\documentclass[twoside]{article}

\usepackage{amsmath,amsfonts}
\usepackage{amssymb,latexsym}
\usepackage{amsxtra}
\usepackage{upref}
\usepackage{exscale}
\usepackage[mathscr]{eucal}

\newcommand{\be}{\begin{equation}}
\newcommand{\ee}{\end{equation}}
\newcommand{\beq}{\begin{eqnarray}}
\newcommand{\eeq}{\end{eqnarray}}
\newcommand{\no}{\nonumber}
\newcommand{\bea}{\begin{array}}
\newcommand{\eea}{\end{array}}
\newcommand{\lb}{\label}
\newcommand{\mcal}{\mathcal}
\newcommand{\mscr}{\mathscr}
\newcommand{\mfrak}{\mathfrak}
\newcommand{\ve}{\varepsilon}

\newcommand{\pp}{\partial}
\newcommand{\im}{\imath}
\newcommand{\ppr}{^{\boldsymbol{\prime}}}
\newcommand{\bprime}{\boldsymbol{\prime}}

\newcommand{\pdag}{^{\dagger}}
\newcommand{\wt}{\widetilde}
\newcommand{\ovv}{\overline}

\newcommand{\feynd}{/\hspace*{-5pt}\hat{\partial}}
\newcommand{\feynD}{\mbox{$/\hspace*{-7pt}\hat{D}$}}
\newcommand{\feynA}{\mbox{$/\hspace*{-7pt}\hat{\mathrm{A}}$}}
\newcommand{\feynVV}{\mbox{$/\hspace*{-6pt}\hat{\mathscr{V}}$}}
\newcommand{\feynV}{\mbox{$/\hspace*{-6pt}\hat{\mathcal{V}}$}}

\newcommand{\feynBV}{\mbox{$\boldsymbol{/\hspace*{-8pt}\hat{\mathcal{V}}}$}}
\newcommand{\feynbv}{\boldsymbol{\scrscr/\hspace*{-5pt}\hat{\mathcal{V}}}}
\newcommand{\trxp}{\mbox{\Large tr$
{\displaystyle\hspace*{-1.5cm}\ph{\int\limits}}_{\int_{C}d^{4}\!x_{p}\;\eta_{p}}$} }
\newcommand{\trxpa}{\mbox{\Large TR$
{\displaystyle\hspace*{-1.5cm}\ph{\int\limits}}_{\int_{C}d^{4}\!x_{p}\;\eta_{p}}^{a(=1,2)}$} }
\newcommand{\trkp}{\mbox{\Large tr$
{\displaystyle\hspace*{-1.5cm}\ph{\int\limits}}_{\int_{C}d^{4}\!k_{p}\;\eta_{p}}$} }
\newcommand{\trkpa}{\mbox{\Large TR$
{\displaystyle\hspace*{-1.5cm}\ph{\int\limits}}_{\int_{C}d^{4}\!k_{p}\;\eta_{p}}^{a(=1,2)}$} }

\newcommand{\trf}{\raisebox{-4pt}{$\mbox{\Large$\mathfrak{tr}$}\atop {\scriptstyle N_{f}}$}}
\newcommand{\trgam}{\raisebox{-4pt}{$\mbox{\Large$\mathfrak{tr}$}\atop {\scriptstyle \hat{\gamma}_{mn}^{(\mu)}}$}}
\newcommand{\trc}{\raisebox{-4pt}{$\mbox{\Large$\mathfrak{tr}$}\atop {\scriptstyle N_{c}}$}}
\newcommand{\trfgam}{\raisebox{-4pt}{$\mbox{\Large$\mathfrak{tr}$}\atop {\scriptstyle N_{f},\hat{\gamma}_{mn}^{(\mu)}}$}}
\newcommand{\trfc}{\raisebox{-4pt}{$\mbox{\Large$\mathfrak{tr}$}\atop {\scriptstyle N_{f},N_{c}}$}}
\newcommand{\trgamc}{\raisebox{-4pt}{$\mbox{\Large$\mathfrak{tr}$}\atop {\scriptstyle \hat{\gamma}_{mn}^{(\mu)},N_{c}}$}}
\newcommand{\trfgamc}{\raisebox{-4pt}{$\mbox{\Large$\mathfrak{tr}$}\atop {\scriptstyle N_{f},\hat{\gamma}_{mn}^{(\mu)},N_{c}}$}}
\newcommand{\TRALL}{\mbox{\Large$
{\displaystyle\mathfrak{T\!r} \hspace*{-1.5cm}
\ph{\int\limits}}_{\scriptscriptstyle N_{f},\hat{\gamma}_{mn}^{(\mu)},N_{c}}^{\scriptscriptstyle a(=1,2)}$} }
\newcommand{\TRFC}{\mbox{\Large$
{\displaystyle\mathfrak{T\!r} \hspace*{-1.5cm}
\ph{\int\limits}}_{\scriptscriptstyle N_{f},N_{c}}^{\scriptscriptstyle a(=1,2)}$} }
\newcommand{\ph}{\phantom}

\newcommand{\scrscr}{\scriptscriptstyle}
\newcommand{\scz}{\scriptsize}

\numberwithin{equation}{section}
\allowdisplaybreaks[4]

\begin{document}

\begin{center}
{\large\bf BCS-like action and Lagrangian from the gradient expansion} \vspace*{0.1cm}\\
{\large\bf of the determinant of Fermi fields in QCD-type,} \vspace*{0.1cm}\\
{\large\bf non-Abelian gauge theories with chiral anomalies} \vspace*{0.3cm}\\
{\bf (Derivation for an effective action of BCS quark pairs} \\
{\bf with the Hopf invariant $\boldsymbol{\Pi_{3}(S^{2})=\mathsf{Z}}$ as a nontrivial topology.)} \vspace*{0.3cm}\\
{\bf Bernhard Mieck}\footnote{e-mail: "bjmeppstein@arcor.de";\newline freelance activity during 2007-2009;
current location : Zum Kohlwaldfeld 16, D-65817 Eppstein, Germany.}
\end{center}

\begin{abstract}
An effective field theory of BCS quark pairs is derived from an ordinary QCD-type path integral with
\(\mbox{SU}_{c}(N_{c}=3)\) non-Abelian gauge fields. We consider the BCS quark pairs
as constituents of nuclei and as the remaining degrees of freedom
in a coset decomposition \(\mbox{SO}(N_{0},N_{0})\,/\,\mbox{U}(N_{0})\otimes \mbox{U}(N_{0})\)
of a corresponding total self-energy matrix taking values as generator within the \(\mbox{so}(N_{0},N_{0})\) Lie algebra.
The underlying dimension (\(N_{0}=N_{f}\cdot 4_{\hat{\gamma}}\cdot N_{c}\))
is determined by the product of isospin- '\(N_{f}=2\)' (flavour- '\(N_{f}=3\)') degrees
of freedom, by the \(4\times 4\) Dirac gamma matrices with factor '\(4_{\hat{\gamma}}\)' and the colour degrees of freedom
'\(N_{c}=3\)'; therefore, the smallest, total self-energy generator has Lie algebra \(\mbox{so}(N_{0},N_{0})\) with \(N_{0}=24\).
We distinguish between a total unitary sub-symmetry \(\mbox{U}(N_{0})\) for purely density related parts of the quarks,
which are taken into account as background fields and as invariant vacuum states in a SSB, and between the BCS terms of quarks as
coset elements \(\mbox{so}(N_{0},N_{0})\,/\,\mbox{u}(N_{0})\). The self-energies are obtained by dyadic products of anomalous
doubled Fermi fields and subsequent HST's where we only use the reproducing property of Gaussian factors
in  Fourier transformations. These HST's are sufficient to achieve a path integral entirely determined by self-energy
matrices for the coset decomposition. Finally, we can compare the derived effective actions of BCS quark pairs with
the effective Skyrme Lagrangian, which is classified by the homotopy group \(\Pi_{3}(\mbox{SU}(2)\,)=\mathsf{Z}\)
for topological solitons as the baryons, and attain the astonishing result that our derived effective actions of BCS quark
pairs are more closely related to the Skyrme-Faddeev field theory with the nontrivial Hopf mapping \(\Pi_{3}(S^{2})=\mathsf{Z}\).\newline

\vspace*{0.1cm}

\end{abstract}

\newpage

\tableofcontents

\newpage

\section{Introduction} \lb{s1}

\subsection{Symmetries and dimensions of self-energies for BCS quark pairs and densities} \lb{s11}

The strong interaction of hadrons in nuclei is described by very different concepts which
range from semi-empirical mass formulas \cite{Henley}-\cite{Ferbel}, nuclear shell models \cite{Hack},
effective Lagrangians of mesons and baryons \cite{Waleck} to QCD- or even string-theory \cite{Muta,Fad1}.
One common part of these models and theories concerns the strong
spin-pairing force of nucleons so that one has already to include spin pairing corrections
in the original Bethe-Weiz\"acker mass formula. This extraordinary pairing force of protons
and neutrons has even provided own models or theories as the 'interacting boson approximation'
or the notion of 'nuclear superfluidity' of spin paired fermionic hadrons \cite{Iachel,Brink}.

In this paper we consider BCS quark pairs as constituents of nuclei and as the remaining
degrees of freedom in a coset decomposition of an ordinary (but subsequently transformed)
QCD path integral with axial gauge fixing. We follow the given coset decomposition
\(\mbox{Osp}(S,S|2L)\,/\,\mbox{U}(L|S)\otimes \mbox{U}(L|S)\) of the ortho-symplectic super-group
\(\mbox{Osp}(S,S|2L)\) for bosonic and fermionic atoms in a trap potential according to
Refs. \cite{mies1,mies2} and 'strictly' derive a transformed path integral with a total
self-energy matrix taking (even)-values in the \(\mbox{so}(N_{0},N_{0})\) Lie algebra.
The dimension \(N_{0}\), (respectively the anomalous doubled case with BCS quark pairs and overall
dimension \((N_{0}\,,\,N_{0})=2\,N_{0}\)), is determined by the product of \(N_{f}=2\)
isospin- (or \(N_{f}=3\) flavour-) degrees of freedom, the \(4_{\hat{\gamma}}\times 4_{\hat{\gamma}}\)
Dirac gamma matrices in 3+1 dimensional spacetime and the \(\mbox{SU}_{c}(N_{c}=3)\) gauge field
degrees of freedom of QCD. This yields in total a unitary sub-symmetry \(\mbox{U}(N_{0})\)
for purely density related parts of the quarks with \(N_{0}=N_{f}\times 4_{\hat{\gamma}}\times N_{c}\).
The self-energy sub-matrix for a single block density section of quarks therefore has the
dimension \(N_{0}=(N_{f}=2)\times 4_{\hat{\gamma}}\times(N_{c=3}=3)=24\) or with
strangeness \((N_{f}=3)\), \(N_{0}=(N_{f}=3)\times 4_{\hat{\gamma}}\times(N_{c=3}=3)=36\).
The anomalous doubling of the fermionic quark fields then leads to the total
\(\mbox{SO}(N_{0},N_{0})\) symmetry for the self-energy which comprises in the block
diagonals the density related \(\mbox{U}(N_{0})\) symmetry and in the off-diagonals blocks
the anti-symmetric sub-matrices for BCS quark pairs; the latter complex, even-valued
BCS parameter fields originate from the coset decomposition
\(\mbox{SO}(N_{0},N_{0})\,/\,\mbox{U}(N_{0})\otimes \mbox{U}(N_{0})\). In advance we
symbolically indicate the density (or subgroup part) and BCS terms of the anomalous
doubling of fermionic quark fields in relation (\ref{s1_1}); this equation also denotes the scheme
of a spontaneous symmetry breaking (SSB) with the coset decomposition
\(\mbox{SO}(N_{0},N_{0})\,/\,\mbox{U}(N_{0})\otimes \mbox{U}(N_{0})\) and the invariant
vacuum or ground states of the density related unitary symmetry \(\mbox{U}(N_{0})\)
\beq \lb{s1_1}
\lefteqn{\Bigg(\bea{c} \mbox{total $\mbox{so}(N_{0},N_{0})$ self-energy} \\
2\,N_{0}\cdot(2\,N_{0}-1)/2\;\;\mbox{ real parameters}\eea\Bigg)^{ab}= } \\ \no &=&
\left(\bea{cc} \left(\bea{c}\mbox{u}(N_{0}) \mbox{ density '11'} \\
(\,N_{0}\,)^{2} \mbox{ real parameters} \eea\right)^{11} &
\left(\bea{c} \im\;{\scrscr\times}\mbox{ (anti-symmetric BCS pairs) '12'} \\
N_{0}\cdot(N_{0}-1)/2 \mbox{ complex parameters}\eea\right)^{12} \vspace*{0.3cm}\\
\left(\bea{c}\im\;{\scrscr\times}\mbox{ (anti-symmetric BCS pairs)$^{\dag}$ '21'} \\
N_{0}\cdot(N_{0}-1)/2 \mbox{ complex parameters}\eea\right)^{21} &
\left(\bea{c} \mbox{u}^{T}(N_{0}) \mbox{ transposed density '22' of '11' block} \\
\mbox{{\it same }} (\,N_{0}\,)^{2} \mbox{ real parameters of '11' part}
\eea\right)^{22} \eea\right)^{ab}_{\mbox{.}}
\eeq
This total self-energy of \(\mbox{so}(N_{0},N_{0})\) with BCS pairing in the off-diagonal
blocks '12' and '21' is achieved by various Hubbard Stratonovich transformations (HST)
from the original, ordinary QCD path integral with axial gauge fixing. We double the
original, odd-valued Fermi- or quark fields
\be\lb{s1_2}
\psi_{M}^{\dag}(x_{p})\;\psi_{M}(x_{p})\rightarrow\frac{1}{2}
\Big(\psi_{M}^{\dag}(x_{p})\;\psi_{M}(x_{p})-\psi_{M}^{T}(x_{p})\;\psi_{M}^{*}(x_{p})\Big)\;,
\ee
and introduce dyadic products of these which then specify the even-valued, hermitian
self-energies. Although the various HST's to self-energies involve intricate manipulations (section \ref{s3}),
we need only to apply the reproducing property of Gaussian integrals in a
Fourier transformation; the Gaussian factor of quartic Fermi fields
(or of quadratic multiples of their dyadic products in a trace relation) is equivalent to the
Gaussian integral of the self-energy matrix with linear coupling to bilinear Fermi fields
or to their corresponding dyadic product in a trace relation. This kind of HST with
Gaussian factors and integrals is also applicable for the eight gauge fields
\(A_{\alpha;\mu}(x_{p})\) of \(\mbox{SU}_{c}(N_{c}=3)\), (\(\alpha=1,\ldots,8\) ,
\(\mu=0,1,2,3\)) \footnote{The semicolon ';' of \(A_{\alpha;\mu}(x_{p})\)
in this paper just separates the internal indices
\(\alpha,\,\beta,\,\gamma=1,\ldots,8\) from the Lorentz-indices \(\kappa,\,\lambda,\,\mu,\,\nu=0,\ldots,3\)
of 3+1 dimensional spacetime, but does not denote something like a covariant derivative.}.
Although there occur three- and four-point vertices of these
gauge fields, we only need the reproducing property of Gaussian integrals in a
Fourier transformation in order to attain an anti-symmetric self-energy matrix
for the quadratic gauge field strength tensor \(\hat{F}_{\alpha}^{\mu\nu}(x_{p})\).
The Gaussian factor of this field strength tensor of gauge fields is transformed
by a Gaussian integral of the corresponding anti-symmetric self-energy matrix
with remaining linear coupling. Since the gauge field strength tensor \(\hat{F}_{\alpha}^{\mu\nu}(x_{p})\)
has only quadratic terms of the original eight gauge fields \(A_{\alpha;\mu}(x_{p})\)
(\(\alpha=1,\ldots,8\)), we can integrate over the remaining quadratic gauge fields in the
linear coupling between the gauge field strength tensor and corresponding self-energy.
This unfinished, quadratic integral of \(A_{\alpha;\mu}(x_{p})\) is modified by the
the 'minimal' coupling principle of a single gauge field to bilinear quark fields;
nevertheless, the remaining Gaussian integral of gauge fields can be performed and results
into the well-known self-interaction of the self-energy matrix for the gauge field
strength tensor. Similar HST's are applied for the quartic interactions of odd-valued
quark fields which are finally kept only in a bilinear, anomalous doubled
kind \((\psi^{*}\,,\,\psi)\,(\hat{H}\,,\,-\hat{H}^{T})\,(\psi\,,\,\psi^{*})^{T}\)
for the doubled one-particle operator \((\hat{H}\), \(-\hat{H}^{T})\); thus they can be
removed by odd-valued, anomalous doubled Gaussian integration properties of Fermi fields.

However, the resulting Fermi determinant contains the total self-energy matrix
\(\mbox{so}(N_{0},N_{0})\) and the 'colour' dressed quark density parts,
combined to a composed gauge field \(\mcal{V}_{\alpha;\mu}(x_{p})\) in place of
the original gauge field \(A_{\alpha;\mu}(x_{p})\), so that further simplification
is not directly obvious. According to Derrick's theorem \cite{raja}, an effective Lagrangian
should have up to order of 'four' derivative terms in order to allow for stable,
static energy configurations in 3+1 spacetime dimensions; the second and fourth order
gradient parts prevent a scaling of the particular configuration to arbitrary small or
large sizes in the three dimensional coordinate space integrations over the static
Hamiltonian density. The intensive, but straightforward HST's in section \ref{s3},
substitute the original QCD path integral (\ref{s2_25}-\ref{s2_27})
\beq \lb{s1_3}
Z[\hat{\mscr{J}},J_{\psi},\hat{J}_{\psi\psi},\hat{\mfrak{j}}^{(\hat{F})}]&=&
\int d[\psi_{M}^{\dag}(x_{p})\,,\,\psi_{M}(x_{p})]\;\;
d[A_{\alpha;\mu}(x_{p})]\;\bigg\{\prod_{\{x_{p},\alpha\}}^{\stackrel{n^{\mu}=(0,\vec{n})}{\vec{n}\cdot\vec{n}=1}}
\delta\big(n^{\mu}\:A_{\alpha;\mu}(x_{p})\big)\bigg\}  \;\times    \\  \no &\times&
\exp\bigg\{\im\;\mscr{A}[\psi,\hat{D}_{\mu}\psi,\hat{F}]-\im\;
\mscr{A}_{S}[\hat{\mscr{J}},J_{\psi},\hat{J}_{\psi\psi},\hat{\mfrak{j}}^{(\hat{F})}]\bigg\} \;;  \\  \lb{s1_4}
\mscr{A}[\psi,\hat{D}_{\mu}\psi,\hat{F}] &=& \int_{C}d^{4}\!x_{p}\bigg\{-\frac{1}{4}\;\hat{F}_{\alpha;\mu\nu}(x_{p})\;
\hat{F}_{\alpha}^{\mu\nu}(x_{p})+ \\  \no &-&
\psi_{N}\pdag(x_{p})\Big[\hat{\beta}\Big(\hat{\gamma}^{\mu}\hat{\pp}_{p,\mu}-\im\;\hat{\ve}_{p}
-\im\;\hat{\gamma}^{\mu}\;A_{\alpha;\mu}(x_{p})\;\hat{t}_{\alpha}+\hat{m}\Big)\Big]_{N;M}\psi_{M}(x_{p})\bigg\}\;; \\  \lb{s1_5}
\hat{\ve}_{p}&=&\hat{\beta}\;\eta_{p}\;\ve_{+}\;;\hspace*{0.6cm}\ve_{+}>0 \;,
\eeq
by the path integral relations (\ref{s3_110}-\ref{s3_116})
\beq  \lb{s1_6}
\hat{\mscr{O}}_{N;M}^{ba}(y_{q},x_{p}) &=&\hspace*{-0.3cm}\bigg\{\hat{\mscr{H}}+
\Big(\hat{T}^{-1}\hat{\mscr{H}}\hat{T}-\hat{\mscr{H}}\Big)+
\hat{T}^{-1}\;\hat{I}\;\hat{S}\;\eta_{q}
\frac{\hat{\mscr{J}}_{N\ppr;M\ppr}^{b\ppr a\ppr}(y_{q},x_{p})}{\mcal{N}}\eta_{p}\;\hat{S}\;\hat{I}\;
\hat{T}\bigg\}_{N;M}^{ba}\hspace*{-0.5cm}(y_{q},x_{p}) \;; \\     \lb{s1_7}
\hspace*{-0.9cm}\mscr{A}_{DET}[\hat{T},\feynV;\hat{\mscr{J}}] &=&\frac{1}{2}\hspace*{0.6cm}\trxpa
\trfgamc\bigg(\ln\Big[\hat{\mscr{O}}_{N;M}^{ba}(y_{q},x_{p})\Big]-
\ln\Big[\hat{\mscr{H}}_{N;M}^{ba}(y_{q},x_{p})\Big]\bigg)  \;; \\  \lb{s1_8}
\mscr{A}_{J_{\psi}}[\hat{T},\feynV;\hat{\mscr{J}}] &=&\frac{1}{2}\int_{C}d^{4}\!x_{p}\;d^{4}\!y_{q}\;\times  \\ \no
\lefteqn{\hspace*{-1.3cm}\times\;J_{\psi;N}^{\dag,b}(y_{q})\;\hat{I}\bigg(\hat{T}(y_{q})\;
\hat{\mscr{O}}_{N\ppr;M\ppr}^{\boldsymbol{-1};b\ppr a\ppr}(y_{q},x_{p})\;\hat{T}^{-1}(x_{p})-
\hat{\mscr{H}}_{N;M}^{\boldsymbol{-1};ba}(y_{q},x_{p})\bigg)\;\hat{I}\;J_{\psi;M}^{a}(x_{p}) \;; }
   \\   \lb{s1_9}
Z[\hat{\mscr{J}},J_{\psi},\hat{J}_{\psi\psi},\hat{\mfrak{j}}^{(\hat{F})}] &=&
\boldsymbol{\Bigg\langle Z\Big[\feynV(x_{p});\hat{\mfrak{S}}^{(\hat{F})},s_{\alpha},\hat{\mfrak{B}}_{\hat{F}},\hat{\mfrak{b}}^{(\hat{F})},
\hat{\mscr{U}}_{\hat{F}},\hat{\mfrak{v}}_{\hat{F}};\sigma_{D};\hat{\mfrak{j}}^{(\hat{F})};
\mbox{\bf Eq. (\ref{s3_59})}\Big]\;\times }   \\ \no &\times&
\int d[\hat{T}^{-1}(x_{p})\;d\hat{T}(x_{p})]\;\;Z_{\hat{J}_{\psi\psi}}[\hat{T}]\;
\exp\Big\{\mscr{A}_{DET}[\hat{T},\feynV;\hat{\mscr{J}}]\Big\}\;
\exp\Big\{\im\;\mscr{A}_{J_{\psi}}[\hat{T},\feynV;\hat{\mscr{J}}]\Big\}\boldsymbol{\Bigg\rangle}\;.
\eeq
Despite intricate appearance of the transformed path integral (\ref{s1_9},\ref{s3_116}), we can attain through
subsequent HST's and a coset decomposition \(\mbox{SO}(N_{0},N_{0})\,/\,\mbox{U}(N_{0})\otimes \mbox{U}(N_{0})\)
a clear separation into BCS terms with coset matrices \(\hat{T}(x_{p})=\exp\{-\hat{Y}(x_{p})\,\}\)
(last line of (\ref{s1_9},\ref{s3_116}))
\beq \lb{s1_10}
\hat{T}_{M;N}^{ab}(x_{p})&=&\Big(\exp\Big\{-\hat{Y}_{M\ppr;N\ppr}^{a\ppr b\ppr}(x_{p})\Big\}\Big)_{M;N}^{ab} \;; \\ \lb{s1_11}
\hat{Y}_{M;N}^{ab}(x_{p}) &=& \left(\bea{cc} 0 & \hat{X}_{M;N}(x_{p}) \\
\hat{X}_{M;N}^{\dag}(x_{p}) &  0 \eea\right)^{a\neq b}\;;\hspace*{0.3cm}\hat{X}_{M;N}^{T}(x_{p})=-\hat{X}_{M;N}(x_{p})\;,
\eeq
and a quark density part with various transformed gauge field parts (first line of (\ref{s1_9},\ref{s3_116}) in
boldface symbols with path integral (\ref{s3_59})). The original gauge field \(A_{\alpha;\mu}(x_{p})\) in
(\ref{s1_3}-\ref{s1_5},\ref{s2_25}-\ref{s2_27}) is exchanged in (\ref{s1_9},\ref{s3_116}) by the gauge field
\(\mcal{V}_{\alpha;\mu}(x_{p})\) (\ref{s3_60}) which is composed of colour dressed quark densities and various
gauge field combinations with auxiliary fields for axial gauge fixing.
This replacement \(\mcal{V}_{\alpha;\mu}(x_{p})\) (\ref{s3_60})
of the original gauge field \(A_{\alpha;\mu}(x_{p})\) is contained in both parts of the final transformed path integral
(\ref{s1_9},\ref{s3_116}), the density and gauge field related, first part (\ref{s3_59}) and the coset part with
\(\hat{T}(x_{p})\) and anti-symmetric BCS generator \(\hat{X}(x_{p})\), \(\hat{X}^{\dag}(x_{p})\). The path integral
(\ref{s1_9},\ref{s3_116}), which is equivalent to the original one (\ref{s1_3}-\ref{s1_5},\ref{s2_25}-\ref{s2_27}),
allows for the separate approximation of the density part (\ref{s3_59}) which yields a mean field solution
\(\langle\feynV(x_{p})\rangle_{\mbox{\scz(\ref{s3_59})}}\)
for the composed gauge field variable \(\mcal{V}_{\alpha;\mu}(x_{p})\); this classical field solution
\(\langle\feynV(x_{p})\rangle_{\mbox{\scz(\ref{s3_59})}}\) of (\ref{s3_59})
has to be inserted into the second path integral part of (\ref{s1_9},\ref{s3_116})
with coset matrix \(\hat{T}(x_{p})\) (\ref{s1_10},\ref{s1_11}).

The remaining Fermi determinant comprises the coset matrices \(\hat{T}(x_{p})\) (\ref{s1_10},\ref{s1_11})
and the mean field \(\langle\feynV(x_{p})\rangle_{\mbox{\scz(\ref{s3_59})}}\) in the thus
approximated, one-particle Hamiltonian operator \(\langle\hat{\mscr{H}}(x_{p})\rangle_{\mbox{\scz(\ref{s3_59})}}\)
(\ref{s1_12}-\ref{s1_14});
however, this Fermi determinant lacks from a simple, obvious reduction to an action
with finite order gradient terms in an effective Lagrangian.
Usually, one considers the exponential trace-logarithm form of the determinant where the gradients
(and matrix potentials) of the one-particle,
mean field operator \(\langle\hat{\mscr{H}}(x_{p})\rangle_{\mbox{\scz(\ref{s3_59})}}\)
are weighted by the coset matrices \(\hat{T}^{-1}\,\langle\hat{\mscr{H}}\rangle_{\mbox{\scz(\ref{s3_59})}}\,\hat{T}\)
relative to the eigenvalue spectrum of \(\langle\hat{\mscr{H}}\rangle_{\mbox{\scz(\ref{s3_59})}}\) which is hence
subtracted for a 'relative' gradient operator
\(\Delta\!\langle\hat{\mscr{H}}\rangle_{\mbox{\scz(\ref{s3_59})}} =
\hat{T}^{-1}\;\langle\hat{\mscr{H}}\rangle_{\mbox{\scz(\ref{s3_59})}}\;\hat{T}-
\langle\hat{\mscr{H}}\rangle_{\mbox{\scz(\ref{s3_59})}}\)
\footnote{In the following the abbreviated trace symbols '\(\mbox{TR}\)',
'\(\mbox{\large$\mfrak{tr}$}\)' denote the summations over the 'Keldysh'
3+1 spacetime contour with inclusion of the anomalous doubled space and over the internal spaces
of isospin (flavour), Dirac gamma matrices and colour degrees of freedom. These spaces are specified in detail in appendix
\ref{sa} (\ref{sa_25}) and in section \ref{s2} (\ref{s2_12}-\ref{s2_14}).}
\beq \lb{s1_12}
\Big\langle\hat{\mscr{H}}_{N;M}^{ba}(y_{q},x_{p})\Big\rangle_{\mbox{\scz(\ref{s3_59})}} &=&
\delta^{(4)}(y_{q}-x_{p})\;\eta_{q}\;\delta_{qp}
\Bigg(\bea{cc} \big\langle\hat{H}_{N;M}(x_{p})\big\rangle_{\mbox{\scz(\ref{s3_59})}} & \\
& \big\langle\hat{H}_{N;M}^{T}(x_{p})\big\rangle_{\mbox{\scz(\ref{s3_59})}}  \eea\Bigg)^{ba} \\ \no &=&\hspace*{-0.25cm}
\delta^{(4)}(y_{q}-x_{p})\;\eta_{q}\;\delta_{qp}\Big[\hat{\mscr{B}}\:\hat{\Gamma}^{\mu}\:\hat{S}\:\hat{\pp}_{p,\mu}+\im\;
\hat{\mscr{B}}\:\hat{\Gamma}^{\mu}\;\hat{\mscr{T}}_{\alpha}\;
\big\langle\mcal{V}_{\alpha;\mu}(x_{p})\big\rangle_{\mbox{\scz(\ref{s3_59})}}+
\hat{\mscr{B}}\:\hat{M}-\im\:\ve_{p}\;\hat{1}_{2N_{0}\times 2N_{0}}\Big] ; \\ \lb{s1_13}
\big\langle\hat{H}(x_{p})\big\rangle_{\mbox{\scz(\ref{s3_59})}} &=&\Big[\hat{\beta}\Big(\,\feynd_{p}+
\im\:\big\langle\feynV(x_{p})\big\rangle_{\mbox{\scz(\ref{s3_59})}}-\im\:\hat{\ve}_{p}+\hat{m}\,\Big)\Big]\;;
\;(\hat{\ve}_{p}=\hat{\beta}\:\ve_{p}=\hat{\beta}\:\eta_{p}\:\ve_{+}\;;\;\ve_{+}>0)\;;  \\ \no &=&
\hat{\beta}\hat{\gamma}^{\mu}\:\hat{\pp}_{p,\mu}+\im\:\hat{\beta}\hat{\gamma}^{\mu}\:\hat{t}_{\alpha}\:
\big\langle\mcal{V}_{\alpha}^{\mu}(x_{p})\big\rangle_{\mbox{\scz(\ref{s3_59})}}+\hat{\beta}\;\hat{m}-
\im\;\ve_{p}\;\hat{1}_{N_{0}\times N_{0}}\;;  \\ \lb{s1_14}
\big\langle\hat{H}^{T}(x_{p})\big\rangle_{\mbox{\scz(\ref{s3_59})}} &=&\Big[\hat{\beta}\Big(\,\feynd_{p}+
\im\:\big\langle\feynV(x_{p})\big\rangle_{\mbox{\scz(\ref{s3_59})}}-\im\:\hat{\ve}_{p}+\hat{m}\,\Big)\Big]^{T}  \\ \no &=&
-(\,\hat{\beta}\hat{\gamma}^{\mu}\,)^{T}\:\hat{\pp}_{p,\mu}+\im\:(\,\hat{\beta}\hat{\gamma}^{\mu}\,)^{T}\:
\hat{t}_{\alpha}^{T}\:\big\langle\mcal{V}_{\alpha}^{\mu}(x_{p})\big\rangle_{\mbox{\scz(\ref{s3_59})}}+
\big(\hat{\beta}\;\hat{m}\big)^{T}-\im\;\ve_{p}\;
\hat{1}_{N_{0}\times N_{0}}\;;
\eeq
\beq \lb{s1_15}
\Delta\!\langle\hat{\mscr{H}}\rangle_{\mbox{\scz(\ref{s3_59})}} &=&
\hat{T}^{-1}\;\langle\hat{\mscr{H}}\rangle_{\mbox{\scz(\ref{s3_59})}}\;\hat{T}-
\langle\hat{\mscr{H}}\rangle_{\mbox{\scz(\ref{s3_59})}} \;;  \\  \lb{s1_16}
\mbox{TR {\large$\mfrak{tr}$}}\Big[\ln\big(\,\hat{T}^{-1}\,
\langle\hat{\mscr{H}}\rangle_{\mbox{\scz(\ref{s3_59})}}\,\hat{T}\,\big)\Big]
&=&\mbox{TR {\large$\mfrak{tr}$}}\bigg[
\ln\Big(\underbrace{\big(\,\hat{T}^{-1}\;\langle\hat{\mscr{H}}\rangle_{\mbox{\scz(\ref{s3_59})}}\;\hat{T}-
\langle\hat{\mscr{H}}\rangle_{\mbox{\scz(\ref{s3_59})}}\,\big)}_{\Delta\!\langle\hat{\mscr{H}}\rangle_{\mbox{\scz(\ref{s3_59})}} }
+\langle\hat{\mscr{H}}\rangle_{\mbox{\scz(\ref{s3_59})}}\Big) \bigg]  \;;  \\  \lb{s1_17}
\mbox{TR {\large$\mfrak{tr}$}}\bigg[\ln\Big(\Delta\!\langle\hat{\mscr{H}}\rangle_{\mbox{\scz(\ref{s3_59})}} +
\langle\hat{\mscr{H}}\rangle_{\mbox{\scz(\ref{s3_59})}}\Big)\bigg] &=&
\mbox{TR {\large$\mfrak{tr}$}}\bigg[\ln\Big(\hat{1}+\Delta\!\langle\hat{\mscr{H}}\rangle_{\mbox{\scz(\ref{s3_59})}}\;
\langle\hat{\mscr{H}}\rangle_{\mbox{\scz(\ref{s3_59})}}^{\boldsymbol{-1}}\Big)+
\ln\big(\langle\hat{\mscr{H}}\rangle_{\mbox{\scz(\ref{s3_59})}}\big)\bigg] \;.
\eeq
Nonetheless, one has to take into account the additional propagation with the inverse of the anomalous doubled
one-particle operator \(\langle\hat{\mscr{H}}\rangle_{\mbox{\scz(\ref{s3_59})}}^{\boldsymbol{-1}}\) so that the
logarithm \(\mbox{TR {\large$\mfrak{tr}$}}[\ln(\hat{1}+\Delta\!\langle\hat{\mscr{H}}\rangle_{\mbox{\scz(\ref{s3_59})}} \;
\langle\hat{\mscr{H}}\rangle_{\mbox{\scz(\ref{s3_59})}}^{\boldsymbol{-1}})]\)
for the gradient expansion is in fact equivalent to the relation
\(\mbox{TR {\large$\mfrak{tr}$}}[\ln(\hat{T}^{-1}\:\langle\hat{\mscr{H}}\rangle_{\mbox{\scz(\ref{s3_59})}}\:\hat{T}\;
\langle\hat{\mscr{H}}\rangle_{\mbox{\scz(\ref{s3_59})}}^{\boldsymbol{-1}})]\)
\beq  \lb{s1_18}
\mbox{TR {\large$\mfrak{tr}$}}\Big[\ln\Big(\hat{1}+\Delta\!\langle\hat{\mscr{H}}\rangle_{\mbox{\scz(\ref{s3_59})}} \;
\langle\hat{\mscr{H}}\rangle_{\mbox{\scz(\ref{s3_59})}}^{\boldsymbol{-1}} \Big)\Big] &=&
\mbox{TR {\large$\mfrak{tr}$}}\Big[\ln\Big(\hat{1}+
\big(\hat{T}^{-1}\;\langle\hat{\mscr{H}}\rangle_{\mbox{\scz(\ref{s3_59})}}\;\hat{T}-
\langle\hat{\mscr{H}}\rangle_{\mbox{\scz(\ref{s3_59})}} \big)\:
\langle\hat{\mscr{H}}\rangle_{\mbox{\scz(\ref{s3_59})}}^{\boldsymbol{-1}}\Big)\Big] \\ \no &=&
\mbox{TR {\large$\mfrak{tr}$}}\Big[\ln\Big(\hat{T}^{-1}\:\langle\hat{\mscr{H}}\rangle_{\mbox{\scz(\ref{s3_59})}}\:\hat{T}\;
\langle\hat{\mscr{H}}\rangle_{\mbox{\scz(\ref{s3_59})}}^{\boldsymbol{-1}}\Big)\Big] \;.
\eeq
The combined occurrence of \(\langle\hat{\mscr{H}}\rangle_{\mbox{\scz(\ref{s3_59})}}\) and of its
inverse \(\langle\hat{\mscr{H}}\rangle_{\mbox{\scz(\ref{s3_59})}}^{\boldsymbol{-1}}\), weighted
by the coset matrices \(\hat{T}^{-1}\), \(\hat{T}\), is suggestive of a gradient expansion with large orders
for slowly varying coset matrices or BCS terms in the logarithm.

We emphasize this point by a gauge transformation of the coset decomposition so that the mean field
operator \(\langle\hat{\mscr{H}}\rangle_{\mbox{\scz(\ref{s3_59})}}\) with
\(\langle\feynV(x_{p})\rangle_{\mbox{\scz(\ref{s3_59})}}\) simplifies to a pure gradient operator
\(\langle\hat{\mscr{H}}_{\hat{\mfrak{W}}}\rangle_{\mbox{\scz(\ref{s3_59})}}=
(\,\langle\hat{H}_{\hat{\mfrak{W}}}\rangle_{\mbox{\scz(\ref{s3_59})}}\:,\:
\langle\hat{H}_{\hat{\mfrak{W}}}^{T}\rangle_{\mbox{\scz(\ref{s3_59})}}\,)\)
with spatially dependent gamma matrices (section \ref{s43})
\be \lb{s1_19}
\langle\hat{H}_{\hat{\mfrak{W}}}\rangle_{\mbox{\scz(\ref{s3_59})}}=
\Big\langle\hat{\beta}_{\hat{\mfrak{W}}}(x_{p})\;\hat{\gamma}_{\hat{\mfrak{W}}}^{\mu}(x_{p})\;
\boldsymbol{\hat{\pp}_{p,\mu}}-\im\:\ve_{p}\Big\rangle_{\mbox{\scz(\ref{s3_59})}} \;.
\ee
Furthermore, the logarithm in (\ref{s1_18}) takes the equivalent gauge transform
\beq \lb{s1_20}
\lefteqn{\hspace*{-1.9cm}
\mbox{TR {\large$\mfrak{tr}$}}\Big[\ln\Big(\hat{1}+
\Delta\!\langle\hat{\mscr{H}}_{\hat{\mfrak{W}}}\rangle_{\mbox{\scz(\ref{s3_59})}} \;
\langle\hat{\mscr{H}}_{\hat{\mfrak{W}}}\rangle_{\mbox{\scz(\ref{s3_59})}}^{\boldsymbol{-1}}\Big)\Big] =
\mbox{TR {\large$\mfrak{tr}$}}\Big[\ln\Big(\hat{1}+\big(\hat{T}^{-1}\;
\langle\hat{\mscr{H}}_{\hat{\mfrak{W}}}\rangle_{\mbox{\scz(\ref{s3_59})}}\;\hat{T}-
\langle\hat{\mscr{H}}_{\hat{\mfrak{W}}}\rangle_{\mbox{\scz(\ref{s3_59})}} \big)\:
\langle\hat{\mscr{H}}_{\hat{\mfrak{W}}}\rangle_{\mbox{\scz(\ref{s3_59})}}^{\boldsymbol{-1}}\Big)\Big] }  \\ \no &=&
\mbox{TR {\large$\mfrak{tr}$}}\Bigg[\ln\Bigg(\hat{T}^{-1}\;
\Bigg(\bea{cc}\langle\hat{H}_{\hat{\mfrak{W}}}\rangle_{\mbox{\scz(\ref{s3_59})}} & \\ &
\langle\hat{H}_{\hat{\mfrak{W}}}\rangle_{\mbox{\scz(\ref{s3_59})}}^{T}\eea\Bigg)\;\hat{T}\;
\Bigg(\bea{cc}
\langle\hat{H}_{\hat{\mfrak{W}}}\rangle_{\mbox{\scz(\ref{s3_59})}}^{\boldsymbol{-1}} & \\ &
\langle\hat{H}_{\hat{\mfrak{W}}}\rangle_{\mbox{\scz(\ref{s3_59})}}^{\boldsymbol{T;-1}}\eea\Bigg)\Bigg)\Bigg]_{\mbox{.}}
\eeq
If one assumes slowly varying finite order gradients of
\(\hat{T}^{-1}\,\langle\hat{\mscr{H}}_{\hat{\mfrak{W}}}\rangle_{\mbox{\scz(\ref{s3_59})}}\,\hat{T}\), one will also
obtain unintented, extraordinary large spatial and time-like variations with
\(\hat{T}^{-1}\,\langle\hat{\mscr{H}}_{\hat{\mfrak{W}}}\rangle_{\mbox{\scz(\ref{s3_59})}}^{\boldsymbol{-1}}\,\hat{T}\)
according to the additional trace operation on the logarithm.
In order to circumvent this problem, we suggest the particular integral representation (\ref{s1_21}-\ref{s1_24})
for the logarithm of an operator \(\hat{\mfrak{O}}\) (and similarly for the inverse)
in order to approximate the total logarithm with the (coset matrix weighted)
combination of \(\langle\hat{\mscr{H}}\rangle_{\mbox{\scz(\ref{s3_59})}}\)
(\(\langle\hat{\mscr{H}}_{\hat{\mfrak{W}}}\rangle_{\mbox{\scz(\ref{s3_59})}}\)) and its inverse
\(\langle\hat{\mscr{H}}\rangle_{\mbox{\scz(\ref{s3_59})}}^{\boldsymbol{-1}}\)
(\(\langle\hat{\mscr{H}}_{\hat{\mfrak{W}}}\rangle_{\mbox{\scz(\ref{s3_59})}}^{\boldsymbol{-1}}\))
to simpler actions in an exponential \cite{Englert}
\beq \lb{s1_21}
\big(\ln\hat{\mfrak{O}}\big)&=&\bigg(\int_{0}^{+\infty}dv\;\;\;
\frac{\exp\{-v\:\hat{1}\}-\exp\{-v\:\hat{\mfrak{O}}\}}{v}\bigg)  \;; \\ \lb{s1_22}
\big(\hat{\mfrak{O}}^{\boldsymbol{-1}}\big)&=& \Big(\int_{0}^{+\infty}dv\;\;\;
\exp\{-v\;\hat{\mfrak{O}}\}\Big) \;;  \\  \lb{s1_23}
\hat{\mfrak{O}}_{\wt{\mscr{J}}} &=&
\bigg(\hat{1}+\Big(\Delta\!\langle\hat{\mscr{H}}\rangle_{\mbox{\scz(\ref{s3_59})}}+
\wt{\mscr{J}}(\hat{T}^{-1},\hat{T})\Big)\;
\Big\langle\hat{\mscr{H}}\Big\rangle_{\mbox{\scz(\ref{s3_59})}}^{\boldsymbol{-1}}\bigg)  \\ \no &=&
\bigg(\hat{T}^{-1}\;\;\big\langle\hat{\mscr{H}}\;\;
\big\rangle_{\mbox{\scz(\ref{s3_59})}}\hat{T}\;\;
\big\langle\hat{\mscr{H}}\big\rangle_{\mbox{\scz(\ref{s3_59})}}^{\boldsymbol{-1}} +
\wt{\mscr{J}}\big(\hat{T}^{-1},\hat{T}\big)\;
\big\langle\hat{\mscr{H}}\big\rangle_{\mbox{\scz(\ref{s3_59})}}^{\boldsymbol{-1}}\bigg) \;;   \\   \lb{s1_24}
\lefteqn{\hspace*{-1.6cm}\mscr{A}_{DET}[\hat{T},\langle\feynV\rangle_{\mbox{\scz(\ref{s3_59})}};\hat{\mscr{J}}\equiv0]= } \\ \no &=&
\frac{1}{2}\int_{0}^{+\infty}dv\;\;\hspace*{0.6cm}\trxpa
\trfgamc\Bigg[\frac{\exp\big\{-v\:\hat{1}\big\}-\exp\bigg\{-v\:
\hat{T}^{-1}\big\langle\hat{\mscr{H}}
\big\rangle_{\mbox{\scz(\ref{s3_59})}}\hat{T}\;
\big\langle\hat{\mscr{H}}\big\rangle_{\mbox{\scz(\ref{s3_59})}}^{\boldsymbol{-1}}\bigg\}}{v}\Bigg]\;.
\eeq
If we suppose positive eigenvalues at order unity or far beyond for the total operator
\(\hat{\mfrak{O}}=\hat{T}^{-1}\,\langle\hat{\mscr{H}}\rangle_{\mbox{\scz(\ref{s3_59})}}\,\hat{T}\,
\langle\hat{\mscr{H}}\rangle_{\mbox{\scz(\ref{s3_59})}}^{\boldsymbol{-1}}\), the inverse factorials \(1/n!\)
of \(\exp\{-v\:\hat{T}^{-1}\,\langle\hat{\mscr{H}}
\rangle_{\mbox{\scz(\ref{s3_59})}}\,\hat{T}\,\langle\hat{\mscr{H}}\rangle_{\mbox{\scz(\ref{s3_59})}}^{\boldsymbol{-1}}\,\}\)
cause a rapid, meaningful expansion and convergence instead of a pure logarithm
\(\ln(\,\hat{T}^{-1}\,\langle\hat{\mscr{H}}\rangle_{\mbox{\scz(\ref{s3_59})}}\,\hat{T}\,
\langle\hat{\mscr{H}}\rangle_{\mbox{\scz(\ref{s3_59})}}^{\boldsymbol{-1}}\,)\)
with reciprocal integer numbers in the expansion.
Therefore, we rely on the integral representation (\ref{s1_21}-\ref{s1_24}) of
the logarithm and of the inverse of an operator \(\hat{\mfrak{O}}\) and apply these relations for reducing
the path integral part to effective actions with coset matrices \(\hat{T}(x_{p})\) and anti-symmetric coset generator
\(\hat{X}(x_{p})\), \(\hat{X}^{\dag}(x_{p})\) for BCS quark pairs. We can even choose the eigenbasis of
the mean field approximated, one-particle operator \(\langle\hat{H}\rangle_{\mbox{\scz(\ref{s3_59})}}\)
or of its anomalous doubled version \(\langle\hat{\mscr{H}}\rangle_{\mbox{\scz(\ref{s3_59})}}\)
instead of the 3+1 dimensional coordinate representation. This particular matrix representation for \(\hat{T}\)
in terms of the eigenbasis of \(\langle\hat{\mscr{H}}\rangle_{\mbox{\scz(\ref{s3_59})}}\) allows to calculate
observables as correlation functions of anomalous quark field combinations
\(\langle\psi_{M}(x_{p})\:\psi_{N}(x_{p})\rangle\), density terms \(\langle\psi_{M}^{*}(x_{p})\:\psi_{N}(x_{p})\rangle\)
and normalized eigenvalue correlations of BCS terms originating from non-Abelian gauge theories as QCD.

Using the above mentioned gauge invariance of coset matrices and generators in the coset decomposition, a kind
of {\it 'interaction'} representation allows to transform the mean field operator
\(\langle\hat{\mscr{H}}\rangle_{\mbox{\scz(\ref{s3_59})}}\) to the already mentioned,
pure gradient terms \(\langle\hat{\mscr{H}}_{\hat{\mfrak{W}}}\rangle_{\mbox{\scz(\ref{s3_59})}}\)
with spatially varying gamma matrices (\ref{s1_19}).
One can take this particular 'interaction' representation in order to extract
a {\it Hopf invariant} with one-form \(\hat{\omega}_{1}(x_{p})\)
\be \lb{s1_25}
N_{I}=\frac{1}{V_{S^{2n-1}}}\int_{S^{2n-1}}\hat{\omega}_{n-1}\wedge (d\hat{\omega}_{n-1})\;\;;\hspace*{0.2cm}
(\hat{\omega}_{n-1}=\mbox{n-1 form with n=2,4,8!}) \;;\;(V_{S^{2n-1}}=\mbox{volume of }S^{2n-1}\mbox{ sphere})\;,
\ee
from the axial current conservation as a nontrivial topology
for the Hopf mapping \(\Pi_{3}(S^{2})=\mbox{\sf Z}\) \cite{Naka,Flanders}.
The zero component of the axial current (with chiral anomaly) contains such a Hopf invariant where one has to consider the
mapping from the 3D spatial coordinate space to the internal \(S^{2}\) sphere with quaternion-valued,
anti-symmetric Pauli matrix \((\hat{\tau}_{2})_{gf}\) as eigenvalues with corresponding complex, even-valued isospin field
\(\bar{f}_{r}(x_{p})\) for BCS quark pairs (Compare for a derivation of the chiral anomaly from the ordinary QCD path integral with
appendix \ref{sc} and with Refs. \cite{Fuji,Fuji2}). Thus our
strictly derived BCS path integral is more closely related to a Skyrme-Faddeev string model \cite{Fad1} with Hopf mapping
\(\Pi_{3}(S^{2})=\mbox{\sf Z}\) than to a topological Skyrme Lagrangian with baryons as winding numbers
following from \(\Pi_{3}(\mbox{SU}(2)\,)=\Pi_{3}(S^{3})=\mbox{\sf Z}\) \cite{Manton,Brown1,Brown2}.
Although finite order gradient expansions have to be taken at least up to fourth order for stable energy
configurations and are questionable concerning the validity of low-momentum approximations, we describe in appendix \ref{sd}
various principles which have necessarily to be regarded for an appropriate expansion in the anomalous doubled Hilbert space
of quantum many particle physics. The remaining coset matrices for the BCS degrees of freedom propagate with anomalous doubled Green functions, containing density related background fields, and are applied to compose an effective (Skyrme-like-)Lagrangian;
this 'analogous' Skyrme-like Lagrangian follows
from the gradient expansion of the self-energy operator within the fermi-determinant and within its inverse of bilinear
source fields, but qualitatively differs by the derived Hopf invariant \(\Pi_{3}(S^{2})=\mbox{\sf Z}\)
from the original Skyrme Lagrangian with homotopy mapping \(\Pi_{3}(\mbox{SU}(2)\,)=\Pi_{3}(S^{3})=\mbox{\sf Z}\).

\subsection{Symmetry breaking source fields for mesons and baryons} \lb{s12}

In order to generate observables from the various, different kinds of path integrals, we introduce the even-valued,
in general nonlocal source matrix \(\hat{\mscr{J}}_{N;M}^{ba}(y_{q},x_{p})\) with two spacetime arguments \(y_{q}\), \(x_{p}\)
on the Keldysh time contour and with anomalous indexing \(a,b=1,2\) and internal space indices \(M,N\) to be specified in
relations (\ref{s2_7}-\ref{s2_14}). This general source field tracks the original observables in terms of quark fields
in the ordinary path integral (\ref{s2_23}-\ref{s2_27}) through multiple transformations (as HST's) to path integrals with
self-energies as the remaining field degrees of freedom. As one considers an observable
\(\langle\Psi_{N}^{\dag,b}(y_{q})\:\Psi_{M}^{a}(x_{p})\rangle\) of bilinear quark fields by differentiating
the original, ordinary QCD path integral with respect to this source \(\hat{\mscr{J}}_{N;M}^{ba}(y_{q},x_{p})\),
one can also generate this same observable for quark fields later after several HST's and a coset decomposition
just by taking the same differentiation of the final transformed generating function.

Besides we incorporate a symmetry breaking, odd-valued source field \(j_{\psi;M}(x_{p})\) (or its anomalous
doubled version \(J_{\psi;M}^{a(=1,2)}(x_{p})\)) which causes non-vanishing observables of quark fields in
odd number as \(\langle\psi_{M}(x_{p})\:\psi_{N}(x_{p})\:\psi_{N\ppr}(x_{p})\rangle\) or
\(\langle\psi_{M}(x_{p})\rangle\), etc..
These source fields \(j_{\psi;M}(x_{p})\) are important for the fermionic degrees in the nuclei and may lead to
an analogous coherent, but fermionic wavefunction as in a BE-condensation for a macroscopic, coherent wavefunction.
Furthermore, one has to include a symmetry breaking source matrix \(\hat{j}_{\psi\psi;N;M}(x_{p})\),
\(\hat{j}_{\psi\psi;N;M}^{\dag}(x_{p})\) (or its combined anomalous doubled version \(\hat{J}_{\psi\psi;N;M}^{ba}(x_{p})\))
for creating an initial configuration of BCS quark pairs for the strong spin pairing force.
However, we do not regard a detailed phase transition from incoherent initial conditions to a coherent
configuartion of BCS quark pairs; this would involve a phase transition with a detailed, experimental
dependence on temperature, density, etc.. Therefore, we just set an initial, coherent configuration of BCS terms
in the coset matrix \(\hat{T}(x_{p})\) and generators \(\hat{X}(x_{p})\,,\,\hat{X}\pdag(x_{p})\)
at intermediate times with wave-packets of appropriate space- and momentum-dependence,
neglecting a detailed phase transition from 'incoherence' to 'coherence' at earlier times \(t_{p}\rightarrow-\infty\).

\section{The path integral with symmetry breaking source fields} \lb{s2}

\subsection{Definitions}\lb{s21} 

The derivation for the effective BCS terms begins with the standard QCD-type Lagrangian (\ref{s2_1}-\ref{s2_6})
of anti-commuting quark field spinors \(\psi(x)\), \(\psi^{\dag}(x)\) and with the non-Abelian \(\mbox{SU}_{c}(N_{c}=3)\)
gauge fields \(A_{\alpha;\mu}(x)\) and the corresponding field strength tensor \(\hat{F}_{\alpha;\mu\nu}(x)\).
We take the notations defined in
\cite{wein1,wein2} ("The Quantum Theory of Fields" Vol. 1-2, S. Weinberg) and label the eight \(\mbox{SU}_{c}(N_{c}=3)\)
gauge field degrees of freedom \(A_{\alpha;\mu}(x)\) by the first Greek letters
\(\alpha,\,\beta,\,\gamma,\,\ldots=1,\,\ldots\,,8\) which are separated from the 3+1 spacetime or Lorentz-indices of the
middle of the Greek alphabet (\(\kappa,\,\lambda,\,\mu,\,\nu=0,1,2,3\)) by a semicolon (spacetime metric tensor
\(\hat{\eta}^{\mu\nu}\) with \(\hat{\eta}^{00}=-1\) and \(\hat{\eta}^{ij}=+\delta^{ij}\) , \(i,j,k,\,\ldots=1,2,3\)).
The covariant derivative \(\hat{D}_{\mu}\) (\ref{s2_3}) is defined in anti-hermitian kind with spacetime derivative \(\hat{\pp}_{\mu}\)
and additional imaginary factor '\(\im\)' which is attached to the eight hermitian \(\mbox{SU}_{c}(N_{c}=3)\) 'colour' generators
\(\hat{t}_{\alpha}\) in the fundamental representation with totally anti-symmetric structure constants \(C_{\alpha\beta\gamma}\).
We assume that all physical quantities are scaled to corresponding
dimensionless objects and list the entire Lagrangian \(\mscr{L}(\psi,\,\hat{D}_{\mu}\psi,\,\hat{F})\) (\ref{s2_1}) with gauge field strength tensor
\(\hat{F}_{\alpha;\mu\nu}(x)\) (\ref{s2_4},\ref{s2_5}) and
coupled fermionic-matter Lagrangian \(\mscr{L}_{M}(\psi,\hat{D}_{\mu}\psi)\) (\ref{s2_2}) according to
Ref. \cite{wein1,wein2} ("The Quantum Theory of Fields" Vol. 2, S. Weinberg),
(\(\hat{m}_{f}=\mbox{diag}\{m_{u},\,m_{d},\,(m_{s})\}\))
\beq \lb{s2_1}
\mscr{L}(\psi,\,\hat{D}_{\mu}\psi,\,\hat{F})&=&-\frac{1}{4}\;\hat{F}_{\alpha;\mu\nu}(x)\;\hat{F}_{\alpha}^{\mu\nu}(x)+
\mscr{L}_{M}(\psi,\hat{D}_{\mu}\psi) \;; \\   \lb{s2_2}
\mscr{L}_{M}(\psi,\hat{D}_{\mu}\psi) &=&-\sum_{f=u,d,(s)}\bar{\psi}_{f}(x)\;\Big[\feynD +\hat{m}_{f}\Big]\;\psi_{f}(x) =
-\hspace*{-0.3cm}\sum_{f=u,d,(s)}\bar{\psi}_{f}(x)\;\Big[\feynd-\im\;\feynA(x) +\hat{m}_{f}\Big]\;\psi_{f}(x)
\\ \no &=&-\sum_{f=u,d,(s)}\bar{\psi}_{f}(x)\;\Big[\hat{\gamma}^{\mu}\;\hat{\pp}_{\mu}-\im\;\hat{\gamma}^{\mu}\;
A_{\alpha;\mu}(x)\;\hat{t}_{\alpha}+\hat{m}_{f}\Big]\;\psi_{f}(x) \;;  \\  \lb{s2_3}
\hat{D}_{\mu}\psi(x) &=&\hat{\pp}_{\mu}\psi(x)-\im\;\hat{t}_{\alpha}\;A_{\alpha;\mu}(x)\;\psi(x)\;;\hspace*{0.5cm}
\feynD(x)=\hat{\gamma}^{\mu}\;\hat{D}_{\mu}(x)\;;\hspace*{0.5cm}\feynd=\hat{\gamma}^{\mu}\;\hat{\pp}_{\mu} \;; \\  \lb{s2_4}
\Big[\hat{D}_{\mu}(x)\:,\:\hat{D}_{\nu}(x)\Big]&=&-\im\;\hat{t}_{\alpha}\;\hat{F}_{\alpha;\mu\nu}(x)  \;; \\   \lb{s2_5}
\hat{F}_{\alpha;\mu\nu}(x) &=& \hat{\pp}_{\mu}A_{\alpha;\nu}(x)-\hat{\pp}_{\nu}A_{\alpha;\mu}(x)+
C_{\alpha\beta\gamma}\;A_{\beta;\mu}(x)\;A_{\gamma;\nu}(x) \;; \\   \lb{s2_6}
\feynA(x) &=& \feynA_{\alpha}(x)\;\hat{t}_{\alpha}=\hat{\gamma}^{\mu}\;A_{\alpha;\mu}(x)\;\hat{t}_{\alpha}\;.
\eeq
In the following we distinguish between three internal spaces whose independent degrees of freedom are marked by the indices
listed in Eqs. (\ref{s2_7}-\ref{s2_9}). The first part of these three independent internal spaces is determined to be the
\(\mbox{SU}_{f}(N_{f}=2)\) isospin (or the extended \(\mbox{SU}_{f}(N_{f}=3)\) flavour) space which is labelled by the Latin
letters \(f,\,g,\,\ldots\) with further sub-indexing \(f_{1},\,g_{1},\,\ldots\) and primes \(f\ppr,\,g\ppr,\,\ldots\)
for '\(\mbox{up}\)', '\(\mbox{down}\)', ('\(\mbox{strange}\)') quarks. The general quark matter Lagrangian is assumed with
non-degenerate isospin (or flavour) masses \(m_{f},\,m_{f\ppr},\ldots,\,m_{g},\,m_{g\ppr},\ldots\) which are diagonal in the
Dirac-gamma matrices and \(\mbox{SU}_{c}(N_{c}=3)\) colour space. We also apply the Clifford-algebra for four-dimensional
spacetime with \(4\times 4\) Dirac matrices \(\hat{\gamma}_{mn}^{\mu}\) of Ref. \cite{wein1} and define the indices
\(m,\,n,\,m\ppr,\,n\ppr,\,m_{1},\,n_{1},\,\ldots\) with further sub-indexing and primes for particular matrix elements of these
\(4\times 4\) gamma matrices as \(\hat{\gamma}_{mn}^{\mu}\) and
\((\hat{\gamma}_{5})_{mn}=-\im\:(\:\hat{\gamma}^{0}\hat{\gamma}^{1}\hat{\gamma}^{2}\hat{\gamma}^{3}\:)_{mn}\). Apart from the Greek indices
\(\alpha,\,\beta,\,\gamma\) for labelling the eight \(\mbox{SU}_{c}(N_{c}=3)\) generators \(\hat{t}_{\alpha}\),
the indices \(r,\,s\) (with further sub-indexing) are ascribed to the \(3\times 3\) fundamental matrix representation
\(\hat{t}_{\alpha;rs}\) of \(\mbox{SU}_{c}(N_{c}=3)\) colour degrees of freedom
\beq \lb{s2_7}
\mbox{isospin (or flavour) index} &:& f,\:f\ppr,\:f_{1},\ldots,\:g,\:g\ppr,\:g_{1},\ldots=\mbox{u(p), d(own), (s(trange))} ; \\ \no
\hspace*{-0.3cm} \mbox{$\mbox{SU}_{f}(2)$ isospin ($\mbox{SU}_{f}(3)$ flavour) mass-matrices} &:&
m_{f},\:m_{f\ppr},\:m_{f_{1}},\ldots,\:m_{g},\:m_{g\ppr},\:m_{g_{1}},\ldots \;; \\  \lb{s2_8}
\mbox{indices for gamma matrices } \hat{\gamma}_{4\times 4}^{\mu}=\hat{\gamma}_{mn}^{\mu} &:&
m,\:m\ppr,\:m_{1},\ldots,\:n,\:n\ppr,\:n_{1},\ldots=1,\ldots,4 \;;\\ \lb{s2_9}
\mbox{indices of $\mbox{SU}_{c}(N_{c}=3)$ colour matrices } \hat{t}_{\alpha;rs} &:&
\alpha,\:\beta,\:\gamma,\:\alpha\ppr,\:\beta\ppr,\:\gamma\ppr,\:\alpha_{1},\:\beta_{1},\:\gamma_{1},\ldots=1,\ldots,8\;; \\ \no &:&
r,\:s,r\ppr,\:s\ppr,\:r_{1},\:s_{1},\ldots=1,2,3 \;.
\eeq
In consequence we attribute the three internal spaces to the Grassmann-valued quark spinors (\ref{s2_10}) by using the labels
\(\{f,m,r\}\), \(\{g,n,s\}\), \(\ldots\) for the isospin- (flavour-) matrices , gamma-matrices and colour-matrix degrees of freedom
\beq   \lb{s2_10}
\psi(x):=\psi_{f,m,r}(x) &;&\bar{\psi}_{g,n,s}(x):=\psi_{g,n\ppr,s}^{\dag}(x)\;\hat{\beta}_{n\ppr n}\;;
\hspace*{0.3cm}\hat{\beta}=\im\;\hat{\gamma}^{0}\;.
\eeq
In order to simplify notations, we combine the isospin- (flavour-) indices \(f,\,g,\,\ldots\), gamma-matrix indices \(m,\,n,\,\ldots\)
and colour-matrix indices \(r,\,s,\,\ldots\) to the collective uppercase indices \(M=\{f,m,r\}\), \(N=\{g,n,s\}\), \(\ldots\) with further
possible sub-indexing. Since we particularly specify on chiral symmetry transformations, the isospin (flavour) index \(f\) or \(g\)
is partially separated from the remaining gamma- and colour-matrix indices which are combined and abbreviated by a bar over
the corresponding uppercase letters as \(\ovv{M},\,\ovv{N},\,\ldots,\,\ovv{M}_{1},\,\ovv{N}_{1},\,\ldots\). Therefore, the entire, collective
indices \(M,\,N,\,M\ppr,\,N\ppr,\,\ldots\) can also be indicated by the combination of the isospin (flavour) index \(f,\,g,\,\ldots\)
and remaining collective index \(\ovv{M},\,\ovv{N},\,\ldots\) for gamma- and colour-matrices
\be \lb{s2_11}
\bea{rclrclrclrcl}
M&:=&\{f,m,r\} & \hspace*{0.6cm} M\ppr&:=&\{f\ppr,m\ppr,r\ppr\} & \hspace*{0.6cm}M_{1}&:=&\{f_{1},m_{1},r_{1}\} & \hspace*{0.6cm} &\ldots &   \\
N&:=&\{g,n,s\} & N\ppr&:=&\{g\ppr,n\ppr,s\ppr\} & N_{1}&:=&\{g_{1},n_{1},s_{1}\} &  &\ldots &   \\
\ovv{M}&:=&\{m,r\} & \ovv{M}\ppr&:=&\{m\ppr,r\ppr\} & \ovv{M}_{1}&:=&\{m_{1},r_{1}\} &  & \ldots& \\
\ovv{N}&:=&\{n,s\} & \ovv{N}\ppr&:=&\{n\ppr,s\ppr\} & \ovv{N}_{1}&:=&\{n_{1},s_{1}\} &  & \ldots& \\
M&:=&\{f,\ovv{M}\} & M\ppr&:=&\{f\ppr,\ovv{M}\ppr\} & M_{1}&:=&\{f_{1},\ovv{M}_{1}\} &  & \ldots& \\
N&:=&\{g,\ovv{N}\} & N\ppr&:=&\{g\ppr,\ovv{N}\ppr\} & N_{1}&:=&\{g_{1},\ovv{N}_{1}\} &  & \ldots&
\eea_{\mbox{.}}
\ee
The final effective BCS related Lagrangian is extracted by performing traces over various combinations of these three internal
spaces. For that reason we have to distinguish between the various traces listed in Eq. (\ref{s2_12}). We denote the symbols
\(\mbox{\large$\mfrak{tr}$}_{\scrscr N_{f}}\), \(\mbox{\large$\mfrak{tr}$}_{\scrscr\hat{\gamma}_{mn}^{(\mu)}}\),
\(\mbox{\large$\mfrak{tr}$}_{\scrscr N_{c}}\) for taking trace operations over isospin
(flavour) matrices, gamma-matrices and \(\mbox{SU}_{c}(N_{c}=3)\) colour matrices according to the above list of indices and labels
in Eqs. (\ref{s2_7}-\ref{s2_11})
\be \lb{s2_12}
\trf\Big[\ldots\Big]\;;\hspace*{0.3cm}\trgam\Big[\ldots\Big]\;; \hspace*{0.3cm}\trc\Big[\ldots\Big]\;.
\ee
In the remainder combinations of the traces (\ref{s2_12}) for these separate internal spaces follow straightforwardly and will
be abbreviated as in Eq. (\ref{s2_13}), as we proceed to the final form of
the effective actions for BCS quark pair condensates
\be\lb{s2_13}
\trfgam\Big[\ldots\Big]\;; \hspace*{0.3cm} \trfc\Big[\ldots\Big]\;; \hspace*{0.3cm}  \trgamc\Big[\ldots\Big]\;; \hspace*{0.3cm}
\trfgamc\Big[\ldots\Big]\; .
\ee
Apart from the above traces (\ref{s2_12},\ref{s2_13}), we point out the overall trace
\(\mbox{$\mfrak{Tr}$}_{\scrscr N_{f},\hat{\gamma}_{mn}^{(\mu)},N_{c}}^{\scrscr a}\) (\ref{s2_14}) of {\it all internal spaces},
which includes an additional summation with \(a,\,b,\,c,\,\ldots=1,2\) over the anomalous doubled space
for BCS pair condensates. In comparison to this overall trace (\ref{s2_14}), the above listed traces
(\ref{s2_12},\ref{s2_13}) only encompass density terms without any possible summations
over anomalous pairings of quark fields
\be\lb{s2_14}
\TRALL\bigg[\ldots\bigg]\;\;\;.
\ee

\subsection{The path integral with fermionic matter- and non-Abelian gauge fields}  \lb{s22}

According to the entire Lagrangian \(\mscr{L}(\psi,\,\hat{D}_{\mu}\psi,\,\hat{F})\) (\ref{s2_1}), we construct the analogous
path integral with fermionic quark matter and non-Abelian gauge fields. However, we introduce the time contour integral
(\ref{s2_15}) with time variable \(x_{p=\pm}^{0}\) on the two branches \(p=\pm\) for time development with \(\mscr{L}\),
\(\mscr{L}_{M}\) (\ref{s2_1},\ref{s2_2}) in forward \(\int_{-\infty}^{+\infty}dx_{+}^{0}\,\ldots\) and backward
\(\int_{+\infty}^{-\infty}dx_{-}^{0}\,\ldots\) direction
\cite{mies1,mies2,mies3,mierep1}. The negative sign of the backward
propagation \(\int_{+\infty}^{-\infty}dx_{-}^{0}\,\ldots=\boldsymbol{-}\int_{-\infty}^{+\infty}dx_{-}^{0}\,\ldots\)
will be frequently taken into account by the symbol \(\eta_{p=\pm}=p=\pm\) (\ref{s2_16}) as a contour time metric
\footnote{This contour time metric \(\eta_{p}=p=\pm\) should not be confused with the spacetime metric tensor
\(\eta^{\mu\nu}\) (\(\eta^{00}=-1\), \(\eta^{ij}=\delta^{ij}\)) for contravariant and covariant components of
vectors and tensors. The indices \(p,\,q,\,p\ppr,\,q\ppr,\,\ldots\) are reserved for the contour time metric
\(\eta_{p},\,\eta_{q},\,\eta_{p\ppr},\,\eta_{q\ppr},\,\ldots\) of the two branches for forward and backward
propagation; on the contrary the indices \(\kappa,\,\lambda,\,\mu,\,\nu\) from the middle of the Greek alphabet
are Lorentz-indices of four-dimensional spacetime, as e.\ g.\ for the metric tensor \(\eta^{\mu\nu},\,\eta^{\kappa\lambda},\,
x^{\mu},\,x_{\nu},\,\ldots\).}
\beq\no
\int_{C}d^{4}\!x_{p}\;\ldots &=&\int_{L^{3}}d^{3}\!\vec{x}\bigg(\int_{-\infty}^{+\infty}dx_{+}^{0}\;\ldots+
\int_{+\infty}^{-\infty}dx_{-}^{0}\;\ldots\bigg) = \int_{L^{3}}d^{3}\!\vec{x}\bigg(\int_{-\infty}^{+\infty}dx_{+}^{0}\;\ldots-
\int_{-\infty}^{+\infty}dx_{-}^{0}\;\ldots\bigg)  \\ \lb{s2_15} &=&
\int_{L^{3}}d^{3}\!\vec{x}\bigg(\sum_{p=\pm}\int_{-\infty}^{+\infty}dx_{p}^{0}\;\eta_{p}\;\ldots\bigg) \;; \\ \lb{s2_16}
\eta_{p} &=&\big\{\underbrace{+1}_{p=+}\;;\;\underbrace{-1}_{p=-}\big\}\;\;\;.
\eeq
An additional contour time label \(p,\,q,\,\ldots\) of the four-dimensional vector \(x^{\mu}=(x^{0},\vec{x})\rightarrow
x_{p}^{\mu}=(x_{p}^{0},\vec{x})\) refers to the two different propagations of zero components \(x_{\pm}^{0}\) of these
contour-time extended four-vectors \(x_{p}\)
\be\lb{s2_17}
\bea{rclrclrcl}
x_{p}^{\mu}&=&\big(x_{p}^{0}\,,\,\vec{x}\big)\;; &\hspace*{0.6cm}x_{+}^{\mu}&=&\big(x_{+}^{0}\,,\,\vec{x}\big)\;; &\hspace*{0.6cm}
x_{-}^{\mu}&=&\big(x_{-}^{0}\,,\,\vec{x}\big)\;;   \\
\hat{\pp}_{p,\mu}&=&\bigg(\frac{\pp}{\pp x_{p}^{0}}\,,\,\frac{\pp}{\pp\vec{x}}\bigg)\;; &
\hat{\pp}_{+,\mu}&=&\bigg(\frac{\pp}{\pp x_{+}^{0}}\,,\,\frac{\pp}{\pp\vec{x}}\bigg)\;; &
\hat{\pp}_{-,\mu}&=&\bigg(\frac{\pp}{\pp x_{-}^{0}}\,,\,\frac{\pp}{\pp\vec{x}}\bigg)\;.
\eea
\ee
We consider four different source fields on the non-equilibrium time contour where two of these are applied for a spontaneous
symmetry breaking of the fermionic matter fields.
The general, anomalous doubled, complex, even-valued source term \(\hat{\mscr{J}}_{g,n,s;f,m,r}^{ba}(y_{q},x_{p})\)
allows to generate observables of bilinear, quartic or higher order, even-numbered quark fields including BCS terms as
\(\langle\psi_{g,n,s}(y_{q})\;\psi_{f,m,r}(x_{p})\rangle\). However, we discern between the source
\(\hat{\mscr{J}}_{N;M}^{ba}(y_{q},x_{p})\) for generating even-numbered quark field observables by differentiating a path
integral and the symmetry breaking, even-valued, complex field \(\hat{J}_{\psi\psi;N;M}^{b\neq a}(x_{p})\) which couples
to \(\psi_{g,n,s}^{*}(x_{p})\;\psi_{f,m,r}^{*}(x_{p})\) and \(\psi_{g,n,s}(x_{p})\;\psi_{f,m,r}(x_{p})\).
In order to create non-vanishing quark fields in odd number, a fermionic, Grassmann-valued source \(J_{\psi;f,m,r}^{a}(x_{p})\)
is incorporated which couples to single quark fields \(\psi_{f,m,r}(x_{p})\), \(\psi_{f,m,r}^{\dag}(x_{p})\) and that is
also extended with its complex value to an anomalous doubled form.
Furthermore, we allow for an anti-symmetric, even-valued,
{\it real} field \(\hat{\mfrak{j}}_{\alpha}^{(\hat{F})\mu\nu}(x_{p})\) which generates the non-Abelian gauge field strength tensor
\(\hat{F}_{\alpha}^{\mu\nu}(x_{p})\). Since the gauge fields are changed to background fields in later steps
of the derivation to the final effective Lagrangian, we do not take into account a symmetry breaking of the gauge field
strength tensor or an anomalous doubling as for the quark fields and simply set the source
\(\hat{\mfrak{j}}_{\alpha}^{(\hat{F})\mu\nu}(x_{p})\) for the field strength to zero at the end of the calculation
\be \lb{s2_18}
\bea{rclrcl}
\mbox{Source fields} &:&   \\
\hat{\mscr{J}}_{g,n,s;f,m,r}^{ba}(y_{q},x_{p})&\in&
\mathsf{C_{even}}\;;&\hspace*{0.6cm} J_{\psi;f,m,r}^{a}(x_{p})&\in&\mscr{C}_{odd}\;; \vspace*{0.16cm} \\
\hat{J}_{\psi\psi;g,n,s;f,m,r}^{b\neq a}(x_{p})&\in&\mathsf{C_{even}}\;;&\hat{\mfrak{j}}_{\alpha;\mu\nu}^{(\hat{F})}(x_{p})&\in&\mathsf{R_{even}}
\eea_{\mbox{.}}
\ee
The precise form of the entire symmetry breaking action \(\mscr{A}_{S}[\hat{\mscr{J}},J_{\psi},\hat{J}_{\psi\psi},\hat{\mfrak{j}}^{(\hat{F})}]\)
is given in Eq. (\ref{s2_22}) for all these four source fields in (\ref{s2_18}). In advance we mention that this source action
\(\mscr{A}_{S}[\hat{\mscr{J}},J_{\psi},\hat{J}_{\psi\psi},\hat{\mfrak{j}}^{(\hat{F})}]\) is specified in its anomalous doubled kind with
metric \(\hat{S}\) (\ref{s2_21}) for the anomalous doubled Fermi fields.
A two-component, anomalous doubled, fermionic field \((\ldots)^{a(=1,2)}\) follows
from the extension with its complex-valued copy
\be\lb{s2_19}
\bea{rcccrcl}
\big(\mbox{field}\big) &\rightarrow&\mbox{anomalous doubled}&\rightarrow&\bigg(\ldots\bigg)^{a(=1,2)}&:=&
\Bigg(\bea{c} \big(\mbox{field}\big)^{a=1} \\ \big(\mbox{field}^{*}\big)^{a=2}\eea \Bigg) \;; \\
\big(\mbox{field}^{\dag}\big) &\rightarrow&\mbox{anomalous doubled}&\rightarrow&\bigg(\ldots\bigg)^{\dag,a(=1,2)}&:=&
\Big( \big(\mbox{field}^{*}\big)^{a=1} \;;\; \big(\mbox{field}\big)^{a=2} \Big)  \;.
\eea
\ee
Since fermionic quark matter fields are only considered for the anomalous doubling, one has to introduce a diagonal
negative sign \(-\hat{1}_{N;M}\) in the '22' block of the metric tensor \(\hat{S}\) for anomalous doubling.
We have to apply for this metric tensor \(\hat{S}\) the Weyl unitary trick where the metric tensor \(\hat{S}\)
is factorized into \(\hat{I}\cdot\hat{I}=\hat{S}\) with the new metric \(\hat{I}=
\delta_{ba}\:\{+\hat{1}_{N;M}\;;\;\im\:\hat{1}_{N;M}\}\). The inverse \(\hat{I}^{-1}\) of \(\hat{I}\) is given by the product
\(\hat{I}^{-1}=\hat{S}\:\hat{I}\) which is frequently used in transformations involving the Weyl unitary trick for the
coset decomposition
\beq \lb{s2_20}
\psi_{M}^{\dag}\psi_{M} &=&\frac{1}{2}\Big(\psi_{N}^{\dag}\:\hat{1}_{N;M}\:\psi_{M}-\psi_{M}^{T}\:\hat{1}_{M;N}\:\psi_{N}^{*}\Big) \\ \no &=&
\frac{1}{2}\Big(\psi_{N}^{*}\:,\:\psi_{N}\Big)^{b}\underbrace{\bigg(\bea{cc} \hat{1}_{N;M} & \\ & -\hat{1}_{N;M}\eea\bigg)^{ba}}_{\hat{S}^{ba}}
\bigg(\bea{c} \psi_{M} \\ \psi_{M}^{*} \eea\bigg)^{a}  \;; \\ \lb{s2_21}
\hat{S}^{ba} &=& \delta_{ba}\Big\{\underbrace{+\hat{1}_{N;M}}_{a=1}\;;\;\underbrace{-\hat{1}_{N;M}}_{a=2}\Big\} \;;  \hspace*{0.6cm}
\hat{I}^{ba} = \delta_{ba}\Big\{\underbrace{+\hat{1}_{N;M}}_{a=1}\;;\;\underbrace{\im\:\hat{1}_{N;M}}_{a=2}\Big\} \;; \hspace*{0.6cm}
\hat{I}\cdot\hat{I}=\hat{S}\;;    \\ \lb{s2_22}
\lefteqn{\hspace*{-1.0cm}\mscr{A}_{S}[\hat{\mscr{J}},J_{\psi},\hat{J}_{\psi\psi},\hat{\mfrak{j}}^{(\hat{F})}]=\frac{1}{2}\int_{C}d^{4}\!x_{p}\Big(
J_{\psi;m,f,r}^{\dag,a}(x_{p})\;\hat{S}^{ab}\;\Psi_{m,f,r}^{b}(x_{p})+
\Psi_{m,f,r}^{\dag,a}(x_{p})\;\hat{S}^{ab}\;J_{\psi;m,f,r}^{b}(x_{p})\Big) +} \\ \no &+&\hspace*{-0.2cm}
\frac{1}{2}\int_{C}d^{4}\!x_{p}\Big(\psi_{g,n,s}(x_{p})\;\hat{j}_{\psi\psi;g,n,s;f,m,r}^{\dag}(x_{p})\;
\psi_{f,m,r}(x_{p})+  \psi_{g,n,s}^{*}(x_{p})\;\hat{j}_{\psi\psi;g,n,s;f,m,r}(x_{p})\;
\psi_{f,m,r}^{*}(x_{p})\Big) + \\ \no &+&\hspace*{-0.2cm}
\int_{C}d^{4}\!x_{p}\;\hat{\mfrak{j}}_{\alpha;\mu\nu}^{(\hat{F})}(x_{p})\;\hat{F}_{\alpha}^{\mu\nu}(x_{p}) +
\frac{1}{2}\int_{C}d^{4}\!x_{p}\;d^{4}\!y_{q}\;\Psi_{g,n,s}^{\dag,b}(y_{q})\;
\hat{\mscr{J}}_{g,n,s;f,m,r}^{ba}(y_{q},x_{p})\;\Psi_{f,m,r}^{a}(x_{p})\;.
\eeq
As already mentioned, the source \(\hat{\mscr{J}}_{g,n,s;f,m,r}^{ba}(y_{q},x_{p})\) differs from the other two source fields
\(\hat{J}_{\psi\psi;N;M}^{b\neq a}(x_{p})\), \(J_{\psi;M}^{a}(x_{p})\) for the quark fields. The latter
are assigned to symmetry breaking processes and therefore are set to equivalent, {\it non-vanishing} values on the two branches of the
time contour at the final end of calculations (\ref{s2_24}); in consequence an entire hermitian action results into a normalized path integral or
generating function. \footnote{In this article we neglect additional effects of symmetry breaking with the gauge field strength tensor
and therefore also set the corresponding source \(\hat{\mfrak{j}}_{\alpha;\mu\nu}^{(\hat{F})}(x_{p})\) to zero at the final end of
transformations. This normalization also allows the treatment of disordered systems with ensemble-averages over random potentials
and interactions \cite{mies3,mierep1}.}
However, this normalization to unity requires a vanishing source term \(\hat{\mscr{J}}_{N;M}^{ba}(y_{q},x_{p})\)
which tests the response to bilinear quark terms (also of the anomalous case) propagating with Lagarangian \(\mscr{L}(\psi,\hat{D}_{\mu}\psi,\hat{F})\)
(\ref{s2_1},\ref{s2_2}) and hermitian, symmetry breaking sources \(\hat{J}_{\psi\psi;N;M}^{b\neq a}(x_{+})=\hat{J}_{\psi\psi;N;M}^{b\neq a}(x_{-})\),
\(J_{\psi;M}^{a}(x_{+})=J_{\psi;M}^{a}(x_{-})\). This response to \(\hat{\mscr{J}}_{N;M}^{ba}(y_{q},x_{p})\neq 0\) can be extended to
higher order functional Taylor-expansion of the path integral \(Z[\hat{\mscr{J}},J_{\psi},\hat{J}_{\psi\psi},\hat{\mfrak{j}}^{(\hat{F})}]\) so that
one obtains the response for higher order correlation functions of quark fields in even number (also with inclusion of BCS related terms)
\beq \lb{s2_23}
Z[\hat{\mscr{J}},J_{\psi},\hat{J}_{\psi\psi},\hat{\mfrak{j}}^{(\hat{F})}] &=&
\overbrace{Z[\hat{\mscr{J}}\equiv 0\,,\,J_{\psi}(x_{+})=J_{\psi}(x_{-})\,,\,\hat{J}_{\psi\psi}(x_{+})=J_{\psi\psi}(x_{-})\,,\,
\hat{\mfrak{j}}^{(\hat{F})}\equiv0]}^{\equiv 1} +
\\ \no &+& \int_{C}d^{4}\!x_{p}\;d^{4}\!y_{q}\;\hat{\mscr{J}}_{N;M}^{ba}(y_{q},x_{p})\;
\bigg(\frac{\delta}{\delta\hat{\mscr{J}}_{N;M}^{ba}(y_{q},x_{p})} Z[\hat{\mscr{J}},J_{\psi},\hat{J}_{\psi\psi},\hat{\mfrak{j}}^{(\hat{F})}]\bigg)\bigg|_{\mbox{\scz conditions}} +
\\ \no &+&\frac{1}{2!} \int_{C}d^{4}\!x_{p_{1}}^{(1)}\;d^{4}\!y_{q_{1}}^{(1)}\;d^{4}\!x_{p_{2}}^{(2)}\;d^{4}\!y_{q_{2}}^{(2)}\;\;
\hat{\mscr{J}}_{N_{1};M_{1}}^{b_{1}a_{1}}(y_{q_{1}}^{(1)},x_{p_{1}}^{(1)})\;\;\hat{\mscr{J}}_{N_{2};M_{2}}^{b_{2}a_{2}}(y_{q_{2}}^{(2)},x_{p_{2}}^{(2)})\;\times \\ \no &\times&
\bigg(\frac{\delta}{\delta\hat{\mscr{J}}_{N_{1};M_{1}}^{b_{1}a_{1}}(y_{q_{1}}^{(1)},x_{p_{1}}^{(1)})}
\frac{\delta}{\delta\hat{\mscr{J}}_{N_{2};M_{2}}^{b_{2}a_{2}}(y_{q_{2}}^{(2)},x_{p_{2}}^{(2)})}
Z[\hat{\mscr{J}},J_{\psi},\hat{J}_{\psi\psi},\hat{\mfrak{j}}^{(\hat{F})}]\bigg)\bigg|_{\mbox{\scz conditions}} +\frac{1}{3!}\ldots \;; \\  \lb{s2_24}
\mbox{conditions} &=& \Big\{\hat{\mscr{J}}\equiv 0\,,\,J_{\psi}(x_{+})=J_{\psi}(x_{-})\,,\,\hat{J}_{\psi\psi}(x_{+})=\hat{J}_{\psi\psi}(x_{-})\,,\,
\hat{\mfrak{j}}^{(\hat{F})}\equiv 0\Big\}\;.
\eeq
According to Ref. \cite{wein1,wein2}, we perform the path integral quantization for the Lagrangian \(\mscr{L}(\psi,\hat{D}_{\mu}\psi,\hat{F})\)
(\ref{s2_1},\ref{s2_2}) with source action \(\mscr{A}_{S}[\hat{\mscr{J}},J_{\psi},\hat{J}_{\psi\psi},\hat{\mfrak{j}}^{(\hat{F})}]\)
(\ref{s2_22}) on the time contour in axial gauge which does not contain the
Gribov ambiguity. This is accomplished by introducing a space-like, four-component, unit vector \(n^{\mu}=(0,\vec{n})\) (\(n_{\mu}n^{\mu}=1\))
into a delta-function with the gauge field \(A_{\alpha;\mu}(x_{p})\)
\beq \lb{s2_25}
Z[\hat{\mscr{J}},J_{\psi},\hat{J}_{\psi\psi},\hat{\mfrak{j}}^{(\hat{F})}]&=&\int d[\psi_{M}^{\dag}(x_{p})\,,\,\psi_{M}(x_{p})]\;\;
d[A_{\alpha;\mu}(x_{p})]\;\bigg\{\prod_{\{x_{p},\alpha\}}^{\stackrel{n^{\mu}=(0,\vec{n})}{\vec{n}\cdot\vec{n}=1}}
\delta\big(n^{\mu}\:A_{\alpha;\mu}(x_{p})\big)\bigg\}  \;\times    \\  \no &\times&
\exp\bigg\{\im\;\mscr{A}[\psi,\hat{D}_{\mu}\psi,\hat{F}]-\im\;\mscr{A}_{S}[\hat{\mscr{J}},J_{\psi},\hat{J}_{\psi\psi},\hat{\mfrak{j}}^{(\hat{F})}]\bigg\} \;;  \\  \lb{s2_26}
\mscr{A}[\psi,\hat{D}_{\mu}\psi,\hat{F}] &=& \int_{C}d^{4}\!x_{p}\bigg\{-\frac{1}{4}\;\hat{F}_{\alpha;\mu\nu}(x_{p})\;
\hat{F}_{\alpha}^{\mu\nu}(x_{p})+ \\  \no &-&
\psi_{N}\pdag(x_{p})\Big[\hat{\beta}\Big(\hat{\gamma}^{\mu}\hat{\pp}_{p,\mu}-\im\;\hat{\ve}_{p}
-\im\;\hat{\gamma}^{\mu}\;A_{\alpha;\mu}(x_{p})\;\hat{t}_{\alpha}+\hat{m}\Big)\Big]_{N;M}\psi_{M}(x_{p})\bigg\}\;; \\  \lb{s2_27}
\hat{\ve}_{p}&=&\hat{\beta}\;\eta_{p}\;\ve_{+}\;;\hspace*{0.6cm}\ve_{+}>0\;.
\eeq
The action \(\mscr{A}[\psi,\hat{D}_{\mu}\psi,\hat{F}]\) of quark matter and gauge fields has to comprise a non-hermitian
\(\hat{\ve}_{p}\) matrix term (\ref{s2_27}) which characterizes a particular time direction in the propagation and which results into
proper time ordering of quark fields and appropriate analytic convergence properties of derived Green functions.

Finally, we end this section of definitions by listing the anomalous doubling of quark fields \(\Psi_{M}^{a}(x_{p})\) with
corresponding anti-commuting source \(J_{\psi;M}^{a}(x_{p})\) (\ref{s2_28},\ref{s2_29}). The symmetry breaking, anomalous doubled source matrix
\(\hat{J}_{\psi\psi;N;M}^{b\neq a}(x_{p})\) (\ref{s2_31})
consists of two anti-symmetric, complex, even-valued sub-matrices \(\hat{j}_{\psi\psi;M;N}(x_{p})\),
\(\hat{j}_{\psi\psi;N;M}^{\dag}(x_{p})\) (\ref{s2_30})
in the off-diagonal '12' and '21' blocks where the anti-symmetric property regards the BCS related
pair condensates of fermionic quark fields. Since the gauge field strength tensor \(\hat{F}_{\alpha}^{\mu\nu}(x_{p})\) is
completely anti-symmetric in its spacetime indices '\(\mu,\nu\)', we have also to require this anti-symmetry for the corresponding, generating, real
source matrix \(\hat{\mfrak{j}}_{\alpha;\mu\nu}^{(\hat{F})}(x_{p})\) (\ref{s2_32})
\beq \lb{s2_28}
\Psi_{f,m,r}^{a(=1,2)}(x_{p})&=&\Psi_{M}^{a(=1,2)}(x_{p})=
\Big\{\underbrace{\psi_{f,m,r}(x_{p})}_{a=1}\;;\;\underbrace{\psi_{f,m,r}^{*}(x_{p})}_{a=2}\Big\}^{T}=
\Big\{\underbrace{\psi_{M}(x_{p})}_{a=1}\;;\;\underbrace{\psi_{M}^{*}(x_{p})}_{a=2}\Big\}^{T} \;; \\  \lb{s2_29}
J_{\psi;f,m,r}^{a(=1,2)}(x_{p}) &=&J_{\psi;M}^{a(=1,2)}(x_{p})=
\Big\{\underbrace{j_{\psi;f,m,r}(x_{p})}_{a=1}\;;\;\underbrace{j_{\psi;f,m,r}^{*}(x_{p})}_{a=2}\Big\}^{T}=
\Big\{\underbrace{j_{\psi;M}(x_{p})}_{a=1}\;;\;\underbrace{j_{\psi;M}^{*}(x_{p})}_{a=2}\Big\}^{T} \;;  \\ \lb{s2_30}
\hat{j}_{\psi\psi;M;N}(x_{p}) &=&-\hat{j}_{\psi\psi;M;N}^{T}(x_{p}) \;; \hspace*{0.6cm}
\hat{j}_{\psi\psi;f,m,r;g,n,s}(x_{p}) = -\hat{j}_{\psi\psi;f,m,r;g,n,s}^{T}(x_{p}) \;; \\ \lb{s2_31}
\hat{J}_{\psi\psi;M;N}^{a\neq b}(x_{p}) &=&
\left(\bea{cc} 0 & \hat{j}_{\psi\psi;M;N}(x_{p})  \\ \hat{j}_{\psi\psi;M;N}^{\dag}(x_{p}) & 0 \eea\right)^{a\neq b}_{M;N}  \;;  \\  \lb{s2_32}
\hat{\mfrak{j}}_{\alpha;\mu\nu}^{(\hat{F})}(x_{p}) &=&-\hat{\mfrak{j}}_{\alpha;\mu\nu}^{(\hat{F}),T}(x_{p})\;;\hspace*{0.6cm}\alpha=1,\ldots,8\;.
\eeq
Apart from the source term \(\mscr{A}_{S}[\hat{\mscr{J}},J_{\psi},\hat{J}_{\psi\psi},\hat{\mfrak{j}}^{(\hat{F})}]\) (\ref{s2_22}), we have also
to implement the anomalous doubling of quark fields into the path integral (\ref{s2_25}) with the action \(\mscr{A}[\psi,\hat{D}_{\mu}\psi,\hat{F}]\)
(\ref{s2_26}) in order to incorporate BCS pair condensates in the time contour propagation. This has to be achieved by
Hubbard-Stratonovich transformations (HST) to self-energies which include the anomalous doubling for the fermionic matter fields.
Eventually, remaining Gaussian integrals of doubled quark fields introduce the Fermi determinant with anomalous doubled
one-particle Hamiltonian and self-energy which allow for the coset decomposition to the
effective, BCS related actions with coset matrices \(\hat{T}(x_{p})\).

\section{HST of Fermi- and non-Abelian gauge fields to self-energies} \lb{s3}

\subsection{HST to the self-energy of the non-Abelian gauge field strength tensor} \lb{s31}

We introduce the real self-energy matrix \(\hat{\mfrak{S}}_{\alpha;\mu\nu}^{(\hat{F})}(x_{p})\) with corresponding anti-symmetry
as of the gauge field strength tensor \(\hat{F}_{\alpha}^{\mu\nu}(x_{p})\) in the two spacetime indices '\(\mu,\nu\)'
for each of the eight \(\mbox{SU}_{c}(N_{c}=3)\) colour generators. In general, HST's are related to Gaussian integrals; therefore,
we start out from the Gaussian identity (\ref{s3_1}) with 'flat', Euclidean integration measure of the real self-energy matrix
\(\hat{\mfrak{S}}_{\alpha;\mu\nu}^{(\hat{F})}(x_{p})\) which couples linearly to the field strength tensor \(\hat{F}_{\alpha}^{\mu\nu}(x_{p})\)
and the real, anti-symmetric source field \(\hat{\mfrak{j}}_{\alpha;\mu\nu}^{(\hat{F})}(x_{p})\) in the action of the exponential
\beq \lb{s3_1}
1&\equiv&\int d[\hat{\mfrak{S}}_{\alpha;\mu\nu}^{(\hat{F})}(x_{p})]\;
\exp\bigg\{\frac{\im}{4}\int_{C}d^{4}\!x_{p}\Big(\hat{\mfrak{S}}_{\alpha}^{(\hat{F})\mu\nu}(x_{p})-
\hat{F}_{\alpha}^{\mu\nu}(x_{p})-2\;\hat{\mfrak{j}}_{\alpha}^{(\hat{F})\mu\nu}(x_{p})\Big) \times \\ \no &\times&
\Big(\hat{\mfrak{S}}_{\alpha;\mu\nu}^{(\hat{F})}(x_{p})-\hat{F}_{\alpha;\mu\nu}(x_{p})-2\;\hat{\mfrak{j}}_{\alpha;\mu\nu}^{(\hat{F})}(x_{p})\Big)\bigg\}\;.
\eeq
The Gaussian integral (\ref{s3_1}) consists of the self-energy matrix \(\hat{\mfrak{S}}_{\alpha;\mu\nu}^{(\hat{F})}(x_{p})\) which is
shifted by \(\hat{F}_{\alpha;\mu\nu}(x_{p})\) and by the source \(\hat{\mfrak{j}}_{\alpha;\mu\nu}^{(\hat{F})}(x_{p})\).
In consequence, one obtains the standard relation of the HST
where the quadratic term of the field strength tensor \(\hat{F}_{\alpha;\mu\nu}(x_{p})\)
is reduced to a linear coupling with the self-energy in a 'Euclidean' Gaussian integral. The quadratic term of the field strength
\(\hat{F}_{\alpha;\mu\nu}(x_{p})\) on the left-hand side of (\ref{s3_2}) is modified by a linear coupling to the symmetry-breaking
source field \(\hat{\mfrak{j}}_{\alpha;\mu\nu}^{(\hat{F})}(x_{p})\) so that we also consider the corresponding part of the source action
\(\mscr{A}_{S}\) (\ref{s2_22})
\beq \lb{s3_2}
\lefteqn{\exp\bigg\{-\im\int_{C}d^{4}\!x_{p}\;\bigg(\frac{1}{4}\;\hat{F}_{\alpha}^{\mu\nu}(x_{p})\;
\hat{F}_{\alpha;\mu\nu}(x_{p})+\hat{\mfrak{j}}_{\alpha;\mu\nu}^{(\hat{F})}(x_{p})\;
\hat{F}_{\alpha}^{\mu\nu}(x_{p})\bigg)\bigg\}= \int d[\hat{\mfrak{S}}_{\alpha;\mu\nu}^{(\hat{F})}(x_{p})] \;\times } \\ \no &\times&
\exp\bigg\{\im\int_{C}d^{4}\!x_{p}
\bigg(\frac{1}{4}\;\hat{\mfrak{S}}_{\alpha}^{(\hat{F})\mu\nu}(x_{p})\;\hat{\mfrak{S}}_{\alpha;\mu\nu}^{(\hat{F})}(x_{p})-\frac{1}{2}
\Big(\hat{F}_{\alpha;\mu\nu}(x_{p})+2\;\hat{\mfrak{j}}_{\alpha;\mu\nu}^{(\hat{F})}(x_{p})\Big)\;\hat{\mfrak{S}}_{\alpha}^{(\hat{F})\mu\nu}(x_{p})+  \\ \no &+&
\hat{\mfrak{j}}_{\alpha;\mu\nu}^{(\hat{F})}(x_{p})\;\hat{\mfrak{j}}_{\alpha}^{(\hat{F})\mu\nu}(x_{p})\bigg)\bigg\}\;.
\eeq
In a subsequent step (\ref{s3_3}), one replaces the field strength tensor \(\hat{F}_{\alpha;\mu\nu}(x_{p})\) in its linear coupling
to the self-energy \(\hat{\mfrak{S}}_{\alpha}^{(\hat{F})\mu\nu}(x_{p})\) by the original expression with the gauge field \(A_{\alpha;\mu}(x_{p})\) (\ref{s2_5},\ref{s2_6}).
Consequently, one gains the actions in the last two lines of (\ref{s3_3})
which combine the self-energy \(\hat{\mfrak{S}}_{\alpha}^{(\hat{F})\mu\nu}(x_{p})\) and its
spacetime derivative to a quadratic and linear part of the gauge field \(A_{\alpha;\mu}(x_{p})\), respectively.
The resulting, quadratic and linear term of \(A_{\alpha;\mu}(x_{p})\) can be removed by Gaussian integration in later steps of the derivation.
We have also to include a non-hermitian \(\hat{\mfrak{e}}_{p}^{(\hat{F})}\) matrix term (\ref{s3_4}) for appropriate time-ordering
and convergence properties of contour time Green functions (similar to \(\hat{\ve}_{p}\) (\ref{s2_27}) for the propagation of quark fields)
\beq  \lb{s3_3}
\lefteqn{\exp\bigg\{-\frac{\im}{2}\int_{C}d^{4}\!x_{p}\;\hat{F}_{\alpha;\mu\nu}(x_{p})\;
\hat{\mfrak{S}}_{\alpha}^{(\hat{F})\mu\nu}(x_{p})\bigg\}=}   \\ \no &=&
\exp\bigg\{-\frac{\im}{2}\int_{C}d^{4}\!x_{p}\;
\Big(\hat{\pp}_{p,\mu}A_{\alpha;\nu}(x_{p})-\hat{\pp}_{p,\nu}A_{\alpha;\mu}(x_{p})+C_{\alpha\beta\gamma}\;A_{\beta;\mu}(x_{p})\;
A_{\gamma;\nu}(x_{p})\Big)\;\hat{\mfrak{S}}_{\alpha}^{(\hat{F})\mu\nu}(x_{p})\bigg\}=
\\ \no &=&   \exp\bigg\{-\frac{\im}{2}\int_{C}d^{4}\!x_{p}\;A_{\beta;\mu}(x_{p})\;\Big[-\im\;\hat{\mfrak{e}}_{p}^{(\hat{F})}+
C_{\alpha\beta\gamma}\;\hat{\mfrak{S}}_{\alpha}^{(\hat{F})\mu\nu}(x_{p})\Big]\;A_{\gamma;\nu}(x_{p})\bigg\}\times  \\ \no &\times&
\exp\bigg\{-\frac{\im}{2}\int_{C}d^{4}\!x_{p}\Big(-A_{\alpha;\nu}(x_{p})\;
\big(\hat{\pp}_{p,\mu}\hat{\mfrak{S}}_{\alpha}^{(\hat{F})\mu\nu}(x_{p})\big)+
A_{\alpha;\mu}(x_{p})\;\big(\hat{\pp}_{p,\nu}\hat{\mfrak{S}}_{\alpha}^{(\hat{F})\mu\nu}(x_{p})\big)\Big)\bigg\} = \\ \no &=&
\exp\bigg\{-\frac{\im}{2}\int_{C}d^{4}\!x_{p}\;A_{\beta;\mu}(x_{p})\;\Big[-\im\;\hat{\mfrak{e}}_{p}^{(\hat{F})}+
C_{\alpha\beta\gamma}\;\hat{\mfrak{S}}_{\alpha}^{(\hat{F})\mu\nu}(x_{p})\Big]\;A_{\gamma;\nu}(x_{p})\bigg\}\times  \\ \no &\times&
\exp\bigg\{-\im\int_{C}d^{4}\!x_{p}\;A_{\alpha;\mu}(x_{p})\;
\big(\hat{\pp}_{p,\nu}\hat{\mfrak{S}}_{\alpha}^{(\hat{F})\mu\nu}(x_{p})\big)\bigg\} \;;  \\ \lb{s3_4} &&
\Big(\hat{\mfrak{e}}_{p}^{(\hat{F})}\Big)_{\beta\gamma}^{\mu\nu} = \eta_{p}\;
\mfrak{e}_{+}^{(\hat{F})}\;\delta_{\beta\gamma}\;\delta^{\mu\nu}\;;
\hspace*{0.6cm}\mfrak{e}_{+}^{(\hat{F})}>0 \;.
\eeq
The combination of Eqs. (\ref{s3_2},\ref{s3_3}) leads to relation (\ref{s3_5}) which transforms the quadratic term of the field
strength tensor and the linear coupling to the source field \(\hat{\mfrak{j}}_{\alpha;\mu\nu}^{(\hat{F})}(x_{p})\) to a Euclidean Gaussian
integral of the corresponding self-energy matrix \(\hat{\mfrak{S}}_{\alpha}^{(\hat{F})\mu\nu}(x_{p})\).
Apart from the linear derivative coupling to the gauge field \(A_{\alpha;\mu}(x_{p})\) (last line of (\ref{s3_5})), the self-energy matrix
is contained as a kind of '{\it inverse variance}' in an action with quadratic gauge fields \(A_{\alpha;\mu}(x_{p})\)
which can be eliminated by Gaussian integration
after adding the coupling to the bilinear quark fields in \(\bar{\psi}(x_{p})\:\feynD(x_{p})\:\psi(x_{p})\)
\beq \lb{s3_5}
\lefteqn{\exp\bigg\{-\im\int_{C}d^{4}\!x_{p}\;\bigg(\frac{1}{4}\;\hat{F}_{\alpha}^{\mu\nu}(x_{p})\;
\hat{F}_{\alpha;\mu\nu}(x_{p})+\hat{\mfrak{j}}^{(\hat{F})}_{\alpha;\mu\nu}(x_{p})\;
\hat{F}_{\alpha}^{\mu\nu}(x_{p})\bigg)\bigg\}=} \\  \no &=&
\int d[\hat{\mfrak{S}}_{\alpha;\mu\nu}^{(\hat{F})}(x_{p})] \; \exp\bigg\{\im\int_{C}d^{4}\!x_{p}\;\bigg(\frac{1}{4}\;
\hat{\mfrak{S}}_{\alpha}^{(\hat{F})\mu\nu}(x_{p})\;\hat{\mfrak{S}}_{\alpha;\mu\nu}^{(\hat{F})}(x_{p})+ \\ \no &-&
\hat{\mfrak{j}}^{(\hat{F})}_{\alpha;\mu\nu}(x_{p})\;\hat{\mfrak{S}}_{\alpha}^{(\hat{F})\mu\nu}(x_{p})+
\hat{\mfrak{j}}^{(\hat{F})}_{\alpha;\mu\nu}(x_{p})\;\hat{\mfrak{j}}_{\alpha}^{(\hat{F})\mu\nu}(x_{p})\bigg)\bigg\}\times   \\ \no &\times&
\exp\bigg\{-\frac{\im}{2}\int_{C}d^{4}\!x_{p}\;A_{\beta;\mu}(x_{p})\;\Big[-\im\;\hat{\mfrak{e}}_{p}^{(\hat{F})}+
C_{\alpha\beta\gamma}\;\hat{\mfrak{S}}_{\alpha}^{(\hat{F})\mu\nu}(x_{p})\Big]\;A_{\gamma;\nu}(x_{p})\bigg\}\times  \\ \no &\times&
\exp\bigg\{-\im\int_{C}d^{4}\!x_{p}\;A_{\alpha;\mu}(x_{p})\;
\big(\hat{\pp}_{p,\nu}\hat{\mfrak{S}}_{\alpha}^{(\hat{F})\mu\nu}(x_{p})\big)\bigg\} \;.
\eeq
Eventually, we can insert the completed HST (\ref{s3_5}) for the quadratic field strength tensor with its linear source-coupling
into the path integral (\ref{s2_25},\ref{s2_26}). The delta-function of the gauge field \(A_{\alpha;\mu}(x_{p})\) in (\ref{s2_25},\ref{s2_26}),
caused by the axial gauge, is taken into account by the standard integral representation with auxiliary, real fields
\(s_{\alpha}(x_{p})\) (\(\alpha,\,\beta,\,\ldots=1,\ldots,8\)). Furthermore, the resulting generating function
\(Z[\hat{\mscr{J}},J_{\psi},\hat{J}_{\psi\psi},\hat{\mfrak{j}}^{(\hat{F})}]\) (\ref{s3_6}) is grouped into integration terms which
consist of the 'gauge field strength' self-energy \(d[\hat{\mfrak{S}}_{\alpha;\mu\nu}^{(\hat{F})}(x_{p})]\), the fermionic
quark fields \(d[\psi_{M}^{\dag}(x_{p}),\,\psi_{M}(x_{p})]\), the auxiliary, real integration variables \(d[s_{\alpha}(x_{p})]\) for the delta-function
of the axial gauge, subsequently followed by remaining Gaussian integration \(d[A_{\alpha;\mu}(x_{p})]\) of the gauge field
\beq   \lb{s3_6}
\lefteqn{Z[\hat{\mscr{J}},J_{\psi},\hat{J}_{\psi\psi},\hat{\mfrak{j}}^{(\hat{F})}]=
\int d[\hat{\mfrak{S}}_{\alpha;\mu\nu}^{(\hat{F})}(x_{p})]\;
\exp\bigg\{\im\int_{C}d^{4}\!x_{p}\;\bigg(\frac{1}{4}\;\hat{\mfrak{S}}_{\alpha}^{(\hat{F})\mu\nu}(x_{p})\;
\hat{\mfrak{S}}_{\alpha;\mu\nu}^{(\hat{F})}(x_{p})+  }   \\ \no &-&
\hat{\mfrak{j}}_{\alpha;\mu\nu}^{(\hat{F})}(x_{p})\;\hat{\mfrak{S}}_{\alpha}^{(\hat{F})\mu\nu}(x_{p})+
\hat{\mfrak{j}}_{\alpha;\mu\nu}^{(\hat{F})}(x_{p})\;\hat{\mfrak{j}}_{\alpha}^{(\hat{F})\mu\nu}(x_{p})\bigg)\bigg\}\;\times \\  \no &\times&
\int d[\psi_{M}^{\dag}(x_{p}),\,\psi_{M}(x_{p})]\;
\exp\bigg\{-\im\int_{C}d^{4}\!x_{p}\;\psi_{N}\pdag(x_{p})\;
\Big[\hat{\beta}\Big(\hat{\gamma}^{\mu}\;\hat{\pp}_{p,\mu}-\im\;\hat{\ve}_{p}+\hat{m}\Big)\Big]_{N;M}\;\psi_{M}(x_{p})\bigg\}\times
\\  \no &\times&
\exp\bigg\{-\frac{\im}{2}\int_{C}d^{4}\!x_{p}\Big(
J_{\psi;M}^{\dag,a}(x_{p})\;\hat{S}^{ab}\;\Psi_{M}^{b}(x_{p})+
\Psi_{M}^{\dag,a}(x_{p})\;\hat{S}^{ab}\;J_{\psi;M}^{b}(x_{p})\Big)\bigg\} \\ \no &\times&
\exp\bigg\{-\frac{\im}{2}\int_{C}d^{4}\!x_{p}\;\Psi_{N}^{\dag,b}(x_{p})\;\hat{J}_{\psi\psi;N;M}^{b\neq a}(x_{p})\;
\Psi_{M}^{a}(x_{p}) \bigg\} \\ \no &\times&
\exp\bigg\{-\frac{\im}{2}\int_{C}d^{4}\!x_{p}\;d^{4}\!y_{q}\;\Psi_{N}^{\dag,b}(y_{q})\;
\hat{\mscr{J}}_{N;M}^{ba}(y_{q},x_{p})\;\Psi_{M}^{a}(x_{p})\bigg\}\;\times\;
\int d[s_{\alpha}(x_{p})]\int d[A_{\alpha;\mu}(x_{p})]\; \times \\ \no &\times&
\exp\bigg\{-\frac{\im}{2}\int_{C}d^{4}\!x_{p}\;A_{\beta;\mu}(x_{p})\;
\Big[-\im\;\hat{\mfrak{e}}_{p}^{(\hat{F})}+C_{\alpha\beta\gamma}\;\hat{\mfrak{S}}_{\alpha}^{(\hat{F})\mu\nu}(x_{p})\Big]\;A_{\gamma;\nu}(x_{p})\bigg\}\times
\\ \no &\times&\exp\bigg\{-\im\int_{C}d^{4}\!x_{p}\;A_{\beta;\mu}(x_{p})\;
\Big[\big(\hat{\pp}_{p,\kappa}\hat{\mfrak{S}}_{\beta}^{(\hat{F})\mu\kappa}(x_{p})\big)-
\psi_{g,\ovv{N}}\pdag(x_{p})\,\big[\hat{\beta}\big(\im\;\hat{\gamma}^{\mu}\;\hat{t}_{\beta}\big)\big]_{\ovv{N};\ovv{M}}\,
\psi_{f,\ovv{M}}(x_{p})\;\delta_{g,f}-s_{\beta}(x_{p})\;n^{\mu}\Big]\bigg\} .
\eeq
We split the Gaussian integration part (\ref{s3_7}) of gauge fields \(A_{\alpha;\mu}(x_{p})\) from the generating
function in (\ref{s3_6}) and separately list the result of the integration. Aside from the inverse square root of the
self-energy matrix for the gauge field strength tensor,
an action appears that is quadratic in spacetime derivatives of the self-energy matrix
\((\,\hat{\pp}_{p}^{\kappa}\hat{\mfrak{S}}_{\beta;\mu\kappa}^{(\hat{F})}(x_{p})\,)\), quartic in the interaction of matter fields
and again quadratic in the auxiliary, real field \(s_{\beta}(x_{p})\) for axial gauge fixing.
The latter action in (\ref{s3_7}) is locally weighted by the inverse of the field strength self-energy matrix as a kind of '{\it variance}',
indicating the self-interaction of the gauge fields. Since the self-interaction of the gauge fields consists of
a quadratic derivative, (symbolically abbreviated by
\((\hat{\pp}\hat{\mfrak{S}}^{(\hat{F})})\:(\hat{\mfrak{S}}^{(\hat{F})})^{\boldsymbol{-1}}\:(\hat{\pp}\hat{\mfrak{S}}^{(\hat{F})})\) ),
one immediately concludes for the asymptotic freedom at high energies according to the spacetime derivative
\((\,\hat{\pp}_{p}^{\kappa}\hat{\mfrak{S}}_{\beta;\mu\kappa}^{(\hat{F})}(x_{p})\,)\) which also occurs in the coupling to the quark fields
\beq \lb{s3_7}
\lefteqn{\int d[A_{\alpha;\mu}(x_{p})]\;
\exp\bigg\{-\frac{\im}{2}\int_{C}d^{4}\!x_{p}\;A_{\beta;\mu}(x_{p})\;
\Big[-\im\;\hat{\mfrak{e}}_{p}^{(\hat{F})}+C_{\alpha\beta\gamma}\;
\hat{\mfrak{S}}_{\alpha}^{(\hat{F})\mu\nu}(x_{p})\Big]\;A_{\gamma;\nu}(x_{p})\bigg\}\times}
\\ \no &\times&\exp\bigg\{-\im\int_{C}d^{4}\!x_{p}\;A_{\beta;\mu}(x_{p})\;
\Big[\big(\hat{\pp}_{p,\kappa}\hat{\mfrak{S}}_{\beta}^{(\hat{F})\mu\kappa}(x_{p})\big)-
\psi_{N}\pdag(x_{p})\;\big[\hat{\beta}\big(\im \;\hat{\gamma}^{\mu}\;\hat{t}_{\beta}\big)\big]_{N;M}\psi_{M}(x_{p})-
s_{\beta}(x_{p})\;n^{\mu}\Big]\bigg\} =  \\ \no &=&
\bigg\{\det\Big[\Big(-\im\;\hat{\mfrak{e}}_{p}^{(\hat{F})}+C_{\alpha\beta\gamma}\;
\hat{\mfrak{S}}_{\alpha}^{(\hat{F})\mu\nu}(x_{p})\Big)_{\beta\gamma}^{\mu\nu}\Big]\bigg\}^{\boldsymbol{-1/2}}  \times \\ \no &\times&
\exp\bigg\{\frac{\im}{2}\int_{C}d^{4}\!x_{p}\Big[\big(\hat{\pp}_{p}^{\lambda}\hat{\mfrak{S}}_{\gamma;\nu\lambda}^{(\hat{F})}(x_{p})\big)-
\psi_{g,\ovv{N}}\pdag(x_{p})\;\big[\hat{\beta}\big(\im \;\hat{\gamma}_{\nu}\;
\hat{t}_{\gamma}\big)\big]_{g,\ovv{N};g\ppr,\ovv{N}\ppr}\;\delta_{g,g\ppr}\;
\psi_{g\ppr,\ovv{N}\ppr}(x_{p})-
s_{\gamma}(x_{p})\;n_{\nu}\Big] \times \\ \no &\times&
\Big[-\im\;\hat{\mfrak{e}}_{p}^{(\hat{F})}+C_{\alpha\beta\ppr\gamma\ppr}\;
\hat{\mfrak{S}}_{\alpha}^{(\hat{F})\mu\ppr\nu\ppr}(x_{p})\Big]_{\gamma\beta}^{\boldsymbol{-1};\nu\mu}\;\times  \\ \no &\times&
\Big[\big(\hat{\pp}_{p}^{\kappa}\hat{\mfrak{S}}_{\beta;\mu\kappa}^{(\hat{F})}(x_{p})\big)-
\psi_{f,\ovv{M}}\pdag(x_{p})\;\big[\hat{\beta}\big(\im\;\hat{\gamma}_{\mu}\;\hat{t}_{\beta}\big)
\big]_{f,\ovv{M};f\ppr,\ovv{M}\ppr}\;\delta_{f,f\ppr}\;
\psi_{f\ppr,\ovv{M}\ppr}(x_{p})-s_{\beta}(x_{p})\;n_{\mu}\Big]\bigg\} \;.
\eeq
We substitute the entire Gaussian integration part (\ref{s3_7}) of the gauge fields \(A_{\alpha;\mu}(x_{p})\) into (\ref{s3_6})
and additionally perform the anomalous doubling of the fermionic quark fields \(\psi_{M}^{\dag}\psi_{M}=\frac{1}{2}
(\Psi_{M}^{\dag,b}\:\hat{S}^{ba}\:\Psi_{M}^{a})\) (\ref{s2_20},\ref{s2_21}) and also include the anomalous doubling
with the transpose of the one-particle Hamiltonian \(\hat{\beta}\,(\hat{\gamma}^{\mu}\hat{\pp}_{p,\mu}-\im\:\hat{\ve}_{p}+\hat{m})_{N;M}\).
The anomalous doubled path integral is listed in detail in Eq. (\ref{s3_8}) where the doubled, fermionic fields
\(\Psi_{M}^{a}(x_{p})\) replace the original quark fields \(\psi_{M}(x_{p})\) in (\ref{s3_6}) with suitable anomalous doubling
of one-particle Hamiltonian and of the quartic interaction term. However, the integration variables of the self-energy
\(\hat{\mfrak{S}}_{\alpha;\mu\nu}^{(\hat{F})}(x_{p})\) of the field strength tensor and of the auxiliary, real field \(s_{\alpha}(x_{p})\),
both related to propagation of the gauge terms,
are separated as background fields from the fermionic quark field integrations \(d[\psi_{M}^{\dag}(x_{p}),\,\psi_{M}(x_{p})]\) in (\ref{s3_8}) with
additional averaging
\(\boldsymbol{\langle}\ldots\boldsymbol{\rangle_{\hat{\mfrak{S}}^{(\hat{F})},s_{\alpha}}^{\mbox{\scz\bf Eq.(\ref{s3_9})}}}\) (\ref{s3_9})
of a 'background generating function'
\beq \lb{s3_8}
\lefteqn{Z[\hat{\mscr{J}},J_{\psi},\hat{J}_{\psi\psi},\hat{\mfrak{j}}^{(\hat{F})}]=
\boldsymbol{\Bigg\langle Z\Big[\hat{\mfrak{S}}^{(\hat{F})},s_{\alpha};\hat{\mfrak{j}}^{(\hat{F})};\mbox{\bf Eq. (\ref{s3_9})}\Big] }\times
\int d[\psi_{M}^{\dag}(x_{p}),\,\psi_{M}(x_{p})]\;\times
\exp\Bigg\{-\frac{\im}{2}\int_{C}d^{4}\!x_{p}\;\times } \\ \no &\times& \Psi_{N}^{\dag,b}(x_{p}) \Bigg(\bea{cc}
[\hat{\beta}(\hat{\gamma}^{\mu}\hat{\pp}_{p,\mu}-\im\;\hat{\ve}_{p}+\hat{m})]_{N;M} &
\hat{j}_{\psi\psi;N;M}(x_{p})  \\  \hat{j}_{\psi\psi;N;M}^{\dag}(x_{p}) &
-[\hat{\beta}(\hat{\gamma}^{\mu}\hat{\pp}_{p,\mu}-\im\;\hat{\ve}_{p}+\hat{m})]_{N;M}^{T}
\eea\Bigg)^{ba}_{N;M}\;\Psi_{M}^{a}(x_{p})\Bigg\} \times   \\ \no &\times&
\exp\Bigg\{-\frac{\im}{2}\int_{C}d^{4}\!x_{p}\;d^{4}\!y_{q}\;
\Psi_{N}^{\dag,b}(y_{q})\;\hat{\mscr{J}}_{N;M}^{ba}(y_{q},x_{p})\;\Psi_{M}^{a}(x_{p})\Bigg\}   \\ \no &\times&
\exp\bigg\{-\frac{\im}{2}\int_{C}d^{4}\!x_{p}\Big(J_{\psi;M}^{\dag,a}(x_{p})\;\hat{S}^{ab}\;
\Psi_{M}^{b}(x_{p})+\Psi_{M}^{\dag,a}(x_{p})\;\hat{S}^{ab}\,J_{\psi;M}^{b}(x_{p})\Big)\bigg\}  \\ \no &\times&
\exp\Bigg\{-\frac{\im}{2}\int_{C}d^{4}\!x_{p}\Big[\big(\hat{\pp}_{p}^{\lambda}\hat{\mfrak{S}}_{\gamma;\nu\lambda}^{(\hat{F})}(x_{p})\big)-
s_{\gamma}(x_{p})\;n_{\nu}\Big]
\Big[-\im\;\hat{\mfrak{e}}_{p}^{(\hat{F})}+C_{\alpha\beta\ppr\gamma\ppr}\;
\hat{\mfrak{S}}_{\alpha}^{(\hat{F})\mu\ppr\nu\ppr}(x_{p})\Big]_{\gamma\beta}^{\boldsymbol{-1};\nu\mu}\times  \\ \no &\times&
\Psi_{g,\ovv{N}}^{\dag,b}(x_{p})   \Bigg(\bea{cc}
[\hat{\beta}(\im \;\hat{\gamma}_{\mu}\;\hat{t}_{\beta})]_{g,\ovv{N};f,\ovv{M}} & 0 \\  0 &
-[\hat{\beta}(\im \;\hat{\gamma}_{\mu}\;\hat{t}_{\beta})]_{g,\ovv{N};f,\ovv{M}}^{T} \eea\Bigg)^{ba}_{g,\ovv{N};f,\ovv{M}}\delta_{g,f}\:
\Psi_{M}^{a}(x_{p})\Bigg\} \times  \\ \no &\times &
\exp\Bigg\{\frac{\im}{8}\int_{C}d^{4}\!x_{p}\;
\Psi_{g\ppr,\ovv{N}\ppr}^{\dag,b\ppr}(x_{p})   \Bigg(\bea{cc}
[\hat{\beta}(\im \;\hat{\gamma}_{\nu}\;\hat{t}_{\gamma})]_{g\ppr,\ovv{N}\ppr;g,\ovv{N}} & 0 \\ 0 &
-[\hat{\beta}(\im \;\hat{\gamma}_{\nu}\;\hat{t}_{\gamma})]_{g\ppr,\ovv{N}\ppr;g,\ovv{N}}^{T}
\eea\Bigg)^{b\ppr b}_{g\ppr,\ovv{N}\ppr;g,\ovv{N}}\delta_{g\ppr,g}\:
\Psi_{g,\ovv{N}}^{b}(x_{p})\;\times \\ \no &\times&
\Big[-\im\;\hat{\mfrak{e}}_{p}^{(\hat{F})}+C_{\alpha\beta\ppr\gamma\ppr}\;
\hat{\mfrak{S}}_{\alpha}^{(\hat{F})\mu\ppr\nu\ppr}(x_{p})\Big]_{\gamma\beta}^{\boldsymbol{-1};\nu\mu}\;\times \\ \no &\times&
\Psi_{f\ppr,\ovv{M}\ppr}^{\dag,a\ppr}(x_{p})   \Bigg(\bea{cc}
[\hat{\beta}(\im \;\hat{\gamma}_{\mu}\;\hat{t}_{\beta})]_{f\ppr,\ovv{M}\ppr;f,\ovv{M}} & 0  \\ 0  &
-[\hat{\beta}(\im \;\hat{\gamma}_{\mu}\;\hat{t}_{\beta})]_{f\ppr,\ovv{M}\ppr;f,\ovv{M}}^{T}
\eea\Bigg)^{a\ppr a}_{f\ppr,\ovv{M}\ppr;f,\ovv{M}}\delta_{f\ppr,f}\:
\Psi_{f,\ovv{M}}^{a}(x_{p})\Bigg\}\boldsymbol{\Bigg\rangle} \;;
\eeq
\beq \lb{s3_9}
\lefteqn{\boldsymbol{\Bigg\langle Z\Big[\hat{\mfrak{S}}^{(\hat{F})},s_{\alpha};\hat{\mfrak{j}}^{(\hat{F})};
\mbox{\bf Eq. (\ref{s3_9})}\Big]\;\Big(\mbox{\bf fields}\Big)\;\Bigg\rangle=} }
\\ \no &=&\boldsymbol{\Bigg\langle}
\int d[\hat{\mfrak{S}}_{\alpha;\mu\nu}^{(\hat{F})}(x_{p})]\;d[s_{\alpha}(x_{p})]\;
\exp\bigg\{\im\int_{C}d^{4}\!x_{p}\;\bigg(\frac{1}{4}\;
\hat{\mfrak{S}}_{\alpha}^{(\hat{F})\mu\nu}(x_{p})\;\hat{\mfrak{S}}_{\alpha;\mu\nu}^{(\hat{F})}(x_{p})-
\hat{\mfrak{j}}_{\alpha;\mu\nu}^{(\hat{F})}(x_{p})\;\hat{\mfrak{S}}_{\alpha}^{(\hat{F})\mu\nu}(x_{p})+  \\ \no &+&
\hat{\mfrak{j}}_{\alpha;\mu\nu}^{(\hat{F})}(x_{p})\;\hat{\mfrak{j}}_{\alpha}^{(\hat{F})\mu\nu}(x_{p})\bigg)\bigg\} \times
\bigg\{\det\Big[\Big(-\im\;\hat{\mfrak{e}}_{p}^{(\hat{F})}+C_{\alpha\beta\gamma}\;
\hat{\mfrak{S}}_{\alpha}^{(\hat{F})\mu\nu}(x_{p})\Big)_{\beta\gamma}^{\mu\nu}\Big]\bigg\}^{\boldsymbol{-1/2}}  \times \\ \no &\times&
\exp\bigg\{\frac{\im}{2}\int_{C}d^{4}\!x_{p}\Big[\big(\hat{\pp}_{p}^{\lambda}\hat{\mfrak{S}}_{\gamma;\nu\lambda}^{(\hat{F})}(x_{p})\big)
-s_{\gamma}(x_{p})\;n_{\nu}\Big] \times \\ \no &\times&
\Big[-\im\;\hat{\mfrak{e}}_{p}^{(\hat{F})}+C_{\alpha\beta\ppr\gamma\ppr}\;
\hat{\mfrak{S}}_{\alpha}^{(\hat{F})\mu\ppr\nu\ppr}(x_{p})\Big]_{\gamma\beta}^{\boldsymbol{-1};\nu\mu}\;
\Big[\big(\hat{\pp}_{p}^{\kappa}\hat{\mfrak{S}}_{\beta;\mu\kappa}^{(\hat{F})}(x_{p})\big)
-s_{\beta}(x_{p})\;n_{\mu}\Big]\bigg\}\;\times\;\boldsymbol{\bigg(\mbox{{\bf fields}}\bigg)\Bigg\rangle} \;.
\eeq
Despite complicated and lengthy appearance, we have moved three and four point vertices of gauge fields
\(A_{\alpha;\mu}(x_{p})\) to a background path integral (\ref{s3_9})
of the field strength self-energy \(\hat{\mfrak{S}}_{\alpha;\mu\nu}^{(\hat{F})}(x_{p})\).
Therefore, it suffices to concentrate onto the anomalous doubled quark fields (\ref{s3_8}) whose quartic interaction contains as two-body potential
the inverse of the gauge field self-energy matrix within a completely local spacetime relation (last three lines of (\ref{s3_8})).

\subsection{HST to the anomalous doubled self-energy of Fermi fields}  \lb{s32}

The HST of Fermi fields is not only more involved by the anomalous doubling, but also because of the 'inverse'
self-energy of the field strength tensor in the quartic interaction (\ref{s3_8}). We recognize that
the combination \(C_{\alpha\beta\gamma}\;\hat{\mfrak{S}}_{\alpha}^{(\hat{F})\mu\nu}(x_{p})\) of anti-symmetric
\(\mbox{SU}_{c}(N_{c}=3)\) structure constants \(C_{\alpha\beta\gamma}\) and anti-symmetric spacetime
indices in \(\hat{\mfrak{S}}_{\alpha}^{(\hat{F})\mu\nu}(x_{p})\) defines a totally real symmetric matrix with doubled
indices of \(C_{\alpha\beta\gamma}\;\hat{\mfrak{S}}_{\alpha}^{(\hat{F})\mu\nu}(x_{p})=
[\mbox{Matrix}]_{\beta\gamma}^{\mu\nu}=[\mbox{Matrix}]_{\gamma\beta}^{\nu\mu}\).
Therefore, we introduce a real, orthogonal diagonalization (\ref{s3_10}) of
\(C_{\alpha\beta\gamma}\;\hat{\mfrak{S}}_{\alpha}^{(\hat{F})\mu\nu}(x_{p})\) with orthogonal matrices
\(\hat{\mfrak{B}}_{\hat{F};\beta(\alpha)}^{\mu(\kappa)}(x_{p})\),
\(\hat{\mfrak{B}}_{\hat{F};(\alpha)\gamma}^{T,(\kappa)\nu}(x_{p})=\hat{\mfrak{B}}_{\hat{F};(\alpha)\gamma}^{\boldsymbol{-1},(\kappa)\nu}(x_{p})\)
and real eigenvalues \(\hat{\mfrak{b}}^{(\hat{F})}_{(\alpha;\kappa)}(x_{p})\)
\footnote{Although we have selected similar indices '\(\alpha=1,\,\ldots\,,8\)' and '\(\kappa=0,1,2,3\)' in anticipation of
the number of colour generators \(\hat{t}_{\alpha}\) and the labeling of spacetime, this summation over \(32\) degrees
of freedom can be realized by arbitrary indexing, as far as only \(32\) independent additions with the eigenvalues
\(\hat{\mfrak{b}}^{(\hat{F})}_{(\alpha;\kappa)}(x_{p})\) and eigenvectors in
\(\hat{\mfrak{B}}_{\hat{F};\beta(\alpha)}^{\mu(\kappa)}(x_{p})\), \(\hat{\mfrak{B}}_{\hat{F};(\alpha)\gamma}^{T,(\kappa)\nu}(x_{p})\)
are concerned; therefore, we need not distinguish between contravariant and covariant spacetime indexing and
set the introduced indices '\((\alpha)=1,\,\ldots\,,8\)' and '\((\kappa)=0,1,2,3\)' into parentheses in order
to emphasize their difference to the other labeling for colour generators and spacetime degrees of freedom
with metric tensor \(\hat{\eta}^{\mu\nu}\) !}.
This induces a change of integration measure from the 'flat', Euclidean self-energy of gauge fields
\(d[\hat{\mfrak{S}}_{\alpha;\mu\nu}^{(\hat{F})}(x_{p})]\) to that of
\(\mbox{SO}(\,(N_{c}^{2}-1=8)\times(3+1)\,)=\mbox{SO}(32)\). Since the self-energy \(\hat{\mfrak{S}}_{\alpha;\mu\nu}^{(\hat{F})}(x_{p})\)
with its diagonalized form (\ref{s3_10}) only occurs as a background field and in a saddle point approximation,
we do not specify details of the \(\mbox{SO}(32)\) integration measure which has to be incorporated
by a delta-function (\ref{s3_12}) into the 'flat', Euclidean integration degrees of freedom \(d[\hat{\mfrak{S}}_{\alpha;\mu\nu}^{(\hat{F})}(x_{p})]\)
\beq\lb{s3_10}
C_{\alpha\beta\gamma}\;\hat{\mfrak{S}}_{\alpha}^{(\hat{F})\mu\nu}(x_{p}) &:=&
\sum_{\scrscr(\alpha)=1,.,8}^{\scrscr(\kappa)=0,.,3}
\hat{\mfrak{B}}_{\hat{F};\beta(\alpha)}^{\mu(\kappa)}(x_{p})\;
\hat{\mfrak{b}}^{(\hat{F})}_{(\alpha;\kappa)}(x_{p})\;\hat{\mfrak{B}}_{\hat{F};(\alpha)\gamma}^{T,(\kappa)\nu}(x_{p}) \;; \\ \lb{s3_11}
d[\hat{\mfrak{S}}_{\alpha}^{(\hat{F})\mu\nu}(x_{p})]&\rightarrow&
d[\hat{\mfrak{S}}_{\alpha}^{(\hat{F})\mu\nu}(x_{p});\hat{\mfrak{B}}_{\hat{F};\beta(\alpha)}^{\mu(\kappa)}(x_{p});
\hat{\mfrak{b}}^{(\hat{F})}_{(\alpha;\kappa)}(x_{p})]  \;;
\eeq
\beq     \lb{s3_12}
\lefteqn{\hspace*{-1.9cm}d[\hat{\mfrak{S}}_{\alpha}^{(\hat{F})\mu\nu}(x_{p});\hat{\mfrak{B}}_{\hat{F};\beta(\alpha)}^{\mu(\kappa)}(x_{p});
\hat{\mfrak{b}}^{(\hat{F})}_{(\alpha;\kappa)}(x_{p})] = d[\hat{\mfrak{S}}_{\alpha}^{(\hat{F})\mu\nu}(x_{p})]\;\;
d[\hat{\mfrak{B}}_{\hat{F};\beta(\alpha)}^{\mu(\kappa)}(x_{p});\hat{\mfrak{b}}^{(\hat{F})}_{(\alpha;\kappa)}(x_{p})]\;\times } \\ \no &\times&
\bigg\{\prod_{\scrscr\{x_{p}\}}\delta\bigg(C_{\alpha\beta\gamma}\;\hat{\mfrak{S}}_{\alpha}^{(\hat{F})\mu\nu}(x_{p}) -
\sum_{\scrscr(\alpha)=1,.,8}^{\scrscr(\kappa)=0,.,3}
\hat{\mfrak{B}}_{\hat{F};\beta(\alpha)}^{\mu(\kappa)}(x_{p})\;
\hat{\mfrak{b}}^{(\hat{F})}_{(\alpha;\kappa)}(x_{p})\;\hat{\mfrak{B}}_{\hat{F};(\alpha)\gamma}^{T,(\kappa)\nu}(x_{p}) \bigg)\bigg\} \;.
\eeq
Moreover, the anomalous doubled gamma- and colour-matrices \(\hat{\beta}\,\hat{\gamma}^{\mu}\), \(\hat{t}_{\alpha}\)
in (\ref{s3_8}) are abbreviated by the matrix symbols \(\hat{\boldsymbol{\Gamma}}_{\beta;\mu;M\ppr;M}^{aa}\) and
\(\hat{\boldsymbol{\Gamma}}_{\gamma;\nu;N\ppr;N}^{bb}\) in order to simplify notation (\ref{s3_13}-\ref{s3_15}).
These anomalous doubled matrices are hermitian (\ref{s3_15})
and are also diagonal in the isospin- (flavour-) degrees of freedom which may be additionally considered by the separate indexing
\(\{M\ppr;M\}\rightarrow\delta_{f\ppr,f}\:\{f\ppr,\ovv{M}\ppr;f,\ovv{M}\}\) according to the definitions in section \ref{s21}
\beq \lb{s3_13}
\hat{\boldsymbol{\Gamma}}_{\beta;\mu;M\ppr;M}^{aa}&=&
\left(\bea{cc} [\hat{\beta}(\im \;\hat{\gamma}_{\mu}\;\hat{t}_{\beta})]_{M\ppr;M} & 0 \\
0  & [\hat{\beta}(\im \;\hat{\gamma}_{\mu}\;\hat{t}_{\beta})]_{M\ppr;M}^{T}
\eea\right)^{aa} \;; \\  \lb{s3_14}
\hat{\boldsymbol{\Gamma}}_{\gamma;\nu;N\ppr;N}^{bb}&=&
\left(\bea{cc} [\hat{\beta}(\im \;\hat{\gamma}_{\nu}\;\hat{t}_{\gamma})]_{N\ppr;N} & 0 \\
0  & [\hat{\beta}(\im \;\hat{\gamma}_{\nu}\;\hat{t}_{\gamma})]_{N\ppr;N}^{T}
\eea\right)^{bb} \;;  \\ \lb{s3_15}
[\hat{\beta}(\im \;\hat{\gamma}_{\mu}\;\hat{t}_{\beta})]_{M\ppr;M}^{\dag} &=&
[\hat{\beta}(\im \;\hat{\gamma}_{\mu}\;\hat{t}_{\beta})]_{M\ppr;M}\;;\hspace*{0.3cm}
[\hat{\beta}(\im \;\hat{\gamma}_{\mu}\;\hat{t}_{\beta})]_{M\ppr;M}=\delta_{f\ppr,f}\;
[\hat{\beta}(\im \;\hat{\gamma}_{\mu}\;\hat{t}_{\beta})]_{f\ppr,\ovv{M}\ppr;f,\ovv{M}}\;.
\eeq
Proceeding with the diagonalization (\ref{s3_10}-\ref{s3_12}) and abbreviations (\ref{s3_13}-\ref{s3_15}),
it is possible to reduce the anomalous doubled, quartic interaction of Fermi fields in (\ref{s3_8})
to relation (\ref{s3_16}). According to the diagonalization of (\ref{s3_10}), one achieves the
sum of \((N_{c}^{2}-1=8)\times (3+1)=32\) interaction terms of anomalous doubled quark fields, each with a different
diagonalized 'two-body potential' \([-\im\,\hat{\mfrak{e}}_{p}^{(\hat{F})}+
\hat{\mfrak{b}}^{(\hat{F})}_{(\alpha;\kappa)}(x_{p})\,]^{\boldsymbol{-1}}\)
of the 32 eigenvalues \(\hat{\mfrak{b}}^{(\hat{F})}_{(\alpha;\kappa)}(x_{p})\). Furthermore, the diagonalizing, orthogonal
eigenvector matrices \(\hat{\mfrak{B}}_{\hat{F};\beta(\alpha)}^{\mu(\kappa)}(x_{p})\),
\(\hat{\mfrak{B}}_{\hat{F};(\alpha)\gamma}^{T,(\kappa)\nu}(x_{p})\) can be shifted into scalar products of anomalous doubled
Fermi fields by defining the potentials \(\hat{\boldsymbol{\mscr{V}}}_{(\alpha);M\ppr;M}^{(\hat{F})a;(\kappa)}(x_{p})\)
(\ref{s3_17},\ref{s3_18}) in terms of (\ref{s3_13}-\ref{s3_15}) and (\ref{s3_10})
\beq \lb{s3_16}
\lefteqn{\exp\bigg\{\frac{\im}{8}\int_{C}d^{4}\!x_{p}\Big(\Psi_{N\ppr}^{\dag,b}(x_{p})\;
\hat{\boldsymbol{\Gamma}}_{\gamma;\nu;N\ppr;N}^{bb}\;\hat{S}^{bb}\;\Psi_{N}^{b}(x_{p})\Big)\;
\Big[-\im\;\hat{\mfrak{e}}_{p}^{(\hat{F})}+C_{\alpha\beta\ppr\gamma\ppr}\;
\hat{\mfrak{S}}_{\alpha}^{(\hat{F})\mu\ppr\nu\ppr}(x_{p})\Big]_{\gamma\beta}^{\boldsymbol{-1};\nu\mu}\;\times } \\ \no &\times&
\Big(\Psi_{M\ppr}^{\dag,a}(x_{p})\;\hat{\boldsymbol{\Gamma}}_{\beta;\mu;M\ppr;M}^{aa}\;\hat{S}^{aa}\;\Psi_{M}^{a}(x_{p})\Big)\bigg\}=
\exp\bigg\{\frac{\im}{8}\int_{C}d^{4}\!x_{p}\sum_{\scrscr(\alpha)=1,.,8}^{\scrscr(\kappa)=0,.,3}
\Psi_{N\ppr}^{\dag,b}(x_{p})\;\hat{\boldsymbol{\mscr{V}}}_{(\alpha);N\ppr;N}^{(\hat{F})b;(\kappa)}(x_{p})\;
\hat{S}^{bb}\;\Psi_{N}^{b}(x_{p})\; \times \\ \no &\times&
\Big[-\im\,\hat{\mfrak{e}}_{p}^{(\hat{F})}+\hat{\mfrak{b}}^{(\hat{F})}_{(\alpha;\kappa)}(x_{p})\Big]^{\boldsymbol{-1}}\;
\Psi_{M\ppr}^{\dag,a}(x_{p})\;
\hat{\boldsymbol{\mscr{V}}}_{(\alpha);M\ppr;M}^{(\hat{F})a;(\kappa)}(x_{p}) \;\hat{S}^{aa}\;\Psi_{M}^{a}(x_{p})\bigg\}  \;;
\eeq\vspace*{-0.7cm}
\beq   \lb{s3_17}
\hat{\boldsymbol{\mscr{V}}}_{(\alpha);M\ppr;M}^{(\hat{F})a;(\kappa)}(x_{p})
&=&\hat{\boldsymbol{\Gamma}}_{\beta;\mu;M\ppr;M}^{aa}\;\hat{\mfrak{B}}_{\hat{F};(\alpha)\beta}^{T;(\kappa)\mu}(x_{p})=
\hat{\boldsymbol{\Gamma}}_{\beta;\mu;f\ppr,\ovv{M}\ppr;f,\ovv{M}}^{aa}\;\delta_{f\ppr,f}\;
\hat{\mfrak{B}}_{\hat{F};\beta(\alpha)}^{\mu(\kappa)}(x_{p}) \;; \\   \lb{s3_18}
\hat{\boldsymbol{\mscr{V}}}_{(\alpha);N\ppr;N}^{(\hat{F})b;(\kappa)}(x_{p})&=&
\hat{\boldsymbol{\Gamma}}_{\gamma;\nu;N\ppr;N}^{bb}\;\hat{\mfrak{B}}_{\hat{F};\gamma(\alpha)}^{\nu(\kappa)}(x_{p})
=\hat{\boldsymbol{\Gamma}}_{\gamma;\nu;g\ppr,\ovv{N}\ppr;g,\ovv{N}}^{bb}\;\delta_{g\ppr,g}\;
\hat{\mfrak{B}}_{\hat{F};\gamma(\alpha)}^{\nu(\kappa)}(x_{p}) \;.
\eeq
This allows to perform the dyadic product of anomalous doubled quark fields to the required density matrix
\(\hat{R}_{M;N\ppr}^{ab}(x_{p})\) with BCS paired terms in the off-diagonal blocks \((a\neq b)\) so that the HST
becomes possible with an anomalous doubled self-energy of quark matter fields in a Gaussian integral
\be \lb{s3_19}
\Psi_{M}^{a}(x_{p})\otimes \Psi_{N\ppr}^{\dag,b}(x_{p})=
\left(\bea{c} \psi_{M}(x_{p}) \\ \psi_{M}^{*}(x_{p}) \eea\right)\otimes
\Big(\psi_{N\ppr}^{*}(x_{p})\,;\,\psi_{N\ppr}(x_{p})\Big)=
\hat{R}_{M;N\ppr}^{ab}(x_{p}) \;.
\ee
Combination of the entire relations (\ref{s3_10}-\ref{s3_19}) for the path integral (\ref{s3_8}) yields Eq. (\ref{s3_20}) with
density matrices (\ref{s3_19}) for the anomalous doubled, quartic interaction part of Fermi fields
\beq \lb{s3_20}
\lefteqn{\exp\bigg\{\frac{\im}{8}\int_{C}d^{4}\!x_{p}\Big(\Psi_{N\ppr}^{\dag,b}(x_{p})\;
\hat{\boldsymbol{\Gamma}}_{\gamma;\nu;N\ppr;N}^{bb}\;\hat{S}^{bb}\;\Psi_{N}^{b}(x_{p})\Big)\;
\Big[-\im\;\hat{\mfrak{e}}_{p}^{(\hat{F})}+C_{\alpha\beta\ppr\gamma\ppr}\;
\hat{\mfrak{S}}_{\alpha}^{(\hat{F})\mu\ppr\nu\ppr}(x_{p})\Big]_{\gamma\beta}^{\boldsymbol{-1};\nu\mu}\;\times } \\ \no &\times&
\Big(\Psi_{M\ppr}^{\dag,a}(x_{p})\;\hat{\boldsymbol{\Gamma}}_{\beta;\mu;M\ppr;M}^{aa}\;\hat{S}^{aa}\;\Psi_{M}^{a}(x_{p})\Big)\bigg\}=
\exp\bigg\{-\frac{\im}{8}\int_{C}d^{4}\!x_{p}\sum_{\scrscr(\alpha)=1,.,8}^{\scrscr(\kappa)=0,.,3} \bigg(
\hat{R}_{M;N\ppr}^{ab}(x_{p})\;\hat{\boldsymbol{\mscr{V}}}_{(\alpha);N\ppr;N}^{(\hat{F})b;(\kappa)}(x_{p})\;\hat{S}^{bb}\;\times \\  \no &\times&
\hat{R}_{N;M\ppr}^{ba}(x_{p})\;\hat{\boldsymbol{\mscr{V}}}_{(\alpha);M\ppr;M}^{(\hat{F})a;(\kappa)}(x_{p})\;\hat{S}^{aa}\;
\Big[-\im\,\hat{\mfrak{e}}_{p}^{(\hat{F})}+\hat{\mfrak{b}}^{(\hat{F})}_{(\alpha;\kappa)}(x_{p})\Big]^{\boldsymbol{-1}}\bigg)\bigg\}\;.
\eeq
However, the dyadic product (\ref{s3_19}) to the anomalous doubled density matrix \(\hat{R}_{M;N\ppr}^{ab}(x_{p})\)
does not regard the \(8\times(3+1)=32\) different potentials
\(\hat{\boldsymbol{\mscr{V}}}_{(\alpha);M\ppr;M}^{(\hat{F})a;(\kappa)}(x_{p})\),
\(\hat{\boldsymbol{\mscr{V}}}_{(\alpha);N\ppr;N}^{(\hat{F})b;(\kappa)}(x_{p})\) (\ref{s3_17},\ref{s3_18})
{\it which even depend on the anomalous indexing with \(a,b=1,2\)}.
Therefore, we take an additional, unitary diagonalization
\(\hat{\mscr{U}}_{\hat{F};\ovv{N};\ovv{M}}^{(\alpha;\kappa)}(x_{p})\) (\ref{s3_22}) of the
hermitian, doubled '11' and '22' potentials (\ref{s3_17},\ref{s3_18})
into real eigenvalues \(\hat{\mfrak{v}}_{\hat{F};\ovv{N}}^{(\alpha;\kappa)}(x_{p})\)
(The indices '\((\alpha;\kappa)\)' in (\ref{s3_22}) are also embraced by
parentheses because a diagonalization is performed for every pair '\((\alpha;\kappa)\)'
of the \(8\times(3+1)=32\) combinations without any additional summation of these !).
Using the unitary properties of \(\hat{\mscr{U}}_{\hat{F};\ovv{N};\ovv{M}}^{(\alpha;\kappa)}(x_{p})\) in the '11' block,
one straightforwardly derives conditions (\ref{s3_23}) for the transposed '22' block and determines
through diagonalization with \(\hat{\mscr{U}}_{\hat{F};\ovv{N};\ovv{M}}^{(\alpha;\kappa)}(x_{p})\) new density matrices
\(\hat{\mscr{R}}_{M;N}^{(\alpha;\kappa)ab}(x_{p})\). These are modified from \(\hat{R}_{M;N}^{ab}(x_{p})\)
by the block diagonal unitary transformations
\(\hat{\mscr{U}}_{\hat{F};\ovv{M};\ovv{N}\ppr}^{(\alpha;\kappa),aa}(x_{p})\),
\(\hat{\mscr{U}}_{\hat{F};\ovv{M}\ppr;\ovv{N}}^{(\alpha;\kappa),bb,\dag}(x_{p})\) with eigenvalues
\(\hat{\mfrak{v}}_{\hat{F};\ovv{N}}^{(\alpha;\kappa)}(x_{p})\)  which do not depend on the anomalous doubling in contrast to the potentials
\(\hat{\boldsymbol{\mscr{V}}}_{(\alpha);M\ppr;M}^{(\hat{F})a;(\kappa)}(x_{p})\),
\(\hat{\boldsymbol{\mscr{V}}}_{(\alpha);N\ppr;N}^{(\hat{F})b;(\kappa)}(x_{p})\) (\ref{s3_17},\ref{s3_18})
\beq \lb{s3_21}
\lefteqn{\hat{\boldsymbol{\mscr{V}}}_{(\alpha);M\ppr;M}^{(\hat{F})a;(\kappa)}(x_{p})=
\hat{\boldsymbol{\Gamma}}_{\beta;\mu;f\ppr,\ovv{M}\ppr;f,\ovv{M}}^{aa}\;\delta_{f\ppr,f}\;
\hat{\mfrak{B}}_{\hat{F};\beta(\alpha)}^{\mu(\kappa)}(x_{p})= } \\ \no &=&
\left(\bea{cc}
\big[\hat{\beta}(\im \;\hat{\gamma}_{\mu}\;\hat{t}_{\beta}\;
\hat{\mfrak{B}}_{\hat{F};\beta(\alpha)}^{\mu(\kappa)}(x_{p})\,)\big]_{f\ppr,\ovv{M}\ppr;f,\ovv{M}} & 0 \\
0 & \big[\hat{\beta}(\im \;\hat{\gamma}_{\mu}\;\hat{t}_{\beta}\;
\hat{\mfrak{B}}_{\hat{F};\beta(\alpha)}^{\mu(\kappa)}(x_{p})\,)\big]_{f\ppr,\ovv{M}\ppr;f,\ovv{M}}^{T}
\eea\right)_{f\ppr,\ovv{M}\ppr;f,\ovv{M}}^{aa}\delta_{f\ppr,f}\;;
\eeq
\be \lb{s3_22}
\big[\hat{\beta}(\im \;\hat{\gamma}_{\mu}\;\hat{t}_{\beta}\,)\big]_{f\ppr,\ovv{M}\ppr;f,\ovv{M}}\;
\hat{\mfrak{B}}_{\hat{F};\beta(\alpha)}^{\mu(\kappa)}(x_{p})=
\hat{\mscr{U}}_{\hat{F};\ovv{M}\ppr;\ovv{N}}^{(\alpha;\kappa),\dag}(x_{p})\;
\underbrace{\hat{\mfrak{v}}_{\hat{F};\ovv{N}}^{(\alpha;\kappa)}(x_{p})}_{\mbox{\scz real}}\;
\hat{\mscr{U}}_{\hat{F};\ovv{N};\ovv{M}}^{(\alpha;\kappa)}(x_{p})\;\delta_{f\ppr,f}\;;
\ee
\be\lb{s3_23}
\bea{rclrcl}
\hat{\mscr{U}}_{\hat{F};\ovv{N};\ovv{M}}^{(\alpha;\kappa)}(x_{p}) &\rightarrow& \hat{\mscr{U}}_{\hat{F};\ovv{N};\ovv{M}}^{(\alpha;\kappa),ab}(x_{p})\;\delta_{ab} \;; & \hat{1}_{\ovv{N};\ovv{M}} &=&\hat{\mscr{U}}_{\hat{F};\ovv{N};\ovv{M}\ppr}^{(\alpha;\kappa),\dag}(x_{p}) \;\;
\hat{\mscr{U}}_{\hat{F};\ovv{M}\ppr;\ovv{M}}^{(\alpha;\kappa)}(x_{p}) \;;  \\
\hat{\mscr{U}}_{\hat{F};\ovv{N};\ovv{M}}^{(\alpha;\kappa),11}(x_{p}) &:=&\hat{\mscr{U}}_{\hat{F};\ovv{N};\ovv{M}}^{(\alpha;\kappa)}(x_{p})\;;
 &\hspace*{1.0cm}
\hat{\mscr{U}}_{\hat{F};\ovv{N};\ovv{M}}^{(\alpha;\kappa),22}(x_{p}) &:=&\hat{\mscr{U}}_{\hat{F};\ovv{N};\ovv{M}}^{(\alpha;\kappa),*}(x_{p}) \;; \\
\hat{\mscr{U}}_{\hat{F};\ovv{M}\ppr;\ovv{N}}^{(\alpha;\kappa),11,\dag}(x_{p}) &:=&
\hat{\mscr{U}}_{\hat{F};\ovv{M}\ppr;\ovv{N}}^{(\alpha;\kappa),\dag}(x_{p}) \;; &
\hat{\mscr{U}}_{\hat{F};\ovv{M}\ppr;\ovv{N}}^{(\alpha;\kappa),22,\dag}(x_{p}) &:=&
\hat{\mscr{U}}_{\hat{F};\ovv{N};\ovv{M}}^{(\alpha;\kappa),T}(x_{p})  \;;
\eea
\ee
\beq \lb{s3_24}
\hat{\boldsymbol{\mscr{V}}}_{(\alpha);f\ppr,\ovv{M}\ppr;f,\ovv{M}}^{(\hat{F})a;(\kappa)}(x_{p})&=&
\hat{\mscr{U}}_{\hat{F};\ovv{M}\ppr;\ovv{N}}^{(\alpha;\kappa),aa,\dag}(x_{p})\;\hat{\mfrak{v}}_{\hat{F};\ovv{N}}^{(\alpha;\kappa)}(x_{p})\;
\hat{\mscr{U}}_{\hat{F};\ovv{N};\ovv{M}}^{(\alpha;\kappa),aa}(x_{p})  \;\delta_{f\ppr,f} \;; \\ \lb{s3_25}
\hat{\boldsymbol{\mscr{V}}}_{(\alpha);g\ppr,\ovv{N}\ppr;g,\ovv{N}}^{(\hat{F})b;(\kappa)}(x_{p})&=&
\hat{\mscr{U}}_{\hat{F};\ovv{N}\ppr;\ovv{M}}^{(\alpha;\kappa),bb,\dag}(x_{p})\;\hat{\mfrak{v}}_{\hat{F};\ovv{M}}^{(\alpha;\kappa)}(x_{p})\;
\hat{\mscr{U}}_{\hat{F};\ovv{M};\ovv{N}}^{(\alpha;\kappa),bb}(x_{p})  \;\delta_{g\ppr,g}  \;; \\ \lb{s3_26}
\hat{R}_{M;N}^{ab}(x_{p})&\rightarrow&\hat{\mscr{R}}_{f,\ovv{M};g,\ovv{N}}^{(\alpha;\kappa)ab}(x_{p})=
\hat{\mscr{U}}_{\hat{F};\ovv{M};\ovv{N}\ppr}^{(\alpha;\kappa),aa}(x_{p})\;\;\hat{R}_{f,\ovv{N}\ppr;g,\ovv{M}\ppr}^{ab}(x_{p})\;\;
\hat{\mscr{U}}_{\hat{F};\ovv{M}\ppr;\ovv{N}}^{(\alpha;\kappa),bb,\dag}(x_{p}) \;.
\eeq
The transformation (\ref{s3_26}) of the density matrices \(\hat{R}_{M;N}^{ab}(x_{p})\) to
\(\hat{\mscr{R}}_{f,\ovv{M};g,\ovv{N}}^{(\alpha;\kappa)ab}(x_{p})\) with block diagonal, unitary matrices
\(\hat{\mscr{U}}_{\hat{F};\ovv{M};\ovv{N}\ppr}^{(\alpha;\kappa),aa}(x_{p})\),
\(\hat{\mscr{U}}_{\hat{F};\ovv{M}\ppr;\ovv{N}}^{(\alpha;\kappa),bb,\dag}(x_{p})\) (\ref{s3_23})
changes the quartic interaction term of Fermi fields in Eq. (\ref{s3_20})
to relation (\ref{s3_27}). The quartic interaction (\ref{s3_27}) with the trace
\(\mbox{$\mathscr{T\!r}$}_{\scrscr N_{f},\hat{\gamma}_{mn}^{(\mu)},N_{c}}^{\scriptscriptstyle a(=1,2)}\)
over \(\hat{\mscr{R}}_{M;N}^{(\alpha;\kappa)ab}(x_{p})\) differs from (\ref{s3_20}) with \(\hat{R}_{M;N\ppr}^{ab}(x_{p})\) by the
\(8\times(3+1)=32\) eigenvalues \(\hat{\mfrak{v}}_{\hat{F};\ovv{M}}^{(\alpha;\kappa)}(x_{p})\) instead of the anomalous indexed potentials
\(\hat{\boldsymbol{\mscr{V}}}_{(\alpha);f\ppr,\ovv{M}\ppr;f,\ovv{M}}^{(\hat{F})a;(\kappa)}(x_{p})\) (\ref{s3_17},\ref{s3_18},\ref{s3_21}-\ref{s3_26}).
The transformation of \(\hat{R}_{M;N}^{ab}(x_{p})\) to \(8\times(3+1)=32\) density matrices \(\hat{\mscr{R}}_{M;N}^{(\alpha;\kappa)ab}(x_{p})\)
also underlines the \(32\) different diagonalized 'two-body' potentials
\([-\im\:\hat{\mfrak{e}}_{p}^{(\hat{F})}+\hat{\mfrak{b}}_{(\alpha;\kappa)}^{(\hat{F})}(x_{p})]^{\boldsymbol{-1}}\)
which cause nontrivial couplings in the anomalous doubled, internal space with index \(a,b=1,2\) (\ref{s3_20}). The unitary transformation
of density matrices gives rise to a change of integration measure with
\(\mbox{U}(\,(N_{c}^{2}-1=8)\times(3+1))=\mbox{U}(32)\) which is not determined in detail
because it is absorbed into the background generating function of gauge fields \cite{jmp1}. An additional delta-function (\ref{s3_28}) in the
measure of the field strength self-energy \(d[\hat{\mfrak{S}}_{\alpha;\mu\nu}^{(\hat{F})}(x_{p})]\) guarantees the equivalence of the two
quartic interactions (\ref{s3_20}) and (\ref{s3_27}) with the transformation (\ref{s3_26}) to the \(8\times(3+1)=32\) density matrices
\(\hat{\mscr{R}}_{M;N}^{(\alpha;\kappa)ab}(x_{p})\)
\beq\lb{s3_27}
\lefteqn{\exp\Bigg\{\frac{\im}{8}\int_{C}d^{4}\!x_{p}\bigg(\Psi_{N\ppr}^{\dag,b}(x_{p})\;
\hat{\boldsymbol{\Gamma}}_{\gamma;\nu;N\ppr;N}^{bb}\;\hat{S}^{bb}\;\Psi_{N}^{b}(x_{p})\bigg)
\Big[-\im\,\hat{\mfrak{e}}_{p}^{(\hat{F})}+C_{\alpha\beta\ppr\gamma\ppr}\;
\hat{\mfrak{S}}_{\alpha}^{(\hat{F})\mu\ppr\nu\ppr}(x_{p})\Big]_{\gamma\beta}^{\boldsymbol{-1};\nu\mu}\times }  \\ \no &\times&
\bigg(\Psi_{M\ppr}^{\dag,a}(x_{p})\;\hat{\boldsymbol{\Gamma}}_{\beta;\mu;M\ppr;M}^{aa}\;\hat{S}^{aa}\;
\Psi_{M}^{a}(x_{p})\bigg)\bigg\}=
\exp\bigg\{-\frac{\im}{8}\int_{C}d^{4}\!x_{p}\;\sum_{\scrscr(\alpha)=1,.,8}^{\scrscr(\kappa)=0,.,3}
\Big[-\im\,\hat{\mfrak{e}}_{p}^{(\hat{F})}+\hat{\mfrak{b}}^{(\hat{F})}_{(\alpha\;\kappa)}(x_{p})\Big]^{\boldsymbol{-1}}\times \\ \no &\times&
\TRALL\bigg[\hat{\mscr{R}}_{f,\ovv{M};g,\ovv{N}}^{(\alpha;\kappa)ab}(x_{p})\;
\hat{\mfrak{v}}_{\hat{F};\ovv{N}}^{(\alpha;\kappa)}(x_{p})\;\hat{S}^{bb}\;
\hat{\mscr{R}}_{g,\ovv{N};f,\ovv{M}}^{(\alpha;\kappa)ba}(x_{p})\;
\hat{\mfrak{v}}_{\hat{F};\ovv{M}}^{(\alpha;\kappa)}(x_{p})\;\hat{S}^{aa}\bigg]\Bigg\}\;;    \\  \lb{s3_28}
\lefteqn{\mbox{\bf Eq. (\ref{s3_12})} \boldsymbol{\rightarrow} d[\hat{\mfrak{S}}_{\alpha;\mu\nu}^{(\hat{F})}(x_{p})]\;\;
d[\hat{\mfrak{B}}_{\hat{F};\beta(\alpha)}^{\mu(\kappa)}(x_{p});\hat{\mfrak{b}}^{(\hat{F})}_{(\alpha;\kappa)}(x_{p})]\;\;
d[\hat{\mscr{U}}_{\hat{F};\ovv{N};\ovv{M}}^{(\alpha;\kappa)}(x_{p});\hat{\mfrak{v}}_{\hat{F};\ovv{N}}^{(\alpha;\kappa)}(x_{p})]
\;\times}  \\  \no &\times&
\bigg\{\prod_{\scrscr\{x_{p}\}}\delta\bigg(C_{\alpha\beta\gamma}\;\hat{\mfrak{S}}_{\alpha}^{(\hat{F})\mu\nu}(x_{p}) -
\sum_{\scrscr(\alpha)=1,.,8}^{\scrscr(\kappa)=0,.,3}
\hat{\mfrak{B}}_{\hat{F};\beta(\alpha)}^{\mu(\kappa)}(x_{p})\;
\hat{\mfrak{b}}^{(\hat{F})}_{(\alpha;\kappa)}(x_{p})\;
\hat{\mfrak{B}}_{\hat{F};(\alpha)\gamma}^{T,(\kappa)\nu}(x_{p}) \bigg)\bigg\} \times \\ \no &\times&
\bigg\{\prod_{\scrscr\{x_{p};(\alpha)=1,.,8\}}^{\scrscr\{(\kappa)=0,.,3\}}
\delta\bigg(\hat{\mscr{U}}_{\hat{F};\ovv{M}\ppr;\ovv{N}}^{(\alpha;\kappa),11,\dag}(x_{p})\;
\hat{\mfrak{v}}_{\hat{F};\ovv{N}}^{(\alpha;\kappa)}(x_{p})\;
\hat{\mscr{U}}_{\hat{F};\ovv{N};\ovv{M}}^{(\alpha;\kappa),11}(x_{p})-
\big[\hat{\beta}\,(\im \;\hat{\gamma}_{\mu}\;\hat{t}_{\beta})\big]_{\ovv{M}\ppr;\ovv{M}}\;
\hat{\mfrak{B}}_{\hat{F};\beta(\alpha)}^{\mu(\kappa)}(x_{p})\bigg)\bigg\} \;.
\eeq
Relations (\ref{s3_27},\ref{s3_28}) of the anomalous doubled, quartic interaction of Fermi fields allow to
apply \(8\times(3+1)=32\) HST's by introducing \(32\) independent, different self-energies
\(\delta\hat{\Sigma}_{M;N}^{(\alpha;\kappa)ab}(x_{p})\) (\ref{s3_29}-\ref{s3_33}) for the matter fields in analogy
to the super-symmetric case of Ref. \cite{mies1}. The anti-hermitian, anti-symmetric BCS related self-energy parts
\(\im\:\delta\hat{\Sigma}_{M;N}^{(\alpha;\kappa)a\neq b}(x_{p})\) are denoted by a tilde '\(\wt{\ph{\Sigma}}\)' over the total self-energy symbol
\(\delta\wt{\Sigma}_{M;N}^{(\alpha;\kappa)ab}(x_{p})\) of the quark matter fields.
The anti-symmetry of \(\delta\wt{\Sigma}_{M;N}^{(\alpha;\kappa)a\neq b}(x_{p})=
\im\:\delta\hat{\Sigma}_{M;N}^{(\alpha;\kappa)a\neq b}(x_{p})\) regards
the BCS pairing of quarks whereas the anti-hermiticity takes into
account a proper coset parametrization with the appropriate number of independent field degrees of
freedom; these have to coincide with the number of independent parameters
for the transformations leaving the path integral (\ref{s3_8},\ref{s3_9}) invariant
\beq \lb{s3_29}
\delta\wt{\Sigma}_{M;N}^{(\alpha;\kappa)ab}(x_{p}) &=&
\left(\bea{cc} \delta\hat{\Sigma}_{M;N}^{(\alpha;\kappa)11}(x_{p}) &
\im\;\delta\hat{\Sigma}_{M;N}^{(\alpha;\kappa)12}(x_{p})    \\
\im\;\delta\hat{\Sigma}_{M;N}^{(\alpha;\kappa)21}(x_{p})   &
\delta\hat{\Sigma}_{M;N}^{(\alpha;\kappa)22}(x_{p})  \eea\right)_{M;N}^{ab} \;; \\   \lb{s3_30}
\delta\hat{\Sigma}_{M;N}^{(\alpha;\kappa)12}(x_{p})
&=&-\Big(\delta\hat{\Sigma}_{M;N}^{(\alpha;\kappa)12}(x_{p})\Big)^{T} \;; \hspace*{0.6cm}
\delta\hat{\Sigma}_{M;N}^{(\alpha;\kappa)21}(x_{p}) =-\Big(\delta\hat{\Sigma}_{M;N}^{(\alpha;\kappa)21}(x_{p})\Big)^{T} \;; \\  \lb{s3_31}
\delta\hat{\Sigma}_{M;N}^{(\alpha;\kappa)21}(x_{p}) &=&\Big(\delta\hat{\Sigma}_{M;N}^{(\alpha;\kappa)12}(x_{p})\Big)^{\dag} \;;
\\  \lb{s3_32} \Big(\delta\hat{\Sigma}_{M;N}^{(\alpha;\kappa)11}(x_{p})\Big)^{\dag}&=&
\delta\hat{\Sigma}_{M;N}^{(\alpha;\kappa)11}(x_{p}) \;; \hspace*{0.6cm}
\Big(\delta\hat{\Sigma}_{M;N}^{(\alpha;\kappa)22}(x_{p})\Big)^{\dag}=\delta\hat{\Sigma}_{M;N}^{(\alpha;\kappa)22}(x_{p}) \;; \\  \lb{s3_33}
\delta\hat{\Sigma}_{M;N}^{(\alpha;\kappa)22}(x_{p})
&=&-\Big(\delta\hat{\Sigma}_{M;N}^{(\alpha;\kappa)11}(x_{p}) \Big)^{T} \;;     \\ \lb{s3_34}
\hat{\Sigma}_{M;N}^{(\alpha;\kappa)11}(x_{p}) &=& \sigma_{D}^{(\alpha;\kappa)}(x_{p})\;
\hat{\mfrak{v}}_{\hat{F};\ovv{M}}^{(\alpha;\kappa)}(x_{p})\;
\delta_{M;N}+\delta\hat{\Sigma}_{M;N}^{(\alpha;\kappa)11}(x_{p}) \;;  \\   \lb{s3_35}
\hat{\Sigma}_{M;N}^{(\alpha;\kappa)22}(x_{p}) &=&
\sigma_{D}^{(\alpha;\kappa)}(x_{p})\;\hat{\mfrak{v}}_{\hat{F};\ovv{M}}^{(\alpha;\kappa)}(x_{p})\;
\delta_{M;N}+\delta\hat{\Sigma}_{M;N}^{(\alpha;\kappa)22}(x_{p}) \;.
\eeq
According to Ref. \cite{mies1}, we catalogue the analogous HST's in (\ref{s3_36},\ref{s3_37}) and briefly describe how half of
the original, anomalous doubled, quartic interaction of fields \(\Psi_{M}^{a}(x_{p})\), \(\Psi_{M}^{\dag,a}(x_{p})\) is transformed
by the off-diagonal blocks \(a\neq b\) or BCS related self-energies \(\delta\wt{\Sigma}_{M;N}^{(\alpha;\kappa)a\neq b}(x_{p})=\im\:
\delta\hat{\Sigma}_{M;N}^{(\alpha;\kappa)a\neq b}(x_{p})\)\footnote{Note the boldface typing
\(\boldsymbol{\im\:\delta\hat{\Sigma}_{M;N}^{(\alpha;\kappa)12}(x_{p})}\),
\(\boldsymbol{\im\:\delta\hat{\Sigma}_{M;N}^{(\alpha;\kappa)21}(x_{p})}\)
in (\ref{s3_36}) !} and how
the other half (\ref{s3_37}) is transformed by including \(8\times (3+1)=32\) real, scalar, diagonal self-energy fields
\(\sigma_{D}^{(\alpha;\kappa)}(x_{p})\) for the quark densities (\ref{s3_34},\ref{s3_35}).
Moreover, we emphasize the anti-hermitian kind of the \(32\)
HST's with \(\im\:\delta\hat{\Sigma}_{M;N}^{(\alpha;\kappa)12}(x_{p})\),  \(\im\:\delta\hat{\Sigma}_{M;N}^{(\alpha;\kappa)21}(x_{p})\)
so that one achieves a suitable coset decomposition with the identical number of independent parameter fields as in the
invariant transformations of the path integral (\ref{s3_8},\ref{s3_9}).
Corresponding to Ref. \cite{mies1}, we perform the HST's of the \(32\) density matrices
\(\hat{\mscr{R}}_{N;M}^{(\alpha;\kappa)a\neq b}(x_{p})\) for half of the quartic interaction\footnote{Pre-factor \(\im/16\) instead
of \(\im/8\) as in (\ref{s3_27}) for the 'half' of the quartic interaction of Fermi fields !} by 'flat', Euclidean Gaussian
integrals of \(32\) self-energies \(\delta\hat{\Sigma}_{M;N}^{(\alpha;\kappa)a\neq b}(x_{p})\) in the off-diagonal, BCS related
sector of the anomalous doubled quark fields
\beq \lb{s3_36}
\lefteqn{\exp\bigg\{\frac{\im}{16}\int_{C}d^{4}\!x_{p}\;
\Big(\Psi_{N\ppr}^{\dag,b}(x_{p})\;\hat{\boldsymbol{\Gamma}}_{\gamma;\nu;N\ppr;N}^{bb}\;\hat{S}^{bb}\;\Psi_{N}^{b}(x_{p})\Big)\;
\Big[-\im\,\hat{\mfrak{e}}_{p}^{(\hat{F})}+C_{\alpha\beta\ppr\gamma\ppr}\;
\hat{\mfrak{S}}_{\alpha}^{(\hat{F})\mu\ppr\nu\ppr}(x_{p})\Big]_{\gamma\beta}^{\boldsymbol{-1};\nu\mu} \times } \\  \no &\times&
\Big(\Psi_{M\ppr}^{\dag,a}(x_{p})\;\hat{\boldsymbol{\Gamma}}_{\beta;\mu;M\ppr;M}^{aa}\;\hat{S}^{aa}\;\Psi_{M}^{a}(x_{p})\Big)\bigg\}=
\int d[\delta\hat{\Sigma}_{M;N}^{(\alpha;\kappa)12}(x_{p});\delta\hat{\Sigma}_{M;N}^{(\alpha;\kappa)21}(x_{p})]\;\times  \\ \no &\times&
\mfrak{P}_{1}\Big(\hat{\mfrak{b}}^{(\hat{F})}_{(\alpha;\kappa)}(x_{p})\Big)\;
\mfrak{Q}_{1}\Big(\hat{\mfrak{v}}^{(\alpha;\kappa)}_{\hat{F};\ovv{M}}(x_{p})\Big)\;
\exp\Bigg\{\frac{\im}{8}\int_{C}d^{4}\!x_{p}\sum_{\scrscr(\alpha)=1,.,8}^{\scrscr(\kappa)=0,.,3}
\Big[-\im\,\hat{\mfrak{e}}_{p}^{(\hat{F})}+\hat{\mfrak{b}}^{(\hat{F})}_{(\alpha;\kappa)}(x_{p})\Big]\;\times  \\ \no &\times&
\TRALL\Bigg[\left(\boldsymbol{\delta\hat{\Sigma}_{f,\ovv{M};g,\ovv{N}}^{(\alpha;\kappa)a\neq b}(x_{p})} \right)
\frac{\left(\bea{cc} 1 & \\ & \boldsymbol{-1} \eea\right)^{bb}}{\hat{\mfrak{v}}_{\hat{F};\ovv{N}}^{(\alpha;\kappa)}(x_{p})}
\left(\boldsymbol{\delta\hat{\Sigma}_{g,\ovv{N};f,\ovv{M}}^{(\alpha;\kappa)b\neq a}(x_{p})} \right)
\frac{\left(\bea{cc} 1 & \\ & \boldsymbol{-1} \eea\right)^{aa}}{\hat{\mfrak{v}}_{\hat{F};\ovv{M}}^{(\alpha;\kappa)}(x_{p})}\Bigg]\Bigg\}
\times \\ \no  &\times&
\exp\Bigg\{\frac{\im}{4}\int_{C}d^{4}\!x_{p}\sum_{\scrscr(\alpha)=1,.,8}^{\scrscr(\kappa)=0,.,3}
\TRALL\Bigg[\left(\delta\hat{\Sigma}_{M;N}^{(\alpha;\kappa)\boldsymbol{a\neq b}}(x_{p})\right)\bigg(\bea{cc} 1 & \\ & -1 \eea\bigg)^{bb}
\left(\hat{\mscr{R}}_{N;M}^{(\alpha;\kappa)\boldsymbol{b\neq a}}(x_{p}) \right)
\bigg(\bea{cc} 1 & \\ & -1 \eea\bigg)^{aa}\Bigg]\Bigg\} =   \\  \no &=&
\int d[\delta\hat{\Sigma}_{M;N}^{(\alpha;\kappa)12}(x_{p});\delta\hat{\Sigma}_{M;N}^{(\alpha;\kappa)21}(x_{p})]\;\;\;
\mfrak{P}_{1}\Big(\hat{\mfrak{b}}^{(\hat{F})}_{(\alpha;\kappa)}(x_{p})\Big)\;
\mfrak{Q}_{1}\Big(\hat{\mfrak{v}}^{(\alpha;\kappa)}_{\hat{F};\ovv{M}}(x_{p})\Big)\;\times \\ \no &\times&
\exp\Bigg\{\frac{\im}{8}\int_{C}d^{4}\!x_{p}\sum_{\scrscr(\alpha)=1,.,8}^{\scrscr(\kappa)=0,.,3}
\Big[-\im\,\hat{\mfrak{e}}_{p}^{(\hat{F})}+\hat{\mfrak{b}}^{(\hat{F})}_{(\alpha;\kappa)}(x_{p})\Big]  \times \\ \no &\times&
\TRALL\Bigg[\left(\boldsymbol{\im\;\delta\hat{\Sigma}_{f,\ovv{M};g,\ovv{N}}^{(\alpha;\kappa)a\neq b}(x_{p})}\right)
\frac{\Bigg(\bea{cc} 1 & \\ & \boldsymbol{+1}\eea\Bigg)^{bb}}{\hat{\mfrak{v}}_{\hat{F};\ovv{N}}^{(\alpha;\kappa)}(x_{p})}
\left(\boldsymbol{\im\;\delta\hat{\Sigma}_{g,\ovv{N};f,\ovv{M}}^{(\alpha;\kappa)b\neq a}(x_{p})}\right)
\frac{\Bigg(\bea{cc} 1 & \\ & \boldsymbol{+1}\eea\Bigg)^{aa}}{\hat{\mfrak{v}}_{\hat{F};\ovv{M}}^{(\alpha;\kappa)}(x_{p})}
\Bigg]\Bigg\}  \times   \\  \no &\times&
\exp\Bigg\{\frac{\im}{4}\int_{C}d^{4}\!x_{p}\sum_{\scrscr(\alpha)=1,.,8}^{\scrscr(\kappa)=0,.,3}\hspace*{0.6cm}
\TRALL\Bigg[\left(\delta\hat{\Sigma}_{M;N}^{(\alpha;\kappa)\boldsymbol{a\neq b}}(x_{p})\right)
\bigg(\bea{cc} 1 & \\ & -1 \eea\bigg)^{bb}
\left(\hat{\mscr{R}}_{N;M}^{(\alpha;\kappa)\boldsymbol{b\neq a}}(x_{p})\right)
\bigg(\bea{cc} 1 & \\ & -1 \eea\bigg)^{aa}\Bigg]\Bigg\} \;.
\eeq
The diagonalized, two-body potential part
\([-\im\:\hat{\mfrak{e}}_{p}^{(\hat{F})}+\hat{\mfrak{b}}_{(\alpha;\kappa)}^{(\hat{F})}(x_{p})]\)
and the eigenvalues \(\hat{\mfrak{v}}_{\hat{F};\ovv{M}}^{(\alpha;\kappa)}(x_{p})\), which replace the
anomalous indexed potentials \(\hat{\boldsymbol{\mscr{V}}}_{(\alpha);f\ppr,\ovv{M}\ppr;f,\ovv{M}}^{(\hat{F})a;(\kappa)}(x_{p})\) (\ref{s3_17},\ref{s3_18}),
have the effect of a kind of 'variance' in the Gaussian integrations of the BCS related quark matter self-energy terms so that the
polynomials \(\mfrak{P}_{1}(\,\hat{\mfrak{b}}^{(\hat{F})}_{(\alpha;\kappa)}(x_{p})\,)\),
\(\mfrak{Q}_{1}(\,\hat{\mfrak{v}}^{(\alpha;\kappa)}_{\hat{F};\ovv{M}}(x_{p})\,)\) have to be incorporated
in the Gaussian integrations of the HST's for a proper normalization.

The other half of the HST's for the quartic interaction\footnote{Note again the pre-factor \(\im/16\) instead
of \(\im/8\) as in (\ref{s3_27}) for the 'half' of the quartic interaction of Fermi fields !}
with the original dyadic product of anomalous doubled Fermi fields is obtained
by the independent, \(32\), real, scalar self-energies \(\sigma_{D}^{(\alpha;\kappa)}(x_{p})\) (\ref{s3_34},\ref{s3_35})
which are related to quark densities.
Since the Gaussian integrals with \(\sigma_{D}^{(\alpha;\kappa)}(x_{p})\) are also weighted by the variance of the
two-body related potential \([-\im\:\hat{\mfrak{e}}_{p}^{(\hat{F})}+\hat{\mfrak{b}}_{(\alpha;\kappa)}^{(\hat{F})}(x_{p})]\),
we have to include an additional polynomial \(\mfrak{P}_{2}(\,\hat{\mfrak{b}}^{(\hat{F})}_{(\alpha;\kappa)}(x_{p})\,)\)
for the normalization of the \(32\) HST's
\beq \lb{s3_37}
\lefteqn{\exp\bigg\{\frac{\im}{16}\int_{C}d^{4}\!x_{p}\;
\Big(\Psi_{N\ppr}^{\dag,b}(x_{p})\;\hat{\boldsymbol{\Gamma}}_{\gamma;\nu;N\ppr;N}^{bb}\;\hat{S}^{bb}\;\Psi_{N}^{b}(x_{p})\Big)\;
\Big[-\im\,\hat{\mfrak{e}}_{p}^{(\hat{F})}+C_{\alpha\beta\ppr\gamma\ppr}\;
\hat{\mfrak{S}}_{\alpha}^{(\hat{F})\mu\ppr\nu\ppr}(x_{p})\Big]_{\gamma\beta}^{\boldsymbol{-1};\nu\mu} \times } \\ \no &&
\Big(\Psi_{M\ppr}^{\dag,a}(x_{p})\;\hat{\boldsymbol{\Gamma}}_{\beta;\mu;M\ppr;M}^{aa}\;\kappa^{aa}\;\Psi_{M}^{a}(x_{p})\Big)\bigg\}=  \\ \no &=&
\int d[\sigma_{D}^{(\alpha;\kappa)}(x_{p})]\;\;\mfrak{P}_{2}\Big(\hat{\mfrak{b}}^{(\hat{F})}_{(\alpha;\kappa)}(x_{p})\Big)\;
\exp\bigg\{\frac{\im}{4}\int_{C}d^{4}\!x_{p}\sum_{\scrscr(\alpha)=1,.,8}^{\scrscr(\kappa)=0,.,3}
\Big[\im\,\hat{\mfrak{e}}_{p}^{(\hat{F})}+\hat{\mfrak{b}}^{(\hat{F})}_{(\alpha;\kappa)}(x_{p})\Big]\;
\Big(\sigma_{D}^{(\alpha;\kappa)}(x_{p})\Big)^{2}\bigg\}\times  \\ \no &\times&
\exp\Bigg\{\frac{\im}{4}\int_{C}d^{4}\!x_{p}\sum_{\scrscr(\alpha)=1,.,8}^{\scrscr(\kappa)=0,.,3}\hspace*{0.6cm}
\TRALL\Bigg[\bigg(\bea{cc} \hat{\mscr{R}}_{g,\ovv{N};f,\ovv{M}}^{(\alpha;\kappa)11}(x_{p}) & 0 \\
0 & \hat{\mscr{R}}_{g,\ovv{N};f,\ovv{M}}^{(\alpha;\kappa)22}(x_{p}) \eea\bigg)\bigg(\bea{cc} 1 &  \\  & -1 \eea\bigg)\;\times \\ \no &\times&
\bigg(\bea{cc} \sigma_{D}^{(\alpha;\kappa)}(x_{p})\;\hat{\mfrak{v}}_{\hat{F};\ovv{M}}^{(\alpha;\kappa)}(x_{p})\;\delta_{\ovv{M};\ovv{N}}
\;\delta_{f,g} & 0 \\
0 &  -\sigma_{D}^{(\alpha;\kappa)}(x_{p})\;\hat{\mfrak{v}}_{\hat{F};\ovv{M}}^{(\alpha;\kappa)}(x_{p})\;\delta_{\ovv{M};\ovv{N}} \;
\delta_{f,g}\eea\bigg)
\bigg(\bea{cc} 1 & \\ & -1 \eea\bigg)\Bigg]\Bigg\} \;.
\eeq
Moreover, one has to introduce 'hinge' fields which allow for the coset decomposition
\(\mbox{SO}(N_{0}\:\boldsymbol{,}\:N_{0})/\mbox{U}(N_{0})\otimes \mbox{U}(N_{0})\), (\(N_{0}=N_{f}\times 4_{\hat{\gamma}}\times N_{c}\))
as the subgroup part in the spontaneous symmetry breaking to BCS related
pair condensates of quark fields. (QCD-type case with up-, down-isospins yields
\(\mbox{SO}(24\,\boldsymbol{,}\,24)/\mbox{U}(24)\otimes\mbox{U}(24)\) and with the inclusion of strangeness
\(\mbox{SO}(36\,\boldsymbol{,}\,36)/\mbox{U}(36)\otimes\mbox{U}(36)\) .)
In compliance with Ref. \cite{mies1}, we therefore take into account
additional Gaussian integrations of self-energy densities \(\delta\hat{\Sigma}_{M;N}^{(\alpha;\kappa)11}(x_{p})\),
\(\delta\hat{\Sigma}_{M;N}^{(\alpha;\kappa)22}(x_{p})\) (\ref{s3_32},\ref{s3_33}) which are normalized to unity by the polynomials
\(\mfrak{P}_{3}(\,\hat{\mfrak{b}}^{(\hat{F})}_{(\alpha;\kappa)}(x_{p})\,)\),
\(\mfrak{Q}_{3}(\,\hat{\mfrak{v}}^{(\alpha;\kappa)}_{\hat{F};\ovv{M}}(x_{p})\,)\).
This normalization to unity is caused by the additional minus sign \(\boldsymbol{-}\delta\hat{\Sigma}_{M;N}^{(\alpha;\kappa)22}(x_{p})\)
in the '22' block density part which can be transformed to a non-vanishing '+' part
\(\boldsymbol{+}\delta\hat{\Sigma}_{M;N}^{(\alpha;\kappa)22}(x_{p})\), but then induces the required anti-hermitian
BCS related parts in the off-diagonal '12', '21' blocks for the HST of half of the quartic interaction (\ref{s3_36})
\beq \lb{s3_38}
\lefteqn{1\equiv
\int d[\delta\hat{\Sigma}_{M;N}^{(\alpha;\kappa)11}(x_{p});\delta\hat{\Sigma}_{M;N}^{(\alpha;\kappa)22}(x_{p})]\;\;
\mfrak{P}_{3}\Big(\hat{\mfrak{b}}^{(\hat{F})}_{(\alpha;\kappa)}(x_{p})\Big)\;
\mfrak{Q}_{3}\Big(\hat{\mfrak{v}}^{(\alpha;\kappa)}_{\hat{F};\ovv{M}}(x_{p})\Big)\;
\exp\Bigg\{\frac{\im}{8}\int_{C}d^{4}\!x_{p}\sum_{\scrscr(\alpha)=1,.,8}^{\scrscr(\kappa)=0,.,3} \times  }   \\ \no &\times&
\Big[-\im\,\hat{\mfrak{e}}_{p}^{(\hat{F})}+\hat{\mfrak{b}}^{(\hat{F})}_{(\alpha;\kappa)}(x_{p})\Big]\,
\hspace*{0.7cm}\TRALL\Bigg[
\left(\delta\hat{\Sigma}_{f,\ovv{M};g,\ovv{N}}^{(\alpha;\kappa)\boldsymbol{aa}}(x_{p})\right)
\frac{1}{\hat{\mfrak{v}}_{\hat{F};\ovv{N}}^{(\alpha;\kappa)}(x_{p})}
\left(\delta\hat{\Sigma}_{g,\ovv{N};f,\ovv{M}}^{(\alpha;\kappa)\boldsymbol{aa}}(x_{p})\right)
\frac{1}{\hat{\mfrak{v}}_{\hat{F};\ovv{M}}^{(\alpha;\kappa)}(x_{p})}\Bigg]\Bigg\}  \times   \\  \no &\times&
\exp\Bigg\{\frac{\im}{4}\int_{C}d^{4}\!x_{p}\sum_{\scrscr(\alpha)=1,.,8}^{\scrscr(\kappa)=0,.,3}
\hspace*{0.7cm}\TRALL\Bigg[\left(\bea{cc} \hat{\mscr{R}}_{N;M}^{(\alpha;\kappa)11}(x_{p}) &  0  \\
0  & \hat{\mscr{R}}_{N;M}^{(\alpha;\kappa)22}(x_{p}) \eea \right)
\bigg(\bea{cc} 1 & \\ & -1 \eea\bigg)\; \times  \\ \no &\times&
\left(\bea{cc} \delta\hat{\Sigma}_{M;N}^{(\alpha;\kappa)11}(x_{p}) & 0  \\
0  &  \boldsymbol{-\delta\hat{\Sigma}_{M;N}^{(\alpha;\kappa)22}(x_{p})} \eea \right)
\bigg(\bea{cc} 1 & \\ & -1 \eea\bigg)\Bigg]\Bigg\}_{\mbox{.}}
\eeq
The combination of all subsequent parts eventually yields the entire HST (\ref{s3_39}) to self-energies
of the quark matter fields
\beq  \lb{s3_39}
\lefteqn{\exp\bigg\{\frac{\im}{8}\int_{C}d^{4}\!x_{p}\;
\Big(\Psi_{N\ppr}^{\dag,b}(x_{p})\;\hat{\boldsymbol{\Gamma}}_{\gamma;\nu;N\ppr;N}^{bb}\;\hat{S}^{bb}\;\Psi_{N}^{b}(x_{p})\Big)\;
\Big[-\im\,\hat{\mfrak{e}}_{p}^{(\hat{F})}+C_{\alpha\beta\ppr\gamma\ppr}\;
\hat{\mfrak{S}}_{\alpha}^{(\hat{F})\mu\ppr\nu\ppr}(x_{p})\Big]_{\gamma\beta}^{\boldsymbol{-1};\nu\mu} \times } \\ \no &&
\Big(\Psi_{M\ppr}^{\dag,a}(x_{p})\;\hat{\boldsymbol{\Gamma}}_{\beta;\mu;M\ppr;M}^{aa}\;\kappa^{aa}\;\Psi_{M}^{a}(x_{p})\Big)\bigg\}=  \\ \no &=&
\int d[\sigma_{D}^{(\alpha;\kappa)}(x_{p})]\;\;\mfrak{P}_{2}\Big(\hat{\mfrak{b}}^{(\hat{F})}_{(\alpha;\kappa)}(x_{p})\Big)\;
\exp\bigg\{\frac{\im}{4}\int_{C}d^{4}\!x_{p}\sum_{\scrscr(\alpha)=1,.,8}^{\scrscr(\kappa)=0,.,3}
\Big[\im\,\hat{\mfrak{e}}_{p}^{(\hat{F})}+\hat{\mfrak{b}}^{(\hat{F})}_{(\alpha;\kappa)}(x_{p})\Big]\;
\Big(\sigma_{D}^{(\alpha;\kappa)}(x_{p})\Big)^{2}\bigg\}\times  \\ \no &\times&
\int d[\delta\wt{\Sigma}_{M;N}^{(\alpha;\kappa)ab}(x_{p})]\;\;\mfrak{P}_{1}\Big(\hat{\mfrak{b}}^{(\hat{F})}_{(\alpha;\kappa)}(x_{p})\Big)\;
\mfrak{P}_{3}\Big(\hat{\mfrak{b}}^{(\hat{F})}_{(\alpha;\kappa)}(x_{p})\Big)\;
\;\mfrak{Q}_{1}\Big(\hat{\mfrak{v}}^{(\alpha;\kappa)}_{\hat{F};\ovv{M}}(x_{p})\Big)\;
\mfrak{Q}_{3}\Big(\hat{\mfrak{v}}^{(\alpha;\kappa)}_{\hat{F};\ovv{M}}(x_{p})\Big)\;\times   \\  \no &\times&
\exp\Bigg\{\frac{\im}{8}\int_{C}d^{4}\!x_{p}\sum_{\scrscr(\alpha)=1,.,8}^{\scrscr(\kappa)=0,.,3}
\Big[-\im\,\hat{\mfrak{e}}_{p}^{(\hat{F})}+\hat{\mfrak{b}}^{(\hat{F})}_{(\alpha;\kappa)}(x_{p})\Big]\; \times \\   \no &\times&
\TRALL\bigg[\delta\wt{\Sigma}_{f,\ovv{M};g,\ovv{N}}^{(\alpha;\kappa)ab}(x_{p})\;
\frac{1}{\hat{\mfrak{v}}_{\hat{F};\ovv{N}}^{(\alpha;\kappa)}(x_{p})}\;
\delta\wt{\Sigma}_{g,\ovv{N};f,\ovv{M}}^{(\alpha;\kappa)ba}(x_{p})\;
\frac{1}{\hat{\mfrak{v}}_{\hat{F};\ovv{M}}^{(\alpha;\kappa)}(x_{p})}\bigg]\Bigg\}
\exp\Bigg\{\frac{\im}{4}\int_{C}d^{4}\!x_{p}\sum_{\scrscr(\alpha)=1,.,8}^{\scrscr(\kappa)=0,.,3}\times
\\ \no &\hspace*{-1.2cm}\times& \hspace*{-0.3cm}
\TRALL\Bigg[\bigg(\bea{cc} \hat{\mscr{R}}_{N;M}^{(\alpha;\kappa)11}(x_{p})  &
\hat{\mscr{R}}_{N;M}^{(\alpha;\kappa)12}(x_{p})  \\   \hat{\mscr{R}}_{N;M}^{(\alpha;\kappa)21}(x_{p})  &
\hat{\mscr{R}}_{N;M}^{(\alpha;\kappa)22}(x_{p})  \eea\bigg)\bigg(\bea{cc} 1 & \\ & -1 \eea\bigg)
\bigg(\bea{cc} \hat{\Sigma}_{M;N}^{(\alpha;\kappa)11}(x_{p})  &
\delta\hat{\Sigma}_{M;N}^{(\alpha;\kappa)12}(x_{p})  \\   \delta\hat{\Sigma}_{M;N}^{(\alpha;\kappa)21}(x_{p})  &
-\hat{\Sigma}_{M;N}^{(\alpha;\kappa)22}(x_{p})  \eea\bigg)\bigg(\bea{cc} 1 & \\ & -1 \eea\bigg)\Bigg]\Bigg\}_{\mbox{.}}
\eeq
The entire HST (\ref{s3_39}) for the quartic interaction of fermionic fields can be inserted into the generating function (\ref{s3_8},\ref{s3_9})
\(Z[\hat{\mscr{J}},J_{\psi},\hat{J}_{\psi\psi},\hat{\mfrak{j}}^{(\hat{F})}]\) so that we obtain relation (\ref{s3_40})
which depends on the \(32\) independent self-energies \(\delta\wt{\Sigma}_{M;N}^{(\alpha;\kappa)ab}(x_{p})\) of quark matter fields
(also with anomalous pairing), on the \(32\) independent, real, diagonal,
scalar self-energy density fields \(\sigma_{D}^{(\alpha;\kappa)}(x_{p})\) and remaining
Gaussian integrations of anti-commuting variables \(d[\psi_{M}^{\dag}(x_{p}),\,\psi_{M}(x_{p})]\)
\beq  \lb{s3_40}
\lefteqn{Z[\hat{\mscr{J}},J_{\psi},\hat{J}_{\psi\psi},\hat{\mfrak{j}}^{(\hat{F})}]=
\boldsymbol{\Bigg\langle Z\Big[\hat{\mfrak{S}}^{(\hat{F})},s_{\alpha},\hat{\mfrak{B}}_{\hat{F}},\hat{\mfrak{b}}^{(\hat{F})},
\hat{\mscr{U}}_{\hat{F}},\hat{\mfrak{v}}_{\hat{F}};\hat{\mfrak{j}}^{(\hat{F})};
\mbox{\bf Eq. (\ref{s3_42})}\Big]\; } \times} \\ \no &\times&
\int d[\sigma_{D}^{(\alpha;\kappa)}(x_{p})]\;
\exp\bigg\{\frac{\im}{4}\int_{C}d^{4}\!x_{p}\sum_{\scrscr(\alpha)=1,.,8}^{\scrscr(\kappa)=0,.,3}
\Big[\im\,\hat{\mfrak{e}}_{p}^{(\hat{F})}+\hat{\mfrak{b}}^{(\hat{F})}_{(\alpha;\kappa)}(x_{p})\Big]\;
\Big(\sigma_{D}^{(\alpha;\kappa)}(x_{p})\Big)^{2}\bigg\} \times   \\ \no &\times&
\int d[\delta\wt{\Sigma}_{M;N}^{(\alpha;\kappa)ab}(x_{p})]\;
\exp\bigg\{\frac{\im}{8}\int_{C}d^{4}\!x_{p}\sum_{\scrscr(\alpha)=1,.,8}^{\scrscr(\kappa)=0,.,3}
\Big[-\im\,\hat{\mfrak{e}}_{p}^{(\hat{F})}+\hat{\mfrak{b}}^{(\hat{F})}_{(\alpha;\kappa)}(x_{p})\Big]\;
\times  \\ \no &\times&
\TRALL\bigg[\delta\wt{\Sigma}_{f,\ovv{M};g,\ovv{N}}^{(\alpha;\kappa)ab}(x_{p})\;
\frac{1}{\hat{\mfrak{v}}_{\hat{F};\ovv{N}}^{(\alpha;\kappa)}(x_{p})}\;
\delta\wt{\Sigma}_{g,\ovv{N};f,\ovv{M}}^{(\alpha;\kappa)ba}(x_{p})\;
\frac{1}{\hat{\mfrak{v}}_{\hat{F};\ovv{M}}^{(\alpha;\kappa)}(x_{p})}\bigg]\bigg\}\times  \\ \no &\times&
\int d[\psi_{M}^{\dag}(x_{p}),\,\psi_{M}(x_{p})]\;\exp\bigg\{-\frac{\im}{2}\int_{C}d^{4}\!x_{p}\;d^{4}\!y_{q}\;
\Psi_{N}^{\dag,b}(y_{q})\;\wt{M}_{N;M}^{ba}(y_{q},x_{p})\;\Psi_{M}^{a}(x_{p}) \bigg\} \times \\ \no &\times&
\exp\bigg\{-\frac{\im}{2}\int_{C}d^{4}\!x_{p}\;\Big(J_{\psi;M}^{\dag,a}(x_{p})\;\hat{S}^{ab}\;\Psi_{M}^{b}(x_{p})+
\Psi_{M}^{\dag,a}(x_{p})\;\hat{S}^{ab}\;J_{\psi;M}^{b}(x_{p})\Big)\bigg\}
\boldsymbol{\Bigg\rangle}\;.
\eeq
Relation (\ref{s3_40}) is simplified by introducing the matrix \(\wt{M}_{N;M}^{ba}(y_{q},x_{p})\) (\ref{s3_41}) and the
background generating function (\ref{s3_42})
of the gauge field propagation. Apart from the sources \(\hat{\mscr{J}}_{N;M}^{ba}(y_{q},x_{p})\),
\(\hat{J}_{\psi\psi;N;M}^{b\neq a}(x_{p})\) and the
one-particle Hamiltonian, the matrix \(\wt{M}_{N;M}^{ba}(y_{q},x_{p})\) (\ref{s3_41}) contains the gauge field strength self-energy
\(\hat{\mfrak{S}}_{\alpha;\mu\nu}^{(\hat{F})}(x_{p})\), the structure constants \(C_{\alpha\beta\gamma}\) of \(\mbox{SU}_{c}(N_{c}=3)\)
and the quark self-energy density fields \(\sigma_{D}^{(\alpha;\kappa)}(x_{p})\) coupled to diagonalizing matrices
\(\hat{\mscr{U}}_{\hat{F};\ovv{M}\ppr;\ovv{M}}^{(\alpha;\kappa),aa}(x_{p})\),
\(\hat{\mscr{U}}_{\hat{F};\ovv{N};\ovv{M}\ppr}^{(\alpha;\kappa),bb,\dag}(x_{p})\)
with eigenvalues \(\hat{\mfrak{v}}_{\hat{F};\ovv{M}\ppr}^{(\alpha;\kappa)}(x_{p})\) of the anomalous indexed potentials
\(\hat{\boldsymbol{\mscr{V}}}_{(\alpha);f\ppr,\ovv{M}\ppr;f,\ovv{M}}^{(\hat{F})a;(\kappa)}(x_{p})\)
(\ref{s3_17},\ref{s3_18}). We assume that these gauge field related terms
create a confining potential for the anomalous doubled self-energies \(\delta\wt{\Sigma}_{M;N}^{(\alpha;\kappa)ab}(x_{p})\) of quark matter fields
where the confinement potential is also partially determined by
the source fields \(J_{\psi;M}^{a}(x_{p})\), \(\hat{J}_{\psi\psi;N;M}^{b\neq a}(x_{p})\).
After using Eqs. (\ref{s3_21}-\ref{s3_25}), we can also link the quark self-energy density fields \(\sigma_{D}^{(\alpha;\kappa)}(x_{p})\)
and the term with the self-energy of the gauge field strength tensor together into a single potential part so that the
relevant terms for a confinement become more obvious
\beq \lb{s3_41}
\lefteqn{\wt{M}_{N;M}^{ba}(y_{q},x_{p})=\hat{\mscr{J}}_{N;M}^{ba}(y_{q},x_{p})+\delta^{(4)}(y_{q}-x_{p})\;\delta_{pq}\,\eta_{q}\times } \\ \no &\times&
\Bigg[\bigg(\bea{cc} \big[\hat{\beta}(\hat{\gamma}^{\mu}\hat{\pp}_{p,\mu}-\im\,\hat{\ve}_{p}+\hat{m})\big]_{N;M} & \hat{j}_{\psi\psi;N;M}(x_{p}) \\
\hat{j}_{\psi\psi;N;M}^{\dag}(x_{p}) & -\big[\hat{\beta}(\hat{\gamma}^{\mu}\hat{\pp}_{p,\mu}-\im\,\hat{\ve}_{p}+\hat{m})\big]_{N;M}^{T}
\eea\bigg)_{N;M}^{ba}
+ \\ \no &+& \Big[\Big(\hat{\pp}_{p}^{\lambda}\hat{\mfrak{S}}_{\gamma;\nu\lambda}^{(\hat{F})}(x_{p})\Big)-s_{\gamma}(x_{p})\;n_{\nu}\Big]\;
\Big[-\im\,\hat{\mfrak{e}}_{p}^{(\hat{F})}+C_{\alpha\beta\ppr\gamma\ppr}\;
\hat{\mfrak{S}}_{\alpha}^{(\hat{F})\mu\ppr\nu\ppr}(x_{p})\Big]_{\gamma\beta}^{\boldsymbol{-1};\nu\mu}\times \\ \no &\times&
\bigg(\bea{cc} \big[\hat{\beta}(\im \;\hat{\gamma}_{\mu}\;\hat{t}_{\beta})\big]_{\ovv{N};\ovv{M}} & 0 \\
0 & -\big[\hat{\beta}(\im \;\hat{\gamma}_{\mu}\;\hat{t}_{\beta})\big]_{\ovv{N};\ovv{M}}^{T} \eea\bigg)_{g,\ovv{N};f,\ovv{M}}^{ba}\delta_{g,f}
+  \\ \no &+&
\sum_{\scrscr(\alpha)=1,.,8}^{\scrscr(\kappa)=0,.,3}\bigg(\frac{\hat{S}^{bb}}{2}\;\sigma_{D}^{(\alpha;\kappa)}(x_{p})\;\delta_{ab}\;
\hat{\mscr{U}}_{\hat{F};\ovv{N};\ovv{M}\ppr}^{(\alpha;\kappa),bb,\dag}(x_{p})\;\;
\hat{\mfrak{v}}_{\hat{F};\ovv{M}\ppr}^{(\alpha;\kappa)}(x_{p})\;\;
\hat{\mscr{U}}_{\hat{F};\ovv{M}\ppr;\ovv{M}}^{(\alpha;\kappa),aa}(x_{p})\;\;\delta_{g,f}+ \\ \no &+& \frac{1}{2}\hat{S}^{bb}\;\;
\hat{\mscr{U}}_{\hat{F};\ovv{N};\ovv{N}\ppr}^{(\alpha;\kappa),bb,\dag}(x_{p})\;
\bigg(\bea{cc} \delta\hat{\Sigma}_{g,\ovv{N}\ppr;f,\ovv{M}\ppr}^{(\alpha;\kappa)11}(x_{p}) &
\delta\hat{\Sigma}_{g,\ovv{N}\ppr;f,\ovv{M}\ppr}^{(\alpha;\kappa)12}(x_{p}) \\
\delta\hat{\Sigma}_{g,\ovv{N}\ppr;f,\ovv{M}\ppr}^{(\alpha;\kappa)21}(x_{p}) &
-\delta\hat{\Sigma}_{g,\ovv{N}\ppr;f,\ovv{M}\ppr}^{(\alpha;\kappa)22}(x_{p}) \eea\bigg)_{N\ppr;M\ppr}^{ba}
\hat{\mscr{U}}_{\hat{F};\ovv{M}\ppr;\ovv{M}}^{(\alpha;\kappa),aa}(x_{p})\;\;\hat{S}^{aa}\bigg)\Bigg]_{g,\ovv{N};f,\ovv{M}}^{ba} = \\ \no &=&
\hat{\mscr{J}}_{N;M}^{ba}(y_{q},x_{p})+\delta^{(4)}\!(y_{q}-x_{p})\;\delta_{pq}\,\eta_{q}
\Bigg[\bigg(\bea{cc} \big[\hat{\beta}(\hat{\gamma}^{\mu}\hat{\pp}_{p,\mu}-\im\,\hat{\ve}_{p}+\hat{m})\big]_{N;M} & \hat{j}_{\psi\psi;N;M}(x_{p}) \\
\hat{j}_{\psi\psi;N;M}^{\dag}(x_{p}) &
\hspace*{-0.6cm}-\big[\hat{\beta}(\hat{\gamma}^{\mu}\hat{\pp}_{p,\mu}-\im\,\hat{\ve}_{p}+\hat{m})\big]_{N;M}^{T}
\eea\bigg)_{N;M}^{ba} \hspace*{-0.3cm} +
\\ \no &+& \hspace*{-0.3cm}
\bigg(\Big[\Big(\hat{\pp}_{p}^{\lambda}\hat{\mfrak{S}}_{\gamma;\nu\lambda}^{(\hat{F})}(x_{p})\Big)-s_{\gamma}(x_{p})\;n_{\nu}\Big]
\Big[-\im\,\hat{\mfrak{e}}_{p}^{(\hat{F})}+C_{\alpha\beta\ppr\gamma\ppr}\;
\hat{\mfrak{S}}_{\alpha}^{(\hat{F})\mu\ppr\nu\ppr}(x_{p})\Big]_{\gamma\beta}^{\boldsymbol{-1};\nu\mu}\hspace*{-0.3cm}+ \hspace*{-0.1cm}\frac{1}{2}\hspace*{-0.1cm}
\sum_{\scrscr(\alpha)=1,.,8}^{\scrscr(\kappa)=0,.,3}\hat{\mfrak{B}}_{\hat{F};\beta(\alpha)}^{\mu(\kappa)}(x_{p})\;
\sigma_{D}^{(\alpha;\kappa)}(x_{p})\bigg)_{\beta}^{\mu} \hspace*{-0.3cm}\times \\ \no &\times&
\bigg(\bea{cc} \big[\hat{\beta}(\im \;\hat{\gamma}_{\mu}\;\hat{t}_{\beta})\big]_{\ovv{N};\ovv{M}} & 0 \\
0 & -\big[\hat{\beta}(\im \;\hat{\gamma}_{\mu}\;\hat{t}_{\beta})\big]_{\ovv{N};\ovv{M}}^{T} \eea\bigg)_{g,\ovv{N};f,\ovv{M}}^{ba}\delta_{g,f} +
\\ \no &+& \hspace*{-0.3cm} \frac{1}{2}\hat{S}^{bb} \sum_{\scrscr(\alpha)=1,.,8}^{\scrscr(\kappa)=0,.,3}
\hat{\mscr{U}}_{\hat{F};\ovv{N};\ovv{N}\ppr}^{(\alpha;\kappa),bb,\dag}(x_{p})\;
\bigg(\bea{cc} \delta\hat{\Sigma}_{g,\ovv{N}\ppr;f,\ovv{M}\ppr}^{(\alpha;\kappa)11}(x_{p}) &
\delta\hat{\Sigma}_{g,\ovv{N}\ppr;f,\ovv{M}\ppr}^{(\alpha;\kappa)12}(x_{p}) \\
\delta\hat{\Sigma}_{g,\ovv{N}\ppr;f,\ovv{M}\ppr}^{(\alpha;\kappa)21}(x_{p}) &
-\delta\hat{\Sigma}_{g,\ovv{N}\ppr;f,\ovv{M}\ppr}^{(\alpha;\kappa)22}(x_{p}) \eea\bigg)^{ba}
\hat{\mscr{U}}_{\hat{F};\ovv{M}\ppr;\ovv{M}}^{(\alpha;\kappa),aa}(x_{p})\;\hat{S}^{aa}\bigg)\Bigg]_{g,\ovv{N};f,\ovv{M}}^{ba}\,.
\eeq
The background path integral for (\ref{s3_40}) is listed in (\ref{s3_42}) and also incorporates the delta-functions (\ref{s3_28}) of the diagonalization (\ref{s3_10},\ref{s3_22})
and the real, auxiliary field \(s_{\alpha}(x_{p})\) for the axial gauge fixing. Furthermore, we have to include the polynomials
\(\mfrak{P}_{1}(\,\hat{\mfrak{b}}_{\alpha;\kappa}^{(\hat{F})}(x_{p})\,)\), \(\mfrak{P}_{2}(\,\hat{\mfrak{b}}_{\alpha;\kappa}^{(\hat{F})}(x_{p})\,)\),
\(\mfrak{P}_{3}(\,\hat{\mfrak{b}}_{\alpha;\kappa}^{(\hat{F})}(x_{p})\,)\),
\(\mfrak{Q}_{1}(\,\hat{\mfrak{v}}_{\hat{F};\ovv{M}}^{(\alpha;\kappa)}(x_{p})\,)\),
\(\mfrak{Q}_{3}(\,\hat{\mfrak{v}}_{\hat{F};\ovv{M}}^{(\alpha;\kappa)}(x_{p})\,)\)
which have been introduced for proper normalization of Gaussian integrals in the HST's
\beq  \lb{s3_42}
\lefteqn{\boldsymbol{\Bigg\langle Z\Big[\hat{\mfrak{S}}^{(\hat{F})},s_{\alpha},\hat{\mfrak{B}}_{\hat{F}},\hat{\mfrak{b}}^{(\hat{F})},
\hat{\mscr{U}}_{\hat{F}},\hat{\mfrak{v}}_{\hat{F}};\hat{\mfrak{j}}^{(\hat{F})};
\mbox{\bf Eq. (\ref{s3_42})}\Big]\;\Big(\mbox{\bf fields}\Big)\;\Bigg\rangle } =}  \\ \no &=&\boldsymbol{\Bigg\langle}
\int d[\hat{\mfrak{S}}_{\alpha}^{(\hat{F})\mu\nu}(x_{p})]\;\;
d[\hat{\mfrak{B}}_{\hat{F};\beta(\alpha)}^{\mu(\kappa)}(x_{p});\hat{\mfrak{b}}_{(\alpha;\kappa)}^{(\hat{F})}(x_{p})]\;\;
d[\hat{\mscr{U}}_{\hat{F};\ovv{N};\ovv{M}}^{(\alpha;\kappa)}(x_{p});\hat{\mfrak{v}}_{\hat{F};\ovv{N}}^{(\alpha;\kappa)}(x_{p})]\;\times
\\  \no &\times& \bigg\{\prod_{\scrscr\{x_{p}\}}\delta\bigg(C_{\alpha\beta\gamma}\;\hat{\mfrak{S}}_{\alpha}^{(\hat{F})\mu\nu}(x_{p}) -
\sum_{\scrscr(\alpha)=1,.,8}^{\scrscr(\kappa)=0,.,3}\hat{\mfrak{B}}_{\hat{F};\beta(\alpha)}^{\mu(\kappa)}(x_{p})\;
\hat{\mfrak{b}}^{(\hat{F})}_{(\alpha;\kappa)}(x_{p})\;\hat{\mfrak{B}}_{\hat{F};(\alpha)\gamma}^{T,(\kappa)\nu}(x_{p})
\bigg)\bigg\} \times \\ \no &\times&
\bigg\{\prod_{\scrscr\{x_{p}\};(\alpha)=1,.,8}^{\scrscr\{(\kappa)=0,.,3\}}
\delta\bigg(\hat{\mscr{U}}_{\hat{F};\ovv{M}\ppr;\ovv{N}}^{(\alpha;\kappa),11,\dag}(x_{p})\;
\hat{\mfrak{v}}_{\hat{F};\ovv{N}}^{(\alpha;\kappa)}(x_{p})\;
\hat{\mscr{U}}_{\hat{F};\ovv{N};\ovv{M}}^{(\alpha;\kappa),11}(x_{p})-
\big[\hat{\beta}\,(\im \;\hat{\gamma}_{\mu}\;\hat{t}_{\beta})\big]_{\ovv{M}\ppr;\ovv{M}}\;
\hat{\mfrak{B}}_{\hat{F};\beta(\alpha)}^{\mu(\kappa)}(x_{p})\bigg)\bigg\} \times
\\ \no &\times& \mfrak{P}_{1}\Big(\hat{\mfrak{b}}_{(\alpha;\kappa)}^{(\hat{F})}(x_{p})\Big)\;
\mfrak{P}_{2}\Big(\hat{\mfrak{b}}_{(\alpha;\kappa)}^{(\hat{F})}(x_{p})\Big)\;
\mfrak{P}_{3}\Big(\hat{\mfrak{b}}_{(\alpha;\kappa)}^{(\hat{F})}(x_{p})\Big)\; \;
\mfrak{Q}_{1}\Big(\hat{\mfrak{v}}_{\hat{F};\ovv{M}}^{(\alpha;\kappa)}(x_{p})\Big)\;
\mfrak{Q}_{3}\Big(\hat{\mfrak{v}}_{\hat{F};\ovv{M}}^{(\alpha;\kappa)}(x_{p})\Big)\;\int d[s_{\alpha}(x_{p})]\;\times   \\ \no &\times&
\exp\bigg\{\im\int_{C}d^{4}\!x_{p}\bigg(\frac{1}{4}\hat{\mfrak{S}}_{\alpha}^{(\hat{F})\mu\nu}(x_{p})\;
\hat{\mfrak{S}}_{\alpha;\mu\nu}^{(\hat{F})}(x_{p})-\hat{\mfrak{j}}^{(\hat{F})}_{\alpha;\mu\nu}(x_{p})\;
\hat{\mfrak{S}}_{\alpha}^{(\hat{F})\mu\nu}(x_{p})+\hat{\mfrak{j}}^{(\hat{F})}_{\alpha;\mu\nu}(x_{p})\;
\hat{\mfrak{j}}^{(\hat{F})\mu\nu}_{\alpha}(x_{p})\bigg)\bigg\}\times
\\ \no &\times& \bigg\{\mbox{det}\Big[\Big(-\im\,\hat{\mfrak{e}}_{p}^{(\hat{F})}+C_{\alpha\beta\gamma}\;
\hat{\mfrak{S}}_{\alpha}^{(\hat{F})\mu\nu}(x_{p})\Big)_{\beta\gamma}^{\mu\nu}\Big]\bigg\}^{\boldsymbol{-1/2}}\times
\exp\bigg\{\frac{\im}{2}\int_{C}d^{4}\!x_{p}\;\Big[\big(\,\hat{\pp}_{p}^{\lambda}\hat{\mfrak{S}}_{\gamma;\nu\lambda}^{(\hat{F})}(x_{p})\,\big)-
s_{\gamma}(x_{p})\;n_{\nu}\Big]\;\times \\ \no &\times&
\Big[-\im\,\hat{\mfrak{e}}_{p}^{(\hat{F})}+C_{\alpha\beta\ppr\gamma\ppr}\;
\hat{\mfrak{S}}_{\alpha}^{(\hat{F})\mu\ppr\nu\ppr}(x_{p})\Big]_{\gamma\beta}^{\boldsymbol{-1};\nu\mu}\;
\Big[\big(\,\hat{\pp}_{p}^{\kappa}\hat{\mfrak{S}}_{\beta;\mu\kappa}^{(\hat{F})}(x_{p})\,\big)-
s_{\beta}(x_{p})\;n_{\mu}\Big]\bigg\}\;  \times  \boldsymbol{\bigg(\mbox{\bf fields}\bigg)\Bigg\rangle}\;.
\eeq

\subsection{Coset decomposition for BCS pair condensate degrees of freedom} \lb{s33}

The HST's of sections \ref{s31}, \ref{s32} have transformed the original path integral (\ref{s2_25}-\ref{s2_27})
to \(Z[\hat{\mscr{J}},J_{\psi},\hat{J}_{\psi\psi},\hat{\mfrak{j}}^{(\hat{F})}]\) (\ref{s3_8},\ref{s3_9})
and finally to (\ref{s3_40}) with matrix
\(\wt{M}_{N;M}^{ba}(y_{q},x_{p})\) (\ref{s3_41}) and background generating functional (\ref{s3_42}) of the gauge field degrees of freedom.
If we disregard the anomalous doubled one-particle Hamiltonian, the quark self-energy densities \(\sigma_{D}^{(\alpha;\kappa)}(x_{p})\) and
the gauge field strength tensor \(\hat{\mfrak{S}}_{\alpha;\mu\nu}^{(\hat{F})}(x_{p})\), the generating function
\(Z[\hat{\mscr{J}},J_{\psi},\hat{J}_{\psi\psi},\hat{\mfrak{j}}^{(\hat{F})}]\) (\ref{s3_40}) mainly consists of the
sum of \((N_{c}^{2}-1=8)\times(3+1)=32\) anomalous doubled self-energies
\(\delta\wt{\Sigma}_{N;M}^{(\alpha;\kappa)ba}(x_{p})\) in \(\wt{M}_{N;M}^{ba}(y_{q},x_{p})\) (\ref{s3_41}) for the
fermionic matter fields; the latter are also dressed by the block diagonal
unitary matrices \(\hat{\mscr{U}}_{\hat{F};\ovv{N};\ovv{N}\ppr}^{(\alpha;\kappa),bb,\dag}(x_{p})\),
\(\hat{\mscr{U}}_{\hat{F};\ovv{M}\ppr;\ovv{M}}^{(\alpha;\kappa),aa}(x_{p})\) with colour degrees of freedom. Therefore, we can introduce a single,
anomalous doubled self-energy \(\delta\wt{\Sigma}_{M;N}^{ab}(x_{p})\) (\ref{s3_44})
with anti-hermitian, BCS related terms in the off-diagonal blocks \(a\neq b\)
for the entire sum of \(32\), colour dressed
self-energies \(\delta\wt{\Sigma}_{M;N}^{(\alpha;\kappa)ab}(x_{p})\) in \(\wt{M}_{N;M}^{ba}(y_{q},x_{p})\) (\ref{s3_41}).
This is accomplished by including the delta-function (\ref{s3_43}) with \(\delta\wt{\Sigma}_{M;N}^{ab}(x_{p})\) for the sum of the \(32\),
anomalous doubled, colour dressed self-energies \(\delta\wt{\Sigma}_{M;N}^{(\alpha;\kappa)ab}(x_{p})\); one has also to reckon the similar symmetries
(\ref{s3_45}) for the density- and anomalous-related blocks of \(\delta\wt{\Sigma}_{M;N}^{ab}(x_{p})\) as
those for the independent, \(32\) self-energies \(\delta\wt{\Sigma}_{M;N}^{(\alpha;\kappa)ab}(x_{p})\) (\ref{s3_29}-\ref{s3_35})
\beq  \lb{s3_43}
\lefteqn{\hspace*{-1.9cm} \int d[\delta\wt{\Sigma}_{M;N}^{ab}(x_{p})]\;
\delta\bigg(\delta\wt{\Sigma}_{f,\ovv{M};g,\ovv{N}}^{ab}(x_{p})-\sum_{\scrscr(\alpha)=1,.,8}^{\scrscr(\kappa)=0,.,3}
\hat{\mscr{U}}_{\hat{F};\ovv{M};\ovv{M}\ppr}^{(\alpha;\kappa),aa,\dag}(x_{p})\;
\delta\wt{\Sigma}_{f,\ovv{M}\ppr;g,\ovv{N}\ppr}^{(\alpha;\kappa)ab}(x_{p})\;
\hat{\mscr{U}}_{\hat{F};\ovv{N}\ppr;\ovv{N}}^{(\alpha;\kappa),bb}(x_{p})\bigg)=} \\ \no &=&
 \int d[\delta\wt{\Sigma}_{M;N}^{ab}(x_{p})]\;\;
\boldsymbol{\delta\bigg[}\bigg(\bea{cc}\delta\hat{\Sigma}_{M;N}^{11}(x_{p}) & \im\;\delta\hat{\Sigma}_{M;N}^{12}(x_{p}) \\
\im\;\delta\hat{\Sigma}_{M;N}^{21}(x_{p}) & \delta\hat{\Sigma}_{M;N}^{22}(x_{p}) \eea\bigg)_{f,\ovv{M};g,\ovv{N}}^{ab} +  \\ \no &-&
\sum_{\scrscr(\alpha)=1,.,8}^{\scrscr(\kappa)=0,.,3}
\hat{\mscr{U}}_{\hat{F};\ovv{M};\ovv{M}\ppr}^{(\alpha;\kappa),aa,\dag}(x_{p})\;
\bigg(\bea{cc}\delta\hat{\Sigma}_{M\ppr;N\ppr}^{(\alpha;\kappa)11}(x_{p}) & \im\;\delta\hat{\Sigma}_{M\ppr;N\ppr}^{(\alpha;\kappa)12}(x_{p}) \\
\im\;\delta\hat{\Sigma}_{M\ppr;N\ppr}^{(\alpha;\kappa)21}(x_{p}) &
\delta\hat{\Sigma}_{M\ppr;N\ppr}^{(\alpha;\kappa)22}(x_{p})\eea\bigg)_{f,\ovv{M}\ppr;g,\ovv{N}\ppr}^{ab}\;
\hat{\mscr{U}}_{\hat{F};\ovv{N}\ppr;\ovv{N}}^{(\alpha;\kappa),bb}(x_{p})\boldsymbol{\bigg]}\;;
\\  \lb{s3_44}
\delta\wt{\Sigma}_{M;N}^{ab}(x_{p}) &=&\left(\bea{cc}\delta\hat{\Sigma}_{M;N}^{11}(x_{p}) & \im\;\delta\hat{\Sigma}_{M;N}^{12}(x_{p}) \\
\im\;\delta\hat{\Sigma}_{M;N}^{21}(x_{p}) & \delta\hat{\Sigma}_{M;N}^{22}(x_{p}) \eea\right)_{M;N}^{ab} \;;
\eeq
\be \lb{s3_45}
\bea{rclrcl}
\Big(\delta\hat{\Sigma}_{M;N}^{11}(x_{p})\Big)^{\dag}&=&\delta\hat{\Sigma}_{M;N}^{11}(x_{p})\;; & \hspace*{0.6cm}
\Big(\delta\hat{\Sigma}_{M;N}^{22}(x_{p})\Big)^{\dag}&=&\delta\hat{\Sigma}_{M;N}^{22}(x_{p})\;; \\
\delta\hat{\Sigma}_{M;N}^{22}(x_{p})&=&-\Big(\delta\hat{\Sigma}_{M;N}^{11}(x_{p})\Big)^{T}\;; &
\delta\hat{\Sigma}_{M;N}^{21}(x_{p})&=&\Big(\delta\hat{\Sigma}_{M;N}^{12}(x_{p})\Big)^{\dag}\;; \\
\delta\hat{\Sigma}_{M;N}^{12}(x_{p})&=&-\Big(\delta\hat{\Sigma}_{M;N}^{12}(x_{p})\Big)^{T}\;; &
\delta\hat{\Sigma}_{M;N}^{21}(x_{p})&=&-\Big(\delta\hat{\Sigma}_{M;N}^{21}(x_{p})\Big)^{T}\;.
\eea
\ee
The transformation (\ref{s3_43}-\ref{s3_45}) to a single, anomalous doubled self-energy for the fermionic fields considerably simplifies the
matrix \(\wt{M}_{N;M}^{ba}(y_{q},x_{p})\) (\ref{s3_41}) in the path integral (\ref{s3_40},\ref{s3_42}).
This is possible according to the analogous symmetries (\ref{s3_45}) of \(\delta\wt{\Sigma}_{M;N}^{ab}(x_{p})\) (\ref{s3_44}) as
those of \(\delta\wt{\Sigma}_{M;N}^{(\alpha;\kappa)ab}(x_{p})\) (\ref{s3_29}-\ref{s3_35}).
After inserting the delta-function (\ref{s3_43}) into (\ref{s3_40}-\ref{s3_42}), we integrate over the original, anti-commuting quark fields
\(d[\psi_{M}^{\dag}(x_{p}),\,\psi_{M}(x_{p})]\) in (\ref{s3_40}) and obtain the square root
of the anomalous doubled Fermi determinant and the bilinear
source term with \(J_{\psi;N}^{\dag,b}(y_{q})\), \(J_{\psi;M}^{a}(x_{p})\) for
the inverse of the simplified matrix \(\wt{M}_{N;M}^{\boldsymbol{-1};ba}(y_{q},x_{p})\) (\ref{s3_47})
\beq \lb{s3_46}
\lefteqn{Z[\hat{\mscr{J}},J_{\psi},\hat{J}_{\psi\psi},\hat{\mfrak{j}}^{(\hat{F})}]=
\boldsymbol{\Bigg\langle Z\Big[\hat{\mfrak{S}}^{(\hat{F})},s_{\alpha},\hat{\mfrak{B}}_{\hat{F}},\hat{\mfrak{b}}^{(\hat{F})},
\hat{\mscr{U}}_{\hat{F}},\hat{\mfrak{v}}_{\hat{F}};\hat{\mfrak{j}}^{(\hat{F})};
\mbox{\bf Eq. (\ref{s3_42})}\Big]\; } \times } \\  \no &\times&
\int d[\sigma_{D}^{(\alpha;\kappa)}(x_{p})]\;\;\exp\bigg\{\frac{\im}{4}\int_{C}d^{4}\!x_{p}
\sum_{\scrscr(\alpha)=1,.,8}^{\scrscr(\kappa)=0,.,3}
\Big[\im\,\hat{\mfrak{e}}_{p}^{(\hat{F})}+\hat{\mfrak{b}}_{(\alpha;\kappa)}^{(\hat{F})}(x_{p})\Big]\;
\Big(\sigma_{D}^{(\alpha;\kappa)}(x_{p})\Big)^{2}\bigg\} \times   \\  \no &\times&
\int d[\delta\wt{\Sigma}_{M;N}^{(\alpha;\kappa)ab}(x_{p})]\;
\exp\bigg\{\frac{\im}{8}\int_{C}d^{4}\!x_{p}\sum_{\scrscr(\alpha)=1,.,8}^{\scrscr(\kappa)=0,.,3}
\Big[-\im\,\hat{\mfrak{e}}_{p}^{(\hat{F})}+\hat{\mfrak{b}}_{(\alpha;\kappa)}^{(\hat{F})}(x_{p})\Big]\;\times \\ \no &\times&
\TRALL\bigg[\delta\wt{\Sigma}_{f,\ovv{M};g,\ovv{N}}^{(\alpha;\kappa)ab}(x_{p})\;
\frac{1}{\hat{\mfrak{v}}_{\hat{F};\ovv{N}}^{(\alpha;\kappa)}(x_{p})}\;
\delta\wt{\Sigma}_{g,\ovv{N};f,\ovv{M}}^{(\alpha;\kappa)ba}(x_{p})
\frac{1}{\hat{\mfrak{v}}_{\hat{F};\ovv{M}}^{(\alpha;\kappa)}(x_{p})}\bigg]\bigg\}\times \\ \no &\times&
\int d[\delta\wt{\Sigma}_{M;N}^{ab}(x_{p})]\;
\delta\bigg(\delta\wt{\Sigma}_{f,\ovv{M};g,\ovv{N}}^{ab}(x_{p})-\sum_{\scrscr(\alpha)=1,.,8}^{\scrscr(\kappa)=0,.,3}
\hat{\mscr{U}}_{\hat{F};\ovv{M};\ovv{M}\ppr}^{(\alpha;\kappa),aa,\dag}(x_{p})\;
\delta\wt{\Sigma}_{f,\ovv{M}\ppr;g,\ovv{N}\ppr}^{(\alpha;\kappa)ab}(x_{p})\;
\hat{\mscr{U}}_{\hat{F};\ovv{N}\ppr;\ovv{N}}^{(\alpha;\kappa),bb}(x_{p})\bigg)\times \\ \no &\times&
\bigg\{\mbox{DET}\Big[\wt{M}_{N;M}^{ba}(y_{q},x_{p})\Big]\bigg\}^{1/2}
\exp\bigg\{\frac{\im}{2}\int_{C}d^{4}\!x_{p}\;d^{4}\!y_{q}\;
J_{\psi;N}^{\dag,b}(y_{q})\;\hat{S}^{bb}\;\wt{M}_{N;M}^{\boldsymbol{-1};ba}(y_{q},x_{p})\;\hat{S}^{aa}\;
J_{\psi;M}^{a}(x_{p})\bigg\}\boldsymbol{\Bigg\rangle}\;;
\eeq
\beq \lb{s3_47}
\lefteqn{\wt{M}_{N;M}^{ba}(y_{q},x_{p}) = \eta_{q}\,
\frac{\hat{\mscr{J}}_{N;M}^{ba}(y_{q},x_{p})}{\mcal{N}}\,\eta_{p}+  }  \\ \no &+&
\delta^{(4)}\!(y_{q}-x_{p})\;\delta_{pq}\,\eta_{q}
\Bigg[\bigg(\bea{cc} \big[\hat{\beta}(\hat{\gamma}^{\mu}\hat{\pp}_{p,\mu}-\im\,\hat{\ve}_{p}+\hat{m})\big]_{N;M} & \hat{j}_{\psi\psi;N;M}(x_{p}) \\
\hat{j}_{\psi\psi;N;M}^{\dag}(x_{p}) & -\big[\hat{\beta}(\hat{\gamma}^{\mu}\hat{\pp}_{p,\mu}-\im\,\hat{\ve}_{p}+\hat{m})\big]_{N;M}^{T}
\eea\bigg)_{N;M}^{ba} \hspace*{-0.3cm} +
\\ \no &+& \hspace*{-0.3cm}
\bigg(\Big[\Big(\hat{\pp}_{p}^{\lambda}\hat{\mfrak{S}}_{\gamma;\nu\lambda}^{(\hat{F})}(x_{p})\Big)-s_{\gamma}(x_{p})\;n_{\nu}\Big]
\Big[-\im\,\hat{\mfrak{e}}_{p}^{(\hat{F})}+C_{\alpha\beta\ppr\gamma\ppr}\;
\hat{\mfrak{S}}_{\alpha}^{(\hat{F})\mu\ppr\nu\ppr}(x_{p})\Big]_{\gamma\beta}^{\boldsymbol{-1};\nu\mu}\hspace*{-0.3cm}+ \hspace*{-0.1cm}\frac{1}{2}\hspace*{-0.1cm}
\sum_{\scrscr(\alpha)=1,.,8}^{\scrscr(\kappa)=0,.,3}\hat{\mfrak{B}}_{\hat{F};\beta(\alpha)}^{\mu(\kappa)}(x_{p})\;
\sigma_{D}^{(\alpha;\kappa)}(x_{p})\bigg)_{\beta}^{\mu} \times \\ \no &\times&\hspace*{-0.3cm}
\bigg(\bea{cc} \big[\hat{\beta}(\im \;\hat{\gamma}_{\mu}\;\hat{t}_{\beta})\big]_{\ovv{N};\ovv{M}} & 0 \\
0 & -\big[\hat{\beta}(\im \;\hat{\gamma}_{\mu}\;\hat{t}_{\beta})\big]_{\ovv{N};\ovv{M}}^{T} \eea\bigg)^{ba}\delta_{g,f} +
 \frac{1}{2}\hat{S}^{bb}
\bigg(\bea{cc} \delta\hat{\Sigma}_{N;M}^{11}(x_{p}) &
\delta\hat{\Sigma}_{N;M}^{12}(x_{p}) \\
\delta\hat{\Sigma}_{N;M}^{21}(x_{p}) &
-\delta\hat{\Sigma}_{N;M}^{22}(x_{p}) \eea\bigg)^{ba} \hat{S}^{aa}\bigg)\Bigg]_{g,\ovv{N};f,\ovv{M}\;.}^{ba}
\eeq
According to Ref.\ \cite{mies1}, we transform the matrix \(\wt{M}_{N;M}^{ba}(y_{q},x_{p})\) (\ref{s3_47}) with hermitian, BCS terms to
\(\wt{\mscr{M}}_{N;M}^{ba}(y_{q},x_{p})\) (\ref{s3_51}) with anti-hermitian, anomalous parts by the subsequent steps (\ref{s3_48}-\ref{s3_50});
additional use is made of the delta-function (\ref{s3_43}) for abbreviating the sum of the \(32\), colour dressed
self-energies \(\delta\wt{\Sigma}_{M;N}^{(\alpha;\kappa)ab}(x_{p})\) by the single,
anomalous doubled one \(\delta\wt{\Sigma}_{M;N}^{ab}(x_{p})\) (\ref{s3_44})
\beq \lb{s3_48}
\wt{M}_{N;M}^{ba}(y_{q},x_{p}) &\rightarrow& \hat{S}\;\Big(\hat{S}\;\wt{M}_{N;M}^{ba}(y_{q},x_{p})\;\hat{S}\Big)\;\hat{S} \; \rightarrow \; \hat{S}\,\hat{I}^{-1}\;\Big(
\underbrace{\hat{I}\;\hat{S}\;\wt{M}_{N;M}^{ba}(y_{q},x_{p})\;\hat{S}\;\hat{I}}_{\wt{\mscr{M}}_{N;M}^{ba}(y_{q},x_{p})}
\Big)\;\hat{I}^{-1}\;\hat{S}  \;;  \\ \no
\wt{\mscr{M}}_{N;M}^{ba}(y_{q},x_{p})&=&\hat{I}\;\hat{S}\;\wt{M}_{N;M}^{ba}(y_{q},x_{p})\;\hat{S}\;\hat{I}  \;; \\  \lb{s3_49}
\mbox{DET}\Big\{\wt{M}_{N;M}^{ba}(y_{q},x_{p})\Big\} &=&
\mbox{DET}\Big\{\hat{S}\;\hat{I}^{-1}\;\wt{\mscr{M}}_{N;M}^{ba}(y_{q},x_{p})\;\hat{I}^{-1}\;\hat{S}\Big\}=
\mbox{DET}\Big\{\wt{\mscr{M}}_{N;M}^{ba}(y_{q},x_{p})\Big\} \;; \\  \lb{s3_50}
\wt{M}_{N;M}^{\boldsymbol{-1};ba}(y_{q},x_{p}) &=& \hat{S}\;\hat{I}\;\wt{\mscr{M}}_{N;M}^{\boldsymbol{-1};ba}(y_{q},x_{p})\;\hat{I}\;\hat{S} \;;
\eeq\vspace*{-0.6cm}
\beq \lb{s3_51}
\lefteqn{\wt{\mscr{M}}_{N;M}^{ba}(y_{q},x_{p})=
\hat{I}\;\hat{S}\;\eta_{q}\,\frac{\hat{\mscr{J}}_{N;M}^{ba}(y_{q},x_{p})}{\mcal{N}}\,\eta_{p}\;\hat{S}\;\hat{I}+
\delta^{(4)}(y_{q}-x_{p})\;\eta_{q}\,\delta_{pq} \;\times } \\ \no &\times&
\Bigg[\bigg(\bea{cc} \big[\hat{\beta}(\hat{\gamma}^{\mu}\hat{\pp}_{p,\mu}-\im\,\hat{\ve}_{p}+\hat{m})\big]_{N;M} &
-\im\;\hat{j}_{\psi\psi;N;M}(x_{p})   \\ -\im\;\hat{j}_{\psi\psi;N;M}^{\dag}(x_{p}) &
\big[\hat{\beta}(\hat{\gamma}^{\mu}\hat{\pp}_{p,\mu}-\im\,\hat{\ve}_{p}+\hat{m})\big]_{N;M}^{T} \eea\bigg)_{N;M}^{ba} +  \\ \no &+&
\mcal{V}_{\beta}^{\mu}(x_{p})\;\bigg(\bea{cc} \big[\hat{\beta}(\im \;\hat{\gamma}_{\mu}\;\hat{t}_{\beta})\big]_{\ovv{N};\ovv{M}} & 0 \\
0 & \big[\hat{\beta}(\im \;\hat{\gamma}_{\mu}\;\hat{t}_{\beta})\big]_{\ovv{N};\ovv{M}}^{T}\eea \bigg)_{g,\ovv{N};f,\ovv{M}}^{ba} \;\delta_{g,f}
+ \frac{1}{2}\;
\bigg(\bea{cc} \delta\hat{\Sigma}_{N;M}^{11}(x_{p}) & \im\;\delta\hat{\Sigma}_{N;M}^{12}(x_{p})  \\
\im\;\delta\hat{\Sigma}_{N;M}^{21}(x_{p}) & \delta\hat{\Sigma}_{N;M}^{22}(x_{p}) \eea\bigg)_{N;M}^{ba}\Bigg]_{\mbox{.}}
\eeq
Furthermore, we have shortened expressions for \(\wt{\mscr{M}}_{N;M}^{ba}(y_{q},x_{p})\) in (\ref{s3_51}) by substituting the
field strength self-energy term of \(\hat{\mfrak{S}}_{\alpha;\mu\nu}^{(\hat{F})}(x_{p})\) and the quark self-energy density
terms of \(\sigma_{D}^{(\alpha;\kappa)}(x_{p})\) with \(\hat{\mfrak{B}}_{\hat{F};\beta\alpha}^{\mu\kappa}(x_{p})\)
by a potential \(\mcal{V}_{\beta}^{\mu}(x_{p})\) (\ref{s3_52}). This real-valued potential
variable \(\mcal{V}_{\beta}^{\mu}(x_{p})\) (\ref{s3_52}) follows from the background path integral (\ref{s3_42}) with
additional averaging of the quark self-energy densities
\be \lb{s3_52}
\mcal{V}_{\beta}^{\mu}(x_{p})=\Big[\Big(\hat{\pp}_{p}^{\lambda}\hat{\mfrak{S}}_{\gamma;\nu\lambda}^{(\hat{F})}(x_{p})\Big)-
s_{\gamma}(x_{p})\;n_{\nu}\Big]\;\Big[-\im\,\hat{\mfrak{e}}_{p}^{(\hat{F})}+C_{\alpha\beta\ppr\gamma\ppr}\;
\hat{\mfrak{S}}_{\alpha}^{(\hat{F})\mu\ppr\nu\ppr}(x_{p})\big]_{\gamma\beta}^{\boldsymbol{-1};\nu\mu} +
\frac{1}{2}\sum_{\scrscr(\alpha)=1,.,8}^{\scrscr(\kappa)=0,.,3}\hat{\mfrak{B}}_{\hat{F};\beta(\alpha)}^{\mu(\kappa)}(x_{p})\;\;
\sigma_{D}^{(\alpha;\kappa)}(x_{p})\;.
\ee
Eventually, the matrix \(\wt{M}_{N;M}^{ba}(y_{q},x_{p})\) is replaced by \(\wt{\mscr{M}}_{N;M}^{ba}(y_{q},x_{p})\) (\ref{s3_51},\ref{s3_52})
in (\ref{s3_46}) so that we achieve the generating function
\(Z[\hat{\mscr{J}},J_{\psi},\hat{J}_{\psi\psi},\hat{\mfrak{j}}^{(\hat{F})}]\) (\ref{s3_53},\ref{s3_54})
which is further reduced by shifting the anti-hermitian, anomalous doubled,
single self-energy (\ref{s3_44}) in \(\wt{\mscr{M}}_{N;M}^{ba}(y_{q},x_{p})\)
with the source matrix \(\im\:\hat{J}_{\psi\psi;N;M}^{b\neq a}(x_{p})\) (\ref{s3_55}).
In consequence the source matrix \(\im\:\hat{J}_{\psi\psi;N;M}^{b\neq a}(x_{p})\)
disappears from \(\wt{\mscr{M}}_{N;M}^{ba}(y_{q},x_{p})\) (\ref{s3_51}), but modifies the delta-function (\ref{s3_43}) into (\ref{s3_56}).
The background path integral of gauge field degrees of freedom is listed again in relation (\ref{s3_54})
for convenience, but additionally includes averaging with the real, scalar quark self-energy densities \(d[\sigma_{D}^{(\alpha;\kappa)}(x_{p})]\)
(boldface symbols in (\ref{s3_54}), last line in Eq. (\ref{s3_54}))
\beq \lb{s3_53}
\lefteqn{Z[\hat{\mscr{J}},J_{\psi},\hat{J}_{\psi\psi},\hat{j}_{\hat{F}}]=
\boldsymbol{\Bigg\langle Z\Big[\hat{\mfrak{S}}^{(\hat{F})},s_{\alpha},\hat{\mfrak{B}}_{\hat{F}},\hat{\mfrak{b}}^{(\hat{F})},
\hat{\mscr{U}}_{\hat{F}},\hat{\mfrak{v}}_{\hat{F}};\sigma_{D};;\hat{\mfrak{j}}^{(\hat{F})};
\mbox{\bf Eq. (\ref{s3_54})}\Big]\; } \times }  \\ \no &\times&
\int d[\delta\wt{\Sigma}_{M;N}^{(\alpha;\kappa)ab}(x_{p})]\;
\exp\bigg\{\frac{\im}{8}\int_{C}d^{4}\!x_{p}\sum_{\scrscr(\alpha)=1,.,8}^{\scrscr(\kappa)=0,.,3}
\Big[-\im\,\hat{\mfrak{e}}_{p}^{(\hat{F})}+\hat{\mfrak{b}}_{(\alpha;\kappa)}^{(\hat{F})}(x_{p})\Big]\;\times   \\ \no &\times&
\TRALL\bigg[\delta\wt{\Sigma}_{f,\ovv{M};g,\ovv{N}}^{(\alpha;\kappa)ab}(x_{p})\;
\frac{1}{\hat{\mfrak{v}}_{\hat{F};\ovv{N}}^{(\alpha;\kappa)}(x_{p})}\;\delta\wt{\Sigma}_{g,\ovv{N};f,\ovv{M}}^{(\alpha;\kappa)ba}(x_{p})\;
\frac{1}{\hat{\mfrak{v}}_{\hat{F};\ovv{M}}^{(\alpha;\kappa)}(x_{p})}\bigg]\bigg\} \times
\int d[\delta\wt{\Sigma}_{M;N}^{ab}(x_{p})]\;\times  \\ \no &\times&
\delta\bigg(\delta\wt{\Sigma}_{f,\ovv{M};g,\ovv{N}}^{ab}(x_{p})+2\,\im\:\hat{J}_{\psi\psi;f,\ovv{M};g,\ovv{N}}^{a\neq b}(x_{p})-
\sum_{\scrscr(\alpha)=1,.,8}^{\scrscr(\kappa)=0,.,3}
\hat{\mscr{U}}_{\hat{F};\ovv{M};\ovv{M}\ppr}^{(\alpha;\kappa),aa,\dag}(x_{p})\;
\delta\wt{\Sigma}_{f,\ovv{M}\ppr;g,\ovv{N}\ppr}^{(\alpha;\kappa)ab}(x_{p})\;
\hat{\mscr{U}}_{\hat{F};\ovv{N}\ppr;\ovv{N}}^{(\alpha;\kappa),bb}(x_{p})\bigg)\times   \\  \no &\times&
\bigg\{\mbox{DET}\Big[\wt{\mscr{M}}_{N;M}^{ba}(y_{q},x_{p})\Big]\bigg\}^{1/2}\;
\exp\bigg\{\frac{\im}{2}\int_{C}d^{4}\!x_{p}\;d^{4}\!y_{q}\;\;
J_{\psi;N}^{\dag,b}(y_{q})\;\hat{I}\;
\wt{\mscr{M}}_{N;M}^{\boldsymbol{-1};ba}(y_{q},x_{p})\;\hat{I}\;J_{\psi;M}^{a}(x_{p})\bigg\}
\boldsymbol{\Bigg\rangle}\;;
\eeq
\beq \lb{s3_54}
\lefteqn{\boldsymbol{\Bigg\langle Z\Big[\hat{\mfrak{S}}^{(\hat{F})},s_{\alpha},\hat{\mfrak{B}}_{\hat{F}},\hat{\mfrak{b}}^{(\hat{F})},
\hat{\mscr{U}}_{\hat{F}},\hat{\mfrak{v}}_{\hat{F}};\sigma_{D};\hat{\mfrak{j}}^{(\hat{F})};
\mbox{\bf Eq. (\ref{s3_54})}\Big]\;\Big(\mbox{\bf fields}\Big)\;\Bigg\rangle } =}  \\ \no &=&\boldsymbol{\Bigg\langle}
\int d[\hat{\mfrak{S}}_{\alpha;\mu\nu}^{(\hat{F})}(x_{p})]\;\;
d[\hat{\mfrak{B}}_{\hat{F};\beta(\alpha)}^{\mu(\kappa)}(x_{p});\hat{\mfrak{b}}_{(\alpha;\kappa)}^{(\hat{F})}(x_{p})]\;\;
d[\hat{\mscr{U}}_{\hat{F};\ovv{N};\ovv{M}}^{(\alpha;\kappa)}(x_{p});\hat{\mfrak{v}}_{\hat{F};\ovv{N}}^{(\alpha;\kappa)}(x_{p})]\;\times
\\  \no &\times& \bigg\{\prod_{\scrscr\{x_{p}\}}\delta\bigg(C_{\alpha\beta\gamma}\;\hat{\mfrak{S}}_{\alpha}^{(\hat{F})\mu\nu}(x_{p}) -
\sum_{\scrscr(\alpha)=1,.,8}^{\scrscr(\kappa)=0,.,3}\hat{\mfrak{B}}_{\hat{F};\beta(\alpha)}^{\mu(\kappa)}(x_{p})\;
\hat{\mfrak{b}}^{(\hat{F})}_{(\alpha;\kappa)}(x_{p})\;\hat{\mfrak{B}}_{\hat{F};(\alpha)\gamma}^{T,(\kappa)\nu}(x_{p})
\bigg)\bigg\} \times \\ \no &\times&
\bigg\{\prod_{\scrscr\{x_{p}\};(\alpha)=1,.,8}^{\scrscr\{(\kappa)=0,.,3\}}
\delta\bigg(\hat{\mscr{U}}_{\hat{F};\ovv{M}\ppr;\ovv{N}}^{(\alpha;\kappa),11,\dag}(x_{p})\;
\hat{\mfrak{v}}_{\hat{F};\ovv{N}}^{(\alpha;\kappa)}(x_{p})\;
\hat{\mscr{U}}_{\hat{F};\ovv{N};\ovv{M}}^{(\alpha;\kappa),11}(x_{p})-
\big[\hat{\beta}\,(\im \;\hat{\gamma}_{\mu}\;\hat{t}_{\beta})\big]_{\ovv{M}\ppr;\ovv{M}}\;
\hat{\mfrak{B}}_{\hat{F};\beta(\alpha)}^{\mu(\kappa)}(x_{p})\bigg)\bigg\} \times
\\ \no &\times& \mfrak{P}_{1}\Big(\hat{\mfrak{b}}_{(\alpha;\kappa)}^{(\hat{F})}(x_{p})\Big)\;
\mfrak{P}_{2}\Big(\hat{\mfrak{b}}_{(\alpha;\kappa)}^{(\hat{F})}(x_{p})\Big)\;
\mfrak{P}_{3}\Big(\hat{\mfrak{b}}_{(\alpha;\kappa)}^{(\hat{F})}(x_{p})\Big)\; \;
\mfrak{Q}_{1}\Big(\hat{\mfrak{v}}_{\hat{F};\ovv{M}}^{(\alpha;\kappa)}(x_{p})\Big)\;
\mfrak{Q}_{3}\Big(\hat{\mfrak{v}}_{\hat{F};\ovv{M}}^{(\alpha;\kappa)}(x_{p})\Big)\;\int d[s_{\alpha}(x_{p})]\;\times   \\ \no &\times&
\exp\bigg\{\im\int_{C}d^{4}\!x_{p}\bigg(\frac{1}{4}\hat{\mfrak{S}}_{\alpha}^{(\hat{F})\mu\nu}(x_{p})\;
\hat{\mfrak{S}}_{\alpha;\mu\nu}^{(\hat{F})}(x_{p})-\hat{\mfrak{j}}^{(\hat{F})}_{\alpha;\mu\nu}(x_{p})\;
\hat{\mfrak{S}}_{\alpha}^{(\hat{F})\mu\nu}(x_{p})+\hat{\mfrak{j}}^{(\hat{F})}_{\alpha;\mu\nu}(x_{p})\;
\hat{\mfrak{j}}^{(\hat{F})\mu\nu}_{\alpha}(x_{p})\bigg)\bigg\}\times
\\ \no &\times& \bigg\{\mbox{det}\Big[\Big(-\im\,\hat{\mfrak{e}}_{p}^{(\hat{F})}+C_{\alpha\beta\gamma}\;
\hat{\mfrak{S}}_{\alpha}^{(\hat{F})\mu\nu}(x_{p})\Big)_{\beta\gamma}^{\mu\nu}\Big]\bigg\}^{\boldsymbol{-1/2}}\times
\exp\bigg\{\frac{\im}{2}\int_{C}d^{4}\!x_{p}\;
\Big[\big(\,\hat{\pp}_{p}^{\lambda}\hat{\mfrak{S}}_{\gamma;\nu\lambda}^{(\hat{F})}(x_{p})\,\big)-
s_{\gamma}(x_{p})\;n_{\nu}\Big]\;\times \\ \no &\times&
\Big[-\im\,\hat{\mfrak{e}}_{p}^{(\hat{F})}+C_{\alpha\beta\ppr\gamma\ppr}\;
\hat{\mfrak{S}}_{\alpha}^{(\hat{F})\mu\ppr\nu\ppr}(x_{p})\Big]_{\gamma\beta}^{\boldsymbol{-1};\nu\mu}\;
\Big[\big(\,\hat{\pp}_{p}^{\kappa}\hat{\mfrak{S}}_{\beta;\mu\kappa}^{(\hat{F})}(x_{p})\,\big)-
s_{\beta}(x_{p})\;n_{\mu}\Big]\bigg\}\;  \times  \\ \no &\times&
\boldsymbol{\int d[\sigma_{D}^{(\alpha;\kappa)}(x_{p})]\;
\exp\bigg\{\frac{\im}{4}\int_{C}d^{4}\!x_{p}\sum_{\scrscr(\alpha)=1,.,8}^{\scrscr(\kappa)=0,.,3}
\Big[\im\,\hat{\mfrak{e}}_{p}^{(\hat{F})}+\hat{\mfrak{b}}_{(\alpha;\kappa)}^{(\hat{F})}(x_{p})\Big]\;
\Big(\sigma_{D}^{(\alpha;\kappa)}(x_{p})\Big)^{2}\bigg\}\times \bigg(\mbox{\bf fields}\bigg)\Bigg\rangle}\;;
\eeq
\be \lb{s3_55}
\frac{1}{2}\bigg(\bea{cc} \delta\hat{\Sigma}_{N;M}^{11}(x_{p}) & \im\;\delta\hat{\Sigma}_{N;M}^{12}(x_{p})   \\
\im\;\delta\hat{\Sigma}_{N;M}^{21}(x_{p}) &   \delta\hat{\Sigma}_{N;M}^{22}(x_{p}) \eea\bigg)-
\bigg(\bea{cc} 0 & \hspace*{-0.2cm}\im\;\hat{j}_{\psi\psi;N;M}(x_{p}) \\
\im\;\hat{j}_{\psi\psi;N;M}^{\dag}(x_{p}) & 0 \eea\bigg)\rightarrow
\frac{1}{2}\bigg(\bea{cc} \delta\hat{\Sigma}_{N;M}^{11}(x_{p}) & \im\;\delta\hat{\Sigma}_{N;M}^{12}(x_{p})   \\
\im\;\delta\hat{\Sigma}_{N;M}^{21}(x_{p}) &   \delta\hat{\Sigma}_{N;M}^{22}(x_{p}) \eea\bigg)_{\mbox{;}}
\ee
\beq \lb{s3_56}
\lefteqn{\delta\bigg(\delta\wt{\Sigma}_{f,\ovv{M};g,\ovv{N}}^{ab}(x_{p})-\sum_{\scrscr(\alpha)=1,.,8}^{\scrscr(\kappa)=1,.,3}
\hat{\mscr{U}}_{\hat{F};\ovv{M};\ovv{M}\ppr}^{(\alpha;\kappa),aa,\dag}(x_{p})\;\;
\delta\wt{\Sigma}_{f,\ovv{M}\ppr;g,\ovv{N}\ppr}^{(\alpha;\kappa)ab}(x_{p})\;\;
\hat{\mscr{U}}_{\hat{F};\ovv{N}\ppr;\ovv{N}}^{(\alpha;\kappa),bb}(x_{p})\bigg) \;\rightarrow}  \\ \no &\rightarrow&
\delta\bigg(\delta\wt{\Sigma}_{f,\ovv{M};g,\ovv{N}}^{ab}(x_{p})+2\,\im\:\hat{J}_{\psi\psi;f,\ovv{M};g,\ovv{N}}^{a\neq b}(x_{p})-
\sum_{\scrscr(\alpha)=1,.,8}^{\scrscr(\kappa)=0,.,3}
\hat{\mscr{U}}_{\hat{F};\ovv{M};\ovv{M}\ppr}^{(\alpha;\kappa),aa,\dag}(x_{p})\;
\delta\wt{\Sigma}_{f,\ovv{M}\ppr;g,\ovv{N}\ppr}^{(\alpha;\kappa)ab}(x_{p})\;
\hat{\mscr{U}}_{\hat{F};\ovv{N}\ppr;\ovv{N}}^{(\alpha;\kappa),bb}(x_{p})\bigg)\;.
\eeq
Furthermore, we move the potential variable \(\mcal{V}_{\beta}^{\mu}(x_{p})\) (\ref{s3_52}), which consists of the field strength self-energy
\(\hat{\mfrak{S}}_{\alpha;\mu\nu}^{(\hat{F})}(x_{p})\), other colour-related
degrees of freedom as \(\hat{\mfrak{B}}_{\hat{F};\beta(\alpha)}^{\mu(\kappa)}(x_{p})\)
and the quark self-energy densities \(\sigma_{D}^{(\alpha;\kappa)}(x_{p})\), to the background functional (\ref{s3_54}).
This transformation becomes obvious, as one expands the anomalous doubled determinant
\(\{\mbox{DET}[\wt{\mscr{M}}_{N;M}^{ba}(y_{q},x_{p})\,]\}^{1/2}\)
and Green function \(\wt{\mscr{M}}_{N;M}^{\boldsymbol{-1};ba}(y_{q},x_{p})\) in (\ref{s3_53},\ref{s3_51}) in terms of
\(\hat{H}_{N;M}(x_{p})=[\hat{\beta}(\,\feynd_{p}+\im\:\feynV(x_{p})-\im\:\hat{\ve}_{p}+\hat{m})]_{N;M}\)
(boldface symbols in Eqs. (\ref{s3_57}-\ref{s3_59}) ).
Since the part \(\hat{H}_{N;M}(x_{p})=[\hat{\beta}(\,\feynd_{p}+\im\:\feynV(x_{p})-\im\:\hat{\ve}_{p}+\hat{m})]_{N;M}\) is contained in
\(\wt{\mscr{M}}_{N;M}^{ba}(y_{q},x_{p})\) in its anomalous doubled kind with the transpose \(\hat{H}_{N;M}^{T}(x_{p})\),
one can simplify the separated actions
for the background functionals by multiplying with an additional factor of two. The particular form of the anomalous doubled
Hilbert space is described in appendix \ref{sa}, which also summarizes the definitions of the doubled spacetime coordinate
states and the appropriate decomposition of the unit operator into complete sets of anomalous doubled spacetime or
momentum-energy states. The decomposition of the unit operator also yields the suitable form of trace operations
of the doubled Hilbert space with inclusion of the time contour integrals for forward and backward propagation
(compare also with chapter 4 in \cite{mies1})
\beq  \lb{s3_57}
\lefteqn{\Big\{\mbox{DET}[\wt{\mscr{M}}_{N;M}^{ba}(y_{q},x_{p})\,]\Big\}^{1/2}=
\exp\bigg\{\frac{1}{2}\hspace*{0.6cm}
\trxpa\bigg[\trfgamc\ln\Big[\wt{\mscr{M}}_{N;M}^{ba}(y_{q},x_{p})\Big]\bigg]\bigg\} =  }  \\ \no &=&
\exp\Bigg\{\frac{1}{2}\hspace*{0.6cm}\trxpa\Bigg[\trfgamc\Bigg(\ln\Big[\wt{\mscr{M}}_{N;M}^{ba}(y_{q},x_{p})\Big]-
\ln\bigg[\bea{cc} \hat{\mscr{H}}_{N;M}^{11}(y_{q},x_{p})  &  0  \\
0  & \hat{\mscr{H}}_{N;M}^{11,T}(y_{q},x_{p})\eea\bigg]\Bigg)\Bigg]\Bigg\} \times  \\ \no &\times&
\underbrace{\boldsymbol{\exp\bigg\{\hspace*{0.6cm}\trxp\bigg[\trfgamc\ln\bigg[\eta_{p}\hat{\beta}\Big(
\feynd_{p}+\im\:\feynV(x_{p})-\im\:\hat{\ve}_{p}+\hat{m}\Big)\bigg]\bigg]\bigg\}}}_{
\mbox{Move to background functional (\ref{s3_54}) $\rightarrow$ (\ref{s3_59}) }} \;;
\eeq
\beq \lb{s3_58}
\lefteqn{\exp\bigg\{\frac{\im}{2}\int_{C}d^{4}\!x_{p}\;d^{4}\!y_{q}\;\;
J_{\psi;N}^{\dag,b}(y_{q})\;\hat{I}\;
\wt{\mscr{M}}_{N;M}^{\boldsymbol{-1};ba}(y_{q},x_{p})\;\hat{I}\;J_{\psi;M}^{a}(x_{p})\bigg\} =
\exp\Bigg\{\frac{\im}{2}\int_{C}d^{4}\!x_{p}\;d^{4}\!y_{q}\;\times } \\ \no &\times&
J_{\psi;N}^{\dag,b}(y_{q})\;\hat{I}\;\Bigg(\bigg(
\wt{\mscr{M}}_{N;M}^{\boldsymbol{-1};ba}(y_{q},x_{p})-
\bigg(\bea{cc} \hat{\mscr{H}}_{N;M}^{11;\boldsymbol{-1}}(y_{q},x_{p})  &  0  \\
0  & \hat{\mscr{H}}_{N;M}^{11,T;\boldsymbol{-1}}(y_{q},x_{p})\eea\bigg)\Bigg)\hat{I}\;J_{\psi;M}^{a}(x_{p})
\Bigg\} \times \\ \no &\times&
\underbrace{\boldsymbol{\exp\bigg\{\im\int_{C}d^{4}\!x_{p}\;d^{4}\!y_{q}\;\;
j_{\psi;N}^{\dag}(y_{q})\;
\Big\langle y_{q}
\Big|\Big[\hat{\beta}\big(\feynd_{p}+\im\:\feynV(x_{p})-\im\:\hat{\ve}_{p}+\hat{m}\big)\Big]_{N;M}^{\boldsymbol{-1}}\Big|x_{p}\Big\rangle\;
j_{\psi;M}(x_{p})\bigg\}}}_{\mbox{Move to background functional (\ref{s3_54}) $\rightarrow$ (\ref{s3_59}) }} \;;
\eeq
\beq \lb{s3_59}
\lefteqn{\boldsymbol{\Bigg\langle Z\Big[\feynV(x_{p});\hat{\mfrak{S}}^{(\hat{F})},s_{\alpha},\hat{\mfrak{B}}_{\hat{F}},\hat{\mfrak{b}}^{(\hat{F})},
\hat{\mscr{U}}_{\hat{F}},\hat{\mfrak{v}}_{\hat{F}};\sigma_{D};\hat{\mfrak{j}}^{(\hat{F})};
\mbox{\bf Eq. (\ref{s3_59})}\Big]\;\Big(\mbox{\bf fields}\Big)\;\Bigg\rangle } =}    \\ \no &=&\boldsymbol{\Bigg\langle}
\int d[\hat{\mfrak{S}}_{\alpha}^{(\hat{F})\mu\nu}(x_{p})]\;\;
d[\hat{\mfrak{B}}_{\hat{F};\beta(\alpha)}^{\mu(\kappa)}(x_{p});\hat{\mfrak{b}}_{(\alpha;\kappa)}^{(\hat{F})}(x_{p})]\;\;
d[\hat{\mscr{U}}_{\hat{F};\ovv{N};\ovv{M}}^{(\alpha;\kappa)}(x_{p});\hat{\mfrak{v}}_{\hat{F};\ovv{N}}^{(\alpha;\kappa)}(x_{p})]\;\times
\\  \no &\times& \bigg\{\prod_{\scrscr\{x_{p}\}}\delta\bigg(C_{\alpha\beta\gamma}\;\hat{\mfrak{S}}_{\alpha}^{(\hat{F})\mu\nu}(x_{p}) -
\sum_{\scrscr(\alpha)=1,.,8}^{\scrscr(\kappa)=0,.,3}\hat{\mfrak{B}}_{\hat{F};\beta(\alpha)}^{\mu(\kappa)}(x_{p})\;
\hat{\mfrak{b}}^{(\hat{F})}_{(\alpha;\kappa)}(x_{p})\;\hat{\mfrak{B}}_{\hat{F};(\alpha)\gamma}^{T,(\kappa)\nu}(x_{p})
\bigg)\bigg\} \times \\ \no  &\times&
\bigg\{\prod_{\scrscr\{x_{p}\};(\alpha)=1,.,8}^{\scrscr\{(\kappa)=0,.,3\}}
\delta\bigg(\hat{\mscr{U}}_{\hat{F};\ovv{M}\ppr;\ovv{N}}^{(\alpha;\kappa),11,\dag}(x_{p})\;
\hat{\mfrak{v}}_{\hat{F};\ovv{N}}^{(\alpha;\kappa)}(x_{p})\;
\hat{\mscr{U}}_{\hat{F};\ovv{N};\ovv{M}}^{(\alpha;\kappa),11}(x_{p})-
\big[\hat{\beta}\,(\im \;\hat{\gamma}_{\mu}\;\hat{t}_{\beta})\big]_{\ovv{M}\ppr;\ovv{M}}\;
\hat{\mfrak{B}}_{\hat{F};\beta(\alpha)}^{\mu(\kappa)}(x_{p})\bigg)\bigg\} \times
\\ \no &\times& \mfrak{P}_{1}\Big(\hat{\mfrak{b}}_{(\alpha;\kappa)}^{(\hat{F})}(x_{p})\Big)\;
\mfrak{P}_{2}\Big(\hat{\mfrak{b}}_{(\alpha;\kappa)}^{(\hat{F})}(x_{p})\Big)\;
\mfrak{P}_{3}\Big(\hat{\mfrak{b}}_{(\alpha;\kappa)}^{(\hat{F})}(x_{p})\Big)\; \;
\mfrak{Q}_{1}\Big(\hat{\mfrak{v}}_{\hat{F};\ovv{M}}^{(\alpha;\kappa)}(x_{p})\Big)\;
\mfrak{Q}_{3}\Big(\hat{\mfrak{v}}_{\hat{F};\ovv{M}}^{(\alpha;\kappa)}(x_{p})\Big)\;\int d[s_{\alpha}(x_{p})]\;\times   \\ \no &\times&
\exp\bigg\{\im\int_{C}d^{4}\!x_{p}\bigg(\frac{1}{4}\hat{\mfrak{S}}_{\alpha}^{(\hat{F})\mu\nu}(x_{p})\;
\hat{\mfrak{S}}_{\alpha;\mu\nu}^{(\hat{F})}(x_{p})-\hat{\mfrak{j}}^{(\hat{F})}_{\alpha;\mu\nu}(x_{p})\;
\hat{\mfrak{S}}_{\alpha}^{(\hat{F})\mu\nu}(x_{p})+\hat{\mfrak{j}}^{(\hat{F})}_{\alpha;\mu\nu}(x_{p})\;
\hat{\mfrak{j}}^{(\hat{F})\mu\nu}_{\alpha}(x_{p})\bigg)\bigg\}\times
\\ \no &\times& \bigg\{\mbox{det}\Big[\Big(-\im\,\hat{\mfrak{e}}_{p}^{(\hat{F})}+C_{\alpha\beta\gamma}\;
\hat{\mfrak{S}}_{\alpha}^{(\hat{F})\mu\nu}(x_{p})\Big)_{\beta\gamma}^{\mu\nu}\Big]\bigg\}^{\boldsymbol{-1/2}}\times
\exp\bigg\{\frac{\im}{2}\int_{C}d^{4}\!x_{p}\;
\Big[\big(\,\hat{\pp}_{p}^{\lambda}\hat{\mfrak{S}}_{\gamma;\nu\lambda}^{(\hat{F})}(x_{p})\,\big)-
s_{\gamma}(x_{p})\;n_{\nu}\Big]\;\times \\ \no &\times&
\Big[-\im\,\hat{\mfrak{e}}_{p}^{(\hat{F})}+C_{\alpha\beta\ppr\gamma\ppr}\;
\hat{\mfrak{S}}_{\alpha}^{(\hat{F})\mu\ppr\nu\ppr}(x_{p})\Big]_{\gamma\beta}^{\boldsymbol{-1};\nu\mu}\;
\Big[\big(\,\hat{\pp}_{p}^{\kappa}\hat{\mfrak{S}}_{\beta;\mu\kappa}^{(\hat{F})}(x_{p})\,\big)-
s_{\beta}(x_{p})\;n_{\mu}\Big]\bigg\}\;\times \\ \no &\times&
\int d[\sigma_{D}^{(\alpha;\kappa)}(x_{p})]\;
\exp\bigg\{\frac{\im}{4}\int_{C}d^{4}\!x_{p}\sum_{\scrscr(\alpha)=1,.,8}^{\scrscr(\kappa)=0,.,3}
\Big[\im\,\hat{\mfrak{e}}_{p}^{(\hat{F})}+\hat{\mfrak{b}}_{(\alpha;\kappa)}^{(\hat{F})}(x_{p})\Big]\;
\Big(\sigma_{D}^{(\alpha;\kappa)}(x_{p})\Big)^{2}\bigg\} \times \\ \no &\times& \boldsymbol{
\exp\bigg\{\hspace*{0.6cm}\trxp\bigg[\trfgamc\ln\bigg[\eta_{p}\hat{\beta}\Big(
\feynd_{p}+\im\:\feynV(x_{p})-\im\:\hat{\ve}_{p}+\hat{m}\Big)\bigg]\bigg]\bigg\} } \times \\ \no &\times&\boldsymbol{
\exp\bigg\{\im\int_{C}d^{4}\!x_{p}\;d^{4}\!y_{q}\;\;
j_{\psi;N}^{\dag}(y_{q})\;
\Big\langle y_{q}\Big|\Big[\hat{\beta}\big(\feynd_{p}+\im\:\feynV(x_{p})-\im\:\hat{\ve}_{p}+\hat{m}\big)\Big]_{N;M}^{\boldsymbol{-1}}\Big|x_{p}\Big\rangle\;
j_{\psi;M}(x_{p})\bigg\} } \boldsymbol{\bigg(\mbox{\bf fields}\bigg)\Bigg\rangle}  \;;
\eeq
\beq \no
\hat{\mscr{H}}_{N;M}^{11}(y_{q},x_{p}) &=& \delta_{pq}\;\delta^{(4)}(y_{q}-x_{p})\;\eta_{q}
\Big[\hat{\beta}\big(\feynd_{p}+\im\:\feynV(x_{p})-\im\:\hat{\ve}_{p}+\hat{m}\big)\Big]_{N;M} \;;  \\ \lb{s3_60}
\mcal{V}_{\beta}^{\mu}(x_{p})\hspace*{-0.3cm}&=&\hspace*{-0.3cm}
\Big[\Big(\hat{\pp}_{p}^{\lambda}\hat{\mfrak{S}}_{\gamma;\nu\lambda}^{(\hat{F})}(x_{p})\Big)-
s_{\gamma}(x_{p})\;n_{\nu}\Big]\;\Big[-\im\,\hat{\mfrak{e}}_{p}^{(\hat{F})}+C_{\alpha\beta\ppr\gamma\ppr}\;
\hat{\mfrak{S}}_{\alpha}^{(\hat{F})\mu\ppr\nu\ppr}(x_{p})\big]_{\gamma\beta}^{\boldsymbol{-1};\nu\mu} +   \\ \no &+&
\frac{1}{2}\sum_{\scrscr(\alpha)=1,.,8}^{\scrscr(\kappa)=0,.,3}\hat{\mfrak{B}}_{\hat{F};\beta(\alpha)}^{\mu(\kappa)}(x_{p})\;\;
\sigma_{D}^{(\alpha;\kappa)}(x_{p})\;.
\eeq
In consistency to the shift of \(\frac{1}{2}\;\delta\wt{\Sigma}_{M;N}^{ab}(x_{p})\) by
\(\im\:\hat{J}_{\psi\psi;M;N}^{a\neq b}(x_{p})\) (\ref{s3_55}), one attains a simplified matrix
\(\wt{\mscr{M}}_{N;M}^{ba}(y_{q},x_{p})\) (\ref{s3_61}) without \(\im\:\hat{J}_{\psi\psi;M;N}^{a\neq b}(x_{p})\) instead of (\ref{s3_51}),
but has a modified delta function (\ref{s3_56}) instead of (\ref{s3_43}) for the sum of 32, colour dressed self-energies
\(\delta\wt{\Sigma}_{M;N}^{(\alpha;\kappa)ab}(x_{p})\). The new delta function (\ref{s3_56}) is reduced to an additional
effective functional (\ref{s3_62}), which has a local spacetime dependence and which follows from integrating by
\(\delta\wt{\Sigma}_{M;N}^{(\alpha;\kappa)ab}(x_{p})\) over the delta function (\ref{s3_56}) and over remaining
Gaussian factors. These Gaussian factors consist of actions with the eigenvalues \(\hat{\mfrak{b}}_{(\alpha;\kappa)}^{(\hat{F})}(x_{p})\),
\(\hat{\mfrak{v}}_{\hat{F};\ovv{N}}^{(\alpha;\kappa)}(x_{p})\) as gauge field variables from the diagonalization of
the interaction potentials (\ref{s3_17},\ref{s3_18}) and (\ref{s3_22})
\beq \lb{s3_61}
\lefteqn{\wt{\mscr{M}}_{N;M}^{ba}(y_{q},x_{p})=\hat{I}\;\hat{S}\;\eta_{q}\,
\frac{\hat{\mscr{J}}_{N;M}^{ba}(y_{q},x_{p})}{\mcal{N}}\,\eta_{p}\;\hat{S}\;\hat{I}+
\delta^{(4)}(y_{q}-x_{p})\;\eta_{q}\,\delta_{pq}\;\times }   \\  \no &\times&
\Bigg[\bigg(\bea{cc} \big[\hat{\beta}\,(\feynd_{p}+\im\;
\feynV(x_{p}) -\im\;\hat{\ve}_{p}+\hat{m})\big]_{N;M}  &  0  \\
0  & \big[\hat{\beta}\,(\feynd_{p}+\im \;
\feynV(x_{p}) -\im\;\hat{\ve}_{p}+\hat{m})\big]_{N;M}^{T} \eea\bigg)_{N;M}^{ba} + \\ \no &+&
\frac{1}{2}\bigg(\bea{cc} \delta\hat{\Sigma}_{N;M}^{11}(x_{p})  & \im\;\delta\hat{\Sigma}_{N;M}^{12}(x_{p}) \\
\im\;\delta\hat{\Sigma}_{N;M}^{21}(x_{p})  & \delta\hat{\Sigma}_{N;M}^{22}(x_{p}) \eea\bigg)_{N;M}^{ba}\Bigg]_{N;M}^{ba} \;;
\eeq
\beq \no
\lefteqn{\bigg\{\prod_{\{x_{p}\}}\boldsymbol{\Delta}
\bigg(\delta\wt{\Sigma}_{f,\ovv{M};g,\ovv{N}}^{ab}(x_{p})+2\:\im\hat{J}_{\psi\psi;f,\ovv{M};g,\ovv{N}}^{a\neq b}(x_{p})\:;\:
\hat{\mscr{U}}_{\hat{F};\ovv{M};\ovv{N}}^{(\alpha;\kappa)}(x_{p})\,;\,\hat{\mfrak{v}}_{\hat{F};\ovv{N}}^{(\alpha;\kappa)}(x_{p})\,;\,
\hat{\mfrak{b}}_{(\alpha;\kappa)}^{(\hat{F})}(x_{p})\bigg)\bigg\} =}    \\ \lb{s3_62} &=&
\int d[\delta\wt{\Sigma}_{M;N}^{(\alpha;\kappa)ab}(x_{p})]\;
\exp\Bigg\{\frac{\im}{8}\int_{C}d^{4}\!x_{p}\sum_{\scrscr(\alpha)=1,.,8}^{\scrscr(\kappa)=0,.,3}
\Big[-\im\:\hat{\mfrak{e}}_{p}^{(\hat{F})}+\hat{\mfrak{b}}_{(\alpha;\kappa)}^{(\hat{F})}(x_{p})\Big]\times \\ \no &\times&
\TRALL\bigg[\delta\wt{\Sigma}_{f,\ovv{M};g,\ovv{N}}^{(\alpha;\kappa)ab}(x_{p})
\frac{1}{\hat{\mfrak{v}}_{\hat{F};\ovv{N}}^{(\alpha;\kappa)}(x_{p})}
\delta\wt{\Sigma}_{g,\ovv{N};f,\ovv{M}}^{(\alpha;\kappa)ba}(x_{p})
\frac{1}{\hat{\mfrak{v}}_{\hat{F};\ovv{M}}^{(\alpha;\kappa)}(x_{p})}\bigg]\Bigg\}\times
\\ \no &\hspace*{-0.3cm}\times&\hspace*{-0.6cm}
\bigg\{\prod_{\{x_{p})\}}
\delta\bigg(\delta\wt{\Sigma}_{f,\ovv{M};g,\ovv{N}}^{ab}(x_{p})+2\:\im\hat{J}_{\psi\psi;f,\ovv{M};g,\ovv{N}}^{a\neq b}(x_{p})-
\sum_{\scrscr(\alpha)=1,.,8}^{\scrscr(\kappa)=0,.,3}
\hat{\mscr{U}}_{\hat{F};\ovv{M};\ovv{M}\ppr}^{(\alpha;\kappa),aa,\dag}(x_{p})
\;\delta\wt{\Sigma}_{f,\ovv{M}\ppr;g,\ovv{N}\ppr}^{(\alpha;\kappa)ab}(x_{p})\;
\hat{\mscr{U}}_{\hat{F};\ovv{N}\ppr;\ovv{N}}^{(\alpha;\kappa),bb}(x_{p})\bigg)\bigg\}_{\mbox{.}}
\eeq
We can therefore remove the 32, colour dressed self-energies \(\delta\wt{\Sigma}_{M;N}^{(\alpha;\kappa)ab}(x_{p})\)
by inserting the effective functional (\ref{s3_62}) into (\ref{s3_53}). Furthermore, it has to be considered that the
background path integral (\ref{s3_54}) has been transformed by separating effective one-particle potentials
\(\hat{H}_{N;M}(x_{p})\), \(\hat{H}_{N;M}^{T}(x_{p})\) with \(\mcal{V}_{\beta}^{\mu}(x_{p})\) (\ref{s3_60}),
also containing the quark self-energy densities \(\sigma_{D}^{(\alpha;\kappa)}(x_{p})\), to the new
background generating function (\ref{s3_59}). These transformations require adjustment of the anomalous doubled determinant (\ref{s3_57})
and bilinear source term (\ref{s3_58}) with the simplified matrix \(\wt{\mscr{M}}_{N;M}^{ba}(y_{q},x_{p})\) (\ref{s3_61}).
The transformations (\ref{s3_55}-\ref{s3_58}) finally yield the generating function (\ref{s3_63}) whose most important
dependence is given by the single, anomalous doubled self-energy \(\delta\wt{\Sigma}_{M;N}^{ab}(x_{p})\) with anti-hermitian
BCS terms. The transformations (\ref{s3_55}-\ref{s3_58}) cause a different background field averaging with the
path integral (\ref{s3_59},\ref{s3_60}) instead of (\ref{s3_54})
\beq \no
\lefteqn{Z[\hat{\mscr{J}},J_{\psi},\hat{J}_{\psi\psi},\hat{\mfrak{j}}^{(\hat{F})}]=
\boldsymbol{\Bigg\langle Z\Big[\feynV(x_{p});\hat{\mfrak{S}}^{(\hat{F})},s_{\alpha},\hat{\mfrak{B}}_{\hat{F}},\hat{\mfrak{b}}^{(\hat{F})},
\hat{\mscr{U}}_{\hat{F}},\hat{\mfrak{v}}_{\hat{F}};\sigma_{D};\hat{\mfrak{j}}^{(\hat{F})};
\mbox{\bf Eq. (\ref{s3_59})}\Big]\;\times }\int d[\delta\wt{\Sigma}_{M;N}^{ab}(x_{p})]\;\times  }  \\ \lb{s3_63} &\times&
\bigg\{\prod_{\{x_{p}\}}\boldsymbol{\Delta}
\bigg(\delta\wt{\Sigma}_{f,\ovv{M};g,\ovv{N}}^{ab}(x_{p})+2\:\im\hat{J}_{\psi\psi;f,\ovv{M};g,\ovv{N}}^{a\neq b}(x_{p})\:;\:
\hat{\mscr{U}}_{\hat{F};\ovv{M};\ovv{N}}^{(\alpha;\kappa)}(x_{p})\,;\,\hat{\mfrak{v}}_{\hat{F};\ovv{N}}^{(\alpha;\kappa)}(x_{p})\,;\,
\hat{\mfrak{b}}_{(\alpha;\kappa)}^{(\hat{F})}(x_{p})\bigg)\bigg\} \times \\ \no &\times&
\exp\Bigg\{\frac{1}{2}\hspace*{0.6cm}\trxpa\Bigg[\trfgamc\Bigg(\ln\Big[\wt{\mscr{M}}_{N;M}^{ba}(y_{q},x_{p})\Big]-
\ln\bigg[\bea{cc} \hat{\mscr{H}}_{N;M}^{11}(y_{q},x_{p})  &  0  \\
0  & \hat{\mscr{H}}_{N;M}^{11,T}(y_{q},x_{p})\eea\bigg]\Bigg)\Bigg]\Bigg\} \times \\ \no &\times&
\exp\Bigg\{\frac{\im}{2}\hspace*{-0.1cm}\int_{C}\hspace*{-0.2cm}d^{4}\!x_{p}\;d^{4}\!y_{q}\;
J_{\psi;N}^{\dag,b}(y_{q})\;\hat{I}\Bigg(\bigg(
\wt{\mscr{M}}_{N;M}^{\boldsymbol{-1};ba}(y_{q},x_{p})-
\bigg(\bea{cc} \hat{\mscr{H}}_{N;M}^{11;\boldsymbol{-1}}(y_{q},x_{p})  &  0  \\
0  & \hspace*{-0.6cm}\hat{\mscr{H}}_{N;M}^{11,T;\boldsymbol{-1}}(y_{q},x_{p})\eea\bigg)\Bigg)\hat{I}\;J_{\psi;M}^{a}(x_{p})
\Bigg\}\hspace*{-0.1cm}\boldsymbol{\Bigg\rangle}_{\mbox{.}}
\eeq
Corresponding to chapters 3 and 4 of Ref.\ \cite{mies1}, the coset
decomposition is accomplished for \(\delta\wt{\Sigma}_{M;N}^{ab}(x_{p})\) by
coset matrices \(\hat{T}(x_{p})\), \(\hat{T}^{-1}(x_{p})\) and by
block diagonal self-energy densities \(\delta\hat{\Sigma}_{D;M\ppr;N\ppr}^{11}(x_{p})\),
\(\delta\hat{\Sigma}_{D;M\ppr;N\ppr}^{22}(x_{p})\) (\ref{s3_64}) which are further diagonalized to
the anomalous doubled eigenvalues \(\delta\hat{\Lambda}(x_{p})\)
with block diagonal 'eigenvector' matrices \(\hat{Q}(x_{p})\), \(\hat{Q}^{-1}(x_{p})\) (\ref{s3_65}) (compare \cite{mies1})
\beq \lb{s3_64}
\delta\hat{\Sigma}_{M;N}^{ab}(x_{p})&=&\left(\bea{cc} \delta\hat{\Sigma}_{M;N}^{11}(x_{p}) & \im\;\delta\hat{\Sigma}_{M;N}^{12}(x_{p}) \\
\im\;\delta\hat{\Sigma}_{M;N}^{21}(x_{p}) & \delta\hat{\Sigma}_{M;N}^{22}(x_{p}) \eea\right) \\  \no &=&
\Big(\hat{T}(x_{p})\Big)_{M;M\ppr}^{aa\ppr}
\left(\bea{cc} \delta\hat{\Sigma}_{D;M\ppr;N\ppr}^{11}(x_{p}) & 0 \\
0 & \delta\hat{\Sigma}_{D;M\ppr;N\ppr}^{22}(x_{p}) \eea\right)_{M\ppr;N\ppr}^{a\ppr b\ppr}
\Big(\hat{T}^{-1}(x_{p})\Big)_{N\ppr;N}^{b\ppr b}  \\ \no &=&
\hat{T}(x_{p})\;\;\hat{Q}^{-1}(x_{p})\;\;\delta\hat{\Lambda}(x_{p})\;\;\hat{Q}(x_{p})\;\;\hat{T}^{-1}(x_{p})\;; \\ \lb{s3_65}
\delta\hat{\Sigma}_{D;M;N}^{aa}(x_{p}) &=&
\hat{Q}_{M;M\ppr}^{aa;-1}(x_{p})\;\;\delta\hat{\Lambda}_{M\ppr;N\ppr}^{aa}(x_{p})\;\;\hat{Q}_{N\ppr;N}^{aa}(x_{p}) \;.
\eeq
We have to require the corresponding symmetries for block diagonal self-energy densities \(\delta\hat{\Sigma}_{D;M;N}^{aa}(x_{p})\)
following from the symmetries of densities \(\delta\hat{\Sigma}_{M;N}^{aa}(x_{p})\) of the original, anomalous doubled, single
self-energy \(\delta\wt{\Sigma}_{M;N}^{ab}(x_{p})\) (\ref{s3_44},\ref{s3_45})
\be  \lb{s3_66}
\bea{rclrcl}
\delta\hat{\Sigma}_{M;N}^{11,\dag}(x_{p}) &=&\delta\hat{\Sigma}_{M;N}^{11}(x_{p})\;;& \hspace*{0.6cm}
\delta\hat{\Sigma}_{M;N}^{22,\dag}(x_{p}) &=&\delta\hat{\Sigma}_{M;N}^{22}(x_{p})\;; \\
\delta\hat{\Sigma}_{M;N}^{22}(x_{p}) &=&-\delta\hat{\Sigma}_{M;N}^{11,T}(x_{p})\;; & && \\
\delta\hat{\Sigma}_{D;M;N}^{11,\dag}(x_{p}) &=&\delta\hat{\Sigma}_{D;M;N}^{11}(x_{p})\;;& \hspace*{0.6cm}
\delta\hat{\Sigma}_{D;M;N}^{22,\dag}(x_{p}) &=&\delta\hat{\Sigma}_{D;M;N}^{22}(x_{p})\;; \\
\delta\hat{\Sigma}_{D;M;N}^{22}(x_{p}) &=&-\delta\hat{\Sigma}_{D;M;N}^{11,T}(x_{p})\;. & &&
\eea
\ee
Proceeding with chapters 3, 4 of Ref.\ \cite{mies1}, the anomalous doubled
eigenvalues \(\delta\hat{\Lambda}^{a}(x_{p})\) (\ref{s3_67},\ref{s3_68})
of block diagonal self-energy densities \(\delta\hat{\Sigma}_{D;M;N}^{aa}(x_{p})\)
(\ref{s3_65},\ref{s3_66}) are given by diagonal matrices
\(\mbox{diag}\{(a=1):\,+\delta\hat{\lambda}_{N_{0}\times N_{0}}(x_{p})\:
\boldsymbol{;}\:(a=2):\,-\delta\hat{\lambda}_{N_{0}\times N_{0}}(x_{p})\}\)
with dimension \(N_{0}=N_{f}\cdot 4_{\hat{\gamma}}\cdot (N_{c}=3)\), (\(N_{0}=24,\:(36)\) for isospin- (flavour-)
degrees of freedom). The block diagonal, diagonalizing 'eigenvector' matrices \(\hat{Q}_{N_{0}\times N_{0}}^{a=b}(x_{p})\) (\ref{s3_69})
consist of the hermitian generator \(\hat{\mscr{F}}_{D;N_{0}\times N_{0}}(x_{p})\) (\ref{s3_70}) with vanishing diagonal
because these degrees of freedom are already contained in the eigenvalues \(\delta\hat{\lambda}_{N_{0}\times N_{0}}(x_{p})\).
Since the '22' self-energy density block \(\delta\hat{\Sigma}_{D;M;N}^{22}(x_{p})\),
(respectively \(\delta\hat{\Sigma}_{M;N}^{22}(x_{p})\)),  is equivalent to the negative, transposed
'11' self-energy density \(-\delta\hat{\Sigma}_{D;M;N}^{11,T}(x_{p})\)
(\(-\delta\hat{\Sigma}_{M;N}^{11,T}(x_{p})\)), we have to require symmetries (\ref{s3_69}) and have to
construct \(\hat{Q}_{N_{0}\times N_{0}}^{22}(x_{p})\) by the negative, transposed generator
\(-\hat{\mscr{F}}_{D;N_{0}\times N_{0}}^{T}(x_{p})\) (\ref{s3_71})
\beq \lb{s3_67}
\delta\hat{\Lambda}^{ab}(x_{p}) &=&\delta_{ab}\;
\mbox{diag}\Big\{\underbrace{\delta\hat{\lambda}_{N_{0}\times N_{0}}(x_{p})}_{a=1}\;\boldsymbol{;}\;
\underbrace{-\delta\hat{\lambda}_{N_{0}\times N_{0}}(x_{p})}_{a=2}\Big\} \;;\hspace*{0.1cm}
N_{0}=N_{f}\cdot 4_{\hat{\gamma}} \cdot (N_{c}=3)\;;  \\ \lb{s3_68}
\delta\hat{\lambda}_{N_{0}\times N_{0}}(x_{p})&=&\mbox{diag}\Big\{\delta\hat{\lambda}_{1}(x_{p}),\;\ldots\;,
\delta\hat{\lambda}_{N_{0}}(x_{p})\Big\}\;;      \\     \lb{s3_69}
\hat{Q}_{N_{0}\times N_{0}}^{ab}(x_{p}) &=&\left(\bea{cc} \hat{Q}_{N_{0}\times N_{0}}^{11}(x_{p}) & 0 \\
0 & \hat{Q}_{N_{0}\times N_{0}}^{22}(x_{p}) \eea\right)^{ab} \;; \\  \no
\Big(\hat{Q}_{N_{0}\times N_{0}}^{22}(x_{p})\Big)^{T} &=&\hat{Q}_{N_{0}\times N_{0}}^{11,\dag}(x_{p}) =
\hat{Q}_{N_{0}\times N_{0}}^{11,-1}(x_{p}) \;; \\  \lb{s3_70}
\hat{Q}_{N_{0}\times N_{0}}^{11}(x_{p}) &=& \exp\Big\{\im\;\hat{\mcal{F}}_{D;N_{0}\times N_{0}}(x_{p})\Big\}\;;
\hspace*{0.3cm}\hat{\mcal{F}}_{D;N_{0}\times N_{0}}^{\dag}(x_{p})=\hat{\mcal{F}}_{D;N_{0}\times N_{0}}(x_{p})\;;  \\  \lb{s3_71}
\hat{Q}_{N_{0}\times N_{0}}^{22}(x_{p}) &=& \exp\Big\{-\im\;\hat{\mcal{F}}_{D;N_{0}\times N_{0}}^{T}(x_{p})\Big\}\;;
\hspace*{0.3cm}\hat{\mcal{F}}_{D;ii}(x_{p})=0\;;(i=1,\ldots,N_{0})\;.
\eeq
In analogy to Ref.\ \cite{mies1}, the coset matrices \(\hat{T}_{M;N}^{ab}(x_{p})\) are specified by the
generator \(\hat{Y}_{M;N}^{ab}(x_{p})\) with anti-symmetric sub-generators \(\hat{X}_{M;N}(x_{p})\), \(\hat{X}_{M;N}^{\dag}(x_{p})\)
of complex commuting variables for the BCS degrees of freedom
\beq \lb{s3_72}
\hat{T}_{M;N}^{ab}(x_{p})&=&\Big(\exp\Big\{-\hat{Y}_{M\ppr;N\ppr}^{a\ppr b\ppr}(x_{p})\Big\}\Big)_{M;N}^{ab} \;; \\ \lb{s3_73}
\hat{Y}_{M;N}^{ab}(x_{p}) &=& \left(\bea{cc} 0 & \hat{X}_{M;N}(x_{p}) \\
\hat{X}_{M;N}^{\dag}(x_{p}) &  0 \eea\right)^{a\neq b}\;;\hspace*{0.3cm}\hat{X}_{M;N}^{T}(x_{p})=-\hat{X}_{M;N}(x_{p})\;.
\eeq
The generator \(\hat{Y}_{2N_{0}\times 2N_{0}}(x_{p})\) (\ref{s3_73}) with anti-symmetric sub-generators \(\hat{X}_{N_{0}\times N_{0}}(x_{p})\),
\(\hat{X}_{N_{0}\times N_{0}}^{\dag}(x_{p})\) is supplementary decomposed into block diagonal matrices
\(\hat{P}_{2N_{0}\times 2N_{0}}(x_{p})=\hat{P}_{N_{0}\times N_{0}}^{aa}(x_{p})\) (\ref{s3_74})
and into \(\hat{Y}_{DD;2N_{0}\times 2N_{0}}(x_{p})\) (\ref{s3_75})
with anti-symmetric, quaternion-valued, diagonal elements \(\hat{X}_{DD;N_{0}\times N_{0}}(x_{p})\),
\(\hat{\ovv{f}}_{N_{0}\times N_{0}}(x_{p})\) (\ref{s3_76},\ref{s3_77}).
The latter, quaternion-eigenvalues refer with the anti-symmetric
Pauli matrix \((\tau_{2})_{gf}\) (with the anti-symmetric parts of
the \(\mbox{SU}_{f}(N_{f}=3)\) Gell-Mann matrices) to isospin- (flavour-)
degrees of freedom where the complex, eigenvalue parameters
\(\ovv{f}_{\ovv{M}}(x_{p})\) (\ref{s3_78}) are labeled by the collective
index \(\ovv{M}=\{m,r\}\) of gamma- \(\hat{\gamma}_{mn}^{(\mu)}\)
and colour-matrices \(\hat{t}_{\alpha;rs}\)
\beq  \lb{s3_74}
\hat{Y}_{2N_{0}\times 2N_{0}}(x_{p}) &=&\hat{P}_{2N_{0}\times 2N_{0}}^{-1}(x_{p})\;\:
\hat{Y}_{DD;2N_{0}\times 2N_{0}}(x_{p})\;\:\hat{P}_{2N_{0}\times 2N_{0}}(x_{p}) \;; \\  \lb{s3_75}
\hat{Y}_{DD;2N_{0}\times 2N_{0}}(x_{p}) &=& \left(\bea{cc} 0 & \hat{X}_{DD;N_{0}\times N_{0}}(x_{p}) \\
\hat{X}_{DD;N_{0}\times N_{0}}^{\dag}(x_{p}) &  0 \eea\right)  \;;  \\  \lb{s3_76}
\hat{X}_{DD;N_{0}\times N_{0}}(x_{p})  &=& \hat{\ovv{f}}_{N_{0}\times N_{0}}(x_{p})
= \hat{\ovv{f}}_{g,\ovv{N};f,\ovv{M}}(x_{p})=\big(\tau_{2}\big)_{gf}\;\delta_{\ovv{N};\ovv{M}}\;\ovv{f}_{\ovv{M}}(x_{p})\;; \\  \lb{s3_77}
\hat{\ovv{f}}_{N_{0}\times N_{0}}(x_{p}) &=&
\mbox{diag}\Big\{\big(\tau_{2}\big)_{gf}\:\ovv{f}_{1}(x_{p}),\,\ldots\,,\big(\tau_{2}\big)_{gf}\:\ovv{f}_{\ovv{M}}(x_{p}),\,\ldots\,,
\big(\tau_{2}\big)_{gf}\:\ovv{f}_{N_{0}/2}(x_{p})\Big\}  \;;  \\  \lb{s3_78}
\ovv{f}_{\ovv{M}}(x_{p}) &=& \big|\ovv{f}_{\ovv{M}}(x_{p})\big|\;\exp\big\{\im\;\phi_{\ovv{M}}(x_{p})\big\}\;;\hspace*{0.3cm}
\big(\ovv{f}_{\ovv{M}}(x_{p})\in\mbox{\sf C}\big)\;; \\ \no &&
(\ovv{M}=1,\,\ldots\,,N_{0}/2)\;;\;(g,f=\mbox{up, down, (strange))}\;; \\ \no \mbox{(inclusion of strangeness} &\rightarrow&
\mbox{with anti-symmetric $\mbox{SU}_{f}(N_{f}=3)$ matrices and $N_{0}/3$ instead of $N_{0}/2$!)}\;.
\eeq
The block diagonal eigenvector-matrices
\(\hat{P}_{N_{0}\times N_{0}}^{aa}(x_{p})\) (\ref{s3_79}) of the coset generators
\(\hat{Y}_{M;N}^{ab}(x_{p})\), \(\hat{X}_{M;N}(x_{p})\), \(\hat{X}_{M;N}^{\dag}(x_{p})\)
(\ref{s3_72},\ref{s3_73}) have to fulfill symmetries
(\ref{s3_80}) with a hermitian, quaternion-valued generator
\(\hat{\mscr{G}}_{D;N_{0}\times N_{0}}(x_{p})\) (\ref{s3_81})
whose negative transposition \(-\hat{\mscr{G}}_{D;N_{0}\times N_{0}}^{T}(x_{p})\)
yields the suitable generator for the '22' block
\(\hat{P}_{N_{0}\times N_{0}}^{22}(x_{p})\) (\ref{s3_82}).
Since \(N_{0}/2\) complex parameters are already contained
in the quaternion-valued, anti-symmetric eigenvalues
\(\hat{\ovv{f}}_{N_{0}\times N_{0}}(x_{p})\) (\ref{s3_77},\ref{s3_78}),
the analogous, quaternion-valued, diagonal matrix elements
of \(\hat{\mcal{G}}_{D;f,\ovv{M};g,\ovv{M}}(x_{p})\)
have to vanish completely in the isospin- (flavour-) indices \(f,\,g=\mbox{up, down, (strange)}\)
\beq \lb{s3_79}
\hat{P}_{2N_{0}\times 2N_{0}}(x_{p}) &=&
\left(\bea{cc} \hat{P}_{N_{0}\times N_{0}}^{11}(x_{p}) & 0 \\
0 & \hat{P}_{N_{0}\times N_{0}}^{22}(x_{p}) \eea\right)^{ab}  \\ \lb{s3_80}
\Big(\hat{P}_{N_{0}\times N_{0}}^{22}(x_{p})\Big)^{T} &=&\hat{P}_{N_{0}\times N_{0}}^{11,\dag}(x_{p}) =
\hat{P}_{N_{0}\times N_{0}}^{11,-1}(x_{p}) \;; \\   \lb{s3_81}
\hat{P}_{N_{0}\times N_{0}}^{11}(x_{p}) &=& \exp\Big\{\im\;\hat{\mcal{G}}_{D;N_{0}\times N_{0}}(x_{p})\Big\}\;;
\hspace*{0.3cm}\hat{\mcal{G}}_{D;N_{0}\times N_{0}}^{\dag}(x_{p})=\hat{\mcal{G}}_{D;N_{0}\times N_{0}}(x_{p})\;;  \\ \lb{s3_82}
\hat{P}_{N_{0}\times N_{0}}^{22}(x_{p}) &=& \exp\Big\{-\im\;\hat{\mcal{G}}_{D;N_{0}\times N_{0}}^{T}(x_{p})\Big\}\;; \\  \lb{s3_83}
\hat{\mscr{G}}_{D;f\ovv{M};g\ovv{M}}(x_{p})&=&0\;;\;\;(\ovv{M}=1,\,\ldots\,,N_{0}/2)\;;\;(g,f=\mbox{up, down, (strange with $N_{0}/3$ !)})\;.
\eeq
We apply the coset decomposition of \(\delta\wt{\Sigma}_{M;N}^{ab}(x_{p})\) in relations (\ref{s3_64}-\ref{s3_78})
to the matrix \(\wt{\mscr{M}}_{N;M}^{ba}(y_{q},x_{p})\) (\ref{s3_61}) on which a similarity transformation is performed by
\(\hat{T}(y_{q})\), \(\hat{T}^{-1}(x_{p})\) and which comprises an effective potential
\(\feynV(x_{p})\) (\ref{s3_60}) abbreviating the sum of the gauge fields \(\hat{\mfrak{S}}_{\alpha;\mu\nu}^{(\hat{F})}(x_{p})\),
\(\hat{\mfrak{B}}_{\hat{F};\beta\alpha}^{\mu\kappa}(x_{p})\) and quark self-energy densities \(\sigma_{D}^{(\alpha;\kappa)}(x_{p})\)
\beq  \no
\lefteqn{\wt{\mscr{M}}_{N;M}^{ba}(y_{q},x_{p})=
\hat{T}_{N;N\ppr}^{bb\ppr}(y_{q})\Bigg\{\hat{T}_{N\ppr;N_{1}}^{-1;b\ppr b_{1}}(y_{q})\:\hat{I}\:\hat{S}\:\eta_{q}\:
\frac{\hat{\mscr{J}}_{N_{1};M_{1}}^{b_{1}a_{1}}(y_{q},x_{p})}{\mcal{N}}\:\eta_{p}\:\hat{S}\:\hat{I}\:\hat{T}_{M_{1};M\ppr}^{a_{1}a\ppr}(x_{p})+
\delta^{(4)}(y_{q}-x_{p})\;\eta_{q}\,\delta_{pq}\;\times  }   \\ \no &\times&  \Bigg[\hat{T}^{-1;b\ppr b_{1}}_{N\ppr;N_{1}}(x_{p})\Bigg(\bea{cc}
\big[\hat{\beta}(\feynd_{p}+\im\:\feynV(x_{p})-\im\:\hat{\ve}_{p}+\hat{m})\big]_{N_{1};M_{1}} & 0 \\ 0 & \hspace*{-0.3cm}
\big[\hat{\beta}(\feynd_{p}+\im\:\feynV(x_{p})-\im\:\hat{\ve}_{p}+\hat{m})\big]_{N_{1};M_{1}}^{T} \eea\Bigg)^{b_{1}a_{1}}
\hspace*{-0.3cm}\hat{T}_{M_{1};M\ppr}^{a_{1}a\ppr}(x_{p})+ \\ \lb{s3_84} &+&
\frac{1}{2}\Bigg(\bea{cc} \delta\hat{\Sigma}_{D;N\ppr;M\ppr}^{11}(x_{p}) & 0 \\ 0 &
\delta\hat{\Sigma}_{D;N\ppr;M\ppr}^{22}(x_{p}) \eea\Bigg)^{b\ppr a\ppr}\Bigg]\Bigg\}\hat{T}_{M\ppr;M}^{-1;a\ppr a}(x_{p}) \;.
\eeq
Taking into account the change of integration measure of the coset decomposition, we attain the generating function (\ref{s3_85})
with matrix \(\wt{\mscr{M}}_{N;M}^{ba}(y_{q},x_{p})\) (\ref{s3_84}) and with an additional polynomial
\(\mfrak{P}_{5}(\delta\hat{\lambda}(x_{p})\,)\) of the eigenvalues \(\delta\hat{\lambda}(x_{p})\) (\ref{s3_67},\ref{s3_68})
following from the Jacobian for the new integration variables
\beq \lb{s3_85}
\lefteqn{Z[\hat{\mscr{J}},J_{\psi},\hat{J}_{\psi\psi},\hat{\mfrak{j}}^{(\hat{F})}]=
\boldsymbol{\Bigg\langle Z\Big[\feynV(x_{p});\hat{\mfrak{S}}^{(\hat{F})},s_{\alpha},\hat{\mfrak{B}}_{\hat{F}},\hat{\mfrak{b}}^{(\hat{F})},
\hat{\mscr{U}}_{\hat{F}},\hat{\mfrak{v}}_{\hat{F}};\sigma_{D};\hat{\mfrak{j}}^{(\hat{F})};
\mbox{\bf Eq. (\ref{s3_59})}\Big]\;\times } }  \\ \no \no &\times&
\int d[\hat{T}^{-1}(x_{p})\;d\hat{T}(x_{p})]\;\;\int d[\delta\hat{\Sigma}_{D}(x_{p})]\;\;\mfrak{P}_{5}(\delta\hat{\lambda}(x_{p})\,)
\;\times  \\   \no   &\times&
\bigg\{\prod_{\{x_{p}\}}\boldsymbol{\Delta}
\bigg(\delta\wt{\Sigma}_{f,\ovv{M};g,\ovv{N}}^{ab}(x_{p})+2\:\im\hat{J}_{\psi\psi;f,\ovv{M};g,\ovv{N}}^{a\neq b}(x_{p})\:;\:
\hat{\mscr{U}}_{\hat{F};\ovv{M};\ovv{N}}^{(\alpha;\kappa)}(x_{p})\,;\,\hat{\mfrak{v}}_{\hat{F};\ovv{N}}^{(\alpha;\kappa)}(x_{p})\,;\,
\hat{\mfrak{b}}_{(\alpha;\kappa)}^{(\hat{F})}(x_{p})\bigg)\bigg\} \times \\ \no &\times& \hspace*{-0.2cm}
\exp\Bigg\{\frac{1}{2}\hspace*{0.6cm}\trxpa\Bigg[\trfgamc\Bigg(\ln\Big[\wt{\mscr{M}}_{N;M}^{ba}(y_{q},x_{p})\Big]-
\ln\bigg[\bea{cc} \hat{\mscr{H}}_{N;M}^{11}(y_{q},x_{p})  &  0  \\
0  & \hat{\mscr{H}}_{N;M}^{11,T}(y_{q},x_{p})\eea\bigg]\Bigg)\Bigg]\Bigg\} \times \\ \no &\times& \hspace*{-0.2cm}
\exp\Bigg\{\frac{\im}{2}\hspace*{-0.05cm}\int_{C}\hspace*{-0.2cm}d^{4}\!x_{p}\;d^{4}\!y_{q}\;
J_{\psi;N}^{\dag,b}(y_{q})\,\hat{I}\Bigg(\bigg(
\wt{\mscr{M}}_{N;M}^{\boldsymbol{-1};ba}(y_{q},x_{p})-
\bigg(\bea{cc} \hat{\mscr{H}}_{N;M}^{11;\boldsymbol{-1}}(y_{q},x_{p})  &  0  \\
0  & \hspace*{-0.6cm}\hat{\mscr{H}}_{N;M}^{11,T;\boldsymbol{-1}}(y_{q},x_{p})\eea\bigg)\Bigg)\hat{I}\,J_{\psi;M}^{a}(x_{p})
\Bigg\}\hspace*{-0.1cm}\boldsymbol{\Bigg\rangle}_{\mbox{.}}
\eeq

\subsection{Separation of 'hinge' fields from BCS pair condensate terms} \lb{s34}

It is the aim to derive an effective Lagrangian with BCS related field degrees of freedom so that we have to remove
the block diagonal self-energy densities or 'hinge' fields of the spontaneous symmetry breaking. This has to be combined
with the coset decomposition \(\mbox{SO}(24\:\boldsymbol{,}\:24)\,/\,\mbox{U}(24)\otimes\mbox{U}(24)\) (for the case
with 'up', 'down' isopsin degrees of freedom) of section \ref{s33} where we have performed a factorization (\ref{s3_84})
of the matrix \(\wt{\mscr{M}}_{N;M}^{ba}(y_{q},x_{p})\) into anomalous field degrees of freedom with coset matrices
\(\hat{T}(y_{q})\), \(\hat{T}^{-1}(x_{p})\) and with block diagonal self-energy densities \(\delta\hat{\Sigma}_{D;N;M}^{aa}(x_{p})\)
or 'hinge' fields. We symbolically abbreviate this factorization (\ref{s3_84}) by Eqs. (\ref{s3_86}-\ref{s3_90}) and introduce
the gradient term \(\hat{T}^{-1}\:\hat{\mscr{H}}\:\hat{T}-\hat{\mscr{H}}\) with anomalous doubled one-particle part
\(\hat{\mscr{H}}_{N;M}^{ba}(y_{q},x_{p})\) (\ref{s3_87},\ref{s3_88}) which also includes the potential part
\(\langle\feynV(x_{p})\rangle\) (\ref{s3_60})
\beq\lb{s3_86}
\lefteqn{\wt{\mscr{M}}_{N;M}^{ba}(y_{q},x_{p})=\Big[\hat{\mscr{H}}+\wt{\mscr{J}}+\frac{1}{2}\hat{T}\;\delta\hat{\Sigma}_{D}\;
\hat{T}^{-1}\Big]_{N;M}^{ba}(y_{q},x_{p}) =
\Big[\hat{\mscr{H}}+\wt{\mscr{J}}+\frac{1}{2}\hat{T}\;\hat{Q}^{-1}\;\delta\hat{\Lambda}\;
\hat{Q}\;\hat{T}^{-1}\Big]_{N;M}^{ba}(y_{q},x_{p})} \\ \no &=&
\hat{T}(y_{q})\Bigg[\hat{\mscr{H}}+\Big(\hat{T}^{-1}\;\hat{\mscr{H}}\;\hat{T}-\hat{\mscr{H}}\Big)+
\wt{\mscr{J}}(\hat{T}^{-1},\hat{T}) + \frac{1}{2}
\Bigg(\bea{cc}\delta\hat{\Sigma}_{D;N_{0}\times N_{0}}^{11} & 0 \\ 0 & \delta\hat{\Sigma}_{D;N_{0}\times N_{0}}^{22}\eea\Bigg)
\Bigg]_{N\ppr;M\ppr}^{b\ppr a\ppr}\hspace*{-0.5cm}(y_{q},x_{p})\;\;\hat{T}^{-1}(x_{p})\;;
\eeq\vspace*{-0.6cm}
\beq \lb{s3_87}
\hat{\mscr{H}}_{N;M}^{ba}(y_{q},x_{p}) &=&\delta_{pq}\;\eta_{q}\;\delta^{(4)}(y_{q}-x_{p})
\Bigg(\bea{cc} \hat{H}_{N;M}(x_{p}) & \\ & \hat{H}_{N;M}^{T}(x_{p}) \eea\Bigg)^{ba} \;;  \\ \lb{s3_88}
\hat{H}_{N;M}(x_{p}) &=& \big[\hat{\beta}\,(\feynd_{p}+\im\:\feynV(x_{p})-\im\,\hat{\ve}_{p}+\hat{m}\,)\big]_{N;M} \;; \\ \lb{s3_89}
\wt{\mscr{J}}_{N;M}^{ba}(y_{q},x_{p}) &=&\hat{I}\;\hat{S}\;\eta_{q}\;\frac{\hat{\mscr{J}}_{N;M}^{ba}(y_{q},x_{p})}{\mcal{N}}\;
\eta_{p}\;\hat{S}\;\hat{I} \;; \\  \lb{s3_90}
\wt{\mscr{J}}_{N;M}^{ba}\big(\hat{T}^{-1}(y_{q})\,,\,\hat{T}(x_{p})\big) &=&
\Big(\hat{T}^{-1}(y_{q})\;\hat{I}\;\hat{S}\;\eta_{q}\;
\frac{\hat{\mscr{J}}_{N\ppr;M\ppr}^{b\ppr a\ppr}(y_{q},x_{p})}{\mcal{N}}\;
\eta_{p}\;\hat{S}\;\hat{I} \;\hat{T}(x_{p})\Big)_{N;M}^{ba}\;.
\eeq
The subsequent steps (\ref{s3_48}-\ref{s3_50}), which have transformed the matrix \(\wt{M}_{N;M}^{ba}(y_{q},x_{p})\) (\ref{s3_47})
to \(\wt{\mscr{M}}_{N;M}^{ba}(y_{q},x_{p})\) (\ref{s3_51}) with anti-hermitian anomalous parts, are inverted by the operations
in relation (\ref{s3_91})
\beq \lb{s3_91}
\wt{\mscr{M}}_{N;M}^{ba}(y_{q},x_{p})\hspace*{-0.2cm}&=&\hspace*{-0.2cm}\bigg(\hat{T}(y_{q})\;\hat{I}\;\hat{S}
\underbrace{\Big(\hat{S}\;\hat{I}^{-1}\;
\wt{\mscr{N}}_{N\ppr;M\ppr}^{b\ppr a\ppr}(y_{q},x_{p};
\delta\hat{\Sigma}_{D})\;\hat{I}^{-1}\;\hat{S}\Big)}_{
\wt{N}_{N\ppr;M\ppr}^{b\ppr a\ppr}\big(y_{q},x_{p};\,\hat{I}^{-1}\:\delta\hat{\Sigma}_{D}\:\hat{I}^{-1}\big)}
\hat{S}\;\hat{I}\;\hat{T}^{-1}(x_{p}) \bigg)_{N;M}^{ba} ; \\ \no \hspace*{-0.8cm}
\wt{N}_{N\ppr;M\ppr}^{b\ppr a\ppr}\big(y_{q},x_{p};\,\hat{I}^{-1}\:\delta\hat{\Sigma}_{D}\:\hat{I}^{-1}\big)
\hspace*{-0.2cm}&=&\hspace*{-0.2cm}
\Big(\hat{S}\;\hat{I}^{-1}\;\wt{\mscr{N}}_{N\ppr;M\ppr}^{b\ppr a\ppr}\big(y_{q},x_{p};
\delta\hat{\Sigma}_{D}\big)\;\hat{I}^{-1}\;\hat{S}\Big)  \;.
\eeq
Using the factorization (\ref{s3_84},\ref{s3_86}), one obtains a new matrix
\(\wt{N}_{N;M}^{ba}(y_{q},x_{p};\,\hat{I}^{-1}\:\delta\hat{\Sigma}_{D}\:\hat{I}^{-1})\) (\ref{s3_92}) under inclusion
of the 'hinge' fields \(\delta\hat{\Sigma}_{D;N_{0}\times N_{0}}^{11}(x_{p})\),
\(-\delta\hat{\Sigma}_{D;N_{0}\times N_{0}}^{22}(x_{p})\) (Note the minus sign before the '22' density part !)
\beq \lb{s3_92}
\wt{N}_{N;M}^{ba}\big(y_{q},x_{p};\hat{I}^{-1}\,\delta\hat{\Sigma}_{D}\,\hat{I}^{-1}\big) \hspace*{-0.2cm}&=&\hspace*{-0.2cm}
\wt{N}_{N;M}^{ba}(y_{q},x_{p}) +\frac{1}{2}\;\Bigg[\hat{S}\Bigg(\bea{cc} \delta\hat{\Sigma}_{D;N_{0}\times N_{0}}^{11} & 0 \\ 0 &
-\delta\hat{\Sigma}_{D;N_{0}\times N_{0}}^{22} \eea\Bigg)\hat{S}\Bigg]_{N;M}^{ba}\hspace*{-0.5cm}(y_{q},x_{p}) \; .
\eeq
We separate the 'hinge' degrees of freedom \(\delta\hat{\Sigma}_{D;N_{0}\times N_{0}}^{aa}(x_{p})\) in
\(\wt{N}_{N;M}^{ba}(y_{q},x_{p};\hat{I}^{-1}\,\delta\hat{\Sigma}_{D}\,\hat{I}^{-1})\) (\ref{s3_92}) and define
new matrices \(\wt{N}_{N;M}^{ba}(y_{q},x_{p})\), \(\hat{\mscr{O}}_{N;M}^{ba}(y_{q},x_{p})\) (\ref{s3_93})
which only comprise BCS related field degrees of freedom in the coset matrices
\(\hat{T}^{-1}(y_{q})\), \(\hat{T}(x_{p})\) with anomalous doubled one-particle and potential part
 \(\feynV(x_{p})\) in \(\hat{\mscr{H}}_{N;M}^{ba}(y_{q},x_{p})\) (\ref{s3_87})
\beq \lb{s3_93}
\wt{N}_{N;M}^{ba}(y_{q},x_{p}) &=&\hat{S}\;\hat{I}^{-1}\;\hat{\mscr{O}}_{N;M}^{ba}(y_{q},x_{p})\;\hat{I}   =
\hat{S}\Bigg\{\hat{\mscr{H}}+\Big[\Big(\hat{T}(y_{q})\;\hat{I}\Big)^{-1}\hat{\mscr{H}}\Big(\hat{T}(x_{p})\;\hat{I}\Big)-
\hat{\mscr{H}}\Big]+ \\ \no &+&
\Big(\hat{T}(y_{q})\;\hat{I}\Big)^{-1}\hat{I}\;\hat{S}\;\eta_{q}\frac{\hat{\mscr{J}}_{N\ppr;M\ppr}^{b\ppr a\ppr}(y_{q},x_{p})}{\mcal{N}}
\eta_{p}\;\hat{S}\;\hat{I}\Big(\hat{T}(x_{p})\;\hat{I}\Big)\Bigg\}_{N;M}^{ba}\hspace*{-0.5cm}(y_{q},x_{p}) \;.
\eeq
Since we have factorized the matrix \(\wt{\mscr{M}}_{N;M}^{ba}(y_{q},x_{p})\) (\ref{s3_86}) in a kind of a 'similarity-transformation', the
determinant (\ref{s3_94}) reduces to an action
\(\mscr{A}_{DET}\ppr[\hat{T},\delta\hat{\Sigma}_{D},\feynV;\hat{\mscr{J}}]\) (\ref{s3_95})
where the block diagonal self-energy densities with additional minus in the '22' part appear as a summand without the coset matrices
\(\hat{T}^{-1}(x_{p})\), \(\hat{T}(x_{p})\) of the BCS field degrees of freedom
\footnote{The actions \(\mscr{A}_{DET}\ppr[\hat{T},\delta\hat{\Sigma}_{D},\feynV;\hat{\mscr{J}}]\) (\ref{s3_95}),
\(\mscr{A}_{J_{\psi}}\ppr[\hat{T},\delta\hat{\Sigma}_{D},\feynV;\hat{\mscr{J}}]\) (\ref{s3_97}) are additionally denoted
by a prime '$\ppr$' in order to point out the missing one-particle potential parts which have been moved to the
background path integral (\ref{s3_59}) (Compare Eqs. (\ref{s3_57}-\ref{s3_60}) ).}
\beq \lb{s3_94}
\mbox{DET}\Big\{\wt{\mscr{M}}_{N;M}^{ba}(y_{q},x_{p})\Big\}^{1/2} &=&
\mbox{DET}\Big\{\hat{T}(y_{q})\;\hat{I}\;\hat{S}\;
\wt{N}_{N\ppr;M\ppr}^{b\ppr a\ppr}(y_{q},x_{p};\,\hat{I}^{-1}\:\delta\hat{\Sigma}_{D}\:\hat{I}^{-1})\;
\hat{S}\;\hat{I}\;\hat{T}^{-1}(x_{p}) \Big\}^{1/2}  \\ \no &=&
\mbox{DET}\Big\{\wt{N}_{N\ppr;M\ppr}^{b\ppr a\ppr}(y_{q},x_{p};\,\hat{I}^{-1}\:\delta\hat{\Sigma}_{D}\:\hat{I}^{-1})\Big\}^{1/2} =
\exp\Big\{\mscr{A}_{DET}\ppr\big[\hat{T},\delta\hat{\Sigma}_{D},\feynV;\hat{\mscr{J}}\big]\Big\}  \;;
\eeq
\be     \lb{s3_95}
\mscr{A}_{DET}\ppr\big[\hat{T},\delta\hat{\Sigma}_{D},\feynV;\hat{\mscr{J}}\big]  = \frac{1}{2}
\hspace*{0.6cm}
\trxpa\trfgamc\hspace*{-0.2cm}\ln\Bigg[\wt{N}_{N;M}^{ba}(y_{q},x_{p})+\frac{1}{2}\hat{S}
\Bigg(\bea{cc} \delta\hat{\Sigma}_{D;N_{0}\times N_{0}}^{11} & 0 \\ 0 & -\delta\hat{\Sigma}_{D;N_{0}\times N_{0}}^{22}\eea\Bigg)
\hat{S}\Bigg]_{\mbox{.}}
\ee
Similarly, we apply the factorization (\ref{s3_86}) for the inverted matrix or Green function
\(\wt{\mscr{M}}_{N;M}^{\boldsymbol{-1};ba}(y_{q},x_{p})\) and obtain relations (\ref{s3_96},\ref{s3_97}) with the new action
\(\mscr{A}_{J_{\psi}}\ppr[\hat{T},\delta\hat{\Sigma}_{D},\feynV;\hat{\mscr{J}}]\) (\ref{s3_97})
of the bilinear source fields \(J_{\psi;N}^{\dag,b}(y_{q })\), \(J_{\psi;M}^{a}(x_{p})\)
\beq \lb{s3_96}
\wt{\mscr{M}}_{N;M}^{\boldsymbol{-1};ba}(y_{q},x_{p})&=&\hat{T}(y_{q})\;\hat{I}^{-1}\;\hat{S}\;
\wt{N}_{N\ppr;M\ppr}^{\boldsymbol{-1};b\ppr a\ppr}(y_{q},x_{p};\hat{I}^{-1}\;\delta\hat{\Sigma}_{D}\;
\hat{I}^{-1})\;\hat{S}\;\hat{I}^{-1}\;\hat{T}^{-1}(x_{p}) \;;
\eeq
\beq \lb{s3_97}
\lefteqn{\mscr{A}_{J_{\psi}}\ppr[\hat{T},\delta\hat{\Sigma}_{D},\feynV;\hat{\mscr{J}}] =
\frac{1}{2}\int_{C}d^{4}\!x_{p}\;d^{4}\!y_{q}\;
J_{\psi;N}^{\dag,b}(y_{q})\;\hat{I}\;\wt{\mscr{M}}_{N;M}^{\boldsymbol{-1};ba}(y_{q},x_{p})\;\hat{I}\;J_{\psi;M}^{a}(x_{p}) } \\ \no &=&
\frac{1}{2}\int_{C}d^{4}\!x_{p}\;d^{4}\!y_{q}\;
\underbrace{J_{\psi;N}^{\dag,b}(y_{q})\big(\hat{I}\;\hat{T}(y_{q})\;\hat{I}^{-1}\big)}_{\wt{J}_{\psi;N\ppr}^{\dag,b\ppr}(y_{q})}
\hat{S}\;\wt{N}_{N\ppr;M\ppr}^{\boldsymbol{-1};b\ppr a\ppr}(y_{q},x_{p};\hat{I}^{-1}\;\delta\hat{\Sigma}_{D}\;\hat{I}^{-1})\;\hat{S}
\underbrace{\big(\hat{I}^{-1}\;\hat{T}^{-1}(x_{p})\;\hat{I}\big)J_{\psi;M}^{a}(x_{p})}_{\wt{J}_{\psi;M\ppr}^{a\ppr}(x_{p})} \;.
\eeq
The defined source field \(\wt{J}_{\psi;M}^{a}(x_{p})\) (\ref{s3_98}) in (\ref{s3_97}), which also encompasses the coset matrices,
fulfills the appropriate property (\ref{s3_101}) for hermitian conjugation; this relation (\ref{s3_101}) follows from the
properties (\ref{s3_99},\ref{s3_100}) of the coset matrices under multiplication with \(\hat{I}\), \(\hat{I}^{-1}\)
\beq \lb{s3_98}
\wt{J}_{\psi;M}^{a}(x_{p}) &=&\Big(\big(\hat{I}^{-1}\;\hat{T}^{-1}(x_{p})\;\hat{I}\big)J_{\psi;M\ppr}^{a\ppr}(x_{p})\Big)_{M}^{a} \;;  \\ \lb{s3_99}
\Big(\wt{J}_{\psi;M}^{a}(x_{p})\Big)^{\dag}&=&\Big(J_{\psi;M\ppr}^{\dag,a\ppr}(x_{p})\;\hat{I}^{-1}\;
\big(\hat{T}^{-1}(x_{p})\big)^{\dag}\;\hat{I}\Big)_{M}^{a} \stackrel{?}{=}
\Big(J_{\psi;M\ppr}^{\dag,a\ppr}(x_{p})\big(\hat{I}\;\hat{T}(x_{p})\;\hat{I}^{-1}\big)\Big)_{M}^{a}
\;; \\ \lb{s3_100}
\hat{I}^{-1}\;\big(\hat{T}^{-1}(x_{p})\big)^{\dag}\;\hat{I} &=&\hat{I}^{-1}
\Bigg[\exp\Bigg(\bea{cc} 0 & \hat{X}(x_{p}) \\ \hat{X}^{\dag}(x_{p}) & 0 \eea\Bigg)\Bigg]^{\dag}\hat{I}  =  \hat{I}^{-1}
\exp\Bigg(\bea{cc} 0 & \hat{X}(x_{p}) \\ \hat{X}^{\dag}(x_{p}) & 0 \eea\Bigg)\hat{I}  \\ \no &=&\hat{I}
\exp\Bigg(\bea{cc} 0 & -\hat{X}(x_{p}) \\ -\hat{X}^{\dag}(x_{p}) & 0 \eea\Bigg)\hat{I}^{-1}= \hat{I}\;\hat{T}(x_{p})\;\hat{I}^{-1} \;; \\ \lb{s3_101}
\Longrightarrow\Big(\wt{J}_{\psi;M}^{a}(x_{p})\Big)^{\dag}&=&\wt{J}_{\psi;M}^{\dag,a}(x_{p}) \;.
\eeq
In correspondence to Eqs. (\ref{s3_94}-\ref{s3_101}), one achieves relations (\ref{s3_102},\ref{s3_103}) where the actions
\(\mscr{A}_{DET}\ppr\big[\hat{T},\delta\hat{\Sigma}_{D},\feynV;\hat{\mscr{J}}\big]\) (\ref{s3_95}),
\(\mscr{A}_{J_{\psi}}\ppr[\hat{T},\delta\hat{\Sigma}_{D},\feynV;\hat{\mscr{J}}]\) (\ref{s3_97}) are derived from the integrations
with bilinear, anomalous doubled, anti-commuting Fermi fields \(d[\psi_{M}^{\dag}(x_{p})\,,\,\psi_{m}(x_{p})]\).
In this manner we have inverted the various steps which have lead from the original path integral to the anomalous doubling of
quark fields and to the self-energies with densities and additional BCS terms
\beq \lb{s3_102}
\lefteqn{Z[\hat{\mscr{J}},J_{\psi},\hat{J}_{\psi\psi},\hat{\mfrak{j}}^{(\hat{F})}]=
\boldsymbol{\Bigg\langle Z\Big[\feynV(x_{p});\hat{\mfrak{S}}^{(\hat{F})},s_{\alpha},\hat{\mfrak{B}}_{\hat{F}},\hat{\mfrak{b}}^{(\hat{F})},
\hat{\mscr{U}}_{\hat{F}},\hat{\mfrak{v}}_{\hat{F}};\sigma_{D};\hat{\mfrak{j}}^{(\hat{F})};
\mbox{\bf Eq. (\ref{s3_59})}\Big]\;\times } }  \\ \no \no &\times&
\int d[\hat{T}^{-1}(x_{p})\;d\hat{T}(x_{p})]\;\;\int d[\delta\hat{\Sigma}_{D}(x_{p})]\;\;\mfrak{P}_{5}(\delta\hat{\lambda}(x_{p})\,)
\;\times  \\ \no &\hspace*{-0.7cm}\times& \hspace*{-0.7cm}
\bigg\{\hspace*{-0.1cm}\prod_{\{x_{p}\}}\boldsymbol{\Delta}
\bigg(\Big(\hat{T}(x_{p})\;\delta\hat{\Sigma}_{D}(x_{p})\;\hat{T}^{-1}(x_{p})\,\Big)_{f,\ovv{M};g,\ovv{N}}^{ab} +
2\:\im\hat{J}_{\psi\psi;f,\ovv{M};g,\ovv{N}}^{a\neq b}(x_{p});\:
\hat{\mscr{U}}_{\hat{F};\ovv{M};\ovv{N}}^{(\alpha;\kappa)}(x_{p})\,;\,\hat{\mfrak{v}}_{\hat{F};\ovv{N}}^{(\alpha;\kappa)}(x_{p})\,;\,
\hat{\mfrak{b}}_{(\alpha;\kappa)}^{(\hat{F})}(x_{p})\bigg)\bigg\}\!\! \times \\ \no &\times&
\exp\bigg\{\mscr{A}_{DET}\ppr[\hat{T},\delta\hat{\Sigma}_{D},\feynV;\hat{\mscr{J}}]\bigg\}\times
\exp\bigg\{\im\;\mscr{A}_{J_{\psi}}\ppr[\hat{T},\delta\hat{\Sigma}_{D},\feynV;\hat{\mscr{J}}]\bigg\} \times \\ \no &\times&
\exp\bigg\{-\frac{1}{2}\hspace*{0.6cm}\trxpa\trfgamc
\ln\Big[\hat{\mscr{H}}_{N;M}^{ba}(y_{q},x_{p})\Big]\bigg\} \times \\ \no &\times& \exp\bigg\{ -\frac{\im}{2}
\hspace*{-0.1cm}\int_{C}\hspace*{-0.2cm}d^{4}\!x_{p}\;d^{4}\!y_{q}\;
J_{\psi;N}^{\dag,b}(y_{q})\;\hat{I}\;
\hat{\mscr{H}}_{N;M}^{\boldsymbol{-1};ba}(y_{q},x_{p})\;\hat{I}\;J_{\psi;M}^{a}(x_{p})
\bigg\}\boldsymbol{\Bigg\rangle}_{\mbox{;}}
\eeq
\beq\lb{s3_103}
\lefteqn{Z[\hat{\mscr{J}},J_{\psi},\hat{J}_{\psi\psi},\hat{\mfrak{j}}^{(\hat{F})}]=
\boldsymbol{\Bigg\langle Z\Big[\feynV(x_{p});\hat{\mfrak{S}}^{(\hat{F})},s_{\alpha},\hat{\mfrak{B}}_{\hat{F}},\hat{\mfrak{b}}^{(\hat{F})},
\hat{\mscr{U}}_{\hat{F}},\hat{\mfrak{v}}_{\hat{F}};\sigma_{D};\hat{\mfrak{j}}^{(\hat{F})};
\mbox{\bf Eq. (\ref{s3_59})}\Big]\;\times } }  \\ \no \no &\times&
\int d[\hat{T}^{-1}(x_{p})\;d\hat{T}(x_{p})]\;\;\int d[\delta\hat{\Sigma}_{D}(x_{p})]\;\;\mfrak{P}_{5}(\delta\hat{\lambda}(x_{p})\,)
\;\times \int d[\psi_{M}^{\dag}(x_{p}),\,\psi_{M}(x_{p})]\;\times  \\   \no  &\hspace*{-0.7cm}\times& \hspace*{-0.7cm}
\bigg\{\hspace*{-0.1cm}\prod_{\{x_{p}\}}\boldsymbol{\Delta}
\bigg(\Big(\hat{T}(x_{p})\;\delta\hat{\Sigma}_{D}(x_{p})\;\hat{T}^{-1}(x_{p})\,\Big)_{f,\ovv{M};g,\ovv{N}}^{ab} +
2\:\im\hat{J}_{\psi\psi;f,\ovv{M};g,\ovv{N}}^{a\neq b}(x_{p})\:;\:
\hat{\mscr{U}}_{\hat{F};\ovv{M};\ovv{N}}^{(\alpha;\kappa)}(x_{p})\,;\,\hat{\mfrak{v}}_{\hat{F};\ovv{N}}^{(\alpha;\kappa)}(x_{p})\,;\,
\hat{\mfrak{b}}_{(\alpha;\kappa)}^{(\hat{F})}(x_{p})\bigg)\bigg\} \!\! \times \\ \no &\times&
\exp\Bigg\{-\frac{\im}{2}\int_{C}d^{4}\!x_{p}\;d^{4}\!y_{q}\;\Psi_{N}^{\dag,b}(y_{q})
\Bigg[\wt{N}_{N;M}^{ba}(y_{q},x_{p})+\frac{1}{2}\;\hat{S}\Bigg(\bea{cc}\delta\hat{\Sigma}_{D;N_{0}\times N_{0}}^{11} & 0 \\ 0 &
-\delta\hat{\Sigma}_{D;N_{0}\times N_{0}}^{22} \eea\Bigg)\hat{S}\Bigg]\Psi_{M}^{a}(x_{p})\Bigg\} \times  \\ \no &\times&
\exp\bigg\{-\frac{\im}{2}\int_{C}d^{4}\!x_{p}\;d^{4}\!y_{q}\;\wt{J}_{\psi;M}^{\dag,a}(x_{p})\;\hat{S}\;\Psi_{M}^{a}(x_{p})+
\Psi_{M}^{\dag,a}(x_{p})\;\hat{S}\;\wt{J}_{\psi;M}^{a}(x_{p})\bigg\} \times \\ \no &\times&
\exp\bigg\{-\frac{1}{2}\hspace*{0.6cm}\trxpa\trfgamc
\ln\Big[\hat{\mscr{H}}_{N;M}^{ba}(y_{q},x_{p})\Big]\bigg\} \times \\ \no &\times& \exp\bigg\{ -\frac{\im}{2}
\hspace*{-0.1cm}\int_{C}\hspace*{-0.2cm}d^{4}\!x_{p}\;d^{4}\!y_{q}\;
J_{\psi;N}^{\dag,b}(y_{q})\;\hat{I}\;
\hat{\mscr{H}}_{N;M}^{\boldsymbol{-1};ba}(y_{q},x_{p})\;\hat{I}\;J_{\psi;M}^{a}(x_{p})
\bigg\}\boldsymbol{\Bigg\rangle}  \;;     \\   \lb{s3_104}  &&
\wt{J}_{\psi;M}^{a}(x_{p}) = \Big(\hat{I}^{-1}\;\hat{T}^{-1}(x_{p})\;\hat{I}\;J_{\psi;M\ppr}^{a\ppr}(x_{p})\Big)_{M}^{a}\;;\hspace{0.6cm}
\wt{J}_{\psi;M}^{\dag,a}(x_{p}) =\Big(J_{\psi;M\ppr}^{\dag,a\ppr}(x_{p})\;\hat{I}\;\hat{T}(x_{p})\;\hat{I}^{-1}\Big)_{M}^{a} \;.
\eeq
We note that the part (\ref{s3_105}) with the 'hinge' fields \(\delta\hat{\Sigma}_{D;N;M}^{11}(x_{p})\),
\(-\delta\hat{\Sigma}_{D;N;M}^{22}(x_{p})\) and anti-commuting, anomalous doubled Fermi fields does not contribute
in the generating functions (\ref{s3_102},\ref{s3_103}); therefore, we have accomplished a {\it projection} onto the
BCS related field degrees of freedom with the coset matrices \(\hat{T}^{-1}(x_{p})\), \(\hat{T}(x_{p})\); it has to
be emphasized that this operation is not invertible according to standard properties of projections which do not allow
the construction of any inverses
\be\lb{s3_105}
\exp\Bigg\{-\frac{\im}{2}\int_{C} d^{4}\!x_{p}\;\underbrace{\Psi_{N}^{\dag,b}(x_{p})\;\frac{1}{2}\;\hat{S}
\Bigg(\bea{cc} \delta\hat{\Sigma}_{D;N;M}^{11}(x_{p}) & 0 \\ 0 & -\delta\hat{\Sigma}_{D;N;M}^{22}(x_{p}) \eea\Bigg)\hat{S}\;\Psi_{M}^{a}(x_{p})}_{\equiv 0}\Bigg\}\equiv 1 \;.
\ee
After removal of part (\ref{s3_105}) from (\ref{s3_103}), we can again perform integrations of anomalous doubled
Grassmann fields and attain the path integral (\ref{s3_106}) with sub-generating function \(Z_{\hat{J}_{\psi\psi}}[\hat{T}]\) (\ref{s3_107})
of the BCS related source field \(\hat{J}_{\psi\psi;N;M}^{b\neq a}(x_{p})\). The matrix \(\wt{N}_{N:M}^{ba}(y_{q},x_{p})\) remains
in the actions \(\mscr{A}_{DET}[\hat{T},\feynV;\hat{\mscr{J}}]\),
\(\mscr{A}_{J_{\psi}}[\hat{T},\feynV;\hat{\mscr{J}}]\) , but without any 'hinge' or block diagonal self-energy density degrees
of freedom
\beq \lb{s3_106}
Z[\hat{\mscr{J}},J_{\psi},\hat{J}_{\psi\psi},\hat{\mfrak{j}}^{(\hat{F})}] &=&
\boldsymbol{\Bigg\langle Z\Big[\feynV(x_{p});\hat{\mfrak{S}}^{(\hat{F})},s_{\alpha},\hat{\mfrak{B}}_{\hat{F}},\hat{\mfrak{b}}^{(\hat{F})},
\hat{\mscr{U}}_{\hat{F}},\hat{\mfrak{v}}_{\hat{F}};\sigma_{D};\hat{\mfrak{j}}^{(\hat{F})};
\mbox{\bf Eq. (\ref{s3_59})}\Big]\;\times }   \\ \no \no &\times&
\int d[\hat{T}^{-1}(x_{p})\;d\hat{T}(x_{p})]\;\;Z_{\hat{J}_{\psi\psi}}[\hat{T}]\;
\exp\Big\{\mscr{A}_{DET}[\hat{T},\feynV;\hat{\mscr{J}}]\Big\}\;
\exp\Big\{\im\;\mscr{A}_{J_{\psi}}[\hat{T},\feynV;\hat{\mscr{J}}]\Big\}\boldsymbol{\Bigg\rangle}\;; \\ \lb{s3_107}
Z_{\hat{J}_{\psi\psi}}[\hat{T}]&=&\int d[\delta\hat{\Sigma}_{D}(x_{p})]\;
\mfrak{P}_{5}\big(\delta\hat{\lambda}(x_{p})\big) \; \times  \\ \no \hspace*{-4.0cm}&\hspace*{-7.3cm}\times&\hspace*{-4.0cm}
\bigg\{\prod_{\{x_{p}\}}\boldsymbol{\Delta}
\bigg(\Big(\hat{T}(x_{p})\;\delta\hat{\Sigma}_{D}(x_{p})\;\hat{T}^{-1}(x_{p})\,\Big)_{f,\ovv{M};g,\ovv{N}}^{ab} \hspace*{-0.1cm}+
2\:\im\hat{J}_{\psi\psi;f,\ovv{M};g,\ovv{N}}^{a\neq b}(x_{p})\:;\:
\hat{\mscr{U}}_{\hat{F};\ovv{M};\ovv{N}}^{(\alpha;\kappa)}(x_{p})\,;\,\hat{\mfrak{v}}_{\hat{F};\ovv{N}}^{(\alpha;\kappa)}(x_{p})\,;\,
\hat{\mfrak{b}}_{(\alpha;\kappa)}^{(\hat{F})}(x_{p})\bigg)\bigg\} ;   \\      \lb{s3_108}
\mscr{A}_{DET}[\hat{T},\feynV;\hat{\mscr{J}}] &=&\frac{1}{2}\hspace*{0.6cm}\trxpa\trfgamc\bigg(
\ln\Big[\wt{N}_{N;M}^{ba}(y_{q},x_{p})\Big] - \ln\Big[\hat{\mscr{H}}_{N;M}^{ba}(y_{q},x_{p})\Big]\bigg) \;;  \\  \lb{s3_109}
\mscr{A}_{J_{\psi}}[\hat{T},\feynV;\hat{\mscr{J}}] &=&\frac{1}{2}\int_{C}d^{4}\!x_{p}\;d^{4}\!y_{q}\;\times  \\ \no
\lefteqn{\hspace*{-1.3cm}\times\;J_{\psi;N}^{\dag,b}(y_{q})\;\hat{I}\bigg(\hat{T}(y_{q})\;\hat{I}^{-1}\;\hat{S}\;
\wt{N}_{N\ppr;M\ppr}^{\boldsymbol{-1};b\ppr a\ppr}(y_{q},x_{p})\;\hat{S}\;\hat{I}^{-1}\;\hat{T}^{-1}(x_{p})-
\hat{\mscr{H}}_{N;M}^{\boldsymbol{-1};ba}(y_{q},x_{p})\bigg)\;\hat{I}\;J_{\psi;M}^{a}(x_{p}) \;. }
\eeq
The matrix \(\wt{N}_{N;M}^{ba}(y_{q},x_{p})\) is further simplified to the related matrix \(\hat{\mscr{O}}_{N;M}^{ba}(y_{q},x_{p})\)
which has the equivalent determinant (\ref{s3_112}) and similar time contour Green function (\ref{s3_113}).
We define the actions \(\mscr{A}_{DET}[\hat{T},\feynV;\hat{\mscr{J}}]\) (\ref{s3_114}),
\(\mscr{A}_{J_{\psi}}[\hat{T},\feynV;\hat{\mscr{J}}]\) (\ref{s3_115}) in terms of \(\hat{\mscr{O}}_{N;M}^{ba}(y_{q},x_{p})\)
(\ref{s3_111}) with gradient part \(\hat{T}^{-1}\,\hat{\mscr{H}}\,\hat{T}-\hat{\mscr{H}}\) and can finally
cease with the effective path integral (\ref{s3_116}) which contains the coset matrices for the BCS terms as the only
remaining field degrees of freedom. The gauge field and quark self-energy density \(\sigma_{D}^{(\alpha;\kappa)}(x_{p})\)
degrees of freedom are incorporated in the averaging with the background functional (\ref{s3_59})
and the sub-generating function \(Z_{\hat{J}_{\psi\psi}}[\hat{T}]\) (\ref{s3_107})
\beq \lb{s3_110}
\wt{N}_{N;M}^{ba}(y_{q},x_{p})&=&\hat{S}\;\hat{I}^{-1}\;\hat{\mscr{O}}_{N;M}^{ba}(y_{q},x_{p})\;\hat{I} \;; \\ \lb{s3_111}
\hat{\mscr{O}}_{N;M}^{ba}(y_{q},x_{p}) &=&\hspace*{-0.3cm}\bigg\{\hat{\mscr{H}}+\Big(\hat{T}^{-1}\hat{\mscr{H}}\hat{T}-\hat{\mscr{H}}\Big)+
\hat{T}^{-1}\;\hat{I}\;\hat{S}\;\eta_{q}
\frac{\hat{\mscr{J}}_{N\ppr;M\ppr}^{b\ppr a\ppr}(y_{q},x_{p})}{\mcal{N}}\eta_{p}\;\hat{S}\;\hat{I}\;
\hat{T}\bigg\}_{N;M}^{ba}\hspace*{-0.3cm}(y_{q},x_{p}) \;; \\  \lb{s3_112}
\mbox{DET}\Big[\wt{N}_{N;M}^{ba}(y_{q},x_{p})\Big] &=&\mbox{DET}\Big[\hat{\mscr{O}}_{N;M}^{ba}(y_{q},x_{p})\Big] \;; \\ \lb{s3_113}
\wt{N}_{N;M}^{\boldsymbol{-1};ba}(y_{q},x_{p}) &=& \hat{I}^{-1}\;\hat{\mscr{O}}_{N;M}^{\boldsymbol{-1};ba}(y_{q},x_{p})\;\hat{I}\;\hat{S} \;;  \\   \lb{s3_114}
\hspace*{-0.9cm}\mscr{A}_{DET}[\hat{T},\feynV;\hat{\mscr{J}}] &=&\frac{1}{2}\hspace*{0.6cm}\trxpa
\trfgamc\bigg(\ln\Big[\hat{\mscr{O}}_{N;M}^{ba}(y_{q},x_{p})\Big]- \ln\Big[\hat{\mscr{H}}_{N;M}^{ba}(y_{q},x_{p})\Big]\bigg)  \;; \\  \lb{s3_115}
\mscr{A}_{J_{\psi}}[\hat{T},\feynV;\hat{\mscr{J}}] &=&\frac{1}{2}\int_{C}d^{4}\!x_{p}\;d^{4}\!y_{q}\;\times  \\ \no
\lefteqn{\hspace*{-1.3cm}\times\;J_{\psi;N}^{\dag,b}(y_{q})\;\hat{I}\bigg(\hat{T}(y_{q})\;
\hat{\mscr{O}}_{N\ppr;M\ppr}^{\boldsymbol{-1};b\ppr a\ppr}(y_{q},x_{p})\;\hat{T}^{-1}(x_{p})-
\hat{\mscr{H}}_{N;M}^{\boldsymbol{-1};ba}(y_{q},x_{p})\bigg)\;\hat{I}\;J_{\psi;M}^{a}(x_{p}) \;; }
   \\   \lb{s3_116}
Z[\hat{\mscr{J}},J_{\psi},\hat{J}_{\psi\psi},\hat{\mfrak{j}}^{(\hat{F})}] &=&
\boldsymbol{\Bigg\langle Z\Big[\feynV(x_{p});\hat{\mfrak{S}}^{(\hat{F})},
s_{\alpha},\hat{\mfrak{B}}_{\hat{F}},\hat{\mfrak{b}}^{(\hat{F})},
\hat{\mscr{U}}_{\hat{F}},\hat{\mfrak{v}}_{\hat{F}};\sigma_{D};\hat{\mfrak{j}}^{(\hat{F})};
\mbox{\bf Eq. (\ref{s3_59})}\Big]\;\times }   \\ \no &\times&
\int d[\hat{T}^{-1}(x_{p})\;d\hat{T}(x_{p})]\;\;Z_{\hat{J}_{\psi\psi}}[\hat{T}]\;
\exp\Big\{\mscr{A}_{DET}[\hat{T},\feynV;\hat{\mscr{J}}]\Big\}\;
\exp\Big\{\im\;\mscr{A}_{J_{\psi}}[\hat{T},\feynV;\hat{\mscr{J}}]\Big\}\boldsymbol{\Bigg\rangle}\;.
\eeq

\section{Infinite order gradient expansion to an effective action} \lb{s4}

\subsection{Separation into path integrals of BCS terms with coset matrices and
density related parts}  \lb{s41}

Although we have performed several involved HST's to self-energies and a coset decomposition in section \ref{s3},
the finally obtained, exact relations (\ref{s3_110}-\ref{s3_116}) are remarkable because they indicate
a clear separation of the original path integral (\ref{s2_15}-\ref{s2_32}) into a density related part
with generating function (\ref{s3_59}) and into BCS degrees of freedom with coset matrices
\(\hat{T}(x_{p})=\exp\{-\hat{Y}(x_{p})\,\}\) (\ref{s3_72}-\ref{s3_83}). The composed gauge field
\(\mcal{V}_{\alpha;\mu}(x_{p})\) (\ref{s3_60}) appears in both parts of the total path integral
(\ref{s3_116}) and replaces the original gauge fields \(A_{\alpha;\mu}(x_{p})\) (\ref{s2_1}-\ref{s2_6})
with auxiliary real field \(s_{\alpha}(x_{p})\) for axial gauge fixing. One can even prove a gauge invariance
between \(\mcal{V}_{\alpha;\mu}(x_{p})\) and the coset matrices \(\hat{T}(x_{p})=\exp\{-\hat{Y}(x_{p})\,\}\)
of (\ref{s3_116}) in a classical consideration where a chosen gauge condition for the composed field
\(\mcal{V}_{\alpha;\mu}(x_{p})\) is achieved by adaption of the auxiliary real field \(s_{\gamma}(x_{p})\)
with a shift of its value (compare Eq. (\ref{s3_60})).
In the quantum mechanical case, we obtain a Ward identity of the derived path integral (\ref{s3_116})
with background field averaging (\ref{s3_59}) and with projection \(\hat{S}\) onto BCS terms in the actions
\(\mscr{A}_{DET}[\hat{T},\feynV;\hat{\mscr{J}}]\), \(\mscr{A}_{J_{\psi}}[\hat{T},\feynV;\hat{\mscr{J}}]\) (cf appendix \ref{sb}).
Although there appears no action of a field strength tensor as \(\hat{F}_{\alpha}^{\mu\nu}(x_{p})\) (\ref{s2_5})
for the composed gauge field \(\mcal{V}_{\alpha}^{\mu}(x_{p})\), a gauge invariance follows because the change of
actions with \(\delta\mcal{V}_{\alpha}^{\mu}(x_{p})\) in a gauge transformation is compensated by the change of
the coset matrices \(\hat{T}(x_{p})=\exp\{-\hat{Y}(x_{p})\,\}\). In this respect the actions
\(\mscr{A}_{DET}[\hat{T},\feynV;\hat{\mscr{J}}]\), \(\mscr{A}_{J_{\psi}}[\hat{T},\feynV;\hat{\mscr{J}}]\)
of the coset matrices replace the action of a quadratic field strength tensor for the composed gauge field.

According to the separation into density and BCS terms, we split the total path integral
\(Z[\hat{\mscr{J}},J_{\psi},\hat{J}_{\psi\psi},\hat{\mfrak{j}}^{(\hat{F})}]\) (\ref{s3_116})
into the density related generating function (\ref{s3_59}) which allows to determine a mean
field solution \(\langle\mcal{V}_{\alpha;\mu}(x_{p})\,\rangle_{\mbox{\scz eq. (\ref{s3_59})}}\);
this assumed particular solution \(\langle\mcal{V}_{\alpha;\mu}(x_{p})\,\rangle_{\mbox{\scz eq. (\ref{s3_59})}}\)
comprises real and imaginary parts where the sign of the imaginary values of
\(\langle\mcal{V}_{\alpha;\mu}(x_{p})\,\rangle_{\mbox{\scz(\ref{s3_59})}}\) has to comply with
the anti-hermitian '\(-\im\:\hat{\ve}_{p}\)' terms (\ref{s2_27}) for stable propagation of the
coset matrices \(\hat{T}(x_{p})=\exp\{-\hat{Y}(x_{p})\,\}\). This definite, fixed mean field solution
\(\langle\mcal{V}_{\alpha;\mu}(x_{p})\,\rangle_{\mbox{\scz(\ref{s3_59})}}\) is inserted into the actions
\(\mscr{A}_{DET}[\hat{T},\langle\feynV\rangle_{\mbox{\scz(\ref{s3_59})}};\hat{\mscr{J}}]\),
\(\mscr{A}_{J_{\psi}}[\hat{T},\langle\feynV\rangle_{\mbox{\scz(\ref{s3_59})}};\hat{\mscr{J}}]\),
where it enters into the one-particle operators \(\hat{H}(x_{p})\), \(\hat{H}^{T}(x_{p})\) or more
precisely into the anomalous doubled version \(\hat{\mscr{H}}(x_{p})\) (\ref{s3_87},\ref{s3_88})
\beq \lb{s4_1}
\lefteqn{Z[\hat{\mscr{J}},J_{\psi},\hat{J}_{\psi\psi},\hat{\mfrak{j}}^{(\hat{F})}] =
\boldsymbol{\Bigg\langle Z\Big[\feynV(x_{p});\hat{\mfrak{S}}^{(\hat{F})},s_{\alpha},
\hat{\mfrak{B}}_{\hat{F}},\hat{\mfrak{b}}^{(\hat{F})},
\hat{\mscr{U}}_{\hat{F}},\hat{\mfrak{v}}_{\hat{F}};\sigma_{D};\hat{\mfrak{j}}^{(\hat{F})};
\mbox{\bf Eq. (\ref{s3_59})}\Big]\;\times }  }   \\ \no  &\times&
\int d[\hat{T}^{-1}(x_{p})\;d\hat{T}(x_{p})]\;\;Z_{\hat{J}_{\psi\psi}}[\hat{T}]\;
\exp\Big\{\mscr{A}_{DET}[\hat{T},\feynV;\hat{\mscr{J}}]\Big\}\;
\exp\Big\{\im\;\mscr{A}_{J_{\psi}}[\hat{T},\feynV;\hat{\mscr{J}}]\Big\}\boldsymbol{\Bigg\rangle}  \\ \no &\approx&
\underbrace{Z\Big[\feynV(x_{p});\hat{\mfrak{S}}^{(\hat{F})},s_{\alpha},\hat{\mfrak{B}}_{\hat{F}},\hat{\mfrak{b}}^{(\hat{F})},
\hat{\mscr{U}}_{\hat{F}},\hat{\mfrak{v}}_{\hat{F}};\sigma_{D};\hat{\mfrak{j}}^{(\hat{F})};
\mbox{\bf Eq. (\ref{s3_59})}\Big]}_{\Longrightarrow\mbox{classical solution }\;
\langle\mcal{V}_{\alpha;\mu}(x_{p})\,\rangle_{\mbox{\scz(\ref{s3_59})}} }\;\times  \\ \no &\times&
\int d[\hat{T}^{-1}(x_{p})\;d\hat{T}(x_{p})]\;\;\;
\Big\langle Z_{\hat{J}_{\psi\psi}}[\hat{T}]\Big\rangle_{\mbox{\scz(\ref{s3_59})}}\;
\exp\Big\{\mscr{A}_{DET}[\hat{T},\langle\feynV\rangle_{\mbox{\scz(\ref{s3_59})}};\hat{\mscr{J}}]\Big\}\;
\exp\Big\{\im\;\mscr{A}_{J_{\psi}}[\hat{T},\langle\feynV\rangle_{\mbox{\scz(\ref{s3_59})}};\hat{\mscr{J}}]\Big\}\;;
\eeq
\beq\lb{s4_2}
\mscr{A}_{DET}[\hat{T},\langle\feynV\rangle_{\mbox{\scz(\ref{s3_59})}};\hat{\mscr{J}}] &=&\hspace*{-0.3cm}
\frac{1}{2}\hspace*{0.6cm}\trxpa
\trfgamc\bigg(\ln\Big[\Big\langle\hat{\mscr{O}}_{N;M}^{ba}(y_{q},x_{p})\Big\rangle_{\mbox{\scz(\ref{s3_59})}}\Big]-
\ln\Big[\Big\langle\hat{\mscr{H}}_{N;M}^{ba}(y_{q},x_{p})\Big\rangle_{\mbox{\scz(\ref{s3_59})}}\Big]\bigg)  ; \\  \lb{s4_3}
\mscr{A}_{J_{\psi}}[\hat{T},\langle\feynV\rangle_{\mbox{\scz(\ref{s3_59})}};
\hat{\mscr{J}}] &=&\frac{1}{2}\int_{C}d^{4}\!x_{p}\;d^{4}\!y_{q}\;\times  \\ \no
\lefteqn{\hspace*{-1.3cm}\times\;J_{\psi;N}^{\dag,b}(y_{q})\;\hat{I}\bigg(\hat{T}(y_{q})\;
\Big\langle\hat{\mscr{O}}_{N\ppr;M\ppr}^{b\ppr a\ppr}(y_{q},x_{p})\Big\rangle_{\mbox{\scz(\ref{s3_59})}}^{\boldsymbol{-1}}
\;\hat{T}^{-1}(x_{p})-
\Big\langle\hat{\mscr{H}}_{N;M}^{ba}(y_{q},x_{p})\Big\rangle_{\mbox{\scz(\ref{s3_59})}}^{\boldsymbol{-1}}\bigg)
\hat{I}\;J_{\psi;M}^{a}(x_{p}) \;; } \;;
\eeq
\beq \lb{s4_4}
\lefteqn{\Big\langle\hat{\mscr{O}}_{N;M}^{ba}(y_{q},x_{p})\Big\rangle_{\mbox{\scz(\ref{s3_59})}} =}  \\ \no &=&
\bigg\{\big\langle\hat{\mscr{H}}\big\rangle_{\mbox{\scz(\ref{s3_59})}}+
\underbrace{\Big(\hat{T}^{-1}
\big\langle\hat{\mscr{H}}\big\rangle_{\mbox{\scz(\ref{s3_59})}}\hat{T}-
\big\langle\hat{\mscr{H}}\big\rangle_{\mbox{\scz(\ref{s3_59})}}\Big)}_{\Delta\!\langle\hat{\mscr{H}}
\rangle_{\mbox{\scz(\ref{s3_59})}} }+
\hat{T}^{-1}\;\hat{I}\;\hat{S}\;\eta_{q}
\frac{\hat{\mscr{J}}_{N\ppr;M\ppr}^{b\ppr a\ppr}(y_{q},x_{p})}{\mcal{N}}\eta_{p}\;\hat{S}\;\hat{I}\;
\hat{T}\bigg\}_{N;M}^{ba}\hspace*{-0.3cm}(y_{q},x_{p}) \;;   \\  \lb{s4_5}
\Delta\!\langle\hat{\mscr{H}}\rangle_{\mbox{\scz(\ref{s3_59})}} &=&\hat{T}^{-1}\;
\langle\hat{\mscr{H}}\rangle_{\mbox{\scz(\ref{s3_59})}}\;\hat{T}-
\langle\hat{\mscr{H}}\rangle_{\mbox{\scz(\ref{s3_59})}} \;.
\eeq
The background gauge field \(\mcal{V}_{\alpha}^{\mu}(x_{p})\) consists of the self-energy field
strength tensor \(\hat{\mfrak{S}}_{\alpha;\mu\nu}^{(\hat{F})}(x_{p})\), its derivative and its inverse
with structure constants \(C_{\alpha\beta\gamma}\) and additionally of the self-energy quark densities
\(\sigma_{D}^{(\alpha;\kappa)}(x_{p})\) which are dressed by the eigenvectors
\(\hat{\mfrak{B}}_{\hat{F};\beta(\alpha)}^{\mu(\kappa)}(x_{p})\) of the self-energy field strength term
\(C_{\alpha\beta\gamma}\;\hat{\mfrak{S}}_{\alpha;\mu\nu}^{(\hat{F})}(x_{p})\)
\beq \lb{s4_6}
\mcal{V}_{\beta}^{\mu}(x_{p})&=&
\Big[\Big(\hat{\pp}_{p}^{\lambda}\hat{\mfrak{S}}_{\gamma;\nu\lambda}^{(\hat{F})}(x_{p})\Big)-
s_{\gamma}(x_{p})\;n_{\nu}\Big]\;\Big[-\im\,\hat{\mfrak{e}}_{p}^{(\hat{F})}+C_{\alpha\beta\ppr\gamma\ppr}\;
\hat{\mfrak{S}}_{\alpha}^{(\hat{F})\mu\ppr\nu\ppr}(x_{p})\Big]_{\gamma\beta}^{\boldsymbol{-1};\nu\mu} +   \\ \no &+&
\frac{1}{2}\sum_{\scrscr(\alpha)=1,.,8}^{\scrscr(\kappa)=0,.,3}\hat{\mfrak{B}}_{\hat{F};\beta(\alpha)}^{\mu(\kappa)}(x_{p})\;
\sigma_{D}^{(\alpha;\kappa)}(x_{p})\;;   \\   \lb{s4_7}
C_{\alpha\beta\gamma}\;\hat{\mfrak{S}}_{\alpha}^{(\hat{F})\mu\nu}(x_{p}) &:=&
\sum_{\scrscr(\alpha)=1,.,8}^{\scrscr(\kappa)=0,.,3}
\hat{\mfrak{B}}_{\hat{F};\beta(\alpha)}^{\mu(\kappa)}(x_{p})\;
\hat{\mfrak{b}}^{(\hat{F})}_{(\alpha;\kappa)}(x_{p})\;\hat{\mfrak{B}}_{\hat{F};(\alpha)\gamma}^{T,(\kappa)\nu}(x_{p})
\;; \\ \lb{s4_8}
d[\hat{\mfrak{S}}_{\alpha}^{(\hat{F})\mu\nu}(x_{p})]&\rightarrow&
d[\hat{\mfrak{S}}_{\alpha}^{(\hat{F})\mu\nu}(x_{p});\hat{\mfrak{B}}_{\hat{F};\beta(\alpha)}^{\mu(\kappa)}(x_{p});
\hat{\mfrak{b}}^{(\hat{F})}_{(\alpha;\kappa)}(x_{p})]  \;; \\   \lb{s4_9}
\lefteqn{\hspace*{-2.8cm}d[\hat{\mfrak{S}}_{\alpha}^{(\hat{F})\mu\nu}(x_{p});
\hat{\mfrak{B}}_{\hat{F};\beta(\alpha)}^{\mu(\kappa)}(x_{p});
\hat{\mfrak{b}}^{(\hat{F})}_{(\alpha;\kappa)}(x_{p})] = d[\hat{\mfrak{S}}_{\alpha}^{(\hat{F})\mu\nu}(x_{p})]\;\;
d[\hat{\mfrak{B}}_{\hat{F};\beta(\alpha)}^{\mu(\kappa)}(x_{p});\hat{\mfrak{b}}^{(\hat{F})}_{(\alpha;\kappa)}(x_{p})]\;\times } \\ \no &\times&
\bigg\{\prod_{\scrscr\{x_{p}\}}\delta\bigg(C_{\alpha\beta\gamma}\;\hat{\mfrak{S}}_{\alpha}^{(\hat{F})\mu\nu}(x_{p}) -
\sum_{\scrscr(\alpha)=1,.,8}^{\scrscr(\kappa)=0,.,3}
\hat{\mfrak{B}}_{\hat{F};\beta(\alpha)}^{\mu(\kappa)}(x_{p})\;
\hat{\mfrak{b}}^{(\hat{F})}_{(\alpha;\kappa)}(x_{p})\;\hat{\mfrak{B}}_{\hat{F};(\alpha)\gamma}^{T,(\kappa)\nu}(x_{p}) \bigg)\bigg\} \;.
\eeq
The anomalous doubled one-particle operator \(\hat{\mscr{H}}(x_{p})\) contains the composed gauge
fields \(\mcal{V}_{\alpha}^{\mu}(x_{p})\), \(\feynV(x_{p})\) or more precisely the anomalous doubled
version \(\hat{\mscr{V}}^{\mu}(x_{p})=\hat{\mscr{T}}_{\alpha}\;\mcal{V}_{\alpha}^{\mu}(x_{p})\),
\(\feynVV(x_{p})\) (\ref{s4_14}) with extended generators \(\hat{\mscr{T}}_{\alpha}\) of \(\mbox{SU}_{c}(N_{c}=3)\)
which are doubled by the transpose \(\hat{t}_{\alpha}^{T}\) in the '22' block.
In analogy we double the Dirac gamma matrices \(\hat{\beta}\), \(\hat{\gamma}^{\mu}\) to their
extended block diagonal forms \(\hat{\mscr{B}}^{aa}\), \(\hat{\Gamma}^{\mu,aa}\) (\ref{s4_13}) according to
the anomalous doubling of Fermi fields
\beq \lb{s4_10}
\hat{\mscr{H}}_{N;M}^{ba}(y_{q},x_{p}) &=&\delta^{(4)}(y_{q}-x_{p})\;\eta_{q}\;\delta_{qp}
\Bigg(\bea{cc} \hat{H}_{N;M}(x_{p}) & \\ & \hat{H}_{N;M}^{T}(x_{p})  \eea\Bigg)^{ba} \;; \\ \lb{s4_11}
\hat{H}(x_{p}) &=&\Big[\hat{\beta}\Big(\,\feynd_{p}+\im\:\feynV(x_{p})-\im\:\hat{\ve}_{p}+\hat{m}\,\Big)\Big]\;;
\;(\hat{\ve}_{p}=\hat{\beta}\:\ve_{p}=\hat{\beta}\:\eta_{p}\:\ve_{+}\;;\;\ve_{+}>0)\;;  \\ \no &=&
\hat{\beta}\hat{\gamma}^{\mu}\:\hat{\pp}_{p,\mu}+\im\:\hat{\beta}\hat{\gamma}^{\mu}\:\hat{t}_{\alpha}\:
\mcal{V}_{\alpha}^{\mu}(x_{p})+\hat{\beta}\;\hat{m}-\im\;\ve_{p}\;\hat{1}_{N_{0}\times N_{0}}\;;  \\ \lb{s4_12}
\hat{H}^{T}(x_{p}) &=&\Big[\hat{\beta}\Big(\,\feynd_{p}+\im\:\feynV(x_{p})-\im\:\hat{\ve}_{p}+\hat{m}\,\Big)\Big]^{T}  \\ \no &=&
-\big(\,\hat{\beta}\hat{\gamma}^{\mu}\,\big)^{T}\:\hat{\pp}_{p,\mu}-\im\:\big(\,\hat{\beta}\hat{\gamma}^{\mu}\,\big)^{T}\:
(-\hat{t}_{\alpha}^{T})\:\mcal{V}_{\alpha}^{\mu}(x_{p})-\big(\hat{\beta}\:(-\hat{m})\,\big)^{T}-\im\;\ve_{p}\;
\hat{1}_{N_{0}\times N_{0}}\;; \\ \lb{s4_13}
\hat{\mscr{B}}\:\hat{\Gamma}^{\mu}&=& \Bigg(\bea{cc} \hat{\beta}\;\hat{\gamma}^{\mu} & \\ &
(\,\hat{\beta}\;\hat{\gamma}^{\mu}\,)^{T} \eea\Bigg)^{ba} \;;\;
\hat{\mscr{B}}= \Bigg(\bea{cc} \hat{\beta} & \\ & \hat{\beta}^{T} \eea\Bigg)^{ba} \;;\;
\hat{\mscr{B}}\hat{M}= \Bigg(\bea{cc} (\hat{\beta}\,\hat{m}) & \\ &
(\hat{\beta}\,(-\hat{m})\,)^{T} \eea\Bigg)^{ba} \!\!\!;  \\ \lb{s4_14}
\feynVV(x_{p})&=&\hat{\Gamma}^{\mu}\;\hat{\mscr{V}}_{\mu}(x_{p})=
\hat{\Gamma}^{\mu}\;\hat{\mscr{T}}_{\alpha}\;\mcal{V}_{\alpha;\mu}(x_{p}) \;; \;\;\;
\hat{\Gamma}^{0}=\Bigg(\bea{cc}\hat{\gamma}^{0} & 0 \\ 0 & \hat{\gamma}^{0,T} \eea\Bigg)\;;\;\;\;
\hat{\vec{\Gamma}}=\Bigg(\bea{cc}\hat{\vec{\gamma}} & 0 \\ 0 & -\hat{\vec{\gamma}}^{T} \eea\Bigg)\;;  \\ \no
\hat{\mscr{T}}_{\alpha}&=& \Bigg(\bea{cc} \hat{t}_{\alpha} & \\ & -\hat{t}_{\alpha}^{T} \eea\Bigg)^{ba}\;;
\hat{\mscr{V}}^{\mu}(x_{p})=\hat{\mscr{T}}_{\alpha}\;\mcal{V}_{\alpha}^{\mu}(x_{p}) \;; \\ \lb{s4_15}
\hat{\mscr{H}}(x_{p}) &=& \hat{S}\Big(\hat{\mscr{B}}\:\hat{\Gamma}^{\mu}\:\hat{\pp}_{p,\mu}+
\hat{\mscr{B}}\:\big(\im\;\feynVV(x_{p})+\hat{M}\big)\Big)-\im\:\ve_{p}\;\hat{1}_{2N_{0}\times 2N_{0}}\;.
\eeq
It remains to expand the actions
\(\mscr{A}_{DET}[\hat{T},\langle\feynV\rangle_{\mbox{\scz(\ref{s3_59})}};\hat{\mscr{J}}]\),
\(\mscr{A}_{J_{\psi}}[\hat{T},\langle\feynV\rangle_{\mbox{\scz(\ref{s3_59})}};\hat{\mscr{J}}]\)
in terms of the gradient operator
\(\Delta\!\langle\hat{\mscr{H}}\rangle_{\mbox{\scz(\ref{s3_59})}}\!\!=
\hat{T}^{-1}\langle\hat{\mscr{H}}\rangle_{\mbox{\scz(\ref{s3_59})}}\hat{T}-
\langle\hat{\mscr{H}}\rangle_{\mbox{\scz(\ref{s3_59})}}\) with mean field solution
\(\langle\feynV(x_{p})\rangle_{\mbox{\scz(\ref{s3_59})}}\) of (\ref{s3_59}) which is
also applied for the anomalous doubled propagator
\(\langle\hat{\mscr{H}}\rangle_{\mbox{\scz(\ref{s3_59})}}^{\boldsymbol{-1}}\)
\beq \no
\Big\langle\hat{\mscr{H}}_{N;M}^{ba}(y_{q},x_{p})\Big\rangle_{\mbox{\scz(\ref{s3_59})}} &=&\hspace*{-0.25cm}
\delta^{(4)}(y_{q}-x_{p})\;\eta_{q}\;\delta_{qp}
\Bigg(\bea{cc} \big\langle\hat{H}_{N;M}(x_{p})\big\rangle_{\mbox{\scz(\ref{s3_59})}} & \\
& \big\langle\hat{H}_{N;M}^{T}(x_{p})\big\rangle_{\mbox{\scz(\ref{s3_59})}}  \eea\Bigg)^{ba} \hspace*{-0.15cm}=
\delta^{(4)}(y_{q}-x_{p})\;\eta_{q}\;\delta_{qp}\times  \\ \lb{s4_16} &\times&
\Big[\hat{S}\Big(\hat{\mscr{B}}\:\hat{\Gamma}^{\mu}\:\hat{\pp}_{p,\mu}+\im\;
\hat{\mscr{B}}\:\hat{\Gamma}^{\mu}\:\hat{\mscr{T}}_{\alpha}\;
\big\langle\mcal{V}_{\alpha;\mu}(x_{p})\big\rangle_{\mbox{\scz(\ref{s3_59})}}+
\hat{\mscr{B}}\:\hat{M}\Big)-\im\:\ve_{p}\;\hat{1}_{2N_{0}\times 2N_{0}}\Big] \;; \\ \lb{s4_17}
\big\langle\hat{H}(x_{p})\big\rangle_{\mbox{\scz(\ref{s3_59})}} &=&\Big[\hat{\beta}\Big(\,\feynd_{p}+
\im\:\big\langle\feynV(x_{p})\big\rangle_{\mbox{\scz(\ref{s3_59})}}-\im\:\hat{\ve}_{p}+\hat{m}\,\Big)\Big]\;;
\;(\hat{\ve}_{p}=\hat{\beta}\:\ve_{p}=\hat{\beta}\:\eta_{p}\:\ve_{+}\;;\;\ve_{+}>0)\;;  \\ \no &=&
\hat{\beta}\hat{\gamma}^{\mu}\:\hat{\pp}_{p,\mu}+\im\:\hat{\beta}\hat{\gamma}^{\mu}\:\hat{t}_{\alpha}\:
\big\langle\mcal{V}_{\alpha}^{\mu}(x_{p})\big\rangle_{\mbox{\scz(\ref{s3_59})}}+\hat{\beta}\;\hat{m}-\im\;\ve_{p}\;\hat{1}_{N_{0}\times N_{0}}\;;  \\ \lb{s4_18}
\big\langle\hat{H}^{T}(x_{p})\big\rangle_{\mbox{\scz(\ref{s3_59})}} &=&\Big[\hat{\beta}\Big(\,\feynd_{p}+
\im\:\big\langle\feynV(x_{p})\big\rangle_{\mbox{\scz(\ref{s3_59})}}-\im\:\hat{\ve}_{p}+\hat{m}\,\Big)\Big]^{T}  \\ \no &=&
-(\,\hat{\beta}\hat{\gamma}^{\mu}\,)^{T}\:\hat{\pp}_{p,\mu}-\im\:(\,\hat{\beta}\hat{\gamma}^{\mu}\,)^{T}\:
(-\hat{t}_{\alpha}^{T})\:\big\langle\mcal{V}_{\alpha}^{\mu}(x_{p})\big\rangle_{\mbox{\scz(\ref{s3_59})}}-
\big(\hat{\beta}\;(-\hat{m})\,\big)^{T}-\im\;\ve_{p}\;\hat{1}_{N_{0}\times N_{0}}\;;   \\   \lb{s4_19}
\Big\langle\hat{\mscr{H}}_{N;M}^{ba}(y_{q},x_{p})\Big\rangle_{\mbox{\scz(\ref{s3_59})}}^{\boldsymbol{-1}} &=&
\delta^{(4)}(y_{q}-x_{p})\;\eta_{q}\;\delta_{qp}
\Bigg(\bea{cc} \big\langle\hat{H}_{N;M}(x_{p})\big\rangle_{\mbox{\scz(\ref{s3_59})}}^{\boldsymbol{-1}} & \\
& \big\langle\hat{H}_{N;M}^{T}(x_{p})\big\rangle_{\mbox{\scz(\ref{s3_59})}}^{\boldsymbol{-1}}  \eea\Bigg)^{ba} \;; \\ \lb{s4_20}
\big\langle\hat{H}(x_{p})\big\rangle_{\mbox{\scz(\ref{s3_59})}}^{\boldsymbol{-1}} &=&
\Big[\hat{\beta}\Big(\,\feynd_{p}+
\im\:\big\langle\feynV(x_{p})\big\rangle_{\mbox{\scz(\ref{s3_59})}}-\im\:\hat{\ve}_{p}+
\hat{m}\,\Big)\Big]^{\boldsymbol{-1}}\;;
\;(\hat{\ve}_{p}=\hat{\beta}\:\ve_{p}=\hat{\beta}\:\eta_{p}\:\ve_{+}\;;\;\ve_{+}>0)\;;  \\ \no &=&\Big[
\hat{\beta}\hat{\gamma}^{\mu}\:\hat{\pp}_{p,\mu}+\im\:\hat{\beta}\hat{\gamma}^{\mu}\:\hat{t}_{\alpha}\:
\big\langle\mcal{V}_{\alpha}^{\mu}(x_{p})\big\rangle_{\mbox{\scz(\ref{s3_59})}}+\hat{\beta}\;\hat{m}-\im\;\ve_{p}\;\hat{1}_{N_{0}\times N_{0}}
\Big]^{\boldsymbol{-1}}\;;  \\ \lb{s4_21}
\big\langle\hat{H}^{T}(x_{p})\big\rangle_{\mbox{\scz(\ref{s3_59})}}^{\boldsymbol{-1}} &=&\Big[\hat{\beta}\Big(\,\feynd_{p}+
\im\:\big\langle\feynV(x_{p})\big\rangle_{\mbox{\scz(\ref{s3_59})}}-\im\:\hat{\ve}_{p}+\hat{m}\,\Big)\Big]^{\boldsymbol{T;-1}}  \\ \no &=&\Big[
-(\,\hat{\beta}\hat{\gamma}^{\mu}\,)^{T}\:\hat{\pp}_{p,\mu}-\im\:(\,\hat{\beta}\hat{\gamma}^{\mu}\,)^{T}\:
(-\hat{t}_{\alpha}^{T})\:\big\langle\mcal{V}_{\alpha}^{\mu}(x_{p})\big\rangle_{\mbox{\scz(\ref{s3_59})}}-
\big(\hat{\beta}\:(-\hat{m})\,\big)^{T}-\im\;\ve_{p}\;\hat{1}_{N_{0}\times N_{0}}\Big]^{\boldsymbol{-1}}\;.
\eeq
The relevant path integral (\ref{s3_116}) is thereby reduced to the relevant actions
\(\mscr{A}_{DET}[\hat{T},\langle\feynV\rangle_{\mbox{\scz(\ref{s3_59})}};\hat{\mscr{J}}]\),
\(\mscr{A}_{J_{\psi}}[\hat{T},\langle\feynV\rangle_{\mbox{\scz(\ref{s3_59})}};\hat{\mscr{J}}]\)
of coset matrices \(\hat{T}(x_{p})=\exp\{-\hat{Y}(x_{p})\,\}\) which are combined to the path integral
\(Z[\hat{\mscr{J}},J_{\psi},\hat{J}_{\psi\psi};\langle\feynV\rangle_{\mbox{\scz(\ref{s3_59})}};\hat{T}]\) (\ref{s4_22})
with averaged action \(\langle Z_{\hat{J}_{\psi\psi}}[\hat{T}]\rangle_{\mbox{\scz(\ref{s3_59})}}\)
for the initial configuration of BCS terms at times \(t_{p=\pm}\rightarrow\,-\infty\).
We neglect the detailed phase transition for the creation of coherent BCS terms from incoherent initial
conditions which involves a detailed dependence of experimental parameters and temperature for the initial
configuration of the nucleus; thus we simply set an initial configuration at intermediate time
\(0>t_{p=\pm}\gg\,-\infty\) by choosing a definite coset matrix \(\hat{T}(x_{p})=\exp\{-\hat{Y}(x_{p})\,\}\)
\beq  \lb{s4_22}
Z[\hat{\mscr{J}},J_{\psi},\hat{J}_{\psi\psi};\langle\feynV\rangle_{\mbox{\scz(\ref{s3_59})}};\hat{T}] &=&
\int d[\hat{T}^{-1}(x_{p})\;d\hat{T}(x_{p})]\;\;\;
\Big\langle Z_{\hat{J}_{\psi\psi}}[\hat{T}]\Big\rangle_{\mbox{\scz(\ref{s3_59})}}\;\times   \\ \no &\times&
\exp\Big\{\mscr{A}_{DET}[\hat{T},\langle\feynV\rangle_{\mbox{\scz(\ref{s3_59})}};\hat{\mscr{J}}]\Big\}\;{\scrscr\times}\;
\exp\Big\{\im\;\mscr{A}_{J_{\psi}}[\hat{T},\langle\feynV\rangle_{\mbox{\scz(\ref{s3_59})}};\hat{\mscr{J}}]\Big\}\;.
\eeq

\subsection{Infinite order gradient expansion of logarithmic and inverted operators}  \lb{s42}

The actions \(\mscr{A}_{DET}[\hat{T},\langle\feynV\rangle_{\mbox{\scz(\ref{s3_59})}};\hat{\mscr{J}}]\),
\(\mscr{A}_{J_{\psi}}[\hat{T},\langle\feynV\rangle_{\mbox{\scz(\ref{s3_59})}};\hat{\mscr{J}}]\) (\ref{s4_2},\ref{s4_3})
of (\ref{s4_1}-\ref{s4_5}) are transformed to relation (\ref{s4_23}) as the remaining path integral of BCS quark pairs
within the coset decomposition \(\mbox{SO}(N_{0},N_{0})\,/\,\mbox{U}(N_{0})\otimes \mbox{U}(N_{0})\),
(\(N_{0}=(N_{f}=2)\times 4_{\hat{\gamma}}\times (N_{c}=3)=24\)).
We abbreviate (\ref{s4_23}) in terms of anomalous doubled Hilbert space states for
the representation of coset operators, as e.\ g.\ \(\hat{T}(x_{p})=\exp\{-\hat{Y}(x_{p})\,\}\)
with coset generator \(\hat{Y}(x_{p})\) of \(\mbox{so}(N_{0},N_{0})\,/\,\mbox{u}(N_{0})\).
(The reader is referred to appendix \ref{sa} for the important specification of doubled Hilbert space
states with linear and anti-linear representations following from the dyadic product of anomalous doubled
Fermi fields).
\beq \lb{s4_23}
\lefteqn{
Z[\hat{\mscr{J}},J_{\psi},\hat{J}_{\psi\psi};\langle\feynV\rangle_{\mbox{\scz(\ref{s3_59})}};\hat{T}] =
\int d[\hat{T}^{-1}(x_{p})\;d\hat{T}(x_{p})]\;\;\;
\Big\langle Z_{\hat{J}_{\psi\psi}}[\hat{T}]\Big\rangle_{\mbox{\scz(\ref{s3_59})}}\;\times } \\ \no &\times&
\exp\bigg\{\frac{1}{2}\hspace*{0.6cm}\trxpa
\trfgamc\bigg(\ln\Big[\hat{1}+\Big(\Delta\!\langle\hat{\mscr{H}}\rangle_{\mbox{\scz(\ref{s3_59})}}+
\wt{\mscr{J}}(\hat{T}^{-1},\hat{T})\Big)\;\Big\langle\hat{\mscr{H}}\Big\rangle_{\mbox{\scz(\ref{s3_59})}}^{\boldsymbol{-1}}\Big]\bigg)\bigg\}\;
\times \\ \no &\times&
\exp\bigg\{\frac{\im}{2}\;\Big\langle \widehat{J_{\psi}}\Big|\hat{\eta}\:\hat{I}\bigg[\hat{T}\:
\big\langle\hat{\mscr{H}}\big\rangle_{\mbox{\scz(\ref{s3_59})}}^{\boldsymbol{-1}}\bigg(
\hat{1}+\Big(\Delta\!\langle\hat{\mscr{H}}\rangle_{\mbox{\scz(\ref{s3_59})}}+
\wt{\mscr{J}}(\hat{T}^{-1},\hat{T})\Big)\;\Big\langle\hat{\mscr{H}}\Big\rangle_{\mbox{\scz(\ref{s3_59})}}^{\boldsymbol{-1}}\bigg)^{\boldsymbol{-1}}\;
\hat{T}^{-1}-\Big\langle\hat{\mscr{H}}\Big\rangle_{\mbox{\scz(\ref{s3_59})}}^{\boldsymbol{-1}}\bigg]\hat{I}\:\hat{\eta}
\Big| \widehat{J_{\psi}}\Big\rangle\bigg\} \;.
\eeq
We have already defined the 'relative' gradient operator
\(\Delta\!\langle\hat{\mscr{H}}\rangle_{\mbox{\scz(\ref{s3_59})}}\) (\ref{s4_24}) in (\ref{s4_5}); it is determined
by the coset matrix weighted '\(\hat{T}^{-1}\ldots\hat{T}\)', anomalous doubled mean field operator
\(\langle\hat{\mscr{H}}\rangle_{\mbox{\scz(\ref{s3_59})}}\) (see Eq. (\ref{s4_1})) relative to its own
eigenvalue spectrum and basis so that one has to subtract the mean field part
\(\langle\hat{\mscr{H}}\rangle_{\mbox{\scz(\ref{s3_59})}}\) from the coset matrix weighted part
\(\hat{T}^{-1}\,\langle\hat{\mscr{H}}\rangle_{\mbox{\scz(\ref{s3_59})}}\,\hat{T}\). Therefore, we have added and
subtracted the anomalous doubled, mean field, one-particle operator \(\langle\hat{\mscr{H}}\rangle_{\mbox{\scz(\ref{s3_59})}}\)
in \(\langle\hat{\mscr{O}}_{N;M}^{ba}(y_{q},x_{p})\rangle_{\mbox{\scz(\ref{s3_59})}}\) (\ref{s4_4},\ref{s4_5})
so that one obtains in combination with \(-\ln\langle\hat{\mscr{H}}\rangle_{\mbox{\scz(\ref{s3_59})}}\) and
\(-\langle\hat{\mscr{H}}\rangle_{\mbox{\scz(\ref{s3_59})}}^{\boldsymbol{-1}}\) within
\(\mscr{A}_{DET}[\hat{T},\langle\feynV\rangle_{\mbox{\scz(\ref{s3_59})}};\hat{\mscr{J}}]\) (\ref{s4_2}) and
\(\mscr{A}_{J_{\psi}}[\hat{T},\langle\feynV\rangle_{\mbox{\scz(\ref{s3_59})}};\hat{\mscr{J}}]\) (\ref{s4_3})
the path integral (\ref{s4_23}) specified by the operator
\(\hat{1}+\Delta\!\langle\hat{\mscr{H}}\rangle_{\mbox{\scz(\ref{s3_59})}}\;
\langle\hat{\mscr{H}}\rangle_{\mbox{\scz(\ref{s3_59})}}^{\boldsymbol{-1}}\) (\ref{s4_25}).
Aside from the source term \(\wt{\mscr{J}}(\hat{T}^{-1},\hat{T})\:
\langle\hat{\mscr{H}}\rangle_{\mbox{\scz(\ref{s3_59})}}^{\boldsymbol{-1}}\) (\ref{s4_26}),
this important operator part simplifies to the combination
\(\hat{T}^{-1}\;\langle\hat{\mscr{H}}\rangle_{\mbox{\scz(\ref{s3_59})}}\:\hat{T}\;
\langle\hat{\mscr{H}}\rangle_{\mbox{\scz(\ref{s3_59})}}^{\boldsymbol{-1}}\) (\ref{s4_25})
\beq  \lb{s4_24}
\Delta\!\big\langle\hat{\mscr{H}}\big\rangle_{\mbox{\scz(\ref{s3_59})}}&=&\hat{T}^{-1}\;\;\big\langle\hat{\mscr{H}}
\big\rangle_{\mbox{\scz(\ref{s3_59})}}\hat{T}-
\big\langle\hat{\mscr{H}}\big\rangle_{\mbox{\scz(\ref{s3_59})}}   \;;  \\  \lb{s4_25}
\hat{1}+\Delta\!\big\langle\hat{\mscr{H}}\big\rangle_{\mbox{\scz(\ref{s3_59})}}\;
\big\langle\hat{\mscr{H}}\big\rangle_{\mbox{\scz(\ref{s3_59})}}^{\boldsymbol{-1}} &=&
\hat{1}+\Big(\hat{T}^{-1}\;\;\big\langle\hat{\mscr{H}}
\big\rangle_{\mbox{\scz(\ref{s3_59})}}\hat{T}-
\big\langle\hat{\mscr{H}}\big\rangle_{\mbox{\scz(\ref{s3_59})}} \Big)\;
\big\langle\hat{\mscr{H}}\big\rangle_{\mbox{\scz(\ref{s3_59})}}^{\boldsymbol{-1}}  \\  \no &=&
\hat{T}^{-1}\;\;\big\langle\hat{\mscr{H}}
\big\rangle_{\mbox{\scz(\ref{s3_59})}}\hat{T}\;\;
\big\langle\hat{\mscr{H}}\big\rangle_{\mbox{\scz(\ref{s3_59})}}^{\boldsymbol{-1}} \;; \\   \lb{s4_26}
\wt{\mscr{J}}\big(\hat{T}^{-1},\hat{T}\big) &=&\hat{T}^{-1}\;\hat{I}\;\hat{S}\;\hat{\eta}\;
\frac{\hat{\mscr{J}}}{\mcal{N}}\;\hat{\eta}\;\hat{S}\;\hat{I}\;\hat{T}\;.
\eeq
However, as we try to reduce the lastly occurring gradient operator combination (\ref{s4_25}) to
lowest order derivatives following from
\(\hat{T}^{-1}\:\langle\hat{\mscr{H}}\rangle_{\mbox{\scz(\ref{s3_59})}}\:\hat{T}\), one has also to
take into account the propagation of \(\langle\hat{\mscr{H}}\rangle_{\mbox{\scz(\ref{s3_59})}}^{\boldsymbol{-1}}\)
which is weighted by \(\hat{T}\:\langle\hat{\mscr{H}}\rangle_{\mbox{\scz(\ref{s3_59})}}^{\boldsymbol{-1}}\:\hat{T}^{-1}\)
according to the trace operations in (\ref{s4_23}). If we restrict to gradient operators up to order of four
for slowly varying coset matrices in
\(\hat{T}^{-1}\:\langle\hat{\mscr{H}}\rangle_{\mbox{\scz(\ref{s3_59})}}\:\hat{T}\) of a small momentum
expansion ('Derrick's theorem' \cite{raja}), one unintentionally causes strongly varying fields
\(\hat{T}\:\langle\hat{\mscr{H}}\rangle_{\mbox{\scz(\ref{s3_59})}}^{\boldsymbol{-1}}\:\hat{T}^{-1}\)
from the inverse mean field operator \(\langle\hat{\mscr{H}}\rangle_{\mbox{\scz(\ref{s3_59})}}^{\boldsymbol{-1}}\).
One can emphasize this point by a gauge transformation of the coset matrix
\(\hat{T}(x_{p})\rightarrow\hat{T}_{\hat{\mfrak{W}}}(x_{p})\) so that the mean field operator
\(\langle\hat{\mscr{H}}\rangle_{\mbox{\scz(\ref{s3_59})}}\) is altered to
\(\langle\hat{\mscr{H}}_{\hat{\mfrak{W}}}\rangle_{\mbox{\scz(\ref{s3_59})}}\) consisting only of pure
gradient terms without any potential parts as \(\langle\feynV\rangle_{\mbox{\scz(\ref{s3_59})}}\)
(compare section \ref{s43}).
In addition we alternatively suggest the exponential integral representations of the logarithm (\ref{s4_27}) and of
the inverse (\ref{s4_28}) for the operator \(\hat{\mfrak{O}}_{\wt{\mscr{J}}}\) (\ref{s4_29}) so that one obtains a meaningful expansion and convergence with \(1/n!\) instead of the reciprocal
integer numbers of a logarithmic expansion \cite{Englert}
\beq  \lb{s4_27}
\big(\ln\hat{\mfrak{O}}\big)&=&\bigg(\int_{0}^{+\infty}dv\;\;\;
\frac{\exp\{-v\:\hat{1}\}-\exp\{-v\:\hat{\mfrak{O}}\}}{v}\bigg) \;; \\ \lb{s4_28}
\big(\hat{\mfrak{O}}^{\boldsymbol{-1}}\big)&=& \Big(\int_{0}^{+\infty}dv\;\;\;
\exp\{-v\;\hat{\mfrak{O}}\}\Big) \;;  \\  \lb{s4_29}
\hat{\mfrak{O}}_{\wt{\mscr{J}}} &=& \bigg(\hat{1}+\Big(\Delta\!\langle\hat{\mscr{H}}\rangle_{\mbox{\scz(\ref{s3_59})}}+
\wt{\mscr{J}}(\hat{T}^{-1},\hat{T})\Big)\;\Big\langle\hat{\mscr{H}}\Big\rangle_{\mbox{\scz(\ref{s3_59})}}^{\boldsymbol{-1}}\bigg)
\\ \no &=&\bigg(\hat{T}^{-1}\;\;\big\langle\hat{\mscr{H}}\;\;
\big\rangle_{\mbox{\scz(\ref{s3_59})}}\hat{T}\;\;
\big\langle\hat{\mscr{H}}\big\rangle_{\mbox{\scz(\ref{s3_59})}}^{\boldsymbol{-1}} +
\wt{\mscr{J}}\big(\hat{T}^{-1},\hat{T}\big)\;
\big\langle\hat{\mscr{H}}\big\rangle_{\mbox{\scz(\ref{s3_59})}}^{\boldsymbol{-1}}\bigg)\;.
\eeq
In consequence one inserts Eqs. (\ref{s4_27}-\ref{s4_29}) into the path integral (\ref{s4_23}) with 'relative'
gradient operator and source term (\ref{s4_24}-\ref{s4_26}) so that we achieve for the actions
\(\mscr{A}_{DET}[\hat{T},\langle\feynV\rangle_{\mbox{\scz(\ref{s3_59})}};\hat{\mscr{J}}]\),
\(\mscr{A}_{J_{\psi}}[\hat{T},\langle\feynV\rangle_{\mbox{\scz(\ref{s3_59})}};\hat{\mscr{J}}]\) (\ref{s4_2},\ref{s4_3})
in (\ref{s4_23},\ref{s4_22}) the exponential integral representations (\ref{s4_30},\ref{s4_31})
with integration variable \(v\in[0,\infty)\)
\beq \lb{s4_30}
\lefteqn{
\mscr{A}_{DET}[\hat{T},\langle\feynV\rangle_{\mbox{\scz(\ref{s3_59})}};\hat{\mscr{J}}]=
\frac{1}{2}\hspace*{0.6cm}\trxpa
\trfgamc\bigg(\ln\Big[\hat{1}+\Big(\Delta\!\langle\hat{\mscr{H}}\rangle_{\mbox{\scz(\ref{s3_59})}}+
\wt{\mscr{J}}(\hat{T}^{-1},\hat{T})\Big)\;\Big\langle\hat{\mscr{H}}\Big\rangle_{\mbox{\scz(\ref{s3_59})}}^{\boldsymbol{-1}}\Big]\bigg) } \\ \no &=&
\frac{1}{2}
\int_{0}^{+\infty}dv\;\;\exp\{-v\}\hspace*{0.6cm}\trxpa
\trfgamc\bigg[\frac{\hat{1}-\exp\bigg\{-v\:\Big(\Delta\!\langle\hat{\mscr{H}}\rangle_{\mbox{\scz(\ref{s3_59})}}+
\wt{\mscr{J}}(\hat{T}^{-1},\hat{T})\Big)\;\Big\langle\hat{\mscr{H}}\Big\rangle_{\mbox{\scz(\ref{s3_59})}}^{\boldsymbol{-1}}\bigg\}}{v}\bigg]\;;
\eeq
\beq  \lb{s4_31}
\lefteqn{
\im\;\mscr{A}_{J_{\psi}}[\hat{T},\langle\feynV\rangle_{\mbox{\scz(\ref{s3_59})}};\hat{\mscr{J}}] = }  \\ \no &=&
\frac{\im}{2}\;\Big\langle \widehat{J_{\psi}}\Big|\hat{\eta}\:\hat{I}\bigg[\hat{T}\:
\big\langle\hat{\mscr{H}}\big\rangle_{\mbox{\scz(\ref{s3_59})}}^{\boldsymbol{-1}}\bigg(
\hat{1}+\Big(\Delta\!\langle\hat{\mscr{H}}\rangle_{\mbox{\scz(\ref{s3_59})}}+
\wt{\mscr{J}}(\hat{T}^{-1},\hat{T})\Big)\;\Big\langle\hat{\mscr{H}}\Big\rangle_{\mbox{\scz(\ref{s3_59})}}^{\boldsymbol{-1}}\bigg)^{\boldsymbol{-1}}\;
\hat{T}^{-1}-\Big\langle\hat{\mscr{H}}\Big\rangle_{\mbox{\scz(\ref{s3_59})}}^{\boldsymbol{-1}}\bigg]\hat{I}\:\hat{\eta}\Big|
\widehat{J_{\psi}}\Big\rangle
\\ \no &=&\frac{\im}{2}\int_{0}^{+\infty}dv\;\exp\{-v\}\;\;
\Big\langle \widehat{J_{\psi}}\Big|\hat{\eta}\:\hat{I}\bigg[\hat{T}\:\Big\langle\hat{\mscr{H}}\Big\rangle_{\mbox{\scz(\ref{s3_59})}}^{\boldsymbol{-1}}
\exp\boldsymbol{\bigg\{}-v\:\Big(\Delta\!\langle\hat{\mscr{H}}\rangle_{\mbox{\scz(\ref{s3_59})}}+
\wt{\mscr{J}}(\hat{T}^{-1},\hat{T})\Big)\;\Big\langle\hat{\mscr{H}}\Big\rangle_{\mbox{\scz(\ref{s3_59})}}^{\boldsymbol{-1}}
\boldsymbol{\bigg\}}\:\hat{T}^{-1}\:
\hat{I}\:\hat{\eta}\Big| \widehat{J_{\psi}}\Big\rangle + \\ \no &-&\frac{\im}{2}
\Big\langle\widehat{J_{\psi}}\Big|\hat{\eta}\:\hat{I}\:
\Big\langle\hat{\mscr{H}}\Big\rangle_{\mbox{\scz(\ref{s3_59})}}^{\boldsymbol{-1}}\:\hat{I}\:\hat{\eta}\Big| \widehat{J_{\psi}}\Big\rangle \;.
\eeq
As one applies relations (\ref{s4_24},\ref{s4_25}) for the 'relative' gradient operator
\(\Delta\!\langle\hat{\mscr{H}}\rangle_{\mbox{\scz(\ref{s3_59})}}\) in the case of a vanishing
source term \(\wt{\mscr{J}}(\hat{T}^{-1},\hat{T})\), we accomplish the exponential integral representations
(\ref{s4_32},\ref{s4_33}) for
\(\mscr{A}_{DET}[\hat{T},\langle\feynV\rangle_{\mbox{\scz(\ref{s3_59})}};\hat{\mscr{J}}\equiv0]\) and for
\(\mscr{A}_{J_{\psi}}[\hat{T},\langle\feynV\rangle_{\mbox{\scz(\ref{s3_59})}};\hat{\mscr{J}}\equiv0]\)
with the exponent of the already described and composed operator
\(\hat{T}^{-1}\;\langle\hat{\mscr{H}}\rangle_{\mbox{\scz(\ref{s3_59})}}\:\hat{T}\;
\langle\hat{\mscr{H}}\rangle_{\mbox{\scz(\ref{s3_59})}}^{\boldsymbol{-1}}\) (\ref{s4_25}). Since the
latter operator has neither a valid small, nor large momentum expansion, the exponentials in
(\ref{s4_32},\ref{s4_33}) give a meaningful representation for the previous form (\ref{s4_2},\ref{s4_3})
of actions within (\ref{s4_1},\ref{s4_4}) or for (\ref{s4_23}) with 'relative' gradients (\ref{s4_24},\ref{s4_25})
\beq \lb{s4_32}
\lefteqn{\mscr{A}_{DET}[\hat{T},\langle\feynV\rangle_{\mbox{\scz(\ref{s3_59})}};\hat{\mscr{J}}\equiv0]= } \\ \no &=&
\frac{1}{2}\int_{0}^{+\infty}dv\;\;\hspace*{0.6cm}\trxpa
\trfgamc\Bigg[\frac{\exp\big\{-v\:\hat{1}\big\}-\exp\bigg\{-v\:
\hat{T}^{-1}\big\langle\hat{\mscr{H}}
\big\rangle_{\mbox{\scz(\ref{s3_59})}}\hat{T}\big\langle\hat{\mscr{H}}\big\rangle_{\mbox{\scz(\ref{s3_59})}}^{\boldsymbol{-1}}\bigg\}}{v}\Bigg]\;;
\eeq
\beq \lb{s4_33}
\lefteqn{\im\;\mscr{A}_{J_{\psi}}[\hat{T},\langle\feynV\rangle_{\mbox{\scz(\ref{s3_59})}};\hat{\mscr{J}}\equiv0]= } \\ \no &=&
\frac{\im}{2}\int_{0}^{+\infty}dv\;
\Big\langle\widehat{J_{\psi}}\Big|\hat{\eta}\:\hat{I}\hat{T}\:
\Big\langle\hat{\mscr{H}}\Big\rangle_{\mbox{\scz(\ref{s3_59})}}^{\boldsymbol{-1}}
\exp\bigg\{-v\:\hat{T}^{-1}\big\langle\hat{\mscr{H}}
\big\rangle_{\mbox{\scz(\ref{s3_59})}}\hat{T}\big\langle\hat{\mscr{H}}\big\rangle_{\mbox{\scz(\ref{s3_59})}}^{\boldsymbol{-1}}\bigg\}\:
\hat{T}^{-1}\:\hat{I}\:\hat{\eta}\Big| \widehat{J_{\psi}}\Big\rangle - \frac{\im}{2}
\Big\langle\widehat{J_{\psi}}\Big|\hat{\eta}\:\hat{I}\:
\Big\langle\hat{\mscr{H}}\Big\rangle_{\mbox{\scz(\ref{s3_59})}}^{\boldsymbol{-1}}\:\hat{I}\:\hat{\eta}\Big| \widehat{J_{\psi}}\Big\rangle \;.
\eeq
The representations (\ref{s4_32},\ref{s4_33}) for the actions
\(\mscr{A}_{DET}[\hat{T},\langle\feynV\rangle_{\mbox{\scz(\ref{s3_59})}};\hat{\mscr{J}}\equiv0]\),
\(\mscr{A}_{J_{\psi}}[\hat{T},\langle\feynV\rangle_{\mbox{\scz(\ref{s3_59})}};\hat{\mscr{J}}\equiv0]\)
allow a straightforward calculation of observables, especially after a gauge transformation to pure
gradient terms
\(\langle\hat{\mscr{H}}\rangle_{\mbox{\scz(\ref{s3_59})}}\rightarrow
\langle\hat{\mscr{H}}_{\hat{\mfrak{W}}}\rangle_{\mbox{\scz(\ref{s3_59})}}\),
\(\hat{T}(x_{p})\rightarrow\hat{T}_{\hat{\mfrak{W}}}(x_{p})\). In order to compute correlation functions,
we consider again the operator \(\hat{\mfrak{O}}_{\wt{\mscr{J}}}\) (\ref{s4_34}), but under inclusion
of the source term \(\wt{\mscr{J}}(\hat{T}^{-1},\hat{T})\). One can track one-point or two-point
correlation functions of the original Grassmann-valued fields \(\psi_{M}(x_{p})\) in the orginal
QCD-type path integral (\ref{s2_25}-\ref{s2_27}) by subsequent differentiation of
\(\hat{\mscr{J}}_{N;M}^{ba}(y_{q},x_{p})\) (\ref{s2_22}-\ref{s2_24}) to the corresponding observables
in terms of the coset matrices from the actions
\(\mscr{A}_{DET}[\hat{T},\langle\feynV\rangle_{\mbox{\scz(\ref{s3_59})}};\hat{\mscr{J}}]\),
\(\mscr{A}_{J_{\psi}}[\hat{T},\langle\feynV\rangle_{\mbox{\scz(\ref{s3_59})}};\hat{\mscr{J}}]\).
Therefore, we perform variations \(\delta_{\hat{\mscr{J}}(\mbox{\scz arg.}1)}\) (\ref{s4_35}),
\(\delta_{\hat{\mscr{J}}(\mbox{\scz arg.}2)}\) (\ref{s4_36}) of the particular operator
\(\hat{\mfrak{O}}_{\wt{\mscr{J}}}\) (\ref{s4_34}) with arguments
'\((\mbox{arg.}1)\)', '\((\mbox{arg.}2)\)' (\ref{s4_37},\ref{s4_38}) for one-point and
two-point correlation functions, respectively
\beq  \lb{s4_34}
\hat{\mfrak{O}}_{\wt{\mscr{J}}} &=&\hat{1}+\Delta\!\langle\hat{\mscr{H}}\rangle_{\mbox{\scz(\ref{s3_59})}}\;
\big\langle\hat{\mscr{H}}\big\rangle_{\mbox{\scz(\ref{s3_59})}}^{\boldsymbol{-1}}+ \wt{\mscr{J}}(\hat{T}^{-1},\hat{T})\;
\langle\hat{\mscr{H}}\rangle_{\mbox{\scz(\ref{s3_59})}}^{\boldsymbol{-1}} \\ \no &=& \hat{1}+
\Big(\hat{T}^{-1}\;\;\big\langle\hat{\mscr{H}}\;\;
\big\rangle_{\mbox{\scz(\ref{s3_59})}}\hat{T}-
\big\langle\hat{\mscr{H}}\big\rangle_{\mbox{\scz(\ref{s3_59})}} \Big)\;
\big\langle\hat{\mscr{H}}\big\rangle_{\mbox{\scz(\ref{s3_59})}}^{\boldsymbol{-1}}+ \wt{\mscr{J}}(\hat{T}^{-1},\hat{T})\;
\langle\hat{\mscr{H}}\rangle_{\mbox{\scz(\ref{s3_59})}}^{\boldsymbol{-1}}   \\ \no &=&
\bigg(\hat{T}^{-1}\;\;\big\langle\hat{\mscr{H}}\;\;
\big\rangle_{\mbox{\scz(\ref{s3_59})}}\hat{T}\;\;
\big\langle\hat{\mscr{H}}\big\rangle_{\mbox{\scz(\ref{s3_59})}}^{\boldsymbol{-1}} +
\wt{\mscr{J}}\big(\hat{T}^{-1},\hat{T}\big)\;
\big\langle\hat{\mscr{H}}\big\rangle_{\mbox{\scz(\ref{s3_59})}}^{\boldsymbol{-1}}\bigg)\;;  \\ \lb{s4_35}
\delta_{\hat{\mscr{J}}(\mbox{\scz arg.}1)} &=& \delta_{(\mbox{\scz arg.}1)}\!\hat{\mfrak{O}}=
\hat{T}^{-1}\;\hat{I}\;\hat{S}\;\hat{\eta}\;
\frac{\delta\hat{\mscr{J}}(\mbox{arg.}1)}{\mcal{N}}\;\hat{\eta}\;\hat{S}\;\hat{I}\;\hat{T}  \;
\big\langle\hat{\mscr{H}}\big\rangle_{\mbox{\scz(\ref{s3_59})}}^{\boldsymbol{-1}} \;;  \\  \lb{s4_36}
\delta_{\hat{\mscr{J}}(\mbox{\scz arg.}2)} &=& \delta_{(\mbox{\scz arg.}2)}\!\hat{\mfrak{O}}=\hat{T}^{-1}\;\hat{I}\;\hat{S}\;\hat{\eta}\;
\frac{\delta\hat{\mscr{J}}(\mbox{arg.}2)}{\mcal{N}}\;\hat{\eta}\;\hat{S}\;\hat{I}\;\hat{T}  \;
\big\langle\hat{\mscr{H}}\big\rangle_{\mbox{\scz(\ref{s3_59})}}^{\boldsymbol{-1}} \;;  \\  \lb{s4_37}
\delta\hat{\mscr{J}}(\mbox{arg.}1) &:=&
\delta\hat{\mscr{J}}_{M_{1};N_{1}}^{a_{1}b_{1}}(x_{p_{1}}^{(1)},y_{q_{1}}^{(1)}) \;; \\   \lb{s4_38}
\delta\hat{\mscr{J}}(\mbox{arg.}2) &:=&
\delta\hat{\mscr{J}}_{M_{2};N_{2}}^{a_{2}b_{2}}(x_{p_{2}}^{(2)},y_{q_{2}}^{(2)}) \;.
\eeq
Suitable integral representations are used for the variation of the logarithmic and inverted operators
of \(\hat{\mfrak{O}}\) which are transformed to exponential integral representations for a meaningful
expansion and convergence according to the \(1\,/\,n!\) reciprocal factorials \cite{Englert}
\beq\lb{s4_39}
\delta\big(\ln\hat{\mfrak{O}}\big) &=&\bigg(\int_{0}^{+\infty}du\;\;\;
\frac{\hat{1}}{\hat{1}\:u+
\hat{\mfrak{O}}}\;\big(\delta\hat{\mfrak{O}}\big)\;\frac{\hat{1}}{\hat{1}\:u+\hat{\mfrak{O}}}\bigg)\;;  \\  \lb{s4_40}
\delta\big(\hat{\mfrak{O}}^{\boldsymbol{-1}}\big) &=&-\Big(\hat{\mfrak{O}}^{\boldsymbol{-1}}\;
\big(\delta\hat{\mfrak{O}}\big)\;\hat{\mfrak{O}}^{\boldsymbol{-1}}\Big) \;;  \\  \lb{s4_41}
\delta\big(\ln\hat{\mfrak{O}}\big) &=&\int_{0}^{+\infty}du\;dv_{1}\;dv_{2}\;\;\exp\{-u(v_{1}+v_{2})\}
\bigg(\exp\{-v_{1}\:\hat{\mfrak{O}}\}\;\big(\delta\hat{\mfrak{O}}\big)\;\exp\{-v_{2}\:\hat{\mfrak{O}}\}\bigg)\;;  \\ \lb{s4_42}
\delta\big(\hat{\mfrak{O}}^{\boldsymbol{-1}}\big) &=&-\int_{0}^{+\infty}dv_{1}\;dv_{2}\;\;
\bigg(\exp\{-v_{1}\:\hat{\mfrak{O}}\}\;\big(\delta\hat{\mfrak{O}}\big)\;\exp\{-v_{2}\:\hat{\mfrak{O}}\}\bigg)\;.
\eeq
As we consider the rather involved appearing, first order variation
\(\delta_{\hat{\mscr{J}}(\mbox{\scz arg.}1)}\) (\ref{s4_35},\ref{s4_37}) of
\(\mscr{A}_{DET}[\hat{T},\langle\feynV\rangle_{\mbox{\scz(\ref{s3_59})}};\hat{\mscr{J}}]\) and of
\(\mscr{A}_{J_{\psi}}[\hat{T},\langle\feynV\rangle_{\mbox{\scz(\ref{s3_59})}};\hat{\mscr{J}}]\), we
finally attain the one-point observable (\ref{s4_45}) from (\ref{s4_43},\ref{s4_44}) which is mainly determined
by exponential integrals of the operator
\(\hat{T}^{-1}\;\langle\hat{\mscr{H}}\rangle_{\mbox{\scz(\ref{s3_59})}}\:\hat{T}\;
\langle\hat{\mscr{H}}\rangle_{\mbox{\scz(\ref{s3_59})}}^{\boldsymbol{-1}}\) (\ref{s4_25}) having neither a valid
small nor large momentum expansion. It has to be noted that relation (\ref{s4_45}) comprises anomalous
parts as \(\langle\psi_{N}(y_{q})\;\psi_{M}(x_{p})\rangle\) with \(b=2\), \(a=1\) and as well density
terms as \(\langle\psi_{N}^{*}(y_{q})\;\psi_{M}(x_{p})\rangle\) by choosing the anomalous indices
\(b=1\), \(a=1\). As one neglects the functional
\(\langle Z_{\hat{J}_{\psi\psi}}[\hat{T}]\rangle_{\mbox{\scz(\ref{s3_59})}}\) by taking appropriate
initial conditions for \(\hat{T}(x_{p})\), one can accomplish the rather involved appearing path integral
(\ref{s4_45}) for the one-point correlation function which becomes more accessible by transforming to the
eigenbasis of \(\langle\hat{\mscr{H}}\rangle_{\mbox{\scz(\ref{s3_59})}}\). This is in particular applicable
for the bulk of the nucleus where surface effects are negligible and where the mean field potential
\(\langle\feynV\rangle_{\mbox{\scz(\ref{s3_59})}}\) is expected to have a constant value. However,
surface effects can additionally be taken into account by appropriate Ward identities for the
coset decomposition (cf appendix \ref{sb})
\beq \lb{s4_43}
\lefteqn{
\delta_{\hat{\mscr{J}}(\mbox{\scz arg.}1)}
\exp\Big\{\mscr{A}_{DET}[\hat{T},\langle\feynV\rangle_{\mbox{\scz(\ref{s3_59})}};\hat{\mscr{J}}]\Big\}=
\delta_{\hat{\mscr{J}}(\mbox{\scz arg.}1)}
\mscr{A}_{DET}[\hat{T},\langle\feynV\rangle_{\mbox{\scz(\ref{s3_59})}};\hat{\mscr{J}}] } \\ \no &=&
\frac{1}{2}\hspace*{0.6cm}
\trxpa\trfgamc
\delta_{\hat{\mscr{J}}(\mbox{\scz arg.}1)}
\ln\Big(\underbrace{\hat{1}+\Delta\!\big\langle\hat{\mscr{H}}\big\rangle_{\mbox{\scz(\ref{s3_59})}} \;
\big\langle\hat{\mscr{H}}\big\rangle_{\mbox{\scz(\ref{s3_59})}}^{\boldsymbol{-1}} }_{\hat{\mfrak{O}}_{\wt{\mscr{J}}=0}}+
\wt{\mscr{J}}(\hat{T}^{-1},\hat{T})\;
\langle\hat{\mscr{H}}\rangle_{\mbox{\scz(\ref{s3_59})}}^{\boldsymbol{-1}} \Big)  \\ \no &=&
\frac{1}{2}\int_{0}^{+\infty}du\;dv_{1}\;dv_{2}\;\exp\Big\{-u\:\big(v_{1}+v_{2}\big)\Big\}\; \times
\hspace*{0.6cm}\trxpa\trfgamc\bigg[\exp\Big\{-v_{1}\:\hat{T}^{-1}\:\langle\hat{\mscr{H}}\rangle_{\mbox{\scz(\ref{s3_59})}}\:
\hat{T}\:\langle\hat{\mscr{H}}\rangle_{\mbox{\scz(\ref{s3_59})}}^{\boldsymbol{-1}}\Big\}\;\times  \\ \no &\times&
\hat{T}^{-1}\:\hat{I}\:\hat{S}\:\hat{\eta}\:
\frac{\delta\hat{\mscr{J}}(\mbox{arg.}1)}{\mcal{N}}\:\hat{\eta}\:\hat{S}\:\hat{I}\:\hat{T}\:
\langle\hat{\mscr{H}}\rangle_{\mbox{\scz(\ref{s3_59})}}^{\boldsymbol{-1}}
\exp\Big\{-v_{2}\:\hat{T}^{-1}\:\langle\hat{\mscr{H}}\rangle_{\mbox{\scz(\ref{s3_59})}}\:
\hat{T}\:\langle\hat{\mscr{H}}\rangle_{\mbox{\scz(\ref{s3_59})}}^{\boldsymbol{-1}}\Big\}\;\bigg]\;;
\eeq
\beq \lb{s4_44}
\lefteqn{
\delta_{\hat{\mscr{J}}(\mbox{\scz arg.}1)}
\exp\Big\{\im\:\mscr{A}_{J_{\psi}}[\hat{T},\langle\feynV\rangle_{\mbox{\scz(\ref{s3_59})}};\hat{\mscr{J}}]\Big\}= \im\:
\delta_{\hat{\mscr{J}}(\mbox{\scz arg.}1)}
\mscr{A}_{J_{\psi}}[\hat{T},\langle\feynV\rangle_{\mbox{\scz(\ref{s3_59})}};\hat{\mscr{J}}] }
\\ \no &=& \frac{\im}{2}\Big\langle\widehat{J_{\psi}}\Big|\hat{\eta}\:\hat{I}\hat{T}\:
\Big\langle\hat{\mscr{H}}\Big\rangle_{\mbox{\scz(\ref{s3_59})}}^{\boldsymbol{-1}}\:
\delta_{\hat{\mscr{J}}(\mbox{\scz arg.}1)}
\Big(\underbrace{\hat{1}+\Delta\!\big\langle\hat{\mscr{H}}\big\rangle_{\mbox{\scz(\ref{s3_59})}} \;
\big\langle\hat{\mscr{H}}\big\rangle_{\mbox{\scz(\ref{s3_59})}}^{\boldsymbol{-1}} }_{\hat{\mfrak{O}}_{\wt{\mscr{J}}=0}}+
\wt{\mscr{J}}(\hat{T}^{-1},\hat{T})\;
\langle\hat{\mscr{H}}\rangle_{\mbox{\scz(\ref{s3_59})}}^{\boldsymbol{-1}} \Big)^{\boldsymbol{-1}}\:
\hat{T}^{-1}\:\hat{I}\:\hat{\eta}\Big|\widehat{J_{\psi}}\Big\rangle  \\ \no &=&
-\frac{\im}{2}\int_{0}^{+\infty}dv_{1}\;dv_{2}\;
\Big\langle\widehat{J_{\psi}}\Big|\hat{\eta}\:\hat{I}\hat{T}\:
\Big\langle\hat{\mscr{H}}\Big\rangle_{\mbox{\scz(\ref{s3_59})}}^{\boldsymbol{-1}}\:
\exp\Big\{-v_{1}\:\hat{T}^{-1}\:\langle\hat{\mscr{H}}\rangle_{\mbox{\scz(\ref{s3_59})}}\:
\hat{T}\:\langle\hat{\mscr{H}}\rangle_{\mbox{\scz(\ref{s3_59})}}^{\boldsymbol{-1}}\Big\}\;\times  \\ \no &\times&
\hat{T}^{-1}\:\hat{I}\:\hat{S}\:\hat{\eta}\:
\frac{\delta\hat{\mscr{J}}(\mbox{arg.}1)}{\mcal{N}}\:\hat{\eta}\:\hat{S}\:\hat{I}\:\hat{T}\:
\langle\hat{\mscr{H}}\rangle_{\mbox{\scz(\ref{s3_59})}}^{\boldsymbol{-1}}
\exp\Big\{-v_{2}\:\hat{T}^{-1}\:\langle\hat{\mscr{H}}\rangle_{\mbox{\scz(\ref{s3_59})}}\:
\hat{T}\:\langle\hat{\mscr{H}}\rangle_{\mbox{\scz(\ref{s3_59})}}^{\boldsymbol{-1}}\Big\}\;
\hat{T}^{-1}\:\hat{I}\:\hat{\eta}\Big|\widehat{J_{\psi}}\Big\rangle   \\ \no &=&
\frac{\im}{2}\int_{0}^{+\infty}dv_{1}\;dv_{2}\;
\hspace*{0.6cm}\trxpa\trfgamc\bigg[\exp\Big\{-v_{1}\:\hat{T}^{-1}\:\langle\hat{\mscr{H}}\rangle_{\mbox{\scz(\ref{s3_59})}}\:
\hat{T}\:\langle\hat{\mscr{H}}\rangle_{\mbox{\scz(\ref{s3_59})}}^{\boldsymbol{-1}}\Big\}\;\times  \\ \no &\times&
\hat{T}^{-1}\:\hat{I}\:\hat{S}\:\hat{\eta}\:
\frac{\delta\hat{\mscr{J}}(\mbox{arg.}1)}{\mcal{N}}\:\hat{\eta}\:\hat{S}\:\hat{I}\:\hat{T}\:
\langle\hat{\mscr{H}}\rangle_{\mbox{\scz(\ref{s3_59})}}^{\boldsymbol{-1}}
\exp\Big\{-v_{2}\:\hat{T}^{-1}\:\langle\hat{\mscr{H}}\rangle_{\mbox{\scz(\ref{s3_59})}}\:
\hat{T}\:\langle\hat{\mscr{H}}\rangle_{\mbox{\scz(\ref{s3_59})}}^{\boldsymbol{-1}}\Big\}\;\times \\ \no &\times&
\hat{T}^{-1}\:\hat{I}\:\hat{\eta}\Big|\widehat{J_{\psi}}\Big\rangle \boldsymbol{\otimes}
\Big\langle\widehat{J_{\psi}}\Big|\hat{\eta}\:\hat{I}\hat{T}\:
\Big\langle\hat{\mscr{H}}\Big\rangle_{\mbox{\scz(\ref{s3_59})}}^{\boldsymbol{-1}}\:\bigg] \;.
\eeq
Referring to the original QCD-type path integral with relations (\ref{s2_22}-\ref{s2_27}), we obtain for
the one-point correlation \(\langle\Psi_{N}^{\dag,b}(y_{q})\;\Psi_{M}^{a}(x_{p})\rangle\) the following
path integral in terms of BCS quark pairs within the coset generator \(\hat{Y}(x_{p})\) of
the coset matrices \(\hat{T}(x_{p})=\exp\{-\hat{Y}(x_{p})\,\}\)
\beq \lb{s4_45}
\lefteqn{-\frac{\im}{2}\Big\langle\Psi_{N}^{\dag,b}(y_{q})\;\Psi_{M}^{a}(x_{p})\Big\rangle =
\int d[\hat{T}^{-1}(x_{p})\;d\hat{T}(x_{p})]\;\;\;
\Big\langle Z_{\hat{J}_{\psi\psi}}[\hat{T}]\Big\rangle_{\mbox{\scz(\ref{s3_59})}}\;\times } \\ \no &\times&
\exp\bigg\{\frac{1}{2}\int_{0}^{+\infty}dv\;\;\hspace*{0.6cm}\trxpa
\trfgamc\bigg[\frac{1}{v}\bigg(e^{-v\:\hat{1}}-e^{-v\:\hat{T}^{-1}\big\langle\hat{\mscr{H}}
\big\rangle_{\mbox{\scz(\ref{s3_59})}}\hat{T}\big\langle\hat{\mscr{H}}\big\rangle_{\mbox{\scz(\ref{s3_59})}}^{\boldsymbol{-1}} }\bigg)\bigg]\bigg\}
\times  \\ \no &\times&
\exp\bigg\{\frac{\im}{2}\int_{0}^{+\infty}dv\;
\Big\langle\widehat{J_{\psi}}\Big|\hat{\eta}\:\hat{I}
\bigg(\hat{T}\:\big\langle\hat{\mscr{H}}\big\rangle_{\mbox{\scz(\ref{s3_59})}}^{\boldsymbol{-1}}
e^{-v\:\hat{T}^{-1}\big\langle\hat{\mscr{H}}
\big\rangle_{\mbox{\scz(\ref{s3_59})}}\hat{T}
\big\langle\hat{\mscr{H}}\big\rangle_{\mbox{\scz(\ref{s3_59})}}^{\boldsymbol{-1}} }\:
\hat{T}^{-1}-\big\langle\hat{\mscr{H}}\big\rangle_{\mbox{\scz(\ref{s3_59})}}^{\boldsymbol{-1}}\bigg)
\:\hat{I}\:\hat{\eta}\Big| \widehat{J_{\psi}}\Big\rangle\bigg\} \times \\ \no &\times&
\boldsymbol{\Bigg\{}\frac{1}{2}\int_{0}^{+\infty}dv_{1}\;dv_{2}
\hspace*{0.6cm}\trxpa\trfgamc\Bigg[\bigg(e^{-v_{1}\:\hat{T}^{-1}\:
\langle\hat{\mscr{H}}\rangle_{\mbox{\scz(\ref{s3_59})}}\:\hat{T}\:
\langle\hat{\mscr{H}}\rangle_{\mbox{\scz(\ref{s3_59})}}^{\boldsymbol{-1}} }\Big|\widehat{y_{q}}\Big\rangle_{N\ppr}^{b\ppr}\;\;
\hat{T}_{N\ppr;N}^{-1;b\ppr b}(y_{q})\:\hat{S}\:\hat{T}_{M;M\ppr}^{aa\ppr}(x_{p})\;\times \\ \no &\times&
_{M\ppr}^{a\ppr}\Big\langle\widehat{x_{p}} \Big|\langle\hat{\mscr{H}}\rangle_{\mbox{\scz(\ref{s3_59})}}^{\boldsymbol{-1}}
e^{-v_{2}\:\hat{T}^{-1}\:\langle\hat{\mscr{H}}\rangle_{\mbox{\scz(\ref{s3_59})}}\:
\hat{T}\:\langle\hat{\mscr{H}}\rangle_{\mbox{\scz(\ref{s3_59})}}^{\boldsymbol{-1}} }\bigg) \times   \\ \no &\times& \bigg(
\int_{0}^{+\infty}du\;\hat{1}\;e^{-u\:(v_{1}+v_{2})}  + \im
\Big(\hat{T}^{-1}\:\hat{I}\:\hat{\eta}\Big|\widehat{J_{\psi}}\Big\rangle \boldsymbol{\otimes}
\Big\langle\widehat{J_{\psi}}\Big|\hat{\eta}\:\hat{I}\hat{T}\:
\Big\langle\hat{\mscr{H}}\Big\rangle_{\mbox{\scz(\ref{s3_59})}}^{\boldsymbol{-1}}\Big)\:\bigg)
\Bigg] \boldsymbol{\Bigg\}}\;.
\eeq
Eventually, we also state the second order variations
\(\delta_{\hat{\mscr{J}}(\mbox{\scz arg.}2)}\delta_{\hat{\mscr{J}}(\mbox{\scz arg.}1)}\) of
\(\mscr{A}_{DET}[\hat{T},\langle\feynV\rangle_{\mbox{\scz(\ref{s3_59})}};\hat{\mscr{J}}]\) and
\(\mscr{A}_{J_{\psi}}[\hat{T},\langle\feynV\rangle_{\mbox{\scz(\ref{s3_59})}};\hat{\mscr{J}}]\)
for two-point correlations so that it becomes possible to compute e.\ g.\ eigenvalue correlations of nuclei
\beq  \lb{s4_46}
\lefteqn{\delta_{\hat{\mscr{J}}(\mbox{\scz arg.}2)}\delta_{\hat{\mscr{J}}(\mbox{\scz arg.}1)}
\exp\Big\{\mscr{A}_{DET}[\hat{T},\langle\feynV\rangle_{\mbox{\scz(\ref{s3_59})}};\hat{\mscr{J}}] +
\im\:\mscr{A}_{J_{\psi}}[\hat{T},\langle\feynV\rangle_{\mbox{\scz(\ref{s3_59})}};\hat{\mscr{J}}]\Big\}= } \\ \no &=&\frac{1}{2}
\Bigg(\hspace*{0.6cm}\trxpa\trfgamc\delta_{\hat{\mscr{J}}(\mbox{\scz arg.}2)}
\big(\ln\hat{\mfrak{O}}_{\wt{\mscr{J}}}\big)+\im\big\langle\widehat{J_{\psi}}\big|\hat{\eta}\:\hat{I}\:\hat{T}\:
\big\langle\hat{\mscr{H}}\big\rangle_{\mbox{\scz(\ref{s3_59})}}^{\boldsymbol{-1}}
\delta_{\hat{\mscr{J}}(\mbox{\scz arg.}2)}\big(\hat{\mfrak{O}}_{\wt{\mscr{J}}}\big)^{\boldsymbol{-1}}\:\hat{T}^{-1}\:\hat{I}\:\hat{\eta}
\big|\widehat{J_{\psi}}\big\rangle\Bigg) \times \\ \no &\times&   \frac{1}{2}
\Bigg(\hspace*{0.6cm}\trxpa\trfgamc\delta_{\hat{\mscr{J}}(\mbox{\scz arg.}1)}
\big(\ln\hat{\mfrak{O}}_{\wt{\mscr{J}}}\big)+\im\big\langle\widehat{J_{\psi}}\big|\hat{\eta}\:\hat{I}\:\hat{T}\:
\big\langle\hat{\mscr{H}}\big\rangle_{\mbox{\scz(\ref{s3_59})}}^{\boldsymbol{-1}}
\delta_{\hat{\mscr{J}}(\mbox{\scz arg.}1)}\big(\hat{\mfrak{O}}_{\wt{\mscr{J}}}\big)^{\boldsymbol{-1}}\:\hat{T}^{-1}\:\hat{I}\:\hat{\eta}
\big|\widehat{J_{\psi}}\big\rangle\Bigg) +  \\ \no &-&\frac{1}{2}\int_{0}^{+\infty}du\hspace*{0.6cm}\trxpa\trfgamc
\bigg[\frac{1}{\hat{1}\:u+\hat{\mfrak{O}}_{\wt{\mscr{J}}}}
\boldsymbol{\bigg\{}
\big(\delta_{\hat{\mscr{J}}(\mbox{\scz arg.}2)}
\hat{\mfrak{O}}_{\wt{\mscr{J}}}\big)\frac{1}{\hat{1}\:u+\hat{\mfrak{O}}_{\wt{\mscr{J}}}}
\;\mbox{{\bf{\LARGE,}}}\;
\big(\delta_{\hat{\mscr{J}}(\mbox{\scz arg.}1)}
\hat{\mfrak{O}}_{\wt{\mscr{J}}}\big)\frac{1}{\hat{1}\:u+\hat{\mfrak{O}}_{\wt{\mscr{J}}}}
\boldsymbol{\bigg\}_{+}}\bigg] +   \\ \no &+&\frac{\im}{2}
\Big\langle\widehat{J_{\psi}}\Big|\hat{\eta}\:\hat{I}\:\hat{T}\:
\big\langle\hat{\mscr{H}}\big\rangle_{\mbox{\scz(\ref{s3_59})}}^{\boldsymbol{-1}}
\big(\hat{\mfrak{O}}_{\wt{\mscr{J}}}\big)^{\boldsymbol{-1}}\:\boldsymbol{\bigg\{}
\big(\delta_{\hat{\mscr{J}}(\mbox{\scz arg.}2)}
\hat{\mfrak{O}}_{\wt{\mscr{J}}}\big)\;\big(\hat{\mfrak{O}}_{\wt{\mscr{J}}}\big)^{\boldsymbol{-1}}
\;\mbox{{\bf{\LARGE,}}}\;
\big(\delta_{\hat{\mscr{J}}(\mbox{\scz arg.}1)}
\hat{\mfrak{O}}_{\wt{\mscr{J}}}\big)\;\big(\hat{\mfrak{O}}_{\wt{\mscr{J}}}\big)^{\boldsymbol{-1}}
\boldsymbol{\bigg\}_{+}}\hat{T}^{-1}\:\hat{I}\:\hat{\eta}\Big|\widehat{J_{\psi}}\Big\rangle = \\ \no &=&-\frac{1}{4}
\Bigg( \Big\langle\Psi_{N_{2}}^{\dag,b_{2}}(y_{q_{2}}^{(2)})\;
\Psi_{M_{2}}^{a_{2}}(x_{p_{2}}^{(2)})\Big\rangle
\Big\langle\Psi_{N_{1}}^{\dag,b_{1}}(y_{q_{1}}^{(1)})\;
\Psi_{M_{1}}^{a_{1}}(x_{p_{1}}^{(1)})\Big\rangle +
\bigg\langle\Psi_{N_{2}}^{\dag,b_{2}}(y_{q_{2}}^{(2)})\;
\Psi_{M_{2}}^{a_{2}}(x_{p_{2}}^{(2)})\;\;\Psi_{N_{1}}^{\dag,b_{1}}(y_{q_{1}}^{(1)})\;
\Psi_{M_{1}}^{a_{1}}(x_{p_{1}}^{(1)})\bigg\rangle \Bigg) \;.
\eeq
This straightforward, but involved appearing second order variation (\ref{s4_46}) of the source term
\(\wt{\mscr{J}}(\hat{T}^{-1},\hat{T})\) (\ref{s4_26}) results into relation (\ref{s4_47})
for the two-point correlator which also considerably simplifies through the transformation
from the coordinate spacetime representation of \(\hat{T}(x_{p})\) to the eigenbasis
of the mean field operator \(\langle\hat{\mscr{H}}\rangle_{\mbox{\scz(\ref{s3_59})}}\)
with mean field potential \(\langle\feynV\rangle_{\mbox{\scz(\ref{s3_59})}}\)
\beq\no
\lefteqn{\bigg\langle\Psi_{N_{2}}^{\dag,b_{2}}(y_{q_{2}}^{(2)})\;
\Psi_{M_{2}}^{a_{2}}(x_{p_{2}}^{(2)})\;\times\;\Psi_{N_{1}}^{\dag,b_{1}}(y_{q_{1}}^{(1)})\;
\Psi_{M_{1}}^{a_{1}}(x_{p_{1}}^{(1)})\bigg\rangle -
\Big\langle\Psi_{N_{2}}^{\dag,b_{2}}(y_{q_{2}}^{(2)})\;
\Psi_{M_{2}}^{a_{2}}(x_{p_{2}}^{(2)})\Big\rangle
\times\Big\langle\Psi_{N_{1}}^{\dag,b_{1}}(y_{q_{1}}^{(1)})\;
\Psi_{M_{1}}^{a_{1}}(x_{p_{1}}^{(1)})\Big\rangle =  }   \\  \lb{s4_47}  &=&4
\int d[\hat{T}^{-1}(x_{p})\;d\hat{T}(x_{p})]\;\;\;
\Big\langle Z_{\hat{J}_{\psi\psi}}[\hat{T}]\Big\rangle_{\mbox{\scz(\ref{s3_59})}}\;\times \\ \no &\times&
\exp\bigg\{\frac{1}{2}\int_{0}^{+\infty}dv\;\;\hspace*{0.6cm}\trxpa
\trfgamc\bigg[\frac{1}{v}\bigg(e^{-v\:\hat{1}}-e^{-v\:\hat{T}^{-1}\big\langle\hat{\mscr{H}}
\big\rangle_{\mbox{\scz(\ref{s3_59})}}\hat{T}\big\langle\hat{\mscr{H}}\big\rangle_{\mbox{\scz(\ref{s3_59})}}^{\boldsymbol{-1}} }\bigg)\bigg]\bigg\}
\times  \\ \no &\times&
\exp\bigg\{\frac{\im}{2}\int_{0}^{+\infty}dv\;
\Big\langle\widehat{J_{\psi}}\Big|\hat{\eta}\:\hat{I}
\bigg(\hat{T}\:\big\langle\hat{\mscr{H}}\big\rangle_{\mbox{\scz(\ref{s3_59})}}^{\boldsymbol{-1}}
e^{-v\:\hat{T}^{-1}\big\langle\hat{\mscr{H}}
\big\rangle_{\mbox{\scz(\ref{s3_59})}}\hat{T}
\big\langle\hat{\mscr{H}}\big\rangle_{\mbox{\scz(\ref{s3_59})}}^{\boldsymbol{-1}} }\:
\hat{T}^{-1}-\big\langle\hat{\mscr{H}}\big\rangle_{\mbox{\scz(\ref{s3_59})}}^{\boldsymbol{-1}}\bigg)
\:\hat{I}\:\hat{\eta}\Big| \widehat{J_{\psi}}\Big\rangle \bigg\}\times \\ \no &\times&\boldsymbol{\Bigg\{}
\frac{1}{2}\int_{0}^{+\infty}dv_{1}\;dv_{2} \;dv_{3}
\hspace*{0.6cm}\trxpa\trfgamc\Bigg[
\Bigg(\:\bigg(e^{-v_{3}\:\hat{T}^{-1}\:\langle\hat{\mscr{H}}\rangle_{\mbox{\scz(\ref{s3_59})}}\:
\hat{T}\:\langle\hat{\mscr{H}}\rangle_{\mbox{\scz(\ref{s3_59})}}^{\boldsymbol{-1}} }
\Big|\widehat{y_{q_{2}}}^{(2)}\Big\rangle_{N_{2}\ppr}^{b_{2}\ppr}\;\times  \\ \no &\times&
\hat{T}_{N_{2}\ppr;N_{2}}^{-1;b_{2}\ppr b_{2}}(y_{q_{2}}^{(2)})\:\hat{S}\:
\hat{T}_{M_{2};M_{2}\ppr}^{a_{2}a_{2}\ppr}(x_{p_{2}}^{(2)})\;
_{M_{2}\ppr}^{a_{2}\ppr}\Big\langle\widehat{x_{p_{2}}^{(2)}}
\Big|\langle\hat{\mscr{H}}\rangle_{\mbox{\scz(\ref{s3_59})}}^{\boldsymbol{-1}}
e^{-v_{2}\:\hat{T}^{-1}\:\langle\hat{\mscr{H}}\rangle_{\mbox{\scz(\ref{s3_59})}}\:
\hat{T}\:\langle\hat{\mscr{H}}\rangle_{\mbox{\scz(\ref{s3_59})}}^{\boldsymbol{-1}} }
\Big|\widehat{y_{q_{1}}}^{(1)}\Big\rangle_{N_{1}\ppr}^{b_{1}\ppr}\;\times  \\ \no &\times&
\hat{T}_{N_{1}\ppr;N_{1}}^{-1;b_{1}\ppr b_{1}}(y_{q_{1}}^{(1)})\:\hat{S}\:
\hat{T}_{M_{1};M_{1}\ppr}^{a_{1}a_{1}\ppr}(x_{p_{1}}^{(1)})\;
_{M_{1}\ppr}^{a_{1}\ppr}\Big\langle\widehat{x_{p_{1}}^{(1)}}
\Big|\langle\hat{\mscr{H}}\rangle_{\mbox{\scz(\ref{s3_59})}}^{\boldsymbol{-1}}
e^{-v_{1}\:\hat{T}^{-1}\:\langle\hat{\mscr{H}}\rangle_{\mbox{\scz(\ref{s3_59})}}\:
\hat{T}\:\langle\hat{\mscr{H}}\rangle_{\mbox{\scz(\ref{s3_59})}}^{\boldsymbol{-1}} } \bigg) +
\bigg(\mbox{sub-indices }\;1\leftrightarrow2\bigg)\:\Bigg) \times \\ \no &\times& \bigg(
\int_{0}^{+\infty}du\;\hat{1}\;e^{-u\:(v_{1}+v_{2}+v_{3})}  + \im
\Big(\hat{T}^{-1}\:\hat{I}\:\hat{\eta}\Big|\widehat{J_{\psi}}\Big\rangle \boldsymbol{\otimes}
\Big\langle\widehat{J_{\psi}}\Big|\hat{\eta}\:\hat{I}\hat{T}\:
\Big\langle\hat{\mscr{H}}\Big\rangle_{\mbox{\scz(\ref{s3_59})}}^{\boldsymbol{-1}}\Big)\:
\bigg)\Bigg] \boldsymbol{\Bigg\}} \;.
\eeq

\subsection{Gauge transformation to the interaction representation of pure gradient terms} \lb{s43}

The gradient term \(\Delta\!\hat{\mscr{H}}_{N;M}^{ba}(y_{q},x_{p})\) (\ref{s4_5},\ref{s4_24}) depends on the
detailed mean field potential \(\langle\feynV(x_{p})\rangle_{\mbox{\scz(\ref{s3_59})}}\) as the appropriate interaction; however, we can perform a gradient expansion
of \(\mscr{A}_{DET}[\hat{T},\langle\feynV\rangle_{\mbox{\scz(\ref{s3_59})}};\hat{\mscr{J}}]\) (\ref{s4_2}),
\(\mscr{A}_{J_{\psi}}[\hat{T},\langle\feynV\rangle;\hat{\mscr{J}}]\) (\ref{s4_3}) with '{\it universal properties}'
by changing to the 'interaction representation' (\ref{s4_48},\ref{s4_49}) of the coset matrices.
We assume general, complex- and even-valued, block diagonal matrices
\(\hat{\mfrak{W}}_{N_{0}\times N_{0}}^{bb}(x_{p})\), \(\hat{\mfrak{W}}_{N_{0}\times N_{0}}^{-1;aa}(x_{p})\)
for this transformation so that the matrix
\(\hat{T}_{\hat{\mfrak{W}}}^{ab}(x_{p})\) (\ref{s4_48}) follows from \(\hat{T}^{ab}(x_{p})\) in the
'interaction representation' by choosing a suitable dependence of \(\hat{\mfrak{W}}_{N_{0}\times N_{0}}^{bb}(x_{p})\),
\(\hat{\mfrak{W}}_{N_{0}\times N_{0}}^{-1;aa}(x_{p})\) on the potential
\(\langle\feynV(x_{p})\rangle_{\mbox{\scz(\ref{s3_59})}}\). Since the '22' part
'\(\hat{\mscr{H}}_{\hat{\mfrak{W}}}^{22}(x_{p})\)' is given as the transpose of the '11' part
'\((\hat{\mscr{H}}_{\hat{\mfrak{W}}}^{11}(x_{p})\,)^{T}\)' (\ref{s4_50},\ref{s4_51}),
one has to require relations (\ref{s4_52}-\ref{s4_54}) where we introduce a general, complex- and even-valued
generator \(\hat{\mfrak{w}}_{N_{0}\times N_{0}}(x_{p})\) (\ref{s4_54}) for the block diagonal transformation with
\(\hat{\mfrak{W}}_{N_{0}\times N_{0}}^{bb}(x_{p})\), \(\hat{\mfrak{W}}_{N_{0}\times N_{0}}^{-1;aa}(x_{p})\)
\beq \lb{s4_48}
\hat{T}^{ab}(x_{p}) &\rightarrow& \hat{T}_{\hat{\mfrak{W}}}^{ab}(x_{p})=
\hat{\mfrak{W}}^{-1;aa}(x_{p})\;\hat{T}^{ab}(x_{p})\;\hat{\mfrak{W}}^{bb}(x_{p})  \;; \\  \lb{s4_49}
\hat{T}^{-1;ab}(x_{p}) &\rightarrow& \hat{T}_{\hat{\mfrak{W}}}^{-1;ab}(x_{p})=
\hat{\mfrak{W}}^{-1;aa}(x_{p})\;\hat{T}^{-1;ab}(x_{p})\;\hat{\mfrak{W}}^{bb}(x_{p})  \;; \\ \lb{s4_50}
\hat{\mscr{H}}^{a=b}(x_{p}) &\rightarrow& \hat{\mscr{H}}_{\hat{\mfrak{W}}}^{a=b}(x_{p})=
\hat{\mfrak{W}}^{-1;aa}(x_{p})\;\hat{\mscr{H}}^{a=b}(x_{p})\;\hat{\mfrak{W}}^{bb}(x_{p})  \;; \\ \lb{s4_51}
\Big(\hat{\mscr{H}}_{\hat{\mfrak{W}}}^{11}(x_{p})\Big)^{T} &=&\hat{\mscr{H}}_{\hat{\mfrak{W}}}^{22}(x_{p})
\Longrightarrow \hat{\mfrak{W}}^{22}(x_{p})=\Big(\hat{\mfrak{W}}^{-1;11}(x_{p})\Big)^{T}  \;;  \\ \lb{s4_52}
\hat{\mfrak{W}}^{11}(x_{p}) &=& \exp\big\{\hat{\mfrak{w}}(x_{p})\big\}\;;\hspace*{0.3cm}
\hat{\mfrak{W}}^{11;-1}(x_{p}) = \exp\big\{-\hat{\mfrak{w}}(x_{p})\big\}\;;   \\ \lb{s4_53}
\hat{\mfrak{W}}^{22}(x_{p}) &=& \exp\big\{-\hat{\mfrak{w}}^{T}(x_{p})\big\}\;;\hspace*{0.3cm}
\hat{\mfrak{W}}^{22;-1}(x_{p}) = \exp\big\{\hat{\mfrak{w}}^{T}(x_{p})\big\}\;;   \\ \lb{s4_54}
\hat{\mfrak{w}}(x_{p}) &:=&\hat{\mfrak{w}}_{N_{0}\times N_{0}}(x_{p})\in\mathsf{C_{even}}  \;.
\eeq
The suitable choice of gauge (\ref{s4_55}) for \(\exp\{\hat{\mfrak{w}}(x_{p})\}\) is determined
in such a dependence on the potential \(\langle\feynV(x_{p})\rangle_{\mbox{\scz(\ref{s3_59})}}\)
and mass term that the interaction representation \(\hat{\mscr{H}}_{\hat{\mfrak{W}}}^{aa}(x_{p})\) reduces to
the contour spacetime gradients \(\boldsymbol{\hat{\pp}_{p,\mu}}\) which are dressed by the matrices
\(\exp\{\pm\hat{\mfrak{w}}_{N_{0}\times N_{0}}(x_{p})\}\),
\(\exp\{\pm\hat{\mfrak{w}}_{N_{0}\times N_{0}}^{T}(x_{p})\}\)
aside from the gamma matrices \(\hat{\beta}\), \(\hat{\gamma}^{\mu}\).
(Note that one has to distinguish between saturated derivatives, as e.g.
\((\hat{\pp}_{p,\mu}\hat{T}(x_{p})\,)\), and unsaturated
gradient operators \(\boldsymbol{\hat{\pp}_{p,\mu}}\) (typed in boldface)
acting further to the right or left beyond the coset matrices!).
Therefore, one finds for the interaction representation of \(\hat{\mscr{H}}_{\hat{\mfrak{W}}}^{aa}(x_{p})\),
in which the potential \(\langle\feynV(x_{p})\rangle_{\mbox{\scz(\ref{s3_59})}}\)
and mass \(\hat{m}\) are removed by the chosen gauge (\ref{s4_55}),
the simplified relations (\ref{s4_56},\ref{s4_57})
\beq \lb{s4_55}
-\Big(\im \:\big\langle\feynV(x_{p})\big\rangle+\hat{m}\Big) &=&
\Big(\feynd_{p}\exp\big\{\hat{\mfrak{w}}(x_{p})\big\}\Big)\;\exp\big\{-\hat{\mfrak{w}}(x_{p})\big\} \;;   \\  \lb{s4_56}
\hat{\mscr{H}}_{\hat{\mfrak{W}}}^{11}(x_{p}) &=&\exp\big\{-\hat{\mfrak{w}}(x_{p})\big\}\;\hat{H}_{p}(x_{p})\;
\exp\big\{\hat{\mfrak{w}}(x_{p})\big\}  \\ \no &=&\exp\big\{-\hat{\mfrak{w}}(x_{p})\big\}\;
\hat{\beta}\:\hat{\gamma}^{\mu}\;\exp\big\{\hat{\mfrak{w}}(x_{p})\big\}\;\boldsymbol{\hat{\pp}_{p,\mu}}-
\im\:\ve_{p}\:\hat{1}_{N_{0}\times N_{0}} \;; \\ \lb{s4_57}
\hat{\mscr{H}}_{\hat{\mfrak{W}}}^{22}(x_{p}) &=&\Big(\hat{\mscr{H}}_{\hat{\mfrak{W}}}^{11}(x_{p})\Big)^{T}=
\exp\big\{\hat{\mfrak{w}}^{T}(x_{p})\big\}\;\hat{H}_{p}^{T}(x_{p})\;
\exp\big\{-\hat{\mfrak{w}}^{T}(x_{p})\big\}  \\ \no &=&\boldsymbol{-\hat{\pp}_{p,\mu}}
\exp\big\{\hat{\mfrak{w}}^{T}(x_{p})\big\}\;(\hat{\beta}\hat{\gamma}^{\mu})^{T}\;
\exp\big\{-\hat{\mfrak{w}}^{T}(x_{p})\big\}-\im\:\ve_{p}\:\hat{1}_{N_{0}\times N_{0}} \;.
\eeq
The successive definitions and steps (\ref{s4_58}-\ref{s4_62}) for the interaction picture of Dirac gamma matrices lead to anomalous doubled, block diagonal matrices
\(\hat{\mscr{B}}_{\hat{\mfrak{W}}}^{aa}(x_{p})\), \(\hat{\Gamma}_{\hat{\mfrak{W}}}^{\mu,aa}(x_{p})\) which fulfill the identical
Clifford algebra for 3+1 spacetime dimensions despite of their local spacetime dependence
\beq \lb{s4_58}
\hat{\mscr{B}}_{\hat{\mfrak{W}}}^{11}(x_{p}) &=&e^{-\hat{\mfrak{w}}(x_{p})}\;\hat{\beta}\;
e^{\hat{\mfrak{w}}(x_{p})}  \;;    \\  \lb{s4_59}
\hat{\mscr{B}}_{\hat{\mfrak{W}}}^{22}(x_{p}) &=&\Big(\hat{\mscr{B}}_{\hat{\mfrak{W}}}^{11}(x_{p})\Big)^{T}=
e^{\hat{\mfrak{w}}^{T}(x_{p})}\;\hat{\beta}\;
e^{-\hat{\mfrak{w}}^{T}(x_{p})}\;;\hspace*{0.3cm}\hat{\beta}^{T}=\hat{\beta} \;;  \\ \lb{s4_60}
\hat{\mscr{B}}_{\hat{\mfrak{W}}}(x_{p})\;\hat{\Gamma}_{\hat{\mfrak{W}}}^{\mu}(x_{p}) &=&
\Bigg(\bea{cc} e^{-\hat{\mfrak{w}}(x_{p})} & \\ & e^{\hat{\mfrak{w}}^{T}(x_{p})}  \eea\Bigg)
\Bigg(\bea{cc} \hat{\beta}\:\hat{\gamma}^{\mu} & \\  & \big(\hat{\beta}\:\hat{\gamma}^{\mu}\big)^{T} \eea \Bigg)
\Bigg(\bea{cc} e^{\hat{\mfrak{w}}(x_{p})} & \\ & e^{-\hat{\mfrak{w}}^{T}(x_{p})}  \eea\Bigg) \;;  \\ \lb{s4_61}
\hat{\mscr{B}}_{\hat{\mfrak{W}}}^{11}(x_{p})\;\hat{\Gamma}_{\hat{\mfrak{W}}}^{\mu,11}(x_{p}) &=&
e^{-\hat{\mfrak{w}}(x_{p})}\;\hat{\beta}\;e^{\hat{\mfrak{w}}(x_{p})}\;\;
e^{-\hat{\mfrak{w}}(x_{p})}\;\hat{\gamma}^{\mu}\;e^{\hat{\mfrak{w}}(x_{p})}  \;; \\  \lb{s4_62}
\hat{\mscr{B}}_{\hat{\mfrak{W}}}^{22}(x_{p})\; \hat{\Gamma}_{\hat{\mfrak{W}}}^{\mu,22}(x_{p}) &=&
\Big(\hat{\mscr{B}}_{\hat{\mfrak{W}}}^{11}(x_{p})\; \hat{\Gamma}_{\hat{\mfrak{W}}}^{\mu,11}(x_{p})\Big)^{T}=
e^{\hat{\mfrak{w}}^{T}(x_{p})}\;\big(\hat{\beta}\;\hat{\gamma}^{\mu}\big)^{T}\;
e^{-\hat{\mfrak{w}}^{T}(x_{p})}\;;\;\;\;\big(\hat{\gamma}^{0}\big)^{T}=\hat{\gamma}^{0}  \;.
\eeq
Application of (\ref{s4_58}-\ref{s4_62}) with the chosen gauge (\ref{s4_55}) reduces the anomalous doubled, one-particle
Hamiltonian to pure gradient terms (\ref{s4_63}) with locally transformed gamma matrices in the interaction picture for
\(\hat{T}_{\hat{\mfrak{W}}}^{ab}(x_{p})\). The anomalous doubled, one-particle Hamiltonian or contour spacetime gradient
\(\hat{\mscr{H}}_{\hat{\mfrak{W}}}(x_{p})\) (\ref{s4_63}) is further decomposed into commutator and anti-commutator parts
(\ref{s4_64}-\ref{s4_66}) with spacetime dependent gamma matrices; in consequence one eventually attains the Hamiltonian
\(\hat{\mscr{H}}_{\hat{\mfrak{W}}}(x_{p})\) (\ref{s4_67}) of the interaction representation with unsaturated gradients
'\(\boldsymbol{\hat{\pp}_{p,\mu}}\)' and saturated derivatives
\((\hat{\pp}_{p,\mu}\hat{\mscr{B}}_{\hat{\mfrak{W}}}^{aa}(x_{p})\:\hat{\Gamma}_{\hat{\mfrak{W}}}^{\mu,aa}(x_{p})\:)\)
of gamma matrices
\beq \lb{s4_63}
\hat{\mscr{H}}_{\hat{\mfrak{W}}}(x_{p}) &=&\Bigg(\bea{cc}
\hat{\mscr{B}}_{\hat{\mfrak{W}}}^{11}(x_{p})\;\hat{\Gamma}_{\hat{\mfrak{W}}}^{\mu,11}(x_{p})
\;\boldsymbol{\hat{\pp}_{p,\mu}}-\im\:\ve_{p}\:
\hat{1}_{N_{0}\times N_{0}}  & 0 \\ 0 & \boldsymbol{-\hat{\pp}_{p,\mu}}\;
\hat{\mscr{B}}_{\hat{\mfrak{W}}}^{22}(x_{p})\;
\hat{\Gamma}_{\hat{\mfrak{W}}}^{\mu,22}(x_{p})-\im\:\ve_{p}\:\hat{1}_{N_{0}\times N_{0}} \eea
\Bigg)  \;;   \\  \lb{s4_64}
\hat{\mscr{H}}_{\hat{\mfrak{W}}}(x_{p}) &=&-\im\:\ve_{p}\;\hat{1}_{2N_{0}\times 2N_{0}} +   \\ \no &+&\frac{1}{2}
\Bigg(\bea{cc} \boldsymbol{\big[}\hat{\mscr{B}}_{\hat{\mfrak{W}}}^{11}(x_{p})\;\hat{\Gamma}_{\hat{\mfrak{W}}}^{\mu,11}(x_{p})\;\boldsymbol{,}\;
\boldsymbol{\hat{\pp}_{p,\mu}}\boldsymbol{\big]_{-}} & \\ &
\boldsymbol{\big[}\hat{\mscr{B}}_{\hat{\mfrak{W}}}^{22}(x_{p})\;\hat{\Gamma}_{\hat{\mfrak{W}}}^{\mu,22}(x_{p})\;
\boldsymbol{,}\;\boldsymbol{\hat{\pp}_{p,\mu}}\boldsymbol{\big]_{-}} \eea\Bigg) +  \\ \no &+&\frac{1}{2}
\Bigg(\bea{cc} \boldsymbol{\big\{}\hat{\mscr{B}}_{\hat{\mfrak{W}}}^{11}(x_{p})\;\hat{\Gamma}_{\hat{\mfrak{W}}}^{\mu,11}(x_{p})\;\boldsymbol{,}\;\boldsymbol{\hat{\pp}_{p,\mu}}\boldsymbol{\big\}_{+}} & \\ &
-\boldsymbol{\big\{}\hat{\mscr{B}}_{\hat{\mfrak{W}}}^{22}(x_{p})\;\hat{\Gamma}_{\hat{\mfrak{W}}}^{\mu,22}(x_{p})\;\boldsymbol{,}\;
\boldsymbol{\hat{\pp}_{p,\mu}}\boldsymbol{\big\}_{+}} \eea\Bigg) \;;
\eeq
\beq \lb{s4_65}
\boldsymbol{\big[}\hat{\mscr{B}}_{\hat{\mfrak{W}}}^{aa}(x_{p})\;\hat{\Gamma}_{\hat{\mfrak{W}}}^{\mu,aa}(x_{p})\;\boldsymbol{,}\;
\boldsymbol{\hat{\pp}_{p,\mu}}\boldsymbol{\big]_{-}} &=&-\big(\hat{\pp}_{p,\mu}
\hat{\mscr{B}}_{\hat{\mfrak{W}}}^{aa}(x_{p})\;\hat{\Gamma}_{\hat{\mfrak{W}}}^{\mu,aa}(x_{p})\big) \;; \\ \lb{s4_66}
\boldsymbol{\big\{}\hat{\mscr{B}}_{\hat{\mfrak{W}}}^{aa}(x_{p})\;\hat{\Gamma}_{\hat{\mfrak{W}}}^{\mu,aa}(x_{p})\;\boldsymbol{,}\;
\boldsymbol{\hat{\pp}_{p,\mu}}\boldsymbol{\big\}_{+}} &=&2\;\hat{\mscr{B}}_{\hat{\mfrak{W}}}^{aa}(x_{p})\;
\hat{\Gamma}_{\hat{\mfrak{W}}}^{\mu,aa}(x_{p})\;\boldsymbol{\hat{\pp}_{p,\mu}} +
\big(\hat{\pp}_{p,\mu}
\hat{\mscr{B}}_{\hat{\mfrak{W}}}^{aa}(x_{p})\;\hat{\Gamma}_{\hat{\mfrak{W}}}^{\mu,aa}(x_{p})\big)  \;;
\eeq
\beq  \lb{s4_67}
\hat{\mscr{H}}_{\hat{\mfrak{W}}}(x_{p}) &=&-\im\:\ve_{p}\;\hat{1}_{2N_{0}\times 2N_{0}} - \frac{1}{2}\;\delta_{ab}\;
\big(\hat{\pp}_{p,\mu}
\hat{\mscr{B}}_{\hat{\mfrak{W}}}^{aa}(x_{p})\;\hat{\Gamma}_{\hat{\mfrak{W}}}^{\mu,aa}(x_{p})\big) +
\frac{1}{2}\;\hat{S}\;\delta_{ab}\;
\big(\hat{\pp}_{p,\mu}
\hat{\mscr{B}}_{\hat{\mfrak{W}}}^{aa}(x_{p})\;\hat{\Gamma}_{\hat{\mfrak{W}}}^{\mu,aa}(x_{p})\big) +  \\ \no  &+&
\delta_{ab}\;\hat{S}\;\hat{\mscr{B}}_{\hat{\mfrak{W}}}^{aa}(x_{p})\;\hat{\Gamma}_{\hat{\mfrak{W}}}^{\mu,aa}(x_{p})\;
\boldsymbol{\hat{\pp}_{p,\mu}} \;.
\eeq
The change to the interaction picture transforms the generating function (\ref{s3_110}-\ref{s3_116}) with \(\hat{T}(x_{p})\) to the
corresponding path integral \(Z[\hat{\mscr{J}}_{\hat{\mfrak{W}}},J_{\psi;\hat{\mfrak{W}}},\hat{J}_{\psi\psi}]\) (\ref{s4_68}) with colour-dressed
source terms \(J_{\psi;\hat{\mfrak{W}}}\), \(\hat{\mscr{J}}_{\hat{\mfrak{W}}}\) and coset matrices \(\hat{T}_{\hat{\mfrak{W}}}(x_{p})\)
\beq \lb{s4_68}
\lefteqn{Z[\hat{\mscr{J}}_{\hat{\mfrak{W}}},J_{\psi;\hat{\mfrak{W}}},\hat{J}_{\psi\psi}]=\int
d[\hat{T}^{-1}(x_{p})\:d\hat{T}(x_{p})]\;\;
\big\langle Z_{\hat{J}_{\psi\psi}}[\hat{\mfrak{W}}\:\hat{T}_{\hat{\mfrak{W}}}\:\hat{\mfrak{W}}^{-1}]\big\rangle\;
\times }   \\
\no &\times& \exp\Big\{\mscr{A}_{DET}[\hat{T}_{\hat{\mfrak{W}}};\hat{\mscr{J}}_{\hat{\mfrak{W}}}]+\im\:
\mscr{A}_{J_{\psi;\hat{\mfrak{W}}}}[\hat{T}_{\hat{\mfrak{W}}};\hat{\mscr{J}}_{\hat{\mfrak{W}}}] \Big\} \;; \\ \lb{s4_69}
\mscr{A}_{DET}[\hat{T}_{\hat{\mfrak{W}}};\hat{\mscr{J}}_{\hat{\mfrak{W}}}] &=&\frac{1}{2}\int_{C}d^{4}\!x_{p}\;\eta_{p}\;\mcal{N}
\TRALL\bigg(\ln\Big[\hat{\mscr{O}}_{\hat{\mfrak{W}};N;M}^{ba}(y_{q},x_{p})\Big]
-\ln\Big[\hat{\mscr{H}}_{\hat{\mfrak{W}};N;M}^{ba}(y_{q},x_{p})\Big]\bigg)
\;;   \\  \lb{s4_70}
\mscr{A}_{J_{\psi;\hat{\mfrak{W}}}}[\hat{T}_{\hat{\mfrak{W}}};\hat{\mscr{J}}_{\hat{\mfrak{W}}}] &=&\frac{1}{2}\int_{C}d^{4}\!x_{p}\;d^{4}\!y_{q} \Big(J_{\psi;N_{1}}^{\dag,b}(y_{q})\;\hat{\mfrak{W}}_{N_{1};N}^{bb}(y_{q})\Big)\;\hat{I}\;\times   \\ \no &\times&
\bigg(\hat{T}_{\hat{\mfrak{W}};N;N\ppr}^{bb\ppr}(y_{q})\; \hat{\mscr{O}}_{\hat{\mfrak{W}};N\ppr;M\ppr}^{\boldsymbol{-1};b\ppr
a\ppr}(y_{q},x_{p})\; \hat{T}_{\hat{\mfrak{W}};M\ppr;M}^{-1;a\ppr a}(x_{p}) -
\hat{\mscr{H}}_{\hat{\mfrak{W}};N;M}^{\boldsymbol{-1};ba}(y_{q},x_{p})\bigg)\;\times  \\ \no &\times&\hat{I}\;
\Big(\hat{\mfrak{W}}_{M;M_{1}}^{-1;aa}(x_{p})\;J_{\psi;M_{1}}^{a}(x_{p})\Big)\;;   \\  \lb{s4_71}
\hat{\mscr{O}}_{\hat{\mfrak{W}};N;M}^{ba}(y_{q},x_{p}) &=& \bigg\{\hat{\mscr{H}}_{\hat{\mfrak{W}}}+
\Big(\hat{T}_{\hat{\mfrak{W}}}^{-1}\:\hat{\mscr{H}}_{\hat{\mfrak{W}}}\;\hat{T}_{\hat{\mfrak{W}}}-\hat{\mscr{H}}_{\hat{\mfrak{W}}}\Big)+  \\ \no &+&
\hat{T}_{\hat{\mfrak{W}}}^{-1}\;\hat{I}\;\hat{S}\;\eta_{q}\;\hat{\mfrak{W}}_{N\ppr;N_{1}}^{-1;b\ppr b\ppr}(y_{q})
\frac{\hat{\mscr{J}}_{N_{1};M_{1}}^{b\ppr a\ppr}(y_{q},x_{p})}{\mcal{N}} \hat{\mfrak{W}}_{M_{1};M\ppr}^{a\ppr a\ppr}(x_{p})\;\eta_{p}\;
\hat{S}\;\hat{I}\;\hat{T}_{\hat{\mfrak{W}}}\bigg\}_{N;M}^{ba}(y_{q},x_{p})\;.
\eeq
It has to be pointed out that the transformation to the interaction picture cannot be incorporated into the coset integration measure
\(d[\hat{T}^{-1}(x_{p})\:d\hat{T}(x_{p})]\nRightarrow d[\hat{T}_{\hat{\mfrak{W}}}^{-1}(x_{p})\:d\hat{T}_{\hat{\mfrak{W}}}(x_{p})]\)
because the generator, determined by the saddle point approximation of
\(\langle\feynV(x_{p})\rangle_{\mbox{\scz(\ref{s3_59})}}\),
consists of a {\it general}, complex-valued matrix  \(\hat{\mfrak{w}}_{N_{0}\times N_{0}}(x_{p})\) instead of the necessarily hermitian
generator \(\hat{\mcal{G}}_{D;N_{0}\times N_{0}}(x_{p})\) for the diagonalizing matrices \(\hat{P}_{N_{0}\times N_{0}}^{aa}(x_{p})\);
the hermitian-conjugation symmetry between \(\hat{X}(x_{p})\), \(\hat{X}^{\dag}(x_{p})\) as sub-generators of \(\hat{Y}(x_{p})\)
does not persist in the interaction picture with \(\hat{Y}_{\hat{\mfrak{W}}}(x_{p})=\hat{\mfrak{W}}^{-1}(x_{p})\;\hat{Y}(x_{p})\;\hat{\mfrak{W}}(x_{p})\)
because of \((\hat{X}_{\hat{\mfrak{W}}}(x_{p})\,)^{\dag}\neq (\hat{X}_{\hat{\mfrak{W}}}^{\dag}(x_{p})\,)\) due to the completely arbitray
complex matrix structure of \(\hat{\mfrak{w}}_{N_{0}\times N_{0}}(x_{p})\) (only restricted by the '\(-\im\:\hat{\ve}_{p}\)' term).
As we specify the operators in \(\mscr{A}_{DET}[\hat{T}_{\hat{\mfrak{W}}};\hat{\mscr{J}}_{\hat{\mfrak{W}}}\equiv0]\),
\(\mscr{A}_{J_{\psi;\hat{\mfrak{W}}}}[\hat{T}_{\hat{\mfrak{W}}};\hat{\mscr{J}}_{\hat{\mfrak{W}}}\equiv0]\)
in correspondence to section \ref{s42}, we obtain the combination
\(\hat{T}_{\hat{\mfrak{W}}}^{-1}\:\hat{\mscr{H}}_{\hat{\mfrak{W}}}\:\hat{T}_{\hat{\mfrak{W}}}\) of pure gradients terms and
also the inverse weight \(\hat{T}_{\hat{\mfrak{W}}}\:\hat{\mscr{H}}_{\hat{\mfrak{W}}}^{\boldsymbol{-1}}\:\hat{T}_{\hat{\mfrak{W}}}^{-1}\)
due to the trace operations. Thus, the transformation to the interaction picture points out the problem
of finite order gradients, having neither a small nor large momentum expansion
\beq \lb{s4_72}
\lefteqn{\mscr{A}_{DET}[\hat{T}_{\hat{\mfrak{W}}};\hat{\mscr{J}}_{\hat{\mfrak{W}}}\equiv0] =\frac{1}{2}\hspace*{0.6cm}
\trxpa\trfgamc\bigg[\ln\Big[\hat{T}_{\hat{\mfrak{W}}}^{-1}\;\hat{\mscr{H}}_{\hat{\mfrak{W}}}\;\hat{T}_{\hat{\mfrak{W}}}\;
\hat{\mscr{H}}_{\hat{\mfrak{W}}}^{\boldsymbol{-1}}\Big]\bigg]  } \\ \no &=&\frac{1}{2}\int_{0}^{+\infty}dv\;\;\hspace*{0.6cm}
\trxpa\trfgamc\bigg[\frac{\exp\{-v\:\hat{1}\}-
\exp\Big\{-v\:\hat{T}_{\hat{\mfrak{W}}}^{-1}\;\hat{\mscr{H}}_{\hat{\mfrak{W}}}\;\hat{T}_{\hat{\mfrak{W}}}\;
\hat{\mscr{H}}_{\hat{\mfrak{W}}}^{\boldsymbol{-1}}\Big\}}{v}\bigg]\;;      \\       \lb{s4_73}
\lefteqn{\mscr{A}_{J_{\psi;\hat{\mfrak{W}}}}[\hat{T}_{\hat{\mfrak{W}}};\hat{\mscr{J}}_{\hat{\mfrak{W}}}\equiv0] =\frac{\im}{2}
\big\langle\widehat{J_{\psi;\hat{\mfrak{W}}} }\big|\hat{\eta}\:\hat{I}\:
\bigg(\hat{T}_{\hat{\mfrak{W}}}\:\hat{\mscr{H}}_{\hat{\mfrak{W}}}^{\boldsymbol{-1}}\;
\Big[\hat{1}+\Delta\!\hat{\mscr{H}}_{\hat{\mfrak{W}}}\;\hat{\mscr{H}}_{\hat{\mfrak{W}}}^{\boldsymbol{-1}}\Big]^{\boldsymbol{-1}}
\hat{T}_{\hat{\mfrak{W}}}^{-1} -\hat{\mscr{H}}_{\hat{\mfrak{W}}}^{\boldsymbol{-1}}\bigg)\:
\hat{I}\:\hat{\eta}\big|\widehat{J_{\psi;\hat{\mfrak{W}}} }\big\rangle }  \\   \no &=&\frac{\im}{2}
\big\langle\widehat{J_{\psi;\hat{\mfrak{W}}} }\big|\hat{\eta}\:\hat{I}\:\bigg(\int_{0}^{+\infty}dv\;\;
\hat{T}_{\hat{\mfrak{W}}}\:\hat{\mscr{H}}_{\hat{\mfrak{W}}}^{\boldsymbol{-1}}\:
\exp\Big\{-v\:\hat{T}_{\hat{\mfrak{W}}}^{-1}\:\hat{\mscr{H}}_{\hat{\mfrak{W}}}\:
\hat{T}_{\hat{\mfrak{W}}}\:\hat{\mscr{H}}_{\hat{\mfrak{W}}}^{\boldsymbol{-1}}\Big\}\:
\hat{T}_{\hat{\mfrak{W}}}^{-1} - \hat{\mscr{H}}_{\hat{\mfrak{W}}}^{\boldsymbol{-1}}\bigg)
\:\hat{I}\:\hat{\eta}\big|\widehat{J_{\psi;\hat{\mfrak{W}}} }\big\rangle  \;.
\eeq

\subsection{Green functions of gauge transformed gradient terms in gradually varying background fields} \lb{s44}

The change to the interaction picture transforms the generating function (\ref{s3_110}-\ref{s3_116}) with \(\hat{T}(x_{p})\) to the
corresponding path integral \(Z[\hat{\mscr{J}}_{\hat{\mfrak{W}}},J_{\psi;\hat{\mfrak{W}}},\hat{J}_{\psi\psi}]\) (\ref{s4_74}) with colour-dressed
source terms \(J_{\psi;\hat{\mfrak{W}}}\), \(\hat{\mscr{J}}_{\hat{\mfrak{W}}}\) and coset matrices \(\hat{T}_{\hat{\mfrak{W}}}(x_{p})\)
\beq \lb{s4_74}
Z[\hat{\mscr{J}}_{\hat{\mfrak{W}}},J_{\psi;\hat{\mfrak{W}}},\hat{J}_{\psi\psi}]&=&\int
d[\hat{T}^{-1}(x_{p})\:d\hat{T}(x_{p})]\;\; Z_{\hat{J}_{\psi\psi}}[\hat{\mfrak{W}}\:\hat{T}_{\hat{\mfrak{W}}}\:\hat{\mfrak{W}}^{-1}]\;\times   \\
\no &\times& \exp\Big\{\mscr{A}_{DET}[\hat{T}_{\hat{\mfrak{W}}};\hat{\mscr{J}}_{\hat{\mfrak{W}}}]+\im\:
\mscr{A}_{J_{\psi;\hat{\mfrak{W}}}}[\hat{T}_{\hat{\mfrak{W}}};\hat{\mscr{J}}_{\hat{\mfrak{W}}}] \Big\} \;; \\ \lb{s4_75}
\mscr{A}_{DET}[\hat{T}_{\hat{\mfrak{W}}};\hat{\mscr{J}}_{\hat{\mfrak{W}}}] &=&\frac{1}{2}\int_{C}d^{4}\!x_{p}\;\eta_{p}\;\mcal{N}
\TRALL\ln\Big[\hat{\mscr{O}}_{\hat{\mfrak{W}};N;M}^{ba}(y_{q},x_{p})\Big]  \;;   \\  \lb{s4_76}
\mscr{A}_{J_{\psi;\hat{\mfrak{W}}}}[\hat{T}_{\hat{\mfrak{W}}};\hat{\mscr{J}}_{\hat{\mfrak{W}}}] &=&\frac{1}{2}\int_{C}d^{4}\!x_{p}\;d^{4}\!y_{q} \Big(J_{\psi;N_{1}}^{\dag,b}(y_{q})\;\hat{\mfrak{W}}_{N_{1};N}^{bb}(y_{q})\Big)\;\times   \\ \no &\times&
\hat{I}\;\hat{T}_{\hat{\mfrak{W}};N;N\ppr}^{bb\ppr}(y_{q})\; \hat{\mscr{O}}_{\hat{\mfrak{W}};N\ppr;M\ppr}^{\boldsymbol{-1};b\ppr
a\ppr}(y_{q},x_{p})\; \hat{T}_{\hat{\mfrak{W}};M\ppr;M}^{-1;a\ppr a}(x_{p})\;\hat{I}\;\times\;
\Big(\hat{\mfrak{W}}_{M;M_{1}}^{-1;aa}(x_{p})\;J_{\psi;M_{1}}^{a}(x_{p})\Big)\;;   \\  \lb{s4_77}
\hat{\mscr{O}}_{\hat{\mfrak{W}};N;M}^{ba}(y_{q},x_{p}) &=& \bigg\{\hat{\mscr{H}}_{\hat{\mfrak{W}}}+
\Big(\hat{T}_{\hat{\mfrak{W}}}^{-1}\:\hat{\mscr{H}}_{\hat{\mfrak{W}}}\;\hat{T}_{\hat{\mfrak{W}}}-\hat{\mscr{H}}_{\hat{\mfrak{W}}}\Big)+  \\ \no &+&
\hat{T}_{\hat{\mfrak{W}}}^{-1}\;\hat{I}\;\hat{S}\;\eta_{q}\;\hat{\mfrak{W}}_{N\ppr;N_{1}}^{-1;b\ppr b\ppr}(y_{q})
\frac{\hat{\mscr{J}}_{N_{1};M_{1}}^{b\ppr a\ppr}(y_{q},x_{p})}{\mcal{N}} \hat{\mfrak{W}}_{M_{1};M\ppr}^{a\ppr a\ppr}(x_{p})\;\eta_{p}\;
\hat{S}\;\hat{I}\;\hat{T}_{\hat{\mfrak{W}}}\bigg\}_{N;M}^{ba}(y_{q},x_{p})\;.
\eeq
According to Derrick's theorem \cite{raja},
a consistent gradient expansion requires terms up to the order of four for stable, static energy configurations and
also the assumption of slowly varying spacetime dependent background fields as \(\langle\feynV(x_{p})\rangle\). We therefore neglect
the 'saturated' derivatives (\ref{s4_78}) of slowly varying gamma matrices in \(\hat{\mscr{H}}_{\hat{\mfrak{W}}}(x_{p})\)
(\ref{s4_63},\ref{s4_64}) so that
the approximated gradient term \(\Delta\!\hat{\mscr{H}}_{\hat{\mfrak{W}};N;M}^{ba}(y_{q},x_{p})\) (\ref{s4_79}) follows in place of the original defined one
\beq \lb{s4_78}
0 &\approx& \big(\hat{\pp}_{p,\mu}
\hat{\beta}_{\hat{\mfrak{W}}}^{aa}(x_{p})\;\hat{\gamma}_{\hat{\mfrak{W}}}^{\mu,aa}(x_{p})\big) \;;  \\ \lb{s4_79}
\lefteqn{\Delta\!\hat{\mscr{H}}_{\hat{\mfrak{W}};N;M}^{ba}(y_{q},x_{p}) =
\Big(\hat{T}_{\hat{\mfrak{W}}}^{-1}\hat{\mscr{H}}_{\hat{\mfrak{W}}}\hat{T}_{\hat{\mfrak{W}}}-\hat{\mscr{H}}_{\hat{\mfrak{W}}}\Big)_{N;M}^{ba}(y_{q},x_{p})
\approx \delta^{(4)}(y_{q}-x_{p})\;\eta_{q}\;\times  }  \\ \no &\times&\hspace*{-0.4cm}
\bigg[\hat{T}_{\hat{\mfrak{W}};N;N\ppr}^{-1;bb\ppr}(x_{p})\;\hat{S}\;\delta_{b\ppr a\ppr}
\Big(\hat{\beta}_{\hat{\mfrak{W}}}^{a\ppr a\ppr}(x_{p})\:\hat{\gamma}_{\hat{\mfrak{W}}}^{\mu,a\ppr a\ppr}(x_{p})\Big)_{N\ppr;M\ppr}^{a\ppr a\ppr}\;
\Big(\hat{\pp}_{p,\mu}\hat{T}_{\hat{\mfrak{W}};M\ppr;M}^{a\ppr a}(x_{p})\Big) +   \\ \no &+&\hspace*{-0.4cm}
\bigg\{\hat{T}_{\hat{\mfrak{W}};N;N\ppr}^{-1;bb\ppr}(x_{p})\;\hat{S}\;\delta_{b\ppr a\ppr}
\Big(\hat{\beta}_{\hat{\mfrak{W}}}^{a\ppr a\ppr}(x_{p})\:\hat{\gamma}_{\hat{\mfrak{W}}}^{\mu,a\ppr a\ppr}(x_{p})\Big)_{N\ppr;M\ppr}^{a\ppr a\ppr}\;
\hat{T}_{\hat{\mfrak{W}};M\ppr;M}^{a\ppr a}(x_{p}) - \hat{S}\;\delta_{ba}
\Big(\hat{\beta}_{\hat{\mfrak{W}}}^{aa}(x_{p})\:\hat{\gamma}_{\hat{\mfrak{W}}}^{\mu,aa}(x_{p})\Big)_{N;M}^{ba} \bigg\}
\boldsymbol{\hat{\pp}_{p,\mu}}\bigg]_{\mbox{.}}
\eeq
The completely labeled gradient term \(\Delta\!\hat{\mscr{H}}_{\hat{\mfrak{W}};N;M}^{ba}(y_{q},x_{p})\) (\ref{s4_79}) is abbreviated by Eq. (\ref{s4_80})
with assumed constant, block diagonal gamma matrices \(\hat{\beta}_{\hat{\mfrak{W}}}^{aa}\), \(\hat{\gamma}_{\hat{\mfrak{W}}}^{\mu,aa}\)
\beq \lb{s4_80}
\Delta\!\hat{\mscr{H}}_{\hat{\mfrak{W}}}(y_{q},x_{p}) &\approx& \delta^{(4)}(y_{q}-x_{p})\;\eta_{q}\;
\bigg[\hat{T}_{\hat{\mfrak{W}}}^{-1}(x_{p})\;\hat{S}\;
\hat{\beta}_{\hat{\mfrak{W}}}^{aa}\;\hat{\gamma}_{\hat{\mfrak{W}}}^{\mu,aa}\;
\Big(\hat{\pp}_{p,\mu}\hat{T}_{\hat{\mfrak{W}}}(x_{p})\Big) +   \\ \no &+&
\bigg\{\hat{T}_{\hat{\mfrak{W}}}^{-1}(x_{p})\;\hat{S}\;\hat{\beta}_{\hat{\mfrak{W}}}^{aa}\;\hat{\gamma}_{\hat{\mfrak{W}}}^{\mu,aa}\;
\hat{T}_{\hat{\mfrak{W}}}(x_{p}) - \hat{S}\;\hat{\beta}_{\hat{\mfrak{W}}}^{aa}\;\hat{\gamma}_{\hat{\mfrak{W}}}^{\mu,aa} \bigg\}
\boldsymbol{\hat{\pp}_{p,\mu}}\bigg]\;.
\eeq
Proceeding as in chapter 4 of \cite{mies1}, one has to determine the anomalous-doubled time contour Green function
\(\hat{G}^{(0)}=[\hat{\mscr{H}}_{\hat{\mfrak{W}}}(x_{p})\,]^{-1}\) (\ref{s4_82}) with the transpose of the '11' block
extended to the '22' block
\beq \lb{s4_81}
\hat{\mscr{H}}_{\hat{\mfrak{W}}}(x_{p}) &\approx& \hat{S}\;\hat{\beta}_{\hat{\mfrak{W}}}^{aa}\;
\hat{\gamma}_{\hat{\mfrak{W}}}^{\mu,aa}\;\boldsymbol{\hat{\pp}_{p,\mu}}-
\im\:\ve_{p}\;\hat{1}_{2N_{0}\times 2N_{0}}  \;;  \\   \lb{s4_82}
\hat{G}^{(0)} &=& \big[\hat{\mscr{H}}_{\hat{\mfrak{W}}}(\hat{x}_{p})\big]^{-1} =
\Bigg(\bea{cc} \hat{g}^{(0)} & \\ & \big[\hat{g}^{(0)}\big]^{T} \eea\Bigg)_{\mbox{.}}
\eeq
In compliance with the interaction picture, one has to calculate relations (\ref{s4_83},\ref{s4_84}) for
\(\hat{g}^{(0)}_{N;M}\), \([\hat{g}^{(0)}]_{N;M}^{T}\) with colour-dressed gamma matrices but which fulfill
the identical Clifford algebra as the original untransformed matrices \(\hat{\beta}\), \(\hat{\gamma}^{\mu}\)
\beq \lb{s4_83}
\hat{g}_{g,n,s;f,m,r}^{(0)} &=&\Big[\hat{\beta}_{\hat{\mfrak{W}}}^{11}\;\hat{\gamma}_{\hat{\mfrak{W}}}^{\mu,11}\;
\boldsymbol{\hat{\pp}_{p,\mu}}-\im\:\ve_{p}\Big]_{g,n,s;f,m,r}^{-1}  \;; \\  \lb{s4_84} \big[\hat{g}^{(0)}\big]_{g,n,s;f,m,r}^{T}&=&
\Big[-\hat{\beta}_{\hat{\mfrak{W}}}^{22}\;\hat{\gamma}_{\hat{\mfrak{W}}}^{\mu,22}\;
\boldsymbol{\hat{\pp}_{p,\mu}}-\im\:\ve_{p}\Big]_{g,n,s;f,m,r}^{-1}\;.
\eeq
We use the overview of the anomalous-doubled Hilbert space, summarized in appendix \ref{sa}, and the definitions
and notations of chapter 4 in \cite{mies1} so that the 'anti-unitary', 'anti-linear' '22' states accompany
as extensions the original states in the '11' block. One therefore gains the Green functions (\ref{s4_85},\ref{s4_86})
which can be combined into the anomalous-doubled one
\(\widehat{\langle x_{p}}^{a}|\hat{G}^{(0)}|\widehat{y_{p}\rangle}^{b}\) (\ref{s4_87}) with the anomalous-doubled
states \(\widehat{\langle x_{p}}^{a}|\), \(|\widehat{y_{p}\rangle}^{b}\) (cf appendix \ref{sa})
\beq \lb{s4_85}
\langle x_{p}|\big(\hat{\beta}\hat{\gamma}^{\mu}\hat{\pp}_{p,\mu}-\im\:\ve_{p}\big)^{-1}|y_{p}\rangle &=&
-\Big(\hat{\gamma}^{\mu}\frac{\pp}{\pp x_{p}^{\mu}}\hat{\beta}-\im\:\ve_{p}\Big)
\frac{\theta_{p}(x_{p}^{0}-y_{p}^{0})}{4\pi\:|\vec{x}-\vec{y}|}\;
\delta\big(|x_{p}^{0}-y_{p}^{0}|-|\vec{x}-\vec{y}|\big)  \;;  \\ \lb{s4_86}
\ovv{\langle x_{p}}|\big(\hat{\beta}\hat{\gamma}^{\mu}\hat{\pp}_{p,\mu}-\im\:\ve_{p}\big)^{T,-1}|\ovv{y_{p}\rangle} &=&
\langle y_{p}|\big(\hat{\beta}\hat{\gamma}^{\mu}\hat{\pp}_{p,\mu}-\im\:\ve_{p}\big)^{-1}|x_{p}\rangle =  \\ \no &=&
-\Big(\hat{\gamma}^{\mu}\frac{\pp}{\pp y_{p}^{\mu}}\hat{\beta}-\im\:\ve_{p}\Big)
\frac{\theta_{p}(y_{p}^{0}-x_{p}^{0})}{4\pi\:|\vec{y}-\vec{x}|}\;
\delta\big(|y_{p}^{0}-x_{p}^{0}|-|\vec{y}-\vec{x}|\big) \;;
\eeq
\be \lb{s4_87}
\widehat{\langle x_{p}}^{a}|\hat{G}^{(0)}|\widehat{y_{p}\rangle}^{b} = -\delta_{ab}
\bigg(\hat{S}^{a}\;\hat{\gamma}_{\hat{\mfrak{W}}}^{\mu,aa}\;\frac{\pp}{\pp x_{p}^{\mu}}\;\hat{\beta}_{\hat{\mfrak{W}}}^{aa}-\im\:\ve_{p}\bigg)
\frac{\delta\big(|x_{p}^{0}-y_{p}^{0}|-|\vec{x}-\vec{y}|\big)}{4\pi\:|\vec{x}-\vec{y}|}
\Bigg(\bea{cc} \theta_{p}(x_{p}^{0}-y_{p}^{0}) & 0 \\ 0 & \theta_{p}(y_{p}^{0}-x_{p}^{0}) \eea\Bigg)^{ab}\;.
\ee
Since we consider time contour Green functions (\ref{s4_85}-\ref{s4_87}), we include the generalized contour Heaviside function
\(\theta_{p}(x_{p}^{0}-y_{p}^{0})\) (\ref{s4_88}) and also obtain the standard relation (\ref{s4_89}) of
non-equilibrium Green functions \cite{Hagen1}. Relation (\ref{s4_89}) incorporates the time ordering of fields and operators
in the path integral (\ref{s4_74}-\ref{s4_77}) according to the time contour integrals with forward and backward propagation
\be \lb{s4_88}
\bea{rclrcl}
\theta_{p=+}(x_{+}^{0}-y_{+}^{0}) &=&\theta(x_{+}^{0}-y_{+}^{0})\;;\hspace*{0.6cm}& x_{+}^{0} &>& y_{+}^{0} \;  \vspace*{0.19cm}\\
\theta_{p=-}(x_{-}^{0}-y_{-}^{0}) &=&\theta(y_{-}^{0}-x_{-}^{0})\;;\hspace*{0.6cm}& y_{-}^{0} &>& x_{-}^{0} \;
\eea
\ee
\beq \lb{s4_89}
\widehat{\langle x_{+}}^{a}|\hat{G}^{(0)}|\widehat{y_{+}\rangle}^{a} + \widehat{\langle x_{-}}^{a}|\hat{G}^{(0)}|\widehat{y_{-}\rangle}^{a}
&=& \widehat{\langle x_{-}}^{a}|\hat{G}^{(0)}|\widehat{y_{+}\rangle}^{a} \;; \;\;\;\;
\widehat{\langle x_{+}}^{a}|\hat{G}^{(0)}|\widehat{y_{-}\rangle}^{a} \equiv 0 \;.
\eeq
Corresponding to the interaction picture, the Green functions (\ref{s4_85}-\ref{s4_87}) are specialized onto the massless case
of the standard Feynman propagator so that the chirality decouples into conserved helicity states as being 'exact' quantum
numbers. (An observer of an arbitrary inertial system cannot 'overrun' the dressed massless BCS-states in the coset matrices
\(\hat{T}_{\hat{\mfrak{W}}}(x_{p})\), \(\hat{T}_{\hat{\mfrak{W}}}^{-1}(x_{p})\) travelling on the light-cone
'\(\delta(|x_{p}^{0}-y_{p}^{0}|-|\vec{x}-\vec{y}|\,)\)' so that the internal angular momentum cannot be projected onto the
opposite momentum direction for a different helicity !). The anomalous-doubled Green function for the massles case (\ref{s4_87})
consists of two time contour Heaviside functions (\ref{s4_88}) \(\theta_{p}(x_{p}^{0}-y_{p}^{0})\), \(\theta_{p}(y_{p}^{0}-x_{p}^{0})\)
on the light-cone '\(\delta(|x_{p}^{0}-y_{p}^{0}|-|\vec{x}-\vec{y}|\,)\)' which result into opposite time contour propagations
\(\eta_{p}(x_{p}^{0}-y_{p}^{0})\:>0\), \(\eta_{p}(y_{p}^{0}-x_{p}^{0})\:>0\) (in the 'contour sense') concerning the '11' and '22'
density blocks. Therefore, we separate the anomalous-doubled, time contour step functions from
\(\widehat{\langle x_{p}}^{a}|\hat{G}^{(0)}|\widehat{y_{p}\rangle}^{a}\) and introduce the Green function 'operator'
\(\widehat{\langle x_{p}}^{a}|\boldsymbol{\hat{\mscr{G}}^{(0)}}|\widehat{y_{p}\rangle}^{a}\) acting onto the anomalous-doubled,
time contour Heaviside function \(\Theta_{p}^{aa}(x_{p}^{0}-y_{p}^{0})\)
\beq \lb{s4_90}
\widehat{\langle x_{p}}^{a}|\hat{G}^{(0)}|\widehat{y_{p}\rangle}^{b} &=&\hspace*{-0.3cm} -\delta_{ab}
\bigg(\hat{S}^{a}\;\hat{\gamma}_{\hat{\mfrak{W}}}^{\mu,aa}\;\frac{\pp}{\pp x_{p}^{\mu}}\;\hat{\beta}_{\hat{\mfrak{W}}}^{aa}-\im\:\ve_{p}\bigg)
\frac{\delta\big(|x_{p}^{0}-y_{p}^{0}|-|\vec{x}-\vec{y}|\big)}{4\pi\:|\vec{x}-\vec{y}|}
\Bigg(\bea{cc} \theta_{p}(x_{p}^{0}-y_{p}^{0}) & 0 \\ 0 & \hspace*{-0.4cm}
\theta_{p}(y_{p}^{0}-x_{p}^{0}) \eea\Bigg)^{ab}  \\ \no &=&\delta_{ab}\;
\widehat{\langle x_{p}}^{a}|\boldsymbol{\hat{\mscr{G}}^{(0)}}|\widehat{y_{p}\rangle}^{a} \;\Theta_{p}^{aa}(x_{p}^{0}-y_{p}^{0}) \;; \\ \lb{s4_91}
\widehat{\langle x_{p}}^{a}|\boldsymbol{\hat{\mscr{G}}^{(0)}}|\widehat{y_{p}\rangle}^{a} &=&
-\delta_{ab} \bigg(\hat{S}^{a}\;\hat{\gamma}_{\hat{\mfrak{W}}}^{\mu,aa}\;\boldsymbol{\frac{\pp}{\pp x_{p}^{\mu}}}\;
\hat{\beta}_{\hat{\mfrak{W}}}^{aa}-\im\:\ve_{p}\bigg)
\frac{\delta\big(|x_{p}^{0}-y_{p}^{0}|-|\vec{x}-\vec{y}|\big)}{4\pi\:|\vec{x}-\vec{y}|} \;;  \\ \lb{s4_92}
\Theta_{p}^{ab}(x_{p}^{0}-y_{p}^{0}) &=&\delta_{ab}
\Bigg(\bea{cc} \theta_{p}(x_{p}^{0}-y_{p}^{0}) & 0 \\ 0 & \theta_{p}(y_{p}^{0}-x_{p}^{0}) \eea\Bigg)^{ab}\;.
\eeq
The anomalous-doubled Heaviside or time contour step function \(\Theta_{p}^{ab}(x_{p}^{0}-y_{p}^{0})\) restricts the possible
terms in the gradient expansion of the action \(\mscr{A}_{DET}[\hat{T}_{\hat{\mfrak{W}}};\hat{\mscr{J}}_{\hat{\mfrak{W}}}]\) (\ref{s4_75})
because the trace operations in (\ref{s4_75}) also involve the contour extended traces of spacetime; in consequence one obtains as the
remaining terms in the gradient expansion of \(\mscr{A}_{DET}[\hat{T}_{\hat{\mfrak{W}}};\hat{\mscr{J}}_{\hat{\mfrak{W}}}]\) (\ref{s4_75})
only those in which the anomalous-doubled, time contour step functions \(\Theta_{p}^{aa}(x_{p}^{0}-y_{p}^{0})\) do not result into
contradictory propagations concerning the time contour extended ordering of gradient terms \(\Delta\!\hat{\mscr{H}}_{\hat{\mfrak{W}}}(x_{p})\) (\ref{s4_80})
(e.g. \(\Theta_{p}^{aa}(x_{p}^{0}-y_{p}^{0})\cdot\Theta_{p}^{aa}(y_{p}^{0}-x_{p}^{0})\equiv 0\) !). This restriction of terms is missing
in the case of the gradient expansion of the action \(\mscr{A}_{J_{\psi;\hat{\mfrak{W}}}}[\hat{T}_{\hat{\mfrak{W}}};\hat{\mscr{J}}_{\hat{\mfrak{W}}}]\) (\ref{s4_76}).
In continuation of principles for a gradient expansion, we state that an anomalous-doubled field \(\Psi_{M}^{a}(x_{p})\) propagates with the
block diagonal. doubled Green function \(\widehat{\langle x_{p}}^{a}|\hat{G}^{(0)}|\widehat{y_{q}\rangle}^{b}\)
\beq \lb{s4_93}
\Psi_{M}^{a(=1,2)}(x_{p}) &=&\Bigg(\bea{c} \psi_{M}(x_{p}) \\ \psi_{M}^{*}(x_{p}) \eea\Bigg)^{a}=\int_{C}d^{4}\!y_{q}\;\mcal{N}^{2}
\widehat{\langle x_{p}}^{a}|\hat{G}^{(0)}_{M;N}|\widehat{y_{q}\rangle}^{b}\;
\Bigg(\bea{c}\psi_{N}(y_{q}) \\ \psi_{N}^{*}(y_{q}) \eea\Bigg)^{b(=1,2)}\;.
\eeq
This principle has to be used in the expansion of
\(\mscr{A}_{J_{\psi;\hat{\mfrak{W}}}}[\hat{T}_{\hat{\mfrak{W}}};\hat{\mscr{J}}_{\hat{\mfrak{W}}}]\) (\ref{s4_76}) where one starts to propagate
with the source field \(J_{\psi;M_{1}}^{a}(x_{p})\) on the right-hand side of the action for a BEC wavefunction. It replaces the wavefunction
\(\Psi_{N}^{b}(y_{q})\) in (\ref{s4_93}). However, the propagation of fields with (\ref{s4_93}) is not directly applicable for the action
\(\mscr{A}_{DET}[\hat{T}_{\hat{\mfrak{W}}};\hat{\mscr{J}}_{\hat{\mfrak{W}}}]\) (\ref{s4_75}) because of the cyclic invariance of traces,
both of the internal state spaces \(\mbox{$\mfrak{tr}$}_{\scrscr N_{f},\hat{\gamma}_{mn}^{(\mu)},N_{c}}\) and the Hilbert state trace of doubled quantum mechanics.

\section{Derivation of a nontrivial topology and the chiral anomaly} \lb{s5}

\subsection{Comparison to the Skyrme model with homotopy group $\Pi_{3}(\mbox{SU}(2))=\mathsf{Z}$} \lb{s51}

The derived actions \(\mscr{A}_{DET}[\hat{T},\feynV;\hat{\mscr{J}}]\), \(\mscr{A}_{J_{\psi}}[\hat{T},\feynV;\hat{\mscr{J}}]\)
(\ref{s3_110}-\ref{s3_116}) consist of coset matrices \(\hat{T}(x_{p})=\exp\{-\hat{Y}(x_{p})\}\) whose manifold is
determined by the coset space \(\mbox{SO}(N_{0},N_{0})\,/\,\mbox{U}(N_{0})\) with
\(N_{0}=(N_{f}=2)\times 4_{\hat{\gamma}}\times (N_{c}=3)=24\). Therefore, the question arises in comparison to effective
nucleon models as the \(\mbox{SU}_{L}(2)\otimes \mbox{SU}_{R}(2)\,/\,\mbox{SU}_{V}(2)\) chiral Skyrme Lagrangian
whether the mapping from 3(+1) spacetime to the \(\mbox{SO}(N_{0},N_{0})\,/\,\mbox{U}(N_{0})\) coset space also allows
for a nontrivial homotopic classification of BCS terms within the \(\mbox{su}(N_{0},N_{0})\,/\,\mbox{u}(N_{0})\)
coset generator \footnote{The isospin chiral symmetry \(\mbox{SU}_{L}(2)\otimes\mbox{SU}_{R}(2)\) is spontaneously
broken by the vector isospin invariance \(\mbox{SU}_{V}(2)\) of the vacuum states with the appearance of three
massless pseudoscalar Nambu-Goldstone bosons or the pions \(\pi^{0}(x_{p})\), \(\pi^{\pm}(x_{p})\) \cite{Nambu,Gold}.}.

Since the so-called '\(\sigma\)-field' \(\sigma(x_{p})\) of the original '\(\sigma\)-model' has turned out to
be dependent on the other three pion fields \(\vec{\pi}(x_{p})=(\pi_{1}(x_{p})\,,\,\pi_{2}(x_{p})\,,\,\pi_{3}(x_{p})\,)\)
as a two pion resonance, one has to introduce the nonlinear condition
\be \lb{s5_1}
\big(\,\sigma(x_{p})\,\big)^{2}+\vec{\pi}(x_{p})\cdot\vec{\pi}(x_{p})=1\; ,
\ee
in order to remove the non-physical, dependent \(\sigma(x_{p})\) field degree of freedom from an effective Lagrangian.
The nonlinear restriction (\ref{s5_1}) is analogous to the four dimensional
\(\vec{x}\cdot\vec{x}-(x^{0})^{2}\), \(\mbox{SO}(3,1)\) Lorentz symmetry, but within the 'internal isospin space'
which gives rise to the chiral symmetry \(\mbox{SU}_{L}(2)\otimes\mbox{SU}_{R}(2)\) for an appropriate effective
Lagrangian in the massless case. A suitable Lagrangian, which incorporates the nonlinear restriction for only three
independent fields, is given by the Skyrme Lagrangian which only comprises quadratic and quartic derivatives of axial
\(\mbox{SU}_{A}(2)\) isospin matrices \(\hat{U}(x_{p})\) corresponding to stable, static energy configurations
in 3+1 spacetime dimensions (compare \cite{Manton,Brown1,Brown2}
concerning the original Skyrme model with Lagrangian (\ref{s5_4})
and 'Derrick's theorem' \cite{raja} for stable, static energy configurations in the 3(+1) spacetime)
\beq \lb{s5_2}
\hat{U}(x_{p}) &=&\exp\big\{\im\:\vec{\tau}\cdot\vec{\varphi}(x_{p})\big\}\;;\hspace*{0.6cm}
\hat{U}^{\dag}(x_{p})\;\hat{U}(x_{p})=\hat{1} \;;  \\ \lb{s5_3}
\hat{L}_{\mu}(x_{p}) &=&\hat{U}^{\dag}(x_{p})\;\big(\hat{\pp}_{p,\mu}\hat{U}(x_{p})\big) \;; \\ \lb{s5_4}
\mscr{L}_{Skyrme} &=&-\frac{f_{\pi}^{2}}{4}\:\:\trf\Big(\hat{L}_{\mu}(x_{p})\;\hat{L}^{\mu}(x_{p})\Big)+
\frac{1}{32\:e^{2}}\;\trf\Big(\big[\hat{L}^{\mu}(x_{p})\:\boldsymbol{,}\:\hat{L}^{\nu}(x_{p})\,\big]_{-}
\big[\hat{L}_{\mu}(x_{p})\:\boldsymbol{,}\:\hat{L}_{\nu}(x_{p})\,\big]_{-}\Big)\;;  \\ \lb{s5_5}
f_{\pi} &\stackrel{\wedge}{=}& \mbox{'pion decay constant'} \;; \;\;\;\;
e \stackrel{\wedge}{=} \mbox{'dimensionless constant' for size of Skyrmion} \;;  \\ \no
\vec{\tau} &\stackrel{\wedge}{=}& \mbox{'isospin Pauli matrices'} \;.
\eeq
The axial \(\mbox{SU}_{A}(2)\) isospin matrices \(\hat{U}(x_{p})=\exp\{\im\:\vec{\tau}\cdot\vec{\varphi}(x_{p})\}\)
include the nonlinear restriction (\ref{s5_1}) due to the reduction to three independent, internal angle field degrees
of freedom \(\vec{\varphi}(x_{p})=(\,\varphi_{1}(x_{p})\,,\,\varphi_{2}(x_{p})\,,\,\varphi_{3}(x_{p})\,)\).
The \(\mbox{SU}_{A}(2)\) isospin matrices \(\hat{U}(x_{p})\) determine the four dependent isospin fields
\((\,\sigma(x_{p})\,,\,\vec{\pi}(x_{p})\,)\)
\beq \lb{s5_6}
\hat{U}(x_{p})&=&\exp\{\im\:\vec{\tau}\cdot\vec{\varphi}(x_{p})\}=\sigma(x_{p}) +\im\;\vec{\pi}(x_{p})\;; \\ \no
\sigma(x_{p})&=&\hat{1}_{2\times 2}\cdot\cos\big(|\vec{\varphi}(x_{p})|\big)\;;\hspace*{0.6cm}
\vec{\pi}(x_{p})=\vec{\tau}\cdot\frac{\vec{\varphi}(x_{p})}{|\vec{\varphi}(x_{p})|}\cdot
\sin\big(|\vec{\varphi}(x_{p})|\big)\;;  \\ \no
|\vec{\varphi}(x_{p})| &=&\sqrt{\varphi_{1}^{2}(x_{p})+\varphi_{2}^{2}(x_{p})+\varphi_{3}^{2}(x_{p})}\:,
\eeq
and give rise to the homotopic classification \(\Pi_{3}(\mbox{SU}(2)\,)=\mathsf{Z}\) for topological solitons
following from the mapping of the compactified three dimensional coordinate space to the internal
\(\mbox{SU}(2)\sim \mbox{S}^{3}\) isospin space or sphere. This homotopic classification
\(\Pi_{3}(\mbox{SU}(2)\,)=\mathsf{Z}\) with nontrivial winding numbers is completely independent from the
Skyrme Lagrangian (\ref{s5_2}-\ref{s5_5}) with the quadratic and quartic derivatives of the \(\mbox{SU}_{A}(2)\)
isospin matrices \(\hat{U}(x_{p})\). The zero component \(B^{0}\) of the integrated, topological current density
\(B^{\mu}(x_{p})\), corresponding to \(\Pi_{3}(\mbox{SU}(2)\,)=\mathsf{Z}\), is assigned to the Baryon number for the
prevailing \(\mbox{SU}_{A}(2)\) isospin field configuration \(\hat{U}(x_{p})\)
\beq \lb{s5_7}
B^{\mu}(x_{p}) &=&\frac{\ve^{\mu\nu\kappa\lambda}}{24\:\pi^{2}}\:\;
\trf\Big[\hat{L}_{\nu}(x_{p})\:\hat{L}_{\kappa}(x_{p})\:\hat{L}_{\lambda}(x_{p})\Big] \;;  \\ \lb{s5_8}
\big(\hat{\pp}_{p,\mu}B^{\mu}(x_{p})\big) &=& 0  \;; \\  \lb{s5_9}
B^{0} &=&\int_{\mathsf{R}^{3}}d^{3}\!x\;\; B^{0}(x_{p}) \;\rightarrow\;\mbox{Baryon number} \;.
\eeq
However, this interpretation and the conserved topological current
\((\hat{\pp}_{p,\mu}B^{\mu}(x_{p})\,)=0\) as a Baryon current is completely independent from the
Skyrme Lagrangian and is argued to occur as an extension of QCD in the large \(N_{c}\) limit
(\(\mbox{SU}(N_{c}\rightarrow\infty)\)) of colour degrees of freedom \cite{Witten1,Witten2,Witten3}.
This causes the question whether
the presented approach of previous sections with fixed \(\mbox{SU}_{c}(N_{c}=3)\) colour symmetry
also contains nontrivial topologies within the derived effective actions
\(\mscr{A}_{DET}[\hat{T},\langle\feynV\rangle_{\mbox{\scz(\ref{s3_59})}};\hat{\mscr{J}}]\),
\(\mscr{A}_{J_{\psi}}[\hat{T},\langle\feynV\rangle_{\mbox{\scz(\ref{s3_59})}};\hat{\mscr{J}}]\)
of coset matrices and generators \(\hat{T}(x_{p})=\exp\{-\hat{Y}(x_{p})\,\}\) with anti-symmetric,
even-valued BCS quark pairs \(\hat{X}_{N;M}(x_{p})\), \(\hat{X}^{\dag}_{N;M}(x_{p})\) (\ref{s3_72}-\ref{s3_78}).

\subsection{Instantons of BCS terms from the Hopf mapping $\Pi_{3}(S^{2})=\mbox{\sf Z}$
and the Hopf invariant} \lb{s52}

The anti-symmetric, even-valued BCS terms \(\hat{X}_{N;M}(x_{p})\), \(\hat{X}^{\dag}_{N;M}(x_{p})\) are given by
a non-compact, internal manifold with 'hyperbolic' trigonometric functions and allow to achieve
a classification of 'instantons' from the derived actions
\(\mscr{A}_{DET}[\hat{T},\langle\feynV\rangle_{\mbox{\scz(\ref{s3_59})}};\hat{\mscr{J}}]\),
\(\mscr{A}_{J_{\psi}}[\hat{T},\langle\feynV\rangle_{\mbox{\scz(\ref{s3_59})}};\hat{\mscr{J}}]\)
instead of the topological solitons as in the Skyrme model. Furthermore, we recognize that the eigenvalues,
as the crucial elements of \(\hat{X}_{N;M}(x_{p})\), \(\hat{X}^{\dag}_{N;M}(x_{p})\), are determined by the
quaternion-valued, complex field variables \(\ovv{f}_{\ovv{M}}(x_{p})\) (\ref{s3_74}-\ref{s3_78}) with
anti-symmetric Pauli matrix \((\hat{\tau}_{2})_{gf}\) of isospin space. Therefore, one can only accomplish
a mapping from the compactified three dimensional coordinate space \(\vec{x}\) to two independent, real
field degrees of freedom within the internal isospin space; in consequence, a valid classification according
to the Hopf fibration \(\Pi_{3}(\mbox{S}^{2})=\mathsf{Z}\) may be expected for the mapping from the compactified
three dimensional coordinate space to the complex, quaternionic eigenvalues with anti-symmetric
Pauli-matrix \((\hat{\tau})_{gf}\). We also anticipate a Hopf classification following from
\(\Pi_{\mbox{\scz S}^{2n-1}}(\mbox{S}^{n})\) with \(n=4\) (instead of \(n=2\) as for the eigenvalues)
because the diagonalizing, eigenvector matrices \(\hat{P}_{N_{0}\times N_{0}}^{aa}(x_{p})\) (\ref{s3_79}-\ref{s3_83})
of \(\hat{X}_{N;M}(x_{p})\), \(\hat{X}_{N;M}^{\dag}(x_{p})\) are specified by four complex-valued fields for
the four independent quaternions with \(2\times 2\) Pauli-matrices (\(\hat{\tau}_{0}\:,\:\vec{\tau}\)) as the
basic entries along the off-diagonal matrix elements of \(\hat{\mscr{G}}_{D;f\ovv{M};g\ovv{M}}(x_{p})\)
(\ref{s3_79}-\ref{s3_83}) \cite{Naka}.
However, it is questionable whether an additional Hopf fibration of \(\Pi_{\mbox{\scz S}^{15}}(\mbox{S}^{8})\)
(or \(n=8\)) can be realized in our model of coset matrices \(\hat{T}(x_{p})\) with BCS terms because this
involves 'octonions' instead of quaternions so that associativity is not preserved.

We verify the assumption of a valid Hopf mapping \(\Pi_{3}(\mbox{S}^{2})=\mathsf{Z}\) from the three dimensional
coordinate space to an anti-symmetric, complex eigenvalue matrix \((\hat{\tau}_{2})_{gf}\) with
fields \(\ovv{f}_{r}(x_{p})\) (\(r=1,\ldots,\,N_{c}=3\)) , (\(\ovv{f}_{r}(x_{p})\in\mathsf{C_{even}}\))
\be \lb{s5_10}
\hat{X}_{N;M}(x_{p}) = \hat{X}_{g,\ovv{N};f,\ovv{M}}(x_{p})=
(\hat{\tau}_{2})_{gf}\;\ovv{f}_{r}(x_{p})\;\delta_{\ovv{N};\ovv{M}} =
(\hat{\tau}_{2})_{gf}\;\ovv{f}_{r}(x_{p})\;\delta_{sr}\;\delta_{nm}\;,
\ee
by using the axial current relation of quark fields with the chiral anomaly (cf. the derivation in appendix \ref{sc})
\beq\no
\lefteqn{\boldsymbol{\hat{\pp}_{p,\mu}\Big(\bar{\psi}(x_{p})\:\hat{\gamma}^{\mu}\:\hat{\gamma}_{5}\:\hat{\mfrak{t}}_{0}\:\psi(x_{p})\Big)=
-\frac{\ve^{\kappa\lambda\mu\nu}}{16\:\pi^{2}}\;
\trfc\Big[\hat{\mfrak{t}}_{0}\;\hat{F}_{\kappa\lambda}(x_{p})\;\hat{F}_{\mu\nu}(x_{p})\Big]  +
\bar{\psi}(x_{p})\:\hat{\gamma}_{5}\:\big\{\hat{\mfrak{t}}_{0}\:\boldsymbol{,}\:\hat{m}\big\}_{\dag}\:\psi(x_{p})   }  +  } \\ \no &+&
j_{\psi}^{\dag}(x_{p})\:\hat{\gamma}_{5}\:\hat{\mfrak{t}}_{0}\:\psi(x_{p})-\psi^{\dag}(x_{p})\:\hat{\gamma}_{5}\:
\hat{\mfrak{t}}_{0}\:j_{\psi}(x_{p}) +  \frac{1}{2}
\bigg[\psi^{T}(x_{p})\Big(\hat{j}_{\psi\psi}^{\dag}(x_{p})\:\hat{\gamma}_{5}\:\hat{\mfrak{t}}_{0}+
\hat{\gamma}_{5}\:\hat{\mfrak{t}}_{0}^{T}\:\hat{j}_{\psi\psi}^{\dag}(x_{p})\Big)\psi(x_{p}) + \\ \lb{s5_11} &-&
\psi^{\dag}(x_{p})\Big(\hat{j}_{\psi\psi}(x_{p})\:\hat{\gamma}_{5}\:\hat{\mfrak{t}}_{0}^{*}+
\hat{\gamma}_{5}\:\hat{\mfrak{t}}_{0}\:\hat{j}_{\psi\psi}(x_{p})\Big)\psi^{*}(x_{p})\bigg] + \frac{1}{2}
\int_{C}d^{4}\!y_{q} \times \\ \no &\times&
\Bigg[\Psi^{\dag}(y_{q},x_{p})\:\hat{\mscr{J}}(y_{q},x_{p})\bigg(\bea{cc}\hat{\gamma}_{5}\:\hat{\mfrak{t}}_{0}
& \\ & -\hat{\gamma}_{5}\:\hat{\mfrak{t}}_{0}^{*}\eea\bigg)\Psi(x_{p})+
\Psi^{\dag}(x_{p})\bigg(\bea{cc} -\hat{\gamma}_{5}\:\hat{\mfrak{t}}_{0}
& \\ & \hat{\gamma}_{5}\:\hat{\mfrak{t}}_{0}^{T} \eea\bigg)\hat{\mscr{J}}(x_{p},y_{q})\:\Psi(y_{q})\Bigg]\;.
\eeq
We have included the various source fields of
\(\mscr{A}_{S}[\hat{\mscr{J}},J_{\psi},\hat{J}_{\psi\psi},\hat{\mfrak{j}}^{(\hat{F})}]\) (\ref{s2_22}) and a finite mass
term \(\hat{m}\) within the axial symmetry variations of the original QCD-type path integral (\ref{s2_25}-\ref{s2_27})
with the constant, hermitian isospin matrix \((\hat{\mfrak{t}}_{0})_{gf}\).
As one performs the zero-mass limit \(\hat{m}\rightarrow 0\) for vanishing source fields, the two chiral
states decouple and result into conserved helicity states of massless fermions moving on the light-cone
so that their projection of spin onto the momentum cannot be overrun by other observers for a different helicity and
becomes conserved. However, this axial current relation also contains the chiral anomaly (Adler-Bell-Jackiw anomaly,
first term on the right-hand side of (\ref{s5_11})) apart from the symmetry violating mass and source field terms.
This anomaly is obtained from the calculation of the Jacobian for quark fields under an axial chiral transformation
with a gauge invariant cut-off regulator \cite{Sred,Das1,Fuji}.
This chiral anomaly has nontrivial instanton numbers and can also be rewritten
in terms of a conserved Chern-Simons current \(K^{\mu}(x_{p})\)
\beq \lb{s5_12}
(N_{I}\in\mathsf{Z}\,,\,\mbox{instanton number}) &=&\frac{\ve^{\mu\nu\kappa\lambda}}{32\:\pi^{2}}
\int_{-\infty}^{+\infty}d^{4}\!x_{p}\;
\trc\Big[\hat{F}_{\mu\nu}(x_{p})\;\hat{F}_{\kappa\lambda}(x_{p})\Big] \;; \\  \lb{s5_13}
\big(\hat{\pp}_{p,\mu}K^{\mu}(x_{p})\,\big)&=&\frac{\ve^{\mu\nu\kappa\lambda}}{8\:\pi^{2}}
\Big(\hat{\pp}_{p,\mu}\trc\Big[\hat{A}_{\nu}(x_{p})\;\big(\hat{\pp}_{p,\kappa}\hat{A}_{\lambda}(x_{p})\big)-
\frac{2\;\im}{3}\;\hat{A}_{\nu}(x_{p})\;\hat{A}_{\kappa}(x_{p})\;\hat{A}_{\lambda}(x_{p})\Big]\Big) \\ \no &=&
\frac{\ve^{\mu\nu\kappa\lambda}}{32\:\pi^{2}}\;\trc\Big[\hat{F}_{\mu\nu}(x_{p})\;\hat{F}_{\kappa\lambda}(x_{p})\Big]\;.
\eeq
If we multiply (\ref{s5_11}) by the factor one-half for a spin '\(\frac{1}{2}\,\hbar\)' angular momentum
and perform the four dimensional spacetime integration in the
massless limit with vanishing source fields, one attains on the right-hand side the integer instanton numbers and on
the left-hand side the integrated, zero-helicity component for vanishing current densities at spatial infinity
from the Gaussian integration law
\beq \lb{s5_14}
\int d^{3}\!\vec{x}\;\;\psi^{\dag}(x_{p})\;\hat{\beta}\;\frac{1}{2}\;\hat{\gamma}^{0}\;
\hat{\gamma}_{5}\;\hat{\mfrak{t}}_{0}\;\psi(x_{p})\;
\Big|_{x_{p}^{0}=-\infty}^{x_{p}^{0}=+\infty} &=&(N_{I}\in\mathsf{Z}\,,\,\mbox{instanton number})\;\cdot\;
(N_{f}=2)\;.
\eeq
The last, integrated relation (\ref{s5_14}) of a conserved helicity number is therefore quantized according to the
instanton numbers from the chiral anomaly.  (Eq. (\ref{s5_14}) contains two isospin degrees of
freedom (\(N_{f}=2\)) with isospin unit matrix \((\hat{\mfrak{t}}_{0})_{gf}=(\hat{1}_{2\times 2})_{gf}\) ).
Since we derive the Hopf invariant
\((V_{S^{2n-1}})^{-1}\int_{\mbox{\scz S}^{2n-1}}\hat{\omega}_{n-1}\:\wedge\:(d\hat{\omega}_{n-1})\) (\(n=2\)) from
relation (\ref{s5_14}) by suitable differentiation of the transformed path integrals with respect to the
source field \(\hat{\mscr{J}}_{N:M}^{ba}(y_{q},x_{p})\), the derived Hopf invariant (for the massless case) classifies the
prevailing field configuration of the coset matrix \(\hat{T}(x_{p})\) with BCS terms according to their
content as a conserved, integrated helicity number.

In consequence, one has to transform relation (\ref{s5_14}) to terms of coset matrices \(\hat{T}(x_{p})\)
and has to find a Hopf invariant
\((V_{S^{2n-1}})^{-1}\int_{\mbox{\scz S}^{2n-1}}\hat{\omega}_{n-1}\:\wedge\:(d\hat{\omega}_{n-1})\) (\(n=2\))
of a one-form \(\hat{\omega}_{1}(x_{p})\) within the actions
\(\mscr{A}_{DET}[\hat{T},\langle\feynV\rangle_{\mbox{\scz(\ref{s3_59})}};\hat{\mscr{J}}]\),
\(\mscr{A}_{J_{\psi}}[\hat{T},\langle\feynV\rangle_{\mbox{\scz(\ref{s3_59})}};\hat{\mscr{J}}]\)
for the specified eigenvalue sort \((\hat{\tau}_{2})_{gf}\:\ovv{f}_{r}(x_{p})\:\delta_{\ovv{N}:\ovv{M}}\)
of the coset generator \(\hat{X}_{N;M}(x_{p})\) (\(r=1,\ldots,\,N_{c}=3\)).
We apply the particular representation of Dirac gamma matrices of \cite{wein1}
\be \lb{s5_15}
\hat{\beta}=\left(\bea{cc} 0 & \hat{1}_{2\times 2} \\ \hat{1}_{2\times 2} & 0 \eea\right)\;;\;\;\;
\hat{\gamma}^{0}=-\im\:\hat{\beta}\;;\;\;\;
\vec{\gamma}=-\im\left(\bea{cc} 0 & \vec{\sigma} \\ -\vec{\sigma} & 0 \eea\right)\;;\;\;\;
\hat{\gamma}_{5} = -\im\:\hat{\gamma}^{0}\:\hat{\gamma}^{1}\:\hat{\gamma}^{2}\:\hat{\gamma}^{3}=
\left(\bea{cc} \hat{1}_{2\times 2} & 0 \\ 0 & -\hat{1}_{2\times 2} \eea\right)\;,
\ee
and perform an anomalous doubling of quark fields \(\psi_{M}(x_{p})\rightarrow\Psi_{M}^{a}(x_{p})=
\{\psi_{M}(x_{p})\,,\,\psi_{M}^{*}(x_{p})\}\) with anomalous doubled 'Gamma' matrices
\(\hat{\mscr{B}}\), \(\hat{\Gamma}^{\mu}\), \(\hat{\Gamma}_{5}\) for the left-hand side of the original,
axial current relation (\ref{s5_11}) with the chiral anomaly
\beq \lb{s5_16}
\lefteqn{\hspace*{-1.3cm}\Big(\hat{\pp}_{p,\mu}\psi^{\dag}(x_{p})\;\hat{\beta}\;
\hat{\gamma}^{\mu}\;\hat{\gamma}_{5}\;\hat{\mfrak{t}}_{0}\;\psi(x_{p})\Big)=
\frac{1}{2}\;\;\hat{\pp}_{p,\mu}\Bigg[\Psi^{\dag,b}(x_{p})\;\hat{S}\left(\bea{cc}
\hat{\beta}\,\hat{\gamma}^{\mu}\,\hat{\gamma}_{5} & 0 \\
0 & \big(\hat{\beta}\,\hat{\gamma}^{\mu}\,\hat{\gamma}_{5}\big)^{T} \eea\right)^{ba}\Psi^{a}(x_{p})\Bigg] = } \\ \no &=&
-\frac{1}{2}\;\;\hat{\pp}_{p,\mu}\Bigg(\hspace*{0.6cm}
\TRALL\bigg[\Big(\hat{S}\:\hat{\mscr{B}}\:\hat{\Gamma}^{\mu}\:\hat{\Gamma}_{5}\Big)^{ba}\;
\Psi^{a}(x_{p})\otimes \Psi^{\dag,b}(x_{p})\bigg] \Bigg) \;; \;\;\;
\big(\,(\hat{\mfrak{t}}_{0})_{gf}=(\hat{1}_{2\times2})_{gf}\big)\;;   \\  \lb{s5_17}
\hat{\mscr{B}} &=& \left(\bea{cc} \hat{\beta} & 0 \\ 0 & \hat{\beta}^{T} \eea \right)\;;\;\;\;
\hat{\mscr{B}}\:\hat{\Gamma}^{\mu}=\left(\bea{cc} \hat{\beta}\:\hat{\gamma}^{\mu} & 0 \\
0 & \big(\hat{\beta}\:\hat{\gamma}^{\mu}\big)^{T} \eea\right)\;;\;\;\;
\hat{\Gamma}_{5}=\left(\bea{cc} \hat{\gamma}_{5} & 0 \\ 0 & \big(\hat{\gamma}_{5}^{T}=\hat{\gamma}_{5}\big)\eea\right)\;.
\eeq
The scalar product of anomalous doubled quark fields is converted to a dyadic product in an anomalous doubled,
'internal space' trace relation so that we can relate the dyadic product of doubled quark fields to the anomalous doubled
self-energy matrix. Furthermore, we can track the scalar or dyadic product of quark fields in (\ref{s5_16}) by
differentiation of the generating functions with respect to the source \(\hat{\mscr{J}}_{N;M}^{ba}(x_{p},x_{p})\)
to corresponding terms of coset matrices; thus, the axial current relation of the original quark fields is transformed
to a relation determined by coset matrices within the path integral (\ref{s3_110}-\ref{s3_116})
\beq \lb{s5_18}
\lefteqn{\hspace*{-1.0cm}
\hat{\pp}_{p,\mu}\Big(\psi^{\dag}(x_{p})\;\hat{\beta}\;\hat{\gamma}^{\mu}\;\hat{\gamma}_{5}\;\hat{\mfrak{t}}_{0}\;
\psi(x_{p})\Big)=
-\frac{1}{2}\;\;\hat{\pp}_{p,\mu}\Bigg(\hspace*{0.6cm}
\TRALL\Big[\Big(\hat{S}\:\hat{\mscr{B}}\:\hat{\Gamma}^{\mu}\:\hat{\Gamma}_{5}\Big)_{N;M}^{ba}\;
\Psi_{M}^{a}(x_{p})\otimes\Psi_{N}^{\dag,b}(x_{p})\Big] \Bigg) } \\ \no &=&\im\;\mcal{N}^{2}\;
\hat{\pp}_{p,\mu}\Bigg(\hspace*{0.6cm}
\TRALL\bigg[\Big(\hat{S}\:\hat{\mscr{B}}\:\hat{\Gamma}^{\mu}\:\hat{\Gamma}_{5}\Big)_{N;M}^{ba}
\bigg(\frac{\pp}{\pp\hat{\mscr{J}}_{N;M}^{ba}(x_{p},x_{p})}
Z[\hat{\mscr{J}},J_{\psi},\hat{J}_{\psi\psi},\hat{\mfrak{j}}^{(\hat{F})}]\bigg)\bigg]\Bigg) \;.
\eeq
We restrict to the 'interaction representation' of the action \(\mscr{A}_{DET}[\hat{T}_{\hat{\mfrak{W}}};\hat{\mscr{J}}]\)
of the determinant and neglect terms from \(\mscr{A}_{J_{\psi}}[\hat{T}_{\hat{\mfrak{W}}};\hat{\mscr{J}}]\)
having arbitrary-valued, anti-commuting source fields \(J_{\psi}(x_{p})\). This allows to extract a third order
derivative term of
\(\Delta\!\hat{\mscr{H}}_{\hat{\mfrak{W}}}\!(\hat{T}^{-1}_{\hat{\mfrak{W}}},\hat{T}_{\hat{\mfrak{W}}})
=\hat{T}^{-1}_{\hat{\mfrak{W}}}\;\hat{\mscr{H}}_{\hat{\mfrak{W}}}\;\hat{T}_{\hat{\mfrak{W}}}  -
\hat{\mscr{H}}_{\hat{\mfrak{W}}}\)
from \(\mscr{A}_{DET}[\hat{T}_{\hat{\mfrak{W}}};\hat{\mscr{J}}]\) by differentiation with respect to
\(\hat{\mscr{J}}_{N;M}^{ba}(x_{p},x_{p})\)
\beq \lb{s5_19}
\lefteqn{Z[\hat{\mscr{J}}_{\hat{\mfrak{W}}},J_{\psi;\hat{\mfrak{W}}},\hat{J}_{\psi\psi}]=\int
d[\hat{T}^{-1}(x_{p})\:d\hat{T}(x_{p})]\;\; \big\langle Z_{\hat{J}_{\psi\psi}}[\hat{\mfrak{W}}\:\hat{T}_{\hat{\mfrak{W}}}\:\hat{\mfrak{W}}^{-1}]\big\rangle\;\times  } \\ \no &\times&
 \exp\Big\{\mscr{A}_{DET}[\hat{T}_{\hat{\mfrak{W}}};\hat{\mscr{J}}_{\hat{\mfrak{W}}}\equiv0]+\im\:
\mscr{A}_{J_{\psi;\hat{\mfrak{W}}}}[\hat{T}_{\hat{\mfrak{W}}};\hat{\mscr{J}}_{\hat{\mfrak{W}}}\equiv0] \Big\}
\times  \\  \no &\times& \exp\bigg\{-\frac{1}{2}\hspace*{0.6cm}\trxpa\trfgamc
\Big[\wt{\mscr{J}}_{\hat{\mfrak{W}}}(\hat{T}^{-1}_{\hat{\mfrak{W}}},\hat{T}_{\hat{\mfrak{W}}})\;
\hat{\mscr{H}}_{\hat{\mfrak{W}}}^{\boldsymbol{-1}}\;
\Big(\Delta\!\hat{\mscr{H}}_{\hat{\mfrak{W}}}(\hat{T}^{-1}_{\hat{\mfrak{W}}},\hat{T}_{\hat{\mfrak{W}}})\;\;
\hat{\mscr{H}}_{\hat{\mfrak{W}}}^{\boldsymbol{-1}}\Big)^{3}\Big]\bigg\} =  \\  \no  &=&
\int d[\hat{T}^{-1}(x_{p})\:d\hat{T}(x_{p})]\;\;
\big\langle Z_{\hat{J}_{\psi\psi}}[\hat{\mfrak{W}}\:\hat{T}_{\hat{\mfrak{W}}}\:\hat{\mfrak{W}}^{-1}]\big\rangle
\;\times   \\
\no &\times& \exp\Big\{\mscr{A}_{DET}[\hat{T}_{\hat{\mfrak{W}}};\hat{\mscr{J}}_{\hat{\mfrak{W}}}\equiv0]+\im\:
\mscr{A}_{J_{\psi;\hat{\mfrak{W}}}}[\hat{T}_{\hat{\mfrak{W}}};\hat{\mscr{J}}_{\hat{\mfrak{W}}}\equiv0] \Big\} \times
\\ \no &\times& \bigg(\bigg(-\frac{1}{2}\bigg)\hspace*{0.6cm}
\trxpa\trfgamc\bigg[\hat{T}_{\hat{\mfrak{W}}}^{-1}\:\hat{I}\:\hat{S}\:
\hat{\eta}\:\hat{\mfrak{W}}^{-1}\;\frac{\hat{\mscr{J}}}{\mcal{N}}\;\hat{\mfrak{W}}\:\hat{\eta}\:\hat{S}\:\hat{I}\;
\hat{T}_{\hat{\mfrak{W}}}\:\hat{\mscr{H}}_{\hat{\mfrak{W}}}^{\boldsymbol{-1}}
\Big(\Delta\!\hat{\mscr{H}}_{\hat{\mfrak{W}}}\!(\hat{T}^{-1}_{\hat{\mfrak{W}}},
\hat{T}_{\hat{\mfrak{W}}})\;\;\hat{\mscr{H}}_{\hat{\mfrak{W}}}^{\boldsymbol{-1}}\Big)^{3}\bigg]\bigg)\;.
\eeq
This yields following equation of the axial current in terms of BCS quark pairs instead of the original
anti-commuting fields
\beq \lb{s5_20}
\lefteqn{\hat{\pp}_{p,\mu}\Big(\psi^{\dag}(x_{p})\;\hat{\beta}\;\hat{\gamma}^{\mu}\;\hat{\gamma}_{5}\;
\hat{\mfrak{t}}_{0}\;\psi(x_{p})\Big)=
\im\:\bigg(-\frac{1}{2}\bigg)\;\hat{\pp}_{p,\mu}\Bigg(\hspace*{0.6cm}\TRALL\bigg[
\Big(\hat{S}\:\hat{\mscr{B}}\:\hat{\Gamma}^{\mu}\:\hat{\Gamma}_{5}\Big)_{N;M}^{ba} \times } \\ \no &\times&
\bigg(\hat{\mfrak{W}}(x_{p})\:\hat{S}\:\hat{I}\:\hat{T}_{\hat{\mfrak{W}}}(x_{p})\;
\Big\langle\widehat{x_{p}}\Big|\hat{\mscr{H}}_{\hat{\mfrak{W}}}^{\boldsymbol{-1}}
\Big(\Delta\!\hat{\mscr{H}}_{\hat{\mfrak{W}}}\!(\hat{T}^{-1}_{\hat{\mfrak{W}}},\hat{T}_{\hat{\mfrak{W}}})\;\;
\hat{\mscr{H}}_{\hat{\mfrak{W}}}^{\boldsymbol{-1}}\Big)^{3}\Big|\widehat{x_{p}}\Big\rangle\;\;
\hat{T}^{-1}_{\hat{\mfrak{W}}}(x_{p})\;\hat{I}\:\hat{S}\:\hat{\mfrak{W}}^{-1}(x_{p})\bigg)_{M;N}^{ab}\bigg]\Bigg)\;.
\eeq
The 'interaction transformed' mean field operator \(\hat{\mscr{H}}^{\boldsymbol{-1}}_{\hat{\mfrak{W}}}\)
(\ref{s4_55}-\ref{s4_57}) of the propagation
and the 'relative' gradient operator \(\Delta\!\hat{\mscr{H}}_{\hat{\mfrak{W}}}(\hat{T}^{-1}_{\hat{\mfrak{W}}},\hat{T}_{\hat{\mfrak{W}}})\) are considered for the bulk of
a nucleus so that we can approximate all 'interaction transformed' Dirac gamma matrices
\(\hat{\mscr{B}}_{\hat{\mfrak{W}}}\), \(\hat{\Gamma}^{\mu}_{\hat{\mfrak{W}}}\),
\(\hat{\Gamma}_{5,\hat{\mfrak{W}}}\) with their original matrices apart from a constant
similarity transformation. According to our ansatz (\ref{s5_21}-\ref{s5_23}) for \(\hat{T}(x_{p})\),
the coset matrix is also not effected by the transformation to the 'interaction picture' because the
transforming block diagonal matrices \(\hat{\mfrak{W}}_{N;M}^{aa}(x_{p})\) need
not be considered for constant similarity transformations in the bulk of a nucleus and decouple from the isospin space
with indices '\(g,\,f\)' for vanishing mass term \(\hat{m}\rightarrow 0\) (compare (\ref{s4_55}-\ref{s4_57}) )
\beq \lb{s5_21}
\Delta\!\hat{\mscr{H}}_{\hat{\mfrak{W}}}(\hat{T}^{-1}_{\hat{\mfrak{W}}},\hat{T}_{\hat{\mfrak{W}}})
&=&\hat{T}_{\hat{\mfrak{W}}}^{-1}\;\hat{\mscr{B}}_{\hat{\mfrak{W}}}(x_{p})\;\hat{S}\;
\hat{\Gamma}^{\mu}_{\hat{\mfrak{W}}}(x_{p})\;\boldsymbol{\hat{\pp}_{p,\mu}}\;\hat{T}_{\hat{\mfrak{W}}}-
\hat{\mscr{B}}_{\hat{\mfrak{W}}}(x_{p})\;\hat{S}\;
\hat{\Gamma}_{\hat{\mfrak{W}}}^{\mu}(x_{p})\;\boldsymbol{\hat{\pp}_{p,\mu}}
  \\ \no &\simeq&
\hat{T}^{-1}\;\hat{\mscr{B}}\;\hat{S}\;\hat{\Gamma}^{\mu}\;\boldsymbol{\hat{\pp}_{p,\mu}}\;\hat{T}-
\hat{\mscr{B}}\;\hat{S}\;\hat{\Gamma}^{\mu}\;\boldsymbol{\hat{\pp}_{p,\mu}} \;;  \\ \lb{s5_22}
\hat{T}_{\hat{\mfrak{W}}}(x_{p}) &=&\exp\big\{-\hat{Y}_{\hat{\mfrak{W}}}(x_{p})\,\big\}\;;\;\;\;
\hat{Y}_{\hat{\mfrak{W}}}(x_{p})=\left(\bea{cc} 0 & \hat{X}_{\hat{\mfrak{W}}}(x_{p})  \\
\hat{X}^{\dag}_{\hat{\mfrak{W}}}(x_{p}) & 0 \eea\right)\;;  \\ \lb{s5_23}
\hat{X}_{\hat{\mfrak{W}};N;M}(x_{p}) &=& \hat{X}_{\hat{\mfrak{W}};g,\ovv{N};f,\ovv{M}}(x_{p}) \simeq
\big(\hat{\tau}_{2}\big)_{gf}\;\ovv{f}_{r}(x_{p})\;\delta_{\ovv{N};\ovv{M}} =\hat{X}_{0}(x_{p})  \;;
\hspace*{0.3cm}(r=1,\ldots,\, N_{c}=3)\;; \\ \lb{s5_24}
\Longrightarrow \hat{T}_{\hat{\mfrak{W}}}(x_{p}) &\simeq&\hat{T}_{0}(x_{p})=
\exp\left\{-\left(\bea{cc} 0 & \hat{X}_{0}(x_{p})  \\ \hat{X}_{0}^{\dag}(x_{p}) & 0 \eea\right)\right\}\;.
\eeq
We introduce the anomalous doubled, diagonal spin matrices \(\hat{\mscr{S}}^{k}\) (\(k=1,2,3\)) instead of
the ordinary Pauli spin matrices \(\hat{\sigma}^{k}\) through inclusion of the Dirac gamma matrix \(\hat{\Gamma}_{5}\)
\be \lb{s5_25}
\im\;\hat{\mscr{S}}^{k}=\hat{\mscr{B}}\;\hat{\Gamma}^{k}\;\hat{\Gamma}_{5}=\im\;\left(\bea{cc}
\big(\hat{\mscr{S}}^{k}\big)^{11} & \\ & \big(\hat{\mscr{S}}^{k}\big)^{22} \eea\right) =\im\;
\left(\bea{cc}\left(\bea{cc} \sigma^{k} & \\ & \sigma^{k} \eea\right) & \\ &
\left(\bea{cc} \sigma^{k} & \\ & \sigma^{k} \eea\right)^{T} \eea\right)_{\mbox{,}}
\ee
and substitute this into the spacetime integrated, zero component of the axial
current relation (\ref{s5_20}) for vanishing spatial current densities at
the surface of the considered three volume. Since the block diagonal, anomalous
doubled spin matrices \(\hat{\mscr{S}}^{k}\) (\ref{s5_25}) commute with the coset matrix \(\hat{T}_{0}(x_{p})\)
of the internal isospin and colour degrees of freedom, the total trace
\(\mfrak{Tr}_{\scrscr N_{f},\hat{\gamma}^{(\mu)},N_{c}}^{a(=1,2)}\) separates into a trace of
Dirac gamma matrices and a trace of anomalous doubled isospin and colour degrees of freedom for the coset matrix
(\ref{s5_26},\ref{s5_27}); furthermore, we point out the trace \(\mfrak{tr}_{\scrscr N_{c}}\) over the completely
diagonal 'colour' terms with \(r=1,\ldots,\,N_{c}=3\) of \(\ovv{f}_{r}(x_{p})\).
Since one also performs the trace over colour and isospin degrees of freedom, the integrated
zero component of the helicity instanton number can only refer to field configurations of 'nucleons in its entity';
one can thereby classify the
total number of nucleons in a field configuration composed of coset matrices \(\hat{T}(x_{p})\)
by the derived helicity instanton number. As we act with the operator \(\boldsymbol{\hat{\pp}_{p,j}}\)
of \(\hat{\mscr{B}}\:\hat{S}\:\hat{\Gamma}^{j}\:\boldsymbol{\hat{\pp}_{p,j}}\) for creating the curl of
\(\hat{T}^{-1}\:\hat{\mscr{B}}\:\hat{S}\:\hat{\Gamma}^{k}\:(\hat{\pp}_{p,k}\hat{T})\)
in relation (\ref{s5_20}) in order to extract
a Hopf invariant of \(\Pi_{3}(\mbox{SU}(2)\,)=\mathsf{Z}\), one retains a single remaining combination and has also
to mind the summation over the different fields \(\ovv{f}_{r}(x_{p})\)
following from the trace over completely diagonal colour degrees of freedom
\beq \lb{s5_26}
\lefteqn{\hspace*{-1.9cm}\int_{-\infty}^{+\infty}d^{4}\!x_{p}\;
\hat{\pp}_{p,\mu}\Big(\psi^{\dag}(x_{p})\;\hat{\beta}\;\hat{\gamma}^{\mu}\;\hat{\gamma}_{5}\;\hat{\mfrak{t}}_{0}\;
\psi(x_{p})\Big)=\int d^{3}\!\vec{x}\;\psi^{\dag}(x_{p})\;\hat{\beta}\;\hat{\gamma}^{0}\;\hat{\gamma}_{5}\;\psi(x_{p})
\Big|_{x_{p}^{0}=-\infty}^{x_{p}^{0}=+\infty}= } \\ \no &=&
\im\:\bigg(-\frac{1}{2}\bigg)\int d^{3}\!\vec{x}\;\hspace*{0.6cm}\TRALL\bigg[\hat{\mscr{B}}\,\hat{\Gamma}^{0}\;
\hat{\mscr{B}}\,\hat{\Gamma}^{i}\,\hat{\Gamma}_{5}\;\hat{\mscr{B}}\,\hat{\Gamma}^{j}\,\hat{\Gamma}_{5}\,
\hat{\mscr{B}}\,\hat{\Gamma}^{k}\,\hat{\Gamma}_{5} \;\times  \\ \no &\times&
\hat{T}^{-1}(x_{p})\;\hat{S}\;\big(\hat{\pp}_{p,i}\hat{T}(x_{p})\big)\Big(\hat{S}\;\hat{\pp}_{p,j}\;
\hat{T}^{-1}(x_{p})\:\hat{S}\:\big(\hat{\pp}_{p,k}\hat{T}(x_{p})\big)\Big)\bigg]\;
\bigg|_{x_{p}^{0}=-\infty}^{x_{p}^{0}=+\infty}\;;
\eeq
\be \lb{s5_27}
\trgam\bigg[\underbrace{\hat{\mscr{B}}\:\hat{\Gamma}^{0}}_{-\im\:\hat{1}}\;
\underbrace{\hat{\mscr{B}}\:\hat{\Gamma}^{i}\:\hat{\Gamma}_{5}}_{\im\:\hat{\mscr{S}}^{i}}\;
\underbrace{\hat{\mscr{B}}\:\hat{\Gamma}^{j}\:\hat{\Gamma}_{5}}_{\im\:\hat{\mscr{S}}^{j}}\;
\underbrace{\hat{\mscr{B}}\:\hat{\Gamma}^{k}\:\hat{\Gamma}_{5}}_{\im\:\hat{\mscr{S}}^{k}}\bigg]=
(-\im)\;\im^{3}\;4\;\im\;\ve^{ijk}\;\hat{1}^{ab}\;\hat{1}_{gf}\;\hat{1}_{sr} \;.
\ee
As we insert the relation (\ref{s5_27}) of the trace over anomalous doubled spin matrices \(\hat{\mscr{S}}^{k}\)
(\ref{s5_25}) into (\ref{s5_26}), one introduces the three dimensional, completely anti-symmetric
'Levi-Civita symbol' '\(\ve^{ijk}\)' and attains Eq. (\ref{s5_28}) with remaining traces over
isospin and colour degrees of freedom
\beq \lb{s5_28}
\lefteqn{\hspace*{-0.6cm}\int_{-\infty}^{+\infty}d^{4}\!x_{p}\;
\hat{\pp}_{p,\mu}\Big(\psi^{\dag}(x_{p})\;\hat{\beta}\;\hat{\gamma}^{\mu}\;\hat{\gamma}_{5}\;\hat{\mfrak{t}}_{0}\;
\psi(x_{p})\Big)=\int d^{3}\!\vec{x}\;\psi^{\dag}(x_{p})\;\hat{\beta}\;\hat{\gamma}^{0}\;\hat{\gamma}_{5}\;\psi(x_{p})
\Big|_{x_{p}^{0}=-\infty}^{x_{p}^{0}=+\infty}= } \\ \no &=&
(-2)\int d^{3}\!\vec{x}\;\ve^{ijk}\;\hspace*{0.6cm}\TRFC
\bigg[\hat{T}^{-1}(x_{p})\;\hat{S}\:
\big(\hat{\pp}_{p,i}\hat{T}(x_{p})\,\big)\;\Big(\hat{S}\;\hat{\pp}_{p,j}\:\hat{T}^{-1}(x_{p})\;\hat{S}\;
\big(\hat{\pp}_{p,k}\hat{T}(x_{p})\big)\Big)\bigg]\;\bigg|_{x_{p}^{0}=-\infty}^{x_{p}^{0}=+\infty}\;.
\eeq
We apply the ansatz (\ref{s5_21}-\ref{s5_24},\ref{s5_29}) for relation (\ref{s5_28}) and
specify the complex fields \(\ovv{f}_{r}(x_{p})\) in terms of their absolute value \(|\ovv{f}_{r}(x_{p})\,|\)
and phase \(\phi_{r}(x_{p})\)
\beq \lb{s5_29}
\hat{T}(x_{p})\stackrel{\wedge}{=}\hat{T}_{0}(x_{p}) &=&
\exp\left\{-\left(\bea{cc} 0 & \hat{X}_{0}(x_{p}) \\ \hat{X}_{0}^{\dag}(x_{p}) \eea\right)\right\} \;;\;\;\;
\hat{X}_{0}(x_{p})\stackrel{\wedge}{=}\big(\hat{\tau}_{2}\big)_{gf}\;\ovv{f}_{r}(x_{p})\;
\delta_{\ovv{N};\ovv{M}}\;;   \\ \no
\ovv{f}_{r}(x_{p}) &=&  \big|\ovv{f}_{r}(x_{p})\,\big|\;\;\exp\big\{\im\;\phi_{r}(x_{p})\,\big\} \;,
\eeq
so that the transformed relation (\ref{s5_28}) of the original axial current conversation with chiral anomaly finally reduces
to the 'curl of the gradient of the phase \(\phi_{r}(x_{p})\)' and to the 'gradient
of the hyperbolic sine of the absolute value \(|\ovv{f}_{r}(x_{p})\,|\)'
\beq \lb{s5_30}
\lefteqn{\hspace*{-1.9cm}\frac{1}{2} \int_{-\infty}^{+\infty}d^{4}\!x_{p}\;\hat{\pp}_{p,\mu}
\Big(\psi^{\dag}(x_{p})\;\hat{\beta}\;\hat{\gamma}^{\mu}\;
\hat{\gamma}_{5}\;\hat{\mfrak{t}}_{0}\;\psi(x_{p})\Big) =
-\frac{\ve^{\kappa\lambda\mu\nu}}{32\:\pi^{2}}\;\int d^{4}\!x_{p}\;
\trfc\Big[\hat{\mfrak{t}}_{0}\;\hat{F}_{\kappa\lambda}(x_{p})\;\hat{F}_{\mu\nu}(x_{p})\Big] = } \\ \no &=&
\im\;(N_{f}=2)\sum_{r=1}^{N_{c}=3}\int d^{3}\!\vec{x}\;\ve^{ijk}\;
\Big(\hat{\pp}_{p,i}\sinh^{4}\big(\big|\ovv{f}_{r}(x_{p})\,\big|\,\big)\Big)\;\;
\Big(\hat{\pp}_{p,j}\hat{\pp}_{p,k}\phi_{r}(x_{p})\Big)\;\;\bigg|_{x_{p}^{0}=-\infty}^{x_{p}^{0}=+\infty}\;.
\eeq
It seems that the anti-symmetric combination \(\ve^{ijk}\;(\hat{\pp}_{p,j}\hat{\pp}_{p,k}\phi_{r}(x_{p})\,)\)
should result into completely vanishing terms; however, one has to take into account the multi-valued
properties of phases \(\phi_{r}(x_{p})\) whose anti-symmetric, second order, spatial gradients commute
everywhere except at line singularities which cause corresponding vortices. If one defines for the
curl of the gradient of the phase \(\phi_{r}(x_{p})\) a line singularity (\ref{s5_31}) along the
\(x^{3}=z\)-axis in standard cylindrical coordinates (\(\rho\,,\,\varphi\,,\,z\)) with locally
orthonormal basis vectors (\(\vec{e}_{\rho}\,,\,\vec{e}_{\varphi}\,,\,\vec{e}_{z}\))
\be \lb{s5_31}
\vec{\nabla}_{\vec{x}}\times\vec{\nabla}_{\vec{x}}\phi_{r}(x_{p})=\vec{e}_{z}\;2\pi\; n_{z,r}\;\;
\delta(x^{1})\;\delta(x^{2}) \;;\;\;\;(n_{z,r}\in\mathsf{Z})\;,
\ee
and applies appropriate boundary conditions (\ref{s5_32}) along the \(x^{3}=z\)-axis, we can verify the quantization
of relation (\ref{s5_30}) by '\(n_{z,r}\)' (\ref{s5_34}) for
the BCS quark pair ansatz (\ref{s5_29}) of the coset matrix \(\hat{T}_{0}(x_{p})\). The phase
\(\phi_{r}(x_{p})\) (\ref{s5_33}) is obtained by Stokes theorem from (\ref{s5_31}) according to the
azimuthal symmetry within the cylinder coordinates
\beq  \lb{s5_32}
\Big(\sinh^{2}\big(\big|\ovv{f}_{r}(t_{p},x^{3}=0)\,\big|\big)=0 &;&
\sinh^{2}\big(\big|\ovv{f}_{r}(t_{p},x^{3}=L_{z})\,\big|\big)=\frac{1}{\sqrt{2\,\pi}}\Big)  \;; \\  \no
\vec{\nabla}_{\vec{x}}\times\vec{\nabla}_{\vec{x}}\phi_{r}(x_{p})&=&\vec{e}_{z}\;2\pi\;n_{z,r}\;\;
\delta(x^{1})\;\delta(x^{2}) \;; \\ \lb{s5_33} (\mbox{Stokes theorem }\Longrightarrow )\;\;
\vec{\nabla}_{\vec{x}}\phi_{r}(t_{p},\vec{x})&=&\vec{e}_{\varphi}\;\frac{1}{\rho}\;
\frac{\pp\phi_{r}(t_{p},\vec{x})}{\pp\varphi}=
\vec{e}_{\varphi}\;\frac{n_{z,r}}{\sqrt{(x^{1})^{2}+(x^{2})^{2}}} \;; \\ \no
\phi_{r}(t_{p},\vec{x}) &=&n_{z,r}\;\;\varphi\;;\;\;\;\varphi\in[0,2\pi) \;;
\eeq
\beq \lb{s5_34}
\lefteqn{\frac{1}{2} \int_{-\infty}^{+\infty}d^{4}\!x_{p}\;\hat{\pp}_{p,\mu}
\Big(\psi^{\dag}(x_{p})\;\hat{\beta}\;\hat{\gamma}^{\mu}\;
\hat{\gamma}_{5}\;\hat{\mfrak{t}}_{0}\;\psi(x_{p})\Big) =
-\frac{\ve^{\kappa\lambda\mu\nu}}{32\:\pi^{2}}\;\int d^{4}\!x_{p}\;
\trfc\Big[\hat{\mfrak{t}}_{0}\;\hat{F}_{\kappa\lambda}(x_{p})\;\hat{F}_{\mu\nu}(x_{p})\Big] = }
\\ \no &=&
\im\;(N_{f}=2)\sum_{r=1}^{N_{c}=3}\int d^{3}\!\vec{x}\;\ve^{ijk}\;
\Big(\hat{\pp}_{p,i}\sinh^{4}\big(\big|\ovv{f}_{r}(x_{p})\,\big|\,\big)\Big)\;\;
\Big(\hat{\pp}_{p,j}\hat{\pp}_{p,k}\phi_{r}(x_{p})\Big)\;\bigg|_{x_{p}^{0}=-\infty}^{x_{p}^{0}=+\infty}    =
\im(N_{f}=2)\sum_{r=1}^{N_{c}=3}n_{z,r}.
\eeq
Eqs. (\ref{s5_30},\ref{s5_34}) can also be related to the Hopf invariant with one-form \(\hat{\omega}_{1}(x_{p})\)
in a direct manner
\be\lb{s5_35}
\int \frac{\hat{\omega}_{1}(x_{p})}{2\,\pi^{2}}\:\wedge\Big(d\hat{\omega}_{1}(x_{p})\Big)=\mbox{integer number}\;,
\ee
by changing the one-forms \(dx^{i}\hat{\pp}_{p,i}\sinh^{4}(|\ovv{f}_{r}(x_{p})\,|)\),
\(dx^{k}\hat{\pp}_{p,k}\phi_{r}(x_{p})\) to corresponding one-forms
\(d\sinh^{4}(|\ovv{f}_{r}(x_{p})\,|)\), \(d\phi_{r}(x_{p})\) {\it which are not 'exact'}
\beq \lb{s5_36}
\lefteqn{\hspace*{-1.9cm}\frac{1}{2} \int_{-\infty}^{+\infty}d^{4}\!x_{p}\;\hat{\pp}_{p,\mu}
\Big(\psi^{\dag}(x_{p})\;\hat{\beta}\;\hat{\gamma}^{\mu}\;
\hat{\gamma}_{5}\;\hat{\mfrak{t}}_{0}\;\psi(x_{p})\Big) =
-\frac{\ve^{\kappa\lambda\mu\nu}}{32\:\pi^{2}}\;\int d^{4}\!x_{p}\;
\trfc\Big[\hat{\mfrak{t}}_{0}\;\hat{F}_{\kappa\lambda}(x_{p})\;\hat{F}_{\mu\nu}(x_{p})\Big] = }
\\ \no &=&
\im\;(N_{f}=2)\sum_{r=1}^{N_{c}=3}\int
\Big(d\sinh^{4}\big(\big|\ovv{f}_{r}(x_{p})\,\big|\,\big)\Big)\;\wedge\;
\Big(\underbrace{dx^{j}\hat{\pp}_{p,j}}_{\equiv d}\;d\phi_{r}(x_{p})\Big)\;
\bigg|_{x_{p}^{0}=-\infty}^{x_{p}^{0}=+\infty}\;;   \\ \no
d&=&d\rho\;\frac{\pp}{\pp\rho}+
d\varphi\;\frac{\pp}{\pp\varphi}+dz\;\frac{\pp}{\pp z}\;.
\eeq
As one substitutes \(d\phi_{r}(x_{p})\), \(d\sinh^{4}(|\ovv{f}_{r}(x_{p})\,|)\)
according to following relations into (\ref{s5_36})
\beq \lb{s5_37}
d\phi_{r}(x_{p}) &=& n_{z,r}\;\theta(\rho-\rho_{z}^{\scrscr(0)})\;d\varphi +
n_{\varphi,r}\;\theta(\rho_{\varphi}^{\scrscr(0)}-\rho)\;dz\;2\,\pi/L_{z} \;; \\ \lb{s5_38}
d\sinh^{4}(|\ovv{f}_{r}(x_{p})\,|) &=&\frac{1}{2\,\pi^{2}}\;\;\frac{d\phi_{r}(x_{p})}{4} \;;\;\;\;
(0<z<L_{z})\;; \\ \no
\rho_{\varphi}^{\scrscr(0)}>\rho_{z}^{\scrscr(0)} &;& n_{z,r}\;,\;n_{\varphi,r}\in\mathsf{Z}\;,
\eeq
one directly achieves a 'Hopf quantization' for \(\Pi_{3}(S^{2})=\mathsf{Z}\)
from (\ref{s5_36}) with the integer numbers \(n_{z,r}\;,\;n_{\varphi,r}\)
\beq \lb{s5_39}
\lefteqn{\frac{1}{2} \int_{-\infty}^{+\infty}d^{4}\!x_{p}\;\hat{\pp}_{p,\mu}
\Big(\psi^{\dag}(x_{p})\;\hat{\beta}\;\hat{\gamma}^{\mu}\;
\hat{\gamma}_{5}\;\hat{\mfrak{t}}_{0}\;\psi(x_{p})\Big) =
-\frac{\ve^{\kappa\lambda\mu\nu}}{32\:\pi^{2}}\;\int d^{4}\!x_{p}\;
\trfc\Big[\hat{\mfrak{t}}_{0}\;\hat{F}_{\kappa\lambda}(x_{p})\;\hat{F}_{\mu\nu}(x_{p})\Big] = }
\\ \no &=&
\im\;(N_{f}=2)\sum_{r=1}^{N_{c}=3}\int
\Big(d\sinh^{4}\big(\big|\ovv{f}_{r}(x_{p})\,\big|\,\big)\Big)\;\wedge\;
\Big(\underbrace{dx^{j}\hat{\pp}_{p,j}}_{\equiv d}\;d\phi_{r}(x_{p})\Big)\;
\bigg|_{x_{p}^{0}=-\infty}^{x_{p}^{0}=+\infty}=\im\;(N_{f}=2)\sum_{r=1}^{N_{c}=3}n_{z,r}\;,\;n_{\varphi,r}\;.
\eeq
Additonally, we note that the integrals (\ref{s5_35},\ref{s5_39}) are invariant under deformations of
the one-forms \(\hat{\omega}_{1}(x_{p})\), \(d\phi_{r}(x_{p})\),
\(d\sinh^{4}(|\ovv{f}_{r}(x_{p})\,|\,)\) (\ref{s5_37},\ref{s5_38}) so that
one has definitely determined a topological invariant, the 'Hopf invariant' \cite{Naka}.

\section{Summary and conclusion} \lb{s6}

The rather involved appearance of section \ref{s3} contains the various HST's from the original
QCD path integral with fermionic quark- and non-Abelian gauge fields to the corresponding self-energies.
Since self-energies comprise the infinite sum of one-particle irreducible terms in a perturbation series,
the path integral representation with self-energy matrices is advantageous to the original representation
with fermionic matter- and non-Abelian gauge fields, especially due to various possible approximations.
The total self-energy for the anomalous doubled dyadic product of quark fields consists of block diagonal
density terms and off-diagonal BCS quark pairs which have been separated in a coset decomposition
\(\mbox{SO}(N_{0},N_{0})\,/\,\mbox{U}(N_{0})\otimes \mbox{U}(N_{0})\) for a SSB with the unitary \(\mbox{U}(N_{0})\)
subgroup symmetry for the invariant ground or vacuum states
(\(N_{0}=(N_{f}=2)\times 4_{\hat{\gamma}}\times(N_{c}=3)=24\)). Bilinear observables follow from differentiating
the prevailing form of the generating function with respect to the source field
\(\hat{\mscr{J}}_{N;M}^{ba}(y_{q},x_{p})\) so that one can track the original observables of quark matter
to their corresponding from in terms of the total self-energy. Since the path integral with self-energy
generator \(\mbox{so}(N_{0},N_{0})\) can be separated by 'hinge' fields into purely density related parts
\(\mbox{u}(N_{0})\) and off-diagonal coset parts \(\mbox{so}(N_{0},N_{0})\,/\,\mbox{u}(N_{0})\) for
BCS quark pairs, we can consider the density related part of the total path integral as background fields
for the BCS terms. The density related path integral particularly allows for a mean field approximation
of a composed gauge field \(\mcal{V}_{\alpha}^{\mu}(x_{p})\) which replaces the original gauge fields
\(A_{\alpha}^{\mu}(x_{p})\) and comprises colour dressed, scalar quark densities and self-energies
for the original field strength tensor \(\hat{F}_{\alpha}^{\mu\nu}(x_{p})\). This mean field solution
\(\langle\feynV(x_{p})\rangle_{\mbox{\scz(\ref{s3_59})}}\) also has a non-hermitian part which has to comply
with the correct sign of the original, imaginary '\(-\im\:\hat{\ve}_{p}\)' terms for a proper convergence
of the generating function. Therefore, we accomplish a path integral which finally only contains coset elements
\(\mbox{so}(N_{0},N_{0})\,/\,\mbox{u}(N_{0})\) propagating in density related, saddle point approximated
gauge fields \(\langle\feynV(x_{p})\rangle_{\mbox{\scz(\ref{s3_59})}}\). The remaining actions
\(\mscr{A}_{DET}[\hat{T},\langle\feynV\rangle_{\mbox{\scz(\ref{s3_59})}};\hat{\mscr{J}}]\),
\(\mscr{A}_{J_{\psi}}[\hat{T},\langle\feynV\rangle_{\mbox{\scz(\ref{s3_59})}};\hat{\mscr{J}}]\)
consist of gradient operators '\(\boldsymbol{\hat{\pp}_{p,\mu}}\)' acting onto coset matrices which do not allow
for simple, finite order gradient expansions; this particular problem is emphasized by a transformation
to an 'interaction representation' so that the anomalous doubled one-particle operator
\(\hat{\mscr{H}}_{\hat{\mfrak{W}}}\) has only unsaturated gradient operators with spatially dependent
Dirac gamma matrices. Since the actions
\(\mscr{A}_{DET}[\hat{T}_{\hat{\mfrak{W}}};\hat{\mscr{J}}_{\hat{\mfrak{W}}}]\)
\(\mscr{A}_{J_{\psi;\hat{\mfrak{W}}}}[\hat{T}_{\hat{\mfrak{W}}};\hat{\mscr{J}}_{\hat{\mfrak{W}}}]\)
mainly encompass the combination of operators
\(\hat{T}_{\hat{\mfrak{W}}}^{-1}\:\hat{\mscr{H}}_{\hat{\mfrak{W}}}\:\hat{T}_{\hat{\mfrak{W}}}\:
\hat{\mscr{H}}_{\hat{\mfrak{W}}}^{\boldsymbol{-1}}\), a small momentum expansion for
\(\hat{T}_{\hat{\mfrak{W}}}^{-1}\:\hat{\mscr{H}}_{\hat{\mfrak{W}}}\:\hat{T}_{\hat{\mfrak{W}}}\)
is always accompanied by a 'large' momentum expansion for
\(\hat{T}_{\hat{\mfrak{W}}}\:\hat{\mscr{H}}_{\hat{\mfrak{W}}}^{\boldsymbol{-1}}\:\hat{T}_{\hat{\mfrak{W}}}^{-1}\)
due to the inverse operator \(\hat{\mscr{H}}_{\hat{\mfrak{W}}}^{\boldsymbol{-1}}\) of pure gradients
in the trace relations. We suggest the specific integral representations (\ref{s4_27}-\ref{s4_29})
of the logarithm and of the inverse of an operator (\ref{s4_29}) which can simplify the computation
of observables.

It is of peculiar interest to investigate the path integral with coset matrices \(\hat{T}(x_{p})\)
and approximated background fields \(\langle\feynV(x_{p})\rangle_{\mbox{\scz(\ref{s3_59})}}\)
for nontrivial topologies especially in comparison to the original Skyrme model with
homotopy group \(\Pi_{3}(\mbox{SU}(2)\,)=\mathsf{Z}\). As one restricts to the anti-symmetric, quaternion-valued
complex eigenvalues of coset generators \(\hat{X}(x_{p})\), \(\hat{X}^{\dag}(x_{p})\)
with Pauli matrix \((\hat{\tau}_{2})_{gf}\), one only has two real angle degrees of freedom
instead of the corresponding three within the Skyrme model. Therefore, the Hopf mapping
\(\Pi_{3}(S^{2})=\mathsf{Z}\) with the Hopf invariant only remains for nontrivial field configurations
of the quaternionic, anti-symmetric complex eigenvalues. In fact, we can extract a Hopf invariant
from the axial current conversation with the chiral anomaly in the massless
limit and can determine nontrivial field configurations of the complex
eigenvalue angles. This is briefly exemplified in cylindrical geometry and can also be illustrated in toroidal coordinates.
The nontrivial Hopf mappings of \(\Pi_{3}(S^{2})=\mathsf{Z}\) can be intuitively understood as one considers
the 'preimage' of the \(S^{2}\) sphere where every point of the \(S^{2}\) sphere corresponds to a
\(S^{1}\) loop in the spatial \(S^{3}\) sphere.
Since there occurs an imaginary factor with the extracted Hopf invariant, the nontrivial
field configurations correspond to 'helicity instantons' following from the axial current conversation
with the chiral anomaly. These 'helicity instantons' thereby classify the prevailing field configuration
of 'nucleons' according to the summations over isospin- and colour-degrees of freedom. For that reason,
our derived path integral with coset matrices \(\hat{T}(x_{p})\) is more closely related to the
Skyrme-Faddeev model with \(\Pi_{3}(S^{2})=\mathsf{Z}\) than to the original Skyrme model
\(\Pi_{3}(\mbox{SU}(2)\,)=\mathsf{Z}\) which is regarded as an effective field theory for QCD in the limit
of infinite colour degrees of freedom \(N_{c}\rightarrow\infty\).

\newpage

\begin{appendix}

\section{Hilbert space of anomalous doubled operators and their representations} \lb{sa}

According to chapter 4 of Ref.\ \cite{mies1},
we describe how the gradient operators \(\boldsymbol{\hat{\pp}_{p,\mu}}\)
act on the coset matrices \(\hat{T}(x_{p})\) in the operator
\(\hat{\mscr{O}}_{N;M}^{ba}(y_{q},x_{p})\) (\ref{s3_111}) with
\(\Delta\!\hat{\mscr{H}}(x_{p})\) and on densities as e.g. the quark self-energy density
\(\hat{\sigma}_{D}^{(\alpha;\kappa)}(x_{p})\) . One can consider some representation as the 3+1 dimensional
spacetime-coordinates where the operators $\hat{T}$, $\hat{Q}$, $\delta\hat{\lambda}$,
$\hat{\sigma}_{D}^{(\alpha;\kappa)}$ are defined to be diagonal in the spacetime variables
and are given by the matrix fields \(\hat{T}_{M;N}^{ab}(x_{p})\),
\(\hat{Q}_{M;N}^{aa}(x_{p})\) and the scalar density fields
\(\delta\hat{\lambda}_{M}(x_{p})\), \(\sigma_{D}^{(\alpha;\kappa)}(x_{p})\).
However, it has to be taken into account
that the square root of the determinant follows
from integration over the bilinear anti-commuting fields which are doubled by their complex conjugates
\(\psi_{M}^{*}(x_{p})\). Consequently a Hilbert space for
\(\psi_{M}(x_{p})\) with 'ket' \(|\psi_{M}\rangle\) has also to be doubled
by its 'dual' space \(\ovv{|\psi_{M}\rangle}=\langle\psi_{M}|\) the 'bra'.
The unsaturated operators \(\boldsymbol{\hat{\pp}_{p,\mu}}\)
are printed in boldface in order to distinguish from the matrix functions as
\((\hat{\pp}_{p,\mu}\hat{T}(x_{p})\,)\) embraced in brackets,
denoting the limited action of the derivative on the prevailing coset matrix \(\hat{T}(x_{p})\).
These 'saturated' gradient operators are not involved in further
derivative actions on matrices or fields outside the parentheses and are therefore not printed in bold type.

The detailed structure of the Hilbert space with its doubled dual part
\(\langle\psi_{M}|=\ovv{|\psi_{M}\rangle}\) is important because the doubled operator
\(\hat{\mscr{H}}(x_{p})\) (\ref{s3_87},\ref{s3_88}) applies the transpose
\(\hat{H}^{T}(x_{p})\) in the '22' block
instead of \(\hat{H}(x_{p})\) as in the '11' part. An operator in quantum mechanics
is defined by the mapping and the space on which it acts. Completely different results can follow
if one considers for one and the same mapping of an operator different spaces where the operator
transforms the prevailing states.
The path integral (\ref{s3_110}-\ref{s3_116}) follows by integration over
the doubled anti-commuting fields \(\Psi_{M}^{a}(x_{p})\) from (\ref{s3_103}).
The corresponding doubled abstract states \(\widehat{|\psi_{M}\rangle}^{a(=1,2)}\)
with internal space label \(M,\;N,\;M\ppr,\;N\ppr,\,\ldots\) are defined in (\ref{sa_1})
\be\lb{sa_1}
\Psi_{M}^{a(=1,2)}(x_{p})=\left(
\bea{c}
\psi_{M}(x_{p}) \\
\psi_{M}^{*}(x_{p})
\eea\right)\propto \widehat{|\Psi_{M}\rangle}^{a(=1,2)}=
\left(\bea{c}
|\psi_{M}\rangle^{a=1} \vspace*{0.15cm} \\
\ovv{|\psi_{M}\rangle}^{a=2}
\eea\right)_{\mbox{.}}
\ee
The appropriate abstract Hilbert space has to be introduced for the definition of the operators
\(\boldsymbol{\hat{\pp}_{p,\mu}}\) in the determinant-action
\(\langle\mscr{A}_{DET}[\hat{T},\feynV;\hat{\mscr{J}}]\rangle_{\feynbv}\) and
\(\langle\mscr{A}_{J_{\psi}}[\hat{T},\feynV;\hat{\mscr{J}}]\rangle_{\feynbv}\) (\ref{s3_114},\ref{s3_115}).
According to the doubling with the dual part \(\ovv{|\psi_{M}\rangle}=\langle\psi_{M}|\),
we have an anti-linear property in the second part \(\widehat{|\psi_{M}\rangle}^{a=2}\)
\be \lb{sa_2}
\widehat{|c\;\Psi_{M}\rangle}=\left(
\bea{c}
c\;|\psi_{M}\rangle \vspace*{0.15cm} \\
c^{*}\;\ovv{|\psi_{M}\rangle}
\eea\right)\;\;;\hspace*{1.0cm}c\in\mbox{\sf C}  \;\;\;.
\ee
Furthermore, we simultaneously have the unitary and 'anti'-unitary representation
of \(U(N_{0})\) (\(N_{0}=N_{f}\times 4_{\hat{\gamma}}\times N_{c}\))
in the '11' and '22' block, respectively. This is in accordance with a theorem of Wigner
that a symmetry in quantum mechanics can have a unitary or anti-unitary realization \cite{wein1}.
The corresponding Hilbert space for 3+1 dimensional spacetime has therefore also to be
doubled with the anti-linear part
\be \lb{sa_3}
\widehat{|x_{p}\rangle}^{a(=1,2)}=
\left(\bea{c}
|x_{p}\rangle^{a=1} \vspace*{0.15cm} \\
\ovv{|x_{p}\rangle}^{a=2}
\eea\right)=\left(\bea{c}
|x_{p}\rangle \vspace*{0.15cm} \\
\langle x_{p}|
\eea\right)\;;
\ee
\be \lb{sa_4}
\bea{rclrclrcl}
\langle x_{p}^{0}|y_{q}^{0}\rangle&=&\delta_{pq}\;\delta_{x_{p}^{0},y_{q}^{0}}\;; &
\hspace*{0.3cm} \langle\vec{x}|\vec{y}\rangle&=&\delta_{\vec{x},\vec{y}}\;; &\hspace*{0.3cm}
\langle x_{p}|y_{q}\rangle &=&\delta_{pq}\;\delta_{x_{p}^{0},y_{q}^{0}}\;\delta_{\vec{x},\vec{y}}\;;
 \\  \langle k_{p}^{0}|p_{q}^{0}\rangle&=&
\delta_{p,q}\;\delta_{k_{p}^{0},p_{q}^{0}} \;; &
\langle\vec{k}|\vec{p}\rangle&=&\delta_{\vec{k},\vec{p}} \;; &
\langle k_{p}|p_{q}\rangle&=&\delta_{pq}\;\delta_{k_{p}^{0},p_{q}^{0}}\;\delta_{\vec{k},\vec{p}}\;; \\
\langle x_{p}^{0}|k_{q}^{0}\rangle&=&\delta_{pq}\;\exp\{-\im\;k_{p}^{0}\cdot x_{p}^{0}\} \;; &
\langle\vec{x}|\vec{k}\rangle&=&\exp\{\im\;\vec{k}\cdot\vec{x}\} \;; &
\langle x_{p}|k_{q}\rangle &=&\delta_{pq}\;\exp\{\im(\vec{k}\cdot\vec{x}-k_{p}^{0}\cdot x_{p}^{0})\}\;;
\eea
\ee
\beq \lb{sa_5} \sum_{a=1,2}
\ph{\,}^{a}\widehat{\langle x_{p}}|\widehat{k_{q}\rangle}^{a}&=&
\langle x_{p}|k_{q}\rangle +
\ovv{\langle x_{p}}|\ovv{k_{q}\rangle} \;; \\ \lb{sa_6}
\ph{\,}^{a=1}\widehat{\langle x_{p}}|\widehat{k_{q}\rangle}^{a=1} &=&\langle x_{p}|k_{q}\rangle =
\delta_{pq}\;\exp\{\im(\vec{k}\cdot\vec{x}-k_{p}^{0}\cdot x_{p}^{0})\} \;; \\ \lb{sa_7}
\ph{\,}^{a=2}\widehat{\langle x_{p}}|\widehat{k_{q}\rangle}^{a=2} &=&
\ovv{\langle x_{p}}|\ovv{k_{q}\rangle} = \langle k_{q}|x_{p}\rangle=
\big(\langle x_{p}|k_{q}\rangle\big)^{*} = \delta_{pq}\;
\exp\{-\im(\vec{k}\cdot\vec{x}-k_{p}^{0}\cdot x_{p}^{0})\}\;.
\eeq
The total unit operators with the unitary and anti-unitary parts are listed in Eqs.
(\ref{sa_8},\ref{sa_9}) for 3+1 dimensional spacetime and momentum-energy states. We have to combine the contour integrals of forward and backward propagation with the contour metric \(\eta_{p=\pm}=\pm\) (\ref{s2_15},\ref{s2_16})
so that the defining relations (\ref{sa_3}-\ref{sa_7}) exactly match with the properties of the unit operators
(\ref{sa_8},\ref{sa_9})
\beq \lb{sa_8}
\hat{1}&=&\left(\bea{cc}
\hat{1}^{11} & \\  & \hat{1}^{22}  \eea\right)=
\sum_{p=\pm}\int_{-\infty}^{+\infty}d^{4}\!x_{p}\;\mcal{N}
 \left(\bea{cc}
|x_{p}\rangle\langle x_{p}| & \\
 & \ovv{|x_{p}\rangle}\;\ovv{\langle x_{p}|} \eea\right)  \\ \no &=&
\int_{C}d^{4}\!x_{p}\;\eta_{p}\;\mcal{N}
 \left(\bea{cc}
|x_{p}\rangle\langle x_{p}| & \\
 & \ovv{|x_{p}\rangle}\;\ovv{\langle x_{p}|} \eea\right) =
\int_{C}d^{4}\!x_{p}\;\eta_{p}\;\mcal{N} \sum_{a=1,2}
 |\widehat{x_{p}\rangle}^{a(=1,2)}\;\ph{\,}^{a(=1,2)}\widehat{\langle x_{p}}|\;; \hspace*{0.3cm}
\mcal{N}=\frac{1}{(\Delta x)^{4}} \;; \\ \lb{sa_9}
\hat{1}&=&\left(\bea{cc}\hat{1}^{11} & \\
& \hat{1}^{22} \eea\right)=
\sum_{p=\pm}\int_{-\infty}^{+\infty}d^{4}\!k_{p}\;\mcal{N}_{k}
 \left(\bea{cc}
|k_{p}\rangle\langle k_{p}| & \\
 & \ovv{|k_{p}\rangle}\;\ovv{\langle k_{p}|} \eea\right)  \\ \no &=&
\int_{C}d^{4}\!k_{p}\;\eta_{p}\;\mcal{N}_{k}
\left(\bea{cc}
|k_{p}\rangle\langle k_{p}| & \\
& \ovv{|k_{p}\rangle}\;\ovv{\langle k_{p}|}
\eea\right) = \int_{C} d^{4}\!k_{p}\;\eta_{p}\;\mcal{N}_{k}\sum_{a=1,2}
|\widehat{k_{p}\rangle}^{a(=1,2)}\;\ph{\,}^{a(=1,2)}\widehat{\langle k_{p}}|
\;;\hspace*{0.3cm}\mcal{N}_{k}=\frac{1}{(\Delta k)^{4}}\;.
\eeq
The 3+1 spacetime '\(x_{p}=\{x_{p}^{0},\vec{x}\}\)' and four-momentum '\(k_{p}=\{k_{p}^{0},\vec{k}\}\)'
representations of the abstract doubled Hilbert space states (\ref{sa_2})
are given in (\ref{sa_10},\ref{sa_11}) where the relations (\ref{sa_3}-\ref{sa_7}) are applied,
including the anti-unitary second part
\beq \lb{sa_10}
\ph{\,}^{a}\widehat{\langle x_{p}}|\widehat{\Psi_{M}\rangle}^{a}&=&
\left(\bea{c}
\langle x_{p}|\psi_{M}\rangle \vspace*{0.15cm}\\
\ovv{\langle x_{p}}|\ovv{\psi_{M}\rangle}
\eea\right)^{a(=1,2)}=
\left(\bea{c} \psi_{M}(x_{p}) \\ \langle\psi_{M}|x_{p}\rangle \eea\right)^{a(=1,2)}=
\left(\bea{c} \psi_{M}(x_{p}) \\ \psi_{M}^{*}(x_{p}) \eea\right)^{a(=1,2)} \;; \\ \lb{sa_11}
\ph{\,}^{a}\widehat{\langle k_{p}}|\widehat{\Psi_{M}\rangle}^{a}&=&
\left(\bea{c} \langle k_{p}|\psi_{M}\rangle \vspace*{0.15cm}\\
\ovv{\langle k_{p}}|\ovv{\psi_{M}\rangle} \eea\right)^{a(=1/2)}=
\left(\bea{c} \psi_{M}(k_{p}) \\ \langle\psi_{M}|k_{p}\rangle \eea\right)^{a(=1,2)}=
\left(\bea{c} \psi_{M}(k_{p}) \\ \psi_{M}^{*}(k_{p}) \eea\right)^{a(=1,2)}\;.
\eeq
The scalar self-energy densities $\hat{\sigma}_{D}^{(\alpha;\kappa)}$ operate on the doubled spacetime state
\(|\widehat{y_{q}\rangle}^{a}\) as in the well-known case of an
annihilation operator on coherent states, but one has to incorporate the contour metric
$\eta_{q}$ and has to consider that the resulting coherent state field
\(\sigma_{D}^{(\alpha;\kappa)}(y_{q})\) only takes real values (\ref{sa_12}).
This additional contour metric \(\eta_{p}\) has to be taken into account
for the one-particle operator $\hat{\mscr{H}}_{N;M}^{ba}(x_{p})$,
the self-energy density \(\hat{\sigma}_{D}^{(\alpha;\kappa)}(x_{p})\)
and \(\delta\hat{\Sigma}_{M;N}^{ab}(x_{p})\) because these operators appear in the
original path integrals and only lead to diagonal matrix elements
in the time contour due to a missing disorder. An ensemble average with a random potential
would include non-diagonal terms in the total self-energy concerning the time contour metric \cite{mies2}.
However, the abstract operator action with the anomalous parts \(\hat{Y}_{M;N}^{ab}(x_{p})\)
in \(\hat{T}_{M;N}^{ab}(x_{p})\) does not involve an additional time contour metric in the
considered case without disorder
\beq \lb{sa_12}
\hat{\sigma}_{D}^{(\alpha;\kappa)}\;|\widehat{y_{q}\rangle}^{a}&=& \eta_{q}\;
\sigma_{D}^{(\alpha;\kappa)}(y_{q})\;|\widehat{y_{q}\rangle}^{a} \;;
\hspace*{0.6cm}\sigma_{D}^{(\alpha;\kappa)}(y_{q})\in\mbox{\sf R} \;; \\ \lb{sa_13}
\hat{T}_{M;N}^{ab}\;|\widehat{y_{q}\rangle}^{b}&=&
\hat{T}_{M;N}^{ab}(y_{q})\;|\widehat{y_{q}\rangle}^{b}\;\;\;.
\eeq
Using the definitions (\ref{sa_3}-\ref{sa_7}) of the abstract doubled Hilbert space, we can
pursue the various steps for calculating matrix elements
\(\ph{\,}^{a}\widehat{\langle x_{p}}|\hat{T}_{M;N}^{ab}|\widehat{y_{q}\rangle}^{b}\)
from the generating operators \(\hat{Y}_{M;N}^{ab}\), \(\hat{X}_{M;N}\),
\(\hat{X}_{M;N}^{\dag}\) in the exponential of \(\hat{T}\). However, the operator
\(\hat{X}_{M;N}\) of the anomalous parts is constructed from two field operators
\(\hat{\psi}_{M}\;\hat{\psi}_{N}\) so that matrix elements
\(\langle x_{p}|\hat{X}_{M;N}|\ovv{y_{q}\rangle}\)
(\ref{sa_15},\ref{sa_16}) result in the expansion of \(\hat{T}_{M;N}^{ab}\)
(\ref{sa_14}), combining also Hilbert states with linear and anti-linear parts
\beq \lb{sa_14}
\lefteqn{\ph{\,}^{a}\widehat{\langle x_{p}}|\hat{T}_{M;N}^{ab}|
\widehat{y_{q}\rangle}^{b}=
\ph{\,}^{a}\widehat{\langle x_{p}}|\big(\hat{1}-\hat{Y}_{M;N}^{ab}\pm\ldots\big)
|\widehat{y_{q}\rangle}^{b}=} \\ \no &=&
\delta_{ab}\;\delta_{M;N}\;\delta_{\vec{x},\vec{y}}\;\delta_{pq}\;
\delta_{x_{p}^{0},y_{q}^{0}}-
\left(\bea{c}
\langle x_{p}| \vspace*{0.15cm} \\ \ovv{\langle x_{p}|} \eea\right)^{T}
\left(\bea{cc}
0 & \hat{X}_{M;N} \\
\hat{X}^{\dag}_{M;N} & 0
\eea\right)\left(\bea{c}
|y_{q}\rangle \vspace*{0.15cm} \\
\ovv{|y_{q}\rangle} \eea\right)\pm\ldots \\ \no &=&
\delta_{\vec{x},\vec{y}}\;\delta_{pq}\;\delta_{x_{p}^{0},y_{q}^{0}}\;
\left[\hat{1}\;\delta_{ab}\;\delta_{M;N}-\left(
\bea{cc} 0 & \hat{X}_{M;N}(x_{p}) \\
\hat{X}_{M;N}^{\dag}(x_{p}) & 0 \eea\right)^{ab}\pm\ldots\right]  \;;
\eeq
\beq \lb{sa_15}
\langle x_{p}|\hat{X}_{M;N}|\ovv{y_{q}\rangle}&=&
\hat{X}_{M;N}(x_{p})\;\delta_{\vec{x},\vec{y}}\;\delta_{pq}\;
\delta_{x_{p}^{0},y_{q}^{0}} \;; \\ \lb{sa_16}
\ovv{\langle x_{p}}|\hat{X}_{M;N}^{\dag}| y_{q}\rangle&=&
\hat{X}_{M;N}^{\dag}(x_{p})\;\delta_{\vec{x},\vec{y}}\;\delta_{pq}\;
\delta_{x_{p}^{0},y_{q}^{0}}  \;; \\ \lb{sa_17}
\Big(\langle x_{p}|\hat{X}_{M;N}|\ovv{y_{q}\rangle}\Big)^{*}&=&
\ovv{\langle y_{q}}|\hat{X}_{M;N}^{\dag}|x_{p}\rangle=
\ovv{\langle x_{p}}|\hat{X}_{M;N}^{\dag}|y_{q}\rangle \;;  \\ \lb{sa_18}
\hat{X}_{M;N}(x_{p})&\propto & \big(\psi_{M}(x_{p})\;
\psi_{N}(x_{p})\big) \;; \\ \lb{sa_19}
\hat{X}_{M;N}^{\dag}(x_{p})&\propto & \big(\psi_{M}(x_{p})\;\psi_{N}(x_{p})\big)^{\dag}=
\big(\psi_{N}^{*}(x_{p})\;\psi_{M}^{*}(x_{p})\big)^{T}= \\ \no &=&
\big(\psi_{M}^{*}(x_{p})\;\psi_{N}^{*}(x_{p})\big) \;;\hspace*{0.6cm}\psi_{M}(x_{p})\;,\:\psi_{N}(x_{p})\:
\in\mscr{C}_{odd}\;\;.
\eeq
Summarizing the effect of the doubling of Hilbert space states with the anti-unitary extension,
we list the matrix elements of the density parts (\ref{sa_20},\ref{sa_21}), always
containing the time contour metric \(\eta_{p}\), and the pair condensates
(\ref{sa_22}), as derived with the properties (\ref{sa_15}-\ref{sa_19}) in the
expansion (\ref{sa_14})
\beq \lb{sa_20}
\ph{\,}^{a}\widehat{\langle x_{p}}|\hat{\mscr{H}}|\widehat{y_{q}\rangle}^{b}&=&\!\!
\delta_{ab}\;\delta_{\vec{x},\vec{y}}\;
\eta_{p}\;\delta_{pq}\;\delta_{x_{p}^{0},y_{q}^{0}} \!\left(
\bea{cc}
\hat{H}_{M;N}(x_{p}) & 0 \\ 0 & \hat{H}_{M;N}^{T}(x_{p})
\eea\right)^{ab}  \;; \\ \lb{sa_21}
\ph{\,}^{a}\widehat{\langle x_{p}}|\hat{\sigma}_{D}^{(\alpha;\kappa)}\;\hat{1}_{2N_{0}\times 2N_{0}}+
\delta\hat{\Sigma}_{M;N}^{ab}|\widehat{y_{q}\rangle}^{b}&=&
\delta_{\vec{x},\vec{y}}\;\eta_{p}\;\delta_{pq}\;\delta_{x_{p}^{0},y_{q}^{0}}\;
\Big(\sigma_{D}^{(\alpha;\kappa)}(x_{p})\big)\;\delta_{ab}\;\delta_{M;N} +
\delta\hat{\Sigma}_{M;N}^{ab}(x_{p})\Big) \;; \\ \lb{sa_22}
\ph{\,}^{a}\widehat{\langle x_{p}}|\hat{T}_{M;N}^{ab}|\widehat{y_{q}\rangle}^{b}&=&
\hat{T}_{M;N}^{ab}(x_{p})\;\;\delta_{\vec{x},\vec{y}}\;
\delta_{pq}\;\delta_{x_{p}^{0},y_{q}^{0}}\;.
\eeq
The source field $J_{\psi;M}^{a}(x_{p})$ for the BEC wave function and
the source matrix \(\wt{\mscr{J}}_{M;N}^{ab}\) for generating observables
are defined in (\ref{sa_23},\ref{sa_24}) for corresponding doubled states and matrices
\beq \lb{sa_23}
J_{\psi;M}^{a}(x_{p})&=&\ph{\,}^{a}\widehat{\langle x_{p}}|
\widehat{J_{\psi;M}\rangle}^{a}=\left(\bea{c}
j_{\psi;M}(x_{p}) \\ j_{\psi;M}^{*}(x_{p}) \eea\right)^{a} \;;  \\  \lb{sa_24}
\wt{\mscr{J}}_{M;N}^{ab}(x_{p},y_{q})&=&
\ph{\,}^{a}\widehat{\langle x_{p}}|\wt{\mscr{J}}_{M;N}^{ab}|
\widehat{y_{q}\rangle}^{b} =
\hat{I}\;\hat{S}\;\eta_{p}\;
\frac{\hat{\mscr{J}}_{M;N}^{ab}(x_{p},y_{q})}{\mcal{N}}\;
\eta_{q}\;\hat{S}\;\hat{I}\;\;\;.
\eeq
The definition of the unit operators (\ref{sa_8},\ref{sa_9}) with time contour integration {\it and}
additional contour metric can be transformed to a trace relation as one can change the
unit operator \(\hat{1}=\sum_{n}|n\rangle\langle n|\) of a complete set of states \(|n\rangle\)
in ordinary quantum mechanics to a trace relation \(\mbox{tr}[\ldots]=\sum_{n}\langle n|\ldots|n\rangle\).
However, we have to distinguish between the  anomalous doubled trace
\(\mbox{TR}_{\scrscr \int_{C}d^{4}x_{p}\;\eta_{p}}^{\scrscr a(=1,2)}[\ldots]\) (\ref{sa_25})
with the anti-unitary second part \(\ovv{\langle x_{p}}|\ldots|\ovv{x_{p}\rangle}\)
and the ordinary trace \(\mbox{tr}_{\scrscr \int_{C}d^{4}x_{p}\;\eta_{p}}[\ldots]\) (\ref{sa_26})
also with time contour integration, but without the anomalous doubled anti-unitary part
\beq \lb{sa_25}
\trxpa\Big[\ldots\Big]&:=&\sum_{p=\pm}\int_{-\infty}^{+\infty}d^{4}\!x_{p}\;\mcal{N}
\sum_{a=1,2}\ph{\,}^{a}\widehat{\langle x_{p}}|\ldots|\widehat{x_{p}\rangle}^{a} =
\int_{C}d^{4}\!x_{p}\;\eta_{p}\;\mcal{N}
\sum_{a=1,2}\ph{\,}^{a}\widehat{\langle x_{p}}|\ldots|\widehat{x_{p}\rangle}^{a} \\ \no &=&
\int_{-\infty}^{+\infty}dx_{+}^{0}\int_{L^{3}} d^{3}\!\vec{x}\;\mcal{N}
\sum_{a=1,2}\ph{\,}^{a}\widehat{\langle x_{+}}|\ldots|\widehat{x_{+}\rangle}^{a} +
\int_{-\infty}^{+\infty} dx_{-}^{0}\int_{L^{3}} d^{3}\!\vec{x}\;  \mcal{N}
\sum_{a=1,2}\ph{\,}^{a}\widehat{\langle x_{-}}|\ldots|\widehat{x_{-}\rangle}^{a} \;; \\ \lb{sa_26}
\trxp\Big[\ldots\Big]&:=&\sum_{p=\pm}\int_{-\infty}^{+\infty}d^{4}\!x_{p}\;\mcal{N}\;
\langle x_{p}|\ldots|x_{p}\rangle=
\int_{C}d^{4}\!x_{p}\;\eta_{p}\;\mcal{N}\;
\langle x_{p}|\ldots|x_{p}\rangle \\ \no &=&
\int_{-\infty}^{+\infty}dx_{+}^{0}\int_{L^{3}} d^{3}\!\vec{x}\;\mcal{N}\;
\langle x_{+}|\ldots|x_{+}\rangle +
\int_{-\infty}^{+\infty}dx_{-}^{0}\int_{L^{3}} d^{3}\!\vec{x}\;\mcal{N}\;
\langle x_{-}|\ldots|x_{-}\rangle \;.
\eeq
In a similar manner one introduces the traces of the four-momentum '\(k_{p}=\{k_{p}^{0},\vec{k}\}\)'
which are needed for a lowest order momentum and gradient expansion of the action
for the determinant
\beq \lb{sa_27}
\trkpa\Big[\ldots\Big]&:=&\sum_{p=\pm}\int_{-\infty}^{+\infty}d^{4}\!k_{p}\;\mcal{N}_{k}
\sum_{a=1,2}\ph{\,}^{a}\widehat{\langle k_{p}}|\ldots|\widehat{k_{p}\rangle}^{a}=
\int_{C}d^{4}\!k_{p}\;\eta_{p}\;\mcal{N}_{k}
\sum_{a=1,2}\ph{\,}^{a}\widehat{\langle k_{p}}|\ldots|\widehat{k_{p}\rangle}^{a} \\ \no &=&
\int_{-\infty}^{+\infty}dk_{+}^{0}\int d^{3}\!\vec{k}\;\mcal{N}_{k}
\sum_{a=1,2}\ph{\,}^{a}\widehat{\langle k_{+}}|\ldots|\widehat{k_{+}\rangle}^{a} +
\int_{-\infty}^{+\infty} dk_{-}^{0}\int d^{3}\!\vec{k}\;  \mcal{N}_{k}
\sum_{a=1,2}\ph{\,}^{a}\widehat{\langle k_{-}}|\ldots|\widehat{k_{-}\rangle}^{a} \;; \\ \lb{sa_28}
\trkp\Big[\ldots\Big]&:=&\sum_{p=\pm}\int_{-\infty}^{+\infty}d^{4}\!k_{p}\;\mcal{N}_{k}\;
\langle k_{p}|\ldots|k_{p}\rangle =
\int_{C}d^{4}\!k_{p}\;\eta_{p}\;\mcal{N}_{k}\;
\langle k_{p}|\ldots|k_{p}\rangle \\ \no &=&
\int_{-\infty}^{+\infty}dk_{+}^{0}\int d^{3}\!\vec{k}\;\mcal{N}_{k}\;
\langle k_{+}|\ldots|k_{+}\rangle +
\int_{-\infty}^{+\infty}dk_{-}^{0}\int d^{3}\!\vec{k}\;\mcal{N}_{k}\;
\langle k_{-}|\ldots|k_{-}\rangle \;.
\eeq

\section{Ward identities corresponding to the gauge symmetries of the coset space} \lb{sb}

\subsection{Ward identities and gauge invariance of the background potential} \lb{sb1}

As we examine the path integral (\ref{s3_59}) and the separated actions
\(\mscr{A}_{DET}[\hat{T},\feynV;\hat{\mscr{J}}]\), \(\mscr{A}_{J_{\psi}}[\hat{T},\feynV;\hat{\mscr{J}}]\)
(\ref{s3_110}-\ref{s3_116}), we recognize the common background potential \(\mcal{V}_{\alpha}^{\mu}(x_{p})\)
(\ref{s3_60}) in \(\hat{H}_{N;M}(y_{q},x_{p})\) (\ref{s3_88}) or in its anomalous doubled version
\(\hat{\mscr{H}}_{N;M}^{ba}(y_{q},x_{p})\) (\ref{s3_87}) with the transpose \(\hat{H}_{N;M}^{T}(y_{q},x_{p})\)
in the '22' block section. The composed background potential \(\mcal{V}_{\alpha}^{\mu}(x_{p})\) substitutes the original
gauge fields \(A_{\alpha}^{\mu}(x_{p})\) of \(\mbox{SU}_{c}(N_{c}=3)\) in axial gauge
\(A_{\alpha}^{\nu}(x_{p})\;n_{\nu}=0\) (\(\vec{n}\cdot\vec{n}=1\)) which is obtained by the standard exponential
integral representation of the delta functions \(\delta(\,A_{\alpha}^{\nu}(x_{p})\;n_{\nu}\,)\)
with auxiliary real fields \(s_{\alpha}(x_{p})\). In consequence, the auxiliary integrals of
\(s_{\alpha}(x_{p})\) for the delta functions \(\delta(\,A_{\alpha}^{\nu}(x_{p})\;n_{\nu}\,)\) fix the
axial gauge so that the original path integrals avoid the multiple (in fact infinite)
weighting of physically equivalent configurations which just differ by a gauge transformation.
Despite of this original axial gauge fixing by \(s_{\alpha}(x_{p})\), a gauge invariance is still
retained for the composed background gauge field \(\mcal{V}_{\alpha}^{\mu}(x_{p})\) (\ref{s3_60}).

The background gauge field \(\mcal{V}_{\alpha}^{\mu}(x_{p})\) (\ref{sb1_1}) consists of the self-energy field
strength tensor \(\hat{\mfrak{S}}_{\alpha}^{(\hat{F})\mu\nu}(x_{p})\), its inverse
with \(\mbox{SU}_{c}(N_{c}=3)\) structure constants \(C_{\alpha\beta\gamma}\) , its derivative and
additionally of the self-energy quark densities \(\sigma_{D}^{(\alpha;\kappa)}(x_{p})\)
which are dressed by the eigenvectors
\(\hat{\mfrak{B}}_{\hat{F};\beta(\alpha)}^{\mu(\kappa)}(x_{p})\) of the self-energy field strength term
\(C_{\alpha\beta\gamma}\;\hat{\mfrak{S}}_{\alpha;\mu\nu}^{(\hat{F})}(x_{p})\)
\beq \lb{sb1_1}
\mcal{V}_{\beta}^{\mu}(x_{p})&=&
\Big[\Big(\hat{\pp}_{p}^{\lambda}\hat{\mfrak{S}}_{\gamma;\nu\lambda}^{(\hat{F})}(x_{p})\Big)-
s_{\gamma}(x_{p})\;n_{\nu}\Big]\;\Big[-\im\,\hat{\mfrak{e}}_{p}^{(\hat{F})}+C_{\alpha\beta\ppr\gamma\ppr}\;
\hat{\mfrak{S}}_{\alpha}^{(\hat{F})\mu\ppr\nu\ppr}(x_{p})\big]_{\gamma\beta}^{-1;\nu\mu} +   \\ \no &+&
\frac{1}{2}\sum_{\scrscr(\alpha)=1,.,8}^{\scrscr(\kappa)=0,.,3}\hat{\mfrak{B}}_{\hat{F};\beta(\alpha)}^{\mu(\kappa)}\!(x_{p})\;\;
\sigma_{D}^{(\alpha;\kappa)}(x_{p})\;;   \\   \lb{sb1_2}
C_{\alpha\beta\gamma}\;\hat{\mfrak{S}}_{\alpha}^{(\hat{F})\mu\nu}(x_{p}) &:=&
\sum_{\scrscr(\alpha)=1,.,8}^{\scrscr(\kappa)=0,.,3}
\hat{\mfrak{B}}_{\hat{F};\beta(\alpha)}^{\mu(\kappa)}(x_{p})\;
\hat{\mfrak{b}}^{(\hat{F})}_{(\alpha;\kappa)}(x_{p})\;\hat{\mfrak{B}}_{\hat{F};(\alpha)\gamma}^{T,(\kappa)\nu}(x_{p})
\;; \\ \lb{sb1_3}
d[\hat{\mfrak{S}}_{\alpha}^{(\hat{F})\mu\nu}(x_{p})]&\rightarrow&
d[\hat{\mfrak{S}}_{\alpha}^{(\hat{F})\mu\nu}(x_{p});\hat{\mfrak{B}}_{\hat{F};\beta(\alpha)}^{\mu(\kappa)}(x_{p});
\hat{\mfrak{b}}^{(\hat{F})}_{(\alpha;\kappa)}(x_{p})]  \;; \\   \lb{sb1_4}
\lefteqn{\hspace*{-2.8cm}d[\hat{\mfrak{S}}_{\alpha}^{(\hat{F})\mu\nu}(x_{p});
\hat{\mfrak{B}}_{\hat{F};\beta(\alpha)}^{\mu(\kappa)}(x_{p});
\hat{\mfrak{b}}^{(\hat{F})}_{(\alpha;\kappa)}(x_{p})] = d[\hat{\mfrak{S}}_{\alpha}^{(\hat{F})\mu\nu}(x_{p})]\;\;
d[\hat{\mfrak{B}}_{\hat{F};\beta(\alpha)}^{\mu(\kappa)}(x_{p});\hat{\mfrak{b}}^{(\hat{F})}_{(\alpha;\kappa)}(x_{p})]\;\times } \\ \no &\times&
\bigg\{\prod_{\scrscr\{x_{p}\}}\delta\bigg(C_{\alpha\beta\gamma}\;\hat{\mfrak{S}}_{\alpha}^{(\hat{F})\mu\nu}(x_{p}) -
\sum_{\scrscr(\alpha)=1,.,8}^{\scrscr(\kappa)=0,.,3}
\hat{\mfrak{B}}_{\hat{F};\beta(\alpha)}^{\mu(\kappa)}(x_{p})\;
\hat{\mfrak{b}}^{(\hat{F})}_{(\alpha;\kappa)}(x_{p})\;\hat{\mfrak{B}}_{\hat{F};(\alpha)\gamma}^{T,(\kappa)\nu}(x_{p}) \bigg)\bigg\} \;.
\eeq
The anomalous doubled one-particle operator \(\hat{\mscr{H}}(x_{p})\) contains the composed gauge
fields \(\mcal{V}_{\alpha}^{\mu}(x_{p})\), \(\feynV(x_{p})\) or more precisely the anomalous doubled
version \(\hat{\mscr{V}}^{\mu}(x_{p})=\hat{\mscr{T}}_{\alpha}\;\mcal{V}_{\alpha}^{\mu}(x_{p})\),
\(\feynVV(x_{p})\) with \(\mbox{SU}_{c}(N_{c}=3)\) generators \(\hat{\mscr{T}}_{\alpha}\) which are doubled by the transpose
\(-\hat{t}_{\alpha}^{T}\) in the '22' block
\beq \lb{sb1_5}
\hat{\mscr{H}}_{N;M}^{ba}(y_{q},x_{p}) &=&\delta^{(4)}(y_{q}-x_{p})\;\eta_{q}\;\delta_{qp}
\Bigg(\bea{cc} \hat{H}_{N;M}(x_{p}) & \\ & \hat{H}_{N;M}^{T}(x_{p})  \eea\Bigg)^{ba} \;; \\ \lb{sb1_6}
\hat{H}(x_{p}) &=&\big[\hat{\beta}(\,\feynd_{p}+\im\:\feynV(x_{p})-\im\:\hat{\ve}_{p}+\hat{m}\,)\big]\;;
\;(\hat{\ve}_{p}=\hat{\beta}\:\ve_{p}=\hat{\beta}\:\eta_{p}\:\ve_{+}\;;\;\ve_{+}>0)\;;  \\ \no &=&
\hat{\beta}\hat{\gamma}^{\mu}\:\hat{\pp}_{p,\mu}+\im\:\hat{\beta}\hat{\gamma}^{\mu}\:\hat{t}_{\alpha}\:
\mcal{V}_{\alpha}^{\mu}(x_{p})+\hat{\beta}\;\hat{m}-\im\;\ve_{p}\;\hat{1}_{N_{0}\times N_{0}}\;;  \\ \lb{sb1_7}
\hat{H}^{T}(x_{p}) &=&\big[\hat{\beta}(\,\feynd_{p}+\im\:\feynV(x_{p})-\im\:\hat{\ve}_{p}+\hat{m}\,)\big]^{T}  \\ \no &=&
-(\,\hat{\beta}\hat{\gamma}^{\mu}\,)^{T}\:\hat{\pp}_{p,\mu}-\im\:(\,\hat{\beta}\hat{\gamma}^{\mu}\,)^{T}\:
(-\hat{t}_{\alpha}^{T})\:\mcal{V}_{\alpha}^{\mu}(x_{p})-(\,\hat{\beta}\,(-\hat{m})\,)^{T}-\im\;\ve_{p}\;
\hat{1}_{N_{0}\times N_{0}}\;; \\ \lb{sb1_8}
\hat{\mscr{B}}\:\hat{\Gamma}^{\mu}&=& \Bigg(\bea{cc} \hat{\beta}\;\hat{\gamma}^{\mu} & \\ &
(\,\hat{\beta}\;\hat{\gamma}^{\mu}\,)^{T} \eea\Bigg)^{ba} \;;\;
\hat{\mscr{B}}= \Bigg(\bea{cc} \hat{\beta} & \\ & \hat{\beta}^{T} \eea\Bigg)^{ba} \;;\;
\hat{\mscr{B}}\:\hat{M}= \Bigg(\bea{cc} (\hat{\beta}\:\hat{m}) & \\ & (\hat{\beta}\:(-\hat{m})\,)^{T} \eea\Bigg)^{ba} ; \\ \lb{sb1_9}
\feynVV(x_{p})&=&\hat{\Gamma}^{\mu}\;\hat{\mscr{V}}_{\mu}(x_{p})=
\hat{\Gamma}^{\mu}\;\hat{\mscr{T}}_{\alpha}\;\mcal{V}_{\alpha;\mu}(x_{p}) \;; \;\;\;
\hat{\Gamma}^{0}=\Bigg(\bea{cc}\hat{\gamma}^{0} & 0 \\ 0 & \hat{\gamma}^{0,T} \eea\Bigg)\;;\;\;\;
\hat{\vec{\Gamma}}=\Bigg(\bea{cc}\hat{\vec{\gamma}} & 0 \\ 0 & -\hat{\vec{\gamma}}^{T} \eea\Bigg)\;;  \\ \no
\hat{\mscr{T}}_{\alpha}&=& \Bigg(\bea{cc} \hat{t}_{\alpha} & \\ & -\hat{t}_{\alpha}^{T} \eea\Bigg)^{ba}\;;
\hat{\mscr{V}}^{\mu}(x_{p})=\hat{\mscr{T}}_{\alpha}\;\mcal{V}_{\alpha}^{\mu}(x_{p}) \;; \\ \lb{sb1_10}
\hat{\mscr{H}}(x_{p}) &=& \hat{S}\Big(\hat{\mscr{B}}\:\hat{\Gamma}^{\mu}\:\hat{\pp}_{p,\mu}+
\hat{\mscr{B}}\:\big(\im\;\feynVV(x_{p})+\hat{M}\big)\Big)-\im\:\ve_{p}\;\hat{1}_{2N_{0}\times 2N_{0}}\;.
\eeq
If we consider the background potential \(\mcal{V}_{\beta}^{\mu}(x_{p})\) (\ref{sb1_1}) as a classical field,
one can always choose a gauge condition, as Lorentz- (or axial-) gauge
\((\,\hat{\pp}_{p,\mu}\mcal{V}_{\beta}^{\mu}(x_{p})\,)=0\;\),
\(\;(n_{\nu}\;\mcal{V}_{\beta}^{\nu}(x_{p})=0)\), respectively. If we assume that the chosen gauge condition of the composed
field \(\mcal{V}_{\beta}^{\mu}(x_{p})\) is violated, we can shift or adapt the auxiliary field \(s_{\alpha}(x_{p})\)
for the axial gauge fixing of the 'original' gauge fields \(A_{\alpha}^{\mu}(x_{p})\) in such a manner that the chosen
gauge condition is again attained (compare (\ref{sb1_1}) with \(s_{\gamma}(x_{p})\) ). Therefore, one can also determine a
gauge condition for \(\mcal{V}_{\alpha}^{\mu}(x_{p})\) in advance, as far as classical fields are concerned.

In the quantum mechanical case, we obtain a Ward identity of the derived path integral (\ref{s3_116})
with background field averaging (\ref{s3_59}) and with projection \(\hat{S}\) onto BCS-terms in the actions
\(\mscr{A}_{DET}[\hat{T},\feynV;\hat{\mscr{J}}]\), \(\mscr{A}_{J_{\psi}}[\hat{T},\feynV;\hat{\mscr{J}}]\).
Although there appears no action of a field strength tensor as \(\hat{F}_{\alpha}^{\mu\nu}(x_{p})\) (\ref{s2_5})
and Grassmann-valued fields \(\psi_{M}(x_{p})\) of quarks
for the composed gauge field \(\mcal{V}_{\alpha}^{\mu}(x_{p})\), as in the original case with field \(A_{\alpha}^{\mu}(x_{p})\),
a gauge invariance follows because the change of
actions with \(\mcal{V}_{\alpha}^{\mu}(x_{p})\) in a gauge transformation is compensated by the change of
the coset matrices \(\hat{T}(x_{p})=\exp\{-\hat{Y}(x_{p})\,\}\). In this respect the actions
\(\mscr{A}_{DET}[\hat{T},\feynV;\hat{\mscr{J}}]\), \(\mscr{A}_{J_{\psi}}[\hat{T},\feynV;\hat{\mscr{J}}]\)
of the coset matrices replace the actions of a quadratic field strength tensor
and of the anti-commuting Fermi fields for the composed gauge field \(\mcal{V}_{\alpha}^{\mu}(x_{p})\).
According to the definitions of Ref. \cite{wein1} and of section \ref{s2},
the gauge transformation (\ref{sb1_11}) or its infinitesimal version (\ref{sb1_12},\ref{sb1_13}) takes the form
of the listed Eqs. (\ref{sb1_11}-\ref{sb1_17}) for the various BCS-fields and one-particle operators with \(\mbox{SU}_{c}(N_{c}=3)\)
Lie group parameters \(v_{\beta}(x_{p})\), \(\delta v_{\beta}(x_{p})\)
\beq \lb{sb1_11}
\psi_{M}(x_{p}) &\rightarrow& \psi_{M}\ppr(x_{p})=\exp\{\im\:\mscr{v}_{\alpha}(x_{p})\:\hat{t}_{\alpha}\}\;
\psi_{M}(x_{p})\;;\;\;\hat{t}_{\alpha}^{\dag}=\hat{t}_{\alpha}\;; \\ \no
\psi_{M}^{*}(x_{p}) &\rightarrow& \psi_{M}^{*\bprime}(x_{p})=\exp\{-\im\:\mscr{v}_{\alpha}(x_{p})\:\hat{t}_{\alpha}^{T}\}\;
\psi_{M}^{*}(x_{p})\;;   \\ \lb{sb1_12}
\psi_{M}(x_{p})&\rightarrow&\psi_{M}\ppr(x_{p})=\psi_{M}(x_{p})+\delta\psi_{M}(x_{p}) \;; \\ \no
\psi_{M}^{*}(x_{p})&\rightarrow&\psi_{M}^{*\bprime}(x_{p})=\psi_{M}(x_{p})+\delta\psi_{M}^{*}(x_{p}) \;; \\ \lb{sb1_13}
\delta\psi_{M}(x_{p}) &=&\delta\psi_{f,m,r}(x_{p})=\im\:\delta v_{\beta}(x_{p})\;
\hat{t}_{\beta;rs}\:\psi_{f,m,s}(x_{p}) \;;  \\ \no
\delta\psi_{M}^{*}(x_{p}) &=&\delta\psi_{f,m,r}^{*}(x_{p})=\im\:\delta v_{\beta}(x_{p})\;
(-\hat{t}^{T}_{\beta;rs})\:\psi_{f,m,s}^{*}(x_{p}) \;.
\eeq
Corresponding to the infinitesimal \(\mbox{SU}_{c}(N_{c}=3)\) Lie group transformations of quark fields,
we list the infinitesimal changes of the composed gauge field \(\feynV(x_{p})\), of the anomalous doubled
one-particle operator \(\hat{\mscr{H}}(x_{p})\), of the coset matrices \(\hat{T}(x_{p})\) of
BCS quark pairs and of the source field \(J_{\psi;M}^{a}(x_{p})\)
\beq\lb{sb1_14}
\feynV(x_{p}) &=&\hat{\gamma}^{\mu}\;\hat{t}_{\alpha}\;\mcal{V}_{\alpha;\mu}(x_{p})\rightarrow
\feynV(x_{p})+\delta\feynV(x_{p}) \;; \\ \no
\delta\feynV(x_{p}) &=&\im\:\delta v_{\beta}(x_{p})\;
[\hat{t}_{\beta}\,\boldsymbol{,}\,\feynV(x_{p})\,]_{-}-\im\:\big(\feynd_{p}\:
\delta v_{\beta}(x_{p})\,\big)\;\hat{t}_{\beta} \;;  \\ \no
\feynVV(x_{p}) &=&\hat{\Gamma}^{\mu}\:\hat{\mscr{T}}_{\alpha}\:\mcal{V}_{\alpha;\mu}(x_{p})\rightarrow
\feynVV(x_{p})+\delta\feynVV(x_{p}) \;;  \\ \no
\delta\feynVV(x_{p})&=&\im\:\delta v_{\beta}(x_{p})\;
[\hat{\mscr{T}}_{\beta}\,\boldsymbol{,}\,\feynVV(x_{p})\,]_{-}-\im\:\hat{\Gamma}^{\mu}
\big(\hat{\pp}_{p,\mu}\:\delta v_{\beta}(x_{p})\,\big)\;\hat{\mscr{T}}_{\beta} \;;  \\ \no
\hat{\mscr{H}}(x_{p}) &\rightarrow& \hat{\mscr{H}}(x_{p}) +\delta\hat{\mscr{H}}(x_{p}) \;;\;\;\;\rightarrow\;
\delta\hat{\mscr{H}}(x_{p})=\im\;\hat{\mscr{B}}\;\hat{S}\;\delta\feynVV(x_{p}) \;;  \\   \lb{sb1_15}
\hat{T}^{ab}(x_{p}) &\rightarrow& \hat{T}^{ab}(x_{p}) +\delta\hat{T}^{ab}(x_{p}) \;; \;\;\;\rightarrow\;
\delta\hat{T}^{ab}(x_{p}) =\im\:\delta v_{\beta}(x_{p})\;
\big(\,[\hat{\mscr{T}}_{\beta}\,\boldsymbol{,}\,\hat{T}(x_{p})\,]_{-}\big)^{ab} \;;  \\ \no
\hat{T}^{-1;ab}(x_{p}) &\rightarrow& \hat{T}^{-1;ab}(x_{p}) - \Big(\hat{T}^{-1}(x_{p})\:
\delta\hat{T}(x_{p})\:\hat{T}^{-1}(x_{p})\Big)^{ab} \;;   \\   \lb{sb1_16}
J_{\psi}^{a}(x_{p}) &\rightarrow& J_{\psi}^{a}(x_{p}) +\delta J_{\psi}^{a}(x_{p})  \;; \;\;\;\rightarrow\;
\delta J_{\psi}^{a}(x_{p}) =\im\:\delta v_{\beta}(x_{p})\;
\big(\hat{\mscr{T}}_{\beta}^{ab\ppr}\:J_{\psi}^{b\ppr}(x_{p})\,\big)^{a} \;;  \\ \no
J_{\psi}^{\dag,b}(x_{p}) &\rightarrow& J_{\psi}^{\dag,b}(x_{p}) +\delta J_{\psi}^{\dag,b}(x_{p})  \;; \;\;\;\rightarrow\;
\delta J_{\psi}^{\dag,b}(x_{p}) = -\im\:\delta v_{\beta}(x_{p})\;
\big(J_{\psi}^{\dag,a\ppr}(x_{p})\:\hat{\mscr{T}}_{\beta}^{a\ppr b}\,\big)^{b} \;;    \\  \lb{sb1_17}
\hat{\mscr{O}} &=&\hat{T}^{-1}\;\hat{\mscr{H}}\;\hat{T} \;;  \\ \no
\delta\hat{\mscr{O}}(x_{p}) &=&\delta\hat{T}^{-1}(x_{p})\;\hat{\mscr{H}}(x_{p})\;\hat{T}(x_{p}) +
\hat{T}^{-1}(x_{p})\;\hat{\mscr{H}}(x_{p})\;\delta\hat{T}(x_{p}) +
\hat{T}^{-1}(x_{p})\;\delta\hat{\mscr{H}}(x_{p})\;\hat{T}(x_{p})   \\ \no &=&
\Big[\hat{T}^{-1}(x_{p})\;\hat{\mscr{H}}(x_{p})\;\hat{T}(x_{p}) \:\boldsymbol{,}\:
\hat{T}^{-1}(x_{p})\;\delta\hat{T}(x_{p})\Big] +
\hat{T}^{-1}(x_{p})\:\im\:\hat{\mscr{B}}\:\hat{S}\:\delta\feynVV(x_{p})\:\hat{T}(x_{p})  \;;  \\ \no
\delta\hat{\mscr{O}}(x_{p}) &=&\im\:\delta v_{\beta}(x_{p})\;
\Big[\hat{\mscr{O}}(x_{p})\:\boldsymbol{,}\:\hat{T}^{-1}(x_{p})\:\big[\hat{\mscr{T}}_{\beta}\,\boldsymbol{,}\,
\hat{T}(x_{p})\,\big]_{-}\Big]_{-}  +   \\ \no &+&
\im\:\delta v_{\beta}(x_{p})\;
\hat{T}^{-1}(x_{p})\:\im\:\hat{\mscr{B}}\:\hat{S}\:\big[\hat{\mscr{T}}_{\beta}\,\boldsymbol{,}\,
\feynVV(x_{p})\,\big]_{-}\:\hat{T}(x_{p}) -\im\:\big(\hat{\pp}_{p,\mu}\delta v_{\beta}(x_{p})\big)\:
\hat{T}^{-1}(x_{p})\:\im\:\hat{\mscr{B}}\:\hat{S}\:\hat{\Gamma}^{\mu}\:\hat{\mscr{T}}_{\beta}\:\hat{T}(x_{p}) \;.
\eeq
The variation of the effective generating functional (\ref{s3_116}) finally yields the conserved Ward identity (\ref{sb1_18})
by expanding up to first order in \(\delta v_{\beta}(x_{p})\) of the gauge parameters where a partial integration
has to be performed for the part \(\delta\feynV(x_{p})\) (\ref{sb1_14}) with derivatives
\((\hat{\pp}_{p,\mu}\delta v_{\beta}(x_{p})\,)\)
\beq \lb{sb1_18}
\lefteqn{\int_{C}d^{4}x_{p}\;\;\frac{\delta Z[\hat{\mscr{J}},J_{\psi},\hat{J}_{\psi\psi},\hat{\mfrak{j}}^{(\hat{F})};
\mbox{(\ref{s3_116})}]}{\delta v_{\beta}(x_{p})}\;\eta_{p}\;\delta v_{\beta}(x_{p}) =}  \\ \no &=&
\boldsymbol{\Bigg\langle} \int_{C}d^{4}x_{p}\;\;\eta_{p}\;\delta v_{\beta}(x_{p})\;
\frac{\delta}{\delta v_{\beta}(x_{p})}
\Bigg(Z\Big[\feynV(x_{p});\hat{\mfrak{S}}^{(\hat{F})},
s_{\alpha},\hat{\mfrak{B}}_{\hat{F}},\hat{\mfrak{b}}^{(\hat{F})},
\hat{\mscr{U}}_{\hat{F}},\hat{\mfrak{v}}_{\hat{F}};\sigma_{D};\hat{\mfrak{j}}^{(\hat{F})};
\mbox{\bf Eq. (\ref{s3_59})}\Big]\;\times \;  \\ \no  &\times&
\int d[\hat{T}^{-1}(x_{p})\;d\hat{T}(x_{p})]\;\;Z_{\hat{J}_{\psi\psi}}[\hat{T}]\;
\exp\Big\{\mscr{A}_{DET}[\hat{T},\feynV;\hat{\mscr{J}}]\Big\}\;
\exp\Big\{\im\;\mscr{A}_{J_{\psi}}[\hat{T},\feynV;\hat{\mscr{J}}]\Big\}\Bigg)  \boldsymbol{\Bigg\rangle}  =  \\ \no &=&
\boldsymbol{\Bigg\langle} Z\Big[\feynV(x_{p});\hat{\mfrak{S}}^{(\hat{F})},
s_{\alpha},\hat{\mfrak{B}}_{\hat{F}},\hat{\mfrak{b}}^{(\hat{F})},
\hat{\mscr{U}}_{\hat{F}},\hat{\mfrak{v}}_{\hat{F}};\sigma_{D};\hat{\mfrak{j}}^{(\hat{F})};
\mbox{\bf Eq. (\ref{s3_59})}\Big]\;\times \;
\int d[\hat{T}^{-1}(x_{p})\;d\hat{T}(x_{p})]\;\;Z_{\hat{J}_{\psi\psi}}[\hat{T}]\;\times \\ \no &\times&
\exp\Big\{\mscr{A}_{DET}[\hat{T},\feynV;\hat{\mscr{J}}] +
\im\;\mscr{A}_{J_{\psi}}[\hat{T},\feynV;\hat{\mscr{J}}]\Big\} \times  \hspace*{0.6cm}
\trxpa\trfgamc\Bigg\{\im\:\delta v_{\beta}(x_{p})\:\eta_{p}\;\times \\ \no &\times&
\Bigg[\bigg(\Big[\hat{\mscr{O}}(x_{p})\:\boldsymbol{,}\:\hat{T}^{-1}(x_{p})\:\big[\hat{\mscr{T}}_{\beta}\,\boldsymbol{,}\,
\hat{T}(x_{p})\,\big]_{-}\Big]_{-}  +
\hat{T}^{-1}(x_{p})\:\im\:\hat{\mscr{B}}\:\hat{S}\:\big[\hat{\mscr{T}}_{\beta}\,\boldsymbol{,}\,
\feynVV(x_{p})\,\big]_{-}\:\hat{T}(x_{p}) \bigg) \times \\ \no &\times&
\bigg(\langle\widehat{x_{p}}|\hat{\mscr{O}}^{-1}|\widehat{x_{p}}\rangle+\frac{\im}{2}
\int_{C}d^{4}y_{q}\;d^{4}y\ppr_{q\ppr}\;
\langle\widehat{x_{p}}|\hat{\mscr{O}}^{-1}|\widehat{y_{q}}\rangle\;
\hat{T}^{-1}(y_{q})\;\hat{I}\;J_{\psi}(y_{q})\;\otimes \;
J_{\psi}^{\dag}(y\ppr_{q\ppr})\;\hat{I}\;\hat{T}(y\ppr_{q\ppr})\:
\langle\widehat{y\ppr_{q\ppr}}|\hat{\mscr{O}}^{-1}|\widehat{x_{p}}\rangle\bigg) + \\ \no &+&
\Bigg(\feynd_{p,\mu}\:
\hat{T}^{-1}(x_{p})\:\im\:\hat{\mscr{B}}\:\hat{S}\:\hat{\Gamma}^{\mu}\:\hat{\mscr{T}}_{\beta}\:\hat{T}(x_{p})
\times \\ \no &\times&
\bigg(\langle\widehat{x_{p}}|\hat{\mscr{O}}^{-1}|\widehat{x_{p}}\rangle+\frac{\im}{2}
\int_{C}d^{4}y_{q}\;d^{4}y\ppr_{q\ppr}\;
\langle\widehat{x_{p}}|\hat{\mscr{O}}^{-1}|\widehat{y_{q}}\rangle\;
\hat{T}^{-1}(y_{q})\;\hat{I}\;J_{\psi}(y_{q})\;\otimes \;
J_{\psi}^{\dag}(y\ppr_{q\ppr})\;\hat{I}\;\hat{T}(y\ppr_{q\ppr})\:
\langle\widehat{y\ppr_{q\ppr}}|\hat{\mscr{O}}^{-1}|\widehat{x_{p}}\rangle\bigg)\Bigg) +  \\ \no &+&\frac{\im}{2}
\int_{C}d^{4}y_{q}\bigg(\langle\widehat{x_{p}}|\hat{\mscr{O}}^{-1}|\widehat{y_{q}}\rangle\;
\hat{T}^{-1}(y_{q})\:\hat{I}\:J_{\psi}(y_{q})\otimes J_{\psi}^{\dag}(x_{p})\:\hat{I}\:\hat{T}(x_{p})\:\hat{\mscr{T}}_{\beta}+
\\ \no &-& \hat{\mscr{T}}_{\beta}\:\hat{T}^{-1}(x_{p})\:\hat{I}\:J_{\psi}(x_{p})\otimes
J_{\psi}^{\dag}(y_{q})\:\hat{I}\:\hat{T}(y_{q})\;
\langle\widehat{y_{q}}|\hat{\mscr{O}}^{-1}|\widehat{x_{p}}\rangle\bigg)\;\Bigg] \Bigg\}\boldsymbol{\Bigg\rangle}\;.
\eeq

\section{Derivation of the chiral anomaly from the change of integration variables} \lb{sc}

\subsection{Axial '$\hat{\gamma}_{5}$' transformations of the actions in the exponentials} \lb{sc1}

In the following we describe the derivation of the chiral anomaly according to \cite{Fuji,Fuji2}. It is finally obtained from the
change of the fermionic path integration variables \(d[\psi\pdag(x_{p})]\;d[\psi(x_{p})]\) where the divergence of the
corresponding Jacobian is determined by a gauge invariant cut-off regulator and has therefore to be regarded as an additional
quantum phenomenon, violating the classical, (massless), conserved '\(\hat{\gamma}_{5}\)' Noether current relation. These chiral anomalies
are also related to nontrivial differential topologies which allow for explicit derivations in a geometrical context \cite{Bertlmann}.
The original derivation of the anomaly is given within perturbation series from the regularization of a triangle diagram with the
axial '\(\hat{\gamma}_{5}\)' matrix \cite{Fuji2}. In this subsection \ref{sc1} we briefly restrict to the change of the original QCD action
\(\mscr{A}[\psi,\hat{D}_{\mu}\psi,\hat{F}]\) and the action
\(\mscr{A}_{S}[\hat{\mscr{J}},J_{\psi},\hat{J}_{\psi\psi},\hat{\mfrak{j}}^{(\hat{F})}]\) of source terms under axial
\(U_{A}(1)\) transformations \(\hat{g}_{A}(x_{p})\), \(\hat{g}_{A}\pdag(x_{p})\) (\ref{sc_5}-\ref{sc_7})
of the Grassmann fields \(\psi_{f,m,r}(x_{p})\) (\ref{sc_1}-\ref{sc_4})
with '\(\hat{\gamma}_{5}\)' Dirac matrix and isospin-(flavour-) generator
\(\hat{\mfrak{t}}_{0,fg}\) and angle \(\alpha(x_{p})\). The finite, axial \(U_{A}(1)\) transformations are listed in relations
(\ref{sc_1}-\ref{sc_7}) with their actions onto the fermionic fields
\beq \lb{sc_1}
\psi_{f,m,r}(x_{p})&=&\delta_{rs}\;\Big[\exp\Big\{\im\;\alpha(x_{p})\;\hat{\mfrak{t}}_{0,f\ppr g\ppr}\;
\big(\hat{\gamma}_{5}\big)_{m\ppr n\ppr}\;\delta_{r\ppr s\ppr}\Big\}\Big]_{f,m,r;g,n,s}\;
\psi_{g,n,s}\ppr(x_{p})\;;  \\ \lb{sc_2}
\psi(x_{p}) &=&\hat{g}_{A}(x_{p})\;\psi\ppr(x_{p})\;; \\ \lb{sc_3}
\psi^{\dag}(x_{p})&=&\psi^{\bprime\dag}(x_{p})\;
\hat{g}_{A}^{\dag}(x_{p})\;;  \\  \lb{sc_4}
\bar{\psi}(x_{p}) &=&\psi^{\dag}(x_{p})\;\hat{\beta}=\psi^{\bprime\dag}(x_{p})\;\hat{g}_{A}^{\dag}(x_{p})\;\hat{\beta}=
\bar{\psi}\ppr(x_{p})\;\hat{g}_{A}(x_{p})\;; \\  \lb{sc_5}
\hat{g}_{A}(x_{p})&=&\cos\Big(\alpha(x_{p})\;\hat{\mfrak{t}}_{0}\Big)+
\im\;\hat{\gamma}_{5}\;\sin\Big(\alpha(x_{p})\;\hat{\mfrak{t}}_{0}\Big) \;;
\hspace*{0.3cm}\hat{\mfrak{t}}_{0}^{\dag}=\hat{\mfrak{t}}_{0}\;; \\  \lb{sc_6}
\hat{g}_{A}^{\dag}(x_{p})&=&\cos\Big(\alpha(x_{p})\;\hat{\mfrak{t}}_{0}\Big)-
\im\;\hat{\gamma}_{5}\;\sin\Big(\alpha(x_{p})\;\hat{\mfrak{t}}_{0}\Big) \;;
\hspace*{0.3cm}\hat{\mfrak{t}}_{0;f,m,r;g,n,s}=\hat{\mfrak{t}}_{0,fg}\;\delta_{mn}\;\delta_{rs}\;; \\   \lb{sc_7}
\hat{\gamma}_{5}^{\dag} &=&\hat{\gamma}_{5}\;;\hspace*{0.3cm}\hat{\gamma}_{5}^{T}=\hat{\gamma}_{5}\;;\hspace*{0.3cm}
\big(\hat{\gamma}_{5}\big)^{2}=\hat{1}_{4\times 4}\;;\hspace*{0.3cm}
\big\{\hat{\beta}\;\boldsymbol{,}\;\hat{\gamma}_{5}\big\}_{+}=0\;;\hspace*{0.3cm}
\big\{\hat{\gamma}^{\mu}\;\boldsymbol{,}\;\hat{\gamma}_{5}\big\}_{+}=0\;.
\eeq
The anti-commuting property (\ref{sc_7}) of the axial '\(\hat{\gamma}_{5}\)' matrix with the Dirac matrices \(\hat{\gamma}^{\mu}\) has to
be emphasized because this particular property causes to transform the fermi field \(\bar{\psi}(x_{p})\) (\ref{sc_4}) with the same matrix
\(\hat{g}_{A}(x_{p})\) as for \(\psi(x_{p})\) (\ref{sc_2}) instead of \(\hat{g}_{A}\pdag(x_{p})\) as for \(\psi\pdag(x_{p})\)
(\ref{sc_3}). We insert the
transformed fields \(\psi\ppr_{g,n,s}(x_{p})\) (\ref{sc_1}-\ref{sc_4})
into the original actions \(\mscr{A}[\psi,\hat{D}_{\mu}\psi,\hat{F}]\) and
\(\mscr{A}_{S}[\hat{\mscr{J}},J_{\psi},\hat{J}_{\psi\psi},\hat{\mfrak{j}}^{(\hat{F})}]\) for a finite angle \(\alpha(x_{p})\)
with the transformation matrices \(\hat{g}_{A}(x_{p})\), \(\hat{g}_{A}\pdag(x_{p})\) (\ref{sc_5}-\ref{sc_7}) and expand to first order \(\delta\alpha(x_{p})\) for infinitesimal values, using the \(U_{A}(1)\) Lie group properties. This results into
\(\delta\mscr{A}[\psi\ppr,\hat{D}_{\mu}\psi\ppr,\hat{F}]\) (\ref{sc_8}-\ref{sc_10}) and
\(\delta\mscr{A}_{S}\ppr[\hat{\mscr{J}},J_{\psi},\hat{J}_{\psi\psi},\hat{\mfrak{j}}^{(\hat{F})}]\) (\ref{sc_11}-\ref{sc_13})
where a partial integration has to be performed
within \(\delta\mscr{A}[\psi\ppr,\hat{D}_{\mu}\psi\ppr,\hat{F}]\) in order to attain only the first order variation
\(\delta\alpha(x_{p})\) without any derivative
\beq \lb{sc_8}
\lefteqn{\mscr{A}[\psi,\hat{D}_{\mu}\psi,\hat{F}]=\int_{C}d^{4}\!x_{p}\bigg\{-\frac{1}{4}\hat{F}_{\alpha;\mu\nu}(x_{p})\;
\hat{F}_{\alpha}^{\mu\nu}(x_{p})+ }  \\ \no &-&
\psi^{\bprime \dag}(x_{p})\;\hat{g}_{A}^{\dag}(x_{p})\Big(\hat{\beta}\:
\hat{\gamma}^{\mu}\:\hat{\pp}_{p,\mu}-\im\, g\:\hat{\beta}\:\hat{\gamma}^{\mu}\:
\hat{t}_{\alpha}\;A_{\alpha;\mu}(x_{p})+\hat{\beta}\:\hat{m}-\im\, \ve_{+}\:
\eta_{p}\Big)\hat{g}_{A}(x_{p})\;\psi\ppr(x_{p})\bigg\} =  \\ \no &=&
\int_{C}d^{4}\!x_{p}\bigg\{-\frac{1}{4}\hat{F}_{\alpha;\mu\nu}(x_{p})\;
\hat{F}_{\alpha}^{\mu\nu}(x_{p}) -\bar{\psi}\ppr(x_{p})\;\hat{g}_{A}(x_{p})
\Big(\feynd_{p}-\im\,g\;\feynA+\hat{m}-\im\,\hat{\beta}\:\ve_{+}\:\eta_{p}\Big)\hat{g}_{A}(x_{p})\;\psi\ppr(x_{p})\bigg\}\;;  \\ \lb{sc_9}
\lefteqn{\mscr{A}[\psi,\hat{D}_{\mu}\psi,\hat{F}] = \mscr{A}[\psi\ppr,\hat{D}_{\mu}\psi\ppr,\hat{F}]+
\delta\mscr{A}[\psi\ppr,\hat{D}_{\mu}\psi\ppr,\hat{F}] \;;} \\  \lb{sc_10}
\lefteqn{\delta\mscr{A}[\psi\ppr,\hat{D}_{\mu}\psi\ppr,\hat{F}]=-\im\hspace*{-0.1cm}\int_{C}\hspace*{-0.2cm}d^{4}\!x_{p}\Big\{
\bar{\psi}\ppr(x_{p})\:\hat{\gamma}_{5}\big\{\hat{\mfrak{t}}_{0}\:\boldsymbol{,}\:
\hat{m}\big\}_{+}\psi\ppr(x_{p})\;\delta\alpha(x_{p})+
\bar{\psi}\ppr(x_{p})\:\hat{\gamma}^{\mu}
\Big(\hat{\pp}_{p,\mu}\delta\alpha(x_{p})\Big)\hat{\gamma}_{5}\:\hat{\mfrak{t}}_{0}\:\psi\ppr(x_{p})\Big\} \hspace*{-0.15cm}= }  \\ \no &=&\im
\int_{C}d^{4}\!x_{p}\Big\{\hat{\pp}_{p,\mu}\Big(\bar{\psi}\ppr(x_{p})\:\hat{\gamma}^{\mu}\:\hat{\gamma}_{5}\:
\hat{\mfrak{t}}_{0}\:\psi\ppr(x_{p})\Big)-
\bar{\psi}\ppr(x_{p})\:\hat{\gamma}_{5}\:
\big\{\hat{\mfrak{t}}_{0}\:\boldsymbol{,}\:\hat{m}\big\}_{+}\:\psi\ppr(x_{p}) \Big\}\;\delta\alpha(x_{p})\;;  \\ \lb{sc_11}
\lefteqn{\mscr{A}_{S}[\hat{\mscr{J}},J_{\psi},\hat{J}_{\psi\psi},\hat{\mfrak{j}}^{(\hat{F})}]=\int_{C}d^{4}\!x_{p}\Bigg\{
\Big(j_{\psi}^{\dag}(x_{p})\:\hat{g}_{A}(x_{p})\:\psi\ppr(x_{p})+
\psi^{\bprime \dag}(x_{p})\:\hat{g}_{A}^{\dag}(x_{p})\:j_{\psi}(x_{p})\Big)+ }  \\ \no &+&\frac{1}{2}
\bigg[\Big(\hat{g}_{A}(x_{p})\:\psi\ppr(x_{p})\Big)^{T}\hat{j}_{\psi\psi}^{\dag}(x_{p})\Big(\hat{g}_{A}(x_{p})\:\psi\ppr(x_{p})\Big)+
\Big(\hat{g}_{A}(x_{p})\:\psi\ppr(x_{p})\Big)^{\dag}\hat{j}_{\psi\psi}(x_{p})
\Big(\hat{g}_{A}(x_{p})\:\psi\ppr(x_{p})\Big)^{*}\bigg] + \\ \no &+& \frac{1}{2}
\int_{C}d^{4}\!y_{q}\;\Psi^{\bprime \dag}(y_{q})\bigg(\bea{cc} \hat{g}_{A}^{\dag}(y_{q}) & \\ &
\hat{g}_{A}^{T}(y_{q}) \eea\bigg)\hat{\mscr{J}}(y_{q},x_{p})
\bigg(\bea{cc} \hat{g}_{A}(x_{p}) & \\ & \hat{g}_{A}^{*}(x_{p}) \eea\bigg)\Psi\ppr(x_{p}) +
\hat{j}_{\hat{F};\alpha;\mu\nu}(x_{p})\;\hat{F}_{\alpha}^{\mu\nu}(x_{p})\Bigg\}\;;  \\  \lb{sc_12}
\lefteqn{\mscr{A}_{S}[\hat{\mscr{J}},J_{\psi},\hat{J}_{\psi\psi},\hat{\mfrak{j}}^{(\hat{F})}] =
\mscr{A}_{S}\ppr[\hat{\mscr{J}},J_{\psi},\hat{J}_{\psi\psi},\hat{\mfrak{j}}^{(\hat{F})}]+
\delta\mscr{A}_{S}\ppr[\hat{\mscr{J}},J_{\psi},\hat{J}_{\psi\psi},\hat{\mfrak{j}}^{(\hat{F})}]  \;; } \\ \lb{sc_13}
\lefteqn{\delta\mscr{A}_{S}\ppr[\hat{\mscr{J}},J_{\psi},\hat{J}_{\psi\psi},\hat{\mfrak{j}}^{(\hat{F})}]=\im\int_{C}d^{4}\!x_{p}\;\,
\delta\alpha(x_{p})\Bigg\{
\Big(j_{\psi}^{\dag}(x_{p})\:\hat{\gamma}_{5}\:\hat{\mfrak{t}}_{0}\:\psi\ppr(x_{p})-
\psi^{\bprime \dag}(x_{p})\:\hat{\gamma}_{5}\:\hat{\mfrak{t}}_{0}\:j_{\psi}(x_{p})\Big)+ }
\\ \no \hspace*{-0.6cm}&\hspace*{-0.6cm}+&\hspace*{-0.6cm}\frac{1}{2}
\bigg[\psi^{\bprime T}(x_{p})\Big(\hat{j}_{\psi\psi}^{\dag}(x_{p})\:\hat{\gamma}_{5}\:\hat{\mfrak{t}}_{0}+
\hat{\gamma}_{5}\:\hat{\mfrak{t}}_{0}^{T}\:\hat{j}_{\psi\psi}^{\dag}(x_{p})\Big)\psi\ppr(x_{p})
- \psi^{\bprime \dag}(x_{p})\Big(\hat{j}_{\psi\psi}(x_{p})\:\hat{\gamma}_{5}\:\hat{\mfrak{t}}_{0}^{*}+
\hat{\gamma}_{5}\:\hat{\mfrak{t}}_{0}\:\hat{j}_{\psi\psi}(x_{p})\Big)\psi^{\bprime *}(x_{p})\bigg]
+ \\ \no &+& \frac{1}{2}
\int_{C}d^{4}\!y_{q}\bigg[\Psi^{\bprime \dag}(y_{q})\:\hat{\mscr{J}}(y_{q},x_{p})
\bigg(\bea{cc} \hat{\gamma}_{5}\:\hat{\mfrak{t}}_{0} & \\ & \hspace*{-0.6cm}-\hat{\gamma}_{5}\:\hat{\mfrak{t}}_{0}^{*} \eea\bigg)
\Psi\ppr(x_{p}) +  \Psi^{\bprime \dag}(x_{p})
\bigg(\bea{cc} -\hat{\gamma}_{5}\:\hat{\mfrak{t}}_{0} & \\ & \hspace*{-0.6cm}
\hat{\gamma}_{5}\:\hat{\mfrak{t}}_{0}^{T} \eea\bigg)\hat{\mscr{J}}(x_{p},y_{q})\;\Psi\ppr(x_{p}) \bigg] \Bigg\}_{.}
\eeq
After reordering the first order variations with \(\delta\alpha(x_{p})\), we achieve the axial current relation
(\ref{sc_14}) of \(\bar{\psi}(x_{p})\:\hat{\gamma}^{\mu}\:\hat{\gamma}_{5}\:\hat{\mfrak{t}}_{0}\:\psi(x_{p})\) whose
conservation is perturbed by the mass term
\(\bar{\psi}(x_{p})\:\hat{\gamma}_{5}\:\big\{\hat{\mfrak{t}}_{0}\:\boldsymbol{,}\:\hat{m}\big\}_{+}\:\psi(x_{p})\) and the source parts
which are also determined by the isospin-(flavour-) generator \(\hat{\mfrak{t}}_{0,fg}\)
\beq \lb{sc_14}
\lefteqn{\delta\mscr{A}[\psi\ppr,\hat{D}_{\mu}\psi\ppr,\hat{F}]-
\delta\mscr{A}_{S}\ppr[\hat{\mscr{J}},J_{\psi},\hat{J}_{\psi\psi},\hat{\mfrak{j}}^{(\hat{F})}]\equiv0\;;\Longrightarrow}
\\ \no &&\hspace*{-1.0cm}
\hat{\pp}_{p,\mu}\Big(\bar{\psi}(x_{p})\:\hat{\gamma}^{\mu}\:\hat{\gamma}_{5}\:\hat{\mfrak{t}}_{0}\:\psi(x_{p})\Big)=
\bar{\psi}(x_{p})\:\hat{\gamma}_{5}\:\big\{\hat{\mfrak{t}}_{0}\:\boldsymbol{,}\:\hat{m}\big\}_{+}\:\psi(x_{p}) +
j_{\psi}^{\dag}(x_{p})\:\hat{\gamma}_{5}\:\hat{\mfrak{t}}_{0}\:\psi(x_{p})-\psi^{\dag}(x_{p})\:
\hat{\gamma}_{5}\:\hat{\mfrak{t}}_{0}\:j_{\psi}(x_{p}) + \\ \no &+& \frac{1}{2}
\bigg[\psi^{T}(x_{p})\Big(\hat{j}_{\psi\psi}^{\dag}(x_{p})\:\hat{\gamma}_{5}\:\hat{\mfrak{t}}_{0}+\hat{\gamma}_{5}\:\hat{\mfrak{t}}_{0}^{T}\:\hat{j}_{\psi\psi}^{\dag}(x_{p})\Big)\psi(x_{p}) -
\psi^{\dag}(x_{p})\Big(\hat{j}_{\psi\psi}(x_{p})\:\hat{\gamma}_{5}\:\hat{\mfrak{t}}_{0}^{*}+
\hat{\gamma}_{5}\:\hat{\mfrak{t}}_{0}\:\hat{j}_{\psi\psi}(x_{p})\Big)\psi^{*}(x_{p})\bigg] + \\  \no &+& \frac{1}{2}
\int_{C}d^{4}\!y_{q}
\Bigg[\Psi^{\dag}(y_{q},x_{p})\:\hat{\mscr{J}}(y_{q},x_{p})\bigg(\bea{cc}\hat{\gamma}_{5}\:
\hat{\mfrak{t}}_{0} & \\ & -\hat{\gamma}_{5}\:\hat{\mfrak{t}}_{0}^{*}\eea\bigg)\Psi(x_{p})+
\Psi^{\dag}(x_{p})\bigg(\bea{cc} -\hat{\gamma}_{5}\:\hat{\mfrak{t}}_{0} & \\ &
\hat{\gamma}_{5}\:\hat{\mfrak{t}}_{0}^{T} \eea\bigg)\hat{\mscr{J}}(x_{p},y_{q})\:\Psi(y_{q})\Bigg]_{.}
\eeq

\subsection{Calculation of the Jacobian with a gauge invariant cut-off regulator} \lb{sc2}

Apart from the transformation of the actions \(\mscr{A}[\psi,\hat{D}_{\mu}\psi,\hat{F}]\),
\(\mscr{A}_{S}[\hat{\mscr{J}},J_{\psi},\hat{J}_{\psi\psi},\hat{\mfrak{j}}^{(\hat{F})}]\) as phases in the exponential, one has also
to consider the change of the Jacobian under transformations (\ref{sc_15},\ref{sc_16}) with matrix '\(\hat{\gamma}_{5}\)'
and the parameter \(\alpha(x_{p})\) for the generator \(\hat{\mfrak{t}}_{0,fg}\) of axial isospin-(flavour-) rotations
\beq\lb{sc_15}
\psi\ppr(x_{p}) &=&\exp\big\{-\im\:\alpha(x_{p})\:\hat{\gamma}_{5}\:\hat{\mfrak{t}}_{0}\big\}\;\psi(x_{p}) \;; \\ \lb{sc_16}
\bar{\psi}\ppr(x_{p}) &=&\bar{\psi}(x_{p})\;\exp\big\{-\im\:\alpha(x_{p})\:\hat{\gamma}_{5}\:\hat{\mfrak{t}}_{0}\big\}\;.
\eeq
We expand the fermionic fields \(\psi(x_{p})\), \(\bar{\psi}(x_{p})\) in terms of orthonormalized eigenfunctions \(\varphi_{L}(x_{p})\)
of the (massless) gauge invariant derivative \(\feynD_{p}\) with eigenvalues \(-\im\,e_{L}^{(p)}\) on the time contour '\(p=\pm\)'
and have to include Dirac spinors \(\chi_{L}^{(p)}\), \(\bar{\chi}_{L}^{(p)}\) in order to regard the anti-commuting character
of the fermi fields \(\psi(x_{p})\), \(\bar{\psi}(x_{p})\) (\ref{sc_17}-\ref{sc_19}).
One might infer that the eigenfunctions \(\varphi_{L}(x_{p})\)
cannot be taken as ordinary, complex-valued, scalar functions, due to the matrix property of \(\feynD_{p}\), but an appropriate
unitary transformation '\(\hat{U}\)' can always be chosen to rotate to scalar, complex-valued eigenfunctions  \(\varphi_{L}(x_{p})\),
leaving the Dirac spinor property of \(\psi(x_{p})\), \(\bar{\psi}(x_{p})\) entirely within \(\chi_{L}^{(p)}\), \(\bar{\chi}_{L}^{(p)}\)
(\ref{sc_20},\ref{sc_21}).
Using the unitary invariance of the 'massless' derivative operator \(\feynD_{p}\) (\ref{sc_22},\ref{sc_23}),
we can eventually compute the expectation
value \(\bar{\psi}(x_{p})\,(\,\feynD_{p}+\hat{m}\,)\,\psi(x_{p})\) (\ref{sc_24}) in terms of
the Dirac spinors \(\chi_{L}^{(p)}\), \(\bar{\chi}_{L}^{(p)}\) and eigenvalues \(-\im\,e_{L}^{(p)}\)
\beq  \lb{sc_17}
\psi(x_{p}) &=&\sum_{L}\varphi_{L}(x_{p})\;\chi_{L}^{(p)}=\sum_{L}\langle x_{p}|L\rangle\;\chi_{L}^{(p)}  \;; \\  \lb{sc_18}
\bar{\psi}(x_{p}) &=& \sum_{L}\bar{\chi}_{L}^{(p)}\;\varphi_{L}^{\dag}(x_{p})=\sum_{L}\bar{\chi}_{L}^{(p)}\;
\langle L|x_{p}\rangle\;;\\ \lb{sc_19}
\chi_{L}^{(p)}\;,\:\bar{\chi}_{L}^{(p)}&:=&\mbox{ Dirac spinors }   \;; \hspace*{0.6cm}
\bar{\chi}_{L}^{(p)}=\chi_{L}^{(p)\dag}\;\hat{\beta}\;;   \\   \lb{sc_20}
\hat{U}\;\feynD_{p}\;\hat{U}^{\dag}\:\big(\hat{U}\varphi_{L}(x_{p})\big) &=&
\hat{U}\Big(\feynd_{p}-\im\,g\:\feynA(x_{p})\Big)\hat{U}^{\dag}\:\big(\hat{U}\varphi_{L}(x_{p})\big) =
-\im\:e_{L}^{(p)}\;\big(\hat{U}\varphi_{L}(x_{p})\big)\;; \\  \lb{sc_21} \mbox{unitary transformation }
\big(\hat{U}\varphi_{L}(x_{p})\big) &\rightarrow&\mbox{for scalar, complex-valued eigenfunctions } \varphi_{L}(x_{p}) \;; \\  \lb{sc_22}
\feynD_{p}\varphi_{L}(x_{p}) &=&\Big(\feynd_{p}-\im\,g\:\feynA(x_{p})\Big)
\varphi_{L}(x_{p})=-\im\:e_{L}^{(p)}\;\varphi_{L}(x_{p})\;; \\    \lb{sc_23}
\delta_{K,L} &=&\int d^{4}\!x_{p}\;\varphi_{K}^{\dag}(x_{p})\;\varphi_{L}(x_{p})\;; \\    \lb{sc_24}
\int d^{4}\!x_{p}\;\bar{\psi}(x_{p})\;\big(\feynD_{p}+\hat{m}\big)\;\psi(x_{p})&=&
\lim_{N_{L}\rightarrow\infty}\sum_{L=1}^{N_{L}}\bar{\chi}_{L}^{(p)}\;\big(-\im\:e_{L}^{(p)}+\hat{m}\big)\;\chi_{L}^{(p)}\;.
\eeq
The change with the Jacobian of \(d[\psi^{\dag}(x_{p})]\;d[\psi(x_{p})]\) to \(d\bar{\chi}_{L}^{(p)}\;d\chi_{L}^{(p)}\) (\ref{sc_25})
follows from the inverse of the transformation matrices \(\langle L|x_{p}\rangle\), \(\langle x_{p}|L\rangle\), due to the
anti-commuting property of the fermionic fields
\beq \no
\lefteqn{\hspace*{-2.3cm}d[\psi^{\dag}(x_{p})]\;d[\psi(x_{p})] \stackrel{\det(\hat{\beta})=1}{=} d[\bar{\psi}(x_{p})]\;d[\psi(x_{p})] =
\prod_{p=\pm}\lim_{N_{L}\rightarrow\infty}\prod_{L=1}^{N_{L}}d\bar{\chi}_{L}^{(p)}\;d\chi_{L}^{(p)}\,
\Big[\det\big(\langle L|x_{p}\rangle\big)\;\det\big(\langle x_{p}|L\rangle\big)\Big]^{-1} }  \\ \lb{sc_25} &=&
\prod_{p=\pm}\lim_{N_{L}\rightarrow\infty}\prod_{L=1}^{N_{L}}d\bar{\chi}_{L}^{(p)}\;d\chi_{L}^{(p)}\;\;
\Big[\det\Big(\int d^{4}\!x_{p}\langle L|x_{p}\rangle\langle x_{p}|K\rangle\Big)\Big]^{-1}  \\ \no &=&
\prod_{p=\pm}\lim_{N_{L}\rightarrow\infty}\prod_{L=1}^{N_{L}}d\bar{\chi}_{L}^{(p)}\;d\chi_{L}^{(p)}\;\;
\Big[\det\big(\delta_{L,K}\big)\Big]^{-1}  =
\prod_{p=\pm}\lim_{N_{L}\rightarrow\infty}\prod_{L=1}^{N_{L}}d\bar{\chi}_{L}^{(p)}\;d\chi_{L}^{(p)}\;.
\eeq
We apply the axial transformations (\ref{sc_1}-\ref{sc_7},\ref{sc_26},\ref{sc_27}) from \(\psi(x_{p})\), \(\bar{\psi}(x_{p})\),
(\(\chi_{L}^{(p)}\), \(\bar{\chi}_{L}^{(p)}\)) to \(\psi\ppr(x_{p})\), \(\bar{\psi}\ppr(x_{p})\) (\ref{sc_28},\ref{sc_30})
and, respectively, to the transformed
Dirac spinors \(\chi_{L}^{\bprime(p)}\), \(\bar{\chi}_{L}^{\bprime(p)}\) and obtain the corresponding first order variation
\(\delta\alpha(x_{p})\) between \(\chi_{L}^{\bprime(p)}\), \(\bar{\chi}_{L}^{\bprime(p)}\) and
\(\chi_{L}^{(p)}\), \(\bar{\chi}_{L}^{(p)}\) (\ref{sc_29},\ref{sc_31})
which has to be substituted in the integration measure (\ref{sc_32}). This
specifies the Jacobian for the change from \(\chi_{L}^{\bprime(p)}\), \(\bar{\chi}_{L}^{\bprime(p)}\) to
\(\chi_{L}^{(p)}\), \(\bar{\chi}_{L}^{(p)}\), and according to relations (\ref{sc_17}-\ref{sc_24}) and especially (\ref{sc_25}),
also the change (\ref{sc_33}) with the Jacobian of (\ref{sc_32}) for the transformation from \(\psi\ppr(x_{p})\), \(\bar{\psi}\ppr(x_{p})\)
to \(\psi(x_{p})\), \(\bar{\psi}(x_{p})\)
\beq\lb{sc_26}
\psi\ppr(x_{p}) &=&\psi(x_{p})-\im\:\hat{\mfrak{t}}_{0}\:\hat{\gamma}_{5}\:\psi(x_{p})\;\delta\alpha(x_{p}) \;; \\  \lb{sc_27}
\bar{\psi}\ppr(x_{p}) &=& \bar{\psi}(x_{p})-\im\:\bar{\psi}(x_{p})\;\hat{\mfrak{t}}_{0}\:\hat{\gamma}_{5}\;\delta\alpha(x_{p})\;; \\ \lb{sc_28}
\psi\ppr(x_{p}) &=&\sum_{L}\varphi_{L}(x_{p})\;\chi_{L}^{\bprime(p)}=\sum_{L}\varphi_{L}(x_{p})\;\chi_{L}^{(p)}-
\im\;\delta\alpha(x_{p})\;\hat{\mfrak{t}}_{0}\:\hat{\gamma}_{5}\sum_{L}\varphi_{L}(x_{p})\;\chi_{L}^{(p)} \;; \\ \lb{sc_29}
\chi_{L}^{\bprime(p)}&=&\chi_{L}^{(p)}-\im\sum_{K}\bigg(\int d^{4}\!x_{p}\;
\varphi_{L}^{\dag}(x_{p})\;\hat{\mfrak{t}}_{0}\;\hat{\gamma}_{5}\;\varphi_{K}(x_{p})\;
\delta\alpha(x_{p})\bigg)\;\chi_{K}^{(p)} \;; \\  \lb{sc_30}
\bar{\psi}\ppr(x_{p}) &=&\sum_{L}\bar{\chi}_{L}^{\bprime(p)}\;\varphi_{L}^{\dag}(x_{p})=
\sum_{L}\bar{\chi}_{L}^{(p)}\;\varphi_{L}^{\dag}(x_{p})-\im\;\delta\alpha(x_{p})\sum_{L}\bar{\chi}_{L}^{(p)}\;\varphi_{L}^{\dag}(x_{p})\;\hat{\mfrak{t}}_{0}\;\hat{\gamma}_{5} \;; \\  \lb{sc_31}
\bar{\chi}_{L}^{\bprime(p)} &=& \bar{\chi}_{L}^{(p)}-\im\sum_{K}\bar{\chi}_{K}^{(p)}
\bigg(\int d^{4}\!x_{p}\;\varphi_{K}^{\dag}(x_{p})\;\hat{\mfrak{t}}_{0}\;\hat{\gamma}_{5}\;\varphi_{L}(x_{p})\;
\delta\alpha(x_{p})\bigg)\;;   \\    \lb{sc_32}
\prod_{p=\pm}\prod_{L=1}^{N_{L}}d\bar{\chi}_{L}^{\bprime(p)}\;d\chi_{L}^{\bprime(p)} &=& \prod_{p=\pm}
\prod_{L=1}^{N_{L}}d\bar{\chi}_{L}^{(p)}\;d\chi_{L}^{(p)}\;\times  \\ \no &\times&
\det\bigg[\delta_{L,K}\;\overbrace{\delta_{f,m,r;g,n,s}}^{\delta_{M;N}}-\im\bigg(\int d^{4}\!x_{p}\;
\delta\alpha(x_{p})\;\varphi_{L}^{\dag}(x_{p})\;
\hat{\mfrak{t}}_{0,fg}\;(\hat{\gamma}_{5})_{mn}\;
\delta_{rs}\;\varphi_{K}(x_{p})\bigg)\Bigg]^{-2}\;;    \\      \lb{sc_33}
d[\psi^{\bprime\dag}(x_{p})]\;d[\psi\ppr(x_{p})] &\stackrel{\det(\hat{\beta})}{=}& d[\bar{\psi}\ppr(x_{p})]\;d[\psi\ppr(x_{p})]
=\prod_{p=\pm}\lim_{N_{L}\rightarrow\infty}d\bar{\chi}_{L}^{\bprime(p)}\;d\chi_{L}^{\bprime(p)} =  \\ \no \lefteqn{\hspace*{-3.7cm}=
\prod_{p=\pm}\bigg(\lim_{N_{L}\rightarrow\infty}\prod_{L=1}^{N_{L}}d\bar{\chi}_{L}^{(p)}\;d\chi_{L}^{(p)}\bigg)
\exp\bigg\{2\im\lim_{N_{L}\rightarrow\infty}\sum_{L=1}^{N_{L}}\int d^{4}\!x_{p}\;\delta\alpha(x_{p})\;
\trfgamc\Big[\varphi_{L}^{\dag}(x_{p})\;\hat{\mfrak{t}}_{0}\;\hat{\gamma}_{5}\;\varphi_{L}(x_{p})\Big]\bigg\} = } \\ \no &=&
d[\bar{\psi}(x_{p})]\;d[\psi(x_{p})]\;\;\prod_{p=\pm}\mfrak{J}_{p} \;.
\eeq
We separate the Jacobian \(\mfrak{J}_{p}\) (on the time contour '\(p=\pm\)') (\ref{sc_33},\ref{sc_34}) from the integration variables
and have to examine a trace relation of the complex-valued, scalar eigenfunctions \(\varphi_{L}(x_{p})\) of the
gauge invariant derivative \(\feynD_{p}\)
\beq \lb{sc_34}
\mfrak{J}_{p}&=&\exp\Big\{2\im\,'\mathsf{exp}'\Big\}=\exp\bigg\{2\im\lim_{N_{L}\rightarrow\infty}\sum_{L=1}^{N_{L}}\int d^{4}\!x_{p}\;\delta\alpha(x_{p})\;
\trfgamc\Big[\varphi_{L}^{\dag}(x_{p})\;\hat{\mfrak{t}}_{0}\;\hat{\gamma}_{5}\;\varphi_{L}(x_{p})\Big]\bigg\}\;.
\eeq
Apparently, the phase '\('\mathsf{exp}'\)' within the Jacobian \(\mfrak{J}_{p}\) (\ref{sc_34}) diverges due to the trace operations
of the eigenfunctions \(\varphi_{L}(x_{p})\); therefore, one transforms to the momentum representation
\(\wt{\varphi}_{L}(k_{p})\), \(\wt{\varphi}_{L}^{\dag}(k_{p}\ppr)\) (\ref{sc_36}) of \(\varphi_{L}(x_{p})\)
\(\varphi_{L}^{\dag}(x_{p})\) and introduces a gauge invariant cut-off regulator
\(\exp\{\feynD_{p}^{2}/M^{2}\}\) of the original operator \(\feynD_{p}\) (\ref{sc_17}-\ref{sc_24}) for the eigenfunctions
\(\varphi_{L}(x_{p})\), modified by the 'regulating' mass parameter '\(M\)'
\beq  \lb{sc_35}
\lefteqn{'\mathsf{exp}'=\lim_{N_{L}\rightarrow\infty}\sum_{L=1}^{N_{L}}\int d^{4}\!x_{p}\;\delta\alpha(x_{p})
\trfgamc\Big[\varphi_{L}^{\dag}(x_{p})\;\hat{\mfrak{t}}_{0}\;\hat{\gamma}_{5}\;\varphi_{L}(x_{p})\Big] = }  \\ \no &=&
\lim_{M^{2}\rightarrow\infty}\sum_{L=1}^{\infty}\int d^{4}\!x_{p}\frac{d^{4}\!k_{p}}{(2\pi)^{4}}
\frac{d^{4}\!k_{p}\ppr}{(2\pi)^{4}}\;\delta\alpha(x_{p})\;
\trfgamc\Big[\wt{\varphi}_{L}^{\dag}(k_{p}\ppr)\:e^{-\im\:k_{p}\ppr\cdot x_{p}}\:\hat{\mfrak{t}}_{0}\:\hat{\gamma}_{5}\:
\exp\big\{\feynD_{p}^{2}/M^{2}\big\}\:e^{\im\:k_{p}\cdot x_{p}}\:\wt{\varphi}_{L}(k_{p})\Big] \;;  \\  \lb{sc_36}  &&
\sum_{L=1}^{\infty}\wt{\varphi}_{L}(k_{p})\;\wt{\varphi}_{L}^{\dag}(k_{p}\ppr)=(2\pi)^{4}\;\delta^{(4)}(k_{p}-k_{p}\ppr)\;\;.
\eeq
Using the completeness relation (\ref{sc_36}) for (\ref{sc_35}), we have exchanged the eigenfunctions \(\varphi_{L}(x_{p})\),
\(\varphi_{L}^{\dag}(x_{p})\) by plane waves \(e^{\im\:k_{p}\cdot x_{p}}\), \(e^{-\im\:k_{p}\cdot x_{p}}\) and obtain the
trace in momentum representation where the divergence is reduced by the operator \(\exp\{\feynD_{p}^{2}/M^{2}\}\)
with mass parameter \(M\)
\beq \lb{sc_37}
'\mathsf{exp}'&=&
\lim_{M^{2}\rightarrow\infty}\int d^{4}\!x_{p}\frac{d^{4}\!k_{p}}{(2\pi)^{4}}\;
\delta\alpha(x_{p})\trfgamc\Big[e^{-\im\:k_{p}\cdot x_{p}}\:\hat{\mfrak{t}}_{0}\:\hat{\gamma}_{5}\:
\exp\big\{\feynD_{p}^{2}/M^{2}\}\:e^{\im\:k_{p}\cdot x_{p}}\Big]  \;;  \\ \lb{sc_38}
\feynD_{p}^{2}&=&\hat{\gamma}^{\mu}\:\hat{\gamma}^{\nu}\:\hat{D}_{p,\mu}\:\hat{D}_{p,\nu}=
\hat{D}_{p,\mu}\hat{D}_{p}^{\mu}+\frac{1}{2}\:\hat{\gamma}^{\mu}\:\hat{\gamma}^{\nu}\:\big[\hat{D}_{p,\mu}\;\boldsymbol{,}\;\hat{D}_{p,\nu}\big]_{-}  \\ \no &=&
\hat{D}_{p,\mu}\hat{D}_{p}^{\mu}-\frac{\im}{2}\:\hat{\gamma}^{\mu}\:\hat{\gamma}^{\nu}\:\hat{F}_{\mu\nu}(x_{p})\;;\hspace*{0.3cm}
\hat{F}_{\mu\nu}(x_{p})=\hat{t}_{\alpha}\;\hat{F}_{\alpha;\mu\nu}(x_{p})  \;.
\eeq
In the case of a true vectorial \(U_{V}(1)\) transformation without '\(\hat{\gamma}_{5}\)' matrix, the following relations
(\ref{sc_41},\ref{sc_45}) do not apply so that the derived anomaly from the change of the integration measure is caused by the
axial '\(\hat{\gamma}_{5}\)' property of the transformation. As we insert Eq. (\ref{sc_38}) for the cut-off regulator into (\ref{sc_37}),
in order to obtain Eq. (\ref{sc_39}), and as we consider relations (\ref{sc_40},\ref{sc_41}), we can expand the exponent of (\ref{sc_39})
with the field strength tensor \(\hat{F}_{\mu\nu}(x_{p})\) up to
quadratic order in \((1/M^{2})\) (\ref{sc_42},\ref{sc_43}) as the only remaining term  of the Gaussian integrations in
the limit \(M^{2}\rightarrow\infty\)
\beq\lb{sc_39}
'\mathsf{exp}'  &=&
\lim_{M^{2}\rightarrow\infty}\int d^{4}\!x_{p}\frac{d^{4}\!k_{p}}{(2\pi)^{4}}\;
\delta\alpha(x_{p}) \;\times  \\  \no &\times&
\trfgamc\Big[e^{-\im\:k_{p}\cdot x_{p}}\:\hat{\mfrak{t}}_{0}\:\hat{\gamma}_{5}\:
\exp\Big\{\Big(\hat{D}_{p,\mu}\hat{D}_{P}^{\mu}-\frac{\im}{2}\hat{\gamma}^{\mu}\hat{\gamma}^{\nu}\:\hat{F}_{\mu\nu}(x_{p})\Big)\Big/M^{2}\Big\}\;
e^{\im\:k_{p}\cdot x_{p}}\Big] \;;   \\ \lb{sc_40}
f\big(\hat{\pp}_{p,\mu}\big)\;e^{\im\:k_{p}\cdot x_{p}}\hspace*{-0.2cm}&=&e^{\im\:k_{p}\cdot x_{p}}\;
f\big(\hat{\pp}_{p,\mu}+\im\:k_{p,\mu}\big)\;; \\ \lb{sc_41}
\trgam\big[\hat{\gamma}_{5}\big]=0 &;&
\hspace*{0.3cm}\trgam\big[\hat{\gamma}_{5}\:\hat{\gamma}_{\mu}\:\hat{\gamma}_{\nu}\big]=0\;;\hspace*{0.3cm}
\trgam\big[\hat{\gamma}_{5}\:\hat{\gamma}_{\mu_{1}}\:\ldots\:\hat{\gamma}_{\mu_{n}}\big]=0\;\;\mbox{ , (n odd) }\;;  \\  \lb{sc_42}
'\mathsf{exp}'&=&\lim_{M^{2}\rightarrow\infty}\int\frac{d^{4}\!k_{p}}{(2\pi)^{4}}\;
\exp\{-k_{p,\mu}\:k_{p}^{\mu}/M^{2}\} \;\times  \\ \no &\times&
\int d^{4}\!x_{p}\;\delta\alpha(x_{p})\;\trfgamc\bigg[\hat{\mfrak{t}}_{0}\;\hat{\gamma}_{5}\,\frac{1}{2!}\,
\bigg(-\frac{\im}{2\,M^{2}}\hat{\gamma}^{\mu}\:\hat{\gamma}^{\nu}\:\hat{F}_{\mu\nu}(x_{p})\bigg)
\bigg(-\frac{\im}{2\,M^{2}}\hat{\gamma}^{\kappa}\:\hat{\gamma}^{\lambda}\:\hat{F}_{\kappa\lambda}(x_{p})\bigg)\bigg]\;;  \\  \lb{sc_43}
'\mathsf{exp}'&=&\lim_{M^{2}\rightarrow\infty}\int\frac{d^{4}\!k_{p}}{(2\pi)^{4}}\;\bigg(-\frac{1}{8\,M^{4}}\;
\exp\Big\{-\big(\vec{k}_{p}\cdot \vec{k}_{p}-(k_{p}^{0})^{2}\big)\big/M^{2}\Big\} \;\times  \\ \no &\times&
\int d^{4}\!x_{p}\;\delta\alpha(x_{p})\;
\trgam\Big[\hat{\gamma}_{5}\:\hat{\gamma}^{\mu}\:\hat{\gamma}^{\nu}\:\hat{\gamma}^{\kappa}\:\hat{\gamma}^{\lambda}\Big]\;
\trfc\Big[\hat{\mfrak{t}}_{0}\;\hat{F}_{\mu\nu}(x_{p})\:\hat{F}_{\kappa\lambda}(x_{p})\Big]\;.
\eeq
The transformation (\ref{sc_44}) to Euclidean integration variables causes the 'instanton' properties (instead of solitons)
of the derived BCS-Hopf invariant in section \ref{s52} and allows to perform the Gaussian integrations. The total trace in (\ref{sc_42})
splits into a trace of Dirac gamma matrices with the peculiar axial '\(\hat{\gamma}_{5}\)' matrix and into a trace of
isospin-(flavour-) and colour matrix degrees of freedom (\ref{sc_43}) where we apply the particular trace relation (\ref{sc_45}) of Dirac
matrices \(\hat{\gamma}^{\mu}\) with the axial '\(\hat{\gamma}_{5}\)' matrix, resulting into the anti-symmetric
Levi-Civita symbol (\ref{sc_45}). Using these properties, one finally achieves
the chiral anomaly (\ref{sc_47},\ref{sc_48}) from insertion into the
Jacobian (\ref{sc_34}) and into the integration measure (\ref{sc_33})
\beq\lb{sc_44}
k_{p}^{0} &\rightarrow&-\im\:\omega_{p} \;\;\;; \\  \lb{sc_45}
\trgam\Big[\hat{\gamma}_{5}\:\hat{\gamma}^{\mu}\:\hat{\gamma}^{\nu}\:\hat{\gamma}^{\kappa}\:\hat{\gamma}^{\lambda}\Big]&=&4\:\im\;
\ve^{\mu\nu\kappa\lambda}\;;\hspace*{0.3cm}\ve^{0123}=+1\;\;\;;   \\ \lb{sc_46}
'\mathsf{exp}' &=&\im\overbrace{\int\frac{d^{3}\!k_{p}}{(2\pi)^{4}}\:\frac{d\omega_{p}}{8\,M^{4}}\;
\exp\Big\{-\big(\vec{k}_{p}\cdot \vec{k}_{p}+
\omega_{p}^{2}\big)\big/M^{2}\Big\}}^{(\sqrt{\pi}\,M)^{4}\cdot (2\pi)^{-4}\cdot (8^{-1}\,M^{-4})
=(128\,\pi^{2})^{-1}} \;\times  \\ \no &\times&
\int d^{4}\!x_{p}\;\delta\alpha(x_{p})\;4\:\im\;\ve^{\mu\nu\kappa\lambda}\;
\trfc\Big[\hat{\mfrak{t}}_{0}\;\hat{F}_{\mu\nu}(x_{p})\:\hat{F}_{\kappa\lambda}(x_{p})\Big] \;;  \\  \lb{sc_47}
'\mathsf{exp}' &=& -\frac{1}{32\:\pi^{2}}
\int d^{4}\!x_{p}\;\delta\alpha(x_{p})\;\ve^{\mu\nu\kappa\lambda}\;
\trfc\Big[\hat{\mfrak{t}}_{0}\;\hat{F}_{\mu\nu}(x_{p})\:\hat{F}_{\kappa\lambda}(x_{p})\Big]\;;  \\  \lb{sc_48}
\prod_{p=\pm}d[\bar{\psi}(x_{p})]\;d[\psi(x_{p})] &=&
\prod_{p=\pm}d[\bar{\psi}\ppr(x_{p})]\;d[\psi\ppr(x_{p})]\;\times  \\ \no &\times&
\exp\bigg\{\frac{\im}{16\:\pi^{2}}\int d^{4}\!x_{p}\;\eta_{p}\;\ve^{\kappa\lambda\mu\nu}
\trfc\Big[\hat{\mfrak{t}}_{0}\;\hat{F}_{\kappa\lambda}(x_{p})\;\hat{F}_{\mu\nu}(x_{p})\Big]\;\delta\alpha(x_{p})\bigg\}\;.
\eeq
As we combine the transformation of the actions \(\mscr{A}[\psi,\hat{D}_{\mu}\psi,\hat{F}]\) and
\(\mscr{A}_{S}[\hat{\mscr{J}},J_{\psi},\hat{J}_{\psi\psi},\hat{\mfrak{j}}^{(\hat{F})}]\) (\ref{sc_14}) in the previous subsection \ref{sc1}
with the transformation of the integration measure, we accomplish the total axial current relation with chiral anomaly given as the first
term on the right-hand side (in the first line of (\ref{sc_49})) which even remains in the massless limit (last term in the first line
of (\ref{sc_49})). The other terms in lines two to four of Eq. (\ref{sc_49}) follow from the source action
\(\mscr{A}_{S}[\hat{\mscr{J}},J_{\psi},\hat{J}_{\psi\psi},\hat{\mfrak{j}}^{(\hat{F})}]\) containing the symmetry
breaking, (odd-valued) fields \(j_{\psi;M}(x_{p})\),  \(j_{\psi;M}\pdag(x_{p})\) and anti-symmetric, (even-valued) matrices
\(\hat{j}_{\psi\psi;M;N}(x_{p})\), \(\hat{j}_{\psi\psi;M;N}\pdag(x_{p})\)
\beq\no
\lefteqn{\boldsymbol{\hat{\pp}_{p,\mu}\Big(\bar{\psi}(x_{p})\:\hat{\gamma}^{\mu}\:\hat{\gamma}_{5}\:\hat{\mfrak{t}}_{0}\:\psi(x_{p})\Big)=
-\frac{\ve^{\kappa\lambda\mu\nu}}{16\:\pi^{2}}\;
\trfc\Big[\hat{\mfrak{t}}_{0}\;\hat{F}_{\kappa\lambda}(x_{p})\;\hat{F}_{\mu\nu}(x_{p})\Big]  +
\bar{\psi}(x_{p})\:\hat{\gamma}_{5}\:\big\{\hat{\mfrak{t}}_{0}\:\boldsymbol{,}\:\hat{m}\big\}_{+}\:\psi(x_{p})   }  +  } \\ \no &+&
j_{\psi}^{\dag}(x_{p})\:\hat{\gamma}_{5}\:\hat{\mfrak{t}}_{0}\:\psi(x_{p})-\psi^{\dag}(x_{p})\:\hat{\gamma}_{5}\:\hat{\mfrak{t}}_{0}\:
j_{\psi}(x_{p}) +  \frac{1}{2}
\bigg[\psi^{T}(x_{p})\Big(\hat{j}_{\psi\psi}^{\dag}(x_{p})\:\hat{\gamma}_{5}\:\hat{\mfrak{t}}_{0}+\hat{\gamma}_{5}\:\hat{\mfrak{t}}_{0}^{T}\:\hat{j}_{\psi\psi}^{\dag}(x_{p})\Big)\psi(x_{p}) + \\ \lb{sc_49} &-&
\psi^{\dag}(x_{p})\Big(\hat{j}_{\psi\psi}(x_{p})\:\hat{\gamma}_{5}\:\hat{\mfrak{t}}_{0}^{*}+\hat{\gamma}_{5}\:\hat{\mfrak{t}}_{0}\:\hat{j}_{\psi\psi}(x_{p})\Big)\psi^{*}(x_{p})\bigg] + \frac{1}{2}
\int_{C}d^{4}\!y_{q} \times \\ \no &\times&
\Bigg[\Psi^{\dag}(y_{q},x_{p})\:\hat{\mscr{J}}(y_{q},x_{p})\bigg(\bea{cc}\hat{\gamma}_{5}\:\hat{\mfrak{t}}_{0} & \\ & -\hat{\gamma}_{5}\:\hat{\mfrak{t}}_{0}^{*}\eea\bigg)\Psi(x_{p})+
\Psi^{\dag}(x_{p})\bigg(\bea{cc} -\hat{\gamma}_{5}\:\hat{\mfrak{t}}_{0} & \\ & \hat{\gamma}_{5}\:\hat{\mfrak{t}}_{0}^{T} \eea\bigg)\hat{\mscr{J}}(x_{p},y_{q})\:\Psi(y_{q})\Bigg]\;.
\eeq

\section{Gradient expansion to an effective Lagrangian with nontrivial topology} \lb{sd}

\subsection{Expansion in the anomalous doubled Hilbert space} \lb{sd1}

The original path integral \(Z[\hat{\mscr{J}},J_{\psi},\hat{J}_{\psi\psi},\hat{\mfrak{j}}^{(\hat{F})}]\) (\ref{s2_25}-\ref{s2_27})
has been changed from matter and gauge fields to self-energies as the appropriate integration variables through subsequent HST's.
Furthermore, the path integral (\ref{s3_63})
has been separated by a coset transformation into block diagonal self-energy densities and BCS-terms which are
the remaining, most important path field integration variables of the final transformed functional (\ref{s3_116}). Moreover, we
have split the gauge field degrees of freedom and scalar quark self-energy densities from the BCS-degrees of freedom by using
a background functional averaging with path integral (\ref{s3_59}). The exact expression (\ref{s3_116}) with background functional
(\ref{s3_59}) is approximated by factorization of the averaging process (\ref{s3_59}) so that the background functional (\ref{s3_59})
acts with its averaging (\ref{sd_2}) for the potential term \(\feynV(x_{p})\) (\ref{s3_60}) individually on the actions
\(\langle\mscr{A}_{DET}[\hat{T},\feynV;\hat{\mscr{J}}]\rangle_{\feynbv}\),
\(\langle\mscr{A}_{J_{\psi}}[\hat{T},\feynV;\hat{\mscr{J}}]\rangle_{\feynbv}\) in the exponents and the BCS-source functional
\(\langle Z_{\hat{J}_{\psi\psi}}[\hat{T}]\rangle_{\feynbv}\)
\beq \lb{sd_1}
\hspace*{-0.6cm}Z[\hat{\mscr{J}},J_{\psi},\hat{J}_{\psi\psi},\hat{\mfrak{j}}^{(\hat{F})}] \hspace*{-0.3cm}&\approx &\hspace*{-0.5cm}
\int d[\hat{T}^{-1}(x_{p})\;d\hat{T}(x_{p})]\;\;\Big\langle Z_{\hat{J}_{\psi\psi}}[\hat{T}]\Big\rangle_{\feynBV}\;
\exp\Big\{\Big\langle \mscr{A}_{DET}[\hat{T},\feynV;\hat{\mscr{J}}]\Big\rangle_{\feynBV}\Big\} \;
\exp\Big\{\im\Big\langle \mscr{A}_{J_{\psi}}[\hat{T},\feynV;\hat{\mscr{J}}]\Big\rangle_{\feynBV}\Big\}_{\mbox{;}}  \\  \lb{sd_2}
\Big\langle\Big(\ldots\Big)\Big\rangle_{\feynBV} &:=&
\boldsymbol{\bigg\langle Z\Big[\feynV(x_{p});\hat{\mfrak{S}}^{(\hat{F})},s_{\alpha},\hat{\mfrak{B}}_{\hat{F}},\hat{\mfrak{b}}^{(\hat{F})},
\hat{\mscr{U}}_{\hat{F}},\hat{\mfrak{v}}_{\hat{F}};\sigma_{D};\hat{\mfrak{j}}^{(\hat{F})};
\mbox{\bf Eq. (\ref{s3_59})}\Big]\;\times \Big(\ldots\Big)\bigg\rangle}\;.
\eeq
Apart from the derivative- '\(\feynd_{p}\)', mass- '\(\hat{m}\)' and '\(-\im\:\hat{\ve}_{p}\)'-terms,
the gradient operator \(\Delta\!\hat{\mscr{H}}_{N;M}^{ba}(y_{q},x_{p})\) (\ref{sd_3}) in (\ref{s3_111})
\(\hat{\mscr{O}}_{N;M}^{ba}(y_{q},x_{p})\) consists of the potential
\(\;\feynV(x_{p})\) (\ref{sd_4},\ref{sd_5},\ref{s3_60})
composed of the gauge and quark self-energy density degrees of freedom within the path integral (\ref{s3_59}).
The transposition  of the '22' part \(\hat{H}^{T}(x_{p})\) (\ref{sd_5})
within the anomalous-doubled, one-particle Hamiltonian \(\hat{\mscr{H}}_{N;M}^{ba}(y_{q},x_{p})\) (\ref{sd_3})
involves the internal symmetry spaces as the
isopspin-(flavour-) matrices, the \(4\times 4\) Dirac gamma matrices and the gauge field generators \(\hat{t}_{\alpha}\),
and also the contour spacetime derivative '\((\hat{\pp}_{p,\mu})^{T}=-\hat{\pp}_{p,\mu}\)'
which results into an additional minus sign. Corresponding to notations and definitions of gamma-matrices \cite{wein1}, one obtains
for the gradient operator \(\Delta\!\hat{\mscr{H}}_{N;M}^{ba}(y_{q},x_{p})\) with the coset matrices \(\hat{T}^{-1}(x_{p})\), \(\hat{T}(x_{p})\)
for the remaining BCS-degrees of freedom following equations
\beq \lb{sd_3}
\lefteqn{\Delta\!\hat{\mscr{H}}_{N;M}^{ba}(y_{q},x_{p})=\Big(\hat{T}^{-1}\hat{\mscr{H}}\hat{T}-\hat{\mscr{H}}\Big)_{N;M}^{ba}(y_{q},x_{p})=
\delta^{(4)}(y_{q}-x_{p})\;\eta_{q}\;\delta_{qp}\times }  \\  \no &\times&
\Bigg[\hat{T}_{N;N\ppr}^{-1;bb\ppr}(x_{p})\Bigg(\bea{cc} \hat{H}_{N\ppr;M\ppr}(x_{p}) & 0 \\ 0 &
\hat{H}_{N\ppr;M\ppr}^{T}(x_{p}) \eea\Bigg)_{N\ppr;M\ppr}^{b\ppr a\ppr}\hat{T}_{M\ppr;M}^{a\ppr a}(x_{p})-
\Bigg(\bea{cc} \hat{H}_{N;M}(x_{p}) & 0 \\ 0 &
\hat{H}_{N;M}^{T}(x_{p}) \eea\Bigg)_{N;M}^{ba} \Bigg]_{\mbox{;}}
\eeq
\beq \lb{sd_4}
\hat{H}(x_{p}) &=& \Big[\hat{\beta}\Big(\feynd_{p}+\im\:\feynV(x_{p})-\im\:\hat{\ve}_{p}+\hat{m}\Big)\Big]\;;
\hspace*{0.3cm}\big(\hat{\ve}_{p}=\hat{\beta}\:\ve_{p}=\hat{\beta}\:\eta_{p}\:\ve_{+}\;;\;\;\;\ve_{+}>0\big)\;; \\ \no &=&
\hat{\beta}\:\hat{\gamma}^{\mu}\:\hat{\pp}_{p,\mu}-\im\:\ve_{p}\:\hat{1}_{N_{0}\times N_{0}}+
\hat{\beta}\,\big(\im\:\feynV(x_{p})+\hat{m}\big)\;;   \\   \lb{sd_5}
\hat{H}^{T}(x_{p}) &=& \big(\hat{\beta}\:\hat{\gamma}^{\mu}\big)^{T}
(-\hat{\pp}_{p,\mu})-\im\:\ve_{p}\:\hat{1}_{N_{0}\times N_{0}}+
\big[\hat{\beta}\,\big(\im\:\feynV(x_{p})+\hat{m}\big)\big]^{T}\;.
\eeq
As the gradient operators \(\hat{\beta}\:\feynd_{p}\) and its transposes \((\hat{\beta}\:\feynd_{p}\,)^{T}\) act onto the
coset matrices \(\hat{T}_{M\ppr;M}^{a\ppr a}(x_{p})\), \(\hat{T}_{N;N\ppr}^{-1;bb\ppr}(x_{p})\) to the right or left,
it has to be distinguished between saturated derivatives as \((\hat{\pp}_{p,\mu}\hat{T}(x_{p})\,)\) whose actions are
limited by outer braces (in the given case for a single coset matrix) and unsaturated derivative 'operators' acting
beyond the outer coset matrices further to the right and left onto other terms in the gradient expansion. Therefore,
we consider two parts \(\delta\hat{\mfrak{h}}(x_{p})=\hat{\mfrak{k}}(x_{p})+\delta\hat{\mfrak{V}}(x_{p})\) (\ref{sd_7},\ref{sd_8}) and
\(\delta\hat{\mscr{K}}^{\boldsymbol{\mu}}(x_{p})\:\boldsymbol{\hat{\pp}_{p,\mu}}\) (\ref{sd_9}) of the total gradient
operator \(\Delta\!\hat{\mscr{H}}\) (\ref{sd_3}-\ref{sd_6}) which consists of 'saturated' derivatives of coset matrices
\((\hat{\pp}_{p,\mu}\hat{T}(x_{p})\,)\) with additional potential matrix \(\hat{\mfrak{V}}(x_{p})\)
and 'unsaturated' gradient '{\it operators}' \(\boldsymbol{\hat{\pp}_{p,\mu}}\)
with Hamiltonian \(\delta\hat{\mscr{K}}^{\boldsymbol{\mu}}(x_{p})\) (\ref{sd_9}), respectively. This difference is
symbolized by boldface letters for the 'unsaturated' gradient '{\it operator}' \(\boldsymbol{\hat{\pp}_{p,\mu}}\)
in the following expansions. We also abbreviate the block diagonal anomalous-doubled potential matrix of the background
potential \(\mcal{V}_{\beta}^{\mu}(x_{p})\) (\ref{s3_60}) and the isospin- (flavour-) masses \(\hat{m}\), \(\hat{m}^{T}\)
by the additional symbol \(\hat{\mfrak{V}}_{N;M}^{ba}(x_{p})\) (\ref{sd_8}). This anomalous-doubled potential
\(\hat{\mfrak{V}}_{N;M}^{ba}(x_{p})\) (\ref{sd_8}) is only contained in the Hamiltonian part \(\delta\hat{\mfrak{h}}(x_{p})\)
(\ref{sd_7}) with saturated gradients and gives rise to effective coupling functions
\(\langle\mcal{V}_{\beta}^{\mu}(x_{p})\;\mcal{V}_{\gamma}^{\nu}(x_{p})\rangle_{\feynbv}\) from the averaging with
the background functional (\ref{s3_59},\ref{sd_2}). Note that the potential matrix
\(\hat{\mfrak{V}}(x_{p})=\mcal{V}_{\alpha;\mu}(x_{p})\;\hat{V}_{\alpha}^{\mu}\) of background field \(\mcal{V}_{\alpha;\mu}(x_{p})\)
is weighted by the coset matrices \(\hat{T}^{-1}(x_{p})\), \(\hat{T}(x_{p})\) so that one achieves the defined difference
\(\delta\hat{\mfrak{V}}(x_{p})=\hat{T}^{-1}(x_{p})\;\hat{\mfrak{V}}(x_{p})\;\hat{T}(x_{p})-\hat{\mfrak{V}}(x_{p})=
\mcal{V}_{\alpha;\mu}(x_{p})\;(\hat{T}^{-1}(x_{p})\:\hat{V}_{\alpha}^{\mu}\:\hat{T}(x_{p})-\hat{V}_{\alpha}^{\mu})\)
because of the original gradient term \(\Delta\!\hat{\mscr{H}}=(\hat{T}^{-1}\:\hat{\mscr{H}}\:\hat{T}-\hat{\mscr{H}})\) and
because of the non-commuting property \([\hat{T}(x_{p})\:,\:\hat{\mfrak{V}}(x_{p})\,]_{-}\neq 0\)
\beq  \lb{sd_6}
\Delta\!\hat{\mscr{H}}_{N;M}^{ba}(y_{q},x_{p}) &=&\delta^{(4)}(y_{q}-x_{p})\;\eta_{q}\;\delta_{qp}
\bigg[\delta\hat{\mfrak{h}}_{N;M}^{ba}(x_{p})+\delta\hat{\mscr{K}}_{N;M}^{\boldsymbol{\mu};ba}(x_{p})\;
\boldsymbol{\hat{\pp}_{p,\mu}}\bigg] \\ \no &=&\delta^{(4)}(y_{q}-x_{p})\;\eta_{q}\;\delta_{qp}
\bigg[\hat{\mfrak{k}}_{N;M}^{ba}(x_{p})+\delta\hat{\mfrak{V}}_{N;M}^{ba}(x_{p})+
\delta\hat{\mscr{K}}_{N;M}^{\boldsymbol{\mu};ba}(x_{p})\;\boldsymbol{\hat{\pp}_{p,\mu}}\bigg] \;; \\ \lb{sd_7}
\delta\hat{\mfrak{h}}_{N;M}^{ba}(x_{p})&=&
\hat{\mfrak{k}}_{N;M}^{ba}(x_{p})+\delta\hat{\mfrak{V}}_{N;M}^{ba}(x_{p}) \;;\;\;\rightarrow
\hat{\mfrak{k}}_{N;M}^{ba}(x_{p}) = \Big[\hat{T}^{-1}(x_{p})\:\hat{\beta}\:\hat{\gamma}^{\mu}\:\hat{S}\:
\big(\hat{\pp}_{p,\mu}\hat{T}(x_{p})\,\big)\Big]_{N;M}^{ba}  \;;  \\ \lb{sd_8}
\delta\hat{\mfrak{V}}_{N;M}^{ba}(x_{p}) &=& \hspace*{-0.3cm}
\Big[\hat{T}^{-1}(x_{p})\:\hat{\mfrak{V}}(x_{p})\:\hat{T}(x_{p})-
\hat{\mfrak{V}}(x_{p})\Big]_{N;M}^{ba}  = \mcal{V}_{\alpha;\mu}(x_{p})\;
\Big[\hat{T}^{-1}(x_{p})\;\hat{V}_{\alpha}^{\mu}\;\hat{T}(x_{p})-\hat{V}_{\alpha}^{\mu}\Big]_{N;M}^{ba} ;  \\ \no
\big[\hat{V}_{\alpha}^{\mu}\big]_{N;M}^{ba} &=& \hat{V}_{\alpha;N;M}^{\mu;ba} = \Bigg(\bea{cc}
\big[\hat{\beta}\,(\im\:\hat{\gamma}^{\mu}\:\hat{t}_{\alpha}+\hat{m})\,\big]_{N;M} & 0  \\ 0 &
\big[\hat{\beta}\,(\im\:\hat{\gamma}^{\mu}\:\hat{t}_{\alpha}+\hat{m})\,\big]_{N;M}^{T} \eea\Bigg)^{ba}\delta_{ab} \;; \\ \no
\hat{\mfrak{V}}_{N;M}^{ba}(x_{p}) &=&\delta_{ab}\Bigg(\bea{cc}
\big[\hat{\beta}\,(\im\:\feynV(x_{p})+\hat{m})\,\big]_{N;M} & 0 \\ 0 &
\big[\hat{\beta}\,(\im\:\feynV(x_{p})+\hat{m})\,\big]_{N;M}^{T}\eea\Bigg)^{ba} \\ \no &=&
\mcal{V}_{\alpha;\mu}(x_{p})\Bigg(\bea{cc}
\big[\hat{\beta}\,(\im\:\hat{\gamma}^{\mu}\:\hat{t}_{\alpha}+\hat{m})\,\big]_{N;M} & 0  \\ 0 &
\big[\hat{\beta}\,(\im\:\hat{\gamma}^{\mu}\:\hat{t}_{\alpha}+\hat{m})\,\big]_{N;M}^{T} \eea\Bigg)^{ba} =
\mcal{V}_{\alpha;\mu}(x_{p})\;\hat{V}_{\alpha;N;M}^{\mu;ba} \;;  \\ \lb{sd_9}
\delta\hat{\mscr{K}}_{N;M}^{\boldsymbol{\mu};ba}(x_{p}) &=&\Big(\hat{T}^{-1}(x_{p})\;
\hat{\beta}\;\hat{\gamma}^{\boldsymbol{\mu}}\;\hat{S}\;
\hat{T}(x_{p})\Big)_{N;M}^{ba}-\delta_{ab}\;\Big(\hat{\beta}\;\hat{\gamma}^{\boldsymbol{\mu}}\;\hat{S}\Big)_{N;M}^{ba}\;.
\eeq
Aside from the background potential matrix \(\hat{\mfrak{V}}(x_{p})\) (\ref{sd_8}), the anomalous-doubled one-particle
operator \(\hat{\mscr{H}}(x_{p})\) (\ref{sd_10}) has the diagonal '\(-\im\:\ve_{p}\:\hat{1}_{2N_{0}\times 2N_{0}}\)' non-hermitian
part which determines the analytic behaviour of Green functions \(\hat{G}^{(0)}\), \(\hat{g}^{(0)}\), \([\hat{g}^{(0)}]^{T}\)
propagating on the non-equilibrium time contour. The anomalous-doubled Green function \(\hat{G}^{(0)}\) (\ref{sd_11}),
consisting of the block diagonal parts \(\hat{g}^{(0)}\), \([\hat{g}^{(0)}]^{T}\) (\ref{sd_12},\ref{sd_13}),
differ from the simple inverses \(\hat{\mscr{H}}^{-1}\), \([\hat{H}(x_{p})]^{-1}\), \([\hat{H}^{T}(x_{p})]^{-1}\)
of the operators (\ref{sd_4},\ref{sd_5}) by the additional averaging with the background functional (\ref{s3_59},\ref{sd_2})
for the potential \(\feynV(x_{p})\) (\ref{s3_60}).
This averaging procedure for products of several inverse operators with \(\hat{\mscr{H}}^{-1}\)
is simplified to corresponding products with independent background averaging of single operators \(\hat{\mscr{H}}^{-1}\)
leading to factors of proper non-equilibrium Green functions. This means that we neglect any correlations between the
potentials \(\mcal{V}_{\beta}^{\mu}(x_{p})\) (\ref{s3_60}), \(\hat{\mfrak{V}}(x_{p})\) (\ref{sd_8}) in the factors of
inverses \(\langle\hat{\mscr{H}}^{-1}{\scrscr\times}\ldots{\scrscr\times}\hat{\mscr{H}}^{-1}\rangle_{\feynbv}\) and
simply reduce to \(\langle\hat{\mscr{H}}^{-1}\rangle_{\feynbv}{\scrscr\times}\ldots{\scrscr\times}\langle\hat{\mscr{H}}^{-1}\rangle_{\feynbv}=
\hat{G}^{(0)}{\scrscr\times}\ldots{\scrscr\times}\hat{G}^{(0)}\)
\beq \lb{sd_10}
\hat{\mscr{H}}(x_{p}) &=&\hat{\beta}\;\hat{\gamma}^{\mu}\;\hat{S}\;\boldsymbol{\hat{\pp}_{p,\mu}}+
\hat{\mfrak{V}}_{N;M}^{ba}(x_{p})-\im\:\ve_{p}\;\hat{1}_{2N_{0}\times 2N_{0}}\;;  \\  \lb{sd_11}
\hat{G}^{(0)} &=& \Big\langle\hat{\mscr{H}}^{-1}\Big\rangle_{\feynBV}=
\Bigg(\bea{cc}  \hat{g}^{(0)}  & 0  \\ 0 & \big[\hat{g}^{(0)}\big]^{T} \eea\Bigg)_{\mbox{;}}   \\   \lb{sd_12}
\hat{g}_{g,n,s;f,m,r}^{(0)} &=&\Big\langle \big[\hat{H}(x_{p})\big]^{-1}_{N;M}\Big\rangle_{\feynBV}  =
\Big\langle\big[\hat{\beta}\big(\feynd_{p}+\im\:\feynV(x_{p})+\hat{m}\big)-\im\:\ve_{p}\big]_{N;M}^{-1}\Big\rangle_{\feynBV}  \\ \no &\approx&\Big[\hat{\beta}\Big(\feynd_{p}+\im\:
\big\langle\feynV(x_{p})\big\rangle_{\feynBV}+\hat{m}\Big)-\im\:\ve_{p}\Big]_{N;M}^{-1} \;;  \\  \lb{sd_13}
\big[\hat{g}^{(0)}\big]_{g,n,s;f,m,r}^{T} &=&
\Big\langle \big[\hat{H}^{T}(x_{p})\big]^{-1}_{N;M}\Big\rangle_{\feynBV}  =
\Big\langle\big[\hat{\beta}\big(\feynd_{p}+\im\:\feynV(x_{p})+\hat{m}\big)-\im\:\ve_{p}\big]_{N;M}^{T;-1}\Big\rangle_{\feynBV} \\ \no
&\approx& \Big[\hat{\beta}\Big(\feynd_{p}+\im\:
\big\langle\feynV(x_{p})\big\rangle_{\feynBV}+\hat{m}\Big)-\im\:\ve_{p}\Big]_{N;M}^{T;-1} \;.
\eeq
Proceeding as in chapter 4 of \cite{mies1}, one has to determine the anomalous-doubled, averaged time contour Green function
\(\hat{G}^{(0)}=\langle\hat{\mscr{H}}^{-1}\rangle_{\feynbv}\) (\ref{sd_11}-\ref{sd_13})
with the transpose of the '11' block extended to the '22' block.
This can be accomplished by a saddle point approximation of the background functional (\ref{s3_59}) with its various gauge
field variables and quark self-energy densities. However, the imaginary parts of \(\hat{\beta}\:\feynV(x_{p})\) (\ref{s3_60}) or
\(\hat{\mfrak{V}}(x_{p})\) (\ref{sd_8}), resulting from a saddle point approximation, have to comply with the imaginary contour sign
'\(-\im\:\ve_{p}\)' for a stable propagation of coset matrices. We use the overview of the anomalous-doubled Hilbert space,
summarized in appendix \ref{sa}, and the definitions and notations of chapter 4 in \cite{mies1} so that the 'anti-unitary',
'anti-linear' '22' states accompany as extensions the original states in the '11' block. It has to be taken into account that
the square root of the determinant follows from integration over the bilinear anti-commuting fields \(\psi_{M}(x_{p})\) which are
doubled by their complex conjugates \(\psi_{M}^{*}(x_{p})\). Consequently a Hilbert space for \(\psi_{M}(x_{p})\) with
'ket' \(|\psi_{M}\rangle\) has also to be doubled by its 'dual' space \(\ovv{|\psi_{M}\rangle}=\langle\psi_{M}|\) the 'bra'.
The corresponding Hilbert space of spacetime variables has therefore also to comprise the anti-linear part
\(\ovv{|x_{p}\rangle}=\langle x_{p}|\) in its section \(a=2\)
\be \lb{sd_14}
|\widehat{x_{p}\rangle}^{a(=1,2)}=\Bigg(\bea{c} |x_{p}\rangle^{a=1} \\ |\ovv{x_{p}\rangle}^{a=2} \eea\Bigg)=
\Bigg(\bea{c} |x_{p}\rangle \\ \langle x_{p}| \eea\Bigg)^{a(=1,2)}_{\mbox{.}}
\ee
The application of rules for an anomalous-doubled Hilbert space, defined in appendix \ref{sa}, leads to the background averaged
matrix representations of Green functions in the '11' part and its transposed '22' part with the particular anti-linear
\(\ovv{\langle x_{p}}|\), \(|\ovv{y_{q}\rangle}\) spacetime states (\ref{sd_15}-\ref{sd_19}).
According to general properties of contour
time Green functions, one obtains generalized time contour Heaviside functions \(\theta_{pq}(x^{0}-y^{0})\) and
background averaged time development operators \(\langle\,\langle\vec{x}|\hat{\mfrak{U}}(x^{0},y^{0})|\vec{y}\rangle\,\rangle_{\feynbv}\).
The matrix representation of the latter time development operator is constructed from standard time path ordering
\(x^{0}>z^{0}>y^{0}\), denoted by the arrow \(\overleftarrow{\exp}\), whereas the generalized Heaviside function
\(\theta_{pq}(x^{0}-y^{0})\) is relevant for the additional appropriate ordering on the '{\it contour extended}' times.
Therefore, one always obtains an inverse propagation of the '22' block relative to the '11' block concerning contour extended
times of the Heaviside functions \(\theta_{pq}(x^{0}-y^{0})\) within the anomalous-doubled Green functions (\ref{sd_17})
\beq \lb{sd_15}
\langle x_{p}|\hat{g}^{(0)}|y_{q}\rangle &=&\Big\langle \langle x_{p}|\big[\hat{\beta}(\feynd_{p}+
\im\,\feynV(x_{p})+\hat{m})-\im\,\ve_{p}\big]^{-1}|y_{q}\rangle\Big\rangle_{\feynBV} \\ \no &=&
 \im\:\theta_{pq}(x^{0}-y^{0})\;
\left\langle\langle\vec{x}|\hat{\mfrak{U}}(x^{0},y^{0})\,|\vec{y}\rangle\right\rangle_{\feynBV} \;; \\  \lb{sd_16}
\ovv{\langle x_{p}}|\big[\hat{g}^{(0)}\big]^{T}|\ovv{y_{q}\rangle}  &=&\langle y_{q}|\hat{g}^{(0)}|x_{p}\rangle =
\im\:\theta_{qp}(y^{0}-x^{0})\;\left\langle\langle \vec{y}|\hat{\mfrak{U}}(y^{0},x^{0})\,|\vec{x}\rangle\right\rangle_{\feynBV} \;; \\
\lb{sd_17}
\ph{\,}^{a}\widehat{\langle x_{p}}|\hat{G}^{(0)}|\widehat{y_{q}\rangle}^{b} &=&\hspace*{-0.1cm}\delta_{ab}\;\im
\Bigg(\bea{cc}\theta_{pq}(x^{0}-y^{0})\;\left\langle\langle\vec{x}|\hat{\mfrak{U}}(x^{0},y^{0})\,|\vec{y}\rangle\right\rangle_{\feynBV} & 0 \\0  & \hspace*{-1.0cm}
\theta_{qp}(y^{0}-x^{0})\;\left\langle\langle\vec{y}|\hat{\mfrak{U}}(y^{0},x^{0})\,|\vec{x}\rangle\right\rangle_{\feynBV} \eea\Bigg);
\\  \lb{sd_18}  \left\langle\langle\vec{x}|\hat{\mfrak{U}}(x^{0},y^{0})\,|\vec{y}\rangle\right\rangle_{\feynBV} &=&
\bigg\langle\overleftarrow{\exp}\bigg\{-\im\int_{y^{0}}^{x^{0}}dz^{0}\;\Big[\hat{\beta}\big(\vec{\gamma}\cdot\vec{\pp}_{\vec{x}}+
\im\,\feynV(z^{0},\vec{x})+\hat{m}\big)-\im\:\ve_{+}\Big]\bigg\}\delta^{(3)}(\vec{x}-\vec{y})\bigg\rangle_{\feynBV}
\\ \no &\approx&
\overleftarrow{\exp}\bigg\{-\im\int_{y^{0}}^{x^{0}}dz^{0}\;\Big[\hat{\beta}\big(\vec{\gamma}\cdot\vec{\pp}_{\vec{x}}+
\im\;\big\langle\feynV(z^{0},\vec{x})\rangle_{\feynBV}+\hat{m}\big)-\im\:\ve_{+}\Big]\bigg\}\delta^{(3)}(\vec{x}-\vec{y})\;;
 \\ \lb{sd_19}
\delta(x^{0}-y^{0})\;\delta^{(3)}(\vec{x}-\vec{y}) &=&
\bigg\langle\Big[\hat{\beta}\big(\hat{\gamma}^{0}\:\hat{\pp}_{x^{0}}+
\vec{\gamma}\cdot\vec{\pp}_{\vec{x}}+\im\:\feynV(x^{0},\vec{x})+
\hat{m}\big)\Big]\hspace*{0.6cm}\langle\vec{x}|\hat{\mfrak{U}}(x^{0},y^{0})\,|\vec{y}\rangle\bigg\rangle_{\feynBV}  \\ \no &\approx&
\Big[\hat{\beta}\big(\hat{\gamma}^{0}\:\hat{\pp}_{x^{0}}+
\vec{\gamma}\cdot\vec{\pp}_{\vec{x}}+\im\:\Big\langle\feynV(x^{0},\vec{x})\Big\rangle_{\feynBV}+
\hat{m}\big)\Big]\hspace*{0.6cm}\Big\langle\langle\vec{x}|\hat{\mfrak{U}}(x^{0},y^{0})\,|\vec{y}\rangle\Big\rangle_{\feynBV}\;.
\eeq
We specify in relations (\ref{sd_20},\ref{sd_21}) the precise from of contour extended Heaviside functions \(\theta_{pq}(x^{0}-y^{0})\)
in terms of the standard Heaviside function \(\theta(x^{0}-y^{0})\) and conclude from the standard relation between
contour extended Heaviside functions the particular expression (\ref{sd_21}) between anomalous-doubled, block diagonal
time contour Green functions. The equation (\ref{sd_21}) also requires a block diagonal, anomalous-doubled
Heaviside function \(\Theta_{pq}^{ab}(x^{0}-y^{0})\) which we define in (\ref{sd_22}-\ref{sd_24})
\beq\no
\theta_{p=+,q=+}(x^{0}-y^{0}) &=&\theta(x^{0}-y^{0})=+1\;\;\;\mbox{for }\;x^{0}\geq y^{0} \;; \\ \no
\theta_{p=-,q=-}(x^{0}-y^{0}) &=&\theta(y^{0}-x^{0})=+1\;\;\;\mbox{for }\;y^{0}\geq x^{0} \;; \\  \lb{sd_20}
\theta_{p=-,q=+}(x^{0}-y^{0}) &\equiv& 1\;\;\;\mbox{for all }\;x^{0},\;y^{0}\;; \\  \no
\theta_{p=+,q=-}(x^{0}-y^{0}) &\equiv& 0\;\;\;\mbox{for all }\;x^{0},\;y^{0}\;; \\ \no
\theta(x^{0}-y^{0}) &=& +1\;\;(\mbox{for }\;x^{0}\geq y^{0})\;\;\mbox{ and }\;=0\;\;(\mbox{for }\;x^{0}<y^{0}) \;; \\ \no
\theta_{++}(x^{0}-y^{0}) + \theta_{--}(x^{0}-y^{0}+0_{+}) &=&
\underbrace{\theta_{-+}(x^{0}-y^{0})}_{\equiv 1}+
\underbrace{\theta_{+-}(x^{0}-y^{0})}_{\equiv 0} \;;   \\   \lb{sd_21}
\ph{\,}^{a}\widehat{\langle x_{+}}|\hat{G}^{(0)}|\widehat{y_{+}\rangle}^{a} +
\ph{\,}^{a}\widehat{\langle x_{-}}|\hat{G}^{(0)}|\widehat{y_{-}\rangle}^{a} &=&
\underbrace{\ph{\,}^{a}\widehat{\langle x_{-}}|\hat{G}^{(0)}|\widehat{y_{+}\rangle}^{a}}_{\neq 0\;
\mbox{\scz for }\;a=1\:, \mbox{\scz but }\;\equiv 0\;\mbox{\scz for }\;a=2} \hspace*{0.2cm}+\hspace*{0.2cm}
\underbrace{\ph{\,}^{a}\widehat{\langle x_{+}}|\hat{G}^{(0)}|\widehat{y_{-}\rangle}^{a}}_{\equiv 0\;
\mbox{\scz for }\;a=1\:, \mbox{\scz but }\;\neq 0\;\mbox{\scz for }\;a=2} \;;
\eeq
\beq   \lb{sd_22}
\widehat{\ph{\,}^{a}\langle x_{p}}|\hat{G}^{(0)}|\widehat{y_{q}\rangle}^{b} &=&\delta_{ab}\;\im\;
\Theta_{pq}^{ab}(x^{0}-y^{0})
\Bigg(\bea{cc}\left\langle\langle\vec{x}|\hat{\mfrak{U}}(x^{0},y^{0})\,|\vec{y}\rangle\right\rangle_{\feynBV} & 0 \\0  &
\left\langle\langle\vec{y}|\hat{\mfrak{U}}(y^{0},x^{0})\,|\vec{x}\rangle\right\rangle_{\feynBV} \eea\Bigg)\;;
 \\   \lb{sd_23}
\Theta_{pq}^{ab}(x^{0}-y^{0}) &=& \delta_{ab}\Bigg(\bea{cc}\theta_{pq}(x^{0}-y^{0}) & 0 \\
0 & \theta_{qp}(y^{0}-x^{0}) \eea\Bigg)^{ab} \;; \\  \lb{sd_24}
\Theta_{pp}^{ab}(x^{0}-y^{0}) &=&\delta_{ab}\Bigg(\bea{cc} \theta_{pp}(x^{0}-y^{0}) & 0 \\
0 & \theta_{pp}(y^{0}-x^{0}) \eea\Bigg) = \delta_{ab} \Bigg\{\bea{c}
\Bigg(\bea{cc} \theta(x^{0}-y^{0}) & 0 \\ 0 & \theta(y^{0}-x^{0}) \eea\Bigg)\;\;\;\mbox{for }\;p=+ \\
\Bigg(\bea{cc} \theta(y^{0}-x^{0}) & 0 \\ 0 & \theta(x^{0}-y^{0}) \eea\Bigg)\;\;\;\mbox{for }\;p=- \eea_{\mbox{.}}
\eeq
The anomalous-doubled Heaviside or time contour step function \(\Theta_{pq}^{ab}(x^{0}-y^{0})\) restricts the possible terms
in the gradient expansion of the action \(\langle\mscr{A}_{DET}[\hat{T},\feynV;\hat{\mscr{J}}]\rangle_{\feynbv}\) because
the trace operations also involve the contour extended traces of spacetime; in consequence one attains as the
remaining terms in the gradient expansion of \(\langle\mscr{A}_{DET}[\hat{T},\feynV;\hat{\mscr{J}}]\rangle_{\feynbv}\)
only those in which the anomalous-doubled time contour step functions \(\Theta_{pp}^{ab}(x^{0}-y^{0})\) do not result
into contradictory propagations concerning the time contour extended ordering of gradient terms \(\Delta\!\hat{\mscr{H}}(x_{p})\)
(e.g. \(\Theta_{pq}^{aa}(x^{0}-y^{0})\;\Theta_{qp}^{aa}(y^{0}-x^{0}) = 0\) !). This restriction of terms is missing
in the case of the gradient expansion of \(\langle\mscr{A}_{J_{\psi}}[\hat{T},\feynV;\hat{\mscr{J}}]\rangle_{\feynbv}\). In
continuation of principles for a gradient expansion, we state that an anomalous-doubled field \(\Psi_{M}^{a}(x_{p})\)
propagates with the block diagonal, doubled Green function
\(\ph{\,}^{a}\widehat{\langle x_{p}}|\hat{G}^{(0)}|\widehat{y_{q}\rangle}^{b}\) (\ref{sd_25})
with background potential \(\mcal{V}_{\beta}^{\mu}(x_{p})\) (\ref{s3_60}). This principle has to be used in the expansion of
\(\langle\mscr{A}_{J_{\psi}}[\hat{T},\feynV;\hat{\mscr{J}}]\rangle_{\feynbv}\) where one starts to propagate with the source
field \(J_{\psi;M}^{a}(x_{p})\) on the right-hand side of the action for a coherent wavefunction. It replaces the
wavefunction \(\Psi_{M}^{a}(x_{p})\) in (\ref{sd_25}). We generalize rule (\ref{sd_25}) for the propagation of arbitrary
fields \((\,\mfrak{w}_{M}(x_{p})\:;\:\mfrak{w}_{M}^{*}(x_{p})\,)^{T,a}\) and list in Eqs. (\ref{sd_26}-\ref{sd_28})
also the propagation for the split parts \(\mfrak{w}_{M}(x_{p})\) and its complex conjugated field \(\mfrak{w}_{M}^{*}(x_{p})\).
The doubled fields \((\,\mfrak{w}_{M}(x_{p})\:;\:\mfrak{w}_{M}^{*}(x_{p})\,)^{T,a}\) can be identified with the fields in the
various steps of propagation in \(\langle\mscr{A}_{J_{\psi}}[\hat{T},\feynV;\hat{\mscr{J}}]\rangle_{\feynbv}\) with
\(\hat{G}^{(0)}\), starting from \(J_{\psi;M}^{a}(x_{p})\) on the right-hand side
\beq  \lb{sd_25}
\Psi_{M}^{a(=1,2)}(x_{p}) &=&\Bigg(\bea{c}\psi_{M}(x_{p}) \\ \psi_{M}^{*}(x_{p})\eea\Bigg)^{a}=
\int_{C}d^{4}\!y_{q}\;\mcal{N}^{2}\;\ph{\,}^{a}\widehat{\langle x_{p}}|\hat{G}_{M;N}^{(0)}|\widehat{y_{q}\rangle}^{b}\;
\Bigg(\bea{c}\psi_{N}(y_{q}) \\ \psi_{N}^{*}(y_{q})\eea\Bigg)^{b}_{\mbox{;}}  \\  \lb{sd_26}
\mfrak{w}_{M}(x_{p}) &=&\int_{C}d^{4}\!y_{q}\;\mcal{N}^{2}\;\langle x_{p}|\hat{g}_{M;N}^{(0)}|y_{q}\rangle\;
\mfrak{w}_{N}(y_{q}) \;;  \\   \lb{sd_27}
\mfrak{w}_{M}^{*}(x_{p}) &=&\int_{C}d^{4}\!y_{q}\;\mcal{N}^{2}\;
\ovv{\langle x_{p}}|\big[\hat{g}\big]_{M;N}^{(0)}|\ovv{y_{q}\rangle}\;
\mfrak{w}_{N}^{*}(y_{q})  =  \int_{C}d^{4}\!y_{q}\;\mcal{N}^{2}\;\mfrak{w}_{N}^{*}(y_{q})\;
\langle y_{q}|\hat{g}_{N;M}^{(0)}|x_{p}\rangle\;;  \\  \lb{sd_28}
\Bigg(\bea{c}\mfrak{w}_{M}(x_{p}) \\ \mfrak{w}_{M}^{*}(x_{p}) \eea\Bigg)^{a} &=&
\int_{C}d^{4}\!y_{q}\;\mcal{N}^{2}\;\ph{\,}^{a}\widehat{\langle x_{p}}|\hat{G}_{M;N}^{(0)}|\widehat{y_{q}\rangle}^{b}\;
\Bigg(\bea{c}\mfrak{w}_{N}(y_{q}) \\ \mfrak{w}_{N}^{*}(y_{q}) \eea\Bigg)^{b} \;.
\eeq
However, the propagation of fields with relations (\ref{sd_25}-\ref{sd_28}) is not directly applicable for the action
\(\langle\mscr{A}_{DET}[\hat{T},\feynV;\hat{\mscr{J}}]\rangle_{\feynbv}\) because of the cyclic invariance of traces,
both of the internal state space and the Hilbert space trace of doubled quantum mechanics. The Hilbert space trace
means a propagation back to the same spacetime point. This property is not included in rules (\ref{sd_25}-\ref{sd_28})
which can only be used directly for the source field \(J_{\psi;M}^{a}(x_{p})\) (as a 'condensate seed') with repeated
propagation of \(\hat{G}^{(0)}\) to the left-hand side \(J_{\psi;N}^{\dag,b}(y_{q})\) in the action
\(\langle\mscr{A}_{J_{\psi}}[\hat{T},\feynV;\hat{\mscr{J}}]\rangle_{\feynbv}\). We circumvent this problem by
introducing anomalous-doubled unit operators of momentum-energy states in the expansion of
\(\langle\mscr{A}_{DET}[\hat{T},\feynV;\hat{\mscr{J}}]\rangle_{\feynbv}\) so that one achieves anomalous-doubled
plane wavefunctions with definite momentum-energy values instead of the source fields \(J_{\psi;N}^{\dag,b}(y_{q})\),
\(J_{\psi;M}^{a}(x_{p})\). The anomalous doubled plane wave states replace the fields in (\ref{sd_25}-\ref{sd_28})
and propagate with the time contour extended, background averaged Green functions (\ref{sd_11}-\ref{sd_24})
from the right-hand side to the left-hand side or vice versa. Since the various gradient terms follow straightforwardly
for \(\langle\mscr{A}_{J_{\psi}}[\hat{T},\feynV;\hat{\mscr{J}}]\rangle_{\feynbv}\) by repeated application of rules
(\ref{sd_25}-\ref{sd_28}) and gradient parts \(\delta\mfrak{h}(x_{p})\),
\(\delta\hat{\mscr{K}}^{\boldsymbol{\mu}}(x_{p})\:\boldsymbol{\hat{\pp}_{p,\mu}}\) (\ref{sd_3}-\ref{sd_9}), we give details of the
expansion for the more involved problem of traces in \(\langle\mscr{A}_{DET}[\hat{T},\feynV;\hat{\mscr{J}}]\rangle_{\feynbv}\).
The gradient expansion of \(\langle\mscr{A}_{J_{\psi}}[\hat{T},\feynV;\hat{\mscr{J}}]\rangle_{\feynbv}\) can then be
obtained from that of \(\langle\mscr{A}_{DET}[\hat{T},\feynV;\hat{\mscr{J}}]\rangle_{\feynbv}\) by replacing
anomalous-doubled plane wave states through the anomalous-doubled source fields \(J_{\psi;N}^{\dag,b}(y_{q})\),
\(J_{\psi;M}^{a}(x_{p})\) which have simpler propagation terms without the projection matrix \(\hat{S}\) for the coset space.
Therefore, we concentrate in the following subsection \ref{sd2} onto the gradient expansion of the action
\(\langle\mscr{A}_{DET}[\hat{T},\feynV;\hat{\mscr{J}}]\rangle_{\feynbv}\) and summarize the effective Lagrangian up to the complete
fourth order gradient with supplementary terms of
\(\langle\mscr{A}_{J_{\psi}}[\hat{T},\feynV;\hat{\mscr{J}}]\rangle_{\feynbv}\) in part \ref{sd3}.

\subsection{Gradient expansion of the anomalous-doubled determinant} \lb{sd2}

In order to allow for stable, static energy configurations in 3+1 spacetime dimensions, one has to expand up to
fourth order gradients so that one cannot scale the particular configuration to arbitrary small or large
sizes in the three dimensional coordinate space integrations over the static Hamiltonian density
('Derrick's theorem' \cite{raja,Naka} !).
Since we have shifted the trace-logarithm (\ref{s3_57}-\ref{s3_59})
of the one-particle operator \(\hat{\mscr{H}}(x_{p})\) with anomalous doubled
potential \(\mcal{V}_{\beta}^{\mu}(x_{p})\),
\(\hat{\mfrak{V}}(x_{p})=\mcal{V}_{\beta;\mu}(x_{p})\cdot\hat{V}_{\beta}^{\mu}\) (\ref{s3_60},\ref{sd_8})
for quark self-energy densities and
gauge degrees of freedom to the background functional (\ref{s3_59}), the trace-logarithm terms
\(\langle\mbox{'traces'} \ln[\hat{\mscr{H}}]\rangle_{\feynbv}\) cancel in the expansion of
\(\langle\mscr{A}_{DET}[\hat{T},\feynV;\hat{\mscr{J}}]\rangle_{\feynbv}\) which is specified in (\ref{sd_29}). The first order
source term with \(\wt{\mscr{J}}(\hat{T}^{-1},\hat{T})\) (\ref{sd_30}) is listed in detail in (\ref{sd_29}) with remaining unspecified
terms ascending from second order \((\,\wt{\mscr{J}}(\hat{T}^{-1},\hat{T})\,)^{n\geq2}\). This source term
\(\wt{\mscr{J}}(\hat{T}^{-1},\hat{T})\) (\ref{sd_29},\ref{sd_30}) allows to track the form of observables from the original
anti-commuting Fermi fields to those in terms of the complex-, even-valued self-energy matrix (\ref{s3_64}-\ref{s3_83}).
However, we concentrate
in this subsection on the derivation of the final effective Lagrangian and set the important source term
\(\wt{\mscr{J}}(\hat{T}^{-1},\hat{T})\) to zero in further steps of the gradient expansion
\beq \lb{sd_29}
\lefteqn{\Big\langle\mscr{A}_{DET}\big[\hat{T},\feynV;\hat{\mscr{J}}\big]\Big\rangle_{\feynBV} =
\frac{1}{2}\;\bigg\langle\hspace*{0.5cm}\trxpa\trfgamc\bigg(\ln
\overbrace{\Big[\hat{\mscr{H}}+\Delta\!\hat{\mscr{H}}+
\wt{\mscr{J}}(\hat{T}^{-1},\hat{T})\Big]}^{\hat{\mscr{O}}_{N;M}^{ba}(y_{q},x_{p})}-
\ln\big[\hat{\mscr{H}}\big]\bigg)\bigg\rangle_{\feynBV} } \\ \no &=&
\frac{1}{2}\sum_{n=1}^{\infty}\frac{(-1)^{n+1}}{n}
\bigg\langle\hspace*{0.5cm}\trxpa\trfgamc\bigg[\Big[\Big(\Delta\!\hat{\mscr{H}}+\wt{\mscr{J}}(\hat{T}^{-1},\hat{T})\Big)
\big(\hat{\mscr{H}}^{-1}\big)\Big]^{n}\bigg]\bigg\rangle_{\feynBV} \\ \no &=&
\frac{1}{2}\sum_{n=1}^{\infty}\frac{(-1)^{n+1}}{n}
\bigg\langle\hspace*{0.5cm}\trxpa\trfgamc\bigg[\Big[\Delta\!\hat{\mscr{H}}\;\;
\hat{\mscr{H}}^{-1}\Big]^{n}\bigg]\bigg\rangle_{\feynBV} + \\ \no &+&
\bigg\langle\hspace*{0.5cm}\trxpa\trfgamc\bigg[\wt{\mscr{J}}(\hat{T}^{-1},\hat{T})\Big)\;\hat{\mscr{H}}^{-1}\;
\bigg(\frac{1}{2}\sum_{n=1}^{\infty}(-1)^{n+1}
\Big[\Delta\!\hat{\mscr{H}}\;
\hat{\mscr{H}}^{-1}\Big]^{n-1}\bigg)\bigg]\bigg\rangle_{\feynBV} +
\Big\langle\mbox{\sf O}\Big[\big(\wt{\mscr{J}}(\hat{T}^{-1},\hat{T})\big)^{2}\Big]\Big\rangle_{\feynBV}\;;
\eeq
\beq \lb{sd_30}
\wt{\mscr{J}}_{N;M}^{ba}\big(\hat{T}^{-1}(y_{q})\,,\,\hat{T}(x_{p})\big) &=&
\Big(\hat{T}^{-1}(y_{q})\;\hat{I}\;\hat{S}\;\eta_{q}\;
\frac{\hat{\mscr{J}}_{N\ppr;M\ppr}^{b\ppr a\ppr}(y_{q},x_{p})}{\mcal{N}}\;
\eta_{p}\;\hat{S}\;\hat{I} \;\hat{T}(x_{p})\Big)_{N;M}^{ba}\;.
\eeq
Apart from the propagation rules (\ref{sd_25}-\ref{sd_28}), one has to include the proper action
of unsaturated gradients \(\boldsymbol{\hat{\pp}_{p,\mu}}\) onto the background potential
\(\mcal{V}_{\alpha;\kappa}(x_{p})\) (\ref{s3_60},\ref{sd_8}),
originating from the inverse operators \(\hat{\mscr{H}}^{-1}\) or their
corresponding background field averaged Green functions \(\hat{G}^{(0)}\) (\ref{sd_10}-\ref{sd_13}).
One has to take into account
the specific commutator relation
\(\boldsymbol{[}\boldsymbol{\hat{\pp}_{p,\kappa}}\:\boldsymbol{,}\:\hat{\mscr{H}}^{-1}\boldsymbol{]_{-}}\)
between the unsaturated derivative \(\boldsymbol{\hat{\pp}_{p,\kappa}}\) and the anomalous doubled inverse
operator \(\hat{\mscr{H}}^{-1}\) (\ref{sd_10}-\ref{sd_13})
in order to incorporate a proper averaging with path integral (\ref{s3_59})
over derivatives of background potentials as
\(\langle\,(\hat{\pp}_{p,\kappa}\mcal{V}_{\alpha;\kappa_{1}}(x_{p})\,)\,\rangle_{\feynbv}\) (or their higher order
products as e.g. \(\langle\,(\hat{\pp}_{p,\kappa}\mcal{V}_{\alpha;\kappa_{1}}(x_{p})\,)\;
(\hat{\pp}_{p,\lambda}\mcal{V}_{\beta;\lambda_{1}}(x_{p})\,)\,\rangle_{\feynbv}\) ).
We have therefore to conclude from Eqs. (\ref{sd_31}) the final commutator
\(\boldsymbol{[}\boldsymbol{\hat{\pp}_{p,\kappa}}\:\boldsymbol{,}\:\hat{\mscr{H}}^{-1}\boldsymbol{]_{-}}\)
which yields an additional saturated derivative of the anomalous doubled background potential
\(\hat{\mfrak{V}}(x_{p})=\mcal{V}_{\alpha;\kappa_{1}}(x_{p})\;\hat{V}_{\alpha}^{\kappa_{1}}\) (\ref{sd_8})
propagating with \(\hat{\mscr{H}}^{-1}\) or approximately with products of
\(\hat{G}^{(0)}=\langle\hat{\mscr{H}}^{-1}\rangle_{\feynbv}\)
\be\lb{sd_31}
\bea{rclrcl}
\boldsymbol{\hat{\pp}_{p,\kappa}}\;\hat{\mscr{H}}\;\hat{\mscr{H}}^{-1}&=& \boldsymbol{\hat{\pp}_{p,\kappa}}\; \hat{1}\;;
& \boldsymbol{\hat{\pp}_{p,\kappa}} &=&
\hat{\mscr{H}}\;\boldsymbol{\hat{\pp}_{p,\kappa}}\;\hat{\mscr{H}}^{-1}+
\boldsymbol{\Big[}\boldsymbol{\hat{\pp}_{p,\kappa}}\;\boldsymbol{,}\;\hat{\mscr{H}}\boldsymbol{\Big]_{-}}\;
\hat{\mscr{H}}^{-1} \;; \\
\boldsymbol{\Big[}\boldsymbol{\hat{\pp}_{p,\kappa}}\;\boldsymbol{,}\;\hat{\mscr{H}}^{-1}\boldsymbol{\Big]_{-}} &=&
-\hat{\mscr{H}}^{-1}\;\boldsymbol{\Big[}\boldsymbol{\hat{\pp}_{p,\kappa}}\;\boldsymbol{,}\;
\hat{\mscr{H}}\boldsymbol{\Big]_{-}}\;\hat{\mscr{H}}^{-1} \;; &
\boldsymbol{\Big[}\boldsymbol{\hat{\pp}_{p,\kappa}}\;\boldsymbol{,}\;\hat{\mscr{H}}\boldsymbol{\Big]_{-}} &=&
 \Big(\hat{\pp}_{p,\kappa}\hat{\mfrak{V}}(x_{p})\Big)=
\big(\hat{\pp}_{p,\kappa}\mcal{V}_{\alpha;\kappa_{1}}\big)\;\hat{V}_{\alpha}^{\kappa_{1}} \;; \\
\boldsymbol{\Big[}\boldsymbol{\hat{\pp}_{p,\kappa}}\;\boldsymbol{,}\;\hat{\mscr{H}}^{-1}\boldsymbol{\Big]_{-}} &=&
-\;\hat{\mscr{H}}^{-1}\;\underbrace{\big(\hat{\pp}_{p,\boldsymbol{\kappa}}
\mcal{V}_{\alpha;\kappa_{1}}\big)\;\hat{V}_{\alpha}^{\kappa_{1}} }_{(\,\hat{\pp}_{p,\boldsymbol{\kappa}}\hat{\mfrak{V}}\,)}
 \;\hat{\mscr{H}}^{-1} \;; &
\boldsymbol{\Big[}\boldsymbol{\hat{\pp}_{p,\kappa}}\;\boldsymbol{,}\;\hat{G}^{(0)}\boldsymbol{\Big]_{-}} &\simeq&
-\;\hat{G}^{(0)}\;\underbrace{\big(\hat{\pp}_{p,\boldsymbol{\kappa}}\mcal{V}_{\alpha;\kappa_{1}}\big)\;
\hat{V}_{\alpha}^{\kappa_{1}} }_{(\,\hat{\pp}_{p,\boldsymbol{\kappa}}\hat{\mfrak{V}}\,)}\;\hat{G}^{(0)} \;;
\eea
\ee
\[
\hat{V}_{\alpha}^{\kappa_{1}}=\Bigg(\bea{cc}
\big[\hat{\beta}(\im\;\hat{\gamma}^{\kappa_{1}}\;\hat{t}_{\alpha}+\hat{m})\,\big] & 0 \\ 0 &
\big[\hat{\beta}(\im\;\hat{\gamma}^{\kappa_{1}}\;\hat{t}_{\alpha}+\hat{m})\,\big]^{T} \eea\Bigg)_{\mbox{.}}
\]
As we use the commutator (\ref{sd_31}) for the expansion (\ref{sd_29}),
one proceeds to the correct expression (\ref{sd_32}) of the gradient expansion
which also consists of derivatives of the background potentials following from
 the inverse operators \(\hat{\mscr{H}}^{-1}\) or their corresponding averaged Green functions (\ref{sd_10}-\ref{sd_13}).
One has even to calculate commutators of \(\hat{\mscr{H}}^{-1}\) with multiple factors of
unsaturated gradient operators in order to accomplish the correct transport functions of background potentials
with their derivatives. However, these commutators of \(\hat{\mscr{H}}^{-1}\) with
higher order gradient operators straightforwardly result from subsequent application of commutators (\ref{sd_31}).
We list relation (\ref{sd_32}), derived from Eq. (\ref{sd_29}), as an intermediate step under single application
of the commutator (\ref{sd_31}); nevertheless, further commutator extensions have to be computed for various factors of gradient operators (\ref{sd_32}) in order to achieve the complete set of transport functions with background potentials
and their derivatives
\beq \lb{sd_32}
\lefteqn{\Big\langle\mscr{A}_{DET}\big[\hat{T},\feynV;\hat{\mscr{J}}\equiv 0\big]\Big\rangle_{\feynBV}   =
\frac{1}{2}\sum_{n=1}^{\infty}\frac{(-1)^{n+1}}{n}
\bigg\langle\hspace*{0.5cm}\trxpa\trfgamc\bigg[\Big(\Delta\!\hat{\mscr{H}}\;
\hat{\mscr{H}}^{-1}\Big)^{n}\bigg]\bigg\rangle_{\feynBV} }  \\ \no &=&
\frac{1}{2}\sum_{n=1}^{\infty}\frac{(-1)^{n+1}}{n}
\bigg\langle\hspace*{0.5cm}\trxpa\trfgamc\bigg[\Big[\Big(\delta\hat{\mfrak{h}}+
\delta\hat{\mscr{K}}^{\boldsymbol{\kappa}}\;\boldsymbol{\hat{\pp}_{p,\kappa}}\Big)
\;\hat{\mscr{H}}^{-1}\Big]^{n}\bigg]\bigg\rangle_{\feynBV}  \\ \no &=&
\frac{1}{2}\sum_{n=1}^{\infty}\frac{(-1)^{n+1}}{n}
\bigg\langle\hspace*{0.5cm}\trxpa\trfgamc\bigg[\Big[\delta\hat{\mfrak{h}}\;\hat{\mscr{H}}^{-1}+
\delta\hat{\mscr{K}}^{\boldsymbol{\kappa}}\;
\boldsymbol{\Big[}\boldsymbol{\hat{\pp}_{p,\kappa}}\;\boldsymbol{,}\;\hat{\mscr{H}}^{-1}\boldsymbol{\Big]_{-}}+
\delta\hat{\mscr{K}}^{\boldsymbol{\kappa}}\;\hat{\mscr{H}}^{-1}\;\boldsymbol{\hat{\pp}_{p,\kappa}}\Big]^{n}\bigg]\bigg\rangle_{\feynBV}
\\ \no &=& \frac{1}{2}\sum_{n=1}^{\infty}\frac{(-1)^{n+1}}{n}
\bigg\langle\hspace*{0.5cm}\trxpa\trfgamc\bigg[\Big[\Big(\delta\hat{\mfrak{h}}-
\delta\hat{\mscr{K}}^{\boldsymbol{\kappa}}\;\hat{\mscr{H}}^{-1}\;\big(\hat{\pp}_{p,\boldsymbol{\kappa}}\mcal{V}_{\alpha;\kappa_{1}}\big)\;
\hat{V}_{\alpha}^{\kappa_{1}}\Big)\;\hat{\mscr{H}}^{-1}+
\delta\hat{\mscr{K}}^{\boldsymbol{\kappa}}\;\hat{\mscr{H}}^{-1}\;\boldsymbol{\hat{\pp}_{p,\kappa}}\Big]^{n}\bigg]\bigg\rangle_{\feynBV} \\ \no &=&
\frac{1}{2}\sum_{n=1}^{\infty}\frac{(-1)^{n+1}}{n}
\bigg\langle\hspace*{0.5cm}\trxpa\trfgamc\bigg[\bigg(\Big[\hat{T}^{-1}\:\hat{S}\:\big(\hat{\beta}\,\feynd_{p}\hat{T}\big)
+\mcal{V}_{\alpha;\kappa_{1}}\:\big(\hat{T}^{-1}\:\hat{V}_{\alpha}^{\kappa_{1}}\:\hat{T}-\hat{V}_{\alpha}^{\kappa_{1}}\big) +
\\ \no &-&\delta\hat{\mscr{K}}^{\boldsymbol{\kappa}}\;
\hat{\mscr{H}}^{-1}\;\big(\hat{\pp}_{p,\boldsymbol{\kappa}}\mcal{V}_{\alpha;\kappa_{1}}\big)\;
\hat{V}_{\alpha}^{\kappa_{1}}\Big]\;\hat{\mscr{H}}^{-1}+
\delta\hat{\mscr{K}}^{\boldsymbol{\kappa}}\;\hat{\mscr{H}}^{-1}\;\boldsymbol{\hat{\pp}_{p,\kappa}}\bigg)^{n}
\bigg]\bigg\rangle_{\feynBV}  \\ \no &=&
\frac{1}{2}\sum_{n=1}^{\infty}\frac{(-1)^{n+1}}{n}
\bigg\langle\hspace*{0.5cm}\trxpa\trfgamc\bigg[\Big[\Delta\ppr\hat{\mfrak{h}}\;\hat{\mscr{H}}^{-1}+
\delta\hat{\mscr{K}}^{\boldsymbol{\kappa}}\;\hat{\mscr{H}}^{-1}\;\boldsymbol{\hat{\pp}_{p,\kappa}}\Big]^{n}
\bigg]\bigg\rangle_{\feynBV}\;.
\eeq
In order to simplify expressions, we introduce the abbreviations (\ref{sd_33}-\ref{sd_38}) for various parts of the gradient operators
\(\Delta\!\hat{\mscr{H}}\) and \(\Delta\!\hat{\mscr{H}}\;\hat{\mscr{H}}^{-1}\simeq \Delta\!\hat{\mscr{H}}\;\hat{G}^{(0)}\), which occur in
the expansion of the logarithm in (\ref{sd_32}), and define the symbol \(\Delta\ppr\hat{\mfrak{h}}\) (\ref{sd_33},\ref{sd_38}) for a part of
\(\Delta\!\hat{\mscr{H}}\;\hat{\mscr{H}}^{-1}\simeq \Delta\!\hat{\mscr{H}}\;\hat{G}^{(0)}\) in the last line of (\ref{sd_32}).
The equations (\ref{sd_6}-\ref{sd_9}) determine the saturated gradient part
\(\hat{\mfrak{k}}(x_{p})=\hat{T}^{-1}(x_{p})\:\hat{S}\:(\hat{\beta}\,\feynd_{p}\hat{T}(x_{p})\,)\) and potential matrix part
\(\delta\hat{\mfrak{V}}(x_{p})=\hat{T}^{-1}(x_{p})\:\hat{\mfrak{V}}(x_{p})\:\hat{T}(x_{p})-\hat{\mfrak{V}}(x_{p})\), which are
combined to \(\delta\hat{\mfrak{h}}(x_{p})=\hat{\mfrak{k}}(x_{p})+\delta\hat{\mfrak{V}}(x_{p})\) (\ref{sd_7}), in addition we discern the
'unsaturated' gradient operator part \(\delta\hat{\mscr{K}}^{\boldsymbol{\kappa}}(x_{p})\;\boldsymbol{\hat{\pp}_{p,\kappa}}\) (\ref{sd_9}).
Moreover, the commutator
\(\boldsymbol{[}\boldsymbol{\hat{\pp}_{p,\kappa}}\:\boldsymbol{,}\:\hat{\mscr{H}}^{-1}\boldsymbol{]_{-}}\) (\ref{sd_31})
in (\ref{sd_32}) gives rise to the potential matrix term
\((\hat{\pp}\ppr\hat{\mscr{V}})\:\hat{\mscr{H}}^{-1}=
(\delta\hat{\mscr{K}}^{\boldsymbol{\kappa}}\:\hat{\mscr{H}}^{-1}\:
(\hat{\pp}_{p,\boldsymbol{\kappa}}\mcal{V}_{\alpha;\kappa_{1}})\;\hat{V}_{\alpha}^{\kappa_{1}})\:\:\hat{\mscr{H}}^{-1}\)
(\ref{sd_33}) which we add to
\(\delta\hat{\mfrak{h}}(x_{p})=\hat{\mfrak{k}}(x_{p})+\delta\hat{\mfrak{V}}(x_{p})\) (\ref{sd_7}) for defining a new
saturated gradient operator part \(\Delta\ppr\hat{\mfrak{h}}(x_{p})\) (\ref{sd_33},\ref{sd_34}).
Nevertheless, we have to distinguish between the presence or
abscence of the anomalous doubled inverse operator \(\hat{\mscr{H}}^{-1}\) or averaged propagator \(\hat{G}^{(0)}\) so that one
has two potential matrix terms \(\hat{\pp}\ppr\hat{\mscr{V}}\) and \(\hat{\pp}\hat{\mscr{V}}(x_{p})\)
from the commutators (\ref{sd_31}) where the missing prime " \(\:\ppr\:\) " of the latter indicates the missing of the inverse
operator \(\hat{\mscr{H}}^{-1}\) or averaged propagator \(\hat{G}^{(0)}\) (compare relations (\ref{sd_33}) and (\ref{sd_34}) !).
This notation with the supplementary
prime is also transferred to the total saturated gradient parts \(\Delta\ppr\hat{\mfrak{h}}\;\hat{\mscr{H}}^{-1}\),
\(\Delta\hat{\mfrak{h}}\;\hat{\mscr{H}}^{-1}\) with the corresponding different potential matrix terms
\(\Delta\ppr\hat{\mfrak{V}}(x_{p})\) (\ref{sd_35}), \(\Delta\hat{\mfrak{V}}(x_{p})\) (\ref{sd_36}).
The potential matrix term \(\Delta\ppr\hat{\mfrak{V}}(x_{p})\)
consists of the sum of \(\delta\hat{\mfrak{V}}(x_{p})\) from \(\Delta\!\hat{\mscr{H}}(x_{p})\) and
\(-\hat{\pp}\ppr\hat{\mscr{V}}\), originating from the commutator
\(\boldsymbol{[}\boldsymbol{\hat{\pp}_{p,\kappa}}\:\boldsymbol{,}\:\hat{\mscr{H}}^{-1}\boldsymbol{]_{-}}=-
(\,\hat{\pp}\ppr\hat{\mscr{V}}\,)\;\hat{\mscr{H}}^{-1}=
-(\,\delta\hat{\mscr{K}}^{\boldsymbol{\kappa}}\;\hat{\mscr{H}}^{-1}\;
(\hat{\pp}_{p,\boldsymbol{\kappa}}\mcal{V}_{\alpha;\kappa_{1}})\:\hat{V}_{\alpha}^{\kappa_{1}}\,)\;\hat{\mscr{H}}^{-1}\),
whereas \(\Delta\hat{\mfrak{V}}(x_{p})\) is defined as the sum of \(\delta\hat{\mfrak{V}}(x_{p})\) from \(\Delta\!\hat{\mscr{H}}(x_{p})\)
and \(-(\hat{\pp}\hat{\mscr{V}}(x_{p})\,)=-(\hat{\pp}_{p,\boldsymbol{\kappa}}\mcal{V}_{\alpha;\kappa_{1}}(x_{p})\,)\:
\delta\hat{\mscr{K}}^{\boldsymbol{\kappa}}(x_{p})\;\hat{V}_{\alpha}^{\kappa_{1}}\)
without any propagator terms \(\hat{\mscr{H}}^{-1}\) or \(\hat{G}^{(0)}\).
Therefore, we have abbreviated the total gradient operator \(\Delta\!\hat{\mscr{H}}\;\hat{\mscr{H}}^{-1}\) of the logarithm in (\ref{sd_32}) by
\(\Delta\ppr\hat{\mfrak{h}}\;\hat{\mscr{H}}^{-1}+\delta\hat{\mscr{K}}^{\boldsymbol{\kappa}}\;\hat{\mscr{H}}^{-1}\;\boldsymbol{\hat{\pp}_{p,\kappa}}\)
(\ref{sd_33},\ref{sd_38}) with a prime, but have to apply the unprimed versions \(\Delta\hat{\mfrak{h}}\) (\ref{sd_34}) and
\(\Delta\hat{\mfrak{V}}(x_{p})\) (\ref{sd_36}) in later steps of transformations where the operators \(\hat{G}^{(0)}\), \(\hat{\mscr{H}}^{-1}\)
are removed according to the assumed rules (\ref{sd_25}-\ref{sd_28}) for propagation of anomalous doubled, generalized fields as
plane wave states or source fields \(J_{\psi;M}^{a}(x_{p})\)
\beq \lb{sd_33}
\Delta\ppr\hat{\mfrak{h}}&=&\hat{\mfrak{k}}+\Delta\ppr\hat{\mfrak{V}}=\hat{\mfrak{k}}+
\delta\hat{\mfrak{V}}-\hat{\pp}\ppr\hat{\mscr{V}}  \;; \\ \no
\hat{\pp}\ppr\hat{\mscr{V}} &=&\delta\hat{\mscr{K}}^{\boldsymbol{\kappa}}\;\hat{\mscr{H}}^{-1}\;
\big(\hat{\pp}_{p,\boldsymbol{\kappa}}\mcal{V}_{\alpha;\kappa_{1}}\,\big)\;\hat{V}_{\alpha}^{\kappa_{1}} =
\delta\hat{\mscr{K}}^{\boldsymbol{\kappa}}\;\hat{\mscr{H}}^{-1}\;
\big(\hat{\pp}_{p,\boldsymbol{\kappa}}\hat{\mfrak{V}}\,\big)   \;; \\ \lb{sd_34}
\Delta\hat{\mfrak{h}}(x_{p})&=&\hat{\mfrak{k}}(x_{p})+\Delta\hat{\mfrak{V}}(x_{p})=\hat{\mfrak{k}}(x_{p})+
\delta\hat{\mfrak{V}}(x_{p})-(\hat{\pp}\hat{\mscr{V}}(x_{p})\,) \;;  \\  \no
(\hat{\pp}\hat{\mscr{V}}(x_{p})\,) &=&
\big(\hat{\pp}_{p,\boldsymbol{\kappa}}\mcal{V}_{\alpha;\kappa_{1}}(x_{p})\,\big)\;
\delta\hat{\mscr{K}}^{\boldsymbol{\kappa}}(x_{p})\;\hat{V}_{\alpha}^{\kappa_{1}}=
\delta\hat{\mscr{K}}^{\boldsymbol{\kappa}}(x_{p})\;\big(\hat{\pp}_{p,\boldsymbol{\kappa}}\hat{\mfrak{V}}(x_{p})\,\big)\;;  \\ \lb{sd_35}
\Delta\ppr\hat{\mfrak{V}}&=&\delta\hat{\mfrak{V}}-\hat{\pp}\ppr\hat{\mscr{V}}  \\ \no &=&
\big(\hat{T}^{-1}\;\hat{\mfrak{V}}\;\hat{T}-\hat{\mfrak{V}}\,\big)-
\delta\hat{\mscr{K}}^{\boldsymbol{\kappa}}\;\hat{\mscr{H}}^{-1}\;
\big(\hat{\pp}_{p,\boldsymbol{\kappa}}\mcal{V}_{\alpha;\kappa_{1}}\,\big)\;\hat{V}_{\alpha}^{\kappa_{1}} \;;  \\ \lb{sd_36}
\Delta\hat{\mfrak{V}}(x_{p})&=&\delta\hat{\mfrak{V}}(x_{p})-(\hat{\pp}\hat{\mscr{V}}(x_{p})\,) \\ \no &=&
\big(\hat{T}^{-1}(x_{p})\;\hat{\mfrak{V}}(x_{p})\;\hat{T}(x_{p})-\hat{\mfrak{V}}(x_{p})\,\big)-
\big(\hat{\pp}_{p,\boldsymbol{\kappa}}\mcal{V}_{\alpha;\kappa_{1}}(x_{p})\,\big)\;
\delta\hat{\mscr{K}}^{\boldsymbol{\kappa}}(x_{p})\;\hat{V}_{\alpha}^{\kappa_{1}} \;;  \\  \lb{sd_37}
\delta\hat{\mfrak{h}}(x_{p})&=&\hat{\mfrak{k}}(x_{p})+\delta\hat{\mfrak{V}}(x_{p})\;;   \\ \no
\hat{\mfrak{k}}(x_{p})&=&\hat{T}^{-1}(x_{p})\:\hat{S}\:\hat{\beta}\:\hat{\gamma}^{\mu}\big(\hat{\pp}_{p,\mu}\hat{T}(x_{p})\,\big)\;;
\\   \lb{sd_38} \Delta\!\hat{\mscr{H}}\;\hat{\mscr{H}}^{-1} &=&\Delta\ppr\hat{\mfrak{h}}\;\hat{\mscr{H}}^{-1}+
\delta\hat{\mscr{K}}^{\boldsymbol{\kappa}}\;\hat{\mscr{H}}^{-1}\;\boldsymbol{\hat{\pp}_{p,\kappa}}\;.
\eeq
The background averaging with (\ref{s3_59},\ref{sd_2}) in (\ref{sd_32}) is split into independent averages
over single factors of the doubled Green function (\ref{sd_10}-\ref{sd_24}) with remaining
background averaging over factors of  \((\,\hat{\pp}_{p,\kappa}\mcal{V}_{\alpha;\kappa_{1}}(x_{p})\,)\) and over
\(\hat{\mfrak{V}}(x_{p})\) (\ref{sd_8}) in \(\Delta\ppr\hat{\mfrak{h}}\) (\ref{sd_33}).
One therefore obtains the described propagation with the doubled non-equilibrium Green functions (\ref{sd_10}-\ref{sd_24})
back to the initial spacetime point. If we assume the appearance of changing or non-diagonal contour time
indices \(p\neq q\) for Green functions as
\(\ph{\,}^{a=1}\widehat{\langle x_{-}}|\hat{G}^{(0)}|\widehat{y_{+}\rangle}^{a=1}\neq 0\)
in the gradient expansion of \(\langle\mscr{A}_{DET}[\hat{T},\feynV;\hat{\mscr{J}}\equiv0]\rangle_{\feynbv}\),
one has also to consider a back
propagation, as e.g. with \(\ph{\,}^{a=1}\widehat{\langle y_{+}}|\hat{G}^{(0)}|\widehat{x_{-}\rangle}^{a=1}\),
which vanishes completely due to the defined, contour extended, doubled Heaviside functions (\ref{sd_20}-\ref{sd_24}).
In consequence one only has a propagation on a fixed branch '\(p\)' of the time contour in the gradient expansion (\ref{sd_29}-\ref{sd_32})
\footnote{In the case of disordered systems, one has to take into account propagations with \(\hat{G}^{(0)}\) of
varying time contour branches because the corresponding gradient operator is non-diagonal in contour time indices
\(p\neq q\) whereas the gradient operator \(\Delta\!\hat{\mscr{H}}(x_{p})\) in this paper is diagonal in
the contour time indices.}
\beq \no   
\lefteqn{\mbox{product of time contour Green functions }\bigg(
\ph{\,}^{a}\widehat{\langle x_{p}}|\hat{G}^{(0)}|\widehat{y_{q}\rangle}^{b}\bigg) \mbox{ in }
\Big\langle\mscr{A}_{DET}[\hat{T},\feynV;\hat{\mscr{J}}\equiv0]\Big\rangle_{\feynBV} \rightarrow
\mbox{remaining terms}}   \\ \no &\rightarrow&\boldsymbol{\delta_{pq}}\;\;
\mbox{product of Green functions on fixed branch '$p$' of time contour }\;
\bigg(\delta_{ab}\;\ph{\,}^{a}\widehat{\langle x_{p}}|\hat{G}^{(0)}|\widehat{y_{p}\rangle}^{a}\bigg) \;.
\eeq
According to the propagation back to the same spacetime point in
\(\langle\mscr{A}_{DET}[\hat{T},\feynV;\hat{\mscr{J}}\equiv0]\rangle_{\feynbv}\), one has to regard at least two
or any higher, even-numbered, non-diagonal factors of gradient terms \(\Delta\!\hat{\mscr{H}}^{a\neq b}(x_{p})\)
(\ref{sd_6}-\ref{sd_9}),
except for the first order case \(n=1\) in (\ref{sd_32}). One has to distinct between the block diagonal or
off-diagonal gradient terms (\ref{sd_39},\ref{sd_40}) by the anti-commutator
\(\frac{1}{2}\hat{S}\,\boldsymbol{\{}\Delta\!\hat{\mscr{H}}\,\boldsymbol{,}\,\hat{S}\boldsymbol{\}_{+}}\) or commutator
\(-\frac{1}{2}\hat{S}\,\boldsymbol{[}\Delta\!\hat{\mscr{H}}\,\boldsymbol{,}\,\hat{S}\boldsymbol{]_{-}}\) parts
with projection matrix \(\hat{S}\) (\ref{s2_21}) of the coset space, respectively
\beq \lb{sd_39}
\Delta\!\hat{\mscr{H}}^{a=b}(x_{p}) &=&\frac{1}{2}\Big(\hat{S}\;\Delta\!\hat{\mscr{H}}(x_{p})\;\hat{S}+
\Delta\!\hat{\mscr{H}}(x_{p})\Big)=
\frac{1}{2}\;\hat{S}\;\boldsymbol{\Big\{}\Delta\!\hat{\mscr{H}}(x_{p})\:\boldsymbol{,}\:\hat{S}\boldsymbol{\Big\}_{+}}\;; \\ \lb{sd_40}
\Delta\!\hat{\mscr{H}}^{a\neq b}(x_{p}) &=&-\frac{1}{2}\Big(\hat{S}\;\Delta\!\hat{\mscr{H}}(x_{p})\;\hat{S}-
\Delta\!\hat{\mscr{H}}(x_{p})\Big)=
-\frac{1}{2}\;\hat{S}\;\boldsymbol{\Big[}\Delta\!\hat{\mscr{H}}(x_{p})\:\boldsymbol{,}\:\hat{S}\boldsymbol{\Big]_{-}} \;.
\eeq
This distinction is necessary because the propagation with only block diagonal operators
\(\Delta\!\hat{\mscr{H}}^{a=b}(x_{p})=
\frac{1}{2}\hat{S}\,\boldsymbol{\{}\Delta\!\hat{\mscr{H}}\,\boldsymbol{,}\,\hat{S}\boldsymbol{\}_{+}}\)
contributes vanishing terms in spacetime integrations over the Hilbert space trace and inserted unit operators
according to back propagation to the initial, contour extended spacetime point. This particular, '{\it vanishing}'
part is symbolically subtracted in (\ref{sd_41}) in order to point out the importance of other, remaining, non-vanishing
combinations with \(\frac{1}{2}\hat{S}\,\boldsymbol{\{}\Delta\!\hat{\mscr{H}}\,\boldsymbol{,}\,\hat{S}\boldsymbol{\}_{+}}\) and
\(-\frac{1}{2}\hat{S}\,\boldsymbol{[}\Delta\!\hat{\mscr{H}}\,\boldsymbol{,}\,\hat{S}\boldsymbol{]_{-}}\)
\beq\lb{sd_41}
\lefteqn{\Big\langle\mscr{A}_{DET}\big[\hat{T},\feynV;\hat{\mscr{J}}\equiv 0\big]\Big\rangle_{\feynBV} =
\frac{1}{2}\sum_{n=1}^{\infty}\frac{(-1)^{n+1}}{n}\bigg\langle\hspace*{0.5cm}\trxpa\trfgamc\bigg[
\Big(\Delta\!\hat{\mscr{H}}\;\hat{G}^{(0)}\Big)^{n}\bigg]\bigg\rangle_{\feynBV}  + } \\ \no &-&
\frac{1}{2}\sum_{n=2}^{\infty}\frac{(-1)^{n+1}}{n}\underbrace{\bigg\langle\hspace*{0.5cm}\trxpa\trfgamc\bigg[
\Big(\frac{1}{2}\:\hat{S}
\boldsymbol{\big\{}\Delta\!\hat{\mscr{H}}\:\boldsymbol{,}\:\hat{S}\boldsymbol{\big\}_{+}}\;\hat{G}^{(0)}\Big)^{n}
\bigg]\bigg\rangle_{\feynBV}}_{\equiv 0,\;
\mbox{\scz vanishing contribution in spacetime integrations !}} \\ \no &=&
\frac{1}{2}\sum_{n=1}^{\infty}\frac{(-1)^{n+1}}{n}\bigg\langle\hspace*{0.5cm}\trxpa\trfgamc\bigg[\bigg(
\hat{T}^{-1}\:\hat{S}\,\big(\hat{\beta}\,\feynd_{p}\hat{T}\big)\:\hat{G}^{(0)}+
\mcal{V}_{\alpha;\kappa_{1}}\:\big(\,\hat{T}^{-1}\:\hat{V}_{\alpha}^{\kappa_{1}}\:\hat{T}-\hat{V}_{\alpha}^{\kappa_{1}}\,\big)\:
\hat{G}^{(0)} + \\ \no &-&
\delta\hat{\mscr{K}}^{\boldsymbol{\kappa}}\:\hat{G}^{(0)}\:\big(\hat{\pp}_{p,\boldsymbol{\kappa}}\mcal{V}_{\alpha;\kappa_{1}}\big)\:
\hat{V}_{\alpha}^{\kappa_{1}}\:\hat{G}^{(0)}+
\delta\hat{\mscr{K}}^{\boldsymbol{\kappa}}\:\hat{G}^{(0)}\;\boldsymbol{\hat{\pp}_{p,\kappa}}\bigg)^{n}\bigg]
\bigg\rangle_{\feynBV} +  \\ \no &-&
\frac{1}{2}\sum_{n=2}^{\infty}\frac{(-1)^{n+1}}{n}\bigg\langle\hspace*{0.5cm}\trxpa\trfgamc\bigg[\bigg(
\frac{\hat{S}}{2} \boldsymbol{\Big\{}
\hat{T}^{-1}\:\hat{S}\,\big(\hat{\beta}\,\feynd_{p}\hat{T}\big)+
\mcal{V}_{\alpha;\kappa_{1}}\:\big(\,\hat{T}^{-1}\:\hat{V}_{\alpha}^{\kappa_{1}}\:\hat{T}-\hat{V}_{\alpha}^{\kappa_{1}}\,\big)
\;\boldsymbol{,}\;\hat{S}\boldsymbol{\Big\}_{+}}\:\hat{G}^{(0)} + \\ \no &-&
\frac{\hat{S}}{2}\boldsymbol{\big\{}
\delta\hat{\mscr{K}}^{\boldsymbol{\kappa}}\;\boldsymbol{,}\;\hat{S}\boldsymbol{\big\}_{+}}\:
\hat{G}^{(0)}\:\big(\hat{\pp}_{p,\boldsymbol{\kappa}}\mcal{V}_{\alpha;\kappa_{1}}\big)\:
\hat{V}_{\alpha}^{\kappa_{1}}\:\hat{G}^{(0)}+\frac{\hat{S}}{2}\boldsymbol{\big\{}\:
\delta\hat{\mscr{K}}^{\boldsymbol{\kappa}}\;\boldsymbol{,}\;\hat{S}\boldsymbol{\big\}_{+}}
\:\hat{G}^{(0)}\;\boldsymbol{\hat{\pp}_{p,\kappa}}\bigg)^{n}\bigg]
\bigg\rangle_{\feynBV} \\ \no &=&
\frac{1}{2}\sum_{n=1}^{\infty}\frac{(-1)^{n+1}}{n}\bigg\langle\hspace*{0.5cm}\trxpa\trfgamc\bigg[\Big[
\Big(\hat{\mfrak{k}}+\delta\hat{\mfrak{V}}-\hat{\pp}\ppr\hat{\mscr{V}}\Big)\;\hat{G}^{(0)}+
\delta\hat{\mscr{K}}^{\boldsymbol{\kappa}}\;\hat{G}^{(0)}\;\boldsymbol{\hat{\pp}_{p,\kappa}}\Big]^{n}\bigg]\bigg\rangle_{\feynBV} +
 \\ \no &-&
\frac{1}{2}\sum_{n=2}^{\infty}\frac{(-1)^{n+1}}{n}\underbrace{\bigg\langle\hspace*{0.5cm}\trxpa\trfgamc\bigg[
\bigg(\frac{1}{2}\:\hat{S}
\boldsymbol{\bigg\{}\Big(\hat{\mfrak{k}}+\delta\hat{\mfrak{V}}-\hat{\pp}\ppr\hat{\mscr{V}}\Big)\;\hat{G}^{(0)}+
\delta\hat{\mscr{K}}^{\boldsymbol{\kappa}}\;\hat{G}^{(0)}\;\boldsymbol{\hat{\pp}_{p,\kappa}}
\:\boldsymbol{,}\:\hat{S}\boldsymbol{\bigg\}_{+}}\bigg)^{n}
\bigg]\bigg\rangle_{\feynBV}}_{\equiv 0,\;
\mbox{\scz vanishing contribution in spacetime integrations !}} \;.
\eeq
Corresponding to the definitions and notations of indices for the internal spaces in section \ref{s2}, we group
again the collective indices \(M_{i}\), \(N_{i}\) which are composed of the isospin- (flavour-) indices
\(f_{i}\), \(g_{i}\), the \(4\times4\) Dirac gamma matrix indices \(m_{i}\), \(n_{i}\) and the colour matrix
indices \(r_{i}\), \(s_{i}\). Furthermore, the total numbers \(\mcal{N}\), \(\mcal{N}_{k}\) of spacetime and
momentum-energy points have to scale the spacetime and momentum-energy integrations in the gradient expansion
of \(\langle\mscr{A}_{DET}[\hat{T},\feynV;\hat{\mscr{J}}\equiv0]\rangle_{\feynbv}\); hence we can perform
a large N-limit with the total number of points \(\mcal{N}\:\mcal{N}_{k}=(\,N_{L}/(2\pi)\,)^{4}\)
of the underlying grids, separated into \(N_{L}\) discrete points for each of the 3+1 dimensions
\be \lb{sd_42}
\bea{rclrcl}
M_{i}&=&\{f_{i}\,,\,m_{i}\,,\,r_{i}\} \hspace*{1.0cm} & N_{i}&=&\{g_{i}\,,\,n_{i}\,,\,s_{i}\} \\
f_{i}&=&\mbox{u(p), d(own), ( s(trange) )}  &  g_{i}&=&\mbox{u(p), d(own), ( s(trange) )}  \\
f_{i}&\stackrel{\mbox{\scz or}}{=}&1,\,\ldots\,,\,N_{f} & g_{i}&\stackrel{\mbox{\scz or}}{=}&1,\,\ldots\,,\,N_{f} \\
m_{i}&=&1,\,\ldots\,,4\;\;\;;[\hat{\gamma}^{\mu}]_{m_{i},n_{j}}  &
n_{i}&=&1,\,\ldots\,,4\;\;\;;[\hat{\gamma}^{\nu}]_{m_{j},n_{i}}   \\
r_{i}&=&1,\ldots,\,N_{c}=3 & s_{i}&=&1,\ldots,\,N_{c}=3  \\
\mcal{N}&=&\delta^{(4)}(0)=\frac{1}{\big(\Delta x\big)^{4}}=\frac{1}{\big(L/N_{L}\big)^{4}}  &
\mcal{N}_{k}&=&\frac{1}{\big(\Delta k\big)^{4}}=\frac{1}{\big(2\pi/L\big)^{4}}  \\
\mcal{N}\;\mcal{N}_{k}&=&\Big(N_{L}/(2\pi)\Big)^{4}  & \sum_{a_{i}=1,2}&&\ldots  \;.
\eea
\ee
The given notations (\ref{sd_42}) are used to label the internal spaces of the trace operations in
\(\langle\mscr{A}_{DET}[\hat{T},\feynV;\hat{\mscr{J}}\equiv0]\rangle_{\feynbv}\) (\ref{sd_43}).
According to the definitions of the anomalous doubled Hilbert states of spacetime, the complete matrix
representation is specified for the gradient expansion of the determinant as a sum over the various orders \(n\)
following from the logarithm. We emphasize the last line in (\ref{sd_43}) with Kronecker deltas over the internal
symmetry spaces and the contour extended delta function of 3+1 dimensional spacetime which causes the propagation
back to the initial spacetime and internal symmetry space state, considering the original traces from the
determinant
\beq  \lb{sd_43}
\lefteqn{\Big\langle\mscr{A}_{DET}\big[\hat{T},\feynV;\hat{\mscr{J}}\equiv 0\big]\Big\rangle_{\feynBV} =
\frac{1}{2}\sum_{n=1}^{\infty}\frac{(-1)^{n+1}}{n}\bigg\langle\hspace*{0.5cm}\trxpa\trfgamc\bigg[
\Big[\Delta\!\hat{\mscr{H}}\;\hat{G}^{(0)}\Big]^{n}\bigg]\bigg\rangle_{\feynBV} = } \\ \no &=&
\frac{1}{2}\sum_{n=1}^{\infty}\frac{(-1)^{n+1}}{n}\int_{C}d^{4}\!x_{p_{n}}^{(n)}\;\eta_{p_{n}}\;\mcal{N}
\sum_{a_{n}=1,2}\sum_{M_{n}}\prod_{i=0}^{n-1}
\bigg(\int_{C}d^{4}\!x_{p_{i}}^{(i)}\;\eta_{p_{i}}\;\mcal{N}
\int_{C}d^{4}\!y_{q_{i}}^{(i)}\;\eta_{q_{i}}\;\mcal{N}\sum_{a_{i}=1,2}\sum_{M_{i}}\sum_{N_{i}}\bigg)\times
\\  \no &\times& \Bigg\langle \ph{\,}^{a_{n}}\widehat{\langle x_{p_{n}}^{(n)}}|
\Delta\!\hat{\mscr{H}}_{M_{n};N_{n-1}}|\widehat{y_{q_{n-1}}^{(n-1)}\rangle}^{a_{n-1}}{\scrscr\times}
\ph{\,}^{a_{n-1}}\widehat{\langle y_{q_{n-1}}^{(n-1)}}|\hat{G}_{N_{n-1};M_{n-1}}^{(0)}|
\widehat{x_{p_{n-1}}^{(n-1)}\rangle}^{a_{n-1}}\;\times \\ \no &\times&
\ph{\,}^{a_{n-1}}\widehat{\langle x_{p_{n-1}}^{(n-1)}}|
\Delta\!\hat{\mscr{H}}_{M_{n-1};N_{n-2}}|\widehat{y_{q_{n-2}}^{(n-2)}\rangle}^{a_{n-2}}{\scrscr\times}
\ph{\,}^{a_{n-2}}\widehat{\langle y_{q_{n-2}}^{(n-2)}}|\hat{G}_{N_{n-2};M_{n-2}}^{(0)}|
\widehat{x_{p_{n-2}}^{(n-2)}\rangle}^{a_{n-2}}\;\times\;\ldots  \\ \no &\times&\ldots\;
\ph{\,}^{a_{2}}\widehat{\langle x_{p_{2}}^{(2)}}|
\Delta\!\hat{\mscr{H}}_{M_{2};N_{1}}|\widehat{y_{q_{1}}^{(1)}\rangle}^{a_{1}}{\scrscr\times}
\ph{\,}^{a_{1}}\widehat{\langle y_{q_{1}}^{(1)}}|\hat{G}_{N_{1};M_{1}}^{(0)}|
\widehat{x_{p_{1}}^{(1)}\rangle}^{a_{1}}\;\times  \\ \no &\times&
\ph{\,}^{a_{1}}\widehat{\langle x_{p_{1}}^{(1)}}|
\Delta\!\hat{\mscr{H}}_{M_{1};N_{0}}|\widehat{y_{q_{0}}^{(0)}\rangle}^{a_{0}}{\scrscr\times}
\ph{\,}^{a_{0}}\widehat{\langle y_{q_{0}}^{(0)}}|\hat{G}_{N_{0};M_{0}}^{(0)}|
\widehat{x_{p_{0}}^{(0)}\rangle}^{a_{0}}\Bigg\rangle_{\feynBV}\;\times \\ \no &\times&
\delta_{p_{n}p_{0}}\;\;\delta^{(4)}(x_{p_{n}}^{(n)}-x_{p_{0}}^{(0)})\;\;\delta_{a_{n}a_{0}}\;\;
\delta_{f_{n}f_{0}}\;\;\delta_{m_{n}m_{0}}\;\;\delta_{r_{n}r_{0}}\;\;;  \\ \no &&
\big(\mbox{summations in (\ref{sd_43}) without }\;a_{n}=a_{n-1}=\ldots=a_{1}=a_{0} !\big)\;\;.
\eeq
The trace operations of the action
\(\langle\mscr{A}_{DET}[\hat{T},\feynV;\hat{\mscr{J}}\equiv0]\rangle_{\feynbv}\) also involve the Kronecker delta
\(\delta_{a_{n}a_{0}}\) of the anomalous doubling. This particular Kronecker delta is substituted by the described
anti-commutator (symbolized in Eq. (\ref{sd_44})) with projection matrix \(\hat{S}\) of the coset space in a similar
kind as for \(\Delta\!\hat{\mscr{H}}^{a=b}(x_{p})\) (\ref{sd_39})
\beq \lb{sd_44}
\lefteqn{\Bigg(\bea{c}\big(\mbox{field}\ppr\big)^{*}  \\ \big(\mbox{field}\ppr\big) \eea\Bigg)^{T,a_{n}}
\underbrace{\frac{1}{2}\Bigg[\hat{S}\Big(\mbox{n-th order term}\Big)\hat{S}+
\Big(\mbox{n-th order term}\Big)\Bigg]}_{\mbox{equivalent to }\delta_{a_{n}a_{0}}}
\Bigg(\bea{c}\big(\mbox{field}\big)  \\ \big(\mbox{field}\big)^{*} \eea\Bigg)^{a_{0}} = } \\ \no &=&
\Bigg(\bea{c}\big(\mbox{field}\ppr\big)^{*}  \\ \big(\mbox{field}\ppr\big) \eea\Bigg)^{T,a_{n}}
\underbrace{\frac{\hat{S}}{2}\boldsymbol{\bigg\{}\Big(\mbox{n-th order term}\Big)\;\boldsymbol{,}\;\hat{S}
\boldsymbol{\bigg\}_{+}} }_{\mbox{equivalent to }\delta_{a_{n}a_{0}}}
\Bigg(\bea{c}\big(\mbox{field}\big)  \\ \big(\mbox{field}\big)^{*} \eea\Bigg)^{a_{0}} \;.
\eeq
The gradient expansion of \(\langle\mscr{A}_{DET}[\hat{T},\feynV;\hat{\mscr{J}}\equiv0]\rangle_{\feynbv}\) is reduced
to that of \(\langle\mscr{A}_{J_{\psi}}[\hat{T},\feynV;\hat{\mscr{J}}\equiv0]\rangle_{\feynbv}\) by factorization of
the anomalous doubled, contour extended delta function of spacetime to plane wave states. Since we perform an
expansion for lowest-order, gradually varying gradient terms, a cutoff momentum \(k_{max}\) is introduced
in the anomalous doubled, contour extended momentum-energy integrations for the unit operator
(compare appendix \ref{sa})
\beq \lb{sd_45}
\lefteqn{\delta_{p_{n}p_{0}}\;\;\delta^{(4)}(x_{p_{n}}^{(n)}-x_{p_{0}}^{(0)})\;\;\delta_{a_{n}a_{0}}=
\sum_{a=1,2}\int_{C}^{k_{max}}d^{4}\!k_{p}\;\eta_{p}\;\mcal{N}_{k}\;
\ph{\,}^{a_{0}}\widehat{\langle x_{p_{0}}^{(0)}}|\widehat{k_{p}\rangle}^{a}{\scrscr \times}
\ph{\,}^{a}\widehat{\langle k_{p}}|\widehat{ x_{p_{n}}^{(n)}\rangle}^{a_{n}} = } \\ \no &=&
\delta_{a_{n}a_{0}}\int_{C}^{k_{max}}d^{4}\!k_{p}\;\eta_{p}\;\mcal{N}_{k}\;
\delta_{pp_{0}}\;\delta_{pp_{n}}\Bigg\{
\bea{c} \exp\big\{\im\,k_{p}\cdot(x_{p_{0}}^{(0)}-x_{p_{n}}^{(n)})\big\}\;;\mbox{ for }\;;a_{0}=a_{n}=1 \;; \\
\exp\big\{-\im\,k_{p}\cdot(x_{p_{0}}^{(0)}-x_{p_{n}}^{(n)})\big\}\;;\mbox{ for }\;;a_{0}=a_{n}=2 \;. \eea
\eeq
One has to achieve a similar propagation as in
\(\langle\mscr{A}_{J_{\psi}}[\hat{T},\feynV;\hat{\mscr{J}}\equiv0]\rangle_{\feynbv}\) due to relations (\ref{sd_25}-\ref{sd_28})
and therefore factorizes the Kronecker deltas of the internal spaces in the last line of the matrix representation
for \(\langle\mscr{A}_{DET}[\hat{T},\feynV;\hat{\mscr{J}}\equiv0]\rangle_{\feynbv}\) (\ref{sd_43})
\be \lb{sd_46}
\delta_{f_{n}f_{0}}=\sum_{\bar{f}=1}^{N_{f}}\delta_{f_{n}\bar{f}}\;\delta_{\bar{f}f_{0}}\;;\hspace*{0.6cm}
\delta_{m_{n}m_{0}}=\sum_{\bar{m}=1}^{4}\delta_{m_{n}\bar{m}}\;\delta_{\bar{m}m_{0}}\;;\hspace*{0.6cm}
\delta_{r_{n}r_{0}}=\sum_{\bar{r}=1}^{N_{c}}\delta_{r_{n}\bar{r}}\;\delta_{\bar{r}r_{0}} \;.
\ee
As we use the factorizations of the delta function and Kronecker deltas (\ref{sd_45},\ref{sd_46}), one can define
anomalous doubled plane wave states
\(\mfrak{W}_{f_{0},m_{0},r_{0}}^{(\bar{f},\bar{m},\bar{r});a_{0}}(x_{p_{0}}^{(0)};k_{p})\),
\((\:\mfrak{W}_{f_{n},m_{n},r_{n}}^{(\bar{f},\bar{m},\bar{r});a_{0}}(x_{p_{n}}^{(n)};k_{p})\:)^{\dag}\)
in terms of (\ref{sd_48},\ref{sd_50}) extended with their particular complex conjugates. The anomalous doubled
plane wave states of four-momentum \(k_{p}\) replace the source fields
\(J_{\psi;N}^{\dag b}(y_{q})\), \(J_{\psi;M}^{a}(x_{p})\) and allow to apply rules (\ref{sd_25}-\ref{sd_28})
for propagation with Green function \(\hat{G}^{(0)}\) averaged by the background functional (\ref{sd_65},\ref{sd_2}).
This is possible because the plane wave states and (\ref{sd_46}) decouple the trace operations leading to
propagation back to the initial spacetime and internal space state
\beq \lb{sd_47}
\mfrak{W}_{f_{0},m_{0},r_{0}}^{(\bar{f},\bar{m},\bar{r});a_{0}}(x_{p_{0}}^{(0)};k_{p}) &=&
\Bigg(\bea{c}\mfrak{w}_{f_{0},m_{0},r_{0}}^{(\bar{f},\bar{m},\bar{r})}(x_{p_{0}}^{(0)};k_{p}) \\
\big(\mfrak{w}_{f_{0},m_{0},r_{0}}^{(\bar{f},\bar{m},\bar{r})}(x_{p_{0}}^{(0)};k_{p})\big)^{*}
\eea\Bigg)^{a_{0}}  \;;  \\   \lb{sd_48}
\mfrak{w}_{f_{0},m_{0},r_{0}}^{(\bar{f},\bar{m},\bar{r})}(x_{p_{0}}^{(0)};k_{p}) &=&
\exp\big\{\im\,k_{p}\cdot x_{p_{0}}^{(0)}\big\}\;\delta_{pp_{0}}\;\delta_{\bar{f}f_{0}}\;
\delta_{\bar{m}m_{0}}\;\delta_{\bar{r}r_{0}}\;;  \\   \lb{sd_49}
\Big(\mfrak{W}_{f_{n},m_{n},r_{n}}^{(\bar{f},\bar{m},\bar{r});a_{0}}(x_{p_{n}}^{(n)};k_{p})\Big)^{\dag}&=&
\Bigg(\bea{c}\big(\mfrak{w}_{f_{n},m_{n},r_{n}}^{(\bar{f},\bar{m},\bar{r})}(x_{p_{n}}^{(n)};k_{p})\big)^{*} \\
\mfrak{w}_{f_{n},m_{n},r_{n}}^{(\bar{f},\bar{m},\bar{r})}(x_{p_{n}}^{(n)};k_{p})
\eea\Bigg)^{T,a_{n}}  \;;  \\  \lb{sd_50}
\mfrak{w}_{f_{n},m_{n},r_{n}}^{(\bar{f},\bar{m},\bar{r})}(x_{p_{n}}^{(n)};k_{p}) &=&
\exp\big\{\im\,k_{p}\cdot x_{p_{n}}^{(n)}\big\}\;\delta_{pp_{n}}\;\delta_{\bar{f}f_{n}}\;
\delta_{\bar{m}m_{n}}\;\delta_{\bar{r}r_{n}}\;.
\eeq
We regroup the integrations of the matrix representation of
\(\langle\mscr{A}_{DET}[\hat{T},\feynV;\hat{\mscr{J}}\equiv 0]\rangle_{\feynbv}\) with inclusion of momentum-energy
integrations and additional internal space summations so that one formally accomplishes an analogous action as
\(\langle\mscr{A}_{J_{\psi}}[\hat{T},\feynV;\hat{\mscr{J}}\equiv 0]\rangle_{\feynbv}\) where the source fields
are substituted by the plane wave fields of (\ref{sd_47}-\ref{sd_50})
\beq \lb{sd_51}
\lefteqn{\Big\langle\mscr{A}_{DET}\big[\hat{T},\feynV;\hat{\mscr{J}}\equiv 0\big]\Big\rangle_{\feynBV} =
\int_{C}^{k_{max}}d^{4}\!k_{p}\;\eta_{p}\;\mcal{N}_{k}\sum_{\bar{f}=1}^{N_{f}}
\sum_{\bar{m}=1}^{4}\sum_{\bar{r}=1}^{N_{c}} \frac{1}{2}\sum_{n=1}^{\infty}\frac{(-1)^{n+1}}{n} \times }
\\ \no &\times&  \int_{C}d^{4}\!x_{p_{n}}^{(n)}\;\eta_{p_{n}}\;\mcal{N}
\sum_{a_{n}=1,2}\sum_{M_{n}}\prod_{i=0}^{n-1}
\bigg(\int_{C}d^{4}\!x_{p_{i}}^{(i)}\;\eta_{p_{i}}\;\mcal{N}
\int_{C}d^{4}\!y_{q_{i}}^{(i)}\;\eta_{q_{i}}\;\mcal{N}\sum_{a_{i}=1,2}\sum_{M_{i}}\sum_{N_{i}}\bigg)\times
\\  \no &\times&
\Bigg\langle \Big(\mfrak{W}_{f_{n},m_{n},r_{n}}^{(\bar{f},\bar{m},\bar{r});a_{n}}(x_{p_{n}}^{(n)};k_{p})\Big)^{\dag}
\;{\scrscr\times}  \ph{\,}^{a_{n}}\widehat{\langle x_{p_{n}}^{(n)}}|
\Delta\!\hat{\mscr{H}}_{M_{n};N_{n-1}}|\widehat{y_{q_{n-1}}^{(n-1)}\rangle}^{a_{n-1}}{\scrscr\times}
\ph{\,}^{a_{n-1}}\widehat{\langle y_{q_{n-1}}^{(n-1)}}|\hat{G}_{N_{n-1};M_{n-1}}^{(0)}|
\widehat{x_{p_{n-1}}^{(n-1)}\rangle}^{a_{n-1}}\;\times \\ \no &\times&
\ph{\,}^{a_{n-1}}\widehat{\langle x_{p_{n-1}}^{(n-1)}}|
\Delta\!\hat{\mscr{H}}_{M_{n-1};N_{n-2}}|\widehat{y_{q_{n-2}}^{(n-2)}\rangle}^{a_{n-2}}{\scrscr\times}
\ph{\,}^{a_{n-2}}\widehat{\langle y_{q_{n-2}}^{(n-2)}}|\hat{G}_{N_{n-2};M_{n-2}}^{(0)}|
\widehat{x_{p_{n-2}}^{(n-2)}\rangle}^{a_{n-2}}\;\times\;\ldots  \\ \no &\times&\ldots\;
\ph{\,}^{a_{2}}\widehat{\langle x_{p_{2}}^{(2)}}|
\Delta\!\hat{\mscr{H}}_{M_{2};N_{1}}|\widehat{y_{q_{1}}^{(1)}\rangle}^{a_{1}}{\scrscr\times}
\ph{\,}^{a_{1}}\widehat{\langle y_{q_{1}}^{(1)}}|\hat{G}_{N_{1};M_{1}}^{(0)}|
\widehat{x_{p_{1}}^{(1)}\rangle}^{a_{1}}\;\times  \\ \no &\times&
\ph{\,}^{a_{1}}\widehat{\langle x_{p_{1}}^{(1)}}|
\Delta\!\hat{\mscr{H}}_{M_{1};N_{0}}|\widehat{y_{q_{0}}^{(0)}\rangle}^{a_{0}}{\scrscr\times}
\ph{\,}^{a_{0}}\widehat{\langle y_{q_{0}}^{(0)}}|\hat{G}_{N_{0};M_{0}}^{(0)}|
\widehat{x_{p_{0}}^{(0)}\rangle}^{a_{0}} {\scrscr\times}\;
\mfrak{W}_{f_{0},m_{0},r_{0}}^{(\bar{f},\bar{m},\bar{r});a_{0}}(x_{p_{0}}^{(0)};k_{p})
\Bigg\rangle_{\feynBV}\;\times \;
\delta_{a_{n}a_{0}}\mbox{ (compare Eq. (\ref{sd_44}))}\;;  \\ \no &&
\big(\mbox{summations in (\ref{sd_51}) without }\;a_{n}=a_{n-1}=\ldots=a_{1}=a_{0}\;!\big)\;\;.
\eeq
Direct application of rules (\ref{sd_25}-\ref{sd_28}) and commutators
\(\boldsymbol{[}\boldsymbol{\hat{\pp}_{p,\kappa}}\;\boldsymbol{,}\;\hat{G}^{(0)}\boldsymbol{]_{-}}\) (\ref{sd_31})
leads to relation (\ref{sd_52}) where the Green functions
\(\hat{G}^{(0)}\), averaged by background path integral (\ref{s3_59},\ref{sd_2}), propagate the plane wave
states \(\mfrak{W}_{f_{0},m_{0},r_{0}}^{(\bar{f},\bar{m},\bar{r});a_{0}}(x_{p_{0}}^{(0)};k_{p})\) on the right-hand
side to the left, analogous 'source field' state
\((\:\mfrak{W}_{f_{n},m_{n},r_{n}}^{(\bar{f},\bar{m},\bar{r});a_{n}}(x_{p_{n}}^{(n)};k_{p})\:)^{\dag}\) under subsequent
action of the gradient operators \(\Delta\!\hat{\mscr{H}}_{M_{i+1};M_{i}}^{a_{i+1}a_{i}}(x_{p})\).
This process of propagation with \(\hat{G}^{(0)}\) is complicated by the action of unsaturated gradients
\(\boldsymbol{\hat{\pp}_{p,\kappa}}\) onto the background potential matrix \(\hat{\mfrak{V}}(x_{p})\) (\ref{sd_8})
within the doubled Green functions. This requires intensive application of commutator relations (\ref{sd_31}) and
results into additional transport coefficients with derivatives
\((\hat{\pp}_{p,\kappa}\mcal{V}_{\alpha;\kappa_{1}}(x_{p})\,)\).
The Kronecker delta \(\delta_{a_{n}a_{0}}\) for the anomalous doubling is taken into account by the
anti-commutator \(\frac{1}{2}\hat{S}\boldsymbol{\{}\mbox{ (n-th order term) }\boldsymbol{,}\:\hat{S}\boldsymbol{\}_{+}}\)
(\ref{sd_44}) for every term '\(n\)' in the expansion of the logarithm.
Every factor with the gradient operator \(\Delta\!\hat{\mscr{H}}\;\hat{G}^{(0)}\) at n-th order
\((\,\Delta\!\hat{\mscr{H}}\;\hat{G}^{(0)}\,)^{n}\) is transformed by the commutator
\(\boldsymbol{[}\boldsymbol{\hat{\pp}_{p,\kappa}}\:\boldsymbol{,}\:\hat{G}^{(0)}\boldsymbol{]_{-}}\) (\ref{sd_31})
so that one gains an extended relation (\ref{sd_52}) for
\(\langle\mscr{A}_{DET}[\hat{T},\feynV;\hat{\mscr{J}}\equiv 0]\rangle_{\feynbv}\);
nevertheless, it is still necessary to transform unsaturated gradients \(\boldsymbol{\hat{\pp}_{p,\kappa}}\)
and Green functions \(\hat{G}^{(0)}\) in (\ref{sd_52}) by further application of extended commutator relations
as \(\boldsymbol{[}\boldsymbol{\hat{\pp}_{p,\kappa}}\:\boldsymbol{,}\:\hat{G}^{(0)}\boldsymbol{]_{-}}\) (\ref{sd_31})
in order to derive the complete, proper set of transport coefficients with the background potential
\(\mcal{V}_{\alpha;\kappa_{1}}(x_{p})\). After substitution of (\ref{sd_33}-\ref{sd_38}) into (\ref{sd_51}), we attain
the action \(\langle\mscr{A}_{DET}[\hat{T},\feynV;\hat{\mscr{J}}\equiv0]\rangle_{\feynbv}\) with a single transformation
of the commutator (\ref{sd_31}) as an intermediate step to be reduced by further applications of (\ref{sd_31}) and
propagation rules (\ref{sd_25}-\ref{sd_28})
\beq \lb{sd_52}
\lefteqn{\Big\langle\mscr{A}_{DET}\big[\hat{T},\feynV;\hat{\mscr{J}}\equiv 0\big]\Big\rangle_{\feynBV} =
\int\limits_{C}^{k_{max}}d^{4}\!k_{p}\;\eta_{p}\;\mcal{N}_{k}\sum_{\bar{f}=1}^{N_{f}}
\sum_{\bar{m}=1}^{4}\sum_{\bar{r}=1}^{N_{c}} \frac{1}{2}\sum_{n=1}^{\infty}\frac{(-1)^{n+1}}{n} \;\times}  \\ \no  &\times&
\int_{C}d^{4}\!x_{p_{n}}^{(n)}\;\mcal{N}\sum_{a_{n}=1,2}\sum_{M_{n}}
\prod_{i=0}^{n-1}\bigg(\int_{C}d^{4}x_{p_{i}}^{(i)}\;\mcal{N}\sum_{a_{i}=1,2}\sum_{M_{i}}\sum_{N_{i}}\bigg) \Bigg\langle
\Big(\mfrak{W}_{f_{n},m_{n},r_{n}}^{(\bar{f},\bar{m},\bar{r});a_{n}}(x_{p_{n}}^{(n)};k_{p})\Big)^{\dag} \times \\ \no &\times&
\frac{\hat{S}}{2}\boldsymbol{\Bigg\{}
\bigg[\Big(\hat{\mfrak{k}}(x_{p_{n}}^{(n)})+\delta\hat{\mfrak{V}}(x_{p_{n}}^{(n)})\Big)_{M_{n};N_{n-1}}^{a_{n}a_{n-1}}\;
\ph{\,}^{a_{n-1}}\widehat{\langle x_{p_{n}}^{(n)}}|\hat{G}_{\scrscr N_{n-1};M_{n-1}}^{(0)}|
\widehat{x_{p_{n-1}}^{(n-1)}\rangle}^{a_{n-1}} +\\  \no &-&
\delta\hat{\mscr{K}}_{M_{n};N_{n-1}}^{\boldsymbol{\kappa};a_{n}a_{n-1}}(x_{p_{n}}^{(n)})\;
\Big(\ph{\,}^{a_{n-1}}\widehat{\langle x_{p_{n}}^{(n)}}|\hat{G}^{(0)}\:
\big(\hat{\pp}_{p,\boldsymbol{\kappa}}\hat{\mfrak{V}}\big)\;\hat{G}^{(0)}|
\widehat{x_{p_{n-1}}^{(n-1)}\rangle}^{a_{n-1}}\Big)_{N_{n-1};M_{n-1}} + \\ \no &+&
\delta\hat{\mscr{K}}_{M_{n};N_{n-1}}^{\boldsymbol{\kappa};a_{n}a_{n-1}}(x_{p_{n}}^{(n)})\;
\ph{\,}^{a_{n-1}}\widehat{\langle x_{p_{n}}^{(n)}}|\hat{G}_{N_{n-1};M_{n-1}}^{(0)}|
\widehat{x_{p_{n-1}}^{(n-1)}\rangle}^{a_{n-1}}\boldsymbol{\hat{\pp}_{p_{n-1},\kappa}^{(n-1)}} \bigg]\times  \\ \no &\times&
\bigg[\Big(\hat{\mfrak{k}}(x_{p_{n-1}}^{(n-1)})+\delta\hat{\mfrak{V}}(x_{p_{n-1}}^{(n-1)})\Big)_{M_{n-1};N_{n-2}}^{a_{n-1}a_{n-2}}\;
\ph{\,}^{a_{n-2}}\widehat{\langle x_{p_{n-1}}^{(n-1)}}|\hat{G}_{\scrscr N_{n-2};M_{n-2}}^{(0)}|
\widehat{x_{p_{n-2}}^{(n-2)}\rangle}^{a_{n-2}} +\\  \no &-&
\delta\hat{\mscr{K}}_{M_{n-1};N_{n-2}}^{\boldsymbol{\lambda};a_{n-1}a_{n-2}}(x_{p_{n-1}}^{(n-1)})\;
\Big(\ph{\,}^{a_{n-2}}\widehat{\langle x_{p_{n-1}}^{(n-1)}}|\hat{G}^{(0)}\:
\big(\hat{\pp}_{p,\boldsymbol{\lambda}}\hat{\mfrak{V}}\big)\;\hat{G}^{(0)}|
\widehat{x_{p_{n-2}}^{(n-2)}\rangle}^{a_{n-2}}\Big)_{N_{n-2};M_{n-2}} + \\ \no &+&
\delta\hat{\mscr{K}}_{M_{n-1};N_{n-2}}^{\boldsymbol{\lambda};a_{n-1}a_{n-2}}(x_{p_{n-1}}^{(n-1)})\;
\ph{\,}^{a_{n-2}}\widehat{\langle x_{p_{n-1}}^{(n-1)}}|\hat{G}_{N_{n-2};M_{n-2}}^{(0)}|
\widehat{x_{p_{n-2}}^{(n-2)}\rangle}^{a_{n-2}}\boldsymbol{\hat{\pp}_{p_{n-2},\lambda}^{(n-2)}} \bigg]\times  \\ \no &\times&
\ldots\ldots \;\times  \\ \no &\times&
\bigg[\Big(\hat{\mfrak{k}}(x_{p_{2}}^{(2)})+\delta\hat{\mfrak{V}}(x_{p_{2}}^{(2)})\Big)_{M_{2};N_{1}}^{a_{2}a_{1}}\;
\ph{\,}^{a_{1}}\widehat{\langle x_{p_{2}}^{(2)}}|\hat{G}_{\scrscr N_{1};M_{1}}^{(0)}|
\widehat{x_{p_{1}}^{(1)}\rangle}^{a_{1}} +\\  \no &-&
\delta\hat{\mscr{K}}_{M_{2};N_{1}}^{\boldsymbol{\mu};a_{2}a_{1}}(x_{p_{2}}^{(2)})\;
\Big(\ph{\,}^{a_{1}}\widehat{\langle x_{p_{2}}^{(2)}}|\hat{G}^{(0)}\:
\big(\hat{\pp}_{p,\boldsymbol{\mu}}\hat{\mfrak{V}}\big)\;\hat{G}^{(0)}|
\widehat{x_{p_{1}}^{(1)}\rangle}^{a_{1}}\Big)_{N_{1};M_{1}} + \\ \no &+&
\delta\hat{\mscr{K}}_{M_{2};N_{1}}^{\boldsymbol{\mu};a_{2}a_{1}}(x_{p_{2}}^{(2)})\;
\ph{\,}^{a_{1}}\widehat{\langle x_{p_{2}}^{(2)}}|\hat{G}_{N_{1};M_{1}}^{(0)}|
\widehat{x_{p_{1}}^{(1)}\rangle}^{a_{1}}\boldsymbol{\hat{\pp}_{p_{1},\mu}^{(1)}} \bigg]\times  \\ \no &\times&
\bigg[\Big(\hat{\mfrak{k}}(x_{p_{1}}^{(1)})+\delta\hat{\mfrak{V}}(x_{p_{1}}^{(1)})\Big)_{M_{1};N_{0}}^{a_{1}a_{0}}\;
\ph{\,}^{a_{0}}\widehat{\langle x_{p_{1}}^{(1)}}|\hat{G}_{\scrscr N_{0};M_{0}}^{(0)}|
\widehat{x_{p_{0}}^{(0)}\rangle}^{a_{0}} +\\  \no &-&
\delta\hat{\mscr{K}}_{M_{1};N_{0}}^{\boldsymbol{\nu};a_{1}a_{0}}(x_{p_{1}}^{(1)})\;
\Big(\ph{\,}^{a_{0}}\widehat{\langle x_{p_{1}}^{(1)}}|\hat{G}^{(0)}\:
\big(\hat{\pp}_{p,\boldsymbol{\nu}}\hat{\mfrak{V}}\big)\;\hat{G}^{(0)}|
\widehat{x_{p_{0}}^{(0)}\rangle}^{a_{0}}\Big)_{N_{0};M_{0}} + \\ \no &+&
\delta\hat{\mscr{K}}_{M_{1};N_{0}}^{\boldsymbol{\nu};a_{1}a_{0}}(x_{p_{1}}^{(1)})\;
\ph{\,}^{a_{0}}\widehat{\langle x_{p_{1}}^{(1)}}|\hat{G}_{N_{0};M_{0}}^{(0)}|
\widehat{x_{p_{0}}^{(0)}\rangle}^{a_{0}}\boldsymbol{\hat{\pp}_{p_{0},\nu}^{(0)}} \bigg]
\;\boldsymbol{,}\;\hat{S}
\boldsymbol{\Bigg\}_{+}}
\mfrak{W}_{f_{0},m_{0},r_{0}}^{(\bar{f},\bar{m},\bar{r});a_{0}}(x_{p_{0}}^{(0)};k_{p})
\Bigg\rangle_{\feynBV}\;;
\\  \no && \big(\mbox{summations in (\ref{sd_52}) without }\;a_{n}=a_{n-1}=\ldots=a_{1}=a_{0}\;!\big)\;\;.
\eeq
Due to Derrick's theorem \cite{raja},
we analyze the orders up to four gradients in (\ref{sd_52}) for a final effective Lagrangian.
As one remembers relation (\ref{sd_41}), we have to exclude the completely diagonal term
of the anomalous doubling '\(a_{n}=a_{n-1}=\ldots=a_{1}=a_{0}\)' in (\ref{sd_52}) because this term only
contributes vanishing measure in the spacetime integrations due to contradictory propagations of the generalized Heaviside functions.
The first order term \(n=1\) (\ref{sd_53}) is not effected by this vanishing term
of contradictory propagations and simply reduces to the part
\(\delta\hat{\mfrak{k}}(x_{p})=\hat{T}^{-1}(x_{p})\:\hat{S}\:(\,\hat{\beta}\,\feynd_{p}\hat{T}(x_{p})\,)\) (\ref{sd_7})
with saturated derivatives because the unsaturated derivative operators \(\boldsymbol{\hat{\pp}_{p,\kappa}}\) with
\(\delta\hat{\mscr{K}}^{\boldsymbol{\kappa}}(x_{p})\) only lead to vanishing four-momentum integrals
due to an anti-symmetric integrand. However, the commutators
\(\boldsymbol{[}\boldsymbol{\hat{\pp}_{p,\kappa}}\:\boldsymbol{,}\:\hat{G}^{(0)}\boldsymbol{]_{-}}\) (\ref{sd_31})
are involved in every factor of \((\,\Delta\!\hat{\mscr{H}}\;\hat{G}^{(0)}\,)^{n}\) at order 'n' where
the unsaturated gradient \(\boldsymbol{\hat{\pp}_{p,\kappa}}\) acts on the background potentials
\(\mcal{V}_{\alpha;\kappa_{1}}(x_{p})\) within the Green functions. This causes the additional part with
'\(-\langle(\,\hat{\pp}_{p,\kappa}\mcal{V}_{\alpha;\kappa_{1}}(x_{p})\,)\rangle_{\feynbv}\;\:
\delta\hat{\mscr{K}}^{\boldsymbol{\kappa}}(x_{p})\;\hat{V}_{\alpha}^{\kappa_{1}}\)' at first order 'n=1'
\beq \lb{sd_53}
\lefteqn{\Big\langle\mscr{A}_{DET}\big[\hat{T},\feynV;\hat{\mscr{J}}\equiv 0;n=1\big]
\Big\rangle_{\feynBV} = k_{max}^{4}\;\mcal{N}_{k}\;\mcal{N}\;\frac{1}{2}\;
\int_{C}d^{4}\!x_{p}\;\times} \\ \no  &\times& \bigg\langle\hspace*{0.5cm}\TRALL\bigg[\frac{\hat{S}}{2}
\boldsymbol{\bigg\{}
\Big(\delta\hat{\mfrak{k}}(x_{p})-\big(\hat{\pp}_{p,\kappa}\mcal{V}_{\alpha;\kappa_{1}}(x_{p})\,\big)\:
\delta\hat{\mscr{K}}^{\boldsymbol{\kappa}}(x_{p})\;\hat{V}_{\alpha}^{\kappa_{1}}\Big)
\;\boldsymbol{,}\;\hat{S}\boldsymbol{\bigg\}_{+}}\bigg]\bigg\rangle_{\feynBV} =
k_{max}^{4}\;\mcal{N}_{k}\;\mcal{N}\;\frac{1}{2}\;
\int_{C}d^{4}\!x_{p}\;\times \\  \no &\times&
\bigg(\hspace*{0.5cm}\TRALL\Big[
\hat{T}^{-1}(x_{p})\;\hat{S}\;\big(\hat{\beta}\:\feynd_{p}\hat{T}(x_{p})\,\big)\Big] -
\Big\langle\big(\hat{\pp}_{p,\kappa}\mcal{V}_{\alpha;\kappa_{1}}(x_{p})\,\big)\Big\rangle_{\feynBV}\:
\hspace*{0.5cm}\TRALL\Big[
\hat{T}^{-1}(x_{p})\;\hat{S}\;\hat{\beta}\:\hat{\gamma}^{\kappa}\;\hat{T}(x_{p})\;
\hat{V}_{\alpha}^{\kappa_{1}}\Big]\bigg)\;.
\eeq
The remaining terms in (\ref{sd_53})
contain the cutoff momentum \(k_{max}^{4}\) to the power of four as a scale for other gradually varying gradients.
As we proceed to higher order gradients as \(n=2\), one also finds terms with \(k_{max}^{6}\) which can be
neglected in a lowest order momentum-energy expansion (with \(k_{max}\ll 1\) in dimensionless units).

The second order term \(n=2\) (\ref{sd_54}) is calculated with detailed description of the various steps.
In order to exclude vanishing contradictory propagation of \(\hat{G}^{(0)}\) with \(a_{2}=a_{1}=a_{0}\),
one has to choose the two commutator terms (\ref{sd_40}) for the off-diagonal blocks
\(\Delta\!\hat{\mscr{H}}_{M_{2};M_{1}}^{a_{2}\neq a_{1}}(x_{p})\),
\(\Delta\!\hat{\mscr{H}}_{M_{1};M_{0}}^{a_{1}\neq a_{0}}(x_{p})\) in the BCS-sector of the anomalous doubled space.
We substitute the total gradient operators \(\Delta\!\hat{\mscr{H}}(x_{p})\) by the part \(\delta\hat{\mfrak{h}}(x_{p})\) (\ref{sd_7})
with unsaturated derivatives and potential matrix \(\hat{\mfrak{V}}(x_{p})\) (\ref{sd_8}) and by the term
\(\delta\hat{\mscr{K}}^{\boldsymbol{\mu}}(x_{p})\:\boldsymbol{\hat{\pp}_{p,\mu}}\) with unsaturated gradient operators.
The latter act onto the anomalous doubled plane wave states of four-momentum \(k_{p}\).
We also incorporate the commutator (\ref{sd_32}) and the relations (\ref{sd_33}-\ref{sd_38}) as abbreviating symbols
in the second order term '$n=2$' (\ref{sd_54}) with application of propagation rules (\ref{sd_25}-\ref{sd_28})
\beq \lb{sd_54}
\lefteqn{\Big\langle\mscr{A}_{DET}\big[\hat{T},\feynV;\hat{\mscr{J}}\equiv 0;n=2\big]
\Big\rangle_{\feynBV} =\sum_{p=\pm}\int\limits^{|k_{p}|<k_{max}/2}\hspace*{-0.4cm}d^{4}\!k_{p}\;\mcal{N}_{k}\;\mcal{N}
\sum_{\bar{M}}\bigg(-\frac{1}{4}\bigg)\int_{-\infty}^{+\infty}d^{4}\!x_{p}\;\eta_{p}\;\times }  \\ \no &\times&
\Bigg\langle\Big(\mfrak{W}_{M_{2}}^{(\bar{M});a_{2}}(x_{p};k_{p})\Big)^{\dag}
\frac{\hat{S}}{2}\:\boldsymbol{\Bigg\{} \frac{\hat{S}}{2}
\boldsymbol{\Big[}\Delta\hat{\mfrak{h}}(x_{p})\;\boldsymbol{,}
\;\hat{S}\boldsymbol{\Big]_{-}} \;
\frac{\hat{S}}{2}
\boldsymbol{\Big[}\Delta\hat{\mfrak{h}}(x_{p})\;\boldsymbol{,}
\;\hat{S}\boldsymbol{\Big]_{-}}  +  \frac{\hat{S}}{2}
\boldsymbol{\Big[} \delta\hat{\mscr{K}}^{\boldsymbol{\kappa}}(x_{p})\;\boldsymbol{,}
\;\hat{S}\boldsymbol{\Big]_{-}}  \times   \\ \no  &\times&  \frac{\hat{S}}{2}
\boldsymbol{\bigg[}\Big(\hat{\pp}_{p,\boldsymbol{\kappa}}\Delta\hat{\mfrak{h}}(x_{p})\Big) -
\delta\hat{\mfrak{h}}(x_{p})\big(\hat{\pp}_{p,\boldsymbol{\kappa}}\hat{\mfrak{V}}(x_{p})\,\big)+
 \\ \no &+&
\delta\hat{\mscr{K}}^{\boldsymbol{\lambda}}(x_{p})
\boldsymbol{\Big\{}
\big(\hat{\pp}_{p,\boldsymbol{\kappa}}\hat{\mfrak{V}}(x_{p})\,\big)\;\boldsymbol{,}\;
\big(\hat{\pp}_{p,\boldsymbol{\lambda}}\hat{\mfrak{V}}(x_{p})\,\big)\boldsymbol{\Big\}_{+}}
\;\boldsymbol{,}
\;\hat{S}\boldsymbol{\bigg]_{-}}\;\boldsymbol{,}\;\hat{S}\boldsymbol{\Bigg\}}_{\boldsymbol{+};M_{2};M_{0}}^{a_{2}=a_{0}}
\mfrak{W}_{M_{0}}^{(\bar{M});a_{0}}(x_{p};k_{p}) +
\\ \no &+&
\Big(\mfrak{W}_{M_{2}}^{(\bar{M});a_{2}}(x_{p};k_{p})\Big)^{\dag}
\frac{\hat{S}}{2}\:\boldsymbol{\bigg\{} \frac{\hat{S}}{2}
\boldsymbol{\Big[}\delta\hat{\mscr{K}}^{\boldsymbol{\kappa}}(x_{p})\;\boldsymbol{,}
\;\hat{S}\boldsymbol{\Big]_{-}}  \frac{\hat{S}}{2}
\boldsymbol{\Big[}\delta\hat{\mscr{K}}^{\boldsymbol{\lambda}}(x_{p})\;\boldsymbol{,}
\;\hat{S}\boldsymbol{\Big]_{-}}  \;\boldsymbol{,}\;
\hat{S}\boldsymbol{\bigg\}}_{\boldsymbol{+};M_{2};M_{0}}^{a_{2}=a_{0}}
\boldsymbol{\hat{\pp}_{p,\kappa}}
\boldsymbol{\hat{\pp}_{p,\lambda}}\mfrak{W}_{M_{0}}^{(\bar{M});a_{0}}(x_{p};k_{p}) + \\ \no &+&
\mbox{'linear' 'unsaturated' gradient operator terms}\Bigg\rangle_{\feynBV} \;.
\eeq
Since the four-momentum integrals vanish for the case with linear, anti-symmetric integrand, one only has an additional
integration of quadratic order \(k_{p,\kappa}\:k_{p,\lambda}\) from the action of unsaturated gradient operators onto
plane wave states
\beq \lb{sd_55}
\lefteqn{\Big\langle\mscr{A}_{DET}\big[\hat{T},\feynV;\hat{\mscr{J}}\equiv 0;n=2\big]
\Big\rangle_{\feynBV} =\sum_{p=\pm}\int\limits^{|k_{p}|<k_{max}/2}\hspace*{-0.4cm}d^{4}\!k_{p}\;\mcal{N}_{k}\;\mcal{N}
\sum_{\bar{M}}\bigg(-\frac{1}{4}\bigg)\int_{-\infty}^{+\infty}d^{4}\!x_{p}\;\eta_{p}\;\times }  \\ \no &\times&
\Bigg\langle\Big(\mfrak{W}_{M_{2}}^{(\bar{M});a_{2}}(x_{p};k_{p})\Big)^{\dag}
\frac{\hat{S}}{2}\:\boldsymbol{\Bigg\{} \frac{\hat{S}}{2}
\boldsymbol{\Big[}\Delta\hat{\mfrak{h}}(x_{p})\;\boldsymbol{,}
\;\hat{S}\boldsymbol{\Big]_{-}} \;
\frac{\hat{S}}{2}
\boldsymbol{\Big[}\Delta\hat{\mfrak{h}}(x_{p})\;\boldsymbol{,}
\;\hat{S}\boldsymbol{\Big]_{-}}  +  \frac{\hat{S}}{2}
\boldsymbol{\Big[} \delta\hat{\mscr{K}}^{\boldsymbol{\kappa}}(x_{p})\;\boldsymbol{,}
\;\hat{S}\boldsymbol{\Big]_{-}}  \times   \\ \no  &\times&  \frac{\hat{S}}{2}
\boldsymbol{\bigg[}\Big(\hat{\pp}_{p,\boldsymbol{\kappa}}\Delta\hat{\mfrak{h}}(x_{p})\Big) -
\delta\hat{\mfrak{h}}(x_{p})\big(\hat{\pp}_{p,\boldsymbol{\kappa}}\hat{\mfrak{V}}(x_{p})\,\big)+
 \\ \no &+&
\delta\hat{\mscr{K}}^{\boldsymbol{\lambda}}(x_{p})
\boldsymbol{\Big\{}
\big(\hat{\pp}_{p,\boldsymbol{\kappa}}\hat{\mfrak{V}}(x_{p})\,\big)\;\boldsymbol{,}\;
\big(\hat{\pp}_{p,\boldsymbol{\lambda}}\hat{\mfrak{V}}(x_{p})\,\big)\boldsymbol{\Big\}_{+}}
\;\boldsymbol{,}
\;\hat{S}\boldsymbol{\bigg]_{-}}\;\boldsymbol{,}\;\hat{S}\boldsymbol{\Bigg\}}_{\boldsymbol{+};M_{2};M_{0}}^{a_{2}=a_{0}}
\mfrak{W}_{M_{0}}^{(\bar{M});a_{0}}(x_{p};k_{p}) +
\Big(\mfrak{W}_{M_{2}}^{(\bar{M});a_{2}}(x_{p};k_{p})\Big)^{\dag} \times  \\ \no &\times&
\frac{\hat{S}}{2}\:\boldsymbol{\bigg\{} \frac{\hat{S}}{2}
\boldsymbol{\Big[}\delta\hat{\mscr{K}}^{\boldsymbol{\kappa}}(x_{p})\;\boldsymbol{,}
\;\hat{S}\boldsymbol{\Big]_{-}}  \frac{\hat{S}}{2}
\boldsymbol{\Big[}\delta\hat{\mscr{K}}^{\boldsymbol{\lambda}}(x_{p})\;\boldsymbol{,}
\;\hat{S}\boldsymbol{\Big]_{-}}  \;\boldsymbol{,}\;
\hat{S}\boldsymbol{\bigg\}}_{\boldsymbol{+};M_{2};M_{0}}^{a_{2}=a_{0}}
\;\Big(\im\,k_{p,\kappa}\;\im\,k_{p,\lambda}\Big)\:\hat{S}^{2}\;
\mfrak{W}_{M_{0}}^{(\bar{M});a_{0}}(x_{p};k_{p})\Bigg\rangle_{\feynBV} \;.
\eeq
Apart from the power of four term \(k_{max}^{4}\) for the parts without action onto plane waves, one therefore
attains an additional order of \(k_{max}^{6}/12\) for the part following from unsaturated gradient operator
actions onto plane wave states
\be \lb{sd_56}
\sum_{p=\pm}\int\limits^{|k_{p}|<k_{max}/2}d^{4}\!k_{p}\;\Big(\im\,k_{p,\kappa}\;\im\,k_{p,\lambda}\Big)=-\eta_{\kappa\lambda}\;
\;\frac{k_{max}^{6}}{12}\;\;\sum_{p=\pm}\;\;.
\ee
After insertion of (\ref{sd_56}) into (\ref{sd_55}), we obtain relation (\ref{sd_57}) with additional, relative
order \(k_{max}^{2}/12\) for the parts resulting from the unsaturated gradient operator action onto
\(\mfrak{W}_{M_{0}}^{(\bar{M});a_{0}}(x_{p};k_{p})\). As already stated, we assume gradually varying
BCS-terms in the coset matrices \(\hat{T}(x_{p})=\exp\{-\hat{Y}(x_{p})\,\}\)
with \(k_{max}\ll 1\) in dimensionless units
\beq \lb{sd_57}
\lefteqn{\Big\langle\mscr{A}_{DET}\big[\hat{T},\feynV;\hat{\mscr{J}}\equiv 0;n=2\big]
\Big\rangle_{\feynBV} =k_{max}^{4}\;\mcal{N}_{k}\;\mcal{N}\;\bigg(-\frac{1}{4}\bigg)
\int_{C}d^{4}\!x_{p}\;\times }  \\ \no &\times& \Bigg\{
\Bigg\langle \hspace*{0.5cm}\TRALL\Bigg[ \frac{\hat{S}}{2}\:\boldsymbol{\bigg\{}
\frac{\hat{S}}{2}
\boldsymbol{\Big[}\Delta\!\hat{\mfrak{h}}(x_{p})\;\boldsymbol{,}
\;\hat{S}\boldsymbol{\Big]_{-}}\frac{\hat{S}}{2}
\boldsymbol{\Big[}\Delta\!\hat{\mfrak{h}}(x_{p})\;\boldsymbol{,}
\;\hat{S}\boldsymbol{\Big]_{-}}\;\boldsymbol{,}\;\hat{S}\boldsymbol{\bigg\}_{+}} + \\ \no &+&
 \frac{\hat{S}}{2}\:\boldsymbol{\Bigg\{}\frac{\hat{S}}{2}
\boldsymbol{\Big[}
\delta\hat{\mscr{K}}^{\boldsymbol{\kappa}}(x_{p})\;\boldsymbol{,}
\;\hat{S}\boldsymbol{\Big]_{-}}\frac{\hat{S}}{2}
\boldsymbol{\bigg[}\Big(\hat{\pp}_{p,\boldsymbol{\kappa}}\Delta\!\hat{\mfrak{h}}(x_{p})\Big)-\delta\hat{\mfrak{h}}(x_{p})\:
\big(\hat{\pp}_{p,\boldsymbol{\kappa}}\hat{\mfrak{V}}(x_{p})\,\big)+    \\ \no &+&
\delta\hat{\mscr{K}}^{\boldsymbol{\lambda}}(x_{p})\;
\boldsymbol{\Big\{}
\big(\hat{\pp}_{p,\boldsymbol{\kappa}}\hat{\mfrak{V}}(x_{p})\,\big)\;\boldsymbol{,}\;
\big(\hat{\pp}_{p,\boldsymbol{\lambda}}\hat{\mfrak{V}}(x_{p})\,\big)\boldsymbol{\Big\}_{+}}
\;\boldsymbol{,}
\;\hat{S}\boldsymbol{\bigg]_{-}}\;\boldsymbol{,}\;\hat{S}\boldsymbol{\Bigg\}}_{\boldsymbol{+}}\Bigg]\Bigg\rangle_{\feynBV} \\ \no &-&
\eta_{\kappa\lambda}\;\frac{k_{max}^{2}}{12}\;\hspace*{0.5cm}\TRALL\Bigg[ \frac{\hat{S}}{2}\:\boldsymbol{\bigg\{}
\frac{\hat{S}}{2}\boldsymbol{\Big[}
\delta\hat{\mscr{K}}^{\boldsymbol{\kappa}}(x_{p})\;\boldsymbol{,}
\;\hat{S}\boldsymbol{\Big]_{-}}\frac{\hat{S}}{2}
\boldsymbol{\Big[}\delta\hat{\mscr{K}}^{\boldsymbol{\lambda}}(x_{p})\;\boldsymbol{,}
\;\hat{S}\boldsymbol{\Big]_{-}}\;\boldsymbol{,}\;\hat{S}\boldsymbol{\bigg\}_{+}} \Bigg] \Bigg\}\;;
\eeq
\be \lb{sd_58}
k_{max} \ll 1\;\;\mbox{ dimensionless units !}\;.
\ee
This assumption allows to simplify the second order term (\ref{sd_57}) to (\ref{sd_59}) without action of unsaturated
gradient operators onto plane wave states
\beq \lb{sd_59}
\lefteqn{\Big\langle\mscr{A}_{DET}\big[\hat{T},\feynV;\hat{\mscr{J}}\equiv 0;n=2\big]
\Big\rangle_{\feynBV} =k_{max}^{4}\;\mcal{N}_{k}\;\mcal{N}\;\bigg(-\frac{1}{4}\bigg)
\int_{C}d^{4}\!x_{p}\;\times }  \\ \no &\times&
\Bigg\langle \hspace*{0.5cm}\TRALL\Bigg[ \frac{\hat{S}}{2}\:\boldsymbol{\bigg\{}
\frac{\hat{S}}{2}
\boldsymbol{\Big[}\Delta\!\hat{\mfrak{h}}(x_{p})\;\boldsymbol{,}
\;\hat{S}\boldsymbol{\Big]_{-}}\frac{\hat{S}}{2}
\boldsymbol{\Big[}\Delta\!\hat{\mfrak{h}}(x_{p})\;\boldsymbol{,}
\;\hat{S}\boldsymbol{\Big]_{-}}\;\boldsymbol{,}\;\hat{S}\boldsymbol{\bigg\}_{+}} + \\ \no &+&
 \frac{\hat{S}}{2}\:\boldsymbol{\Bigg\{}\frac{\hat{S}}{2}
\boldsymbol{\Big[}
\delta\hat{\mscr{K}}^{\boldsymbol{\kappa}}(x_{p})\;\boldsymbol{,}
\;\hat{S}\boldsymbol{\Big]_{-}}\frac{\hat{S}}{2}
\boldsymbol{\bigg[}\Big(\hat{\pp}_{p,\boldsymbol{\kappa}}\Delta\!\hat{\mfrak{h}}(x_{p})\Big)-\delta\hat{\mfrak{h}}(x_{p})\:
\big(\hat{\pp}_{p,\boldsymbol{\kappa}}\hat{\mfrak{V}}(x_{p})\,\big)+    \\ \no &+&
\delta\hat{\mscr{K}}^{\boldsymbol{\lambda}}(x_{p})\;
\boldsymbol{\Big\{}
\big(\hat{\pp}_{p,\boldsymbol{\kappa}}\hat{\mfrak{V}}(x_{p})\,\big)\;\boldsymbol{,}\;
\big(\hat{\pp}_{p,\boldsymbol{\lambda}}\hat{\mfrak{V}}(x_{p})\,\big)\boldsymbol{\Big\}_{+}}
\;\boldsymbol{,}
\;\hat{S}\boldsymbol{\bigg]_{-}}\;\boldsymbol{,}\;\hat{S}\boldsymbol{\Bigg\}}_{\boldsymbol{+}}\Bigg]\Bigg\rangle_{\feynBV}  \;.
\eeq
The anti-commutator \(\frac{1}{2}\hat{S}\:\boldsymbol{\{}\ldots\:\boldsymbol{,}\,\hat{S}\boldsymbol{\}_{+}}\) (\ref{sd_44})
for \(\delta_{a_{2}a_{0}}\) can be removed in (\ref{sd_59}) so that the second order term finally reduces to (\ref{sd_60}) because
the traces of (\ref{sd_59}) regard only terms \(a_{2}=a_{0}\) in any case of values for \(a_{2}\), \(a_{0}\)
\beq \lb{sd_60}
\lefteqn{\Big\langle\mscr{A}_{DET}\big[\hat{T},\feynV;\hat{\mscr{J}}\equiv 0;n=2\big]
\Big\rangle_{\feynBV} =k_{max}^{4}\;\mcal{N}_{k}\;\mcal{N}\;\bigg(-\frac{1}{4}\bigg)
\int_{C}d^{4}\!x_{p}\;\times }  \\ \no &\times&
\Bigg\langle \hspace*{0.5cm}\TRALL\bigg[
\frac{\hat{S}}{2}
\boldsymbol{\Big[}\Delta\!\hat{\mfrak{h}}(x_{p})\;\boldsymbol{,}
\;\hat{S}\boldsymbol{\Big]_{-}}\frac{\hat{S}}{2}
\boldsymbol{\Big[}\Delta\!\hat{\mfrak{h}}(x_{p})\;\boldsymbol{,}
\;\hat{S}\boldsymbol{\Big]_{-}}\bigg] + \hspace*{0.5cm}\TRALL\bigg[
\frac{\hat{S}}{2}\boldsymbol{\Big[}\delta\hat{\mscr{K}}^{\boldsymbol{\kappa}}(x_{p})\;\boldsymbol{,}
\;\hat{S}\boldsymbol{\Big]_{-}} \times \\ \no &\times& \frac{\hat{S}}{2}
\boldsymbol{\Big[}\Big(\hat{\pp}_{p,\boldsymbol{\kappa}}\Delta\!\hat{\mfrak{h}}(x_{p})\Big)-\delta\hat{\mfrak{h}}(x_{p})\:
\big(\hat{\pp}_{p,\boldsymbol{\kappa}}\hat{\mfrak{V}}(x_{p})\,\big) +
\delta\hat{\mscr{K}}^{\boldsymbol{\lambda}}(x_{p})\;
\boldsymbol{\Big\{}
\big(\hat{\pp}_{p,\boldsymbol{\kappa}}\hat{\mfrak{V}}(x_{p})\,\big)\;\boldsymbol{,}\;
\big(\hat{\pp}_{p,\boldsymbol{\lambda}}\hat{\mfrak{V}}(x_{p})\,\big)\boldsymbol{\Big\}_{+}}
\;\boldsymbol{,}
\;\hat{S}\boldsymbol{\Big]_{-}}\bigg]\Bigg\rangle_{\feynBV}
\\   \no &\approx& k_{max}^{4}\;\mcal{N}_{k}\;\mcal{N}\;\bigg(-\frac{1}{4}\bigg)
\int_{C}d^{4}\!x_{p}\;\times \\  \no &\times&
\Bigg\langle \hspace*{0.5cm}\TRALL\bigg[
\frac{\hat{S}}{2}
\boldsymbol{\Big[}\delta\hat{\mfrak{h}}(x_{p})\;\boldsymbol{,}
\;\hat{S}\boldsymbol{\Big]_{-}}\frac{\hat{S}}{2}
\boldsymbol{\Big[}\delta\hat{\mfrak{h}}(x_{p})\;\boldsymbol{,}
\;\hat{S}\boldsymbol{\Big]_{-}}  +
\frac{\hat{S}}{2}\boldsymbol{\Big[}\delta\hat{\mscr{K}}^{\boldsymbol{\kappa}}(x_{p})\;\boldsymbol{,}
\;\hat{S}\boldsymbol{\Big]_{-}}  \frac{\hat{S}}{2}
\boldsymbol{\Big[}\Big(\hat{\pp}_{p,\boldsymbol{\kappa}}\delta\hat{\mfrak{h}}(x_{p})\Big)
\;\boldsymbol{,}
\;\hat{S}\boldsymbol{\Big]_{-}}\bigg]\Bigg\rangle_{\feynBV} \;.
\eeq
If one disregards surface phenomena at the boundary of the nucleus, we can further assume a constant effective
background potential \(\feynV(x_{p})\) so that one can reduce the second order gradient expansion (\ref{sd_60})
to the simplified relation in the last two lines of (\ref{sd_60}) for the bulk of a nucleus.

In the remainder we restrict to the case of constant background potentials \(\feynV(x_{p})\) so that
we can eventually specify the third order term (\ref{sd_61}) of gradients. Since the completely block-diagonal
propagation of Green functions \(\hat{G}^{(0)}\) vanishes for
\(\mscr{A}_{DET}[\hat{T},\feynV;\hat{\mscr{J}}\equiv 0]\rangle_{\feynbv}\) due to contradictory propagations with
time contour extended Heaviside functions, one has to consider two off-diagonal terms
\(\Delta\!\hat{\mscr{H}}^{a_{3}\neq a_{2}}(x_{p})\), \(\Delta\!\hat{\mscr{H}}^{a_{2}\neq a_{1}}(x_{p})\)
and one block-diagonal gradient operator
\(\Delta\!\hat{\mscr{H}}^{a_{1}= a_{0}}(x_{p})\). According to (\ref{sd_39},\ref{sd_40}),
this amounts to two commutator terms and one anti-commutator part. Since there are three possibilities of combinatorial
ordering of two commutators and one anti-commutator terms, one has to start from relation (\ref{sd_61}) for the
third order part \(n=3\) with additional, combinatorial factor '3'
\beq \lb{sd_61}
\lefteqn{\Big\langle\mscr{A}_{DET}\big[\hat{T},\feynV;\hat{\mscr{J}}\equiv 0;n=3\big]
\Big\rangle_{\feynBV} =\sum_{p=\pm}\int\limits^{|k_{p}|<k_{max}/2}d^{4}\!k_{p}\;\mcal{N}_{k}\;\mcal{N}
\sum_{\bar{M}}3\;\frac{1}{2}\;\frac{1}{3}
\int_{-\infty}^{+\infty}d^{4}\!x_{p}\;\eta_{p}\;\times }  \\ \no &\times&
\Bigg\langle
\Big(\mfrak{W}_{M_{3}}^{(\bar{M});a_{3}}(x_{p};k_{p})\Big)^{\dag}
 \frac{\hat{S}}{2}\:\boldsymbol{\bigg\{}
\frac{\hat{S}}{2}
\boldsymbol{\Big[}\Delta\!\hat{\mscr{H}}_{M_{3};M_{2}}^{a_{3}\neq a_{2}}(x_{p})\;\boldsymbol{,}
\;\hat{S}\boldsymbol{\Big]_{-}}
\frac{\hat{S}}{2}
\boldsymbol{\Big[}\Delta\!\hat{\mscr{H}}_{M_{2};M_{1}}^{a_{2}\neq a_{1}}(x_{p})\;\boldsymbol{,}
\;\hat{S}\boldsymbol{\Big]_{-}}\;\times   \\ \no &\times&
\frac{\hat{S}}{2}
\boldsymbol{\Big\{}\Delta\!\hat{\mscr{H}}_{M_{1};M_{0}}^{a_{1}=a_{0}}(x_{p})\;\boldsymbol{,}
\;\hat{S}\boldsymbol{\Big\}_{+}}\;\boldsymbol{,}\;\hat{S}\boldsymbol{\bigg\}_{+}}
\mfrak{W}_{M_{0}}^{(\bar{M});a_{0}}(x_{p};k_{p})\Bigg\rangle_{\feynBV} \;.
\eeq
Equation (\ref{sd_62}) follows from further transformation to the parts \(\delta\hat{\mfrak{h}}(x_{p})\),
\(\delta\hat{\mscr{K}}^{\boldsymbol{\mu}}(x_{p})\:\boldsymbol{\hat{\pp}_{p,\mu}}\) with saturated
and unsaturated derivatives where we only keep the order \(k_{max}^{4}\) and
neglect higher orders of \(k_{max}^{\geq6}\) following from action of unsaturated gradient operators
onto plane wave states
\beq \lb{sd_62}
\lefteqn{\Big\langle\mscr{A}_{DET}\big[\hat{T},\feynV;\hat{\mscr{J}}\equiv 0;n=3\big]
\Big\rangle_{\feynBV} \approx k_{max}^{4}\;\mcal{N}_{k}\;\mcal{N}\;\frac{1}{2}
\int_{C}d^{4}\!x_{p}\;\times }  \\ \no &\times&\Bigg[
\Bigg\langle \hspace*{0.5cm}\TRALL\Bigg[ \frac{\hat{S}}{2}\:\boldsymbol{\bigg\{}
\bigg(\frac{\hat{S}}{2}\boldsymbol{\Big[}\delta\hat{\mfrak{h}}(x_{p})\;\boldsymbol{,}
\;\hat{S}\boldsymbol{\Big]_{-}}\bigg)^{2}
\frac{\hat{S}}{2}
\boldsymbol{\Big\{}\delta\hat{\mfrak{h}}(x_{p})\;\boldsymbol{,}
\;\hat{S}\boldsymbol{\Big\}_{+}}\;\boldsymbol{,}\;\hat{S}\boldsymbol{\bigg\}_{+}}\Bigg]
\Bigg\rangle_{\feynBV} +  \\  \no &+&\Bigg\langle\hspace*{0.5cm} \TRALL\Bigg[\frac{\hat{S}}{2}\:\boldsymbol{\bigg\{}
\frac{\hat{S}}{2}\boldsymbol{\Big[}\delta\hat{\mfrak{h}}(x_{p})\;\boldsymbol{,}\;
\hat{S}\boldsymbol{\Big]_{-}}\frac{\hat{S}}{2}\boldsymbol{\Big[}
\delta\hat{\mscr{K}}^{\boldsymbol{\mu}}(x_{p})\;\boldsymbol{,}
\;\hat{S}\boldsymbol{\Big]_{-}}\frac{\hat{S}}{2}
\boldsymbol{\Big[}\Big(\hat{\pp}_{p,\mu}\delta\hat{\mfrak{h}}(x_{p})\Big)\;\boldsymbol{,}
\;\hat{S}\boldsymbol{\Big]_{+}}\;\boldsymbol{,}\;\hat{S}\boldsymbol{\bigg\}_{+}}\Bigg]
\Bigg\rangle_{\feynBV} +
\\  \no &+&\Bigg\langle \hspace*{0.5cm}\TRALL\Bigg[\frac{\hat{S}}{2}\:\boldsymbol{\bigg\{}
\frac{\hat{S}}{2}\boldsymbol{\Big[}\delta\hat{\mscr{K}}^{\boldsymbol{\nu}}(x_{p})\;\boldsymbol{,}\;
\hat{S}\boldsymbol{\Big]_{-}}\bigg(\hat{\pp}_{p,\nu}\frac{\hat{S}}{2}\boldsymbol{\Big[}
\delta\hat{\mfrak{h}}(x_{p})\;\boldsymbol{,}
\;\hat{S}\boldsymbol{\Big]_{-}}\frac{\hat{S}}{2}
\boldsymbol{\Big\{}\delta\hat{\mfrak{h}}(x_{p})\;\boldsymbol{,}
\;\hat{S}\boldsymbol{\Big\}_{+}}\bigg)\;\boldsymbol{,}\;\hat{S}\boldsymbol{\bigg\}_{+}}\Bigg]
\Bigg\rangle_{\feynBV} +
\\  \no &+&\Bigg\langle \hspace*{0.5cm}\TRALL\Bigg[\frac{\hat{S}}{2}\:\boldsymbol{\bigg\{}
\frac{\hat{S}}{2}\boldsymbol{\Big[}\delta\hat{\mscr{K}}^{\boldsymbol{\nu}}(x_{p})\;\boldsymbol{,}\;
\hat{S}\boldsymbol{\Big]_{-}}\bigg(\hat{\pp}_{p,\nu}\frac{\hat{S}}{2}\boldsymbol{\Big[}
\delta\hat{\mscr{K}}^{\boldsymbol{\mu}}(x_{p})\;\boldsymbol{,}
\;\hat{S}\boldsymbol{\Big]_{-}}\frac{\hat{S}}{2}
\boldsymbol{\Big\{}\Big(\hat{\pp}_{p,\mu}\delta\hat{\mfrak{h}}(x_{p})\Big)\;\boldsymbol{,}
\;\hat{S}\boldsymbol{\Big\}_{+}}\bigg)\;\boldsymbol{,}\;\hat{S}\boldsymbol{\bigg\}_{+}}\Bigg]
\Bigg\rangle_{\feynBV} \Bigg]\;.
\eeq
The anti-commutator (\ref{sd_44}) for the Kronecker delta \(\delta_{a_{3}a_{0}}\) in the anomalous doubled
space is removed as in (\ref{sd_59},\ref{sd_60}) because the traces of (\ref{sd_62}) already reduce to
the diagonal parts \(a_{3}=a_{0}\) in any case. Therefore, one accomplishes relation (\ref{sd_63}) for the third
order gradient term under approximation of \(k_{max}\ll 1\) with \(k_{max}^{4}\) as the lowest order
four-momentum scale
\beq \lb{sd_63}
\lefteqn{\Big\langle\mscr{A}_{DET}\big[\hat{T},\feynV;\hat{\mscr{J}}\equiv 0;n=3\big]
\Big\rangle_{\feynBV} \approx k_{max}^{4}\;\mcal{N}_{k}\;\mcal{N}\;\frac{1}{2}
\int_{C}d^{4}\!x_{p}\;\times }  \\ \no &\times&\Bigg[
\Bigg\langle \hspace*{0.5cm}\TRALL\Bigg[
\bigg(\frac{\hat{S}}{2}\boldsymbol{\Big[}\delta\hat{\mfrak{h}}(x_{p})\;\boldsymbol{,}
\;\hat{S}\boldsymbol{\Big]_{-}}\bigg)^{2}
\frac{\hat{S}}{2}
\boldsymbol{\Big\{}\delta\hat{\mfrak{h}}(x_{p})\;\boldsymbol{,}
\;\hat{S}\boldsymbol{\Big\}_{+}}\Bigg]
\Bigg\rangle_{\feynBV} +  \\  \no &+&\Bigg\langle \hspace*{0.5cm}\TRALL\Bigg[
\frac{\hat{S}}{2}\boldsymbol{\Big[}\delta\hat{\mfrak{h}}(x_{p})\;\boldsymbol{,}\;
\hat{S}\boldsymbol{\Big]_{-}}\frac{\hat{S}}{2}\boldsymbol{\Big[}
\delta\hat{\mscr{K}}^{\boldsymbol{\mu}}(x_{p})\;\boldsymbol{,}
\;\hat{S}\boldsymbol{\Big]_{-}}\frac{\hat{S}}{2}
\boldsymbol{\Big[}\Big(\hat{\pp}_{p,\mu}\delta\hat{\mfrak{h}}(x_{p})\Big)\;\boldsymbol{,}
\;\hat{S}\boldsymbol{\Big\}_{+}}\Bigg]
\Bigg\rangle_{\feynBV} +
\\  \no &+&\Bigg\langle \hspace*{0.5cm}\TRALL\Bigg[
\frac{\hat{S}}{2}\boldsymbol{\Big[}\delta\hat{\mscr{K}}^{\boldsymbol{\nu}}(x_{p})\;\boldsymbol{,}\;
\hat{S}\boldsymbol{\Big]_{-}}\bigg(\hat{\pp}_{p,\nu}\frac{\hat{S}}{2}\boldsymbol{\Big[}
\delta\hat{\mfrak{h}}(x_{p})\;\boldsymbol{,}
\;\hat{S}\boldsymbol{\Big]_{-}}\frac{\hat{S}}{2}
\boldsymbol{\Big\{}\delta\hat{\mfrak{h}}(x_{p})\;\boldsymbol{,}
\;\hat{S}\boldsymbol{\Big\}_{+}}\bigg)\Bigg]
\Bigg\rangle_{\feynBV} +
\\  \no &+&\Bigg\langle \hspace*{0.5cm}\TRALL\Bigg[
\frac{\hat{S}}{2}\boldsymbol{\Big[}\delta\hat{\mscr{K}}^{\boldsymbol{\nu}}(x_{p})\;\boldsymbol{,}\;
\hat{S}\boldsymbol{\Big]_{-}}\bigg(\hat{\pp}_{p,\nu}\frac{\hat{S}}{2}\boldsymbol{\Big[}
\delta\hat{\mscr{K}}^{\boldsymbol{\mu}}(x_{p})\;\boldsymbol{,}
\;\hat{S}\boldsymbol{\Big]_{-}}\frac{\hat{S}}{2}
\boldsymbol{\Big\{}\Big(\hat{\pp}_{p,\mu}\delta\hat{\mfrak{h}}(x_{p})\Big)\;\boldsymbol{,}
\;\hat{S}\boldsymbol{\Big\}_{+}}\bigg)\Bigg]
\Bigg\rangle_{\feynBV} \Bigg]\;.
\eeq
The fourth order derivative part stabilizes the static energy configurations so that we have to extract from
relation (\ref{sd_52}) all terms with four gradients (under the assumption of a constant background potential);
but in addition, these gradient terms of order four encompass various combinatorial factors.
In general the n-th order gradient term of \(\langle\mscr{A}_{DET}[\hat{T},\feynV;\hat{\mscr{J}}\equiv0;n]\rangle_{\feynbv}\)
with $2m$ off-diagonal operators '\(-\frac{1}{2}\hat{S}\,
\boldsymbol{[}\Delta\!\hat{\mscr{H}}_{M_{i+1};M_{i}}^{a_{i+1}\neq a_{i}}(x_{p})\;\boldsymbol{,}\;
\hat{S}\boldsymbol{]_{-}}\)' and \(n-2m\) block diagonal gradient parts '\(\frac{1}{2}\hat{S}\,
\boldsymbol{\{}\Delta\!\hat{\mscr{H}}_{M_{j+1};M_{j}}^{a_{j+1}= a_{j}}(x_{p})\;\boldsymbol{,}\;
\hat{S}\boldsymbol{\}_{+}}\)' is included by the coefficient \(C_{(2m)}^{(n)}\)-times which is determined
by the binomial coefficient \({n \choose 2m}\). However, these \(C_{(2m)}^{(n)}={n \choose 2m}\) combinations are also realized
in various patterns under the trace operations so that we have to introduce combinatorial sub-factors \(C_{(2m),l}^{(n)}\)
whose sum \(\sum_{l}C_{(2m),l}^{(n)}\) is equivalent to the total number of combinations \(C_{(2m)}^{(n)}\) for the n-th order
gradient term with \(2m\) gradients in the BCS- or off-diagonal, anomalous-doubled sectors. Since one has to exclude the
completely block diagonal terms according to opposite propagations of Green functions, we have one combination
\(C_{(2m=4)}^{(n=4)}={4\choose 4}=1\) for four off-diagonal parts \(\Delta\!\hat{\mscr{H}}_{M_{i+1};M_{i}}^{a_{i+1}\neq a_{i}}(x_{p})\)
and six combinations \(C_{(2m=2)}^{(n=4)}={4\choose 2}=6\) for the case of two BCS-sector parts and two block diagonal terms.
The latter six combinations differ by the pattern (\ref{sd_64}) with \(C_{(2m=2),l=1}^{(n=4)}=2\) realizations under the trace
\be\lb{sd_64}
\Delta\!\hat{\mscr{H}}_{M_{4};M_{3}}^{a_{4}\neq a_{3}}(x_{p})\;\;\;
\Delta\!\hat{\mscr{H}}_{M_{3};M_{2}}^{a_{3}= a_{2}}(x_{p})\;\;\;
\Delta\!\hat{\mscr{H}}_{M_{2};M_{1}}^{a_{2}\neq a_{1}}(x_{p})\;\;\;
\Delta\!\hat{\mscr{H}}_{M_{1};M_{0}}^{a_{1}= a_{0}}(x_{p})\;,
\ee
and four combinations \(C_{(2m=2),l=2}^{(n=4)}=4\) for the pattern (\ref{sd_65})
\be\lb{sd_65}
\Delta\!\hat{\mscr{H}}_{M_{4};M_{3}}^{a_{4}\neq a_{3}}(x_{p})\;\;\;
\Delta\!\hat{\mscr{H}}_{M_{3};M_{2}}^{a_{3}\neq a_{2}}(x_{p})\;\;\;
\Delta\!\hat{\mscr{H}}_{M_{2};M_{1}}^{a_{2}= a_{1}}(x_{p})\;\;\;
\Delta\!\hat{\mscr{H}}_{M_{1};M_{0}}^{a_{1}= a_{0}}(x_{p})\;,
\ee
yielding the total number of six combinations \(C_{(2m=2)}^{(n=4)}={4\choose 2}=6\)
\be \lb{sd_66}
\sum_{l=1,2} C_{(2m=2),l}^{(n=4)}=\underbrace{C_{(2m=2),l=1}^{(n=4)}}_{=2}+\underbrace{C_{(2m=2),l=2}^{(n=4)}}_{=4}=
C_{(2m=2)}^{(n=4)}={4\choose 2}=6 \;.
\ee
The various combinations with different sub-patterns for the ordering of block diagonal density parts and  BCS-sectors
therefore comprise the corresponding anti-commutator '+' and commutator '-' operators (\ref{sd_39},\ref{sd_40}) which
we abbreviate by the common symbol \(\boldsymbol{[}\ldots\;\boldsymbol{,}\;\ldots\boldsymbol{\}_{I,J,K,L=\pm}}\)
with indices \(I,J,K,L=\pm\) specifying the anti-commutator '+' or commutator '-'
instead of the symbols \(\boldsymbol{\{}\ldots\;\boldsymbol{,}\;\ldots\boldsymbol{\}_{I,J,K,L=+}}\) (\ref{sd_39}) and
\(\boldsymbol{[}\ldots\;\boldsymbol{,}\;\ldots\boldsymbol{]_{I,J,K,L=-}}\) (\ref{sd_40}), respectively
\beq\lb{sd_67}
\lefteqn{\Big\langle\mscr{A}_{DET}\big[\hat{T},\feynV;\hat{\mscr{J}}\equiv 0;n=4\big]
\Big\rangle_{\feynBV} \approx k_{max}^{4}\;\mcal{N}_{k}\;\mcal{N}\;\bigg(-\frac{1}{8}\bigg)
\int_{C}d^{4}\!x_{p}\; \sum_{m=1,2}\sum_{l=1}^{3-m}C_{(2m),l}^{(n=4)} \times } \\ \no &\times&
\Bigg\langle \hspace*{0.6cm}\TRALL\Bigg[
\frac{\hat{S}}{2}\boldsymbol{\Big[}\delta\hat{\mfrak{h}}(x_{p})\;\boldsymbol{,}
\;\hat{S}\boldsymbol{\Big\}_{I_{m_{l}}}}\;
\frac{\hat{S}}{2}
\boldsymbol{\Big[}\delta\hat{\mfrak{h}}(x_{p})\;\boldsymbol{,}
\;\hat{S}\boldsymbol{\Big\}_{J_{m_{l}}}}\;
\frac{\hat{S}}{2}\boldsymbol{\Big[}\delta\hat{\mfrak{h}}(x_{p})\;\boldsymbol{,}
\;\hat{S}\boldsymbol{\Big\}_{K_{m_{l}}}}\;
\frac{\hat{S}}{2}
\boldsymbol{\Big[}\delta\hat{\mfrak{h}}(x_{p})\;\boldsymbol{,}
\;\hat{S}\boldsymbol{\Big\}_{L_{m_{l}}}}  +  \\ \no &+&
\frac{\hat{S}}{2}\boldsymbol{\Big[}\delta\hat{\mfrak{h}}(x_{p})\;\boldsymbol{,}
\;\hat{S}\boldsymbol{\Big\}_{I_{m_{l}}}}\;
\frac{\hat{S}}{2}
\boldsymbol{\Big[}\delta\hat{\mfrak{h}}(x_{p})\;\boldsymbol{,}
\;\hat{S}\boldsymbol{\Big\}_{J_{m_{l}}}}\;
\frac{\hat{S}}{2}\boldsymbol{\Big[}\delta\hat{\mscr{K}}^{\boldsymbol{\lambda}}(x_{p})\;\boldsymbol{,}
\;\hat{S}\boldsymbol{\Big\}_{K_{m_{l}}}}\;\Big(\hat{\pp}_{p,\lambda}
\frac{\hat{S}}{2}
\boldsymbol{\Big[}\delta\hat{\mfrak{h}}(x_{p})\;\boldsymbol{,}
\;\hat{S}\boldsymbol{\Big\}_{L_{m_{l}}}}\Big) +  \\ \no &+&
\frac{\hat{S}}{2}\boldsymbol{\Big[}\delta\hat{\mfrak{h}}(x_{p})\;\boldsymbol{,}
\;\hat{S}\boldsymbol{\Big\}_{I_{m_{l}}}}\;
\frac{\hat{S}}{2}
\boldsymbol{\Big[}\delta\hat{\mscr{K}}^{\boldsymbol{\mu}}(x_{p})\;\boldsymbol{,}
\;\hat{S}\boldsymbol{\Big\}_{J_{m_{l}}}}\;\bigg(\hat{\pp}_{p,\mu}
\frac{\hat{S}}{2}\boldsymbol{\Big[}\delta\hat{\mfrak{h}}(x_{p})\;\boldsymbol{,}
\;\hat{S}\boldsymbol{\Big\}_{K_{m_{l}}}}\;
\frac{\hat{S}}{2}
\boldsymbol{\Big[}\delta\hat{\mfrak{h}}(x_{p})\;\boldsymbol{,}
\;\hat{S}\boldsymbol{\Big\}_{L_{m_{l}}}}\bigg) +  \\ \no &+&
\frac{\hat{S}}{2}\boldsymbol{\Big[}\delta\hat{\mfrak{h}}(x_{p})\;\boldsymbol{,}
\;\hat{S}\boldsymbol{\Big\}_{I_{m_{l}}}}
\frac{\hat{S}}{2}
\boldsymbol{\Big[}\delta\hat{\mscr{K}}^{\boldsymbol{\mu}}(x_{p})\;\boldsymbol{,}
\;\hat{S}\boldsymbol{\Big\}_{J_{m_{l}}}}\bigg(\hat{\pp}_{p,\mu}
\frac{\hat{S}}{2}\boldsymbol{\Big[}\delta\hat{\mscr{K}}^{\boldsymbol{\lambda}}(x_{p})\;\boldsymbol{,}
\;\hat{S}\boldsymbol{\Big\}_{K_{m_{l}}}}\Big(\hat{\pp}_{p,\lambda}
\frac{\hat{S}}{2}
\boldsymbol{\Big[}\delta\hat{\mfrak{h}}(x_{p})\;\boldsymbol{,}
\;\hat{S}\boldsymbol{\Big\}_{L_{m_{l}}}}\Big)\bigg) +  \\ \no &+&
\frac{\hat{S}}{2}\boldsymbol{\Big[}\delta\hat{\mscr{K}}^{\boldsymbol{\nu}}(x_{p})\;\boldsymbol{,}
\;\hat{S}\boldsymbol{\Big\}_{I_{m_{l}}}}\bigg(\hat{\pp}_{p,\nu}
\frac{\hat{S}}{2}
\boldsymbol{\Big[}\delta\hat{\mfrak{h}}(x_{p})\;\boldsymbol{,}
\;\hat{S}\boldsymbol{\Big\}_{J_{m_{l}}}}\;
\frac{\hat{S}}{2}\boldsymbol{\Big[}\delta\hat{\mfrak{h}}(x_{p})\;\boldsymbol{,}
\;\hat{S}\boldsymbol{\Big\}_{K_{m_{l}}}}\;
\frac{\hat{S}}{2}
\boldsymbol{\Big[}\delta\hat{\mfrak{h}}(x_{p})\;\boldsymbol{,}
\;\hat{S}\boldsymbol{\Big\}_{L_{m_{l}}}} \bigg) +  \\ \no &+&
\frac{\hat{S}}{2}\boldsymbol{\Big[}\delta\hat{\mscr{K}}^{\boldsymbol{\nu}}(x_{p})\;\boldsymbol{,}
\;\hat{S}\boldsymbol{\Big\}_{I_{m_{l}}}}\;\bigg(\hat{\pp}_{p,\nu}
\frac{\hat{S}}{2}
\boldsymbol{\Big[}\delta\hat{\mfrak{h}}(x_{p})\;\boldsymbol{,}
\;\hat{S}\boldsymbol{\Big\}_{J_{m_{l}}}}\;
\frac{\hat{S}}{2}\boldsymbol{\Big[}\delta\hat{\mscr{K}}^{\boldsymbol{\lambda}}(x_{p})\;\boldsymbol{,}
\;\hat{S}\boldsymbol{\Big\}_{K_{m_{l}}}}\;\Big(\hat{\pp}_{p,\lambda}
\frac{\hat{S}}{2}
\boldsymbol{\Big[}\delta\hat{\mfrak{h}}(x_{p})\;\boldsymbol{,}
\;\hat{S}\boldsymbol{\Big\}_{L_{m_{l}}}}\Big)\bigg) +  \\ \no &+&
\frac{\hat{S}}{2}\boldsymbol{\Big[}\delta\hat{\mscr{K}}^{\boldsymbol{\nu}}(x_{p})\;\boldsymbol{,}
\;\hat{S}\boldsymbol{\Big\}_{I_{m_{l}}}}\;\bigg(\hat{\pp}_{p,\nu}
\frac{\hat{S}}{2}
\boldsymbol{\Big[}\delta\hat{\mscr{K}}^{\boldsymbol{\mu}}(x_{p})\;\boldsymbol{,}
\;\hat{S}\boldsymbol{\Big\}_{J_{m_{l}}}}\;\bigg(\hat{\pp}_{p,\mu}
\frac{\hat{S}}{2}\boldsymbol{\Big[}\delta\hat{\mfrak{h}}(x_{p})\;\boldsymbol{,}
\;\hat{S}\boldsymbol{\Big\}_{K_{m_{l}}}}\;
\frac{\hat{S}}{2}
\boldsymbol{\Big[}\delta\hat{\mfrak{h}}(x_{p})\;\boldsymbol{,}
\;\hat{S}\boldsymbol{\Big\}_{L_{m_{l}}}}\bigg)\bigg) +  \\ \no &+&
\frac{\hat{S}}{2}\boldsymbol{\Big[}\delta\hat{\mscr{K}}^{\boldsymbol{\nu}}(x_{p})\;\boldsymbol{,}
\hat{S}\boldsymbol{\Big\}_{I_{m_{l}}}}\Bigg(\hat{\pp}_{p,\nu}
\frac{\hat{S}}{2}
\boldsymbol{\Big[}\delta\hat{\mscr{K}}^{\boldsymbol{\mu}}(x_{p})\;\boldsymbol{,}
\hat{S}\boldsymbol{\Big\}_{J_{m_{l}}}}\bigg(\hat{\pp}_{p,\mu}
\frac{\hat{S}}{2}\boldsymbol{\Big[}\delta\hat{\mscr{K}}^{\boldsymbol{\lambda}}(x_{p})\;\boldsymbol{,}
\hat{S}\boldsymbol{\Big\}_{K_{m_{l}}}}\Big(\hat{\pp}_{p,\lambda}
\frac{\hat{S}}{2}
\boldsymbol{\Big[}\delta\hat{\mfrak{h}}(x_{p})\;\boldsymbol{,}
\hat{S}\boldsymbol{\Big\}_{L_{m_{l}}}}\Big)\bigg)\Bigg) \Bigg]\Bigg\rangle_{\feynBV}\;;
\eeq
\be \lb{sd_68}
\bea{rclrclcrclrclrclrclrcl}
m&=&1; & l&=&1;&\rightarrow &C_{(2m=2),l=1}^{(n=4)}&=&2; & 'I_{m_{l}}\hspace*{-0.2cm}&=&\hspace*{-0.2cm}-';
  & 'J_{m_{l}}\hspace*{-0.2cm}&=&\hspace*{-0.2cm}+'; & 'K_{m_{l}}\hspace*{-0.2cm}&=&\hspace*{-0.2cm}-';  &
L_{m_{l}}\hspace*{-0.2cm}&=&\hspace*{-0.2cm}+';  \\
m&=&1; & l&=&2;&\rightarrow &C_{(2m=2),l=2}^{(n=4)}&=&4; & 'I_{m_{l}}\hspace*{-0.2cm}&=&\hspace*{-0.2cm}-';
  & J_{m_{l}}\hspace*{-0.2cm}&=&\hspace*{-0.2cm}-'; & K_{m_{l}}\hspace*{-0.2cm}&=&\hspace*{-0.2cm}+';  &
L_{m_{l}}\hspace*{-0.2cm}&=&\hspace*{-0.2cm}+';  \\
m&=&2; & l&=&1;&\rightarrow &C_{(2m=4),(l=1)}^{(n=4)}&=&1; & 'I_{m_{l}}\hspace*{-0.2cm}&=&\hspace*{-0.2cm}-';
  & 'J_{m_{l}}\hspace*{-0.2cm}&=&\hspace*{-0.2cm}-'; & K_{m_{l}}\hspace*{-0.2cm}&=&\hspace*{-0.2cm}-';  &
'L_{m_{l}}\hspace*{-0.2cm}&=&\hspace*{-0.2cm}-'\;.
\eea
\ee

\subsection{Lagrangian of the effective action with the bilinear, fermionic source fields} \lb{sd3}

The gradient expansion of \(\langle\mscr{A}_{J_{\psi}}[\hat{T},\feynV;\hat{\mscr{J}}]\rangle_{\feynbv}\)
takes a considerably simpler form than that of \(\langle\mscr{A}_{DET}[\hat{T},\feynV;\hat{\mscr{J}}]\rangle_{\feynbv}\)
because it suffices to use for the propagation of anomalous doubled wavefunctions and fields, starting on the
right-hand side with \(J_{\psi;M}^{a}(x_{p})\), the anomalous doubled Green function of the background potential
\(\langle\feynV\rangle_{\mbox{\scz(\ref{s3_59})}}\) following from a saddle point approximation
\beq \lb{sd_69}
\Big\langle\mscr{A}_{J_{\psi}}[\hat{T},\feynV;\hat{\mscr{J}}]\Big\rangle_{\feynBV} &=&
\frac{1}{2}\int_{C}d^{4}\!x_{p}\;d^{4}\!y_{q}\;\times  \\ \no
\lefteqn{\hspace*{-1.3cm}\times\;J_{\psi;N}^{\dag,b}(y_{q})\;\hat{I}\bigg(\hat{T}(y_{q})\;
\Big\langle\hat{\mscr{O}}_{N\ppr;M\ppr}^{\boldsymbol{-1};b\ppr a\ppr}(y_{q},x_{p})\Big\rangle_{\feynBV}\;\hat{T}^{-1}(x_{p})-
\hat{\mscr{H}}_{N;M}^{\boldsymbol{-1};ba}(y_{q},x_{p})\bigg)\;\hat{I}\;J_{\psi;M}^{a}(x_{p}) \;; }
 \\  \lb{sd_70}\hspace*{-0.6cm}
\Big\langle\hat{\mscr{O}}_{N;M}^{ba}(y_{q},x_{p})\Big\rangle_{\feynBV}
&=&\hspace*{-0.3cm}\bigg\{\langle\hat{\mscr{H}}\rangle_{\feynbv}+
\Big(\hat{T}^{-1}\langle\hat{\mscr{H}}\rangle_{\feynbv}\hat{T}-\langle\hat{\mscr{H}}\rangle_{\feynbv}\Big)+
\hat{T}^{-1}\;\hat{I}\;\hat{S}\;\eta_{q}
\frac{\hat{\mscr{J}}_{N\ppr;M\ppr}^{b\ppr a\ppr}(y_{q},x_{p})}{\mcal{N}}\eta_{p}\;\hat{S}\;\hat{I}\;
\hat{T}\bigg\}_{N;M}^{ba}\hspace*{-0.73cm}(y_{q},x_{p})_{.}
\eeq
Repeated application of the propagation rules (\ref{sd_25}-\ref{sd_28}) then determines the various coefficients
with the background potential \(\langle\feynV(x_{p})\rangle_{\mbox{\scz(\ref{s3_59})}}\). In analogy to the
lowest order momentum expansion of \(\langle\mscr{A}_{DET}[\hat{T},\feynV;\hat{\mscr{J}}]\rangle_{\feynbv}\)
in section \ref{sd2}, we neglect any resulting derivative terms of the fermionic source fields
\((\hat{\pp}_{p,\kappa}J_{\psi;M}^{a}(x_{p})\,)\approx0\) from the action of 'unsaturated' gradients
\(\boldsymbol{\hat{\pp}_{p,\kappa}}\) and take into account the commutator for the action of these
onto the background potentials within the Green functions
\beq \lb{sd_71}
\lefteqn{\Big\langle\mscr{A}_{J_{\psi}}[\hat{T},\feynV;\hat{\mscr{J}}\equiv0]\Big\rangle_{\feynBV} =
\frac{1}{2}\int_{C}d^{4}\!x_{p}\;d^{4}\!y_{q}\;J_{\psi;N}^{\dag,b}(y_{q})\;\hat{I}\bigg(\hat{T}(y_{q})\;\times}
\\ \no &\times&
\bigg(\sum_{n=1}^{\infty}(-1)^{n}
\bigg[\langle\hat{\mscr{H}}\rangle_{\feynbv}^{\boldsymbol{-1}}\Big(\Delta\!\langle\hat{\mscr{H}}\rangle_{\feynbv}\:\;
\langle\hat{\mscr{H}}\rangle_{\feynbv}^{\boldsymbol{-1}}\Big)^{n}\bigg]_{N\ppr;M\ppr}^{b\ppr a\ppr}
\hspace*{-0.4cm}(y_{q},x_{p})\bigg)\;
\;\hat{T}^{-1}(x_{p})\bigg)_{N;M}^{ba}\;\hat{I}\;J_{\psi;M}^{a}(x_{p}) \\  \no &=&
\frac{1}{2}\sum_{n=1}^{\infty}(-1)^{n}\int_{C}d^{4}\!x_{p}\;d^{4}\!y_{q}\;J_{\psi;N}^{\dag,b}(y_{q})\;\hat{I}\bigg(\hat{T}(y_{q})\;\times  \\ \no &\times&
\bigg[\langle\hat{\mscr{H}}\rangle_{\feynbv}^{\boldsymbol{-1}}
\Big(\delta\hat{\mfrak{h}}\;\:\langle\hat{\mscr{H}}\rangle_{\feynbv}^{\boldsymbol{-1}}+
\delta\hat{\mscr{K}}^{\boldsymbol{\kappa}}\boldsymbol{\hat{\pp}_{p,\kappa}}\;\:
\langle\hat{\mscr{H}}\rangle_{\feynbv}^{\boldsymbol{-1}}\Big)^{n}\bigg]_{N\ppr;M\ppr}^{b\ppr a\ppr}
\hspace*{-0.4cm}(y_{q},x_{p})\;\hat{T}^{-1}(x_{p})\bigg)_{N;M}^{ba}\;\hat{I}\;J_{\psi;M}^{a}(x_{p}) \;;  \\ \lb{sd_72} &&
\big(\hat{\pp}_{p,\kappa}J_{\psi;M}^{a}(x_{p})\,\big) \approx 0  \;;\;\;\;\;
\boldsymbol{\Big[}\boldsymbol{\hat{\pp}_{p,\kappa}}\;\boldsymbol{,}\;\langle\hat{\mscr{H}}\rangle_{\feynbv}^{\boldsymbol{-1}}
\boldsymbol{\Big]_{-}} \simeq
-\;\langle\hat{\mscr{H}}\rangle_{\feynbv}^{\boldsymbol{-1}}
\;\underbrace{\big(\hat{\pp}_{p,\boldsymbol{\kappa}}\mcal{V}_{\alpha;\kappa_{1}}\big)\;
\hat{V}_{\alpha}^{\kappa_{1}} }_{(\,\hat{\pp}_{p,\boldsymbol{\kappa}}\hat{\mfrak{V}}\,)}\;
\langle\hat{\mscr{H}}\rangle_{\feynbv}^{\boldsymbol{-1}} \;.
\eeq
Since the action \(\langle\mscr{A}_{J_{\psi}}[\hat{T},\feynV;\hat{\mscr{J}}]\rangle_{\feynbv}\) does not contain
a back-propagation to the same spacetime point as \(\langle\mscr{A}_{DET}[\hat{T},\feynV;\hat{\mscr{J}}]\rangle_{\feynbv}\),
due to the missing of corresponding traces, any combinatorial factors cannot occur in the expansion of
\(\langle\mscr{A}_{J_{\psi}}[\hat{T},\feynV;\hat{\mscr{J}}]\rangle_{\feynbv}\). Therefore, we can simply list the
final approximated gradient expansion of \(\langle\mscr{A}_{J_{\psi}}[\hat{T},\feynV;\hat{\mscr{J}}]\rangle_{\feynbv}\)
in relation (\ref{sd_73}) which can be further reduced by the assumption of constant background potentials
\((\hat{\pp}_{p,\kappa}\mcal{V}_{\alpha;\mu}(x_{p})\,)\approx0\)
\beq \lb{sd_73}
\lefteqn{\Big\langle\mscr{A}_{J_{\psi}}[\hat{T},\feynV;\hat{\mscr{J}}\equiv0]\Big\rangle_{\feynBV} = \frac{1}{2}
\sum_{n=1}^{\infty}(-1)^{n}\int_{C}d^{4}\!x_{p}\;d^{4}\!y_{q}\;J_{\psi;N}^{\dag,b}(y_{q})\;\hat{I}\;
\hat{T}_{N;N\ppr}^{bb\ppr}(y_{q})\;\times }   \\ \no &\hspace*{-0.2cm}\times&\hspace*{-0.4cm}
\boldsymbol{\Bigg[}\bigg[\langle\hat{\mscr{H}}\rangle_{\feynbv}^{\boldsymbol{-1}}
\bigg(\left(\delta\hat{\mfrak{h}}-\delta\hat{\mscr{K}}^{\boldsymbol{\kappa_{n}}}\;
\langle\hat{\mscr{H}}\rangle_{\feynbv}^{\boldsymbol{-1}}
\;\big(\,\hat{\pp}_{p,\boldsymbol{\kappa_{n}}}\hat{\mfrak{V}}\,\big)\right)
\langle\hat{\mscr{H}}\rangle_{\feynbv}^{\boldsymbol{-1}} +
\delta\hat{\mscr{K}}^{\boldsymbol{\kappa_{n}}}\;\langle\hat{\mscr{H}}\rangle_{\feynbv}^{\boldsymbol{-1}}\;
\boldsymbol{\hat{\pp}_{p,\kappa_{n}}}\bigg)\bigg] \times \\ \no &\hspace*{-0.2cm}\times&\hspace*{-0.4cm}
\bigg[\langle\hat{\mscr{H}}\rangle_{\feynbv}^{\boldsymbol{-1}}
\bigg(\left(\delta\hat{\mfrak{h}}-\delta\hat{\mscr{K}}^{\boldsymbol{\kappa_{n-1}}}\;
\langle\hat{\mscr{H}}\rangle_{\feynbv}^{\boldsymbol{-1}}
\;\big(\,\hat{\pp}_{p,\boldsymbol{\kappa_{n-1}}}\hat{\mfrak{V}}\,\big)\right)
\langle\hat{\mscr{H}}\rangle_{\feynbv}^{\boldsymbol{-1}} +
\delta\hat{\mscr{K}}^{\boldsymbol{\kappa_{n-1}}}\;\langle\hat{\mscr{H}}\rangle_{\feynbv}^{\boldsymbol{-1}}\;
\boldsymbol{\hat{\pp}_{p,\kappa_{n-1}}}\bigg)\bigg] \times \\ \no &\hspace*{-0.2cm}\times&\vdots
\\ \no &\hspace*{-0.2cm}\times&\hspace*{-0.4cm}
\bigg[\langle\hat{\mscr{H}}\rangle_{\feynbv}^{\boldsymbol{-1}}
\bigg(\left(\delta\hat{\mfrak{h}}-\delta\hat{\mscr{K}}^{\boldsymbol{\kappa_{2}}}\;
\langle\hat{\mscr{H}}\rangle_{\feynbv}^{\boldsymbol{-1}}
\;\big(\,\hat{\pp}_{p,\boldsymbol{\kappa_{2}}}\hat{\mfrak{V}}\,\big)\right)
\langle\hat{\mscr{H}}\rangle_{\feynbv}^{\boldsymbol{-1}} +
\delta\hat{\mscr{K}}^{\boldsymbol{\kappa_{2}}}\;\langle\hat{\mscr{H}}\rangle_{\feynbv}^{\boldsymbol{-1}}\;
\boldsymbol{\hat{\pp}_{p,\kappa_{2}}}\bigg)\bigg] \times \\ \no &\hspace*{-0.2cm}\times&\hspace*{-0.4cm}
\bigg[\langle\hat{\mscr{H}}\rangle_{\feynbv}^{\boldsymbol{-1}}
\bigg(\left(\delta\hat{\mfrak{h}}-\delta\hat{\mscr{K}}^{\boldsymbol{\kappa_{1}}}\;
\langle\hat{\mscr{H}}\rangle_{\feynbv}^{\boldsymbol{-1}}
\;\big(\,\hat{\pp}_{p,\boldsymbol{\kappa_{1}}}\hat{\mfrak{V}}\,\big)\right)
\langle\hat{\mscr{H}}\rangle_{\feynbv}^{\boldsymbol{-1}} +
\delta\hat{\mscr{K}}^{\boldsymbol{\kappa_{1}}}\;\langle\hat{\mscr{H}}\rangle_{\feynbv}^{\boldsymbol{-1}}\;
\boldsymbol{\hat{\pp}_{p,\kappa_{1}}}\bigg)\bigg]\boldsymbol{\Bigg]_{N\ppr;M\ppr}^{b\ppr a\ppr}
\hspace*{-1.0cm}(y_{q},x_{p})}\;\hat{T}_{M\ppr M}^{-1;a\ppr a}(x_{p})\;\hat{I}\;J_{\psi;M}^{a}(x_{p}) \,.
\eeq

\end{appendix}

\end{document}